\documentclass[12pt,fleqn,oneside]{book}

\usepackage{amsmath} %utk masuk begin{align}
\usepackage[scr=rsfs]{mathalpha}
\usepackage{amsxtra}%supaya \eqref
\usepackage{amsopn}%supaya dapat declare benda baru \citep
\usepackage{amsbsy}
\usepackage{epigraph}
\usepackage{amstext}
\usepackage{amsfonts}
\usepackage{amssymb}
\usepackage{amsthm}
\usepackage{float}
\usepackage{shadow}% kotak shadow
\usepackage{curves}
\usepackage{graphicx}
\usepackage{color}%masukan color
\usepackage{lipsum}
\usepackage{imakeidx}
\usepackage{hyperref}
\usepackage{enumitem}
\usepackage{thmtools}
\hypersetup{
    colorlinks=true,
    linkcolor=blue,
    filecolor=magenta,      
    urlcolor=cyan,
    citecolor=red,
    pdftitle={Overleaf Example},
    pdfpagemode=FullScreen,
    }
\makeindex
\usepackage{titletoc} 
\usepackage{framed}
\usepackage{titlesec}
\usepackage{subfigure}
\definecolor{shadecolor}{rgb}{0.95,0.95,0.95}
\renewenvironment{leftbar}[1][\hsize]
{%
    \MakeFramed{\hsize#1\advance\hsize-\width\FrameRestore}%
}
{\endMakeFramed}

\usepackage{fancyhdr}

\titleformat{\chapter}[display]
  {\normalfont\huge\bfseries\centering}{\chaptertitlename\ \thechapter}{20pt}{\Huge}
\titlespacing*{\chapter}{0pt}{-50pt}{40pt}

\makeatletter
\renewcommand*{\@dotsep}{999}
\makeatother

\newtheorem{theorem}{Theorem}

\usepackage[sort, numbers]{natbib}
\def\be{\begin{equation}}
\def\ee{\end{equation}}
\def\bea{\begin{eqnarray}}
\def\eea{\end{eqnarray}}

\def\rateWt{\omega_{t}}

\def\mP{\mathcal{P} }
\def\mQ{\mathcal{Q} }
\def\Stot{S^{\rm tot}_t }
\def\emStot{\exp(-S^{\rm tot}_t) }

\def\Xon{X_{[0,n]}}
\def\Xom{X_{[0,m]}}
\def\Xot{X_{[0,t]}}
\def\xot{x_{[0,t]}}
\def\Xos{X_{[0,s]}}
\def\Mom{M_{[0,m]}}
\def\X{\mathcal{X}}
\def\D{\mathcal{D}}
\def\la{\langle}
\def\ra{\rangle}

\def\T{\mathcal{T}}

\newcommand{\bmu     }{\mbox{\boldmath$\mu$}}

\newcommand{\bD     }{\mbox{\boldmath$D$}}

\def\beq{\begin{equation}}
\def\eeq{\end{equation}}
\def\bea{\begin{eqnarray}}
\def\eea{\end{eqnarray}}

\def\D{\mathrm{d}}

\theoremstyle{plain}

\theoremstyle{plain}
\theoremstyle{Example}
\theoremstyle{Question}

\theoremstyle{Task}
\theoremstyle{Solution Task}

\theoremstyle{Exercise}
\theoremstyle{definition}

\theoremstyle{solution}

\theoremstyle{remark}

\begin{document}
\setcounter{secnumdepth}{5}  
\setcounter{tocdepth}{4}    %How many levels to include in TOC
\addtocontents{toc}{\protect\sloppy}    %Keeps long chapters within page # margins
\pagenumbering{roman}

\thispagestyle{plain}
\begin{center}
    \Large
    %\textbf{\Huge Martingales for Physicists}%:\\ a Tutorial Review}
    
    \textbf{\Huge Martingales for physicists:  A treatise on stochastic thermodynamics and beyond}%:\\ a Tutorial Review}
        
    \vspace{1cm}
    \textbf{\'{E}dgar Rold\'{a}n$^{1,\star}$, Izaak Neri$^{2,\star}$, Raphael Chetrite$^{3,\star}$, Shamik Gupta$^4$, Simone Pigolotti$^5$, Frank J\"ulicher$^6$, Ken Sekimoto$^7$}
     \vspace{0.5cm}

    \large
$^1$ICTP - The Abdus Salam International Centre for Theoretical Physics, Strada Costiera 11, 34151 Trieste, Italy \\
$^2$Department of Mathematics, King's College London, Strand, London WC2R 2LS, United Kingdom\\
$^3$Universit\'e C\^ote d'Azur, CNRS, LJAD, Parc Valrose, 06108 NICE Cedex 02, France\\ $^4$Department of Theoretical Physics,Tata Institute of Fundamental Research, Homi Bhabha Road, Mumbai 400005, India\\
$^5$Biological Complexity Unit, Okinawa Institute for Science and Technology and Graduate University, Onna, Okinawa 904-0495, Japan\\
$^6$Max-Planck Institute for the Physics of Complex Systems, N\"{o}thnitzer Stra\ss e 38, 01187 Dresden, Germany\\
$^7$Gulliver Laboratoire, CNRS-UMR7083, ESPCI, Paris, France \newline
 \\
$^{\star}$Equal contribution and corresponding authors\\ (\href{mailto:edgar@ictp.it}{edgar@ictp.it}, \href{mailto:izaak.neri@kcl.ac.uk}{izaak.neri@kcl.ac.uk}, \href{mailto:raphael.chetrite@unice.fr}{raphael.chetrite@unice.fr})

          %\Large
    \vspace{0.9cm}
    \textbf{We review  the theory of martingales as applied to stochastic thermodynamics  and  stochastic processes in physics more generally.  }
\end{center}
%%%%%%%

\tableofcontents

\listoftheorems

\chapter*{Notation and definitions}\label{appendix:-1} 

We  introduce the main notation used in this work (see also the List of Symbols below).   

We denote stochastic, physical processes by  $X_t$, where  $t\geq0$ is a discrete ($t=0,1,2,\dots$) or continuous  ($t\geq 0$) time index, and where  $X_t$ takes values in the set  $\mathcal{X}$, which we call the state space.   Depending on the definition of $\mathcal{X}$, the random variable $X_t$ can be scalar or vectorial, and discrete or continuous. For example, if $\mathcal{X}=\mathbb{R}$, then $X_t$ is a one-dimensional process on the real line.  We denote the path (trajectory) of $X$ in the time interval $[s,t]$ by $X_{[s,t]} = \left\{X_{u}\right\}_{u\in[s,t]}$.   

Elements of the set $\mathcal{X}$ are denoted by $x\in \mathcal{X}$.  We use small letters to distinguish them from the stochastic process $X$.   Also, we use $x_{[s,t]} = \left\{x_{u}\right\}_{u\in[s,t]}$ for a deterministic trajectory, in contrast with the stochastic trajectory $X_{[s,t]}$.

Random variables are associated with their probability $\mP$. For example, $\mP(X_t>0)$ is the probability that $X_t$ is positive.  We also use $\mP$ for a probability density of  random variable. In particular, when $\mathcal{X}$ is discrete, then the probability density of the trajectory $X_{[0,t]}$ reads    
\begin{equation}
    \mP(x_{[0,t]}) \equiv\mP(X_0=x_0,\cdots , X_t=x_t),
\end{equation} 
for all $x_{[0,t]}\in \mathcal{X}^{t+1}$, and when $\mathcal{X}$ is continuous, then the probability density is defined by  
\begin{equation}
    \mP(x_{[0,t]}) \equiv\mathcal{P}\left(X_{0}\in\left[x_{0},x_{0}+dx_{0}\right],,....,X_{t}\in\left[x_{t},x_{t}+dx_{t}\right]\right),
\end{equation}
for all $x_0,x_1,\ldots,x_t\in\mathcal{X}$.
Probability densities are normalized, i.e.,
\begin{equation}
    \sum_{x_0\in \mathcal{X}}\cdots \sum_{x_t\in \mathcal{X}} \mP(x_{[0,t]}) =1 %=\sum_{x_0\in \mathcal{X}}\cdots \sum_{x_t\in \mathcal{X}} \mP(X_0=x_0,\cdots , X_t=x_t) =1,
\end{equation}
 for discrete-time processes with discrete state space $\mathcal{X}$,  and 
\begin{equation}
    \int_{x_0\in \mathcal{X}}\cdots \int_{x_t\in \mathcal{X}} \mP(x_{[0,t]})\mathcal{D}x_{[0,t]} =1, 
    %= \int_{x_0\in \mathcal{X}}\cdots  \int_{x_t\in \mathcal{X}} \mathcal{P}\left(X_{0}\in\left[x_{0},x_{0}+dx_{0}\right],,....,X_{t}\in\left[x_{t},x_{t}+dx_{t}\right]\right) =1,
\end{equation}
for discrete-time processes with continuous state space $\mathcal{X}$, 
where we have introduced the notation $\mathcal{D}x_{[0,t]} = dx_0\cdots dx_t$. 

We write expected values (averages) with respect to the path probability $\mathcal{P}$  as~$\langle \cdot \rangle$. For example, for $\mathcal{X}$ discrete, the expectation (also called "average") value of $X_t$ is given by
\begin{equation}
    \langle X_t \rangle  =   \sum_{x_t\in \mathcal{X}} x_t\,  \mP (x_{[0,t]}) = \sum_{x\in \mathcal{X}} x\,  \rho_t (x).
    \label{eq:3}
\end{equation}
If $X_t$ is continuous, we have
\begin{equation}
    \langle X_t \rangle  =   \int_{x_t\in \mathcal{X}} \mathcal{D}x_{[0,t]} x_t\,  \mP (x_{[0,t]})= \int_{x\in \mathcal{X}}dx\, x\,  \rho_t (x).
    \label{eq:4}
\end{equation}
We use $\rho_t(x)$ to denote the instantaneous probability density for both  continuous and discrete random variables, see Eqs.~\eqref{eq:3} and~\eqref{eq:4}. For discrete $X_t$, we formally define the instantaneous probability density by
\begin{equation}
    \rho_t(x) = \langle \delta_{X_t,x}\rangle  
    \equiv \sum_{x_t\in \mathcal{X}}  \mP (x_{[0,t]}) \delta_{x_t,x} = \sum_{x_t\in \mathcal{X}}  \mP (x_{t}) \delta_{x_t,x}\;\;,
\end{equation}
where $\delta_{i,j}$ is Kronecker's delta. For $X_t$ continuous, we have
\begin{equation}
    \rho_t(x) = \langle \delta (X_t-x)\rangle  
    \equiv \int_{x_t\in \mathcal{X}} \mathcal{D}x_{[0,t]}\mP (x_{[0,t]}) \delta(x_t-x)  = \int_{x_t\in \mathcal{X}}  dx_t  \mP (x_{t}) \delta(x_t-x) \;\;, %
\end{equation}
where $\delta(x)$ is the Dirac delta function.  The instantaneous density is  normalized as $\sum_{x\in\mathcal{X}} \rho_t(x) = 1$ for discrete $X_t$ and as $\int_{x\in\mathcal{X}} dx \rho_t(x) = 1$ for continuous $X_t$, for all $t\geq 0$.   

A key concept in martingale theory is the expectation of an observable at a time~$t$ conditioned on its history up to a previous time $s\leq t$. A simple example of  conditional expectation is that of the physical process $X_t$ itself. If $X_t$ is discrete, such conditional expectation is given by
\begin{equation}
    \langle X_t | X_{[0,s]}\rangle = \sum_{x\in \mathcal{X}} x\,  \mP(X_t=x|X_0, X_1,\dots,X_s) ,\label{eq:7}
\end{equation}
whereas if $X_t$ is a continuous random variable, 
\begin{equation}
    \langle X_t | X_{[0,s]}\rangle = \int_{x\in \mathcal{X}} dx\, x\,  \mP(X_{t}\in\left[x,x+dx\right]|X_0, X_1,\dots,X_s) .  \label{eq:8}
\end{equation}
For a discrete  random variable $Y$, we use $\rho_{Y}(y)= \mathcal{P}(Y=y)$ to denote the probability.  
For a continuous random variable $Y$, we denote the probability density by \begin{equation}
\rho_{Y}(y) \equiv \frac{\mathcal{P}(Y\in\left[y,y+dy\right])}{dy}.
\end{equation}   
Analogously, we use the notation $\rho_{Y}(y|X_0=x)$ for a conditional probability density, in this case conditioned on $X_0=x$.

%\blue{ Autonomous notation for $X$ one time density and path proability. Different notation with $\rho$ for the rest.}
%{\bf [IN: I do not understand the last sentence]}

%\begin{equation}
%    \mP(X_0)
%\end{equation}
 %Often  we  consider averages of physical observables $Z_t$, which are  stochastic processes given by functionals of the physical process $X_t$, i.e. $Z_t = Z[X_{[0,t]}]$. 

%In general, we will use $\mathsf{Pr}$ to denote the probability of a given event, e.g. $\mathsf{Pr}(X_t\leq x)$ for the cumulative distribution of $X_t$.\\

%We denote by $\mathcal{T}$ any stopping time, defined as the time at which the process $X_t$ first satisfies a certain criterion. 

%For simplicity, we will set the Boltzmann constant $k_{\rm B}=1$ in this Review.
\vspace{-1cm}
\chapter*{List of Symbols (Part I)}
\begin{tabular}{ll}
    $\mathcal{X}$ & State space, continuous or discrete\\
   $\delta(x)$ & Dirac's delta function  for $\mathcal{X}$ continuous\\
    $\delta_{x,y}$ &Kronecker's delta function for $\mathcal{X}$  discrete: $\delta_{x,x}=1$ and $\delta_{x,y}=0$ for $y\neq x$\\
    $dx$ &Lebesgue measure (counting measure) for $\mathcal{X}$ continuous (discrete) \\
   $t$ & Time (continuous or discrete)\\
    $X_{t}$ & Value of the physical process at time $t$, $X_{t}\in \mathcal{X}$ \\
    $X_{\left[s,t\right]}$ & Stochastic trajectory $X_{[s,t]}$  in  $[s,t]$, with $s\leq t$ \\
     $\mathcal{P}(x_{[0,t]})=\mathcal{P}_{[0,t]}\left(x_{\left[0,t\right]}\right)$ & Path  probability for the stochastic trajectory $\Xot$ to be equal to $x_{[0,t]}$, \\ & i.e. path probability for $x_{[0,t]}$ to occur in the interval~$[0,t]$ \\
 $\mathcal{P}_{[r,s]}\left(x_{\left[0,t\right]}\right)$ & Path  probability marginal of $\mathcal{P}(x_{[0,t]} )$ on the time interval~$[r,s]$, \\
 & with $0\leq r \leq s \leq t$ \\
 $\T$ & Stopping time\\
 $\left\langle   \;\cdot\;\right\rangle$ or $\left\langle
      \;\cdot\;\right\rangle_\mP $ &  Expectation (average) with respect to the path probability~$\mathcal{P}$, \\ 
      & see e.g. Eqs.~(\ref{eq:3}-\ref{eq:4})        for explicit expressions of the average~$\langle X_t\rangle $\\ 
 $Z_{t}\equiv Z\left[X_{\left[0,t\right]}\right]$ & Functional  of $X_{\left[0,t\right]}$ ($X-$adapted  observable) \\
    $\langle Z_u \vert   X_{[s,t]}\rangle$ &Conditional expectation of $Z_u$ with respect to  the filtration\\
    &  generated by $X_{[s,t]}$. For example, $\langle X_t \vert   X_{[0,s]}\rangle$,   is the  average of $X_t$ \\
    & conditioned on the process tracing a specific trajectory\\
    &  $X_{[0,s]}$ in the interval~$[0,s]$, with $0\leq s\leq t$ \\
  $\rho_Y(y)$ & Probability density  $\rho_{Y}(y) \equiv \mathcal{P}(Y\in\left[y,y+dy\right])/dy$  for a continuous\\
  &  random variable $Y$.\\
  &  Probability $\rho_{Y}(y)\equiv \mathcal{P}(Y=y)$ for a discrete  random variable $Y$ \\
 $\rho_t (x)$ or $\rho^{\mathcal{P}}_t (x)$ & Instantaneous       density (or probability) of the process,\\
 & given by $\rho_t (x)= \langle\delta(X_t -x)\rangle$. \\
 & One-point marginal of the path-probability $\mathcal{P}\left(x_{\left[0,t\right]}\right)$\\ 
$\rho_{\rm st}(x)$ & Stationary probability density (or probability) of the process\\
$\mP(x_t\vert x_s)$ & Conditional probability density for the process to be at $X_t=x_t$ \\
& at  time $t$ given that at time $s$ the value of the process\\
&  $X_s=x_s$, with $x_s,x_t\in \mathcal{X}$ \\
$\mathcal{L}_{t}$ & Markovian generator of a generic Markov   process\\ 
$\mathcal{L}_{t}^{\dagger}$ & Adjoint of Markovian generator w.r.t the canonical scalar  product\\
$\pi_{t}$ & Accompanying density,  solution of $\mathcal{L}_{t}^{\dagger}\pi_{t}=0$\\
\end{tabular}

\chapter*{List of Symbols (Part II)}
\begin{tabular}{ll}
$T$ & Temperature of the thermal bath\\
$k_{\rm B}=1$ & Boltzmann's constant, set equal to one in this  Review\\ 
$\bmu_{t}(x)$ & Mobility matrix\\
$F_{t}(x)=-\left(\nabla V_{t}\right)(x)+f_{t}(x)$ & Force vector, with  $V_{t}(x)$ potential and\\
&  $f_{t}(x)$ a non-conservative force\\
$\mathbf{D}_{t}(x)$ & Diffusion matrix\\
$B_t$ & Wiener process \\ 
$\dot{B}_t$ & Gaussian white noise\\
$\displaystyle\dot{X}_{t}=\nu_{t}(X_{t})+\sqrt{2\mathbf{D}_{t}(X_{t})} \dot{B}_{t}$ & Ito-Langevin equation (overdamped dynamics),\\
& with~$\nu_{t}(x)= (\bmu_{t}F_{t}+\nabla\, \mathbf{D}_{t})(x)$ \\
$\mathbf{D}_{t}(x)=\displaystyle\frac{T}{2}(\bmu_{t}(x)+\left[\bmu_{t}(x)\right]^{\dagger})$ & Einstein's relation for isothermal processes,\\
&  with $^\dagger$ denoting matrix transposition. \\
& For symmetric mobility matrix, it reads $\mathbf{D}_{t}(x)=T\bmu_{t}(x)$ \\
$\omega_{t}(x,y)$ & Transition rate  at time $t$   from state $x$ to state $y$ \\
& for a  Markov-jump process in continuous time\\
$w_{t}(x,y)$ & Transition probability at time $t$   from state $x$ to state $y$\\
&  for a  Markov-jump process  in discrete time \\
$J_{t,\rho}(x)$ & Instantaneous probability  current associated with the
      density $\rho_t$. \\
      & For a diffusion process, $J_{t,\rho}(x)=(\bmu_{t}F_{t}\rho_t)(x)-(\mathbf{D}_{t}\nabla\rho_t)(x)$. \\
      & For a jump process,  $J_{t,\rho}(x,y)=\rho_t(x)\omega_{t}(x, y)-\rho_t(y)\omega_{t}(y, x)$ \\
$W_{t}$ & Stochastic work done on the system in the time interval~$\left[0,t\right]$\\
&  along  a stochastic trajectory $\Xot$ \\
$Q_{t}$ & Stochastic heat absorbed by the system in the time interval~$\left[0,t\right]$\\
&  along  a stochastic trajectory $\Xot$\\
$Q_{t}+W_t = V_t(X_t)-V_0(X_0)$ & First law of stochastic thermodynamics\\ & along a stochastic trajectory $\Xot$\\
$S^{\rm sys}_{t}=-\ln\left(\rho_{t}\left(X_{t}\right)\right)$ & Stochastic system entropy at time $t$. \\
& The system entropy change along a stochastic trajectory $\Xot$\\ 
& in   $[0,t]$ reads $\Delta S^{\rm sys}_t = S^{\rm sys}_t - S^{\rm sys}_0 =  \ln (\rho_0(X_0)/\rho_t(X_t)) $\\
$S^{\rm env}_{t}$ & Stochastic environmental entropy change\\ &  along a stochastic trajectory $\Xot$  in   $[0,t]$\\
$S^{\rm tot}_{t}$ & Stochastic total entropy production\\ &  along a stochastic trajectory $\Xot$  in   $[0,t]$\\
\end{tabular}

\chapter*{List of Symbols (Part III)}
\begin{tabular}{ll}
$\Theta_{t}$ & Time reversal operator. \\ & In this Review, it is applied to a trajectory $x_{[0,t]}$ \\
& as follows        $\left[\Theta_{t}\left(x_{\left[0,t\right]}\right)\right]_{s}\equiv       x_{t-s}$\\
$\Lambda_t^{\mathcal{P},\mathcal{Q}}=\ln\displaystyle\left[\frac{\mathcal{P}\left(X_{\left[0,t\right]}\right)}{\mathcal{Q}\left(X_{\left[0,t\right]}\right)}\right]$ & $\Lambda-$entropic functional (associated   with a pair\\
&  of path probabilities  $\mathcal{P}$ and $\mathcal{Q}$) \\ 
& evaluated over the stochastic trajectory $\Xot$ \\
$\Sigma_t^{\mathcal{P},\mathcal{Q}}=\ln\displaystyle\left[\frac{\mathcal{P}\left(X_{\left[0,t\right]}\right)}{\mathcal{Q}^{(t)}\left(\Theta_t (X_{\left[0,t\right]})\right)}\right]$ & $\Sigma-$entropic functional (associated   with a pair\\
&  of path probabilities  $\mathcal{P}$ and $\mathcal{Q}$) \\
&  evaluated over the stochastic trajectory $\Xot$\\ 
$\Sigma_{[r,s];t}^{\mP,\mQ}=\ln\displaystyle\left[\frac{\mP_{[r,s]}(X_{[0,t]})}{\mQ^{(t)}_{[t-s,t-r]}\left(\Theta_{t}(\Xot)\right)}\right]$ & Generalized  $\Sigma$-entropic functional\\
&  over the subset time interval $[r,s]\subseteq[0,t]$\\
$D_{\rm KL}[\rho_X(x)\vert\vert \sigma_X(x)]$ &  Kullback-Leibler divergence between the normalized\\
& distributions $\rho_X(x)$ and $\sigma_X(x)$ of the random variable $X\in\mathcal{X}$. \\
& For the distributions of a random variable  with support $\mathcal{X}$ it \\ & is given by
 $D_{\rm KL}[\rho_X(x)\vert\vert \sigma_X(x)]= \int_{\mathcal{X}} dx\,\rho_X(x)\ln\left[\displaystyle\frac{\rho_X(x)}{\sigma_X(x)}\right]$
\end{tabular}

\pagenumbering{arabic}
\setcounter{chapter}{0}

\chapter{Introduction}
\label{ch:1}

\epigraph{ \textit{Before leaving, M.~M. asked me to go to her casino, to take some
money and to play, taking her as my partner. I did so. I took all the
gold I found, and playing the martingale, doubling my stakes continuously,  I won every day during the rest of the carnival.}} {Giacomo Casanova, {\it History of My Life} (1789).}

\section{Why this {Treatise}?}

Models based on stochastic processes have proven to be  useful in non-equilibrium statistical physics. As a consequence, an extensive set of techniques from stochastic processes have become mainstream in non-equilibrium statistical physics, one notable example being large-deviation theory  \cite{touchette2009large}. Nevertheless, few works in statistical physicists use martingales.  

Martingales play a central role in the theory of stochastic processes and find important applications in statistics and mathematical finance. In contrast, applications of martingale theory in physics are limited.   This is somewhat surprising, given that unbiased random walks and Brownian motion are martingales. These processes are of paramount importance in physics and many of their important properties can be easily derived using that they are martingales.

An explanation for the absence of martingales in contemporary statistical physics is that  martingales are not presented in textbooks and classic references used by physicists to study stochastic processes~\cite{van1992stochastic, risken1989fokker, gardiner1985handbook,haken1983springer,hanggi1982stochastic,zwanzig2001nonequilibrium,redner2001guide}.    For physicists, learning martingale theory is a quest, which can be achieved through an exhaustive reading of mathematical textbooks, just like Don Quixote reading cavalric romances, until losing their mind to become a knight errant~\cite{cervantes1605ingenioso}. 

This {Treatise} gives an overview of the aspects of martingale theory that we think are important for physics. In particular, we build on recent works that develop martingales in statistical physics, see e.g. Refs.~\cite{Chetrite:2011, neri2017statistics,Pigolotti:2017,neri2019integral, neri2020second,manzano2021thermodynamics, ge2021martingale}.
{We emphasize this work is a treatise rather than a review, inasmuch  we discuss a topic in depth by providing a thorough overview of   published results  but also include extensive novel material.} 
We shall show that  martingales are ubiquitous in  nonequilibrium  physics (e.g., in stochastic thermodynamics),  that martingales provide  fundamental insights into central concepts in nonequilibrium physics (e.g.,  on the second law of thermodynamics), and that martingales constitute a powerful tool for mathematical derivations (e.g., for splitting probabilities and extreme value statistics).   The Review is aimed at readers with a basic knowledge on nonequilibrium statistical mechanics and  stochastic processes. It covers mathematical definitions and properties in a comprehensive way, explains how to apply such results to nonequilibrium physics, and  discusses applications of martingales in interdisciplinary fields.

 \section{How to read this {Treatise}}
 
This {Treatise} is organized as follows.   Chapter~\ref{ch:1} presents historical remarks on the origin of martingales, and provides a few illustrative examples of martingales in physics. Chapter~\ref{ch:2a} introduces mathematical definitions and key examples of martingales. Chapter~\ref{ch:2b} revisits the  concept of Markov processes and its importance in statistical physics, and discusses its relation with martingales.  
Chapter~\ref{ch:3} presents martingale properties and theorems. 
Chapter~\ref{sec:martST1}  introduces martingale theory in stochastic thermodynamics through paradigmatic examples of stochastic processes. Chapter~\ref{ch:thermo2}  elaborates advanced knowledge in stochastic thermodynamics; it provides mathematical rigor on how martingales can be identified and applied in  the study of a broad class of nonequilibrium processes (stationary and non-stationary), in particular martingales related to path probability ratios. Chapter~\ref{sec:martST2} further elaborates  the connection between thermodynamics and martingales by presenting universal properties of entropy production in nonequilibrium stationary states. Chapter~\ref{sec:martST3} reviews recent work that applied martingale theory to non-stationary isothermal processes, revealing fluctuation theorems at stopping times. Chapter~\ref{TreeSL} presents a tree-like hierarchy of second law that descend from martingale properties of probability ratios. %\textcolor{red}{Can we add one sentence on 7.3 : the tree of second law ? Moreover, 7.1 is more on advanced thermo sto that on Martingale. [DONE] }
 Chapter~~\ref{sec:PQ} discusses  martingales in the context of progressive quenching in physics. Finally, Chapter~\ref{ch:popdyn} and Chapter~\ref{chapter:finance} review, respectively, applications of martingales in  population dynamics and quantitative finance. Chapter~\ref{ch:conclusion} briefly reviews applications of martingales in quantum collapse, and presents the conclusion of this review.

Key concepts, results, and theorems that we think are essential in this {Treatise} are highlighted in gray boxes. Sections with advanced content, {most of which novel material, and often} not recommended for a first read unless for  intrepid readers, are highlighted with a superscript$^{\spadesuit}$ at the beginning of their title. We recommend to consult the List of Symbols placed after the Table of Contents. Lengthy mathematical proofs and supplemental material are relegated to the Appendices. As martingales are ``fair" games, we do not guarantee potential readers will become wealthy after reading this {Treatise}, but  to acquire rich knowledge after a patient and dedicated read.

\begin{figure}[h!]
\centering
\includegraphics[width=\textwidth]{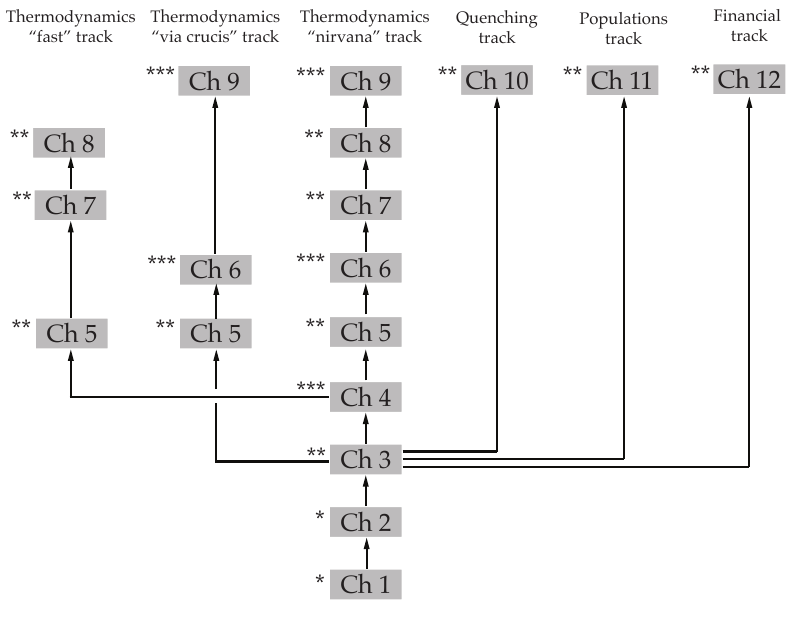}
\caption{Roadmap to this {Treatise}. The branches illustrate the different tracks that readers may take when reading this {Treatise}. We classify the each chapter's level of difficulty as: *---easy, **---normal, and ***---advanced.   }\label{fig:2}
\end{figure}

Depending on the reader's interests and background, it may be preferable to focus on selected chapters of this {Treatise}. Below and in Fig.~\ref{fig:2} we provide possible roadmaps: 
    \begin{itemize}
        \item {\em To know what are martingales and their properties}. We recommend to read  Ch.~\ref{ch:2a} to get a primer on martingales,  Ch.~\ref{ch:2b} to establish connections with Markov processes and  Ch.~\ref{ch:3} to learn about the key theorems and mathematical relations in martingale theory.
        \item {\em To learn foundations of stochastic thermodynamics}.  We recommend  to first read Ch.~\ref{sec:martST1} and then if sufficiently audatious  Ch.~\ref{ch:thermo2}  (at valiant heart nothing is impossible). 
        \item {\em For readers with basic notions on  stochastic thermodynamics wanting to learn its connection to martingales}. We recommend  to first read Ch.~\ref{sec:martST1} to refresh key concepts and learn the martingale structure of the second law in Langevin stationary processes. Further, we recommend  to  read Ch.~\ref{sec:martST2} and Ch.~\ref{sec:martST3} (together with  Ch.~\ref{ch:3} as a mathematical background) to learn how martingale theory can unveil new universal properties in stochastic thermodynamics.
         \item {\em For readers with advanced notions on  stochastic thermodynamics wanting to reach the  ``nirvana" on martingality}. We recommend first to read  Ch.~\ref{sec:martST1} and Ch.~\ref{ch:thermo2}  to get the detailed fundamentals on the martingale structure of stochastic thermodynamics, both in stationary and non-stationary setups. Next, we suggest  to read Ch.~\ref{sec:martST2} and Ch.~\ref{sec:martST3}  (with  Ch.~\ref{ch:3} as a mathematical complement) to learn how martingale theory can unveil new universal properties in stochastic thermodynamics. After this acquired knowledge, the nirvana on martingality can be acquired through a dedicated read of Ch.~\ref{TreeSL}
          \item  {\em For fans of the second law of thermodynamics}. We recommend  to first read Ch.~\ref{sec:martST1} and then Ch.~\ref{sec:martST2}, Ch.~\ref{sec:martST3}, and Ch.~\ref{TreeSL}.
          \item {\em For biophysicists wishing to learn the basics of martingales and their applications.} We recommend to start with Ch.~\ref{ch:2a} to get an informal primer on martingales, Ch.~\ref{sec:markovCont} to learn basics of continuous-time Markov processes, then Ch.~\ref{sec:martST1} to learn foundations of stochastic thermodynamics and/or Ch.~\ref{ch:popdyn} to get familiarized with applications to population dynamics.
         
         \item {\em For experts in martingales that want to learn stochastic thermodynamics}.  We recommend  to first read Ch.~\ref{sec:martST1} and then  Ch.~\ref{ch:thermo2}.
         
         \item {\em To learn applications of martingales other than thermodynamics}. We recommend to read Ch.~\ref{ch:2a}-\ref{ch:3} to familiarized with the mathematical properties and examples of martingales, before exploring applications of martingale theory in other domains, in particular progressive quenching (Ch.~\ref{sec:PQ}) and population dynamics (Ch.~\ref{ch:popdyn}).
         \item {\em For those looking for arbitrage opportunities in the stock market}. We highly recommend to read Ch.~\ref{chapter:finance} where we revisit how martingale theory is  applied in quantitative finance.
    \end{itemize}

\section{History  of Martingales}

\subsection{Etymological origin of the word ``Martingale"}

The word ``martingale" presents numerous etymologies that spread across disciplines including: gambling, mathematics, finance, geography, technology and vernacular language~\cite{Mansuy:2009,Bienvenu:2009,Krickeberg:2009}. The origin of this word dates back to the 16th-17th Centuries at the foundations of probability theory in France. One of its first appearances in literature is  Casanova's memories from 1754. Its etymology remains obscure; it is mentioned in early French, Spanish and Catalan dictionaries, which highlight the  Mediterranean roots of martingales. Some of the  usages of the word martingale, in roughly inverse chronological order, are:

\begin{itemize}
\item The word ``martingale" has a formal meaning in probability theory. Martingales are stochastic processes without drift, i.e. their expected value in the future is given by the last value of a sequence of past observations. Research on the mathematical properties of martingales were mainly developed by Doob in the 20th Century, and applied to derive key results in the theory of stochastic processes, as we discuss in the following.
 \item In mathematical finance, martingale processes have been used for decades as paradigmatic models of fair markets in which there exists no arbitrage opportunities. Martingale theory has been notoriously boosted in financial research. Krickeberg famously stated: ``I was never tempted to get involved in the applications of martingales to the theory and, worse, the practice of financial speculations that have contributed in no small measure to the present crisis of the world's money markets and economy". 
\item In game theory and gambling, martingales represent fair games of chance in which any player may win or lose with equal probability,
%has neither the certainty of winning, nor of losing,
irrespective of the previous outcomes of the game. Such ``fair" games of chance motivated the origin of probability theory in the 17th Century. Even earlier, the book of Fra Luca Paccioli (1494) already discussed fair games in the spirit of what today are known as martingales.
\item Giacomo Casanova's memories~\cite{Casanova} provide the arguably first literary reference of the word:  ``J'y fus [au casino de Venise], j'ai pris tout l'or que j'ai trouv\'e, et portant avec la force qu'en terme de jeu on dit \`a la martingale, j'ai gagn\'e trois et quatre fois par jour pendant tout le rest de carnaval"\footnote{``I went [to Venice's casino], taking all the gold I could get, and by means of what in gambling is called the martingale I won three or four times a day during the rest of the carnival".} . 
The dictionary of the Acad\'emie Fran\c{c}aise describes Casanova's gaming strategy as ``betting all that was lost".
\item Abb\'e Prevost describes the martingale as the celebrated playing strategy where the gambler doubles his/her stake at each loss in order to quit with a sure profit, provided that he/she wins once. In  casino's roulette this is called the ``Double Up" strategy.  Alexandre Dumas describes this strategy in {\em La Femme au collier de velours} as ``introuvable comme l'\^{a}me" (unreachable like the soul) being put at work during the last days of the life of an old gambler who spent all his life looking for the martingale.
\item Martingales have also an equestrian meaning, which is in nowadays registered in e.g. Oxford's English dictionary as ``a strap or set of straps running from the noseband or reins to the girth of a horse, used to prevent the horse from raising its head too high". Similarly, the Spanish word {\em alm\'artaga}, which refers also to a horse harness is also considered among one of the possible etymological roots of the word martingales.  
\item Mistral's Proven\c{c}al dictionary cites {\em martegalo} as the demonym of the residents of Martigues, a French city located northwest of Marseille, currently nested within the Proven\c{c}e-Alpes-C\^ote d'Azur region. The isolated location of the Martigues area, at the merger of three boroughs, brought according to Mistral's dictionary a ``proverbial reputation of naivety".  In the same  dictionary, we  find the Proven\c{c}al expression {\em jouga a la martegalo}, which means to play in an absurd --and thus not necessarily fair-- way.
\item The word {\em martegalo} is used in sailing as a rope attached above the bowsprit needed to secure the flying jib, and  sailors called {\em martegaux} were famous for net fishing in the south of Italy and Andalusia.
\item Cotgrave's dictionary relates martingales to a sailor's dance consisting of a repetitive and rough stamping of the ground with the heels.  This is mentioned in Charles IX trip to Brignoles (1564) with his court where ``the citizens tried to please him through [...] the dances of the area [...] dances named volte or martingale". 
\item In Rabelais' series of novels {\em Gargantua}, the  character  Panurge wears the martingale pants, which contain an orifice at the back. In Rabelais' words ``a drawbridge [...] that makes excretion easier". 
\item Letters from the 17th Century of a  prophetess nicknamed La Martingale have been reported, containing doubtful prophecies (e.g. for the fate of Louis XIV) often accompanied by requests for donations. 
\item In vernacular language, martingale has been used to refer to prostitutes, courtesans, streetwalkers, etc. This meaning can be found in old slang dictionary and also in Scarron's {\em Virgile Travesti}. 
\item In Italian language,  "martingala" has yet another meaning: a sort of half-belt which tightens the back of a jacket or a coat.
\end{itemize}

\begin{figure}
\centering
{\includegraphics[width=0.9\textwidth]{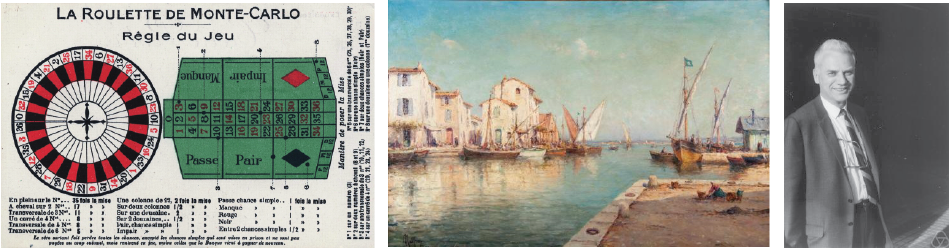}}
\caption{Some celebrated ``Martingales". Left: rules of game in the roulette of casino in Montecarlo. Middle: Painting of the village of Martigues (France). Right: Joseph L. Doob, mathematician who pioneered the development of martingale processes in probability theory.}\label{fig:1}
\end{figure}

\subsection{Martingales in probability theory}

The true explosion of the concept of martingales in mathematics dates back to the works by Joseph Leo Doob in the 1940s. Doob proved many fundamental inequalities and limit theorems associated with martingales. These results deeply changed the field of probability theory. In the following years, finding a suitable martingale became the ``skeleton key" to solve a challenging new problem in probability theory. 

Two important precursors of Doob in martingale theory are:

\hspace{0.5cm}-Jean Ville, who  introduced for the first time the concept of martingale in mathematics in his PhD Thesis ``Etude critique de la notion de collectif" (1939)~\cite{Ville}. His thesis includes the first  proofs of the so-called Doob's maximal inequality.  Doob, who took part to Ville's PhD Thesis committee, recognized that Ville's thesis  was a major inspiration for his  work.

\hspace{0.5cm}-Paul Lévy, whose work is in some way related to martingale theory. For instance, his  book ``Stochastic Processes and Brownian Motion" (1948) deeply influenced  probability theory. Levy's writing style is informal and focused on explanations rather than on mathematical proofs, in contrast with Doob's rigorous and dry mathematical style. 

In 1953, Doob published  the influential book ``Stochastic Processes"~\cite{Doob}, which contains the mathematical foundations of what today is called martingale theory. In the second half of the 20th century, martingales have provided a new perspective on a plethora of problems in probability theory, for example:

\hspace{0.5cm}-Stroock and Varadhan introduced in 1969 the ``martingale problem"~\cite{stroock2007multidimensional}, which enable to characterize the distribution of a stochastic process through a martingale condition. In particular, for Markovian processes, this martingale can be expressed in terms of the infinitesimal generator. This problem is particularly well suited to characterize the limit of a family of Markov processes. 
 
\hspace{0.5cm}-In stochastic calculus, martingales are stochastic processes that form good integrators.   Indeed,  the theory of integration with respect to a Wiener process has been  extended to integrals that use  general martingales as integrators~\cite{kunita1967square}.

\subsection{Martingales in gambling}

Martingales originated in a class of betting strategies that were popular in the 18th-century France. These strategies can be summarized by the principle: ``if you lose, double your wager size."  
Consider a betting involving two gamblers X and Y.
Suppose X starts to toss the coin, taken to be fair, with a betting amount of 50. If the outcome is a head, X retains this amount, otherwise loses it to Y. The coin is tossed, and it falls on the tail. Using the martingale strategy, X now increases the betting amount to 100. The coin is tossed, but again it falls on the tail, and so X again doubles the betting amount. So by the time X tosses the coin for the third time, the total amount that X has lost to Y is 350. The coin is tossed, and now, to X's merriment, the coin falls on the head, and so X gets from Y an amount of 400. In the process, X has retained the initial amount of 50.  The amount of the winning trade in the above martingale betting strategy exceeds the combined losses of all the previous trades, and the difference is the amount of the original trade. It is evident that the  strategy would result in a profit for a gambler,  but as we will see in this {Treatise}  this assumes that the gambler has infinite (i.e. unbounded)  wealth to keep on betting and doubling the betting amount until he wins. 
 Note also that the casino knows that
bankruptcy is a possible outcome in case of infinite wealth. To
avoid such possibility, a casino often uses table limits to control the maximum
 bets that a player can play. Most  casinos in Las Vegas Strip usually offer
 tables with a 10000\$ limit. We note however that  such limits do not exist
in financial markets, and investing in the stock market with Casanova's strategy  could imply a huge bankrupcy!
 
The strategy of doubling up on a loss is what had been the betting strategy of Casanova mentioned earlier. 
Denoting by $S_i$ the total accumulated score up to the $i$-th toss included, and given the outcomes of $i >1$ tosses, the expectation value of $S_{i+1}$ reads
\begin{align}
    \langle S_{i+1}|S_1,S_2,\ldots,S_i\rangle \ge S_i,
    \label{eq:conditional-average}
\end{align}
which makes the stochastic process $S_i$ a {\em submartingale}. 

\subsection{Martingales in finance}

Quantitative finance employs mathematical and statistical tools to anticipate the value of  financial assets as stocks and options. From early days,  physics models such as random walks have been invoked to discuss stock pricing. Jules Augustin Fr\'{e}d\'{e}ric Regnault, an assistant to a French stock broker, was one of the first to propose a modern theory of stock pricing in his 1863 treatise \textit{Calcul des Chances et Philosophie de la Bourse}, in which he writes ``\textit{l'\'{e}cart des cours est en raison directe de la racine carr\'{e}e des temps}", which translates as ``price deviation is directly proportional to the square root of time". Louis Jean-Baptiste Alphonse Bachelier, a French mathematician who lived at the turn of the 20th century, was the first to propose as part of his PhD thesis \textit{Th\'{e}orie de la sp\'{e}culation} a mathematical model for Brownian motion and how it may be used for discussing stock pricing. His contributions make him arguably the forefather of mathematical theory of finance. However, it is the American economist Eugene Francis ``Gene" Fama whom some people argue is the father of finance, owing to his ground-breaking work in the area, and in particular, for proposing the so-called efficient-market hypothesis. This hypothesis states that in an efficient market, it would not be possible to make definite predictions about future price on the basis of the information available today,  so that the best prediction that one can make for the expected future price discounted to the present time is today’s price itself. The hypothesis forms a cornerstone of modern financial theory, and in the light of the present review, an implication of the hypothesis is that asset price is a martingale. We will explore this connection in more detail in Chapter~\ref{chapter:finance}, in which, among others, we will discuss the very-influential Black-Scholes model used widely by options market participants round the world. This model, named after American economists Fischer Black and Myron Scholes, provides a theoretical estimate of the price of European-style option. The model was introduced in the 1973 paper by Black and Scholes titled ``The Pricing of Options and Corporate Liabilities," and published in the Journal of Political Economy. Robert C. Merton published his own article in this area, ``Theory of Rational Option Pricing," in The Bell Journal of Economics and Management Science, in which he coined  the term ``Black–Scholes theory of option pricing." For their work, Black and Merton were awarded the Nobel Prize in Economic Sciences for the year 1997 (Scholes because of his death in 1995 was considered ineligible for the prize).

\subsection{Martingales in stochastic thermodynamics}

Stochastic thermodynamics describes the non-equilibrium behavior of mesoscopic systems~\cite{LNP,seifert2012stochastic,peliti2021stochastic}. The application of martingale theory to stochastic thermodynamics has a short yet fruitful history, see, e.g.,~Refs.~\cite{Chetrite:2011,neri2017statistics, Pigolotti:2017, chetrite2019martingale, neri2019integral,neri2020second, cheng2020infima,yang2021nonequilibrium, ge2021martingale, manzano2021thermodynamics,faggionato2022martingale, neri2022extreme}.  Classical fluctuation relations  of stochastic thermodynamics, such as the integral fluctuation relation and Jarzynski's equality,  can be understood with martingale theory, and martingale theory generalises these  fluctuation relations, providing  a better understanding of fluctuations in mesoscopic systems.    In particular,   with martingale theory we obtain fluctuation relations at random times and for the extreme values of stochastic processes, while stochastic thermodynamics usually deals with fluctuations at fixed time.     Also, martingale theory implies   versions of  the second law of the thermodynamics for mesoscopic systems that are stronger than those obtained in "standard" stochastic thermodynamics, providing us with a better understanding of the implications of the second law at mesoscopic scales.   In particular, the martingale versions of the second law reveal how  the  observer's knowledge  about a system's history affects the second law of thermodynamics.      This body of work forms the core of this review (Chapers~\ref{sec:martST1}-\ref{TreeSL}), and now  we   provide some  ``historical" remarks.  

The link between fluctuation relations in stochastic thermodynamics and martingales was first highlighted in Ref.~\cite{Chetrite:2011}. Reference~\cite{neri2017statistics} rediscovered the link between martingales and fluctuation relations in stochastic thermodynamics within the setup of stationary processes, and moreover used the mathematical properties of martingales to derive universal relations for the statistics of  extreme-values and stopping-times  of entropy production.     The results from~Ref.~\cite{neri2017statistics} were rederived in Ref.~\cite{Pigolotti:2017} within the context of Langevin processes by using  It\^{o} calculus and  random-time transformations, and Ref.~\cite{neri2019integral} shows how most of  the results of Ref.~\cite{neri2017statistics} follow readily from one relation, namely, the  integral fluctuation relation for entropy production at stopping times.   The integral fluctuation relation for entropy production at stopping times is thus a key result of martingale theory for stochastic thermodynamics, and Ref.~\cite{neri2019integral} also introduces the ensuing second law of thermodynamics at stopping times.    This latter version of the second law of thermodynamics describes how classical limits in thermodynamics can be overcome by stopping at a cleverly chosen moment.   %Ref.~\cite{yang2021nonequilibrium} explored generalized Einstein relations in the context of fluctuation-dissipation theorems in stationary states.
%Universal expressions for various statistics of entropy production in Langevin nonequilibrium processes were found by combining Ito calculus and  martingale theory~\cite{Pigolotti:2017}. 
Some of these results have been experimentally verified in  single-electron boxes~\cite{singh2019extreme} and granular systems~\cite{cheng2020infima}. 

Martingales theory has also plays a role for trade-off inequalities between the rate of entropy production, speed, and  precision.    Reference~\cite{roldan2015decision} derives a bound relating first-passage times of current-like observables to the average rate of dissipation.     A more in-depth analysis in  Ref.~\cite{neri2022universal} shows  that this bound can be interpreted as a tradeoff between dissipation, speed, and precision within a first-passage setup, and that the bound is related to the so-called thermodynamic uncertainty relations~\cite{barato2015thermodynamic, pietzonka2016universal,gingrich2016dissipation}.   Moreover,   using martingale theory, Ref.~\cite{neri2022universal}  shows that the bound is tight for currents proportional to the entropy production, and hence is optimal in this case.   

%This work also revealed a fluctuation relation for the decision-time distributions for correct and wrong decisions on the direction of the arrow of time (forward or backwards). Such result was proved rigorously in Refs.~\cite{dorpinghaus2018testing} invoking Doob's martingale theorems~\cite{Doob}, which have been key in the development of martingale approaches in stochastic thermodynamics.

More recently,  martingales have been  employed to describe fluctuations of generic nonequilibrium Markov process driven by arbitrary external protocols~\cite{chetrite2019martingale,chun2019universal,neri2020second,manzano2021thermodynamics,faggionato2022martingale}. Reference~\cite{neri2020second} derives in this setup a  second law of thermodynamics at stopping times and a Jarzynski equality at stopping times. Reference~\cite{manzano2021thermodynamics}  also provides  Jarzynski-like relations and generalized second laws at stopping times, albeit using nonequilibrium free energies instead of equilibrium free energies, and illustrates the result in an experimentally-realized ``gambling" demon which stops the dynamics of a process following specific criteria. Further applications of martingales in stochastic thermodynamics have been reported e.g. in quantum systems~\cite{manzano2019quantum}, molecular motors~\cite{guillet2020extreme}, periodically-driven systems~\cite{faggionato2022martingale}, and  photoelectric devices~\cite{manzano2022survival}.

\section{"Warm-up" on Martingales}\label{sec:warmup}

As a first encounter with martingales we discuss simple random walks, which are possibly the simplest example of martingale processes. 
\begin{figure}[h!]
\centering
{\includegraphics[width=0.6\textwidth]{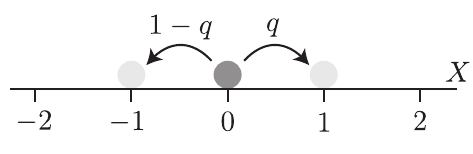}}
\caption{Illustration of a discrete-time biased random walk. A  particle (gray circle) moves in a one dimensional lattice $X$ (black line).  At every (discrete) time step, the particle jumps either forward (positive $X$ direction) with probability $q$ or backward (negative $X$ direction) with probability $1-q$.  }\label{fig:BRW}
\end{figure}

We denote by  $X_t$ the position of a one-dimensional, discrete-time, biased, random walk at  times $t= 0,1,2,\dots,$ with initial condition $X_0=0$ (see Fig.~\ref{fig:BRW} for an illustration). For $t\geq 1$, the position of the walker is given by 
\begin{equation}
    X_t \equiv  \sum_{s=1}^t \xi_s,
    %X_t = X_0 + \sum_{i=1}^t \xi_i
    \label{eq:RW1}
\end{equation}
where $\xi_s$ are independent increments which take the value $+1$ with probability $q<1$ and $-1$ with probability $1-q$. The average (expectation) value of $X$ at time $t$ reads $\langle X_t \rangle = \sum_{s=1}^t \langle\xi_s \rangle = (2q-1)t$.

What is the expected value of $X_t$ at time $t>0$ given its history $X_{[0,s]}$ up to a previous time $s<t$?  This {\em conditional expectation} is formally defined as
\begin{equation}
    \langle X_t | X_{[0,s]}\rangle = \sum_{x\in \mathcal{X}} x\,  \mathcal{P}(X_t=x|X_0, X_1,\dots,X_s) , \label{eq:condExpa}
\end{equation}
where $P(X_t|X_0, X_1,\dots,X_s)$ is the conditional probability  of $X_t$  given  $X_{[0,s]} =X_0, X_1,\dots,X_s $.  For our example, 
\begin{equation}
    \langle X_t |X_{[0,s]} \rangle =    X_s +  \sum_{r=s+1}^t \langle \xi_r\rangle  = X_s + (2q-1)(t-s).
\end{equation}
If $q=1/2$, then the   random walk is unbiased and   satisfies the {\em martingale} property expressed by
\begin{equation}
    \langle X_t| X_{[0,s]}\rangle =    X_s.
\end{equation}
If $q\geq 1/2$,  then $X_t$ is a, so-called, {\em submartingale} for which $ \langle X_t | X_{[0,s]} \rangle \geq      X_s$, whereas if $q\leq 1/2$, then $X_t$ is a {\em  supermartingale} $ \langle X_t | X_{[0,s]} \rangle \leq      X_s$. In words, a martingale  is a fair,  unbiased process whereas a submartingale (supermartingale) is a biased process with positive (negative) drift.

Interestingly, we can transform a biased random walk (sub or supermartingale) into a martingale. For example, the position of the walker in the comoving frame $Y_t=X_t - vt$, with $v=(2q-1)$ the net drift, is  a martingale.

We can construct other martingales from $X_t$, as we discuss now.  
A useful trick is to use the multiplicative structure
\begin{equation}
    M_t  \equiv  \prod_{s=1}^t \eta_s,
    \label{eq:mi}
\end{equation}
where  the $\eta_s$ are independent random variables with $\langle \eta_s\rangle=1 $, and thus $\langle \eta_s\eta_t\rangle=\langle \eta_s\rangle\langle \eta_t\rangle= 1$ for all $s\neq t$.    As one can readily verify, processes of the form~\eqref{eq:mi}  are martingales, i.e., $\langle M_t|M_{[0,s]}\rangle = M_s$, for any $t\geq s \geq 0$.
A possible choice is 
\begin{equation}
    \eta_s \equiv  \frac{\exp(y \xi_s)}{\langle \exp(y\xi_s)\rangle} =\frac{\exp(y\xi_s)}{q\exp(y)+(1-q)\exp(-y)}, 
    \label{eq:etai}
\end{equation}
where  $y\in \mathbb{R}$ is a real number. Plugging \eqref{eq:etai} into  \eqref{eq:mi},  we obtain
\begin{equation}
M_{t} = \frac{\exp(y X_t)}{[q\exp(y)+ (1-q)\exp(-y)]^t}\,,
\label{eq:1p7}
\end{equation}
which are martingales for all $y\in \mathbb{R}$.  As we motivate later in this {Treatise}, a ``popular" choice in stochastic thermodynamics is $y=\ln [(1-q)/q]$, yielding the exponential process
\begin{equation}
M_{t} = \exp\left[  -X_t\ln \left(\frac{q}{1-q}\right)\right]=\left(\frac{1-q}{q}\right)^{X_t}.  
\label{eq:1p7new}
\end{equation}

Let us investigate some consequences of the family of martingales $M_t$, given by Eq.~\eqref{eq:1p7}.
Expanding $M_{t}$ in small values of   $y$ yields 
\begin{eqnarray}
M_{t}&=& 1+ \left. \frac{\partial M_t}{\partial y}\right|_{y=0}y  + \left.\frac{\partial^2 M_t}{\partial y^2}\right|_{y=0}\frac{y^2}{2!} + O(y^3) \nonumber\\
&=& 1+ y (X_t - vt) + \frac{y^2}{2} [ (X_t -v t)^2 - \sigma^2 t] + O(y^3), \label{eq:martexp}
\end{eqnarray}
where 
\begin{equation}
    v \equiv \langle X_t \rangle/t = (2q-1),
\end{equation} and 
\begin{equation}
    \sigma^2 \equiv  4q(1-q).
\end{equation} Because $M_{t}$ is a martingale for all values of  $y\in \mathbb{R}$, also  $M_t^{(k)}=\left.\frac{\partial^k M_t}{\partial y^k}\right|_{y=0}$  are martingales, as for any integer $k\geq 1$ they can be written as the difference between two martingales.   As a result, all the coefficients in the expansion~\eqref{eq:martexp} are martingales, in particular,
\begin{eqnarray}
M_{t}^{(1)} &=& X_t - vt \label{eq:1p9}\\
M_{t}^{(2)} &=& (X_t-vt)^2 - \sigma^2 t ,\label{eq:1p99}
\label{eq:17}
\end{eqnarray}
and so forth, are martingales.    Notice that the higher-order derivatives give $M_t^{(k)}$ in terms of powers of $X_t$ up to degree $k$.  The martingale property of $M_t^{(k)}$  can be used to obtain exact expressions for  the centered moments of $X_t$, e.g., $\langle X_t\rangle =vt$ and $\langle (X_t-\langle X_t\rangle) ^2\rangle =\sigma^2 t$. %This illustrates the opportunities of martingales as moment generating functions of stochastic processes. 

Martingales  are also useful for studying stochastic processes at  {\em stopping times}.   Stopping times generalise  first-passage times~\cite{redner2001guide}. Put simply, a stopping time is the first time when   a process satisfies a certain prescribed condition, provided that the condition is fulfilled at a finite time; otherwise the stopping time is infinite.   An example of  a stopping time $\T$ is the first time when the biased random walk $X_t$, starting at $X_0=0$, reaches any of two absorbing boundaries located at $L>0$  and $-L<0$, i.e., the first exit time from the interval $(-L,L)$.  
In some cases, such as in the present example of a  biased random walk, it is possible to use martingales to determine analytically the  absorption probabilities and the mean first-passage time~\cite{feller1957introduction,redner2001guide}.   The  absoroption probabilities $P_+$ and $P_-=1-P_+$ for the walker  at the positive and negative  boundaries, respectively, are given by
\begin{equation}
    P_+ =      \frac{1-\left(\displaystyle   \frac{1-q}{q}\right)^{L}}{1-\left(\displaystyle   \frac{1-q}{q}\right)^{2L}} , \quad {\rm and} \quad  P_- =      \frac{\left(\displaystyle   \frac{1-q}{q}\right)^L-\left(\displaystyle   \frac{1-q}{q}\right)^{2L}}{1-\left(\displaystyle   \frac{1-q}{q}\right)^{2L}} ,
    \label{eq:ppluspminusintro1}
\end{equation}
for $q\neq 1/2$ (biased random walk), and
\begin{equation}
    P_+ =       P_- =    1/2 ,
    \label{eq:ppluspminusintro2}
\end{equation}
for $q= 1/2$ (unbiased random walk);
see Appendix~\ref{appendix:0} for an explicit derivation of  the Eqs.~(\ref{eq:ppluspminusintro1}-\ref{eq:ppluspminusintro2}).         Note that the average value of the ``exponential" martingale given by Eq.~\eqref{eq:1p7new} evaluated at the first exit time $\T$ out of the $(-L,L)$ reads
\begin{equation}
\langle M_\T \rangle = P_- \left(\frac{1-q}{q}\right)^{-L} + P_+ \left(\frac{1-q}{q}\right)^{L} = P_+ + P_- = 1  ,\label{eq:DoobOpt1}
\end{equation}
i.e., it is equal to the initial value of the martingale $M_0=((1-q)/q)^0=1$. In other words,  using exit times out of a  symmetric interval, the process $M_\T$   can on average neither win nor lose~\eqref{eq:1p7new} .

The mean first-passage time   for $q\neq 1/2$  is given by (see  Appendix~\ref{appendix:0})
\begin{equation}
\langle \T \rangle =  \frac{L}{1-2q}-\frac{2L}{1-2q}\frac{1-\left(\frac{1-q}{q}\right)^{L}}{1-\left(\frac{1-q}{q}\right)^{2L}} = \frac{L}{1-2q} - \frac{2L}{1-2q}P_+ =\frac{ L}{v} (P_+-P_-)  ,\label{eq:TAvDoobIllu1}
\end{equation}
where we have used $v=2q-1$, and  for $q=1/2$  the mean first-passage time reads
\begin{equation}
\langle \T \rangle =  L^2 .\label{eq:TAvDoobIllu2}
\end{equation}
   Combining the Eqs.~(\ref{eq:ppluspminusintro1}) and (\ref{eq:TAvDoobIllu1}), and using $v=2q-1$ we obtain that 
\begin{equation}
\langle M^{(1)}_{\T}\rangle = L(P_+-P_-) - v \langle \T \rangle = 0,  \label{eq:DoobOpt2}
\end{equation} 
which holds for all values of $q\in[0,1]$. This further illustrates  the {\em fairness} of martingales, as the average value of $M^{(1)}$ at the first exit time equals to its initial value $M^{(1)}_0=0$.
 
 Perhaps more striking (and less intuitive) is the fact that Eq.~\eqref{eq:DoobOpt2} also holds for the mean escape time of the unbiased random walk $M_t^{(1)}$ from {\em asymmetric} intervals $(-L_-,L_+)$ with $L_+,L_->0$ any two integer threshold values, with $L_+\neq L_-$. In other words, one cannot ``win" neither ``lose" with the martingale $M_t^{(1)}$ irrespective of the chosen stopping strategy.  For example, for $L_-\gg L_+$, the many trajectories that escape the interval through the positive boundary,  $L_+$, are balanced by the few trajectories that escape the interval through the negative boundary, $L_-$.   This points out to the flaw in Casanova's gambling strategy as the wins on most days are balanced by a few big losses.
 \begin{figure}[h]
\centering
\includegraphics[width=\textwidth]{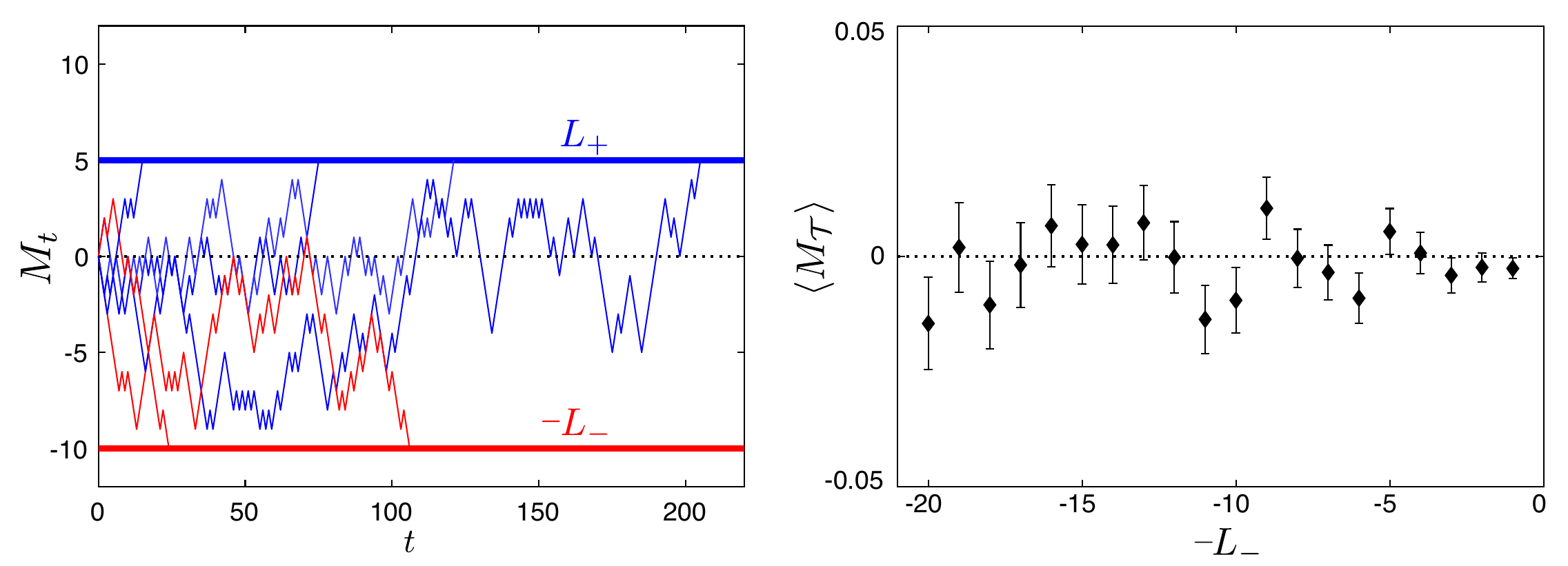}
\caption{Gambling with martingales and stopping conditions. Left: sample trajectories~(lines) of the martingale %$M_n$
given by the one-dimensional, discrete-time,  random walk described by Eq.~\eqref{eq:RW1}, and sketched in Fig.~\ref{fig:BRW}, with equal forward and backward jump probabilities $p = q = 1/2$. %The circles illustrate the instances at which the random walk exits the interval $(-L_-,L_+)$ with $L_-=3$ and $L_+=5$ in this case.   
Right: average value  $\langle M_\T\rangle$ of the random walk evaluated at the stopping time $\T$ given by the first time $M_t$ reaches either $-L_-$ or $L_+$, for fixed $L_+=5$ and different values of $-L_-$. The symbols are obtained from $N=10^6$ simulations and the errorbars from the standard error of the mean. The horizontal dotted lines set to $M_0=0$ illustrate  Doob's optional stopping theorem $\langle M_\T \rangle = M_0=0$. }\label{fig:1}
\end{figure}
This result is illustrated in the left panel in Fig.~\ref{fig:1} for the choice $L_+=5$ and $-L_-=-10$. Consider a gambler that expects to obtain profit by ``stopping" an unbiased random walk whenever it escapes the interval $(-L_-,L_+)$. Let $L_+$   the wealth gained by the gambler if the random walk first reaches the positive threshold, and $-L_-$  the wealth lost by the gambler if instead the random walk  first reaches the negative threshold. The gambler may expect that he/she could get a net profit from the fact that the random walk will reach the positive threshold more often than the negative one, even if the dynamics of the process is unbiased. However, because the unbiased random walk is a martingale, the probability for first reaching the positive threshold $P_+ = L_+/(L_+ + L_-)$ whereas $P_-= L_-/(L_+ + L_-)$ for first reaching the negative one~\cite{redner2001guide}. As a result, the net wealth after many repetitions of this gambling strategy $\langle M_{\T}\rangle = P_+ L_+  - P_- L_- = 0 = M_0$ equals to its initial value, i.e. it is a {\em fair}  strategy that leads to no net win neither to net loss on average. The validity of this property for arbitrary values of the negative threshold value $-L_-$  for is further illustrated with numerical simulations in the right panel in Fig.~\ref{fig:1} 

Equations~(\ref{eq:DoobOpt1}) and (\ref{eq:DoobOpt2}) are two examples of  the so-called Doob's  optional stopping theorem.        Loosely said, Doob's optional stopping theorem states that the martingale condition also holds when stopping a process at a clever moment, viz., 
 \begin{equation}
 \langle M_{\T} \rangle  =\langle M_{0} \rangle  \label{eq:DoobOpt3}
 \end{equation}
 holds, where $M$ is a martingale and $\T$ a stopping time.     Drawing an analogy with fair games, Eq.~(\ref{eq:DoobOpt3}) states that it is not possible to  win on average with a martingale, as its expected outcome at the end of the  game equals to its expected initial value.
In this {Treatise}, we will use repeatedly Doob's optional  stopping theorem to simplify   first-passage-time calculations. For example, as we will show in this {Treatise}, using Doob's optional stopping theorem, Eqs.~(\ref{eq:DoobOpt1}) and (\ref{eq:DoobOpt2}) can be used to shortcut   analytical calculations for e.g. splitting probabilities $P_-$ and $P_+$ and  mean first-passage times.

 \section{Martingales in biophysics}
 \label{motors}
 We discuss briefly how martingales can be a useful concept  in  biophysics.
For this purpose we discuss a minimal model of the motion of a molecular machine (motor) on a filament.
 
 \begin{figure}[h!]
    \centering
    \includegraphics[width=0.7\textwidth]{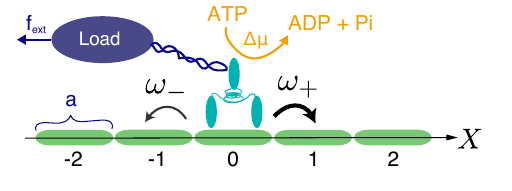}
    \caption{ Illustration of the minimal stochastic model of molecular motor motion, given by a Markov-jump process in a discrete lattice (a 1D biased random walk). The transition rates are given by $\omega_+=\nu\exp(A/2)$ and $\omega_-=\nu\exp(-A/2)$,  for forward and backward stepping respectively. See text for further details }
    \label{fig:motorsch1}
\end{figure}
 
 A molecular motor binds to a linear filament,
which provides a periodic, one-dimensional, lattice of binding sites. The filament has a polar asymmetry, which specifies the direction of motion. The motor catalyzes the hydrolysis of a
fuel, Adenosinetriphosphate (ATP) to the diphosphate form (ADP), releasing inorganic
phosphate (P). This reaction provides an amount $\Delta\mu=\mu_{\rm ATP}-(\mu_{\rm ADP}
+\mu_P)$ of chemical free energy.
As the system is driven out of thermodynamic equilibrium, it will step stochastically from binding site
to binding site with a bias in a direction given by the filament polarity. In the presence of an external force
$f_{\rm ext}$ it can perform mechanical work $f_{\rm ext} a$ per step, where $a$ is the spacing
between binding sites.

For simplicity, we describe the molecular motor stepping process as a continuous-time Markov-jump process (a biased random walk), using a discrete position variable $X_t=x\in \mathbb{Z}$
which describes the discrete binding sites.
Transitions from site $x$ to $x+1$ occur at a rate  $\omega(x,x+1)=\omega_+$, and  transitions in the opposite direction occur 
at a rate $\omega(x,x+1)=\omega_-$. We can write 
\begin{equation}
\omega_\pm=\nu \exp(\pm A/2),
\end{equation}
 where
\begin{equation}
\nu=\sqrt{\omega_+\omega_-} , \quad \text{and} \quad  A=\ln(\omega_+/\omega_-) 
\end{equation}
are the {\it kinetic} rate  and {\it affinity} of the motor, respectively.
In the simplest case of a motor that tightly couples ATP hydrolysis and stepping in a one-to-one manner,
thermodynamics requires that the ratio between forward and backward rates is 
\begin{equation}
\frac{\omega_+}{\omega_-}=\exp[\beta(\Delta\mu-a f_{\rm ext})],
\label{eq:affmotor}
\end{equation} 
and thus 
\begin{equation}
A=\beta(\Delta\mu-a f_{\rm ext})  , \quad \text{with}\quad  \beta= T^{-1}.
\end{equation}
The kinetic rate $\nu$ depends on ATP concentration, the external force, and on internal time scales of the motor molecule.

The probability $\rho_t(x)$ to find the motor at position $x=X_t$ at time $t$ obeys the Master equation
\begin{equation}
\frac{\partial \rho_t(x)}{\partial t}=\omega_+ \rho_t({x-1})-(\omega_+ +\omega_-)\rho_t(x)+\omega_-\rho_t(x+1)  .
\end{equation}
For the initial condition $\rho_{0}(x)=\delta_{x,0}$, the solution is given by
\begin{equation}
\rho_t(x)=\left(\frac{\omega_+}{\omega_-}\right)^{-x/2}\exp \left [-(\omega_+ + \omega_-) t  \right ] I_x(2\sqrt{\omega_+ \omega_-} t), %\exp \left [\frac{Ax}{2}-2\nu t \cosh A \right ] I_x(2\nu t) ,
\label{probex}
\end{equation}
where $I_x(y)$ denotes the modified Bessel function of the first kind.
This can be seen using the relations $dI_x(y)/dy=I_{x-1}(y)-(x/y)I_x(y)$ and $dI_x(y)/dy=I_{x+1}(y)+(x/y)I_x(y)$
which follow from the generating function  
\begin{equation}
\sum_{x=-\infty}^{\infty}z^x I_x(y)=\exp\left [y(z+z^{-1})/2\right ].
\label{Gen}
\end{equation}

The position of the motor is described by the stochastic variable $X_t$ where we choose $X_0=0$. 
%{(Xhy do we need this ?)}The average velocity is  $v=a \langle X_t \rangle /t=a(\omega_+-\omega_-)=2a\nu \sinh(A/2)$. 
Interestingly, the stochastic process [cf. Eq.~\eqref{eq:1p7new}]
\begin{equation}
    M_t = \exp(-AX_t)=\exp\left[-\left(\ln \frac{\omega_+}{\omega_-}\right)X_t \right],
    \label{eq:Mtmotor}
\end{equation}
is a martingale with respect to $X_t$. This can be proved by noting first that 
\begin{equation}
\left\langle \left. M_{t+dt}\right|X_{\left[0,t\right]}\right\rangle  =M_{t}\left\langle \left. \frac{M_{t+dt}}{M_{t}}\right|X_{\left[0,t\right]}\right\rangle 
  =M_{t}\left\langle \left.\left(\frac{w_{-}}{w_{+}}\right)^{X_{t+dt}-X_{t}}\right|X_{\left[0,t\right]}\right\rangle. \label{eq:velimirchaos2}
\end{equation}
Then, the central argument here is then that by definition of the transition rate we have the equality
\begin{eqnarray}
\left\langle \left.\left(\frac{w_{-}}{w_{+}}\right)^{X_{t+dt}-X_{t}}\right|X_{\left[0,t\right]}\right\rangle  & \!=\!&\left(\frac{w_{-}}{w_{+}}\right)w_{+}dt+\left(\frac{w_{-}}{w_{+}}\right)^{-1}w_{-}dt+\left(\frac{w_{-}}{w_{+}}\right)^{0}\left(1-\left(w_{+}+w_{-}\right)dt\right)+O(dt^2)\nonumber\\
 & =& 1+O(dt^2).\label{eq:nisekspress}
\end{eqnarray}
Combining~\eqref{eq:velimirchaos2} with~\eqref{eq:nisekspress} we get $
\left\langle \left.M_{t+dt}\right|X_{\left[0,t\right]}\right\rangle =M_{t}+O(dt^2)
$,  therefore using the {\em tower rule} (Appendix~\ref{app:towerproperty})  we have for any $0\leq s < t $ :  
\begin{equation}
\left\langle \left.M_{t+dt}\right|X_{\left[0,s\right]}\right\rangle =\left\langle\left\langle \left.M_{t+dt}\right|X_{\left[0,t\right]}\right\rangle\left. \right| X_{\left[0,s\right]} \right\rangle = \left\langle \left.M_{t}\right|X_{\left[0,s\right]}\right\rangle +O(dt^2) ,
\end{equation}
which implies that ${\frac{d}{dt}}\left\langle \left.M_{t}\right|X_{\left[0,s\right]}\right\rangle=0$,   and then  the martingale property 
\begin{equation}
\left\langle \left.M_{t}\right|X_{\left[0,s\right]}\right\rangle =M_s. 
\end{equation}
%as
%\begin{eqnarray}
%\left\langle M_t \vert X_{[0,s]}\right\rangle &=&\left\langle \frac{M_t}{M_s} \left\vert\right. X_{[0,s]}\right\rangle M_s\\
%&=&\left\langle M_{t-s}\right\rangle M_s \label{eq:nisburek1}\\
%&=& M_s.  \label{eq:nisburek2}
%\end{eqnarray}
%Equation~\eqref{eq:nisburek1} follows from the fact that $X_t$ is Markovian, time homogeneous and using $X_0=0$. {\bf [IN: I don't think the explanation is correct!]}
%On the other hand, Eq.~\eqref{eq:nisburek2} follows from the ``integral fluctuation relation" 
%\begin{eqnarray}
%\hspace{-1cm}\langle M_{t-s}\rangle&=&\sum_{x=-\infty}^{\infty}\rho_{t-s}(x)\left(\frac{\omega_+}{\omega_-}\right )^{-x}\%\
%&=&\exp(-(\omega_+ +\omega_-)(t-s))\sum_{x=-\infty}^{\infty}\left (\frac{\omega_+}{\omega_-}\right )^{-x/2}I_x(2\sqrt{\omega_+\omega_-}(t-s))\\
%&=&1,
%\end{eqnarray}
%where the last equality follows from Eq.~\eqref{Gen}.
Analogously, one can retrieve the martingale $M_t$ in Eq.~\eqref{eq:Mtmotor} by taking the continuous-time limit of the process~\eqref{eq:1p7} for the choice $z=A=\ln(\omega_+/\omega_-)$. 

%Knowing  the explicit form of the probability density of $X_t$, Eq.~\eqref{probex}, is however not necessary to prove the martingale property of $M_t= \exp(-AX_t)$.  A more fundamental proof follows   

We also note that $\exp(-AX_t)$ is not the {\em only} martingale associated with $X_t$. In fact, an inifinite number of  martingales can be defined as functions of $X_t$ and can be constructed similarly as for the discrete-time case in Sec.~\ref{sec:warmup} (see Eq.~\eqref{eq:1p7}). The {Treatise} will shed light on how to construct martingales from $X_t$ and why this is useful.

%The stochastic entropy produced until time $t$ {(ref equation)} obeys 
%\begin{equation}
%S^{\rm tot}_t =  \ln \frac{{\cal P}(X_{[0,t]})}{{\cal P}(\Theta_t ( X_{[0,t]}))} =  \ln\left [ \left (\frac{\omega_+}{\omega_-}\right )^{X_t}\right ]= AX_t ,\label{eq:Stot}
%\end{equation}
 
\section{Martingales on a ring}
\label{Ring} 
The martingales given by Eqs.~\eqref{eq:1p7new} and \eqref{eq:Mtmotor} in Secs.~\ref{sec:warmup}  and \ref{motors}, respectively,  can be expressed as 
\begin{equation}
    M_t = \exp (-\text{Entropy production in } [0,t]). \label{eq:MartGeneral}
\end{equation}
  For a single step of the random walker the entropy flow into the environment is given by $(X_t-X_{t-1})\log q/(1-q)$, which measures the degree of irreversibility via the ratio of the probabilities of a forward and a backward step, as will be discussed in Chapter 5. 
In the example of the molecular motor, the entropy flow associated with a step is proportional to  $\pm A=\pm \beta(\Delta\mu-a f_{\rm ext})$ which is the heat dissipated to the environment in a forward or backward step.   The martingality of  the process (\ref{eq:MartGeneral}) lies at the root of the use of martingales in stochastic  thermodynamics, and we discuss this extensively in this~{Treatise}.   % , as they appear also in e.g. stationary Langevin processes, and non-stationary Markovian processes that may be continuous or discrete.

In this subsection, we   review the  connection between stochastic thermodynamics and martingales by discussing the paradigmatic    example of a driven particle on a ring. This example, besides its simplicity, is illuminating because it reveals the martingale structure of stochastic entropy production in a simple yet nontrivial way.
\begin{figure}[h!]
\centering
{\includegraphics[width=0.6\textwidth]{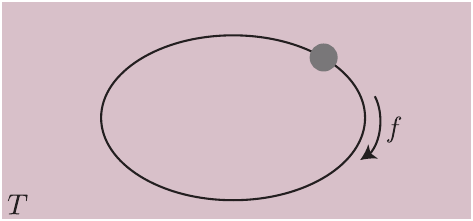}}
\caption{Illustration of a minimal stochastic model of nonequilibrium dynamics. A Brownian particle (gray circle) immersed in a thermal bath at temperature $T$ (red box) is subject to move along a ring under the action of an external constant force $f$.}\label{fig:ring}
\end{figure}

We consider the dynamics of a driven overdamped Brownian particle on a ring, see Fig.~\ref{fig:ring} for an illustration. A constant, homogeneous external force $f$ is applied to the particle along the ring. The particle moves with mobility $\mu$ within  a thermal bath that is at temperature $T$. The dynamics of the position $X$ of the particle is assumed to obey a one-dimensional overdamped Langevin equation
\begin{equation}
    \dot{X}_t  = \mu f + \sqrt{2\mu T}\dot{B}_t,
    \label{eq:le1d}
\end{equation}
where $\dot{B}_t$ is a zero-mean Gaussian white noise $\langle\dot{B}_t\rangle=0$ with autocorrelation $\langle\dot{B}_t\dot{B}_s\rangle=\delta(t-s)$, see Sec.~\ref{sec:diff} for further details about this class of processes.   Here, and throughout the {Treatise},  we have set the Boltzmann constant equal to one. We also assume that the initial state is drawn from the stationary distribution in the ring which is here uniform because $f$ is constant.

In a small  interval $[t,t+dt]$ of time, the particle moves by a stochastic amount $dX_t$. The work done on the particle in  $[t,t+dt]$ by the external force $f$ is stochastic and given by
\begin{equation}
    d W =f dX_t.
\end{equation}
 In this example the particle has no internal degrees of freedom and its
internal energy $U=U_0$  is constant and does not change in time, i.e., $dU_t=0$. We thus obtain from the first law of stochastic thermodynamics $dU_t = d Q_t  + d W_t$  the following expression for  the heat absorbed by the particle in $[t,t+dt]$, viz.,
\begin{equation}
 d Q_t = -d W_t= -f dX_t. \label{eq:dQSimple}
\end{equation}
Using the Langevin equation~\eqref{eq:le1d} in Eq.~(\ref{eq:dQSimple}), we  obtain a  stochastic differential equation for the heat, viz., 
\begin{equation}
    -\frac{\dot{Q}_t}{T}  = \frac{\mu f^2}{T} +    \sqrt{\frac{2 \mu  f^2}{T}} \dot{B}_t  = v^Q + \sqrt{2v^Q} \dot{B}_t,
    \label{eq:Qsde}
\end{equation}
where we have defined the expected heat rate
\begin{equation}
    v^Q \equiv  -\langle \dot{Q}_t  \rangle/T =   \frac{\mu f^2}{T}.
\end{equation}
 Furthermore, changing variables in Eq.~\eqref{eq:Qsde} and applying Ito's lemma (see Appendix~\ref{app:Ito-lemma}),  we find that the exponential $\exp\left(\frac{Q_t}{T}\right) $  satisfies the stochastic differential equation
\begin{equation}
    \frac{d}{dt} \exp\left(\frac{Q_t}{T}\right) = - \sqrt{2 v^Q} \exp\left(\frac{Q_t}{T}\right) \dot{B}_t\;.
     \label{eq:expStotLang}
\end{equation}

%Equivalently, the stochastic entropy production along a single trajectory, which in this case reads $S^{\rm tot}_t=-Q_t/T$, follows the stochastic differential equation
%\begin{equation}
%    \frac{\text{d}}{\text{d}t} e^{-S^{\rm tot}_t} = - \sqrt{2 v^Q} e^{-S^{\rm tot}_t}\cdot  \xi_t.
%    \label{eq:expStotLang}
%\end{equation}
Since the dissipated heat divided by the temperature is the entropy produced in this process, Eq.~\eqref{eq:expStotLang} reveals that the exponential of the negative entropy production a martingale.   This follows from~\eqref{eq:expStotLang} which shows that $\exp\left(Q_t/k_{\rm B}T\right)$ has no drift term, and hence is a  martingale. Because $Q_0=0$,  we find that $\left\langle \exp\left(Q_t/k_{\rm B}T\right) \right\rangle = 1 $ at all times, which is often referred to as the ``integral fluctuation relation (or theorem)" for the absorbed heat, and this relation is thus closely related to the martingality of  $\exp\left(\frac{Q_t}{T}\right) $.

For the present  example, the stochastic heat and its exponential can be determined analytically. Solving~\eqref{eq:Qsde} we get 
\begin{equation}
    Q_t = -T v^Q t - T\sqrt{2 v^Q} B_t,
    \label{fatigue10}
\end{equation}
where $B_t$ is the value of the Wiener process at time $t>0$. Because $v^Q\geq 0$ and $\langle B_t\rangle  =0$, we retrieve the second law of thermodynamics for this example, viz, 
 $\langle Q_t \rangle\leq 0$.   In other words, on average the particle dissipates heat into the environment  %  (see Sec.~\ref{sec:41}) vanishes because the stationary probability density for the particle is uniform and constant in time. 
 Moreover, the relation \eqref{fatigue10} implies that the integral of Eq.~\eqref{eq:expStotLang} is given by
\begin{equation}
    \exp\left(\frac{Q_t}{T}\right) =  \exp\left(   -v^Q t - \sqrt{2v^Q} B_t\right) .
    \label{GBM}
\end{equation}
In other words, $ \exp\left(Q_t/T\right)$ is a geometric Brownian motion with zero drift and volatility $\sqrt{2v^Q}$, a process that has been widely used e.g. in modelling stock fluctuations in quantitative finance; see Ch.~\ref{chapter:finance}.

\chapter{Martingales: Definitions and  examples} \label{ch:2a}

\epigraph{The name ``supermartingale" was spoiled for me by the fact that every evening the exploits of ``Superman" were played on the radio by one of my children.   
\\ \vspace{0.5cm} {\em A conversation with Joe Doob}, J.~L.~Snell, Stat. Sci. {\bf 12} (4) (1947).}

In this Chapter, through examples of martingales,  we  convince ourselves that martingales are ubiquitous.    
This chapter is organised into  two main parts. Section~\ref{MartDiscrete} defines and provides examples of martingales in discrete time, and Sec.~\ref{MartCont} does the same for martingales in continuous time.

For the sake of clarity, in  Chapters~\ref{ch:2a},~\ref{ch:2b} and~\ref{ch:3} we  use   the  symbol $n\in \mathbb{N}\cup \left\{0\right\}$ for a discrete time index (see Sec.~\ref{MartDiscrete}) and  $t\in \mathbb{R}^+$ for a  continuous time index (see e.g. Sec.~\ref{MartCont}).  On the other hand, in the other chapters of this {Treatise} we will use $t$  indiscriminately for both continuous and discrete time.

\section{Martingales  in discrete time}  \label{MartDiscrete}

\subsection{Martingales,  submartingales and supermartingales}   
Martingales are stochastic processes that have no net drift.    
 Formally, we define discrete-time martingales relative to  a stochastic process $X_n\in \mathcal{X}$ as follows.   

     {Let $M_n\in \mathbb{R}$ be a discrete-time stochastic process given by a real-valued function defined on the set of trajectories $X_{[0,n]} = (X_0, X_1, \ldots, X_n)$. We   assume that $M_n$ is integrable, i.e., $\langle  |M_n| \rangle<\infty$ for all $n$.}
{
\begin{leftbar}
    We say that  $M_n$ is a discrete-time {\bf martingale} relative to $X_n$ if  $M_n$ has no drift, i.e., 
    \begin{equation}
    \langle M_n|\Xom\rangle = M_m, \label{eq:defM}
    \end{equation}
    for all $0\leq m\leq n$.
 \end{leftbar} }
   
   %We say that  $M_n\in \mathcal{X}$ is a discrete time {\it martingale} relative to $X_n$ if the following three criteria are met:  
     
   %\begin{itemize}
   %\item  $M_n$ is a real-valued function defined on the set of trajectories $X_{[0,n]} = (X_0, X_1, \ldots, X_n)$;
   %\item  $M_n$ is integrable, i.e., $\langle  |M_n| \rangle<\infty$ for all $n$;
    %\item $M_n$ has no drift, i.e., 
    %\begin{equation}
    %\langle M_n|\Xom\rangle = M_m, \label{eq:defM}
    %\end{equation}
   % for all $0\leq m\leq n$.
   %\end{itemize}   
    
\begin{figure}
\centering
{\includegraphics[width=0.7\textwidth]{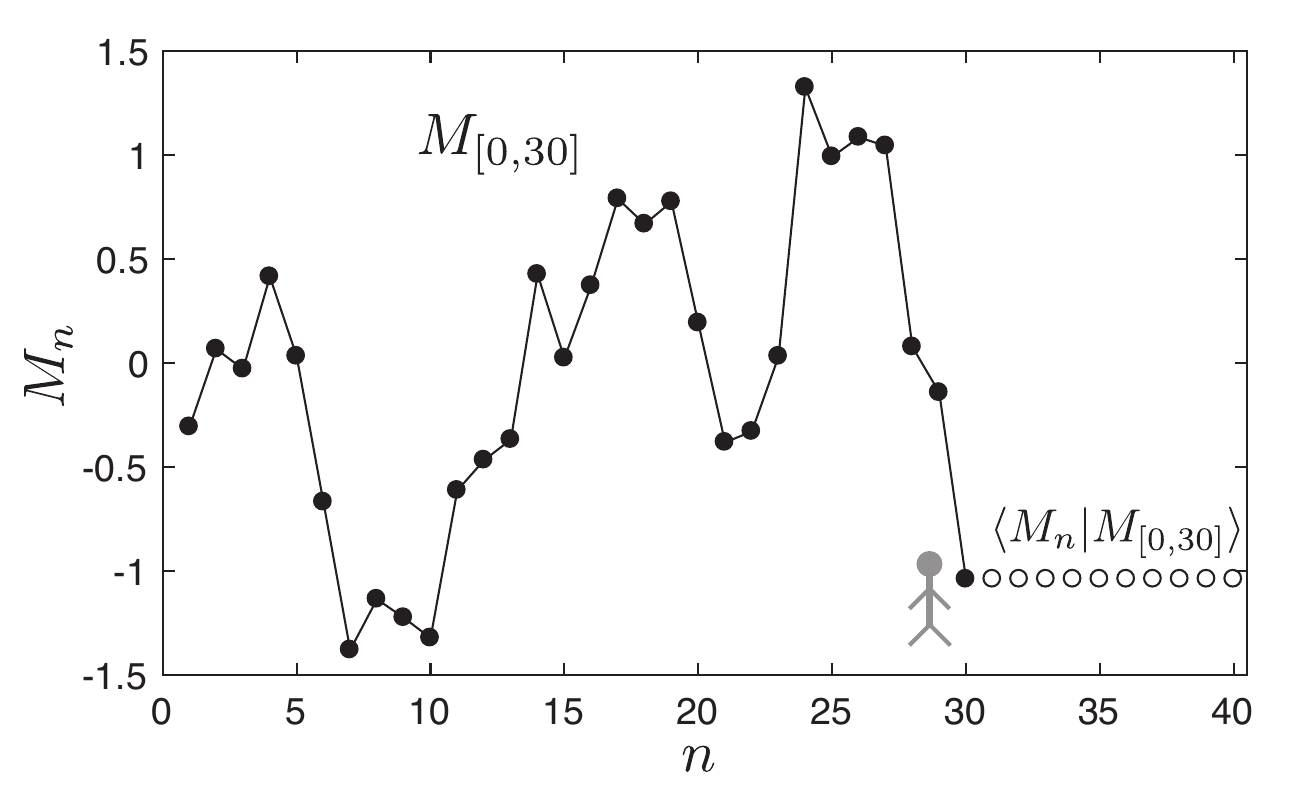}}
\caption{Illustration of a martingale process $M_n$ in discrete time $n\in\mathbb{N}$. Here $M_n$ is given by the cumulative sum of $n$ independent  Gaussian random numbers with zero mean and standard deviation equal to $1/2$. The
  filled circles with lines illustrate a specific trajectory of the process  up to time $m=30$. The  unfilled circles denote the  expected values of the martingale $\langle M_n | \Mom \rangle = M_{m}$ at  future times $n>m$, conditioned on  the  past sequence $M_{[0,m]}$ shown in the figure. The grey cartoon illustrates an observer that collects the values of the sequence $M_{[0,m]}$ and makes predictions about its future expected value.}
\label{fig:martillus}
\end{figure}

    Note that conditional expectations are defined as in Eq.~(\ref{eq:condExpa}).  We require that $M_n$ is integrable, as otherwise the conditional expectation is not well defined.     See Section 9.7 of Ref.~\cite{williams1991probability} for a list of useful properties of conditional expectations and Fig.~\ref{fig:martillus} for an illustration of the martingale concept.

    As done in Fig.~\ref{fig:martillus}, it is often assumed that  $X_n=M_n$ and  thus $\langle M_n | M_{[0,m]}\rangle =M_m$ for $0\leq m\leq n$.

{We  define submartingales (supermartingales) as processes with a nonnegative (nonpositive) drift. Specifically, consider a   real-valued function $S_n$ defined on the set of trajectories $X_{[0,n]}$, and let us  assume that $S_n$ is integrable, i.e., $\langle  |S_n| \rangle<\infty$.    We say that $S_n$ is a {\it submartingale} (supermartingale) relative to $X_n$ if  it has a nonnegative (nonpositive) drift, i.e., 
\begin{equation}
\langle S_n|\Xom\rangle \geq S_m \quad (\langle S_n|\Xom\rangle \leq S_m) \label{eq:SCond}
\end{equation}
for all $0\leq m\leq n$.
}
With these definitions, martingales are particular cases of submartingales. In what follows, when we refer to  martingales (or submartingales) we imply that they are defined with respect to a process $X_n$.   

   {The condition (\ref{eq:defM}) can be complicated to verify in concrete examples of stochastic processes.   However, in discrete time there exists a simpler, equivalent condition for martingality, which is a consequence of the tower property of conditional expectations,~see Appendix~\ref{appendix:A}. 
   %, an equivalent definition to Eq.~\eqref{eq:defM} of a martingale in discrete time can  be formulated through the so-called tower property (or tower rule), which is a central property in martingale theory. The so-called elementary tower rule  is a mathematical property concerning the conditional expectation of a random variable $Z$ condition to another variable $X$ to take a specific value. It is formulated as follows
  % \begin{equation} \label{eq:tower0main}
%\la \,\la{Z}|{X}\ra\,\ra= \la{Z}\ra,
%\end{equation}
%which holds for any two random variables $Z,X$ having arbitrary joint probability distributions, see Appendix~\ref{appendix:A} for further details. This result follows from the property of iterated conditioning (see Appendix~\ref{sec:condexpshamik}) and can be also extended to functionals, see Appendix~\ref{app:towerproperty} and below.
The tower property states that  for any (integrable) functional  $Z_p = Z[X_{[0,p]}]$ it holds that 
   \begin{equation}
\langle \langle Z_p | X_{[0,n]}
 \rangle | X_{[0,m]} \rangle  =  \langle Z_p | X_{[0,m]} \rangle, \label{eq:Tower1}
\end{equation}
for all $0\le m\le n$.  Using this tower property, we get the following simpler ``one-step-ahead" martingale criterion~\cite{shreve2005stochastic} . }

\begin{leftbar}
\textbf{One-step-ahead criterion for martingality.} 
%The tower property~\eqref{eq:Tower1} implies that 
The martingale property~\eqref{eq:defM} is equivalent to the simpler condition
\begin{equation}
 \langle M_{n+1}|\Xon\rangle = M_n, \label{eq:defMtower}
\end{equation}
for all $n$. 
\end{leftbar} 

%The tower property associated with a functional $Z_p = Z[X_{[0,p]}]$ is given by
%\begin{equation}
%\langle \langle Z_p | X_{[0,n]}
 %\rangle | X_{[0,m]} \rangle  =  \langle Z_p | X_{[0,m]} \rangle \qquad 0\le m\le n, \label{eq:Tower1}
%\end{equation}
%which holds for an arbitrary value of $p$, see Appendix~\ref{app:towerproperty} for further details and its derivation. 

%{Even though Eq.~\eqref{eq:defMtower} seems a weaker condition than~\eqref{eq:defM}, it is equivalent to the martingale condition in discrete time~\eqref{eq:defM}.}
{The equivalence between the conditions  \eqref{eq:defM} and \eqref{eq:defMtower}  follows from the tower property of conditional expectations.  Indeed, for all $m<n$ it holds that  \[
\left\langle \left.M_{n}\right|X_{\left[0,m\right]}\right\rangle =\left\langle \left.\left\langle \left.M_{n}\right|X_{\left[0,n-1\right]}\right\rangle \right|X_{\left[0,m\right]}\right\rangle =\left\langle \left.M_{n-1}\right|X_{\left[0,m\right]}\right\rangle ,
\]
which iterates up to
\[
\left\langle \left.M_{n}\right|X_{\left[0,m\right]}\right\rangle =\left\langle \left.M_{m}\right|X_{\left[0,m\right]}\right\rangle =M_{m}.
\]}

\subsection{$^{\spadesuit}$Backward martingales, submartingales and supermartingales}\label{sec:back}
In the definition of the martingale, Eq.~(\ref{eq:defM}), we have that $n\geq m$, and hence the martingale definition uses a part of the trajectory that happened in the past.     We can also define martingales  conditioned on a part of the trajectory that takes place in the future.   In this way, we obtain   {\em backward} martingales.  

{Let $M_{n}$  be a real-valued function defined on the set of trajectories $X_{[n,\infty]} = (X_{n}, X_{n+1}, \ldots, X_{\infty})$.   In addition, we assume that $M_{n}$ is integrable, i.e., $\langle  |M_{n}| \rangle<\infty$ for all $n\in \mathbb{N}$.   
\begin{leftbar}
 We say that  $M_{n}\in \mathcal{X}$ is a  {\bf backward martingale}  relative to $X_{n}$ if  $M_{n}$ has no drift  when conditioned on events in the future, i.e., 
    \begin{equation}
    \langle M_{\ell}|X_{[m,n]}\rangle = M_{m}, \label{eq:defMB}
    \end{equation}
    for all $0\leq \ell \leq m\leq n$.
\end{leftbar}   
}
Backward submartingales and backward supermartingales are  defined by replacing the equality in Eq.~(\ref{eq:defMB}) by $\geq$ and $\leq$, respectively.
To distinguish martingales from backward martingales, we sometimes call the former  {\em forward} martingales.  %As we will see in later chapters,    a stochastic process can be a backward martingale but not a forward martingale.  

\subsection{Examples of martingales in discrete time}  \label{sec:Examples}

\begin{itemize}

\item {\bf Gambler's fortune in a fair game of chance}:  A gambler's fortune  in a fair game of chance is a martingale \cite{Doob}. Let us consider the example of a coin toss.   The game consists of a series of coin flips with equally likely outcomes $X_n\in\left\{{\rm Head},{\rm Tail}\right\}$.  
  Each betting round, the gambler guesses the outcome  of the coin toss through a betting system.   The gambler's guess is denoted by  $Y_n\equiv Y\left(X_{[0,n-1]}\right)\in\left\{{\rm Head},{\rm Tail}\right\} $, where $Y\left(X_{[0,n-1]}\right)$ means that $Y_n$ depends  on $X_{[0,n-1]}$; such processes $Y_n$ are called  {\it predictable} processes.   If the gambler guesses right, i.e., $Y_n=X_n$, they wins $1$ euro, otherwise, if the gambler guesses wrong, i.e., $Y_n\neq X_n$, they loses $1$ euro.     
 The gambler's fortune $F_n$ after $n$ betting rounds satisfies \begin{eqnarray}
F_n \equiv  F_{0} + \sum^n_{m=1} \left( 2\delta_{X_m, Y_m} -1\right), \label{eq:toinCoss}
\end{eqnarray}
 and is  a martingale process. Here, $\delta_{i,j}$ denotes the Kronecker delta function.     The  martingale property $\langle F_n|X_{[0,m]}\rangle  = F_m$ reflects the fairness of this game.

\item {\bf Sums of independent random variables}:    Let   $X_i$ ($i\in\mathbb{N}\cup\left\{0\right\}$) be a sequence of independent and identically distributed ---denoted \textbf{iid} here and in the following--- random variables with  finite variance.   The sum
\begin{eqnarray}
\tilde{X}_n \equiv \sum^n_{i=0}X_i \label{eq:zeroMeanSum}
\end{eqnarray}
has conditional average
\begin{eqnarray}
\langle \tilde{X}_n |X_{[0,m]}\rangle =  \sum^m_{i=0}X_i + \sum^n_{i=m+1} \langle   X_i  \rangle \, .
\end{eqnarray}
Therefore, $\tilde{X}_n$
is  a martingale, submartingale, or supermartingale, if $X_i$ has zero mean, positive mean, or negative mean, respectively.   Indeed, it holds that
\begin{eqnarray}
\langle \tilde{X}_n |X_{[0,m]}\rangle  
\begin{cases}
=\tilde{X}_m, \quad \text{if} \quad \langle X_i\rangle =0, \\
\geq  \tilde{X}_m, \quad \text{if} \quad \langle X_i\rangle \geq 0, \\
\leq  \tilde{X}_m, \quad \text{if} \quad \langle X_i\rangle \leq 0. 
\end{cases}
\end{eqnarray}
Moreover, because the square root is a concave function, we have
\begin{eqnarray}
\langle |\tilde{X}_n | \rangle = \left\langle \sqrt{\tilde{X}_n^2}  \right\rangle  \leq  \sqrt{\left\langle \tilde{X}_n^2 \right\rangle } <\infty,
\end{eqnarray}  
where in the second inequality we used that $X_i$, and thus also $\tilde{X}_n$, has a finite variance.

The sum $\tilde{X}_n$ also  obeys a strong law of large numbers, which states that $\overline{X}_n = \tilde{X}_n/n$  converges almost surely to its mean value $\mu = \langle X_i\rangle$~\cite{Feller}.   In addition, $\tilde{X}_n$ satisfies the  central limit theorem, which states that $(\tilde{X}_n-\mu \: n)/\sqrt{n}$ converges in distribution to a standard, normally distributed random variable.    In Secs.~\ref{subsec:LLN} and~\ref{subsec:CLT}  we consider extensions of these properties  to martingale processes.  

\item{\bf Conditional-expectation process (closed Martingale}: Let $X_i$ ($i\in\mathbb{N}\cup\left\{0\right\}$) be  a sequence of integrable, possibly correlated, random variables.

We consider the conditional expectation
\begin{equation}
     C^{\ell}_{m,n} \equiv \langle X_\ell | X_{[m,n]} \rangle ,
    \label{eq:cepdef}
\end{equation}
which depends on three integers $0 \leq m \leq n$ and $\ell \geq 0$. 
We can interpret $ C^{\ell}_{m,n}$ as a forward matingale or a backward martingale: 
\begin{itemize}
\item If we keep $\ell$ fixed and set $m=0$, then  the process  $C^{\ell}_{0,n}$ is a forward martingale for values of  $n$ in  $0\leq n \leq \ell$.  Indeed, 
\begin{equation}
    \langle C^{\ell}_{0,n} | X_{[0,n']} \rangle = C^{\ell}_{0,n'},
\end{equation} 
for all $0\leq n' \leq n \leq \ell$.   This relation follows from the  {\bf tower property} of conditional expectations { [see Eq.~\eqref{eq:Tower1}]}, 
\begin{equation}
    \langle C^{\ell}_{0,n} | X_{[0,n']} \rangle = \langle \langle X_{\ell} |X_{[0,n]} \rangle | X_{[0,n']}\rangle  =  \langle X_{\ell} | X_{[0,n']}\rangle = C^{\ell}_{0,n'}, 
    \label{eq:towerruleexample}
\end{equation}
where we have used the definition~\eqref{eq:cepdef} in the first and in the third equalities, and the tower property in the second equality. A proof of the tower property can be found in Appendix~\ref{app:towerproperty}.  
\item  Alternatively, for fixed $\ell$ and $n$,  the process $C^{\ell}_{m,n}$ with $m$ such that $0\leq \ell\leq m \leq n$, is a backward martingale.  Indeed, 
\begin{equation}
    \langle C^{\ell}_{m,n} | X_{[m',n]} \rangle = C^{\ell}_{m',n},
\end{equation}
for all $0\leq \ell \leq m \leq m' \leq n$.  Also this result follows from the  {\bf tower property} of conditional expectations  { [see Eq.~\eqref{eq:Tower1}]}, 
\begin{equation}
    \langle C^{\ell}_{m,n} | X_{[m',n]} \rangle = \langle \langle X_{\ell} |X_{[m,n]} \rangle | X_{[m',n]}\rangle  =  \langle X_{\ell} | X_{[m',n]}\rangle = C^{\ell}_{m',n}, 
    \label{eq:towerruleexample}
\end{equation}
where here also we have used the definition~\eqref{eq:cepdef} in the first and in the third equalities, and the tower property in the second equality.
\end{itemize}

\item {\bf Martingale transform}: Let $M_n$ be a martingale  relative to $X_n$, and let $D_n$ be a  process  determined by $X_{[0,n]}$.   The martingale transform 
\begin{equation}
(D\cdot M)_n \equiv  D_0M_0 + \sum^{n}_{k=1}D_{k-1} (M_k-M_{k-1}) 
\label{eq:transform}
\end{equation}
with $(D\cdot M)_0 = M_0$, 
is  a martingale if $|D_n|\leq c$, with $c$ a positive constant.  
{Indeed, it holds that \begin{align}
\left\langle \left.(D\cdot M)_{n}\right|X_{\left[0,n-1\right]}\right\rangle  & =\left\langle \left.(D \cdot M)_{n-1}+D_{n-1}\left(M_{n}-M_{n-1}\right)\right|X_{\left[0,n-1\right]}\right\rangle ,\nonumber \\
 & =(D\cdot M)_{n-1}+D_{n-1}\left\langle \left.\left(M_{n}-M_{n-1}\right)\right|X_{\left[0,n-1\right]}\right\rangle, \nonumber\\
 & =(D \cdot M)_{n-1}+0,\nonumber\\
 & =(D \cdot M)_{n-1}. \label{eq:DMmart}
\end{align}
In the second equality we have used that $D_{n-1}$ is fully determined by $X_{[0,n-1]}$, and the third equality follows from the martingale property of $M$. 
By virtue of the one-step-ahead condition  (\ref{eq:defMtower}),  Eq.~\eqref{eq:DMmart}  implies that $D\cdot M$ is a martingale.   Note that the use of $D_{k-1}$ in the definition (\ref{eq:transform}) is important to guarantee the martingality of $(D\cdot M)_n$.
%puis for all $0\leq m<n$ 
}

\item {\bf Ratios of path probability densities}: Martingales play an important role in stochastic thermodynamics~\cite{neri2019integral}, as well as, in statistics~\cite{tartakovsky2014sequential}. One reason is that several quantities of central interest in these fields are expressed as ratios of probability densities, and ratios of probability densities are  martingales.

Specifically,  consider two probability densities $\mathcal{P}(x_{[0,n]})$ and $\mathcal{Q}(x_{[0,n]})$, defined on the same set of trajectories $x_{[0,n]}\in \mathcal{X}^n$.  We assume that  $Q(x_{[0,n]}) = 0$ if $P(x_{[0,n]}) = 0$ for all $n\in \mathbb{N}$, and we say that  $\mathcal{Q}$  is locally,  {\bf absolutely continuous}  with respect of $\mathcal{P}$ when this condition holds.  For $\mQ$ that are locally, absolutely continuous with respect of $\mP$, the process 
\begin{eqnarray} 
R_n \equiv \frac{\mathcal{Q}(X_{[0,n]})}{\mathcal{P}(X_{[0,n]})}\, , \label{eq:RForm}
\end{eqnarray} 
with the convention that $0/0 = 0$, exists and is a martingale.   Notice that in  Eq.~(\ref{eq:RForm}) we evaluate the probability density $\mathcal{P}(x_{[0,n]})$ on the random realisation $X_{[0,n]}$ of the trajectory $x_{[0,n]}$, and analogously for $\mQ$. 

The fact that $R_n$ is a martingale can  be proven as follows:
 \begin{eqnarray} \label{eq:RnMartingale} %Ken added a label
\langle R_n | X_{[0,m]}\rangle  &=& \sum_{ x_{m+1}}\dots \sum_{ x_{n}} \mathcal{P}(x_{[m+1,n]}| X_{[0,m]})\: \frac{\mathcal{Q}(X_{[0,m]}, x_{[m+1,n]})}{\mathcal{P}(X_{[0,m]}, x_{[m+1,n]})} \nonumber\\ 
&=&  \sum_{x_{m+1}}\dots \sum_{x_n} \frac{\mathcal{P}(X_{[0,m]}, x_{[m+1,n]})}{\mathcal{P}(X_{[0,m]})}\: \frac{\mathcal{Q}(X_{[0,m]}, x_{[m+1,n]})}{\mathcal{P}(X_{[0,m]}, x_{[m+1,n]})}  \nonumber\\
&=&  \frac{\sum_{x_{m+1}}\dots \sum_{x_n}  \mathcal{Q}(X_{[0,m]}, x_{[m+1,n]})}{\mathcal{P}(X_{[0,m]})}\nonumber\\ 
&=& \frac{\mathcal{Q}(X_{[0,m]})}{\mathcal{P}(X_{[0,m]})}=R_m, \label{eq:Der}
\end{eqnarray}
where in the second equality we have used the definition of a conditional probability distribution, and in the last step we have used  that $\mathcal{Q}(X_{[0,m]})$ is the marginal  probability distribution of $\mathcal{Q}(X_{[0,n]})$ for $m<n$.

If instead $R_n$ is the ratio of a sequence of  densities  $\mQ^{(n)}(X_{[0,n]})$ and $\mP^{(n)}(X_{[0,n]})$  that depend explicitly  on time $n$, then the marginalisation condition, used in the last step of the derivation of Eq.~(\ref{eq:Der}), does not hold in general, and in this case $R_n$ is in general not a martingale~\footnote{For the example of a  Markov chain, which we introduce below in Chapter~\ref{ch:2b}, explicit time dependence occurs if the transition matrix  in Eq.~\eqref{eq:transmat} has a supplementary dependence on $n$, i.e., the path probability Eq.~\eqref{eq:PDefMarkov}  reads 
\begin{eqnarray}
 \mathcal{Q}^{(n)}(x_{[0,n]}) = \rho_0(x_0) \prod^{n}_{j=1}w^{(n)}(x_{j-1},x_{j}). \label{eq:PDefMarkov}
 \end{eqnarray}
 Note that this latter property is different than the time-inhomogeneity of  a Markov chain for which the transition matrix has a supplementary dependency on the present time $j$ and the path probability Eq.~\eqref{eq:PDefMarkov}  reads 
\begin{eqnarray}
 \mathcal{Q}(x_{[0,n]}) = \rho_0(x_0) \prod^{n}_{j=1}w^{(j)}(x_{j-1},x_{j}). \label{eq:PDefMarkov}
 \end{eqnarray}
 This time inhomogeneity is not a problem for the last step of the  derivation in Eq.~(\ref{eq:Der}), which remains valid.}.   This observation plays an important role in stochastic  thermodynamics, as we discuss in  detail in Sec.~\ref{sec:6p2}.

\item {\bf Random walker on $\mathbb{Z}$}:    Let $X_n$ denote the position of a biased random walker on $\mathbb{Z}$ with $X_0 = 0$.  The random walker makes one step in the positive direction with a probability $q$ and one step in the negative direction with a probability $1-q$.   This model was introduced in Sec.~\ref{sec:warmup}, and  see Fig.~\ref{fig:BRW} for an illustration.

The position of the walker relative to its mean, i.e.,
\begin{eqnarray}
M_n\equiv X_n - n(2q-1), 
\end{eqnarray}
 is a martingale because it is a sum of independent random variables with zero mean, as in Eq.~(\ref{eq:zeroMeanSum}).
As shown in Sec.~\ref{sec:warmup} (see Eq.~\eqref{eq:1p7}), the exponential 
\begin{eqnarray}
\mathcal{E}_n(y)\equiv  \frac{\exp({y}  X_n)}{ [q\exp({y} )+(1-q)\exp(-{y} )]^n   }   \label{eq:stochExpExample1}
\end{eqnarray}
is a martingale process for all values of ${y} \in \mathbb{R}$. This statement is also proven in  Appendix~\ref{appendix:A} by expressing $\mathcal{E}_n({y} )$ as a ratio of two probability densities.     Using  ${y}  = \ln [(1-q)/q]$, we obtain that 
\begin{eqnarray}
  \mathcal{E}_n\left(\ln \frac{1-q}{q}\right) =\exp\left[X_n\ln \left(\frac{1-q}{q}\right)\right] = \left(\frac{1-q}{q}\right)^{ X_n},    \label{eq:stochExpExample1x}
\end{eqnarray}
which coincides with the martingale given by Eq.~\eqref{eq:1p7new}. 

Since $\mathcal{E}_n({y} )$ is a martingale, it holds that 
\begin{eqnarray}
\langle \mathcal{E}_n({y} ) \rangle  = \langle \mathcal{E}_0({y} ) \rangle  = 1.
\end{eqnarray}
Therefore,  the generating function of $X_n$ is  given by 
\begin{eqnarray}
g_n({y} ) =  \langle  \exp({y}   X_n) \rangle = \left( q\exp({y} )+ (1-q)\exp(-{y} ) \right)^{n},
\end{eqnarray}
which can also be verified with a direct computation. Expanding (\ref{eq:stochExpExample1}) in ${y} $, we obtain 
\begin{eqnarray}
\mathcal{E}_n({y} ) ={1} +  \sum^{\infty}_{j=1}{y} ^j M^{(j)}_n(X_n)
\end{eqnarray}
and hence the processes $M^{(j)}_n(X_n)$ are martingales, viz., the processes
\begin{eqnarray}
M^{(1)}_n(X_n) &=&X_n-n(2q-1), \label{g1} \\ 
M^{(2)}_n(X_n) &=& (X_n - n(2q-1))^2 -  4 n q(1-q), \label{g2}\\
\dots \nonumber\\
M^{(k)}_n(X_n) &=& \left.\frac{\partial^{(k)} \mathcal{E}_n({y} )}{\partial {y} ^k}\right|_{{y} =0},
\end{eqnarray}
and so forth  are martingales (cf.~Eqs.~(\ref{eq:1p9}) and~(\ref{eq:1p99})).

\item {\bf Random walker on $\mathbb{R}$}: 
We consider a random walker moving on the real line.   The position $X_n$ of the random walker satisfies \begin{eqnarray}
X_n \equiv X_{n-1} + a + Y_n  \label{eq:randomWalkR}
\end{eqnarray}
for all $n\geq 1$ and $X_0=0$.  The increments $Y_n$ are iid random variables with zero mean and finite variance, and not necessarily drawn from a  Gaussian distribution.   If $a=0$, then $X_n$ is a martingale.   On the other hand, if $a>0$ or $a<0$, then $X_n$ is a submartingale or a supermartingale, respectively. See Fig.~\ref{fig:rws} for illustrations.
\begin{figure}[H]
\centering
{\includegraphics[width=\textwidth]{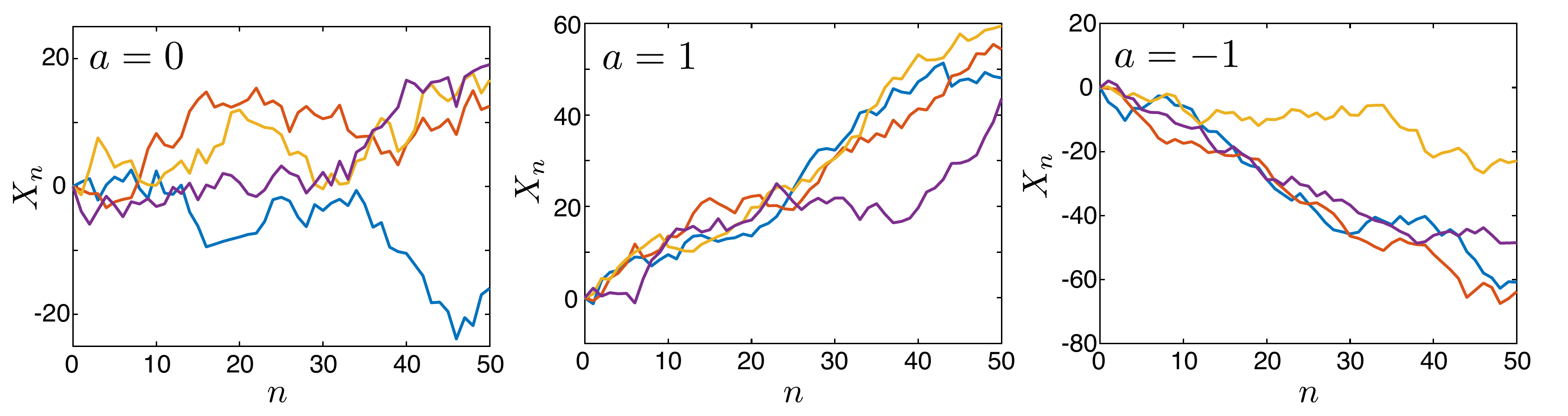}}
\caption{Illustration of  a martingale (left), submartingale (middle) and supermartingale (right). Sample trajectories of  discrete-time random walks on the real line, $X_n$, as a function of time $n$, as  given by Eq.~\eqref{eq:randomWalkR} with $Y_n$ extracted from a Gaussian distribution with zero mean and standard deviation equal to two. The different panels are obtained for three different values of the bias parameter $a$: $a=0$ (left), $a=1$ (middle), and $a=-1$ (right), which correspond respectively to martingale, submartingale and supermartingale processes.  }\label{fig:rws}
\end{figure}

\item {\bf Martingales in branching processes}:  Branching processes are simple models for reproduction \cite{asmussen1983branching, grimmett2001probability, haccou2005branching}.    Consider a population of  constituents, which may be, e.g.,  nuclei, molecules, viruses, cells, or animals,  that  multiply themselves.    We denote  the number of members in the population at time $n$ by  $X_n \in \mathbb{N}\cup \left\{0\right\}$, with the initial condition $X_0 = 1$.   At each time step reproduction takes place, and thus each time step corresponds with one generation. We assume that all members live for exactly one generation.  We denote by $Y_{i,n}\in \mathbb{N}\cup \left\{0\right\}$ the number of progeny of the $i$-th member of the population at generation $n$, see Fig.~\ref{fig:branch} for an explanation. It  holds then that 
\begin{eqnarray}
X_{n} \equiv \sum^{X_{n-1}}_{i=1}Y_{i,n}. 
\end{eqnarray} 
We assume that the $Y_{i,n}$ are iid drawn random variables from a distribution    $\rho_Y(y)$  with $y\in \mathbb{N}\cup \left\{0\right\}$.   We denote by $\mu =  \sum^{\infty}_{y=0}\rho_Y(y)y$      the mean value of $Y$ and by $g(s) = \sum^{\infty}_{y=0}\rho_Y(y)s^y$ the generating function of $Y$.  One can verify that  $\langle X_n \rangle = \mu^n$ and that the   extinction probability  $\eta$, which is the probability that the parent generates a finite population, is the smallest nonnegative root of the equation $\eta = g(\eta)$  \cite{athreya2004branching, grimmett2001probability}; see Fig.~\ref{fig:branch} for a  derivation.

The normalised population size 
\begin{eqnarray}
W_n \equiv \frac{X_n}{\langle X_n\rangle}
\end{eqnarray} 
is a martingale.   Indeed, 
\begin{eqnarray}
\langle W_n |X_{[0,n-1]}\rangle = \frac{\langle X_n |X_{n-1}\rangle  }{\langle X_n \rangle } = \frac{\mu X_{n-1}}{\mu^n} = \frac{X_{n-1}}{\langle X_{n-1} \rangle} = W_{n-1}, \label{eq:WM}
 \end{eqnarray} 
 where in the first equality we have used the Markov nature of the process and in the second equality we have used that $X_{n}$ is the sum of $X_{n-1}$ independent random variables with mean $\mu$.   {It follows  from (\ref{eq:WM}) and the tower property of conditional expectations that $W_n$ is a martingale [see discussion around Eq.~(\ref{eq:defMtower})].}
 
More surprising is that the process   \cite{grimmett2001probability}
\begin{eqnarray} 
V_n  \equiv \eta^{X_n},
\end{eqnarray}
with $\eta$  the extinction probability, is  a martingale.    Indeed, it holds that 
\begin{eqnarray} 
\langle V_n  | X_{[0,n-1]}\rangle =  \langle \eta^{\sum^{X_{n-1}}_{j=1}Y_j}|X_{[0,n-1]}  \rangle  = \prod^{X_{n-1}}_{j=1}\langle \eta^{Y_j} \rangle  = \left(g(\eta)\right)^{X_{n-1}} = \eta^{X_{n-1}} = V_{n-1},\nonumber
\end{eqnarray}
{and thus according to the one-step-ahead condition given by Eq.~(\ref{eq:WM})  $V_n$ is a martingale.}

\begin{figure}[H]
\centering
{\includegraphics[width=0.35\textwidth]{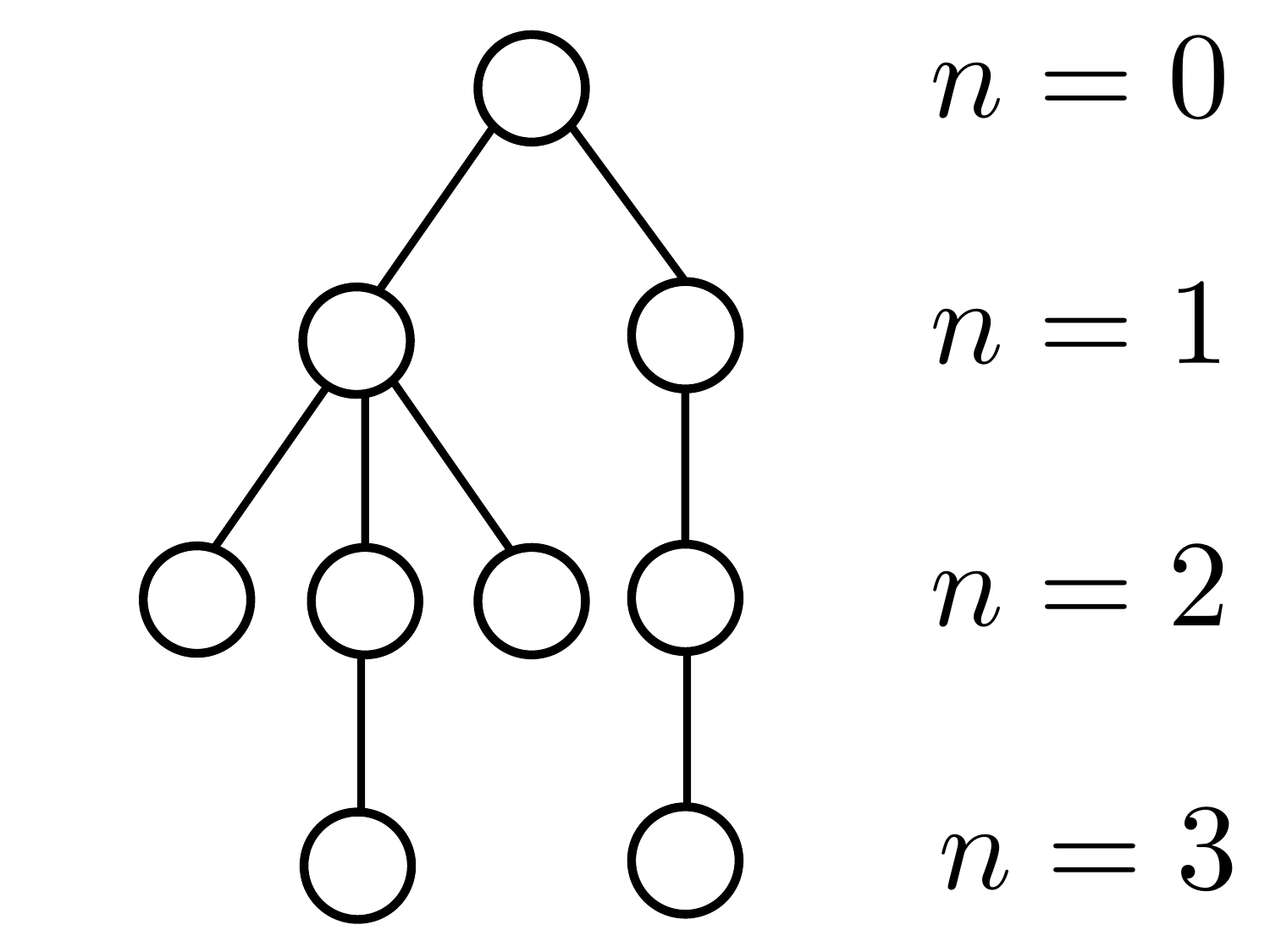}}
\caption{An example of one realisation of a branching process.   In this example the parent generates a finite population of three generations.   For the example shown, $\left\{Y_{1,0} = 2\right\}$, $\left\{Y_{1,1}=3, Y_{2,1} = 1\right\}$, $\left\{ Y_{1,2}=0, Y_{2,2}=1, Y_{3,2}=0,Y_{4,2}=1\right\}$, and $\left\{Y_{1,3}=0, Y_{2,3}=0\right\}$.   }\label{fig:branch}
\end{figure}
\begin{figure}[h]
\centering
{\includegraphics[width=0.45\textwidth]{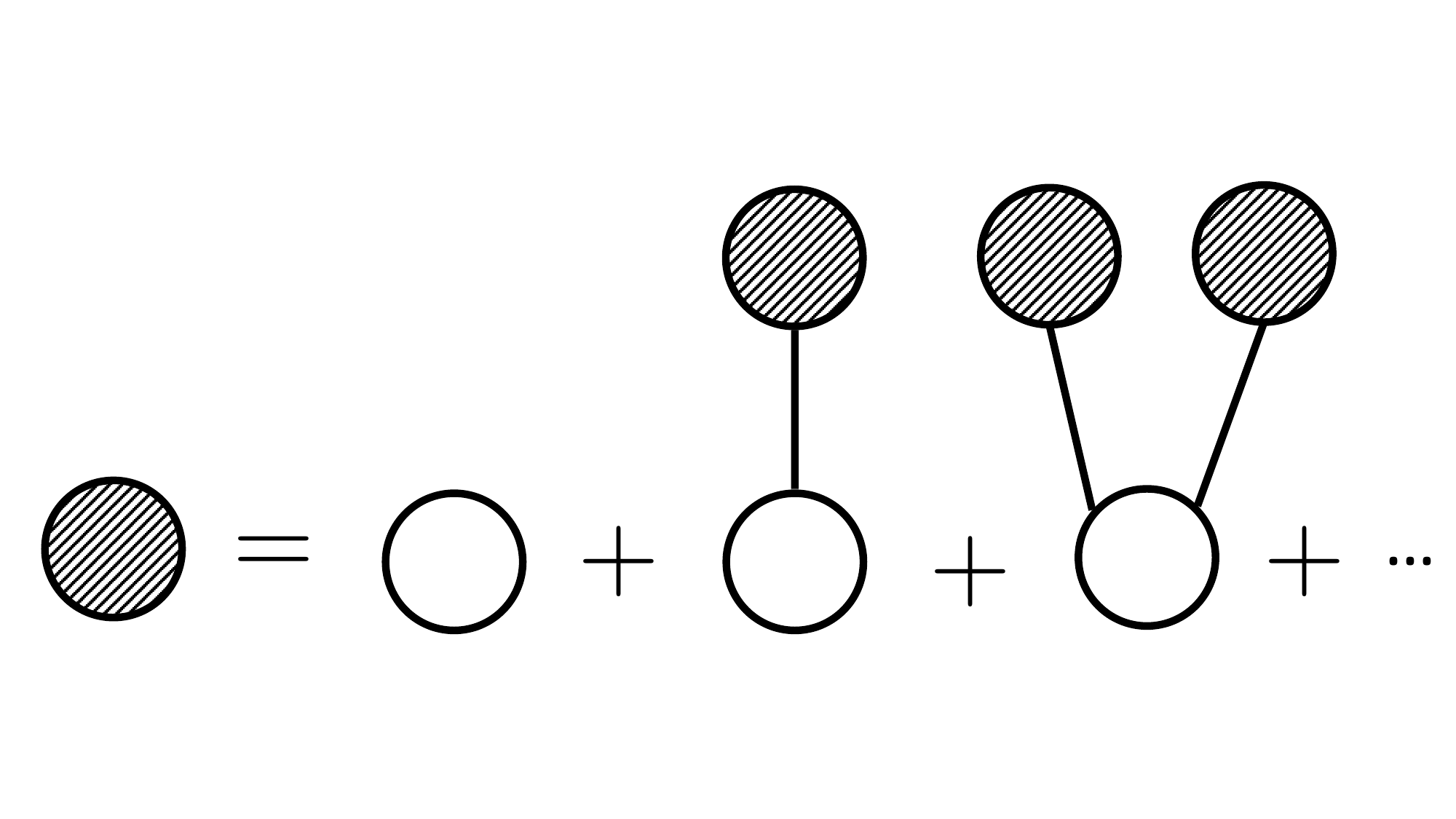}}
\caption{ A graphical illustration of the self-consistent equation $\eta = g(\eta) = \sum_{y=0}\rho_Y(y) \eta^y$ for the extinction probability of a branching process.  The extinction probability  $\eta$, denoted by the filled circle, is equal to the probability that the parent, denoted by an unfilled circle, has no progeny ($Y=0$), plus the probability that the parent has one child ($Y=1$) and this child  generates a finite population, plus the probability that the parent has two children ($Y=2$), both of which generate finite populations, etc.     }\label{fig:branch}
\end{figure}

\item {\bf Martingales in elephant random walks}:  So far, we have considered examples of martingales in processes $X$ that are Markovian.  We consider now an example of a martingale in a non-Markovian process $X$, namely, the elephant random walk.    

{Elephant random walks were introduced in Ref.~\cite{schutz2004elephants} as examples of non-Markovian processes with long-range memory that can exhibit   anomalous diffusion.   A diffusing particle exhibits anomalous diffusion when its mean squared displacement grows as a power law, i.e., $\langle X^2_n\rangle \sim n^{\alpha}$ with an exponent $\alpha\neq 1$~\cite{metzler2004restaurant}.   Anomalous diffusion has been observed, amongst others, in the motion of lipid granules  in the  cytoplasm~\cite{tolic2004anomalous}, in colloidal particles in an optically controlled medium~\cite{Douglass}, and  active particles~\cite{golestanian2009anomalous}.   Although Markov processes can exhibit anomalous diffusion transiently, i.e., within a finite time window, asymptotically they inevitably transition to a regime with standard diffusion, see Ref.~\cite{hart2008scale}.     The elephant random walk describes how superdiffusion emerges in a microscopic random walk model due to the presence of long-range temporal correlations.      }

References~\cite{bercu2017martingale,laulin2022new} identify  martingale processes associated with   elephant random walks, and use these martingales  to characterise  properties of  elephant random walks.  {Here we review some of their findings in a minimal example.}

Let us consider an elephant random walk  located at the position $X_n\in \mathbb{Z}$ at time $n=\{0,1,\dots\}$. The initial position of the walker is $X_0=0$.  At time $n=1$ the walker moves to $X_1=Y_1$, where $Y_1$ equals $+1$ with probability $1/2$ and $-1$ with probability $1/2$. In  the next steps, $n\geq 1$, the motion of the walker is as follows, 
\begin{equation}
    X_{n+1} \equiv X_n + Y_{n+1},
    \label{eq:Xdyn}
\end{equation}
where $Y_{n+1}$  is obtained  by the following rule.   We select uniformly at random an integer $k\in \{1,\dots , n\}$ and we then reverse  with probability $p$ the  sign  of the corresponding $Y_k$, i.e.,
\begin{equation}
Y_{n+1}  \equiv
\left\{
\begin{array}{ll}
Y_k \quad {\rm
with\,\, probability}\quad p,\\
-Y_k \quad  {\rm with\,\, probability}\quad 1-p.\\
               \end{array}
        \right.  \label{eq:Ydyn}
\end{equation}
In other words, 
\begin{equation}
    Y_{n+1} = \sigma_n Y_{\beta_n},
    \label{eq:Ysigma}
\end{equation}
where $\sigma_n=1$ ($\sigma_n=-1$) with probability $p$  ($1-p$) and $\beta_n$ is drawn from a discrete uniform distribution in $\left\{1,\dots,n\right\}$.

The parameter $p$ is called the {\it memory} parameter of the elephant random walk.  For $p\in [0,3/4)$ a central limit theorem applies, and the elephant random walk is diffusive ($X_{n}\sim \sqrt{n}$), while for $p\in (3/4,1)$ the elephant random walk is superdiffusive ($X_n\sim n^{\alpha}$ with $\alpha>1/2$).    These results can be derived with martingale theory, as we discuss in Chapter~\ref{ch:3}.  

The process $X_n$ is, up to a time-dependent constant, related to a martingale process.   Indeed,  using Eqs.~(\ref{eq:Xdyn}-\ref{eq:Ysigma}) we find that
\begin{eqnarray}
    \langle Y_{n+1} | Y_{[0,n]} \rangle &=& \langle  \sigma_n \rangle \langle Y_{\beta_n}|  Y_{[0,n]}\rangle =    \langle \sigma_n \rangle   \frac{\sum_{k=1}^{n}\langle Y_{k}|  Y_{[0,n]}\rangle}{n} \nonumber \\
    &=&  (2p-1)\frac{\sum_{k=1}^n Y_k}{n} = (2p-1)\frac{X_n}{n},
    \label{eq:Ymart}
\end{eqnarray}
where we have used that $Y_0=0$. From Eqs.~(\ref{eq:Xdyn}-\ref{eq:Ymart}) it follows that the position of the elephant random walker is not a martingale, except for the case $p=1/2$ when the elephant random walk is a simple random walk.  In fact, (\ref{eq:Xdyn}) and (\ref{eq:Ymart}) imply that 
\begin{equation}
    \langle X_{n+1} | Y_{[0,n]} \rangle =  \frac{n+2p-1}{n} X_n = \gamma_n X_n,
    \label{eq:Ymart2}
\end{equation}
where  $\gamma_n =(n+2p-1)/n $. Yet, from this result, we obtain a martingale with multiplicative structure.  Indeed, let us introduce the quantity
\begin{equation}
    a_n \equiv \prod_{k=1}^{n-1} \gamma_k^{-1} = \frac{\Gamma(n+1)\Gamma(2p)}{\Gamma(n+2p)},
    \label{eq:an}
\end{equation}
where $\Gamma$ is the Gamma function; notice that asymptotically, 
\begin{equation}
a_n\sim \Gamma(2p)n^{1-2p}. \label{eq:anAsympto}
\end{equation}
Defining 
\begin{equation}
    M_n \equiv  a_n X_n, \label{eq:dfM}
\end{equation}
we obtain from the definition (\ref{eq:dfM}) and Eqs.~(\ref{eq:Ymart})   that
\begin{equation}
    \langle M_{n+1} | Y_{[0,n]}\rangle = a_{n+1}  \langle X_{n+1} | Y_{[0,n]}\rangle = a_{n+1} \gamma_n X_n = a_{n} X_n = M_n,
\end{equation}
and hence also $\langle M_{n+1}|X_{[0,n]}\rangle = M_n$.  Thus, {according to the one-step-ahead condition  (\ref{eq:defMtower}), }  $M_n$ is a martingale.  Figure~\ref{fig:elephantsShamik} shows  a couple of trajectories drawn from the elephant random walk $X_n$ and their associated martingale process $M_n$ given by Eq.~\eqref{eq:dfM}. As illustrated in Fig.~\ref{fig:elephantsShamik},  martingalization not only reduces the persistence of the elephant random walks, rendering them driftless, but also reduces the amplitude of their fluctuations.

\begin{figure}[H]
\centering
{\includegraphics[width=\textwidth]{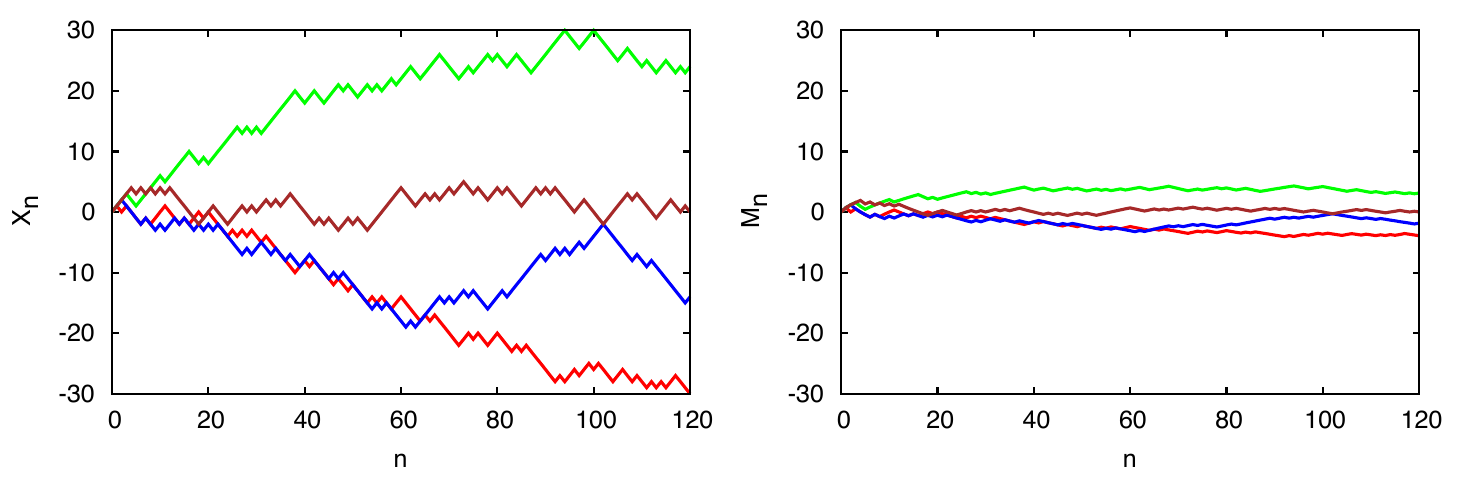}}
 \caption{Left: Example trajectories of the elephant random walk, $X_n$, whose dynamics is given by Eqs.~\eqref{eq:Xdyn} and~\eqref{eq:Ydyn}, as a function of time $n$,  with parameter $p=0.7$. Right: Martingale process, $M_n$, constructed using Eq.~\eqref{eq:dfM}, and associated with the trajectories in the left panel. The trajectories in the right panel are examples of martingales in  a non-Markovian process $X$.  Observe the reduced size of  fluctuations in $M_n$ when compared with $X_n$. {Lines are linear interpolation between the discrete values $X_n$ and serve as a guide to the eye.}   }
\label{fig:elephantsShamik}
\end{figure}

\item { {\bf Run-and-tumble motion:} The run-and-tumble  process  is an example of  a ``false friend" of the martingale.   This process   has  zero average drift, but  nevertheless is {\it not} a martingale.  The position of a one-dimensional run-and-tumble particle with initial position $X_0=0$ may be described as 
\begin{equation}
    X_n = X_{n-1} + \sigma v_{n},
    \label{eq:RnT}
\end{equation}  where the instantaneous normalized velocity $v_{n}=\{-1,1\}$ is a Markovian dichotomous noise process, and $\sigma>0$ the step size. More precisely, the initial value of the normalized velocity is drawn at random $P(\sigma_0=\pm 1)=1/2$, and in the subsequent steps it flips its sign (``tumbles") with probability $q$, i.e., $P(v_n|v_{n-1}) =q \delta_{v_n,-v_{n-1}}+(1-q)\delta_{v_n,v_{n-1}}$ for all $n\geq 1$. 
See Refs.~\cite{tailleur2008statistical,evans2018run,malakar2018steady} for generalizations and extensions.  }
\begin{figure}[H]
\centering
{\includegraphics[width=\textwidth]{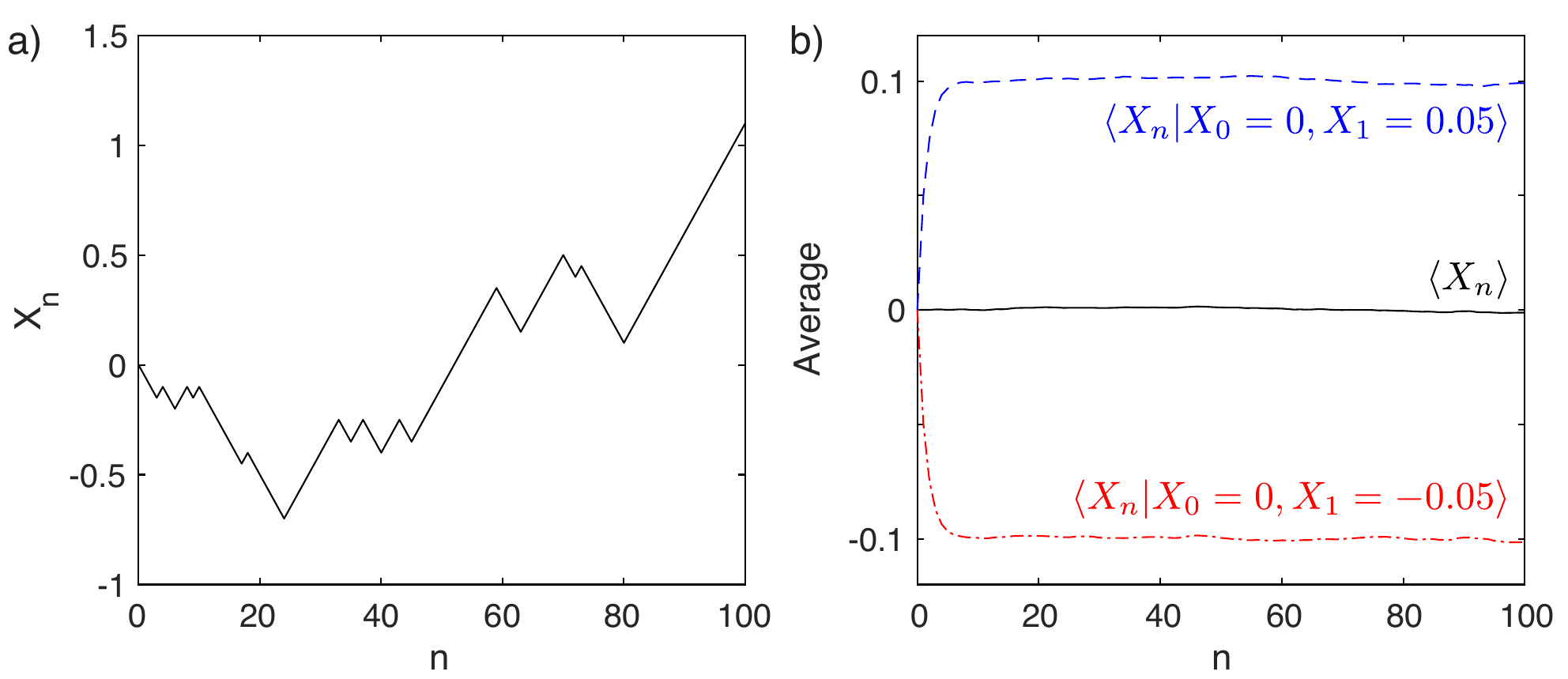}}
 \caption{{Run and tumble motion. a) Example trajectory of the position of a discrete-time run-and-tumble particle described by Eq.~\eqref{eq:RnT}. Lines are linear interpolation between the discrete values $X_n$ and serve as a guide to the eye.  b) Average position as a function time (black solid line), and conditional average of the position over trajectories with a  given history $X_0,X_1$ up to the first jump (blue dashed line, and red dash-dotted line).
 Results are obtained from numerical simulations with parameters: tumble probability $q=1/4$; jump amplitude $\sigma=0.05$, and  averages are done over $10^5$ numerical simulations. }  }
\label{fig:RnT}
\end{figure}
{Figure~\ref{fig:RnT}a shows an example trajectory of the position of a run-and-tumble particle described by Eq.~\eqref{eq:RnT}, which has a zig-zag-like structure. The uncondioned average $\langle X_n\rangle=X_0=0$ vanishes because we have fixed the initial position to $X_0=0$ (black line in Fig.~\ref{fig:RnT}b). On the other hand, the  average of the position conditioned over its history $X_{[0,1]}=X_0,X_1$ up to the first step $n=1$, reveals that $X_n$ is not a martingale. Indeed, $\langle X_n\vert X_0,X_1\rangle $  for $n>1$ is time-dependent for the two possible values of ($X_0= 0,X_1=\sigma$; and $X_0= 0, X_1=-\sigma$), see blue dashed line and red dash-dotted line in Fig.~\ref{fig:RnT}b. Thus we conclude $\langle X_n | X_0,X_1\rangle\neq X_1 $, which implies that $X_n$ is not a martingale.}

%{This example shows that there exist stochastic processes that have zero drift and are not martingales. Furthermore, as we will show in Ch.~\ref{sec:martingaleProblemCont} and~\ref{sec:dynkinCont}, for possible to construct martingales associated with any Markov process. For the run-and-tumble  process, $(X_n,\sigma_n)$ is a two-dimensional Markovian process, from which one could in principle construct a martingale that depends on both $X_n$ and~$\sigma_n$.  } 

\end{itemize}

 \section{Martingales   in continuous time}\label{MartCont}      
  \subsection{Martingales, submartingales, supermartingales} 
   We consider  martingales $M_t$ in continuous time $t\in \mathbb{R}^+$.   Just as for the discrete-time case, martingales in continuous time are processes that have no drift.   

   {Let $M_t$ be a real-valued functional defined on the set of trajectories  of  $\Xot = \left\{X_s\right\}_{s\in[0,t]}$.   In addition, assume that $M_t$ is integrable, i.e., $\langle |M_t|\rangle <\infty$. 
\begin{leftbar}
 We say that a process $M_t$ is a {\bf martingale} with respect to the process $X_t\in \mathcal{X}$ if  $M_t$ has no drift, i.e.,  it holds with probability one that   
    \begin{equation}
    \langle M_t|\Xos\rangle = M_s \label{eq:mart}
    \end{equation}
    for all $0\leq s\leq t$.   
\end{leftbar}}
   
In continuous time the condition $\langle M_t|\Xos\rangle = M_s$ holds with probability one, as  we omit  events that occur with  zero probability.  
 Also, conditional expectations $\langle M_t|X_{[0,s]}\rangle$  in continuous time should be understood as conditional expectations with respect to the filtration generated by $X_{[0,s]}$, see Appendix~\ref{app:probability} for a brief introduction and further references.    
   
   Similarly, we define submartingales (supermartingales) as processes with a nonnegative (nonpositive) drift.
  { We say that $S_t$ is a {\it submartingale} (supermartingale) relative to $X_t$ if it is an integrable stochastic process that has  a nonnegative (nonpositive) drift, i.e., it holds with probability one that
    \begin{equation}
    \langle S_t|\Xos\rangle \geq S_s  \quad (\langle S_t|\Xos\rangle \leq S_s ) \label{eq:submart}
    \end{equation}
    for all $0\leq s\leq t$.}

We  define backward (sub){martingales} by conditioning on a future part of the trajectory, analogously to the discrete time case considered in Sec.~\ref{sec:back}.     
\subsection{Key examples} 
\label{sec:examplescont}

\begin{itemize}
\item {\bf The Brownian motion (Wiener process) $B_t$ }:  The Brownian motion is   a one dimensional stochastic process that satisfies the following four conditions \cite{williams1979diffusions, oksendal2003stochastic}:  
      \begin{itemize}
   \item  $B_0=0$;
   \item  the increments $B_t-B_s$ are normally distributed with mean zero and variance~$|t-s|$;
    \item  for $0\leq t_1<t_2<\ldots <t_n<\infty$ it holds that the increments $B_{t_1}$, $B_{t_2}-B_{t_1}$, $\ldots$, $B_{t_n}-B_{t_{n-1}}$ are independent; 
    \item  the process $B_{t}$ is continuous  with probability one.     
   \end{itemize} 
  Brownian motion $B_t$ is   a paradigmatic physical example of a martingale.  The left panel of  Fig.~\ref{fig:WienerandDoleans} shows a few examples of  Brownian  trajectories. 
\begin{figure}[h!]
\centering
{\includegraphics[width=\textwidth]{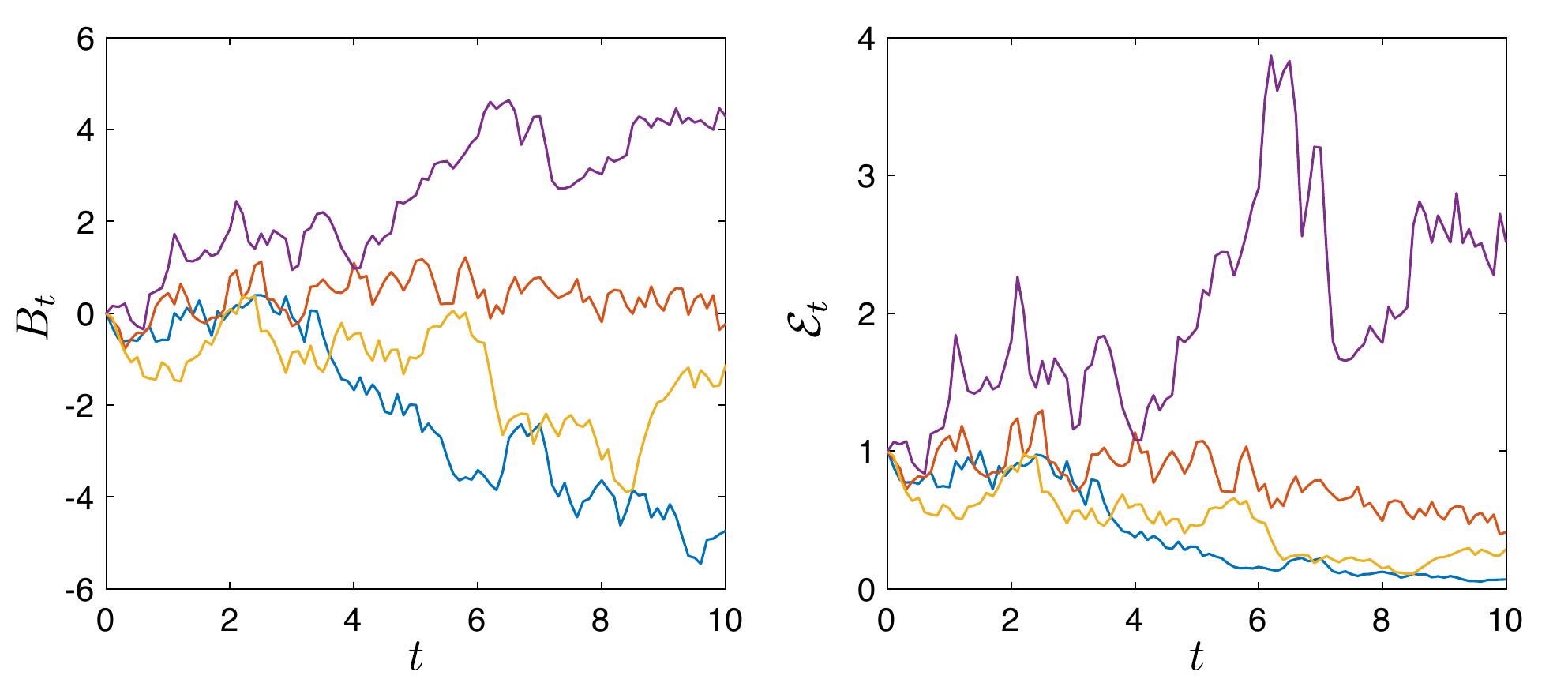}}
\caption{ Example trajectories for two continuous martingales, namely, the one-dimensional Brownian motion $B_t$ (left panel) and the stochastic exponential $\mathcal{E}_t(zB)$ of $zB_t$, as defined by Eq.~\eqref{eq:BExp}, for the parameter $z=0.4$ (right panel).  Notice that $B_t$ can take negative values, whereas $\mathcal{E}_t(zB)$ is a positive martingale.  Trajectories have been generated with   the Euler numerical integration scheme with time discretisation step $0.1$.}
\label{fig:WienerandDoleans}
\end{figure}

\item {\bf Counting processes}:   
Let $N_t\in \mathbb{N}$ be a Poisson process with rate $\lambda$, i.e., $N_t$ denotes the number of ticks in the interval $[0,t]$ of a Poisson point process of constant rate $\lambda$.   Then the process 
\begin{equation}
M_t \equiv N_t - \lambda t \label{eq:countingM}
\end{equation}
is a martingale.   Indeed, since a Poisson process is  Markovian and time-homogeneous, it holds that 
\begin{equation}
\langle M_t|N_{[0,s]}\rangle = \langle N_t|N_{[0,s]}\rangle -\lambda t = \langle N_t|N_s\rangle  -\lambda t  = N_s -\lambda s = M_s .
\end{equation}
 Figure~\ref{fig:poissonmartingale} illustrates four randomly generated trajectories  of both the  counting  (Poisson) process $N_t$  and the corresponding martingale $M_t$, given by Eq.~\eqref{eq:countingM}, as a function of time. 
\begin{figure}[h!]
\centering
{\includegraphics[width=1\textwidth]{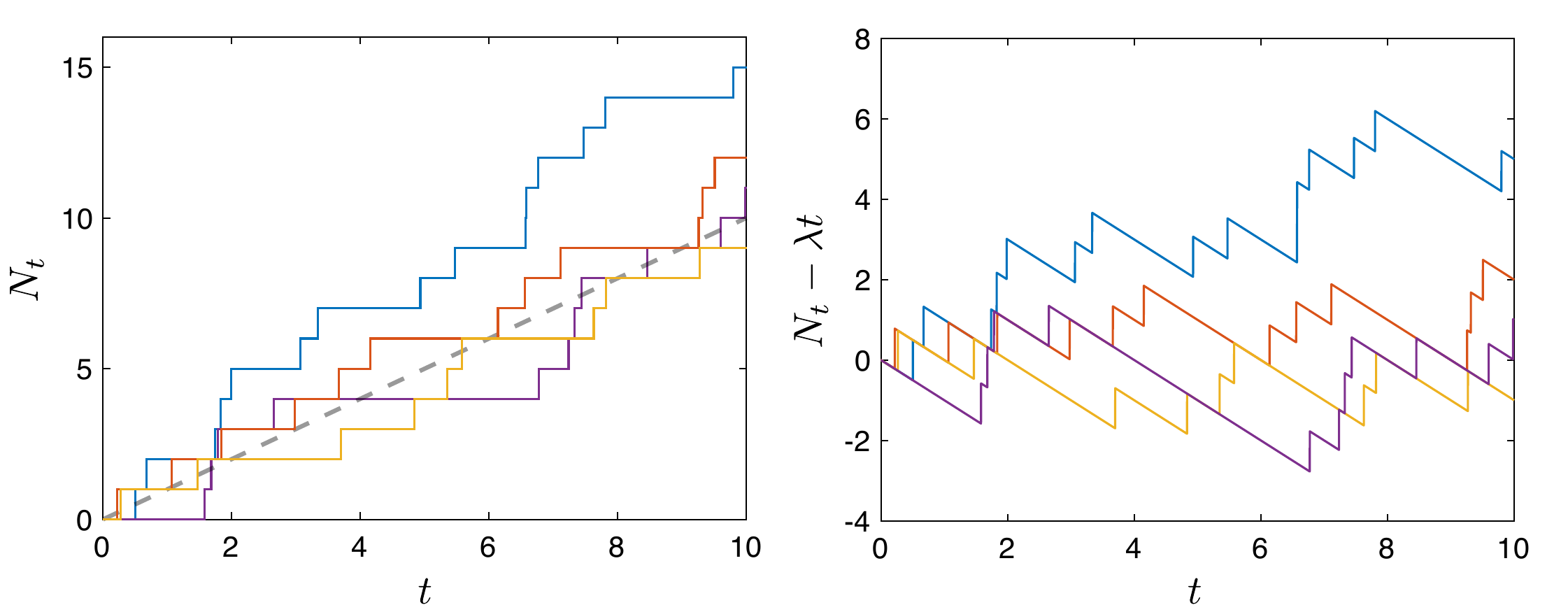}}
\caption{Randomly generated trajectories of a Poisson process $N_t$ with rate parameter $\lambda=1$  (left) and  the corresponding trajectories of the  martingale $N_t-\lambda t$ (right). The black dashed line in the left panel denotes  the deterministic process $\lambda t$.  }
\label{fig:poissonmartingale}
\end{figure}

 \item   {\bf Radon-Nikodym density processes (a.k.a. path probability ratios)}:  We consider an extension of the probability ratio (\ref{eq:RForm})  that applies to processes in continuous time.   These processes are martingales and  are important for the applications discussed in  this review.    
 
 Let $X_t$ be a stochastic process whose statistics are described by one of the two probability measures $\mP$ or $\mQ$.   Probability measures are functions that assign probabilities to measurable sets $\Phi$ of trajectories through~\cite{tao2011introduction}  
 \begin{equation}
 \mathcal{P}\left(\Phi\right)  \equiv \langle \mathbf{1}_{\Phi}(\Xot)\rangle_\mP  , 
\end{equation}
  where 
\begin{eqnarray}
{\bf 1}_{\Phi} \left(X_{[0,t]}\right) \equiv \left\{\begin{array}{ccc} 1, & {\rm if}& X_{[0,t]}\in \Phi, \\ 0, & {\rm if}& X_{[0,t]}\notin \Phi ,  \end{array}\right.  \label{eq:indicator}
\end{eqnarray}
 is the indicator function that equals $1$ when $\Xot\in\Phi$    and equals zero otherwise.     
 
  We define a density process ${R_{t} \equiv R(\Xot)\in \mathbb{R}^+}$  such that  
\begin{eqnarray}
 \langle   f(\Xot) \rangle_{\mQ} = \langle  { f(\Xot) R_{t} }\rangle_{\mP}, \label{eq:radonDef2}
\end{eqnarray}
holds for all nonnegative, measurable functions $f$.
We denote the process $R_t$ by 
 \begin{eqnarray}
 R_{t} = \frac{\mQ(\Xot)}{\mP(\Xot)}, \label{eq:radon}
 \end{eqnarray}
and call it the {\bf Radon-Nikodym density process} (a.k.a.~{\bf path probability ratio}) of $\mQ$ with respect to $\mP$, as $R_t$ plays the role of the  density of         $\mQ$ with respect of $\mP$.  
Note that in Eq.~(\ref{eq:radon}) the numerator $\mQ(\Xot)$  does not exist separately from the denominator  $\mP(\Xot)$, which distinguishes probability ratios in discrete time, as defined by Eq.~(\ref{eq:RForm}), from those in continuous time.

According to the Radon-Nikodym theorem~\cite{lip2013}, the process $R$ exists as long as $\mQ$ is locally, absolutely continuous with respect to  $\mP$, which means that  
\begin{equation}
\mP(\Phi) = 0 \Rightarrow       \mQ(\Phi) = 0 
\end{equation}
for all measurable sets $\Phi$ defined on the set of  trajectories $\Xot$ and for finite $t$. Provided the absolute continuity conditions are satisfied, the process $R_t$ given by Eq.~\eqref{eq:radon} is a martingale with respect to $\mP$.

An alternative way to represent probability measures $\mathcal{P}$ is through the Onsager-Machlup method, see, e.g.,~Refs.~\cite{Onsager1,Onsager2, cugliandolo2017rules}.   In this approach, we consider a family of equivalent probability measures $\mathscr{P}$ that are mutually absolutely continuous, i.e., if $\mathcal{P},\mathcal{Q}\in \mathscr{P}$ then \begin{equation}
\mP(\Phi) = 0 \Leftrightarrow   \mQ(\Phi) = 0 .
\end{equation}
The probability measures  in this family, e.g., $\mathcal{P}$ and $\mathcal{Q}$, can  be represented as 
\begin{equation}
    \mathcal{P}\left(x_{[0,t]}\right) = \mathcal{N}^{-1}\exp\left(-\mathcal{A}_{\mathcal{P}}(x_{[0,t]})\right), \label{eq:onsagerMachlupApproach1}
\end{equation}
and 
\begin{equation}
    \mathcal{Q}\left(x_{[0,t]}\right) =\mathcal{N}^{-1}\exp\left(-\mathcal{A}_{\mQ}(x_{[0,t]})\right), \label{eq:onsagerMachlupApproach2}
\end{equation}
where $\mathcal{A}_{\mathcal{P}}$ and $\mathcal{A}_{\mathcal{Q}}$ are functionals  (often called "action" functionals) defined on the trajectories of the process, and $\mathcal{N}$ is a common prefactor.   Even though the prefactor $\mathcal{N}$ is ill-defined, the Onsager-Machlup representation is convenient as we   obtain  Radon-Nikodym density processes between any two probability measures in the equivalence class $\mathscr{P}$  from  ratios
\begin{equation}
 R_t  =     \frac{\mathcal{Q}\left(X_{[0,t]}\right)}{\mathcal{P}\left(X_{[0,t]}\right)} = \exp\left( -\mathcal{A}_{\mathcal{Q}}(X_{[0,t]})+\mathcal{A}_{\mathcal{P}}(X_{[0,t]})\right).  \label{eq:onsagerMachlupApproach3}
\end{equation} 
In other words, the Onsager-Machlup representation 
 allows us to represent the numerator and denominator of $R_t$ independently in terms of the so-called actions $\mathcal{A}_{\mathcal{Q}}$ and $\mathcal{A}_P$.

 As suggested before, often we will use the physics' slang {\bf path probability} for $\mathcal{P}\left(x_{[0,t]}\right)$ and {\bf path probability ratio} for $R_t$, even though $\mathcal{P}$ is not really a probability, but rather a representation of the measure $\mathcal{P}$ in terms of the action.

\item The {\bf stochastic exponential of $z B_{t}$}:   The exponential
\begin{eqnarray}
\mathcal{E}_t(zB) \equiv \exp\left(z B_{t} - \frac{1}{2}z^2 t\right) \label{eq:BExp}
\end{eqnarray}
is a martingale for all $z\in \mathbb{R}$, as shown in Appendix~\ref{appendix:A}; see  the right  panel of Fig.~\ref{fig:WienerandDoleans}  for an illustration of trajectories of $\mathcal{E}_t(zB)$.     {Note that (\ref{eq:BExp}) can be obtained from the continuous-time limit  of the martingale (\ref{eq:stochExpExample1}) for $q=1/2$ by making the substitutions $n = t/\Delta t $, $ X_n = B_t/\sqrt{\Delta t}$,  and    $y=z\sqrt{\Delta t}$, and by subsequently  taking the limit $\Delta t\rightarrow 0$.}

Expanding the exponential (\ref{eq:BExp}) around $z=0$, we obtain \cite{williams1979diffusions}
\begin{eqnarray}
\exp\left(z x - \frac{1}{2}z^2 t\right) = \sum^{\infty}_{n\geq 0}\frac{z^n}{n!} H_n(t,x) .\label{eq:herm}
\end{eqnarray}
Setting $x = B_{t}$ in Eq.~\eqref{eq:herm}, it follows from the martingale property of the exponential (\ref{eq:BExp}) that the  functions 
$H_n(t,B_{t})$ are  martingales for all $n\in \mathbb{N}$.   Thus, the processes,
 \begin{eqnarray}
 H_1(t,B_{t}) &=& B_{t},\label{trieste1} \\ 
 H_2(t,B_{t}) &=& B_{t}^2-t, \label{trieste2}\\ 
 H_3(t,B_{t}) &=& B_{t}^3 - 3tB_{t}, \label{trieste3}
 \end{eqnarray}
and so forth, are martingales. 
\item The {\bf It\^{o} integral}:  Let $Z_t = Z(B_{[0,t]})$ be a function defined on the space of trajectories of the Brownian motion.  Let {$P=[ t_1<t_2<\ldots<t_{n}]$, with $t_1=0$ and $t_n=t$,} be a finite partition  of the interval $[0,t]$, and define its norm $\|P\|$ be given by the maximum spacing  $t_i-t_{i-1}$ between two consecutive values.  The It\^{o} integral is defined by the limit~\cite{oksendal2003stochastic, protter2005stochastic}
\begin{equation}
   I_t =  \int^{t}_{0} Z_s dB_s = \lim_{\|P\|\rightarrow0}\sum^{n-1}_{i=0}Z_{t_i} \left(B_{t_{i+1}}-B_{t_i}\right), \label{eq:Ito}
\end{equation} 
where the convergence should be understood in probability.  
The  Brownian motion is recovered as  the special case when the diffusion coefficient  $Z_t$  is  constant.      We can also express It\^{o} integrals as stochastic differential equations, i.e., 
\begin{equation}
    \frac{dI_t}{dt} =  Z_t\frac{dB_t}{dt},
\end{equation}
or even more briefly as  
\begin{equation}
    \dot{I}_t = Z_t \dot{B}_t,   
    \label{eq:izaakmartingality}
\end{equation}
where the dot represents a derivative towards time. 
It\^{o} integrals of the form~\eqref{eq:Ito} are  martingales when~\cite{chung1990introduction, oksendal2003stochastic}
\begin{equation}
 \int^{t}_0 \langle Z^2_s \rangle {\rm d}s <\infty. \label{eq:AverageD}
\end{equation}  
Consequently, one has 
\begin{equation}
    \langle I_t\rangle= 0. 
\end{equation}
However, there exist It\^{o} integrals that are not martingales and this leads to the concept of a local martingale, which we introduce later in this review.

Martingales play an important role in the theory of stochastic integration.  In fact, 
It\^{o} integrals also  exist when the integrator   is a   martingale \cite{chung1990introduction}, viz., 
\begin{equation}
   I_t =  \int^{t}_{0} Z_s dM_s \equiv \lim_{\|P\|\rightarrow0}\sum^{n-1}_{i=0}Z_{t_i} \left(M_{t_{i+1}}-M_{t_i}\right) \label{eq:Ito2}
\end{equation}  
where $M_s$ is now a martingale process, not necessarily Brownian motion $B_s$.   Notice that the It\^{o} integral is the continuous-time version of the martingale transform~(\ref{eq:transform}). 
    The integral given by Eq.~(\ref{eq:Ito}) is  a special case of Eq.~(\ref{eq:Ito2}) for an integrator that is a Brownian motion.   In fact, the {\it martingale representation theorem} states that   square integrable, continuous martingales can  be written as  It\^{o} integrals for which the integrator $M_s$ is a Brownian motion \cite{oksendal2003stochastic}, and hence the  generic form of the It\^{o} integral Eq.~(\ref{eq:Ito2}) is mainly relevant for  martingales that admit jumps.     The requisite for martingality~\eqref{eq:AverageD} for the special case of an It\^{o} integral with respect to the Brownian motion, reads  for the generic It\^{o} integral Eq.~(\ref{eq:Ito2})  as
\begin{equation}
\int^{t}_0 \langle Z^2_s \rangle\: {\rm d}[M_s,M_s] <\infty,
\end{equation}
where $[M_s]$ is the quadratic variation process, defined by 
\begin{equation}
[M_t,M_t] \equiv \lim_{\|P\|\rightarrow0}\sum^{n-1}_{i=0} \left(M_{t_i}-M_{t_{i-1}}\right)^2. \label{eq:quadrVar}
\end{equation}

For illustration purposes, let us consider two canonical examples of It\^{o} integrals.   When  the integrator $  M_t = B_t  $ is a Brownian motion, then
\begin{equation}
 [M_t,M_t] = t. 
\end{equation}
Note that  the quadratic variation can be obtained informally by using  the notation $d[Z,Z]_s = \left(dZ_s\right)^2$  and  the rules of It\^{o} calculus. 
\begin{leftbar}  {\bf Rules of It\^{o} calculus}  (see Appendix~\ref{app:ItoFormula}):
\begin{equation}
    \left( dB_t\right)^2 = dt,\quad  dB_tdt = 0,\quad \text{and}\quad (dt)^2 = 0. \label{eq:Itosanrules}
\end{equation}  
\end{leftbar}
A second canonical example of an integrator is a shifted  Poisson process of rate $\lambda$, i.e., $  M_t=N_t-\lambda t  $, for which 
 \begin{equation}
 [M_t,M_t] = N_t. \label{eq:martingalePoisson2}
\end{equation}
Equation~(\ref{eq:martingalePoisson2}) follows from taken the limit  $\|P\|\rightarrow0$ in the right-hand side of Eq.~(\ref{eq:quadrVar}), leading to a  sum of three kind of terms of the form $(M_{t_i}-M_{t_{i-1}})^2$; (i)  there are no jumps between  $t_i$ and $t_{i-1}$, in which case $(M_{t_i}-M_{t_{i-1}})\rightarrow 0$  when $\|P\|\rightarrow0$; (ii)  there is exactly one  jump between  $t_i$ and $t_{i-1}$, in which case $(M_{t_i}-M_{t_{i-1}})\rightarrow 1$  when $\|P\|\rightarrow0$; (iii)
there are multiple jumps between $t_i$ and $t_{i-1}$, in which case  $(M_{t_i}-M_{t_{i-1}})^2$ converges to a nontrivial limit.      However, the number of such terms converges to zero when  $\|P\|\rightarrow0$.

\item {\bf It\^{o} process} with nonnegative drift:  The stochastic differential equation 
\begin{equation}
    \dot{J}_t = b_t +  \sqrt{2D_t} \dot{B}_t,
\end{equation}
where $b_t \equiv b_t(X_{[0,t]})\geq 0$  is a drift term and $D_t  \equiv D_t(X_{[0,t]})\geq 0$ satisfying  $\int^t_0 D^2_u du <\infty$, is solved by 
\begin{equation}
   J_t  = \int^{t}_0 b_u du + \int^{t}_0   \sqrt{2D_u} d{B}_u. 
\end{equation}
The process $J_t$ is a submartingale when   $b_t\geq 0$.  

\item The {\bf multidimensional It\^{o} integral}: 
the multidimensional It\^{o} integral $I_t$ solves
\begin{eqnarray}
\dot{I}_t 
\equiv   \sum^{d}_{a=1}Z_{a,t}  \dot{B}_{a,t} \label{eq:IMulti}
\end{eqnarray}   
where $B_{a,t}$ with $a=1,2,\ldots, d$ are a set of $d$ independent Brownian motions and $Z_{a,t} \equiv Z_{a,t}(B_{1,[0,t]}, B_{2,[0,t]},\ldots, B_{d,[0,t]})$, is a martingale if 
 \begin{eqnarray}
 \sum^{d}_{a=1} \Big\langle  \int^{t}_{0} Z^2_{a,s} ds \Big\rangle  < \infty.   \label{eq:condMulti} 
 \end{eqnarray}

\item The Dol\'{e}ans-Dade {\bf stochastic exponential} of an It\^{o} integral $I_t$:      Let $X_t$ be a possibly high dimensional It\^{o} process, and   let $S_t\in \mathbb{R}$ be an It\^{o} process that solves 
\begin{equation}
\dot{S}_t = D_t + \sqrt{2 D_t } \dot{B}_t ,  \label{eq:St}
\end{equation}
where $D_t \equiv  D(\Xot)$ is a functional     defined on the trajectories of $X$ and $B_t$ is a Brownian motion process that may be correlated with $X$.  
Applying It\^{o}'s formula for the variable change $S\to \exp(-S)$, see Eq.~(\ref{eq:itoFormula})  in Appendix~\ref{app:ItoFormula} and below in Eq.~\eqref{eq:itoFormula0} for the one-dimensional case, we obtain 
\begin{equation}
\frac{d}{dt}\exp(-S_t) =  -\exp(-S_t)  (  \dot{S}_t -   D_t ) = -\exp(-S_t) \sqrt{2 D_t }\dot{B}_t, \label{eq:expst}
\end{equation}
and hence $\exp(-S_t)$ is an It\^{o} integral.  If we identify  in the above equation  the It\^{o} integral
\begin{equation}
\dot{I}_t \equiv -\sqrt{2 D_t }\dot{B}_t,
\end{equation}
then Eq.~(\ref{eq:expst}) reads 
\begin{equation}
\frac{d}{dt}\exp(-S_t) =  \exp(-S_t) \dot{I}_t.   \label{eq:expst2}
\end{equation}
We call the solution to an equation of the form (\ref{eq:expst2}) the {\bf Dol\'{e}ans-Dade stochastic exponential} of $I_t$, and we denote it by $\mathcal{E}_t(I)=\exp(-S_t)$.    
Stochastic exponentials  play an important role in stochastic thermodynamics and quantitative finance, as we will see in  Ch.~\ref{sec:martST1}, and  Ch.~\ref{chapter:finance}, respectively.

\item  {{\bf Position of a tagged particle in the symmetric exclusion process}:   We present  an example  in continuous time of a ``false friend"  of the martingale, i.e., a process with   zero average drift that is  {\it not} a martingale.      Consider   the position of a tagged particle in the symmetric exclusion process (SEP) on $\mathbb{Z}$~\cite{Mallick}.      This  is a continuous-time  random walk of a particle that moves in  a crowded environment.   }
\begin{figure}[h!]
\centering
{\includegraphics[width=0.5\textwidth]{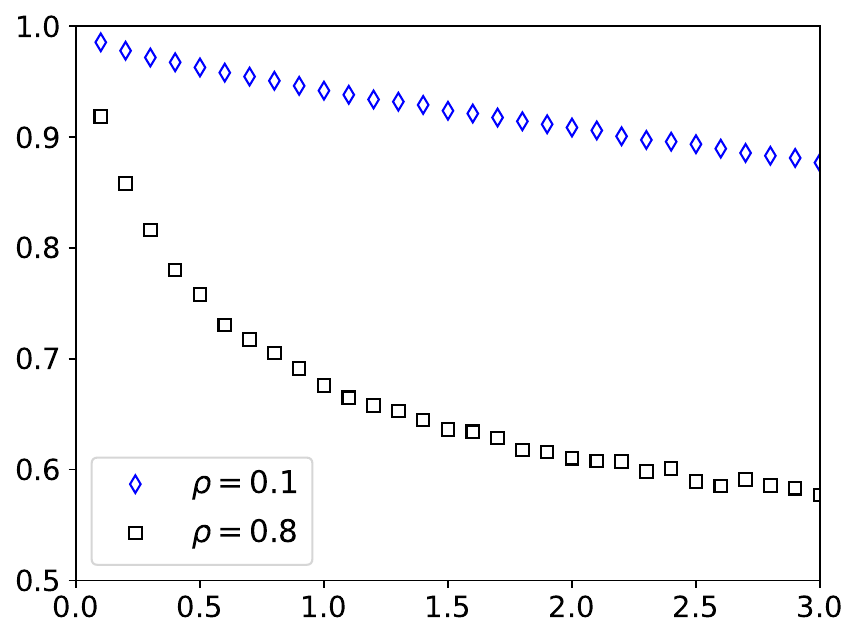}}
  \put(-320,80){$\langle X_t|X_0,X_{0^+}=1\rangle$}
  \put(-110,-10){$t$}
\caption{{\it The position of a tagged particle in the symmetric exclusion process on $\mathbb{Z}$ is not a martingale.}  The average position  $\langle X_t| X_0=0, X_{0^+} = 1\rangle$ as a function of $t$, conditioned on the event that $X_t$ makes a jump to the right at time $t=0$, in the symmetric exclusion process on $\mathbb{Z}$.  The total particle occupation probability  $\rho$ is given in the legend.  Results are empirical means from repeated simulations.}
\label{fig:bias}
\end{figure}

{
  In the initial configuration, each site of $\mathbb{Z}$ is  occupied  with probability~$\rho$ by  a particle, and it is empty with a probability~$1-\rho$.       Subsequently, each particle moves at a rate $1/2$  to its right, or with a rate $1/2$ to its left neighbour.   If the neighbouring site is occupied by a particle, then the jump is blocked and the particle stays in its original position.  }
  
  {Interestingly, although the particle position $X_t$  of a tagged particle is on average driftless,  it is not a martingale.   }
{
Indeed, in Fig.~\ref{fig:bias} we plot $\langle X_t| X_0=0, X_{0^+} = 1\rangle $ as a function of time.  If it were a martingale, then one would have $\langle X_t| X_0=0, X_{0^+} = 1\rangle =1$, independent of $t$.   Note that this is indeed approximately the case for small $\rho$, but for large enough $\rho$, there is a clear drift towards the left, as the particle leaves a hole in its trail wen jumping to the right at time $t=0$.    
 }

  \end{itemize}

 \subsection{On stochastic calculus{: It\^{o}, Stratonovich, and beyond} }
 \label{sec:stochasticcalculus}
 In Sec.~\ref{sec:examplescont} we have reviewed the prominent role of It\^{o} integrals in martingale theory. 
  In physics it is  often common to use the  Stratonovich integral as defined in the books~\cite{gardiner1985handbook,risken1996fokker,pavliotis2014stochastic} and the original references~\cite{fisk1963quasi,stratonovich1968conditional}
  \begin{equation}
   S_t =  \int^{t}_{0} Z_s\circ  dB_s = \lim_{\|P\|\rightarrow0}\sum^{n-1}_{i=0} \left(\frac{Z_{t_i}+ Z_{t_{i+1}}}{2}\right) \left(B_{t_{i+1}}-B_{t_i}\right),  \label{eq:Strat}
\end{equation}
{where we recall that the limit $\|P\|\rightarrow0$ means the limit of small norm $\|P\|$ of  a finite partition $P= [0=t_1<t_2<\ldots<t_{n}=t]$ of the interval $[0,t]$. }
%The Stratonovich-Fisk convention~\eqref{eq:Strat} has an analogous structure to Riemann's midpoint (trapezoidal) rule in calculus.
% Therefore, } 
The  Stratonovich-Fisk convention has the advantage that it  allows  us to use the standard rules of differential calculus, {e.g., the chain rule for derivatives and  the fundamental theorem of calculus, } see Appendix~\ref{app:Ito-lemma}. However, the Stratonovich integral  has the inconvenience of  {\em not} being a martingale as it contains a spurious drift term, see Appendix~\ref{app:strat} for details. 
{We  recall readers the definition of the It\^{o}  integral given by Eq.~\eqref{eq:Ito}, copied here for convenience,
\begin{equation}
   I_t =  \int^{t}_{0} Z_s dB_s = \lim_{\|P\|\rightarrow0}\sum^{n-1}_{i=0}Z_{t_i} \left(B_{t_{i+1}}-B_{t_i}\right), \label{eq:Ito-copy}
\end{equation} 
which differs to the Stratonovich convention on the time point at which the process in the integrand $Z$ is evaluated. The fact that in It\^{o} convention the summation rule is done by evaluating $Z$ at the beginning of each interval of the partition is crucial for  It\^{o} processes of the type~\eqref{eq:Ito-copy} to be martingales.
}

More generally, we  define the $\alpha-$discretization convention,
with $0\leq\alpha\leq1$, via the infinitesimal rules~\footnote{See~\cite{hottovy2012noise} for generalizations where $\alpha\in\mathbb{R}$ and even space dependent.} 
  \begin{equation}
   Y_t =  \int^{t}_{0} Z_s\circ_{\alpha} dB_s = \lim_{\|P\|\rightarrow0}\sum^{n-1}_{i=0} \left((1-\alpha)Z_{t_i}+\alpha Z_{t_{i+1}}\right) \left(B_{t_{i+1}}-B_{t_i}\right).  \label{eq:deltadis}
\end{equation} 
%\begin{equation}
%Z_t\circ_{\alpha}\dot{B}_{t}\equiv g_{t}\left(X_{t+\alpha dt}\right)\left(\frac{B_{t+dt}-B_{t}}{dt}\right).
%\end{equation}
Apart for the Ito corresponding   to $\alpha=0$, and the Stratonovich-Fisk corresponding to  $\alpha=1/2$,  
another $\alpha-$discretization scheme that is widely used in the literature is the anti-It\^o convention, corresponding to $\alpha=1$.   However, only in the case of $\alpha=0$  stochastic integrals are martingales.

In the case of It\^{o} convention, we will  omit the $\circ_0$ symbol throughout the {Treatise}. In the following, we use the symbol $\circ$ to denote the Stratonovich-Fisk convention.  {As we will show in the subsequent chapters, in physics (e.g. stochastic thermodynamics) it is customary to consider stochastic It\^{o} (Stratonovich) integrals  of the type $\int_0^t Z_s  dX_s$ ($\int_0^t Z_s \circ dX_s$) for $Z_s=Z[X_{[0,t]}]$ a functional of the trajectory $X_{[0,t]}$.  }

\subsubsection{It\^{o}'s formula}
\label{sec:itoslemma}
A useful result in It\^{o}'s calculus regards the change of variables, see Appendix~\ref{app:ItoFormula} for details.
Let $X_t\in \mathbb{R}$ be a stochastic process that solves a one-dimensional It\^{o} stochastic  differential equation of the form 
\begin{equation}
    \dot{X}_t = b_t(X_{[0,t]}) + \sigma_t(X_{[0,t]})\dot{B}_t, \label{eq:XDot0}
\end{equation}
where $B_t$ is the one-dimensional Brownian motion (see Sec.~\ref{sec:examplescont}), and where $b_t$ and $\sigma_t$ satisfy suitable  integrability conditions (see Appendix~\ref{app:ItoFormula}). 

\begin{leftbar}
Let $g_t(x)$ be a twice continously differentiable function in $t\in \mathbb{R}^+$ and $x\in\mathbb{R}$ that may depend explicitly on time $t$. Then the process 
\begin{equation}
    Y_t = g_t(X_t),
\end{equation}
with $X_t$ described by the It\^{o} stochastic differential equation~\eqref{eq:XDot0},
solves the stochastic differential equation \cite{oksendal2003stochastic}
\begin{equation}
    \dot{Y}_t = \left(\frac{\partial g_t}{\partial t}\right)(X_t) +  \left[\left(\frac{\partial g_t}{\partial x}\right)(X_t)\right]\dot{X}_t +
    \left[\frac{1}{2} \left(\frac{\partial^2 g_t}{\partial x^2}\right)(X_t)  \right]
    \sigma_t^2(X_{[0,t]})  , \label{eq:itoFormula0}
\end{equation}
which is  known as {\bf It\^{o}'s lemma} (or {\bf It\^{o}'s formula}). 
\end{leftbar}
It\^{o}'s formula may be understood from a Taylor expansion of $g_{t+dt}(X_{t+dt})$, viz.,
 \begin{eqnarray}
g_{t+dt}(X_{t+dt})-g_t(X_t)&=&\left[\left(\frac{\partial g_t}{\partial t}\right){(X_t)} \right]dt+\left[\left(\frac{\partial
g}{\partial x}\right){(X_t)}  \right]dX_t+\left[\frac{1}{2}\left(\frac{\partial^2 g_t}{\partial t^2}\right){(X_t)}\right] (dt)^2\nonumber\\
&+&\left[\frac{1}{2}\left(\frac{\partial^2 g_t}{\partial x^2}\right){(X_t)}\right] (d X_t)^2+\left[\frac{1}{2}\left(\frac{\partial^2 g_t}{\partial t \partial x} \right){(X_t)}\right]  dt 
dX_t+\ldots
\label{appeq:dF}
\end{eqnarray}
Using $dX_t = \dot{X}_tdt$,  the rules of It\^{o} calculus [Eqs.~\eqref{eq:Itosanrules}], 
and neglecting contributions of orders higher than $dt$, we get Eq.~\eqref{eq:itoFormula0}.
Similarly, one can show that if instead one has a Stratonovich stochastic differential equation
\begin{equation}
    \dot{X}_t = b_t(X_{[0,t]}) + \sigma_t(X_{[0,t]})\circ\dot{B}_t, \label{eq:XDot0s}
\end{equation}
the process $Y_t = g_t(X_t)$ obeys the standard ``chain rule"
\begin{equation}
    {\dot{Y}_t = \left(\frac{\partial g_t}{\partial t}\right)(X_t) +  \left[\left(\frac{\partial g_t}{\partial x}\right)(X_t)\right]\circ\dot{X}_t  .} \label{eq:itoFormula0s}
\end{equation}
{Indeed, this follows from using }
{
\begin{eqnarray}
\left[\left(\frac{\partial
g}{\partial x}\right){(X_t)} \right] dX_t +  \left[\frac{1}{2}\left(\frac{\partial^2 g_t}{\partial x^2}\right){(X_t)}\right] (
d X_t)^2 &=& \frac{1}{2}\left[\left(\frac{\partial g_t}{\partial x}\right)(X_t) +\left(\frac{\partial g_t}{\partial x}\right)(X_{t+dt}) \right]dX_t 
  \nonumber\\ 
&=& \left(\frac{\partial g_t}{\partial x}\right)(X_t) \circ dX_t.
\end{eqnarray}
}
We refer readers to Appendix~\ref{app:Ito-lemma} for further details, generalizations and extensions to, e.g., $d>1$ dimensions. 

{\subsubsection{From It\^{o} to Stratonovich and back}As we will show in the subsequent chapters, in statistical physics it is important to convert  It\^{o} integrals of the type $\int_0^t g_s(X_s)  dX_s$ into Stratonovich integrals of the type $\int_0^t g_s(X_s)\circ dX_s$, and vice versa. The theorem below provides a rigorous answer for such  conversions  in the case when $X_t$ is a one-dimensional stochastic process and $g_t(x)$ a smooth function. 
\begin{leftbar}
    \begin{theorem}[{\bf Conversion from Stratonovich to It\^{o} integrals in one dimension}] \label{Strato2Ito} 
  Let $X_t\in \mathbb{R}$ be the solution of the It\^{o} stochastic differential equation
\begin{equation}
    \dot{X}_t = b_t(X_t) + \sigma_t(X_t)\dot{B}_t, \label{eq:XDot0-2}
\end{equation}
with $B_t$ the one-dimensional Brownian motion and  $b_t$ and $\sigma_t$ two functions satisfying suitable  integrability conditions (see Appendix~\ref{app:ItoFormula}). 
 Then the following identity between the Stratonovich and It\^{o} products holds
 \begin{equation}
g_t(X_{t})\circ d{X}_{t} =g_t(X_{t})d{X}_{t}+\frac{\sigma_t^2(X_{t})}{2}\left[ \left(\frac{\partial g_t}{\partial x}\right)(X_t)\right] dt,\label{eq:stratotoito}
\end{equation}
which is valid for any function $g_t(x)$ that may depend explicitly on time and is continously differentiable function in $t\in \mathbb{R}^+$ and $x\in\mathbb{R}$. 
    %$\\$ [{\bf IN comment: If $S_n$ can have negatie values, then  $\mathbb{P}\left[{\rm sup}_{n'\leq n}S_{n'}  \geq \lambda\right] \leq  \frac{\langle {\rm  max}\left\{S_n,0\right\} \rangle}{\lambda}$, which is the formula used in Figure 3.1. Should we present this formula here?  It appears to be more useful when applying to random walks.}]
    \label{th:0}
\end{theorem}
\end{leftbar}
The relation~\eqref{eq:stratotoito} implies that under the assumptions of Theorem~\ref{Strato2Ito},  one has the following rule to convert a Stratonovich integral into an It\^{o} integral:
 \begin{equation}
\int_{0}^{t}g_s(X_{s})\circ\dot{X}_{s}ds  =\int_{0}^{t}g_s(X_{s})\dot{X}_{s}ds+\int_{0}^{t}\frac{\sigma_s^2(X_{s})}{2}\left[(\partial_x g_s)(X_{s})\right]ds,\label{eq:stratotoito2}
\end{equation}
which holds for any $t\geq 0$. Equations~\eqref{eq:stratotoito} and~\eqref{eq:stratotoito} can be generalized to e.g. processes $X_t$ following $(d>1)-$dimensional stochastic differential equations, see Eq.~\eqref{eq:StratotoItoLang} and  Appendix~\ref{app:strat}.}

\chapter{Martingales and Markov processes} \label{ch:2b}

\epigraph{Time, dear friend, time brings round opportunity; opportunity is the martingale of man. The more we have ventured the more we gain, when we know how to wait.   
\\ \vspace{0.5cm} {\em The three musketeers}, A.~Dumas (1844).}

 As discussed in Chapter~\ref{ch:2a},  not all  martingales are defined in Markov processes.   Nevertheless, in this Chapter we focus on martingales associated with Markov processes, as they play a central role in physics.   In fact, most mesoscopic, physical processes, whether they are  an object in a fluid,  transport processes, or chemical reactions, are described by Markov processes.

This chapter is organised into two main parts.   Section~\ref{Sec:markov} is devoted to    martingales in discrete time Markov processes, and  Sec.~\ref{sec:markovCont} reviews the theory of martingales in  continuous-time Markov processes.
%Much of the theory for discrete time martingales carries over to continuous time, but there are also some important differences between both cases, and we will focus on those in Sec.~\ref{MartCont}. 

 \section{Markov processes and martingales in discrete time} \label{Sec:markov}   
We review the theory of discrete-time martingales $M_n$ defined with respect to Markov chain $X_n$.   First,  in Sec.~\ref{sec:defMarkovDiscrete} we revisit   the definition of Markov chains. 
Next, in Sec.~\ref{sec:martingaleProblem}, we consider   the martingale problem, which is one of the central results in the theory of Markov processes.   Subsequently,  we consider important examples of martingales in Markov processes.   In Sec.~\ref{sec:Dynkin}, we define Dynkin's martingales (also referred to L\'{e}vy's Martingales~\cite{bremaud2013markov}).  Then, in  Sec.~\ref{sec:Multiplicative} we define multiplicative martingales, which are  simple examples of martingales that are not Dynkin's martingales.   In Sec.~\ref{subsec:Ratios}, we consider  martingales that are  ratios of path probability densities of Markov chains, which  play a prominent   role in physics, in particular, in nonequilibrium thermodynamics (see Ch.~\ref{sec:martST1}-\ref{TreeSL}).

\subsection{Definition of Markov chains}\label{sec:defMarkovDiscrete}

\begin{leftbar} 

A {\bf discrete-time  Markov chain} is  a stochastic process such that its future values conditioned on its current value are statistically independent of its past values.  For processes on  discrete state space $\mathcal{X}$ this implies 
\begin{equation}
\mathcal{P}\left(X_n=x_n|X_{[0,n-1]}\right) = \mathcal{P}\left(X_n=x_n|X_{n-1}\right),
\label{eq:2.34}
\end{equation}
which motivates us to introduce the  {\bf transition matrix}   of a 
 time-homogeneous discrete-time  Markov chain  $X_n$ in discrete state space $\mathcal{X}$  as 
 \begin{eqnarray}
 w(x,y) \equiv \mathcal{P}\left(X_{n}=y|X_{n-1}=x\right),\quad \forall x,y\in\mathcal{X}.
 \label{eq:transmat}
 \end{eqnarray}
For processes in  continuous state space $\mathcal{X}$, we define their transition matrix as 
\begin{eqnarray}
 w(x,y)  \equiv \frac{\mathcal{P}\left(X_{n}\in\left[y,y+dy\right]|X_{n-1}=x\right)}{dy},\quad \forall x,y\in\mathcal{X}.
 \label{eq:transmat2}
 \end{eqnarray} 
 Equations~(\ref{eq:2.34})-(\ref{eq:transmat2}) imply that the probability (density)  for a sequence $x_{[0,n]}=(x_0,x_1,\dots, x_n)$ to occur in the discrete-time Markov chain is  given by
 \begin{eqnarray}
 \mathcal{P}(x_{[0,n]}) = \rho_0(x_0) \prod^{n}_{j=1}w(x_{j-1},x_{j}), \label{eq:PDefMarkov}
 \end{eqnarray}
 where  $\rho_0(x)$  is the probability (density) of the initial state $X_0$.
 \end{leftbar}
In general, Markov chains are inhomogeneous, i.e.,  their transition probabilities $w_n(x,y)$ may depend explicitly on time $n$.   However, for clarity we postpone the discussion of time-inhomogeneous Markov processes to the Sec.~\ref{sec:markovCont} on   Markov processes in continuous time, while in discrete time we  focus on time-homogeneous processes, i.e., we use $w_n(x,y)=w(x,y)$ throughout Sec.~\ref{Sec:markov}.

 \subsection{{Constructing martingales from Markov processes }}\label{sec:martingaleProblem}
Martingales play a prominent role in the theory of Markov processes \cite{norris1998markov, ethier2009markov}.
One reason is due to the following theorem (Theorem 4.1.3. in \cite{norris1998markov}): 

\begin{leftbar}
\begin{theorem}[\textbf{Characterisation of Markov processes with martingales}]
 $\\$ Let $X_n$ be a stochastic process that takes values in $\mathcal{X}$.   The following two statements are equivalent: 
  \begin{itemize}
  \item $X_n$ is a Markov chain with transition matrix $w(x,y)$; 
  \item for all real-valued, bounded functions $f$ defined on $\mathcal{X}$ it holds that the process 
  \begin{eqnarray}
  M_n=f(X_n) - f(X_0)  - \sum^{n-1}_{m=0} \sum_{x\in\mathcal{X}} \left(w(X_m,x) - \delta_{x,X_m}\right) f(x) \label{eq:markovChain}
  \end{eqnarray}
   is a martingale with respect to $X_n$.
  \end{itemize}
  \label{th:1}
\end{theorem} 
Taken together, Eq.~\eqref{eq:equivalencemart} implies that  $M_{n}$  is a martingale if and only if
$X_{n}$ is Markovian. 
\end{leftbar}
%Theorem~\ref{th:1} follows  from the identity $M_{n+1}-M_{n}=f(X_{n+1}) - \sum_{x\in\mathcal{X}} w(X_n,x) f(x)$,   which together with ~\eqref{eq:transmat} implies that $\langle M_{n+1}|X_{n}\rangle = M_n$.  Since $\langle |M_n|\rangle<\infty$, we consider bounded functions $f$.   
 
 {Theorem~\ref{th:1} follows from the identity
$M_{n+1}-M_{n}=f\left(X_{n+1}\right)-\sum_{x\in\mathcal{X}}w\left(X_{n},x\right)f(x),$
which implies 
\[
\left\langle \left.M_{n+1}\right|X_{\left[0,n\right]}\right\rangle =M_{n}+\left\langle \left.f\left(X_{n+1}\right)\right|X_{\left[0,n\right]}\right\rangle -\sum_{x\in\mathcal{X}}w\left(X_{n},x\right)f(x).
\]
 Due to this last relation, we obtain 
\begin{equation}
\left\langle \left.M_{n+1}\right|X_{\left[0,n\right]}\right\rangle =M_{n}\iff \left\langle \left.f\left(X_{n+1}\right)\right|X_{\left[0,n\right]}\right\rangle =\sum_{x\in\mathcal{X}}w\left(X_{n},x\right)f(x).
\label{eq:equivalencemart}
\end{equation}
%Now, by tower property (ref), the l.h.s of the equivalence is equivalent
%to the Martingale property : for all $m\leq n$ $\left\langle \left.M_{n}\right|X_{\left[0,m\right]}\right\rangle =M_{m}.$
The right-hand-side of the second equality of the equivalence~\eqref{eq:equivalencemart}
is  a function of  $X_{n}$ only, which implies  that $X_{n}$ is a Markov chain with transition matrix $w$. }

 \subsection{Dynkin's  martingales}\label{sec:Dynkin}

Processes of the form~\eqref{eq:markovChain} are called Dynkin's additive martingales, and we can   also express them as \begin{eqnarray}
  M_n= \sum^{n-1}_{m=0} \left(f(X_{m+1})-\sum_{x\in\mathcal{X}}w(X_m,x) f(x)\right). \label{eq:markovChain2}
  \end{eqnarray}
  Put simply,  Dynkin's additive martingales, as defined by Eq.~\eqref{eq:markovChain2}, are the cumulative differences between the  function $f(X_{m+1})$ evaluated on the process $X$ at time $m+1$ minus the expected value of $f$ at time $m+1$ when conditioned on its value at the previous time step.  In what follows, we discuss two key examples of  Dynkin's martingales.     
  
 \subsubsection{Processes without memory}
Let us consider the case when $X_n$ is  an i.i.d.~sequence.   This is the particular case of a Markov chain with transition probability  $w(y,x)=\mathcal{P}(x)$, where $\mathcal{P}(x)$ is the law of the variables in an i.i.d.~sequence.  In this case,  Dynkin's martingale takes the form  \begin{eqnarray}
  M_n= \sum^{n-1}_{m=0} \left(f(X_{m+1})-\langle f (X_m)\rangle \right). \label{eq:markovChain3}
  \end{eqnarray} 
  Specializing to the case $f(x)=\ln(x)$, we obtain the martingale 
  \begin{eqnarray}
  M_n= \sum^{n-1}_{m=0} \left(\ln(X_{m+1})-\langle \ln (X_m)\rangle \right)=\ln\left( \prod_{m=0}^{n-1}  X_{m+1}  \right) - n \langle \ln  X_0 \rangle .
  \label{eq:markovChain35}
  \end{eqnarray}
  On the other hand, for the choice $f(x)= x$  we obtain the additive martingale  
  \begin{eqnarray}
  M_n= \sum^{n-1}_{m=0} X_{m+1} - n \langle X_0\rangle,
  \label{eq:markovChain36}
  \end{eqnarray}
which coincides with the martingale~\eqref{eq:1p9} when $X\in\{1,-1\}$ with probabilities $\mathcal{P}(1)=q$ and $\mathcal{P}(-1)=1-q$.

  \subsubsection{Harmonic functions}

We say that $h(x)$ is a {\it harmonic} function if it is  a bounded function for which 
\begin{eqnarray}
\sum_{x\in \mathcal{X}}  w(y,x) h(x) &=& h(y), \quad \forall y\in\mathcal{X} .\label{eq:harm}
\end{eqnarray}   
Hence, harmonic functions are   the right eigenvectors associated with the Perron root   of  $w$;  notice that these are different from the   left eigenvectors of the Perron root, which represent    the  stationary probability distributions.    For an unbiased random walk, Eq.~(\ref{eq:harm}) is a discrete version of the  equation $\partial^2_x h(x) = 0$, which clarifies why we call $h$ a harmonic function.  
Analogously,  we say that $s(x)$ is a  {\em subharmonic} function if it is a bounded function for which 
\begin{eqnarray}
\sum_{x\in \mathcal{X}} w(y,x)s(x) \geq s(y), \quad \forall y\in\mathcal{X}.
\end{eqnarray}

 Theorem~\ref{th:1} implies that  processes of the form $h(X_n)$, with $h$  a harmonic function, are martingales.   Indeed,  plugging Eq.~\eqref{eq:harm} in Eq.~\eqref{eq:markovChain}, Theorem~\ref{th:1} implies that $h(X_n)$ is a martingale. We can also prove this result directly:  
\begin{eqnarray}
\langle  h(X_n) |X_{[0,n-1]}\rangle =  \langle  h(X_n) |X_{n-1}\rangle = \sum_{x\in \mathcal{X}} w(X_{n-1},x)h(x)  = h(X_{n-1}),
\end{eqnarray} 
where the first equality follows from the Markov property, the second from the definition of the transition matrix~\eqref{eq:transmat}, and the third equality from the definition of harmonic functions~\eqref{eq:harm}.  Analogously, processes of the form $s(X_n)$ with $s$ a subharmonic function are  submartingales, see Ref.~\cite{bremaud2013markov}.  

For ergodic Markov processes, the trivial function $h(x)=1$ is the only harmonic function \cite{bremaud2013markov}.  Indeed, for ergodic processes, the Perron root of the operator $w(x,y)$  is nondegenerate, and hence the  left eigenvector of $w(x,y)$  associated with the Perron root is unique.    On the other hand, for nonergodic processes, the Perron root  is degenerate, and we can construct  nontrivial harmonic functions. 

As an  example of a nontrivial harmonic function, consider the process $X_n$ with initial condition $X_0\in\mathcal{X}\setminus \left( \mathcal{X}_1\cup\mathcal{X}_2\right)$ that "stops" as soon as $X_n$ reaches the  absorbing set  $\mathcal{X}_1 \cup  \mathcal{X}_2$.     In other words, the transition matrix is  given by 
  \begin{eqnarray}
  \widetilde{w}(y,x) =  \left\{ \begin{array}{ccc} \delta_{x,y},  &{\rm if}&  y \in \mathcal{X}_1\cup\mathcal{X}_2, \\ w(x,y), &{\rm if}& y \in  \mathcal{X}\setminus \left( \mathcal{X}_1\cup\mathcal{X}_2\right). \end{array} \right.
  \end{eqnarray}
    We assume that  $\mathcal{X}_1\cap \mathcal{X}_2 = \phi$.    In this case, the process is nonergodic  as the states in the sets $\mathcal{X}_1$ and $\mathcal{X}_2$ are absorbing.  
    Let 
  \begin{eqnarray}
  \mathcal{T}_{\mathcal{X}_1} = {\rm min}\left\{n\geq 0: X_n\in \mathcal{X}_1\right\}, \quad  {\rm and}\quad  \mathcal{T}_{\mathcal{X}_2} = {\rm min}\left\{n\geq 0: X_n\in \mathcal{X}_2\right\}, 
    \end{eqnarray} 
    be the first-passage  times when $X_n$ hits the sets $\mathcal{X}_1$ or $\mathcal{X}_2$, respectively. 
     {Let us now define  the splitting probability that the process $X_t$ hits the set $\mathcal{X}_1$ before hitting the set $\mathcal{X}_2$ given that the state at time $k$ was $X_k=x$,}
    \begin{eqnarray}
   {h_{\mathcal{X}_1,\mathcal{X}_2}(x) = \mathcal{P}\left(\mathcal{T}_{\mathcal{X}_1}<\mathcal{T}_{\mathcal{X}_2}| X_k = x\right) .}\label{eq:hitprob}
    \end{eqnarray} 
    {Note that, because the transition rates are considered to be time homogeneous, the splitting probabilities~\eqref{eq:hitprob} are independent of $k$.}
    It holds that the splitting probability $h_{\mathcal{X}_1,\mathcal{X}_2}(x)$ is a harmonic function related of  $\widetilde{w}(y,x)$ \cite{doyle1984random}.   Indeed, using the Markovianity  of $X$, we find iteration \begin{equation}
\mathcal{P}\left(\mathcal{T}_{\mathcal{X}_1}<\mathcal{T}_{\mathcal{X}_2}| X_0 = x\right) = \sum_{y\in \mathcal{X}}w(x,y) \mathcal{P}\left(\mathcal{T}_{\mathcal{X}_1}<\mathcal{T}_{\mathcal{X}_2}| X_0 = y\right),
    \end{equation}
   and hence $  h_{\mathcal{X}_1,\mathcal{X}_2}(x)$   solves the Dirichlet problem
      \begin{eqnarray}
           h_{\mathcal{X}_1,\mathcal{X}_2}(x) = \left\{ \begin{array}{ccc} 1, &{\rm if}& x\in \mathcal{X}_1, \\ 0, &{\rm if}& x\in \mathcal{X}_2, \\  \sum_{y\in \mathcal{X}} w(x,y) h_{\mathcal{X}_1,\mathcal{X}_2}(y),  &{\rm if}& x\in \mathcal{X}\setminus (\mathcal{X}_1\cup \mathcal{X}_2). \end{array} \right. 
    \end{eqnarray}
Consequently, the splitting probability $  h_{\mathcal{X}_1,\mathcal{X}_2}(x)$ is an example of a nontrivial harmonic function, and  it is a martingale.    

\subsubsection{Doob's $h$-transform }

An interesting application of positive harmonic functions $h$ is the construction of path probability ratios associated with $h$ though the, so-called, Doob's $h$-transform, which we introduce below. 

  Let $X$ be a Markov process, and let $h$ be a positive  and harmonic function.   Then there exists a Markov process with path probability density  $\mathcal{P}_h$  such that 
\begin{eqnarray}
\langle f(X_{[0,n]}) |X_0\rangle_h  = \frac{1}{h(X_0)} \langle h(X_n) f(X_{[0,n]})  |X_0\rangle, 
\end{eqnarray} 
for all $n\in \mathbb{N}\cup \left\{0\right\}$, where $\langle \cdot \rangle_h$  denotes the expectation with respect to $\mathcal{P}_h$.    We call $\mathcal{P}_h$ the Doob $h$-transform of $\mathcal{P}$.  
\begin{leftbar}
\begin{theorem}[\textbf{Doob's $h$-transform}] $\\$
Let $\mathcal{P}$ be the path probability density  of a Markov chain with transition matrix $w(y,x)$.
If $h$ is a nonnegative harmonic function associated with this Markov chain, then there exists a Markov chain $\mathcal{P}_h$ with transition matrix $w_h(y,x)$ such that 
\begin{eqnarray}
 \frac{\mathcal{P}_h(x_{[0,n]}|x_0)}{\mathcal{P}(x_{[0,n]}|x_0)} =\frac{h(x_n)}{h(x_0)}.
\end{eqnarray} 
\end{theorem}
\end{leftbar}
\begin{proof}
Since $h$ is a positive and harmonic function, it holds that  
\begin{eqnarray}
w_h(y,x) = \frac{h(x)}{h(y)} w(y,x)
\end{eqnarray}
is  a transition matrix.   Indeed,  $w_h(y,x)\geq 0$ and 
\begin{eqnarray}
\sum_{x\in\mathcal{X}} w_h(y,x) =\sum_{x\in\mathcal{X}}   \frac{h(x)}{h(y)} w(y,x) = 1.
\end{eqnarray} 
Using that 
\begin{eqnarray}
\mathcal{P}_h(X_{[0,n]}|X_0) =   \prod^{n}_{j=1}w_h(X_{j-1},X_j), \quad  \mathcal{P}(X_{[0,n]}|X_0) =   \prod^{n}_{j=1}w(X_{j-1},X_j),
\end{eqnarray}
we obtain   
\begin{eqnarray}
\frac{\mathcal{P}_h(X_{[0,n]}|X_0)}{\mathcal{P}(X_{[0,n]}|X_0)} = \prod^{n}_{j=1}\frac{w_h(X_{j-1},X_j)}{w(X_{j-1},X_j)} = \frac{h(X_n)}{h(X_0)},
\end{eqnarray}
which completes the proof.
\end{proof} 

Doob's $h$-transform can be used to map the statistics of a conditioned process, provided by the measure $\mathcal{P}_h$, on the statistics provided by an unconditioned process,  given by $\mathcal{P}$, see e.g. Refs.~\cite{majumdar2015effective,chetrite2015nonequilibrium} for some explicit examples.   For example, if  $h$ is the splitting probability~(\ref{eq:hitprob}), {$h(x)=h_{\mathcal{X}_1,\mathcal{X}_2}(x)$}, then  
\begin{eqnarray}
w_h(y,x) &=& \frac{h(x)}{h(y)} w(y,x)    \nonumber\\
&=&{ \frac{h_{\mathcal{X}_1,\mathcal{X}_2}(x)}{h_{\mathcal{X}_1,\mathcal{X}_2}(y)} w(y,x)}
\nonumber\\
&=&\frac{\mathcal{P}\left(\mathcal{T}_{\mathcal{X}_1}<\mathcal{T}_{\mathcal{X}_2}| X_{n}=x \right) }{\mathcal{P}\left(\mathcal{T}_{\mathcal{X}_1}<\mathcal{T}_{\mathcal{X}_2}| X_{n-1}=y\right)}\mathcal{P}\left(X_n = x|X_{n-1}=y\right) \nonumber\\ 
 &=&\frac{\mathcal{P}\left(X_{n}=x , \mathcal{T}_{\mathcal{X}_1}<\mathcal{T}_{\mathcal{X}_2}|X_{n-1}=y\right)  }{\mathcal{P}\left(\mathcal{T}_{\mathcal{X}_1}<\mathcal{T}_{\mathcal{X}_2}| X_{n-1}=y\right)} \nonumber \\ 
 &=&  \mathcal{P} \left(X_n=x|X_{n-1}=y, \mathcal{T}_{\mathcal{X}_1}<\mathcal{T}_{\mathcal{X}_2} \right). 
\end{eqnarray}
Hence, $\mathcal{P}_h$ is the probability distribution of a  Markov process that describes the statistics conditioned on the event $\mathcal{T}_{\mathcal{X}_1}<\mathcal{T}_{\mathcal{X}_2}$.

 \subsection{Multiplicative martingales}\label{sec:Multiplicative}
Given a real-valued, bounded function $f$, the product 
\begin{equation}
M_{n}=\prod_{j=0}^{n-1}\frac{f(X_{j+1})}{\sum_{x\in\mathcal{X}}w(X_j,x) f(x)}
\label{eq:marfomu}
\end{equation}
is martingale.     The  martingality of $M_n$ follows  from the identity 
 \begin{equation}
M_{n+1}=M_{n}\frac{f(X_{n+1})}{\sum_{x\in\mathcal{X}}w(X_n,x) f(x)},
\end{equation}
 together with  
 \begin{equation}
    \langle  f(X_{n+1})|X_{[0,n]}\rangle = \sum_{x\in\mathcal{X}}w(X_n,x)f(x). 
 \end{equation}
 
 For the particular case of  i.i.d. sequences  with transition probability  $w(y,x)=\mathcal{P}(x)$, the multiplicative martingale~\eqref{eq:marfomu} takes the form
 \begin{equation}
     M_n = \frac{\prod_{j=0}^{n-1}f(X_{j+1})}{\langle f(X_0)\rangle^n}.
 \end{equation}
 Setting $f(x)=\exp(y x)$ and assuming that $X=\{1,-1\}$ is a binary random variable with $\mathcal{P}(1)=q$,  we recover the martingale given by Eq.~\eqref{eq:1p7}. 

\subsection{Ratios of path probability densities}\label{subsec:Ratios}
We consider the ratio $R_n$ (see Eq.~(\ref{eq:RForm})) of two path   probability densities $\mathcal{P}$ and  $\mathcal{Q}$ of two time-homogeneous Markov processes in discrete time.       To obtain an explicit expression for $R_n$, we denote $\mathcal{P}(x_{[0,n]})$ as in Eq.~(\ref{eq:PDefMarkov}), and  we write 
\begin{equation}
\mathcal{Q}\left(x_{[0,n]}\right) =  \rho^{\mathcal{Q}}\left(x_{0}\right)  \prod^{n}_{j=1}w^{\mathcal{Q}}(x_{j-1},x_j).
\end{equation}
 When (i) $w^{\mathcal{Q}}(y,x)= 0$ for all $x,y\in \mathcal{X}$ for which $w(y,x)= 0$, and (ii)  $\rho^{\mathcal{Q}}(x)=0$ for all $x\in\mathcal{X}$ for which $\rho(x)=0$, then the probability density $\mathcal{Q}$ is locally absolutely continuous with respect of $\mathcal{P}$, such that  the ratio 
\begin{equation}
R_n = \frac{\rho^{\mathcal{Q}}\left(X_{0}\right)}{\rho\left(X_{0}\right) }\prod^{n}_{j=1}\frac{w^{\mathcal{Q}}   (X_{j-1},X_j)}{w   (X_{j-1},X_j)}
\end{equation}
exists, and is a $\mathcal{P}$-martingale. 

\section{Martingales in continuous-time Markov processes  } \label{sec:markovCont}
The second section of this Chapter deals with martingales that are defined relative to a  Markov process $X_t$ that runs in  continuous time.  These are arguably the most important examples of martingales for physics, as the lion's share of models that describe physical processes at the mesoscopic scale   are  continuous time Markov processes, see e.g.~\cite{schnakenberg1976network, LNP, seifert2012stochastic}.

The present section is organised as follows.   In Sec.~\ref{sec:defMarkovCont}, we introduce the  mathematical quantities defining Markov processes in continuous time.  In the following two sections, we define two main classes of Markov processes in continuous time, namely,  Markov jump processes in Sec.~\ref{eq:jumpprocconta} and diffusion processes in Sec.~\ref{sec:diff}.    In Sec.~\ref{sec:martingaleProblemCont} we formulate the martingale problem for Markov processes in  continuous time.    The last three sections are devoted to examples of martingales that are defined relative to a continuous-time Markov process, namely,  Dynkin's martingales in Sec.~\ref{sec:dynkinCont}, exponential martingales in Sec.~\ref{sec:exponCont}, and Radon-Nikodym derivative processes in Sec.~\ref{sec:radonNikodym}.

\subsection{  Markov processes in continuous time: three definitions}\label{sec:defMarkovCont}
 We discuss three complementary ways to define Markov processes~\cite{norris1998markov}.   The first approach is based on the path probabilities   $\mP(x_{[0,t]})$.   
The second approach is based on the observation that for  Markov processes on a discrete state space $\mathcal{X}$
  \begin{equation}
     \mP(X_{t}=y|X_{[0,s]}) = \mathcal{P}(X_{t}=y | X_{s})
     \label{eq:markovpropertydef}
 \end{equation} 
 for any $t\geq s\geq 0$, and therefore to determine a Markov process it is  sufficient to define the transition function $\mathcal{P}(X_{t}=y  | X_{s}=x)$ that gives the transition probability between two states $X_s$ and $X_{t}$ at times $s$ and $t$, respectively.   A third way to define Markov processes is with the  generator or adjoint generator of the process; the former determines the evolution with respect of time $s$ with $t\geq s\geq 0$
of the  transition function $\mathcal{P}(X_{t}=y  | X_{s}=x)$, and the latter determines the evolution in time of   the instantaneous probability density of $X$.

\subsubsection{Path probabilities}
Let us start with a description of Markov processes through path probabilities.  
 \begin{leftbar}
 The  \textbf{measures}    $\mathcal{P}\left(x_{[0,t]}\right)$   specify the probability   to observe sets of  paths $x_{[0,t]}$ in the time window $[0,t]$.  
In general, it is not possible to present an explicit expression for $\mathcal{P}\left(x_{[0,t]}\right)$.  However, we can express the density of $\mathcal{P}\left(x_{[0,t]}\right)$ relative to another equivalent measure $\mathcal{Q}\left(x_{[0,t]}\right)$ through the  Radon-Nikodym derivative process, see Eq.~(\ref{eq:radonDef2}), or we can use the Onsager-Machlup approach to represent each member $\mathcal{P}$ of an equivalence class $\mathscr{P}$ of mutually absolutely continuous measures in terms of the action fuctional $\mathcal{A}_\mathcal{P}(x_{[0,t]})$, see Eqs.~(\ref{eq:onsagerMachlupApproach1}-\ref{eq:onsagerMachlupApproach3}).
At the end of this section, we present a couple of examples of    Radon-Nikodym derivatives of   jump processes and diffusions.  
\end{leftbar}

\subsubsection{Transition functions}
According to Eq.~(\ref{eq:markovpropertydef}), Markov processes can also be specified with their transition function (again, for discrete state space $\mathcal{X}$)   
\begin{equation}
\mathcal{P}_{s,t}(x,y)\equiv 
\mathcal{P}(X_t=y|X_{s}=x),
\end{equation}
for $t\geq s\geq 0$, see  Refs.~\cite{chung2006markov, barato2015formal}.
For continuous state space $\mathcal{X}$, we define transition function as \begin{equation}
\mathcal{P}_{s,t}(x,y) \equiv \frac{\mathcal{P}(X_t\in\left[y,y+dy\right]|X_{s}=x)}{dy}.
\end{equation}

\begin{leftbar}
The \textbf{transition function} 
operates on bounded, real-valued functions $\phi$ defined on $\mathcal{X}$ through 
\begin{equation}
\mathcal{P}_{s,t}[\phi](x) \equiv \langle \phi(X_{t})|X_{s}=x\rangle =   \int_{\mathcal{X}} dy  \:  \mathcal{P}_{s,t}(x,y)\phi(y). \label{eq:PtTPrime}
\end{equation}

 The transition function 
   satisfies the Chapman-Kolmogorov  condition 
\begin{equation}
\int_{\mathcal{X}} dy\: \mathcal{P}_{s,t}(x,y)\mathcal{P}_{t,t'}(y,z)=\mathcal{P}_{s,t'}(x,z),
\label{eq:2.71}
\end{equation}
for all $s\leq t\leq t'$.  
\end{leftbar}

\subsubsection{Generators}
A third approach to define a Markov process is through either its generator $\mathcal{L}_t$ or the adjoint generator $\mathcal{L}^\dagger_t$ that describes the evolution in time of the instantaneous probability density $\rho_t$.   Since the latter is used more often in physics, we introduce it first.

\begin{leftbar}

The   \textbf{instantaneous density} of a continuous-time Markov process $X_t\in \mathcal{X}$ is defined as
\begin{equation}
    \rho_t (x) \equiv \langle\delta(X_t -x)\rangle,
\end{equation} 
where  $\langle \cdot \rangle$ is the average  over repeated realizations of the Markov process $X$. 
 The instantaneous density $\rho_t$  is the solution of the Fokker-Planck or Master equation
\begin{equation}
\partial_{t}\rho_{t}=\mathcal{L}_{t}^{\dagger}\rho_{t},
\label{eq:FP}
\end{equation}
where  $\mathcal{L}_{t}^{\dagger}$ is the   adjoint of the generator $\mathcal{L}_{t}$ that expresses the evolution in time of the   transition function, 
 \begin{equation}
\partial_{t}\mathcal{P}_{s,t}=\mathcal{P}_{s,t}\mathcal{L}_{t}. 
\label{eq:KFE}
\end{equation}
The  explicit time dependence in  $\mathcal{L}^\dagger_{t}$  is  relevant for  Markov processes with time-dependent, external driving. An invariant density $\rho_{\rm st}(x)$ is a time-independent distribution that solves for all $t\geq 0$ 
\begin{equation}
0=\mathcal{L}_{t}^{\dagger}\rho_{\rm st}
\label{eq:FPI}
\end{equation}
and we say that $\rho_{\rm st}$ is a {\bf stationary probability density} if in addition to Eq.~\eqref{eq:FPI} one has the normalization condition
\begin{equation}
\int_{\mathcal{X}} \rho_{\rm st}(x)dx=1.
\end{equation}
For the special the case of time-homogeneous dynamics, we have  $\mathcal{L}^\dagger_{t}=\mathcal{L}^\dagger$.

\end{leftbar}

Let us clarify some of the mathematical notation used in Eqs.~(\ref{eq:FP})-(\ref{eq:FPI}):

\begin{itemize}

  \item    The generator $\mathcal{L}_t$ is a linear operator that acts on the Hilbert space $L^2(\mathcal{X})$  of functions $\phi:\mathcal{X}\rightarrow \mathbb{R}$  with finite norm $\int_{x\in\mathcal{X}} dx \: \phi^2(x)$, 
and endowed with the inner product
\begin{equation}
    \left( \phi_1,\phi_2\right) =\int_{x\in\mathcal{X}} dx \: \phi_1(x)\phi_2(x).
    \label{eq:sp1axs}
\end{equation} 
In~Eq.~\eqref{eq:sp1axs},  $dx$ refers to the Lebesgue measure if the space  $\mathcal{X}$
is continuous and to the counting measure if $\mathcal{X}$ is discrete. For the latter,  integrals are finite sums, i.e.,
\begin{equation}
\int_{x\in\mathcal{X}} dx \: \phi_1(x)\phi_2(x)  = \sum_{x\in\mathcal{X}} dx \: \phi_1(x)\phi_2(x),
\end{equation}
and operators  $\mathcal{L}_t$ are matrices.   
 We will follow this convention throughout this {Treatise}.  
 \item  
  
  The operator $\mathcal{L}^\dagger_t $  can be seen as  the adjoint of the operator $\mathcal{L}_t $  on the   Hilbert space $L^2(\mathcal{X})$.  In other words, the  $\mathcal{L}_{t}^{\dagger}\phi_2$ is the function such that
 \begin{equation}
   \left( \phi_2,\mathcal{L}_{t}\phi_1\right)=\big( \mathcal{L}_{t}^{\dagger}\phi_2,\phi_1\big) 
\end{equation}   
for all  functions $\phi_1$ in the domain of $\mathcal{L}_{t}$.     Consequently, the right-hand side of Eq.~(\ref{eq:FP}) is the function  
\begin{equation}
\big(\mathcal{L}_{t}^{\dagger}\rho_{t}\big)(x)=  \int_{\mathcal{X}} dy  \rho_{t}(y) \mathcal{L}_{t}(y,x) .
\label{eq:2p755}
\end{equation} 
In the particular case where $\mathcal{X}$ is finite, $\mathcal{L}^\dagger_t$ is the matrix transpose of $\mathcal{L}_t$. 

 \item 
 We underline that  the left-hand side of Eq.~\eqref{eq:KFE} should be understood as  acting on  scalar functions $\phi(x)$ as in Eq.~(\ref{eq:PtTPrime}), as for the right-hand side
\begin{equation}
 (\mathcal{P}_{s,t}[\mathcal{L}_{t} \phi])(x)=  \int_{\mathcal{X}} dy  \:  \mathcal{P}_{s,t} (x,y) \: \int_{\mathcal{X}} dz  \mathcal{L}_t(y,z) \phi(z) .
\label{eq:2p75}
\end{equation} 

  \item Note that time-homogeneous and stationary Markov processes are, in general, not equivalent.  Indeed, a time-homogeneous Markov processes is nonstationary when its distribution $\rho_t$ is nonstationary, and a stationary Markov process is time-inhomogeneous when the generator $\mathcal{L}_t$ depends on time $t$.   Indeed, a Markov process may obey detailed balance with a certain potential $V(x)$ and  have time-dependent rates.  
\end{itemize}

Although the most general Markov process 
consist of   a mixture of diffusions and  random jumps, see Ref.~\cite{applebaum2009levy,bass2007sdes},  
in this {Treatise}, we will focus on two  paradigmatic  classes of Markov processes in nonequilibrium physics, namely  pure jump processes (for which the continuous part is absent) and pure diffusion processes (for which the jump part is absent).

\subsection{Markov jump processes}
\label{eq:jumpprocconta}

\begin{figure}[H]
\centering
{\includegraphics[width=\textwidth]{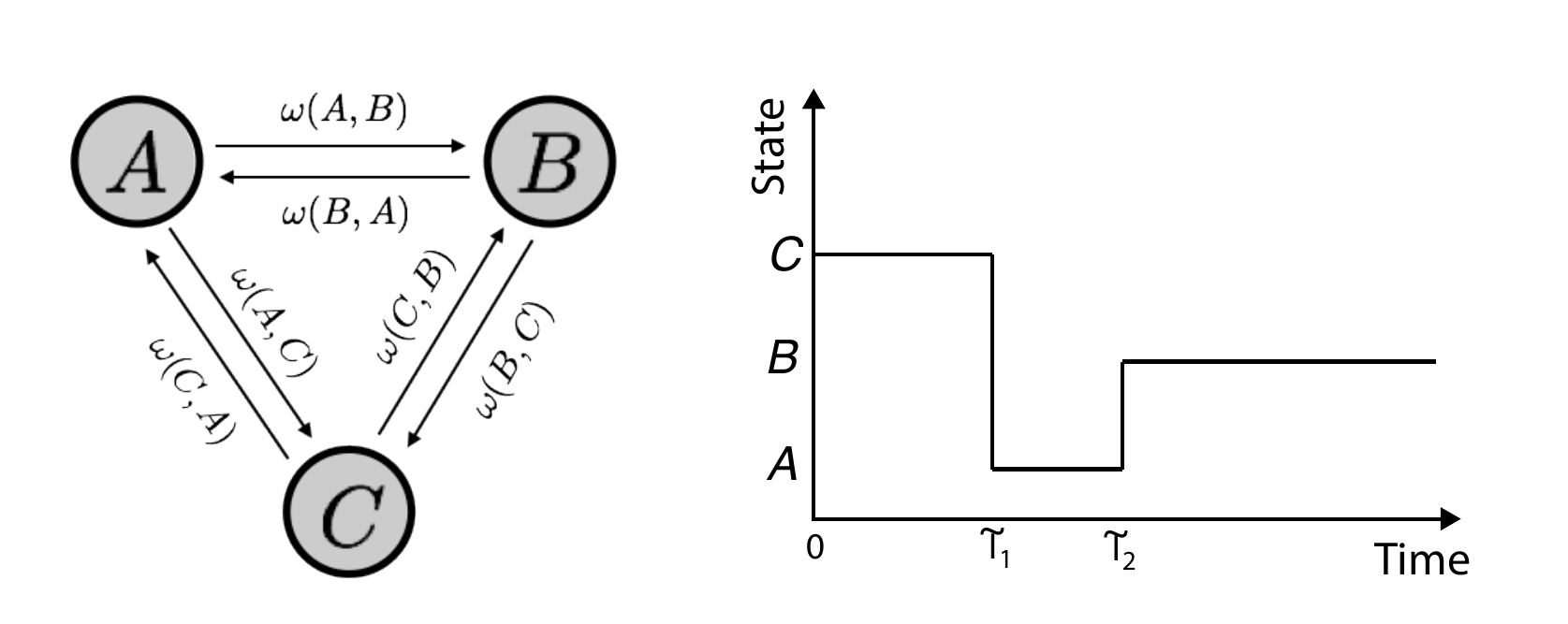}}
\caption{Left: Illustration of a 3-state continuous-time Markov jump process between the  states (gray circles) $A$, $B$, and $C$. The transition rate values between each pair of states are indicated on the arrows. Right: Example trajectory of the process, where the system jumps at a random time $\T_1$ from the initial state $C$ to state $A$ and at a later time $\T_2$ from state $A$ to state $B$. See text for further details.}
\label{fig:3s}
\end{figure}

Markov jump processes are Markov processes for which the process $X_t$  changes its state in a  purely discontinuous manner.    Figure~\ref{fig:3s} shows an example of a minimal model of a continuous-time Markov jump process in a discrete set of states together with an illustration of a single trajectory of the process.

\subsubsection{Mathematical form}

The trajectories of a Markov jump process are piecewise constant functions, with jump times $\T_i$, with $\T_0=0$ and with $i\in \left\{1,2,\ldots,N_{t}\right\}$, and where   $N_t$  is the number of times the process has jumped in the time interval $[0,t]$.   In between two jump times, the process  $X_t$ does not change its value.   We denote the value of $X_t$ right before the $i$-th jump by 
\begin{equation}
    X_{\T_i^{-}} = \lim_{\epsilon\rightarrow 0, \epsilon>0}X_{\T_i-\epsilon} 
\end{equation} 
and right after the $i$-th jump by 
\begin{equation}
   X_{\T_i^{+}} = \lim_{\epsilon\rightarrow 0, \epsilon>0}X_{\T_i+\epsilon},
   \end{equation} 
so that 
\begin{equation}
X_t=  X_{\T_{i-1}^{+}}=X_{\T_i^{-}} \quad {\rm if} \quad  t\in [\mathcal{T}_{i-1},\mathcal{T}_{i})  . \label{eq:defsXMJP}
\end{equation}

The \textbf{transition rate} $\rateWt(x,y)$ of the jump process is  the rate for the transition $x\rightarrow y$ at time $t$, i.e. the average number of jumps from state $x$ to state $y$ occurring at time $t$.

Other observables that we often use for Markov jump processes are the number of times $N_t(x,y)$ that $X_t$ has jumped form $x$ to $y$ in the time window $[0,t]$ and the residence time $\tau_t(x)$ that the process $X_t$ has spend in the the $x$-th state.  For discrete sets $\mathcal{X}$, these quantities are formally defined as 
\begin{equation}
N_t(x,y)\equiv  \sum^{N_t}_{j=1} \delta_{X_{\T^-_j},x}  \delta_{X_{\T^+_j},y}    \label{eq:NtDef} 
\end{equation}
and 
\begin{equation}
\tau_t(x) \equiv \int^{t}_0  \delta_{X_s,x}  ds, \label{eq:tauDef}
\end{equation}
where $\delta$ is here the Kronecker delta function;  analogous definitions can be written down for continuous sets $\mathcal{X}$.    Occasionally, we also use  $\dot{N}_t(x,y)$, for which it should be understood that 
\begin{equation}
    N_t(x,y) = \int^{t}_0 \dot{N}_s(x,y) ds. 
\end{equation}

\begin{leftbar}  
The {\bf Markov generator} associated with the Fokker-Planck equation~\eqref{eq:FP}  is  given by 
\begin{equation}
\left(\mathcal{L}_{t}\phi\right)(x)\equiv\int_{\mathcal{X}}\,dy \, \rateWt(x,y)\left(\phi(y)-\phi(x)\right),
\label{eq:genps}
\end{equation}
where    $\mathcal{X}$ can be either discrete or continuous. {If $\mathcal{X}$ is discrete, the integral in~\eqref{eq:genps} must be read as a sum.}
\end{leftbar}

{When $X_t$ has no explosions~\cite{norris1998markov}, i.e., the total number of jumps $\sum_{x\neq y}N_t(x,y)$ is with probability one finite, then the generator $\mathcal{L}_t$ uniquely defines the Markov jump process. }

\begin{leftbar}
The  {\bf Fokker-Planck equation}~\eqref{eq:FP} associated with a Markov jump process reads  
\begin{equation}
\partial_{t}\rho_t(x)=-\int_{\mathcal{X}}{\rm d}y\: J_{t,\rho}(x,y). \label{eq:fokkerplanck}
\end{equation} 
where the probability current reads
\begin{equation}
   J_{t,\rho}(x,y) = \rho_t(x)\omega_t(x,y)-\rho_t(y)\omega_t(y,x).
\end{equation}

{The Fokker-Planck equation~\eqref{eq:fokkerplanck} can be also written as
\begin{equation}
\partial_{t}\rho_t(x)=\int_{\mathcal{X}}{\rm d}y \Big[\rho_t(y)\omega_t(y,x)-\rho_t(x)\omega_t(x,y)\Big],
\end{equation}
which for $\mathcal{X} $ discrete reads
\begin{equation}
\partial_{t}\rho_t(x)=\sum_{y\in \mathcal{X}}\Big[\rho_t(y)\omega_t(y,x)-\rho_t(x)\omega_t(x,y)\Big].
\label{eq:FPCTMC}
\end{equation}
Equation~\eqref{eq:FPCTMC} provides the familiar form for the Master equation of a continuous-time Markov chain where we identify the first term in the right hand side as probability influxes to state $x$ and the second term as probability outfluxes from state $x$.
}
\end{leftbar}

\subsubsection{Physical setup: isothermal case}
To add physical content to the dynamics of  a Markov jump process, we use the principle of local detailed balance~\cite{seifert2012stochastic, maes2021local}.   Consider a mesoscopic system, say a molecular motor, that is pushed by an external force of magnitude $f_{t}$ and is in contact with  one thermal bath at temperature $T$, and $m$ particle reservoirs characterised by the chemical potentials $\mu^{(a)}$, where $a=1,2,\ldots,m$.  
We  assume that all  particle reservoirs are at temperature~$T$. 
For isothermal processes, the principle of local detailed balance implies that the ratio of transition rates  satisfies
\begin{equation}
        \frac{\omega_t(x,y)}{\omega_t(y,x)}=\exp\left(\frac{-(V_t(y)-V_t(x)) + f_{t}\:r(x,y)+ \sum^{n}_{
    a=1}\mu^{(a)}\:n_{a}(x,y)}{T}\right),
        \label{eq:EDBx}
\end{equation}
where $V_t(x)$ is a thermodynamic potential, %$T$ is the temperature of the environment, 
$r(x,y)$ is the distance moved when the system jumps from $x$ to $y$, and $n_{a}(x,y)$ is the number of particles exchanged with the $a$-th particle reservoir when the system jumps from $x$ to $y$.  The plus sign in front of  $f_{t}$ indicates that a negative  force  opposes forward motion, and the plus sign in front of $\mu^{(a)}$ indicates that   $n_a > 0$ when the system binds particles and $n_a < 0$ when the system releases particles. Generalization to particle reservoirs at different temperatures can be found in e.g.~\cite{van2015ensemble}.

\subsection{Diffusion processes}
\label{sec:diff}
A continuous-time Markov process is  a diffusion process if its trajectories $X_t$ are continuous functions of $t$~\cite{revuz2013continuous}.  We  determine  diffusion processes  through   stochastic differential equations, which we first discuss in their mathematical form, and subsequently, we discuss their physical interpretation.

\subsubsection{Mathematical form}

\begin{leftbar}
A $d$-dimensional  {\bf It\^{o} process} 
   $X_t = (X^1_t, X^2_t,\ldots,X^d_t)\in \mathbb{R}^d$ solves the stochastic differential equation
\begin{equation}
\dot{X}_{t}=b_{t}(X_{t})+ \sigma_t (X_t) \dot{B}_{t},
\label{eq:diff}
\end{equation}
where  $b_{t}= (b_t^1, b_t^2, \dots , b_t^d)^\dagger\in \mathbb{R}^d$ 
is a smooth, vectorial function;  $\sigma_{t}$ is a smooth matrix  ---not necessarily square--- defined on $\mathcal{X}$ with size $d\times n$, and $n$ arbitrary which is the number of noises. In other words,  $B_{t} = (B_t^1, B_t^2, \dots , B_t^n)^\dagger\in \mathbb{R}^n$  is a vector of $n$  independent Brownian processes.  We call 
\begin{equation}
    D_t(x)=\frac{\sigma_t (x) \sigma_t^{\dagger} (x)}{2}
\end{equation} the  {\bf diffusion matrix} which is nonnegative and of size $d\times d$. 

\end{leftbar}

The generator associated with the diffusion process given by   Eq.~(\ref{eq:diff}) takes the form 
\begin{equation}
\ensuremath{\mathcal{L}_{t}}=\ensuremath{b_{t}\nabla}+\mathbf{D}_{t} \ensuremath{\nabla\,\nabla},\label{eq:gpd}
\end{equation}
 where $\nabla =  (\partial_{x^1}, \partial_{x^2}, \dots ,\partial_{x^d})^\dagger$ is the gradient vector.
On a scalar function $\phi$, the generator Eq.~\eqref{eq:gpd} acts as \begin{equation}
\left(\mathcal{L}_{t}\phi\right)(x)=\ensuremath{b_{t}(x)\left(\nabla \phi \right)}+\mathbf{D}_{t}(x) \ensuremath{\left(\nabla \nabla \phi\right)}.\quad
\label{eq:gpd2}
\end{equation}

\begin{leftbar}
The {\bf Fokker-Planck} Eq.~(\ref{eq:FP}) {associated with} a $d$-dimensional It\^{o} process~\eqref{eq:diff} takes the form 
\begin{equation}
\partial_{t}\rho_{t}+\nabla \cdot J_{t,\rho}=0,
\label{ce} 
\end{equation} where the probability current 
\begin{equation}
\ensuremath{J_{t,\rho}(x)= b_{t}(x)\rho_{t}(x)}-\ensuremath{\nabla \cdot \left( \mathbf{D}_{t}(x)\rho_{t}(x)\right)}.
\label{curr}
\end{equation}
\end{leftbar}

\subsubsection{Physical setup: Langevin equations}

In physics, Eqs.~(\ref{eq:diff}) is often written in a different form that highlights  physically relevant quantities, such as, the potential and the external force, and which is commonly called the    \textbf{Langevin equation}, see e.g.~\cite{risken1996fokker, gardiner1985handbook, LNP}.  The Langevin equations are  mathematically equivalent to~(\ref{eq:diff}), and  when describing multi-dimensional diffusions in a physics context  we   consider Langevin equations, as these equations are  useful for describing  physical process, e.g., the dynamics of a set of interacting mesoscopic systems moving in multi-dimensions under external driving.

\begin{leftbar}
 The \textbf{Langevin equation}  is   the It\^{o} process Eq.~(\ref{eq:diff})  for $d=n$ written in terms of physical meaningful quantities~\cite{risken1996fokker, gardiner1985handbook, LNP}.   In particular, we write
\begin{equation}
\dot{X}_{t}= \bmu_{t}(X_t)F_{t}(X_t) + \left(\nabla\, \mathbf{D}_{t}\right) (X_t) +\sqrt{2\mathbf{D}_{t}(X_t)} \dot{B}_{t},\label{eq:LE}
\end{equation}
 where $ \bmu_{t}(x)$ is the mobility matrix that may depend on time and space,  and is not necessarily symmetric.  
 The force vector $F_{t}(x)$ can be decomposed as 
\begin{equation}
F_{t}(x)\equiv-\left(\nabla V_{t}\right)(x)+f_{t}(x),
\label{F}
\end{equation}
 where $V_{t}(x)$ is a time-dependent potential and $f_{t}(x)$ is a non-conservative force.  The potential is  controlled by an external agent through a deterministic protocol $\lambda_t$, such that $V_{t}(x)=V(x,\lambda_t)$.   
 The \textbf{Markovian generator}~\eqref{eq:gpd} associated with the Langevin Eq.~\eqref{eq:LE} takes  the form
\begin{equation}
\mathcal{L}_{t}=(\bmu_{t}F_{t})\nabla+\nabla\,\mathbf{D}_{t}\,\nabla,
\label{eq:gpdL}
\end{equation}
and the \textbf{probability current}   Eq.~\eqref{curr} takes the form
\begin{equation}
\ensuremath{J_{t,\rho}(x)\equiv(\bmu_{t}F_{t})(x)\rho_{t}(x)}-\ensuremath{\mathbf{D}_{t}(x)\nabla\rho_{t}(x)}. 
\label{currL}
\end{equation}
  \end{leftbar}
 The noise vector $B_t$ in the Langevin equation~\eqref{eq:LE} consists of $d$ independent standard Brownian motions.   The $d$-dimensional vector $X_t$ may contain both position and momentum variables, such as {\bf underdamped Langevin equations}, in which case the diffusion matrix $\mathbf{D}_t$ is  singular; this is the reason why in Eq.~(\ref{eq:diff})  
the matrix $\mathbf{D}_{t}(x)$  is  nonnegative instead of positive. 
The term $\left(\nabla\, \mathbf{D}_{t}\right)^{i}(x)=\sum^d_{j=1}\left(\partial_{j}\mathbf{D}_{t}(x)\right)^{ij}$ is a spurious drift term which comes from the $x$ dependence of the diffusion matrix; its physical origin is discussed below.   

\begin{figure}
\centering
{\includegraphics[width=0.9\textwidth]{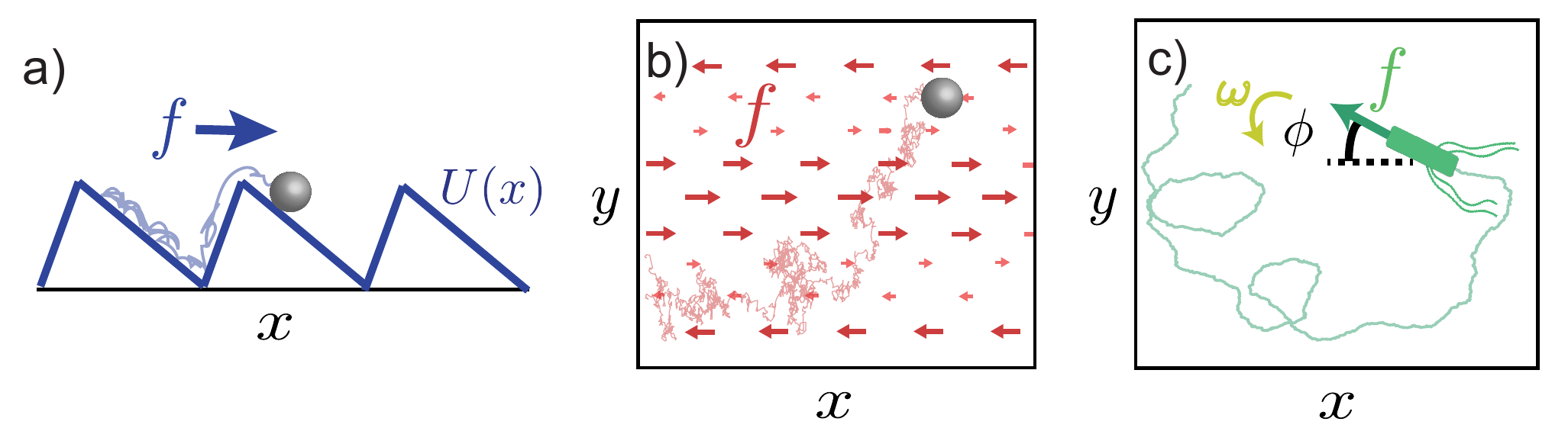}}
\caption{Illustration of three diffusion processes described by a Langevin equation of the type~\eqref{eq:LE}.
(a) A  Brownian particle  moves in a tilted periodic 1D sawtooth potential,
 $\dot{X}_t=\mu[f-\partial_X V(X_t)]+\sqrt{2D}\dot{B}_t$, with  $V(x)=(U_0\,x)/x^*$ for $x\in[0,x^*]$ and $V(x)=U_0 (1-x)/(1-x^*)$ for
 $x\in[x^*,1]$. (b)~Motion of a colloid in a 2D force field: $\dot{X}_t=\mu
 f\cos(2\pi Y_t)+\sqrt{2D}\dot{B}_{x,t}$ and $\dot{Y}_t=\sqrt{2D}\dot{B}_{x,t}$. (c)
 Chiral active Brownian motion described by the position coordinates
 $\dot{X}_t=\mu f \cos(\phi_t)+\sqrt{2D}\dot{B}_{x,t}$,
 $\dot{Y}_t=\mu f \sin(\phi_t)+\sqrt{2D}\dot{B}_{y,t}$ and the orientation angle
 $\dot{\phi}_t=\mu_\phi
 \omega+\sqrt{2D_\omega}\dot{B}_{\phi,t}$. See Ref.~\cite{Pigolotti:2017} for further details.
}
\label{fig:3l}
\end{figure}

In many physical situations, the mobility matrix $ \bmu_{t}$ and  the  diffusion matrix $\mathbf{D}_{t}$ depend on space.   Examples are, among others,  the Landau-Lifshitz-Bloch dynamics of a Brownian
spin~\cite{chetrite2009fluctuation} and the diffusion of water molecules near soft-matter phase boundaries~\cite{belousov2021statistical}. For simplicity, we provide in Fig.~\ref{fig:3l} three paradigmatic examples of diffusions that are relevant to physics.

For isothermal systems the mobility matrix $ \bmu_{t}(x)$ is related to the diffusion matrix $\mathbf{D}_{t}(x)$ by  {\bf Einstein's relation}
\begin{equation}
\mathbf{D}_{t}(x)=\frac{T}{2}\left(\bmu_{t}(x)+\left[\bmu_{t}(x)\right]^{\dagger}\right),
\label{ER}
\end{equation}
where $T$ is the temperature of the environment and we have used units for which the Boltzmann constant is equal to one.       Einstein's relation~\eqref{ER} states
that  friction (dissipation) and  noise (fluctuation) are two conjugated effects of the interaction
with the thermal bath.  For this reason, Eq.~\eqref{ER}   is  also called \textbf{fluctuation-dissipation} theorem~\cite{kubo1966fluctuation}.  When the Einstein relation \eqref{ER} holds,  we say that Eq.~\eqref{eq:LE} is an {\bf isothermal} Langevin equation, and when in addition the mobility matrix is symmetric, i.e. $\bmu_{t}=\left[\bmu_{t}\right]^{\dagger}$, then one retrieves
\begin{equation}
\mathbf{D}_{t}(x)=T\bmu_{t}(x)
\label{ER2}
\end{equation}
and  we say that Eq.~\eqref{eq:LE} is  an  {\bf isothermal overdamped} Langevin.  Note that   Einstein's relations do not apply  if the system interacts with multiple   thermal reservoirs at different temperatures, or if the system interacts with  a  thermal reservoir that is not at equilibrium.

The presence of the ``spurious" drift\footnote{The expression of the spurious drift depends of the convention chosen in the equation (\ref{eq:LE}).  In some references, it is claimed that the spurious term disappears in the anti-It\^{o} convention ($\alpha=1$ in Eq.~\eqref{eq:deltadis}) of the isothermal overdamped version of  (\ref{eq:LE}), and therefore  the convention $\alpha=1$ is often called the  isothermal convention.   Note however that the spurious drift disappears in the anti-It\^{o} convention {\em only for the case}   $\mathbf{D}_{t}(x)=g_{t}(x)\mathbf{D}_{t},$ with $g$ a scalar function  and $\mathbf{D}_t$ a space homogeneous matrix, i.e if all the $x$ dependence is in the scalar part.   The latter condition holds  in one dimension, but is not generally true for $d\geq 2$. See also~\cite{lanccon2001drift,volpe2010influence}. 
{An alternative perspective is to consider the Langevin equation~ (\ref{eq:LE}) as the zero correlation time limit of Eq.~(\ref{eq:LE})  but with a colored, Orsntein-Uhlenbeck noise, see~\cite{wong1965convergence,stratonovich1966new}. Note that the limiting equation has also in general  a non-vanishing spurious drift,  except for the one-dimensional  case if we choose to write the Eq.~(\ref{eq:LE})  in the Stratonovich convention. Such spurious drift is in general different to the spurious drift in Eq.~(\ref{eq:LE}). For the expression and the proof of such spurious drift in the general case, we refer to  Theorem 7.2 on page 497 of the book~\cite{ikeda2014stochastic}.}}term $\left(\nabla \mathbf{D}_{t}\right)(X_{t})$  in Eq.~(\ref{eq:LE}) {may appear exotic to readers, however we note that this term} ensures  {\bf thermodynamic consistency} {in the following sense: }%and  comes from the $x$ dependence of the diffusion matrix  
% \textst{Component-wise, this term reads as  $\left(\nabla\, \mathbf{D}_{t}\right)^{i}(x)=\sum^d_{j=1}\left(\partial_{j}\mathbf{D}_{t}(x)\right)^{ij}$.  }
  if we consider a  time-independent  potential $V_{t}=V$ and if we assume that the  Einstein relation holds, then  in  the absence of a
non-conservative force ($f=0$),   the term  $\left(\nabla \mathbf{D}_{t}\right)(X_{t})$ ensures that the stationary distribution of Eq.~(\ref{eq:FPI}) is the Boltzmann distribution 
\begin{equation}
  \rho_{\rm st}(x) =   \frac{\exp(-V(x)/T)}{Z},
\end{equation}
where $Z$ is the partition function,
  see Refs.~\cite{klimontovich1990ito,lau2007state} for details.     Note that if the  diffusion matrix $\mathbf{D}_t$ depends explicitly on time, then the process is stationary but time-inhomogeneous.

{In the following Chapters it will be useful to consider the following identity relating Stratonovich and Ito integrals associated with 
\begin{equation}
    \int_{0}^{t}g_s(X_{s})\circ\dot{X}_{s}ds =\int_{0}^{t}g_s(X_{s})\dot{X}_{s}ds+\int_{0}^{t}\textbf{D}_{s}(X_{s})\left[\left(\nabla g_s\right)(X_{s})\right]ds,
    \label{eq:StratotoItoLang}
\end{equation}
which is valid  for any  function $g_t(x)$ that is smooth on $x$ and $t$. This relation is a generalization of Eq.~\eqref{eq:stratotoito2} (see also Theorem~\ref{Strato2Ito}) to $d$ dimensions and a special case of the relation~\eqref{eq:stratoToIto} given in Appendix~\ref{app:strat}.}

 \subsection{$^{\spadesuit}$Stroock-Varadhan martingale problem } \label{sec:martingaleProblemCont}
 Martingales play a prominent role in the theory of continuous-time Markov processes because, among others, it is possible to  characterize  Markov processes using martingales, see Refs.~\cite{stroock2007multidimensional, ethier2009markov}.     This is proved rigorously in the Theorem on page 182 in~Ref.~\cite{ ethier2009markov}. Here we give an informal version of the theorem: 

\begin{leftbar}
\begin{theorem}[\textbf{Characterisation of Markov processes with martingales}]$\\$Let $X_t$ be a continuous-time stochastic process that takes values in $\mathcal{X}$. 
 The following two statements are equivalent: 
  \begin{itemize}
  \item $X_t$ is a Markov process with generator $\mathcal{L}_t$, i.e. its instantaneous density obeys $\partial_{t}\rho_{t}=\mathcal{L}_{t}^{\dagger}\rho_{t}$, see Eq.~\eqref{eq:FP}.
  {\item  The process 
  \begin{equation}
M_{t}=g_{t}(X_{t})-g_{0}(X_{0})-\int_{0}^{t}ds\left(\partial_{s}g_{s}+\mathcal{L}_{s}g_{s}\right)(X_{s}).
\label{eq:markovProtc}
\end{equation}
  is a martingale with respect to $X_t$ for all family of real-valued bounded functions $g_t(x)$ defined on $\mathcal{X}$.}
  \end{itemize}
  \label{th:3}
\end{theorem}
 {The martingale $M_{t}$ in  Eq.~\eqref{eq:markovProtc} is called {\bf Dynkin's martingale} associated with the function $g$.
Written in an infinitesimal way, the relation \eqref{eq:markovProtc} gives the {\bf generalized It\^{o} formula}
\begin{equation}
d\left(g_{t}(X_{t})\right)=\left(\partial_{t}g_{t}+\mathcal{L}_{t}g_{t}\right)(X_{t})dt+dM_{t}.
\label{eq:markovProtc2}
\end{equation}}
\label{Th:markov}
\end{leftbar}

Theorem~\ref{Th:markov} is useful in  at least two ways. First, given a Markov process  $X_t$, we can  construct   an arbitrary number of martingales by using different choices of the function~$g$ in   Eq.~\eqref{eq:markovProtc}.  Second, if one proves that the right hand side of~\eqref{eq:markovProtc}  is a martingale for all bounded functions $g$, then it is guaranteed that $X_t$ is a Markov process with  generator $\mathcal{L}_{t}$; { note that in continuous time it is non}. 
Despite its simplicity, Theorem~\ref{th:3}  is one of the most important results of probability theory, as it has no counterpart in the theory of ordinary or partial differential equations.   It was introduced in the late 1960s by D.W.~Stroock and S.R.S.~Varadhan, and it contributed to the boost of  martingales in  modern probability theory.

We  sketch the proof of the equivalence in the first direction, i.e., we show that  $M_t$ given by Eq.~\eqref{eq:markovProtc}  is  a martingale when $X_t$ is a Markov process.  Indeed, starting from Eq.~\eqref{eq:markovProtc}, we obtain:

\begin{equation}
M_{t}-M_{s}=g_{t}(X_{t})-g_{s}(X_{s})-\int_{s}^{t}du\left(\partial_{u}g_{u}+\mathcal{L}_{u}g_{u}\right)(X_{u}). \label{eq:stroock}
\end{equation}

Taking the expectation value  of Eq.~(\ref{eq:stroock}) conditioned on  $X_{[0,s]}$ yields
\begin{eqnarray}
\left\langle \left. \left(M_{t}-M_{s}\right)\right|X_{\left[0,s\right]}\right\rangle  &=& \left\langle \left.g_t(X_{t})\right|X_{\left[0,s\right]}\right\rangle -g_s(X_{s})-\int_{s}^{t}du\left\langle \left.\left(\partial_{u}g_{u}+\mathcal{L}_{u}g_u\right)(X_{u})\right|X_{\left[0,s\right]}\right\rangle \nonumber \\
 &=& \left\langle \left.g_t(X_{t})\right|X_s\right\rangle -g_s(X_{s})-\int_{s}^{t}du\left\langle \left.\left(\partial_{u}g_{u}+\mathcal{L}_{u}g_u\right)(X_{u})\right|X_{s}\right\rangle \nonumber \\
 &=& \left(\mathcal{P}_{s,t}\left[g_t\right]-g_s-\int_{s}^{t}du\mathcal{P}_{s,u}\left[\left(\partial_{u}+\mathcal{L}_{u}\right)g_u\right]\right)(X_{s})\nonumber \\
 &=& \left(P_{s,t}\left[g_t\right]-f_s-\int_{s}^{t}du  \: \partial_{u} \left( \mathcal{P}_{s}^{u} [g_u]\right)\right)(X_{s})\nonumber \\
 &=&0.  \label{eq:theabove}
\end{eqnarray}
The second equality of Eq.~(\ref{eq:theabove}) follows from the Markov property, the third equality comes first from the definition~Eq.~\eqref{eq:PtTPrime} of the transition function. In particular, in this equality,  we have used for the first term the relation
\begin{equation}
    \left\langle \left.g_t(X_{t})\right|X_{s}\right\rangle = \int dy \,  \mathcal{P}_{s,t} (X_s,y) g_t(y)  =  \mathcal{P}_{s,t} [g_t] (X_s),
\end{equation}
which follows from the convention given by Eq.~\eqref{eq:2p75}, 
and we have proceeded analogously for the third term.
Finally, the fourth equality follows from the forward Kolmogorov equation \eqref{eq:KFE} $\partial_{u}P_{s,u}=P_{s,u}\mathcal{L}_{u}$,  fulfilled by the transition probability.

 \subsection{Dynkin's Martingales }\label{sec:dynkinCont}

For each function $g_t$, 
Equation~\eqref{eq:markovProtc} provides us  a recipe to construct  a martingale $M_t$ associated with a given a Markov process $X_t$. Hence we can use Eq.~\eqref{eq:markovProtc} to either systematically construct martingales in  Markov processes, or to show whether a given process $g_t(X_t)$ is a martingale or not.   We call  martingales of the form   \eqref{eq:markovProtc}  \textbf{Dynkin's additive martingales}.
Note that $g_t$ does not necessarily need to be a bounded function  to be a martingale, but it is  sufficient  to guarantee that $\langle |M_t|\rangle<\infty$.   We illustrate some examples below.

\subsubsection{Dynkin's Martingales associated with jump processes}
If   $X_t$ is a Markov  jump process, as defined in Sec.~\ref{eq:jumpprocconta}  with generator given by Eq.~\eqref{eq:genps},  then  Dynkin's Martingales~\eqref{eq:markovProtc} takes the  expression
\begin{equation}
M_{t}=g_{t}(X_{t})-g_{0}(X_{0})-\int_{0}^{t}ds\left(\left(\partial_{s}g_{s}\right)(X_{s})+\int_{\mathcal{X}}dy\,\omega_{s}\left(X_{s},y\right)\left(g_{s}(y)-g_{s}(X_{s})\right)\right),
\end{equation}
where we recall that $g_t(x)$ here is an arbitrary real function of $t\geq 0$ and  of $x\in\mathcal{X}$. To illustrate how martingales can be constructed with Dynkin's formula \eqref{eq:markovProtc}, we  give some explicit  examples.
\begin{enumerate}
\item \textbf{A Dynkin Martingale associated with the Poisson process.} For a Poisson process $N_t$ with time-dependent transition rate $\lambda_t$, i.e $\mathcal{X}=\mathbb{N}$, $N_0=0$,  and $\omega_t(n,n+1)=\lambda_t$ with $n\in\mathbb{N}$,  the associated Dynkin's martingale is given by
\begin{equation}
M_{t}=g_{t}(N_{t})-g_{0}(0)-\int_{0}^{t}ds\Big(\left(\partial_{s}g_{s}\right)(N_{s})+\lambda_{s}\left(g_{s}(N_{s}+1)-g_{s}(N_{s})\right)\Big).\label{eq:DM-1-2}
\end{equation}
In particular, the Dynkin martingale associated with the function $g_t(n)=n$ takes the simple expression
\begin{equation}
M_{t}=N_{t}-\int_{0}^{t}ds\lambda_{s}, \label{eq:rateDepDynkin1}
\end{equation}which  is a generalization of the martingale Eq.~(\ref{eq:countingM}) for time-independent rates $\lambda_t=\lambda$.
Analogously, using $g_t(n)=n^2$, we obtain the martingale  
\begin{equation}
M_{t}=N_{t}^{2}-\int_{0}^{t}ds\:\lambda_{s}\left(2N_{s}+1\right).\label{eq:DM-1-2-1}
\end{equation}
\item {\bf  A Dynkin Martingale associated with a three-state model.} Let $X_{t}$ be a 3-state continuous-time Markov jump process defined on $\mathcal{X}=\left\{ A,B,C\right\} $
and with time independent transition rates $\omega(X,Y)$ between states $X\in \mathcal{X}$ and  $Y\in \mathcal{X}$ (see Fig.~\ref{fig:3s} for an illustration). For $X_{0}=A$, and for the choice $g_{t}(x)=\delta_{C,x}$, with $\delta_{i,j}$  the Kronecker delta, the associated Dynkin martingale is given by
\begin{equation}
M_{t}=\delta_{C,X_{t}}-\int_{0}^{t}ds\,\omega(X_{s},C)+\left(\omega(C,B)+\omega(C,A)\right)\int_{0}^{t}ds\,\delta_{C,X_{s}}.
\label{eq:mart3s}
\end{equation}
The second term in~\eqref{eq:mart3s} is the accumulated {\em inflow probability}~\cite{baiesi2015inflow} to state $C$. On the other hand, the third term in~\eqref{eq:mart3s} depends on the escape rate from state~$C$, $\lambda(C) =\omega(C,B)+\omega(C,A)$, and also on the  empirical occupation probability $(1/t)\int_{0}^{t}ds\delta_{.}(X_{s})$ of state  $C$.  
\end{enumerate}

\subsubsection{Dynkin's Martingales associated with Langevin dynamics}

If $X_t$ is a Markov diffusion process defined by the Langevin equation~\eqref{eq:LE}, then using the explicit expression \eqref{eq:gpdL} of the generator, we obtain  Dynkin's martingales of the form: 
\begin{equation}
M_{t}=g_{t}(X_{t})-g_{0}(X_{0})
-\int_{0}^{t}ds
\left(
\partial_{s}g_{s}+(\bmu_{s}F_{s})\nabla g_{s}
+\nabla\left(\mathbf{D}_{s}\,\nabla g_{s}\right)
\right)(X_{s}). 
\label{eq:DMDP}
\end{equation}
We now provide few illuminating physical examples of the Martingales~\eqref{eq:DMDP}.

\begin{enumerate}
\item \textbf{Martingales associated with  Brownian motion.}  For a Wiener process $X_t=B_t$ with $B_0=0$,     Dynkin's martingales are given by
\begin{equation}
M_{t}=g_{t}(B_{t})-g_{0}(0)
-\int_{0}^{t}ds
\left(
\partial_{s}g_{s}+\frac{1}{2}\partial_{xx} g_{s}
\right)(B_{s}). 
\label{eq:DMW}
\end{equation} 
For example,  the Dynkin martingale associated to the functions $g_t(x)=x$, $g_t(x)=x^2$ and $g_t(x)=x^3$ are, respectively, the martingales $ B_{t}$, $B_{t}^2-t$ and $B_{t}^3 - 3tB_{t}$ given   by, respectively,  Eqs.~\eqref{trieste1},~\eqref{trieste2} and~\eqref{trieste3}, the    key examples  presented in Sec.~\ref{sec:examplescont} of   martingales associated with the Wiener process.

\item \textbf{Martingales associated with Brownian motion in two dimensions. } For $X_t=(B_{x,t},B_{y,t})^T$  a two-dimensional Brownian motion   with $B_{x,t}$ and  $B_{y,t}$ two independent Wiener processes, and initial condition $X_0=(0,0)^T$,     Dynkin's martingales are given by
\begin{equation}
M_{t}=g_{t}(B_{x,t},B_{y,t})-g_{0}(0,0)
-\int_{0}^{t}ds
\left(
\partial_{s}g_{s}+\frac{1}{2}\partial_{xx} g_{s}+\frac{1}{2}\partial_{yy} g_{s}
\right)(B_{x,t},B_{y,t}). 
\label{eq:DMW}
\end{equation} 
For example,   Dynkin's martingale associated with the function   $g_t(x,y)=x^2 y^2$  is given by
\begin{equation}
    M_t=B_{x,t}^2 B_{y,t}^2-\int_{0}^{t}ds\left( B_{x,s}^2+B_{y,s}^2\right).
\end{equation}

\end{enumerate}

\subsubsection{Dynkin's martingales associated with %space-time
harmonic functions}
 Dynkin's martingale construction implies  that processes of the form $h_t(X_t)$, with $h_t(x)$  a  harmonic function are martingales, for generic Markovian $X_t$.  
We say that $h_t(x)$ is a space time {\it harmonic} function if 
\begin{eqnarray}
\partial_{s}h_{s}+\mathcal{L}_{s}h_{s}=0. 
\label{eq:harmtc}
\end{eqnarray}  

As a key example, consider the one-dimensional Brownian motion $B_t$ (Wiener process), whose Markovian generator is   $\mathcal{L}=\frac{1}{2}\partial_{xx}$. An example of a space-time harmonic function associated with this generator is $h_{t}(x)=x^{2}-t$. Then we recover again that $h_t(B_t) = B_{t}^2-t$ is a martingale with respect to the Wiener process.  

Similarly,  $h_{t}(x)=\exp\left(z x-\frac{1}{2}z^{2}t\right)$  is an space-time harmonic function for  $z$ any real number.  From this result, we recover  that  the stochastic exponential of $B_t$ introduced in~\eqref{eq:BExp},  $h_t(B_t)=\exp\left(z B_t-\frac{1}{2}z^{2}t\right)$,   is a martingale with respect to the Wiener process for all values of $z$.
 
The harmonic martingales martingale $h_t(X_t)$, with $h_t(x)$  a space-time harmonic function,  play a crucial role in Doob's conditioning theory, which has applications in control theory~\cite{chetrite2015nonequilibrium}.

  \subsection{Exponential martingales}\label{sec:exponCont}
  
  %We briefly review the concept of multiplicative martingales. 
  For all real-valued, bounded functions $g$, the continuous version of the discrete time multiplicative martingale \eqref{eq:marfomu} is the exponential martingale 
\begin{equation}
M_{t}=\left(g_{0}(X_{0})\right)^{-1}\exp\left[-\int_{0}^{t}ds\left(g_{s}^{-1}\left(\partial_{s}g_{s}\right)+g_{s}^{-1}\mathcal{L}_{s}\left[g_{s}\right]\right)(X_{s})\right]g_{t}(X_{t}). \label{eq:multmartcont}
\end{equation}
A canonical example of an exponential martingale is the stochastic exponential ~\eqref{eq:BExp}, corresponding to the choices $g_{t}(x)=\exp\left(zx\right)$ and $X_t=B_t$ the Wiener process. Multiplicative  martingales~\eqref{eq:multmartcont} have important applications in nonequilibrium physics. As a matter of fact, it was shown in~\cite{Chetrite:2011} that the martingale condition $\langle M_t | X_{[0,s]} \rangle = M_s$ for the martingales~\eqref{eq:multmartcont} is a non perturbative version of the fluctuation-dissipation theorem. Notably, these martingales are also related to the conditioning theory on rare events~\cite{chetrite2013nonequilibrium,chetrite2015nonequilibrium,chetrite2018peregrinations}.  

{To prove that $M_t$ in Eq.~\eqref{eq:multmartcont} is a martingale  we first differentiate the process
\begin{equation}
dM_{t}=-M_{t}\left(f_{t}^{-1}(\partial_{t}f_{t})+f_{t}^{-1}L_{t}\left[f_{t}\right]\right)\left(X_{t}\right)dt+M_{t}f_{t}^{-1}\left(X_{t}\right)d\left(f_{t}\left(X_{t}\right)\right).
\label{eq:dynkin11}
\end{equation}
Next, we apply the generalized It\^{o} formula~\eqref{eq:markovProtc2} to the function $f_t(X_t)$, which reads
\begin{equation}
d\left(f_{t}(X_{t})\right)=\left(\partial_{t}g_{t}+\mathcal{L}_{t}g_{t}\right)(X_{t})dt+dM_{t}^{g},
\label{eq:dynkin12}
\end{equation}
where we note that $M_{t}^{g}$ is the Dynkin's martingale associated with $g_t$ [see Eq.~\eqref{eq:markovProtc}].
Combining Eqs.~(\ref{eq:dynkin11}-\ref{eq:dynkin12}) we obtain
\begin{equation}
dM_{t}=M_{t}g_{t}^{-1}\left(X_{t}\right)dM_{t}^{g}.
\label{eq:dynkin13}
\end{equation}
Equation~\eqref{eq:dynkin13} implies that $M_t$ is a martingale because $M_t^{g}$ is a martingale and the fact that $M_t$ is an It\^{o} integral of the form~\eqref{eq:Ito2}. Or equivalently, because $M_g$ is a stochastic exponential that we will introduce later in Sec.~\ref{sec:stochExp}.}

  \subsection{Path probability ratios} \label{sec:radonNikodym} 
   As a last step in our "world tour"  on the relation between  Markov processes and martingales in continuous time, we present explicit expressions for the  path probability ratios $R_t$, as defined in Eqs.~(\ref{eq:radonDef2}-\ref{eq:radon}).  To this aim, we use the Onsager-Machlup approach that represents  measures $\mathcal{P}$ that belong to a class $\mathscr{P}$ of mutually absolutely continuous measures   with action functionals $\mathcal{A}(X_{[0,t]})$ (see the discussion around Eqs.~(\ref{eq:onsagerMachlupApproach1}-\ref{eq:onsagerMachlupApproach3})).

 \subsubsection{Markov jump processes} 
 
We consider a Markov jump process with transition rates   $\rateWt(x,y)$, that determine its  generator through Eq.~(\ref{eq:genps}).   In addition, we  assume that the  initial distribution  is   $\rho_0(x)$.    Recall that the trajectories $\Xot$  of Markov Jump processes  are   piecewise constant functions with  consecutive states $X_i$ and jump times $\mathcal{T}_i$, see  Eq.~(\ref{eq:defsXMJP}).  
  
The action functional associated with  a Markov jump process is 
\begin{equation}
\mathcal{A}(X_{[0,t]}) = -\ln \left(     \rho_0(X_0)\right)  -\sum^{N_t}_{j=1} \ln \left(  \omega_{\mathcal{T}_{j}}(X_{\mathcal{T}_{j}^{-}}, X_{\mathcal{T}_{j}^{+}})\right)+ \int^{t}_0 ds \lambda_s(X_s),\label{eq:RFormCont}
\end{equation}
where  we have used 
\begin{equation}
\lambda_t(x) = \int_{y\in\mathcal{X}} dy\: \omega_{t}(x,y),
\end{equation}
for the exit rate from state $x$.
 
%{\textcolor{red}{With the title of the section ''Path probability ratios'' : we must add one sentence for explain that the ratio martingale will be given as the exponential of the difference of the 2 actions. }}

\subsubsection{Diffusion processes}
We consider  diffusion processes $X_t\in \mathbb{R}^d$ defined through their generator, given by Eq.~(\ref{eq:gpdL}) with  mobility matrix $\bmu_t\in \mathbb{R}^{d^2}$,  force vector $F_t\in \mathbb{R}^d$, and diffusion matrix $\bD_t\in \mathbb{R}^{d^2}$, all of which are time-dependent.    Also, we consider  initial distributions  $\rho_0(x)$.  Note that     absolute continuity of the measures $\mathcal{P}$ in $\mathscr{P}$ requires, among others,  that all measures  have the same diffusion matrix $\bD_t$. 

If the diffusion matrix $\mathbf{D}_t$ is independent of $X$, then the action takes the form  \cite{stratonovich1971probability,
langouche2013functional}
\begin{eqnarray}
\mathcal{A}(X_{[0,t]}) &=&     -\ln \rho_0(X_0)  - \frac{1}{2} \int^t_0  \left(\mathbf{D}^{-1}_s   \bmu_{s} F_{s}\right)(X_s) \circ \dot{X}_{s} ds
\nonumber\\ 
&&\quad \quad  \quad  + \frac{1}{4}\int^t_0 \left(\bmu_s F_s\right) \cdot  \left(\mathbf{D}^{-1}_s \bmu_s F_s\right)(X_s)  ds + \frac{1}{2}\int^t_0 \nabla\cdot \left( \bmu_s F_s \right)(X_s)ds, \nonumber \\ \label{eq:additiveA}
\end{eqnarray}
where the last term in Eq.~(\ref{eq:additiveA}) appears due to the   Stratonovich convention used in the first integral. We use the Stratonovich convention here as this convention will prove to be useful in physics because of its properties under time reversal.   An alternative  form of the action $\mathcal{A}$, more commonly used in physics, reads
\begin{eqnarray}
\mathcal{A}(X_{[0,t]}) &=&     -\ln \rho_0(X_0)  +  \int^t_0 \frac{1}{4}  \left( \dot{X}_{s}-\bmu_s (X_s) F_s (X_s)  \right) \mathbf{D}^{-1}_s  \left( \dot{X}_{s}-\bmu_s (X_s) F_s (X_s)  \right)  ds
\nonumber\\ 
&+&  \frac{1}{2} \int^t_0  \nabla\cdot \left( \bmu_s F_s \right)(X_s) ds,
 \label{eq:additiveA2}
\end{eqnarray}
which is equivalent to Eq.~(\ref{eq:additiveA}), as in Eq.~(\ref{eq:onsagerMachlupApproach1}) the    $\dot{X}_s\mathbf{D}^{-1}_s \dot{X}_{s}$ can be absorbed into the normalisation constant $\mathcal{N}$; note that this is possible as  $\mathbf{D}_t$ is the same for all measures $\mP\in \mathscr{P}$.   However,  a complication with  Eq.~(\ref{eq:additiveA2}) is that the  mathematical meaning of $\dot{X}_s\mathbf{D}^{-1}_s \dot{X}_{s}$ is not clear, even though this term, whatever it signifies, disappears when taking the ratio between $\mP$ and another measure $\mQ\in  \mathscr{P}$. 

The action can also be expressed as 
\begin{equation}
    \mathcal{A}(X_{[0,t]}) = -\ln \rho_0(X_0) + \int^{t}_0 ds 
    \: \mathscr{L}_s(X_{s}, \dot{X}_s),
    \label{action}
\end{equation}
in terms of a Lagrangian
\begin{equation}
\mathscr{L}_s(X_{s}, \dot{X}_s) \equiv  
\frac{1}{4}  \left( \dot{X}_{s}-\bmu_s (X_s) F_s (X_s)  \right) \mathbf{D}^{-1}_s  \left( \dot{X}_{s}-\bmu_s(X_s) F_s (X_s)  \right)  
+  \frac{1}{2} \nabla\cdot \left( \bmu_s F_s \right)(X_s).
 \label{eq:LagrangianFormula}
\end{equation}

If $\bD_t$ depends on $X_t$, then the mathematical meaning of the action $\mathcal{A}$ is less simple,  see e.g.,~Refs.~\cite{arnold2000symmetric, langouche2013functional, lau2007state}. In this case,  we  directly consider  the Radon-Nikodym derivative process $R_t$ of $\mathcal{Q}$ with respect to $\mathcal{P}$  that reads, see  Refs.~\cite{stroock2007multidimensional, barato2015formal} 
  \begin{eqnarray} 
R_t &=& \frac{\mathcal{Q}(X_{[0,t]})}{\mathcal{P}(X_{[0,t]})} =   \frac{\rho^\mathcal{Q}_0(X_0)}{\rho^\mathcal{P}_0(X_0)} \exp\left( \frac{1}{2} \int^t_0  \left( (\bmu_{s}^{\mathcal{Q}}F_{s}^{\mathcal{Q}}-\bmu_{s}^{\mathcal{P}}F_{s}^{\mathcal{P}}) \right)(X_s)\cdot \mathbf{D}^{-1}_s(X_s)\circ \dot{X}_s\right) ,  \nonumber \\ 
&\times&\exp\left(-\int^t_0 \frac{1}{4} (\bmu_{s}^{\mathcal{Q}}F_{s}^{\mathcal{Q}}-\bmu_{s}^{\mathcal{P}}F_{s}^{\mathcal{P}})(X_s) \cdot \mathbf{D}_s^{-1}(X_s) 
\left(
\bmu_{s}^{\mathcal{Q}}F_{s}^{\mathcal{Q}}+\bmu_{s}^{\mathcal{P}}F_{s}^{\mathcal{P}}-2\nabla\cdot\mathbf{D}_s
\right)(X_s)
ds\right)
\nonumber\\ 
&\times& \exp\left(-\int^t_0 \frac{1}{2} \nabla \cdot \left[ \bmu_{s}^{\mathcal{Q}}F_{s}^{\mathcal{Q}}-\bmu_{s}^{\mathcal{P}}F_{s}^{\mathcal{P}}\right](X_s)ds \right). 
\label{eq:RFormContx}
\end{eqnarray}
Here, it should be understood that  $\rho^{\mathcal{P}}_0$ ($\rho^{\mathcal{Q}}_0$),  $\mathbf{D}^{\mathcal{P}}_s$ ($\mathbf{D}^{\mathcal{Q}}_s$),  $\bmu_{s}^{\mathcal{P}}$ ($\bmu_{s}^{\mathcal{Q}}$), and  $F_{s}^{\mathcal{P}}$ ($F_{s}^{\mathcal{Q}}$ ) determine $\mathcal{P}$ ($\mathcal{Q}$).

\chapter{Martingales:  Mathematical properties}
\label{ch:3}

\epigraph{There exist only two kinds of modern mathematics books: one which you cannot read beyond the first page and one which you cannot read beyond the first sentence.\\ \vspace{0.1cm} Cheng Ning Yang. Physics Nobel prize 1957.}

 The properties of {martingales} are rich, encompassing various branches of mathematics, see, e.g.,~the textbooks~\cite{williams1979diffusions,williams1991probability,lip2013}.   Instead of giving  a complete overview of martingale theory, we   focus  on those   properties that we think are important for physics.  To this aim, we are guided by  recent works on martingales in physics, which we review in later chapters.  
 
 This Chapter is divided into two main sections: Sec.~\ref{sec:discretetime} deals with martingales in discrete time, and Sec.~\ref{sec:continuoustime} deals with martingales in continuous time.   

\section{Discrete time}\label{sec:discretetime}
\subsection{Relating  submartingales to martingales}

  If $M_n$ is a positive martingale, then $-\ln M_n$ is a submartingale. 
This is because $-\ln(x)$ is convex (its second order derivative is nonnegative), and a convex function of a submartingale is a submartingale,  as formulated by the following theorem. 
\begin{leftbar}

  \begin{theorem}[{\bf Convex nondecreasing functions of submartingales}]$\\$ \label{subM}
  Let $S_n$ be a submartingale and let $f(x)$ be a real-valued function defined on $\mathbb{R}$ that is nondecreasing, convex, and $\langle |f(S_n)|\rangle < \infty $ for all $n\in\mathbb{N}$.   Then, the process $f(S_n)$ is a submartingale.
  \end{theorem}
  \end{leftbar}
  \begin{proof} 
  We verify condition (\ref{eq:SCond}) that appears in the definition of the submartingale.   First, we apply 
 Jensen's inequality to the average of the convex function $f$, leading to
 \begin{eqnarray}
 \langle  f(S_n) | X_{[0,m]}\rangle  \geq f\left( \langle  S_n| X_{[0,m]}\rangle \right) .\label{eq:CFirst}
 \end{eqnarray}
Subsequently, we use that $S_n$ is a submartingale, 
 \begin{eqnarray}
\langle  S_n| X_{[0,m]}\rangle \geq   S_m
 \end{eqnarray}
 and  that $f$ is nondecreasing
  \begin{eqnarray}
 f\left( \langle  S_n|X_{[0,m]}\rangle \right) \geq   f(S_m). \label{eq:CLast}
 \end{eqnarray}
 Relation (\ref{eq:CFirst}) together with (\ref{eq:CLast}) implies that $f(S_n)$ has a nonnegative drift and is thus a submartingale.
    \end{proof} 
   
We will use Theorem~\ref{subM}  in Chapters~\ref{sec:martST1} and \ref{sec:martST2} to derive the submartingale property of entropy production and the second law of thermodynamics. 

If $S$ is a submartingale and {its} mean value is a constant independent of time, i.e., 
\begin{equation}
    \langle S_t\rangle = c, \label{eq:constCond}
\end{equation}
 then $S_t$ is a martingale.  Indeed,  the process
\begin{equation}
    \langle S_t|X_{[0,s]}\rangle - S_s , \quad t\geq s,
\end{equation}
for fixed $s$, 
  is a nonnegative process with  zero expectation, and thus  
     \begin{equation}
        \langle S_t|X_{[0,s]}\rangle  = S_s. 
\end{equation}

 \subsection{Doob's decomposition theorem} 
  We call a stochastic process $A_n$ {\bf predictable} if $A_n$ is a function of $X_{[0,n-1]}$ and we say that a  process is {\bf increasing} if with probability one $0 = A_0 \leq A_1 \leq A_2 \cdots$.    
    
It is always possible to decompose a submartingale into a martingale and an increasing process that is predictable, and this decomposition is unique  (see Theorem 2.13 in Ref.~\cite{lip2013}).

 \begin{leftbar}
 \begin{theorem}[\textbf{Doob's decomposition}]  \label{doobD}
 Let $Y_n$ be a discrete-time process that is a function of the set of trajectories $X_{[0,n]} = (X_0, X_1, \ldots, X_n)$, and integrable (i.e., $\langle  |Y_n| \rangle<\infty$ for all $n$). Then  it can be uniquely decomposed as
 \begin{eqnarray}
   Y_n = Y_0 + M_n + \underbrace{\sum_{k=0}^{n-1} v_k}_{\displaystyle  A_n},
  \end{eqnarray}
  where we have introduced the {\bf conditional velocity}
  \begin{equation}
     v_k =  \left \langle  \left(Y_{k+1}-Y_k\right)  | X_{[0,k]}  \right \rangle .
  \end{equation}
   The predictable process $A_n$ is  called the  {\bf compensator} and $M_n$, defined as
   \begin{equation}
 M_{n+1} - M_n  = Y_{n+1}-Y_n - v_n,
  \end{equation}
  is a martingale with respect to the underlying process $X_n$, i.e., $\langle M_n|X_{[0,m]}\rangle = M_m$, for $m\leq n$.
  \end{theorem}
  \end{leftbar}

If $Y_n$ is a submartingale (supermartingale), then  $v_k\geq 0$ ($v_k\leq 0$), and  the compensator $A_n$ is increasing (decreasing).  In Theorem~\ref{doobD} "unique" means that   if there exist two Doob  decompositions  $S_n = M_n+A_n$ and  $S_n = M'_n+A'_n$, then   for all $n \in \mathbb{N}$ one has $\mP(M_n = M'_n,A_n=A'_n) = 1$ .

{\bf Example 1: Doob decomposition for the square of a stochastic process}\\  We consider the Doob decomposition of the square $Y_n=Z_n^2$  of a discrete-time process $Z_n$, namely,
\begin{eqnarray}
Z^2_n =Z_0^2 +  M_n +  \sum_{k=0}^{n-1} v_k, \label{eq:doobDecomp1}
\end{eqnarray}
where  
\begin{eqnarray}
\begin{cases}
  v_k \equiv  \left \langle \left( Z^2_{k+1}-Z^2_k\right)  |X_{[0,k]} \right\rangle, \\
   M_{n+1}   \equiv   M_n + Z^2_{n+1}-Z^2_n - v_n.
  \end{cases}
\end{eqnarray}
The  velocity $v_n$   is called  the {\em  angle bracket}  process of $Z_n$. Now, if additionally $Z_n$ is a martingale with respect to $X_n$, then the angle bracket process satisfies 
\begin{eqnarray}
    v_k &=& \left \langle \left( Z^2_{k+1}-Z^2_k\right)  | X_{[0,k]}   \right\rangle\\
    &=&   \left \langle \left( Z_{k+1}-Z_k\right)^2  | X_{[0,k]}   \right\rangle + 2 \left \langle  Z_{k+1} Z_k   | X_{[0,k]}  \right\rangle - 2 \left \langle   Z_k^2   | X_{[0,k]}  \right\rangle \\
    &=& \left \langle \left( Z_{k+1}-Z_k\right)^2  | X_{[0,k]}   \right\rangle ,
\end{eqnarray}
where we have used the martingale property of $Z_n$ in the third equality. Processes of the type $\left \langle \left( Z_{k+1}-Z_k\right)^2  | X_{[0,k]}   \right\rangle$ are often called {\em sharp bracket} processes and the associated  compensator, which is also called the {\em  conditional variance} of the $Z_n$, reads
\begin{equation}
V_n = \sum_{k=0}^{n-1} v_k = \sum_{k=0}^{n-1} \left \langle 
\left( Z_{k+1}-Z_k \right)^2  | X_{[0,k]}   \right\rangle.
\label{eq:condvar}
\end{equation} Note that   if $Z_n$ is a martingale, then by virtue of Theorem~\ref{subM}  $Z_n^2$ is a submartingale  with respect to $X_n$.

{\bf Example 2: Doob decomposition for a function of a Markov chain $X_{n}$. }
Theorem \ref{th:1}  directly gives   the Doob decomposition of $f\left(X_{n}\right)$ for all real-valued bounded functions $f$, viz.,
 \begin{eqnarray}
  f(X_n)=  f(X_0)  + M_n+ \underbrace{ \sum^{n-1}_{m=0} \sum_{x\in\mathcal{X}} \left(w(X_m,x) - \delta_{x,X_m}\right) f(x)}_{\displaystyle  A_n} \label{eq:funcmarkovChain}, 
  \end{eqnarray}
  where $M_n$ is  Dynkin's additive martingale, as defined in (\ref{eq:markovChain}),  and  $w(x,y)$ is the transition matrix of the Markov chain $X_{n}$.

\subsection{Extreme values}

\subsubsection{Doob's maximum inequality}
Let $A$ be a positive random variable.   Markov's inequality states that 
\begin{eqnarray}
\mathcal{P}\left(A\geq \lambda\right) \leq  \frac{\langle A \rangle}{\lambda}. \label{eq:PA}
\end{eqnarray}  
Doob's maximum inequality is a refinement of Markov's inequality that  involves the supremum of a  submartingale. More precisely the following theorem holds:
%Remarkably, for positive submartingales we can replace  in the righ-hand side of Eq.~(\ref{eq:PA}) $S_n$ by the  supremum of  $S_n$: 
\begin{leftbar}
    \begin{theorem}[{\bf Doob's maximum inequality}] \label{DoobMAx} $\\$
  Let $S_n$ be a  submartingale.   Then, 
  \begin{eqnarray}
\mathcal{P}\left({\rm sup}_{m\leq n}S_{m}  \geq \lambda\right) \leq  \frac{\langle {\rm  max}\left\{S_n,0\right\} \rangle}{\lambda}, \label{eq:Doobmax}
\end{eqnarray}
where $\lambda\geq 0$. %$\\$ [{\bf IN comment: If $S_n$ can have negatie values, then  $\mathbb{P}\left[{\rm sup}_{n'\leq n}S_{n'}  \geq \lambda\right] \leq  \frac{\langle {\rm  max}\left\{S_n,0\right\} \rangle}{\lambda}$, which is the formula used in Figure 3.1. Should we present this formula here?  It appears to be more useful when applying to random walks.}]
\end{theorem}
\end{leftbar}
\begin{proof}
We consider the sequence of sets 
\begin{eqnarray}
\Phi_1 &=& \left\{S_1>\lambda\right\},\\ 
\Phi_2 &=&\left\{S_1\leq \lambda, S_2>\lambda\right\},  \\ 
&\vdots& \nonumber\\ 
\Phi_k &=&\left\{S_1\leq \lambda, S_2 \leq \lambda, \hdots S_{k-1}\leq \lambda, S_k>\lambda\right\},
\end{eqnarray}
with $\lambda > 0$ and $k\geq 2$.
Doob's maximum inequality follows from the following inequalities: 
\begin{eqnarray}
\langle S_n \rangle &\geq& \sum^n_{k=1} \mathcal{P}\left(\Phi_k\right) \langle S_n|\Phi_k \rangle \label{eq:doobMax1} \\ 
 &\geq& \sum^n_{k=1} \mathcal{P}\left(\Phi_k\right) \langle S_k|\Phi_k \rangle \label{eq:doobMax2} \\ 
 &\geq& \lambda  \sum^n_{k=1} \mathcal{P}\left(\Phi_k\right)  = \lambda \:\mathcal{P}\left( {\rm sup}_{n'\leq n}S_{n'}  \geq \lambda \right).\label{eq:doobMax3} 
\end{eqnarray} 
The first inequality (\ref{eq:doobMax1})  follows from the fact that $S_n$ is nonnegative.   The second inequality 
(\ref{eq:doobMax2})  holds because  $S$ is a submartingale.   Finally, the last inequality is a consequence of the definition of the sets $\Phi_k$.
\end{proof}    

Note that in discrete time, the supremum can be replaced by the maximum, whereas in continuous time this will not be the case.     

In Chapter~\ref{sec:martST2} we use Doob's maximum inequality to derive the infimum law for entropy production.

\subsubsection{Application: Extreme values of random walkers}

\begin{figure}[h!]
\centering
\includegraphics[width=\textwidth]{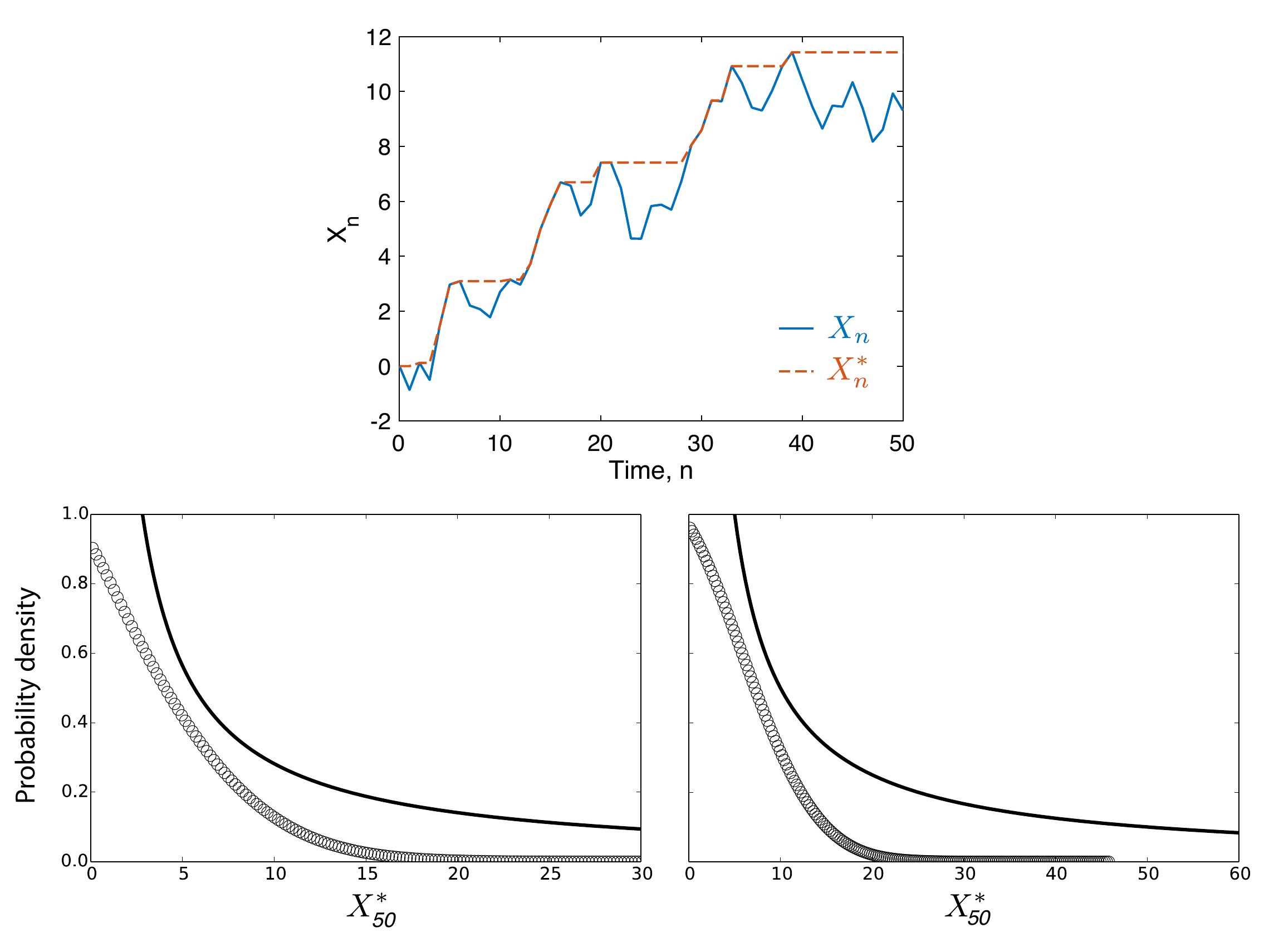}
\caption{Top: Example trajectory $X_n$ (blue line) of a discrete-time random walker on~$\mathbb{R}$ given by $X_n = X_{n-1} + a + Y_n $, with $X_0=0$, $a=0.1$, and $Y_n$ ($n\geq 1$) a Gaussian random number with zero mean and unit standard deviation. The red dashed line is the running maximum $X^\ast_n = {\rm max}_{n'\leq n}X_{n'}$ associated with the trajectory, see Eq.~\eqref{eq:runmax}.
Bottom: Distributions $\rho_{X^\ast_n}$ of the maximum $X^\ast_n$  of a random walker on the real line with parameters $n=50$, $a=0$ (bottom left) and $a=0.1$ (bottom right) and  with $Y$ a Gaussian random variable with zero mean and unit variance.    Markers are  the numerical results of the distribution $\rho_{X^\ast_n}$ and the solid line is the martingale bound given by Eq.~(\ref{eq:Doobmax}) with $\langle {\rm max}\left\{X_n,0\right\}\rangle = \sqrt{n/2\pi}$ (left) and $\langle {\rm max}\left\{X_n,0\right\}\rangle = an+O(\sqrt{n}\exp(-a^2n/2))$ (right).  }\label{fig:martDistri}
\end{figure}

Doob's maximum inequality can be used to bound the cumulative distribution of extreme values of stochastic processes, which have attracted considerable attention in various scientific disciplines such as statistical physics~\cite{bouchaud1997universality,krug2007records,eliazar2019gumbel,majumdar2020extreme}, climate science~\cite{cheng2014non,schar2016percentile}, and finance~\cite{poon2004extreme}.  
Here, for illustrative purposes, we consider the discrete-time random walk  $X_n = X_{n-1} + a + Y_n $, as defined in Eq.~(\ref{eq:randomWalkR}), for different values of $a$ and  with the noise variable  $Y_n$  a random variable with zero mean and finite variance, see Fig.~\ref{fig:martDistri}.

It is in general difficult to obtain an exact expression for the cumulative distribution of the finite-time maximum 
\begin{equation}
    X^\ast_n = {\rm max}_{n'\leq n}X_{n'}. 
    \label{eq:runmax}
\end{equation} 
For example, for the special case of $a=0$ the cumulative distribution of the maximum is described by the  {Pollaczek-Spitzer}  formula~\cite{pollaczek1952fonctions,spitzer1956combinatorial,spitzer1957wiener, majumdar2010universal}.   The quantity $q_n(\lambda) = 1-\mP\left[X^\ast_n \geq \lambda\right]$  denotes the probability that the process stays below the threshold $\lambda$, and therefore we call it  the survival probability.  The  {Pollaczek-Spitzer}  formula   provides a formula for the double inverse Laplace transform of the survival probability in terms of the Fourier transform  $\phi(k) = \int^{\infty}_{-\infty} \rho_Y(y) \exp({\rm i}ky)dy$ of the distribution $\rho_Y$ of the increment~\cite{majumdar2010universal}, viz.,
\begin{equation}
  \int^{\infty}_0 \left[\sum^{\infty}_{n=0}q_n(x_0)s^n\right]  \exp(-px_0) dx_0 = \frac{1}{p\sqrt{1-s}}\exp\left(-\frac{p}{\pi}\int^{\infty}_0 \frac{\ln \left(1-s\phi(k)\right)}{p^2+k^2}  dk\right). \label{eq:doubleLapla}
 \end{equation}

Although it is in general difficult   to take the inverse of the double Laplace transform in Eq.~(\ref{eq:doubleLapla}), one can readily bound the distribution of the maximum of a random walker with the martingale bound Eq.~(\ref{eq:Doobmax}).  Indeed,  it is often easy to determine $\langle {\rm max}\left\{X_n,0\right\} \rangle$, as illustrated in   Fig.~\ref{fig:martDistri}.   In  Fig.~\ref{fig:martDistri}  we compare numerically obtained results for the distribution of $X^\ast_n$ with analytical results from the  martingale bound $\langle {\rm max}\left\{X_n,0\right\} \rangle/x^\ast$     in Theorem~\ref{DoobMAx}.    In particular, we consider  the case when $Y$ is a random variable drawn from a standard Gaussian distribution  with $a=0$ (left) and with $a>0$ (right).

\subsection{Convergence theorems}
\label{subsec:LLN}

% {\color{red}  Let $S_n = \sum^n_{i=1} X_i$ be the sum of $n$ independent  random variables $X_i$.    If  $X_i\in [a_i,b_i]$, then  Hoeffding's inequality states that \cite{hoeffding1994probability}
%\begin{eqnarray}
%P\left[|S_n- \langle S_n|\rangle\geq s \right] \leq 2\exp\left(-\frac{2s^2}{\sum^n_{j=1}(b_j-a_j)^2}\right)
%\end{eqnarray}
%for all $s>0$.   For martingales, this result has been generalised  by Azuma with the following theorem \cite{azuma1967weighted}: 
 %\begin{theorem}[Azuma's inequality] $\\$
 %Let $M_n$ a martingale process with bounded increments, $|M_k-M_{k-1}|\leq c_k$.    Then 
% \begin{eqnarray}
% P\left[|M_n-  M_0|\geq s \right]  \leq 2 \exp\left(-\frac{s^2}{2\sum^n_{j=1}c^2_j}\right).
% \end{eqnarray}
%\end{theorem}  
%Azuma's inequality is a consequence of  Lemma 1 in \cite{azuma1967weighted} and implies that $\frac{M_n}{n}$ converges almost surely to $0$ \cite{azuma1967weighted}.}

%The law of large numbers for martingales is~\cite{lip2013}
%\begin{leftbar}
%Let $M_n$ be a martingale. Then, $M_n/n$ converges almost surely to $0$.
%\end{leftbar}
%{\bf [IN: Is this result correct?  I  cannot find it in (48)?]} 

{A fundamental result in martingale theory is that, under  a set of  conditions  specified in the martingale convergence theorems, the  fluctuations in the  trajectories of a martingale decrease as a function of $n$, yielding the convergence to an asymptotic limit, i.e.,
\begin{equation}
\lim_{n\rightarrow \infty} M_n   = M_{\infty}. \label{eq:convT}
\end{equation}   
The martingale convergence theorem is a fundamental property of martingales that follows from the fact that martingales represent a gambler's fortune in a fair game of chance.    Consequently,  a martingale process cannot keep fluctuating as otherwise a gambler could exploit a buy low and sell high strategy to make profit out of a fair game of chance.     Note that this is more than a simple analogy as the martingale convergence theorem is proved with  Doob's upward crossing lemma, which precisely bounds the profit a gambler can make out of the buy low and sell high strategy.  }%As we will see in Sec.~\ref{sec:stop},  stopping a martingale at a random time yields similar conclusions.      }

There exist two versions of the martingale convergence theorem, one that holds for submartingales bounded from above, and another that holds for uniformly integrable martingales.  

{Now, let us  get to the specifics.}
Let $x_n$, with $n\in\mathbb{N}$, be a nondecreasing deterministic sequence of real numbers that is bounded from above [i.e., ${\rm sup}_n x_n < \infty$], then elementary math gives $\lim_{n\rightarrow \infty}x_n = x_{\infty} \in \mathbb{R}$.     The following theorem (Theorem 2.6 in Ref.~\cite{lip2013}) generalises the previous result to submartingale  processes.
\begin{leftbar}
  \begin{theorem}[\textbf{Submartingale convergence theorem}] \label{convergenceTheorem}$\\$
  $\\$ Let $S_n$ be a submartingale for which
  \begin{eqnarray}
  {\rm sup}_n \langle  {\rm max}\left\{S_n,0\right\}\rangle <\infty.
  \end{eqnarray}
  Then  there exists a $S_{\infty}$ for which
    \begin{eqnarray}
\langle  {\rm max}\left\{S_\infty,0\right\}\rangle <\infty.
\end{eqnarray}
such that  
   \begin{eqnarray}
\mathcal{P}\left(\lim_{n\rightarrow \infty} S_n   = S_{\infty}\right)  = 1.
\label{eq:convMart}
 \end{eqnarray} 
\end{theorem}
\end{leftbar}

Next we discuss  the second version of the martingale convergence theorem that holds for uniformly integrable processes.
We say that a stochastic process $A_n$ is {\it uniformly integrable} if  
\begin{eqnarray}
\lim_{m\rightarrow \infty} {\rm sup}_{n\in \mathbb{N} \cup \left\{0\right\}}   \langle  |A_n| \: {\bf 1}_{|A_n|\geq m}\rangle  = 0.\label{eq:UIDef}
\end{eqnarray} 
Note that because $m$ in Eq.~(\ref{eq:UIDef}) is independent of $n$,  Eq.~(\ref{eq:UIDef}) implies that $A_n$ cannot escape to infinity.    Uniform integrability is important since it  allows us to swap expectation values with limits, i.e., 
\begin{eqnarray}
\left\langle \lim_{n\rightarrow \infty}A_n \right\rangle = \lim_{n\rightarrow \infty} \langle A_n \rangle,
\end{eqnarray} 
if  $\lim_{n\rightarrow \infty}A_n$ exists with probability one.

The properties of uniformly integrable martingales can be characterised with the  following theorem  (Theorem 2.7 in Ref.~\cite{lip2013}), which states that uniformly integrable martingales and conditional expectations processes are equivalent:
\begin{leftbar}
  \begin{theorem}[\textbf{Convergence theorem for  uniformly integrable martingales}] \label{convergenceTheorem2}$\\$ Let $M_n$ be a martingale defined on $n\in \mathbb{N}\cup\left\{0\right\}$.   The following conditions are equivalent: 
      \begin{itemize}
      \item the process $M_n$ is uniformly integrable;
      \item ${\rm sup}_n\langle |M_n| \rangle<\infty$ and thus $M_{\infty} = \lim_{n\rightarrow \infty}M_n$ exists.   In addition, $M_n$ is {\bf regular}, which means that  with probability one it holds that 
      \begin{eqnarray}
  M_n =     \langle  M_{\infty}|X_{[0,n]}\rangle .
      \end{eqnarray}
      \item $M_{\infty} = \lim_{n\rightarrow \infty}M_n$ exists and 
      \begin{eqnarray}
      \lim_{n\rightarrow \infty} \langle |M_{\infty}-M_n|\rangle = 0. \label{eq:UICond}
      \end{eqnarray} 
      \end{itemize}
      \end{theorem}
      \end{leftbar}
Uniform integrability extends thus the martingale sequence from the natural numbers $\mathbb{N}$ to the  natural numbers extended with infinity $\mathbb{N}\cup \left\{\infty\right\}$.   

{
Several fundamental results in probability theory can be derived from Doob's martingale convergence  theorem.   
A notable example is 
 {\it L\'{e}vy's upwards theorem}, which states that 
\begin{equation}
\lim_{n\rightarrow \infty}\langle  A |X_{[0,n]}\rangle = \langle  A |X_{[0,\infty)}\rangle
\end{equation}
holds for  integrable random variables $A$, where convergence should be understood either with probability one or in the $L^1$ norm.     In addition, L\'{e}vy's upwards theorem implies {\it Kolmogorov's zero-one law}, which states that tail events $\Phi$, which are events independent of any finite sequence $X_1,X_2,\ldots,X_n$, i.e., 
 \begin{equation}
 \mathbb{P}\left[\Phi, X_{[0,n]}=x_{[0,n]}\right] =  \mathbb{P}\left[\Phi\right]  \mathbb{P}\left[X_{[0,n]}=x_{[0,n]}\right] 
 \end{equation}
 occur either with probability one, $ \mathbb{P}\left[\Phi\right]=1$, or with probability zero, $ \mathbb{P}\left[\Phi\right]=0$.     This law is used, e.g., in percolation theory~\cite{bollobas2006percolation}, to show that an infinite, percolating  cluster exists either with probability zero or one~\cite{haggstrom2006uniqueness}.  }

   \subsection{Stopping times} \label{sec:stop}
  
Martingales can be used to study  stochastic processes at random times, and this has been up to now one of its main uses in stochastic thermodynamics.    Therefore, in this section we introduce the concept of a stopping time.    
 
  \subsubsection{Definition and examples}

  Put simply, a stopping time is the time when a specific criterion  is met for the first time.   Importantly, the stopping criterion obeys causality, and this makes stopping times  suitable for modelling physical processes.       
    \begin{leftbar}
     A {\bf stopping time}  is a   nonnegative  random variable $\T=\T(X_{[0,\infty)} )\in \mathbb{N}\cup \left\{0,\infty\right\}$ 
     that is statistically independent of the part of the trajectory $X_{[\T+1,\infty)}$  that comes after the stopping time.
    \end{leftbar} 
    Note that this definition can be generalized to continuous time. 
Examples of stopping times are:
     \begin{itemize}
       
         \item The $m$-th time a stochastic process visits a subset of  $\mathcal{X}$.     In the particular case of $m=1$ we obtain first-passage times.  
             \item The first time a functional $f(X_{[0,n]})\in \mathbb{R}$ defined on the trajectories of  $X$ exits an interval $(-\ell_-,\ell_+)$.     Since the main observables of   stochastic thermodynamics are functionals, this example is of particular importance.   In the specific case of $f(X_{[0,n]})=X_n$, this stopping time  equals the first escape time of $X_n$ from the interval $(-\ell_-,\ell_+)$.  
         \item $ \T_1\wedge n = \min(\T_1,n)$, where $\T_1$ is the first time that  a prescribed condition is met for the stochastic process of interest, and $n\in \mathbb{N}$ determines a finite time horizon.  
     \end{itemize}
   
    On the other hand, the following quantities  {\em are not} stopping times:
     \begin{itemize}
         \item The time when a random walker leaves indefinitely  a subset of  $\mathcal{X}$;
         \item The time  a stochastic process attains a  minimum or maximum value (which may be a local minimum or maximum);
         \item The occupation time spent in a given  subset of  $\mathcal{X}$.
     \end{itemize}

     \begin{figure}[h!]
\centering
{\includegraphics[width=\textwidth]{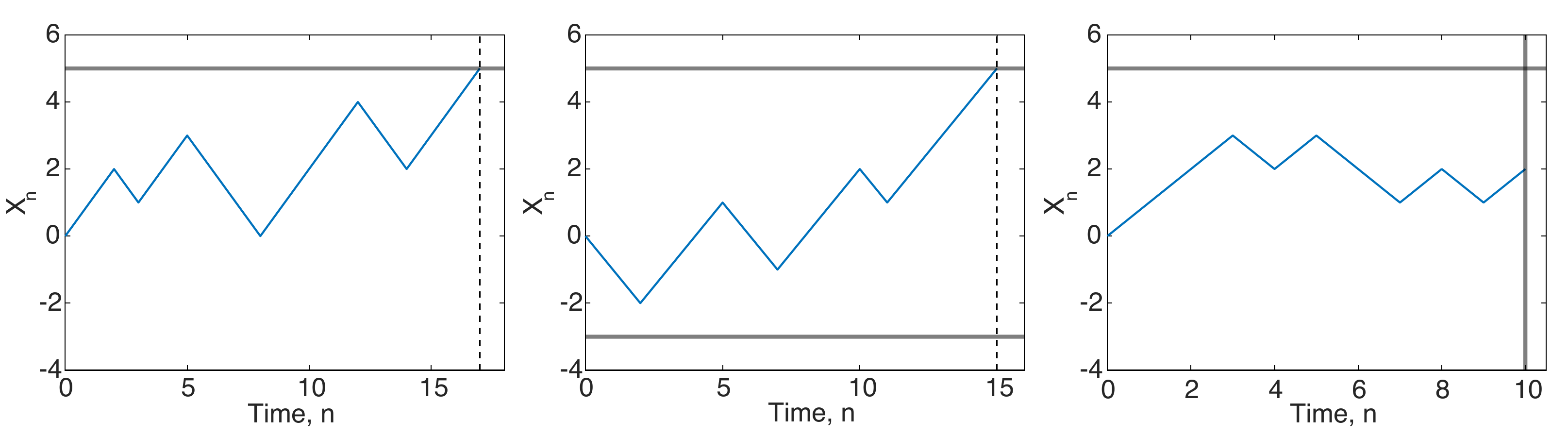}}
\caption{Three examples of stopping times evaluated over trajectories of a discrete-time biased random walk on $\mathbb{Z}$, with forward jump probability $q=0.7$ and backward jump probability $1-q=0.3$. Left: first passage time to reach the absorbing boundary $\ell_+=5$. Middle: first escape time from the interval $[-3,5]$. Right: $\min(\T_1,10)$ with $\T_1$ the first passage time to reach  the absorbing boundary $\ell_+=5$. In the three examples, {the blue zigzag lines are a linear interpolation between the discrete values $X_n$ and a guide to the eye}, the thick horizontal lines illustrate the boundaries of the stopping-time problem, and the dashed vertical lines denote  the time when the stopping condition takes place.   }\label{fig:stoppingtimeexample}
\end{figure}

  \subsubsection{Doob's optional stopping theorems}

Consider a gambler  who participates in a fair game of chance.  Can (s)he make on average profit by leaving the game at an  intelligently chosen moment $\T$?   In other words, is it possible that $\langle M_\T\rangle > \langle M_0 \rangle$?    
  
The optional stopping theorem states that  $\langle M_\T\rangle = \langle M_0 \rangle$, given certain conditions on the stopping time $\T$ and the martingale $M$. Loosely said, these conditions impose that the   gambler does not  have access to an infinite budget. Indeed, if the gambler has access to an infinite budget, then strategies to make profit out of a fair game of chance exist, and this leads to  paradoxes, the most well known being the St.~Petersburg paradox~\cite{bernstein1996against}.

We illustrate the optional stopping theorem with the example of a gambler's wealth  $F_n$ in a fair coin toss game, see Eq.~(\ref{eq:toinCoss}). We assume that $F_0 = f_{\rm init}$. If
\begin{equation}
 \T^{(1)}  = {\rm min} \left\{n\geq0 : F_n = f_{\rm init}+m_+\right\}
 \label{eq:Izaakstoppedonce}
 \end{equation}
with $m_+\in \mathbb{N}$,  then 
 \begin{equation}
\langle F_{\T^{(1)}}\rangle = f_{\rm init} + m_+ \geq f_{\rm init} = F_0, \label{eq:case1}
 \end{equation}
 which implies that the gambler is earning money on average  and that the optional stopping theorem does not apply.   However, if 
 \begin{equation}
 \T^{(2)}  = {\rm min} \left\{n\geq0 : F_n = f_{\rm init}+m_+ \  {\rm or} \   F_n = f_{\rm init}-m_- \right\}
 \label{eq:Izaakstoppedtwice}
 \end{equation}
 with $m_+,m_-\in \mathbb{N}$, then 
  \begin{equation}
\langle F_{\T^{(2)}}\rangle  = f_{\rm init} = F_0 . \label{eq:case2}
 \end{equation} 
The difference between the stopping times $\T^{(1)}$ [Eq.~(\ref{eq:Izaakstoppedonce})] and $\T^{(2)}$ [Eq.~(\ref{eq:Izaakstoppedtwice})] is that in the first case the gambler has access to an infinite budget  ($F_{n}$ can take arbitrary large negative values) whereas in the second case the gambler has  a finite budget  ($F_{n}$ is bounded between $f_{\rm init}-m_- $ and $f_{\rm init}+m_+$).

 In what follows, we consider several versions of Doob's optional stopping theorem.  
  Amongst Doob's theorems, the first important result that we review is the following (Theorem 2.1, Chapter  VII in Ref.~\cite{Doob}).
\begin{leftbar}

\begin{theorem}[{\bf Doob's optional sampling theorem}]\label{optionalSampling} $\\$
 Let $M_n$ be a martingale (submartingale) and  let $\T$  be  a stopping time, both with respect to the  process $X_n$.   Then the stopped process $M_{\T\wedge n}$, with $\T\wedge n = {\rm min}\left\{\T,n\right\}$ a finite stopping time,    is also a martingale (submartingale), i.e.,
 \begin{equation}
     \langle M_{\T\wedge n} | X_{[0,m]}\rangle  = M_{\T\wedge m} \quad  \left(     \langle M_{\T\wedge n} | X_{[0,m]}\rangle  \geq  M_{\T\wedge m}\right),
 \end{equation}
 for $0\leq m\leq n$.
\end{theorem}
\end{leftbar}
\begin{proof}  

The process $M_{\T\wedge n}$  is integrable, since it is a finite sum of integrable random variables.    Because of the tower property of conditional expectations, it is sufficient to show that 
\begin{eqnarray}
\langle M_{\T\wedge n} | X_{[0,n-1]}\rangle  = M_{\T\wedge (n-1)}.  
\end{eqnarray} 
It holds that 
\begin{eqnarray}
\langle M_{\T\wedge n}  |X_{[0,n-1]} \rangle &=& M_{\T\wedge (n-1)}   +  \langle  (M_n-M_{n-1})\mathbf{1}_{\T\geq n}| X_{[0,n-1]}\rangle \\ 
&=& M_{\T\wedge (n-1)}   +  \mathbf{1}_{\T\geq n} \langle  (M_n-M_{n-1})| X_{[0,n-1]}\rangle  = M_{\T\wedge (n-1)},
\end{eqnarray} 
where we used the indicator function Eq.~(\ref{eq:indicator}) for 
\begin{equation}
\Phi = \left\{X_{[0,\infty]}: \T(X_{[0,\infty]})\geq n\right\}.
\end{equation}
The proof in the case of submartingales is analogous.
\end{proof}

Applying the optional sampling theorem to uniform integrable martingales, see definition (\ref{eq:UIDef}), we obtain  Doob's optional stopping theorem (Theorem 2.9 in  \cite{lip2013}). 

\begin{leftbar}
\begin{theorem}[{\bf Doob's Optional stopping, version I }] \label{Doob1x} $\\$
Let $M_n$ be a uniformly integrable martingale and let $\T_1$ and $\T_2$ be two stopping times with $P\left(\T_2\geq \T_1\right)=1$, then 
\begin{eqnarray}
\langle M_{\T_2}| X_{[0,\T_1]}\rangle = M_{\T_1}. 
\label{eq:OTcomplicated}
\end{eqnarray}
For the particular case of $\T_1=0$  { and $\T_2 = \T$}, we obtain
  \begin{eqnarray}
\langle M_\T| X_0\rangle = M_0, \label{eq:optStop}
\end{eqnarray}
i.e., the  average of a uniformly integrable martingale conditioned on the initial state $X_0$ equals its initial value $M_0$.
\end{theorem} 
\end{leftbar}
%{For the case in which $\T_1<\tau$ with $\tau$ a deterministic upper bound to the stopping time $\T_1$, one has for any $t\geq \tau$
%\begin{eqnarray}
%\langle M_{t}| X_{[0,\T_1]}\rangle = M_{\T_1}. 
%\label{eq:OTcomplicated2}
%\end{eqnarray}
%}
For simplicity we give here the proof of the particular case~\eqref{eq:optStop}. 
%, we refer to~\cite{lip2013} for the proof of~\eqref{eq:OTcomplicated}.  
\begin{proof} 
According to Theorem~\ref{optionalSampling} it holds that 
\begin{eqnarray}
\lim_{n\rightarrow \infty}\langle M_{\T\wedge n}| X_0\rangle = M_0. \label{eq:optStop1}
\end{eqnarray}
Since $M_{\T\wedge n}$ is a uniformly integrable, it holds that 
\begin{eqnarray}
\lim_{n\rightarrow \infty}\langle M_{\T\wedge n}| X_0\rangle = \langle \lim_{n\rightarrow \infty} M_{\T\wedge n}| X_0\rangle    .\label{eq:optStop2}
\end{eqnarray}
In addition,
\begin{eqnarray}
 \langle \lim_{n\rightarrow \infty} M_{\T\wedge n}| X_0\rangle =  \langle   M_{\T\wedge \infty}| X_0\rangle =  \langle   M_{\T }| X_0\rangle.  \label{eq:optStop3}
\end{eqnarray}
Equations (\ref{eq:optStop1})- (\ref{eq:optStop3}) imply (\ref{eq:optStop}), which is what we meant to prove.
\end{proof}

An alternative version of Doob's optional stopping theorem, corresponds to the case of a      gambler  that has a finite budget (Theorem 4.1.1 in \cite{norris1998markov}).  
\begin{leftbar}
\begin{theorem}[\textbf{ Doob's Optional stopping, version II} ] \label{Doob2} $\\$
Let $M_n$ be a martingale and let $\T$ be a stopping time. 
If $\mathcal{P}(\T<\infty)=1$ and if there exists a constant $m$  such that  $|M_n|\leq m$ for all $n\leq \T$, then 
\begin{eqnarray} 
\langle M_\T\rangle = \langle M_0\rangle. \label{eq:doobStop1}
\end{eqnarray} 
\end{theorem} 
\end{leftbar}
The two versions of Doob's optional stopping theorem   are related to each other, and in fact one can derive Theorem~\ref{Doob2} from Theorem~\ref{Doob1x}, see for example the proofs in the appendix of Ref.~\cite{neri2019integral}.   %, averaging over the distribution $\mP(X_0)$,

The optional stopping theorem is one of the key properties that characterise martingales, and in fact,    it is a defining property of martingales, see Ref.~\cite{revuz2013continuous}.  Indeed, as we will show, the condition Eq.~(\ref{eq:defM}) can be written in terms of the stopping time
\begin{equation}
\T = m \mathbf{1}_{\Phi}(X_{[0,m]}) + n \mathbf{1}_{\Phi^c}(X_{[0,m]}) , \label{eq:TDefx}
\end{equation} 
where $\Phi$ is a measurable subset of the set of trajectories $x_{[0,m]}$,  where $\Phi^c$ is the complement of $\Phi$, {where $\mathbf{1}_{\Phi}(x_{[0,m]})$ is the indicator function that returns the value $1$ when $x_{[0,m]}\in \Phi$ and $0$ when $x_{[0,m]}\notin \Phi$}, and where $m\leq n$.    
\begin{leftbar}
\begin{theorem}\label{Theorem:Equiv}
A stochastic process $M_n=M[X_{[0,n]}]$  is a martingale if and only if for every bounded stopping time $\T$, 
\begin{equation}
    \langle |M_\T|\rangle<\infty,
\end{equation}
and 
\begin{equation}
    \langle M_\T\rangle = \langle M_0\rangle. \label{eq:MTx}
\end{equation}
\end{theorem}
\end{leftbar}
\begin{proof}
We show the if part, as the only if part readily follows from the optional stopping theorem.  

Applying the optional stopping theorem to the stopping time $\T$ defined in Eq.~(\ref{eq:TDefx}) yields, 
\begin{equation}
 \langle M_0\rangle = \langle M_\T\rangle = \langle M_n \mathbf{1}_{\Phi^c}(X_{[0,m]})\rangle + \langle M_m \mathbf{1}_{\Phi}(X_{[0,m]})\rangle, \label{eq:MTv1}
 \end{equation}
 and applying Eq.~(\ref{eq:MTx}) to the stopping time $n$ yields,
 \begin{equation}
 \langle M_0\rangle = \langle M_n\rangle = \langle M_n \mathbf{1}_{\Phi^c}(X_{[0,m]})\rangle + \langle M_n \mathbf{1}_{\Phi}(X_{[0,m]})\rangle.\label{eq:MTv2}
 \end{equation}
 Equations (\ref{eq:MTv1}) and (\ref{eq:MTv2}) imply that 
 \begin{equation}
\langle M_n \mathbf{1}_{\Phi}(X_{[0,m]})\rangle =  \langle M_m \mathbf{1}_{\Phi}(X_{[0,m]})\rangle 
 \end{equation}
 for all subsets $\Phi$ of the set of trajectories $x_{[0,m]}$.  By the tower property of conditional expectations, we  can rewrite this equation as 
  \begin{equation}
\langle \langle M_n|X_{[0,m]}\rangle \mathbf{1}_{\Phi}(X_{[0,m]})\rangle =  \langle M_m \mathbf{1}_{\Phi}(X_{[0,m]})\rangle ,
 \end{equation}
 for all subsets $\Phi$ of the set of trajectories $x_{[0,m]}$, and therefore by the definition of  conditional expectations it holds with probability one  that
 \begin{equation}
   \langle M_n|X_{[0,m]} \rangle = M_m. \end{equation}
\end{proof}

  \subsubsection{First-passage problems of random walks with martingales} 
  \label{sec:FPTRW}
  
  We use the optional stopping theorem to derive the statistics of  first-passage times in a stochastic process.   We consider the random-walk example    $X_n$ discussed in Section~\ref{sec:Examples}. Here, $X_n$ is a discrete-time,  biased random walker on $\mathbb{Z}$ with $X_0 = 0$; it moves  one step in the positive (negative) direction with probability $q$ ($1-q$).  We consider the first-passage time 
\begin{eqnarray}
\T^{(2)} = {\rm min}\left\{n\geq 0: X_n = -x_- \ {\rm or} \  X_n = x_+ \right\},
  \end{eqnarray} 
  where the constants $x_-,x_+ \in \mathbb{N}$,  such that $-x_-$ and $x_+$ are absorbing sites.  In other words, $\T^{(2)}$ is the first escape time of the walker from the interval $(x_-, x_+)$.
  Using  Doob's optional stopping theorem, version II,  we derive exact results for the statistics of $\T^{(2)}$.   
  Let us  consider the martingale~(\ref{eq:stochExpExample1x}), denoted here as 
  \begin{equation}
  M_n = \eta^{X_n} \ \text{with}\ \eta = (1-q)/q.\label{eq:martingaleshamik}
  \end{equation} 
  Applying Theorem~\ref{Doob2} to the martingale $M_n$ given by Eq.~\eqref{eq:martingaleshamik}, we obtain 
\begin{eqnarray} 
\langle M_{\T^{(2)}}\rangle =   P_+ \eta^{x_+} + (1- P_+) \eta^{-x_-} = 1,\label{eq:stoppaa}
\end{eqnarray}
where $P_+ = \mathcal{P}(X_{\T^{(2)}}=x_+)$, and we have used the fact that $P_- = \mathcal{P}(X_{\T^{(2)}}=-x_-)=1-P_+$ (i.e., $X_n$ escapes the interval at finite time with probability one).   Solving Eq.~\eqref{eq:stoppaa} towards $P_+$
 we obtain 
 \begin{eqnarray}
 P_+ =  \frac{1-\eta^{-x_-}}{\eta^{x_+}-\eta^{-x_-}}.
 \end{eqnarray}   
 Second, we apply Theorem~\ref{Doob2} to the  the martingale $X_t-(2q-1)t$, see Eq.~(\ref{g1}),  obtaining 
\begin{eqnarray}
  \langle  \T^{(2)} \rangle  &=&  \frac{\langle X_{\T^{(2)}}\rangle}{(2q-1)}   . \label{eq:expl}
\end{eqnarray} 
Using   
\begin{eqnarray}
\langle X_{\T^{(2)}}\rangle =P_+ x_+ - (1-P_+)x_- =  \frac{1-\eta^{-x_-}}{\eta^{x_+}-\eta^{-x_-}}x_+ - \frac{\eta^{x_+}-1}{\eta^{x_+}-\eta^{-x_-}}x_- 
\end{eqnarray} 
in (\ref{eq:expl}),
we obtain the following explicit expression for  the mean first-passage time
\begin{equation}
  \displaystyle\langle  \T^{(2)} \rangle  =  \frac{1}{2q-1}\left(\displaystyle\frac{1-\eta^{-x_-}}{\eta^{x_+}-\eta^{-x_-}}x_+ - \displaystyle\frac{\eta^{x_+}-1}{\eta^{x_+}-\eta^{-x_-}}x_-\right)    . \label{eq:expl}
\end{equation}
   Analogously, the optional stopping theorem can be used to derive an explicit expression for  the second moment $\langle  (\T^{(2)})^2 \rangle$ of the first-passage time and its generating function, see e.g.,~the appendices of Ref.~\cite{neri2022universal}.

 \subsection{$^{\spadesuit}$Martingale central limit theorem} \label{subsec:CLT}  
  Central limit theorems refer  to a collection of results that describe how the sum of a large number of random variables converges to a normal distribution.   The study of central limit theorems initiated in the beginning of the 19th century with the work of Pierre-Simon Laplace, who was the first to observe the universal character of the Gaussian distribution, see Ref.~\cite{Laplace}.     The central limit theorem has been extended and refined in various ways ever since, see Ref.~\cite{fischer2011history} for an overview of the history of central limit theorems.    The idea  underlying  the different central limit theorems is  however the same, viz.,  the statistics of the sum of a large number of variables  converges to a normal distribution if the variables are weakly correlated and the sum is not dominated by a few large outliers.     
 
 Let  us consider a sum of $n$ real-valued random variables $Y_j$ given by
 \begin{equation}
     \tilde{X}_n = \sum^n_{j=1}Y_j. \label{eq:SnSum}
 \end{equation}
 Central limit theorems determine under which conditions the statistics of a rescaled and shifted version of $ \tilde{X}_n $ are described by the normal distribution
%\begin{equation}
%    \mathcal{N}(x) = \frac{1}{\sqrt{2\pi}}\exp(-x^2/2),
%\end{equation}
i.e., 
\begin{eqnarray}
\lim_{n\rightarrow \infty} 
\left \langle \delta \left(\frac{
  \tilde{X}_n-\mu_n}{\sigma_n} -x\right)\right \rangle 
=  \frac{1}{\sqrt{2\pi}}\exp(-x^2/2),\label{eq:d}
\end{eqnarray}
where $\delta(x)$ is the Dirac delta distributon,  $\mu_n$ is the average shift, and $\sigma_n$ determines the scaling of $  \tilde{X}_n-\mu_n$ with $n$.

The version of the central limit theorem that is best known holds for sums $\tilde{X}_n$ of iid random variables $Y_j$ with fixed mean $\mu$ and finite variance $\sigma^2$, as defined in Eq.~(\ref{eq:zeroMeanSum}).      This central limit theorem states that Eq.~(\ref{eq:d})  holds for the standard "norming" (see Theorem 27.1  of  Ref.\cite{billingsley2008probability})
\begin{eqnarray}
\mu_n = \mu \: n, \quad {\rm and} \quad \sigma_n = \sigma \sqrt{n}.
\end{eqnarray}

A natural extension of the  central limit theorem for iid random variables  considers
sums of random variables  $Y_j$ that are independent, but not identically, distributed, random variables.    Assuming that the $Y_j$ are independent random variables with 
with mean $\mu_j$ and finite variance $\sigma_j$, then 
 Eq.~(\ref{eq:d}) applies for (see Theorem 27.2  of  Ref.\cite{billingsley2008probability})
\begin{equation}
    \mu_n = \sum^n_{j=1}\mu_j, \quad {\rm and} \quad  \sigma^2_n = \sum^n_{j=1}\sigma^2_j,   \label{eq:independSigmaj}
\end{equation}
as long as the Lindeberg condition 
\begin{equation}
\lim_{n\rightarrow \infty}\frac{1}{\sigma^2_n}\sum^n_{j=1} \int_{|y_j|\geq \epsilon \sigma_n}  dy_j \: y^2_j \:  \rho_{Y}\left(y_j\right) =  0 \label{eq:lindeberg}
\end{equation}
holds for all $\epsilon>0$.       Notice that the Lindeberg condition compares the total accumulated variance $\sigma^2_n$, which is a measure for the number of variables contained in the sum, with the  statistical weight accumulated in the tails of the distribution determined by  $\sum^n_{j=1}\int_{|y_j|\geq \epsilon \sigma_n} dy_j \: y^2_j \:  \rho_{Y}\left(y_j\right)$.     The central limit theorem holds as long as the former is infinitely larger than the latter.

Martingales are natural candidates to extend the  central limit theorem to the case of dependent, albeit uncorrelated, random variables $Y_j$.     Indeed, a martingale $M_n$ can be written as the sum of  martingale differences 
 \begin{equation}
      Y_j = M_j - M_{j-1}   .
  \end{equation}
The martingale condition implies that
  \begin{eqnarray}
    \langle Y_{i_1} Y_{i_2}\ldots Y_{i_k}  \rangle = 0  \label{eq:uncor}
 \end{eqnarray}
  holds for any $k$-tuple of distinct indices $(i_1,i_2,\ldots,i_k)$.
  Therefore, $M_n$ is  a sum $\tilde{X}_n$ of $n$ random variables $Y_j$ with vanishing autocorrelation function. 

Central limit theorems for martingales have been derived originally by  L\'{e}vy~\cite{levy1935proprietes, levy1938theorie}, and many extensions has been derived since, see e.g.,~the book ~\cite{hall2014martingale} for an overview.  We consider here the version of the martingale central limit theorem of Ref.~\cite{brown1971martingale}, as  for clarity we do not want to deal with the  more general case of double indexed sequences considered in Ref.~\cite{hall2014martingale}. 
%\begin{leftbar}
%\begin{theorem}[\textbf{Martingale central limit theorem}]\label{MartCLT}
%$\\$ Let $M_n$ be a zero mean martingale  with respect to the process $X_n$, and 
%\begin{equation}
%V_n=\sum_{k=0}^{n-1} \left \langle  \left( M_{k+1}-M_k \right)^2  | X_{[0,k]}  \right\rangle
%\label{eq:cvmart}
%\end{equation}
%  be the conditional variance of $M_n$ defined  as in Eq.~\eqref{eq:condvar}. Assume first that there exist a $\sigma>0$ such that 
% \begin{eqnarray}
%\lim_{n\rightarrow \infty}\mathcal{P}\left[\left|  V_n   -\sigma^2\right|>\epsilon\right]  = 0
%\label{eq:cond1}
% \end{eqnarray} 
% holds 
%for all $\epsilon>0$.  In addition, assume also that 
%\begin{eqnarray}
%\lim_{n\rightarrow \infty}\mathcal{P}\left[ \sum^{n}_{j=1}\Big\langle \left(M_j-M_{j-1}\right)^2 I_{|M_j-M_{j-1}| > \epsilon_1  } | X_{[0,j-1]}\Big\rangle  > \epsilon_2    \right] = 0   \label{eq:cond2}
%\end{eqnarray}
%holds for all $\epsilon_1, \epsilon_2>0$.  

%Then $M_n$ converges in distribution to a normal distribution with variance $\sigma^2$
%\begin{eqnarray}
%\lim_{n\rightarrow \infty} 
%\left \langle \delta \left(M_n %-x\right)\right \rangle 
%= \frac{1}{\sqrt{2\pi\sigma^2}} %\exp\left(-\frac{x^2}{2\sigma^2}\right).
%\lim_{n\rightarrow \infty}p_{\frac{M_n}{\sqrt{n}}}(x) = \frac{1}{\sqrt{2\pi}}e^{-\frac{x^2}{2\sigma^2}}. \label{eq:cond2}
%\end{eqnarray} 
%\end{theorem}}  
%\end{leftbar}

\begin{leftbar}
\begin{theorem}[\textbf{Martingale central limit theorem}]\label{MartCLT}
$\\$ Let $M_n = \sum^n_{j=1}Y_j$ be a zero mean martingale, and let $V_n$ be its conditional variance, as defined in   Eq.~\eqref{eq:condvar}.  Assume  that  for all $\epsilon>0$, 
\begin{equation}
\lim_{n\rightarrow \infty}\mathcal{P}\left(\left|  V_n   -\sigma^2_n\right|>\epsilon\right)  = 0
\label{eq:cond1}
\end{equation}
where  $\sigma^2_n = \langle V_n\rangle$, 
and assume that the Lindeberg condition  Eq.~(\ref{eq:lindeberg}) holds.   Then the central limit theorem Eq.~(\ref{eq:d}) applies for $\tilde{X}_n=M_n$, $\mu_n=0$, and $\sigma^2$ given by the expected value of the conditional variance.   
\end{theorem}
\end{leftbar}

Note that the martingale central limit theorem also relies on the Lindeberg condition, but now the expected value $\langle V_n\rangle$  of the conditional variance  plays the role of $\sigma_n$, instead of the sum of the variances  Eq.~(\ref{eq:independSigmaj}) as was  the case for independent random variables.   

     Just as is the case for  sum of  iid random variables, in the continuous-time limit a properly rescaled martingale process converges to a Wiener process, see Theorem 3 in Ref.~\cite{brown1971martingale}.  In addition,   martingales  obey a law of iterated logarithm, which determines that the absolute value of the maximum of   $M_n$ grows as $\sqrt{2\sigma^2_n \log\log \sigma^2_n}$, see Ref.~\cite{hall2014martingale}.

\subsection{Elephant random walks: convergence and central limit }

We apply the martingale convergence Theorem~\ref{convergenceTheorem} to  the martingale $M_n$ of Eq.~(\ref{eq:dfM}), associated with the elephant random walk  $X_n$ defined in (\ref{eq:Xdyn}). 
As shown in Ref.~\cite{bercu2017martingale}, the conditional variance $V_n$  of the martingale $M_n$, as defined in Eq.~(\ref{eq:condvar}), is bounded from above  by 
\begin{equation}
\nu_n = \sum^n_{k=1}a_n^2 > V_n.   
\end{equation}
The asymptotic behaviour of the sequence $\nu_n$  depends on the memory parameter $p$, namely, 
\begin{equation}
   \nu_n \sim \left\{\begin{array}{ccc} \frac{(\Gamma(2p))^2}{3-4p} n^{3-4p},&{\rm if}& p\in [0,3/4),  \\ \frac{\pi}{4}\log n,&{\rm if}& p=3/4, \\ b, &{\rm if}&  p \in (3/4,1],\end{array}\right.  \label{eq:nun}
\end{equation}
where $b$ is a finite number that can be expressed in terms of a generalised hypergeometric function, see Ref.~\cite{bercu2017martingale}.

\begin{figure}[h!]
{\includegraphics[width=\textwidth]{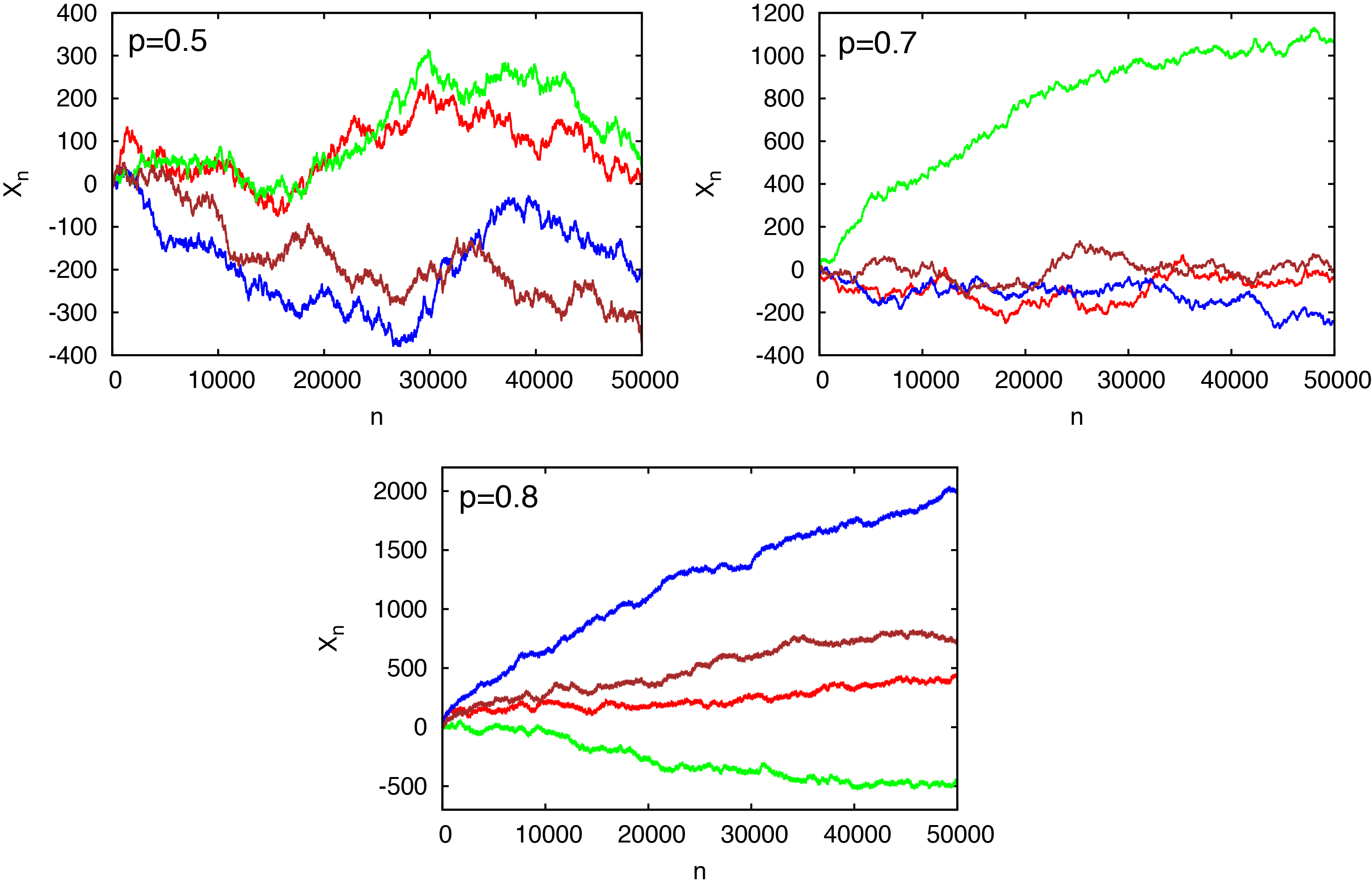}}
\caption{ Illustration of the implication of the martingale central limit theorem for the trajectories of the elephant random walk $X_n$ with memory parameter $p$.   Plots show four trajectories of the elephant random walk for three values of $p$.  Upper left panel: $p=0.5$, corresponding to a simple random walk without memory. The standard central limit theorem applies, and  in the asymptotic (or continuum) limit $X_n$ converges to a standard Brownian motion with $\langle X_nX_m\rangle = m$  for $m<n$.  Upper right panel:  $p=0.7$, corresponding to a diffusive random walk  with memory.  The martingale central limit theorem applies, and in the asymptotic limit  $X_n$ converges to a Brownian motion with a nontrivial memory kernel, such that,  $\langle X_nX_m\rangle =5 n^{0.6}m^{0.4}$ for $m<n$.    Lower panel: $p=0.8$, corresponding with the superdiffusive regime.  The martingale central limit theorem does not apply, and  the asymptotic limit takes the form  $X_n\sim Y n^{0.6}$ with $Y$ a time-independent random variable. }\label{fig:elephant}
\end{figure}

%It follows readily from Eq.~(\ref{eq:nun}) and Theorem~\ref{convergenceTheorem} that  the martingale $M_n$ converges almost surely to a finite random variable $M_{\infty}$ when $p>3/4$.    
{It follows  from Eq.~(\ref{eq:nun}) that $\sup_n\langle M_n\rangle$ is finite, as $\sup_n\langle M_n\rangle\leq \sup_n\sqrt{\langle M^2_n\rangle} =\sup_n\sqrt{\sum^{n-1}_{k=0}\langle (M_{k+1}-M_k)^2\rangle} \sim b $, where we have used (\ref{eq:doobDecomp1}) and (\ref{eq:condvar}).   Therefore, 
Theorem~\ref{convergenceTheorem} applies and   the martingale $M_n$ converges almost surely to a finite random variable $M_{\infty}$ when $p>3/4$.  }{As shown in Ref.~\cite{bercu2017martingale},  $M_\infty$ has a sub-Gaussian distribution with a $p$-dependent kurtosis $\mathcal{K}(p)$ that decreases monotonically as a function of $p$, such that $\mathcal{K}(3/4)=3$ and $\mathcal{K}(1)=1$.}  Consequently, according to Eqs.~(\ref{eq:anAsympto}) and (\ref{eq:dfM}), the elephant random walk process converges almost surely to 
\begin{equation}
X_n \sim n^{2p-1}\frac{M_{\infty}}{\Gamma(2p)},
\end{equation}
which is superdiffusive for $p>3/4$.  Note that for $p\rightarrow 3/4$ it approaches the diffusive regime $p\in [0,3/4)$.    We refer the reader to Fig.~\ref{fig:elephant} where we plot example trajectories of $X_n$ for $p=0.5$, $p=0.7$ and $p=0.8$.

We discuss the implications of the martingale central limit, Theorem~\ref{MartCLT}, on the elephant random walk. The martingale $M_n$, given by Eq.~(\ref{eq:dfM}),  satisfies the martingale central limit theorem when $p\leq 3/4$~\cite{bercu2017martingale, baur2016elephant, coletti2017central}.  Indeed, as indicated by   Eq.~(\ref{eq:nun}), the conditional variance $V_n$ grows indefinitely  for $p\leq 3/4$.  This argument can be made rigorous, and in Ref.~\cite{bercu2017martingale} Bercu has shown that the martingale   $M_n$ satisfies  the martingale central limit if  $p\leq 3/4$.        Using Eqs.~(\ref{eq:anAsympto}) and (\ref{eq:dfM}), it follows that also $X_n$ obeys a central limit theorem with $\mu_n=0$ and $\sigma_n =\sqrt{n/(3-4p)}$ or $\sigma_n = \sqrt{n\log(n)}$ for $p\in[0,3/4)$ or $p=3/4$, respectively.   For $p>3/4$, the conditional variance $V_n$ converges to a finite limit, and hence the Lindeberg condition is not satisfied.  In this case, the correlations in the process are too strong to generate enough data in the process, as quantified by $V_n$.  The distinction between the diffusive regime, where the central limit theorem applies, and the superdiffusive regime, with strong memory effects, is also apparent in the continuum limit of the model, see Ref.~\cite{baur2016elephant}.   For $p\in [0,3/4)$, $X_{\lfloor nt \rfloor}/\sqrt{n}$, with $\lfloor a \rfloor$  the floor function,   converges for large $n$ to a Wiener process $B_t$ with zero mean and autocovariance 
$\langle B_tB_s\rangle = s(t/s)^{2p-1}/(3-4p)$ for $0<s\leq t$, while in the superdiffusive regime, $X_{\lfloor nt \rfloor}/n^{2p-1}$ converges to $t^{2p-1}Y$   with $Y$ a real-valued random variable independent of time.

  \section{Continuous time}\label{sec:continuoustime}

 \subsection{Properties of continuous-time martingales  that carry over from discrete time} 
   %Fundamental  properties of martingales, such as Doob's optional stopping theorems and Doob's maximum inequality,   carry over to the continuous-time case if we assume  that the trajectories of the martingale  are right continuous, i.e., the process  is  continuous with occasional jumps.    Fortunately, according to Doob's regularity theorem, see Theorem 3.1 in Ref.~\cite{lip2013},   (sub)martingales can be considered  right-continuuous when the mean value  $\langle S_t \rangle$ is right continuous, i.e., $\lim_{\epsilon\rightarrow 0^+}\langle S_{t+\epsilon}\rangle = \langle S_t\rangle$.      Indeedm in  this case there exists a process $\tilde{S}_t$ that is right continuous and for which  $\mP(\tilde{S}_{t}= S_{t}) = 1$ for all $t\geq0$.     So, Doob's regularity theorem implies that when working with martingales or submartingales we can assume that we work on its right continuous modification, and hence  Doob's optional stopping theorems  and maximum inequality apply to this modification.  
 Fundamental  properties of martingales, such as Doob's optional stopping theorems and Doob's maximum inequality,   carry over to the continuous-time case if we assume  that the trajectories of the martingale  are right continuous, i.e., the process  is  continuous with occasional jumps.    Fortunately, according to Doob's regularity theorem, see Theorem 3.1 in Ref.~\cite{lip2013},   (sub)martingales can be considered  right-continuuous when the mean value  $\langle S_t \rangle$ is right continuous, i.e., $\lim_{\epsilon\rightarrow 0^+}\langle S_{t+\epsilon}\rangle = \langle S_t\rangle$.      {Indeed,} in  this case there exists a process $\tilde{S}_t$ that is right continuous and for which  $\mP(\tilde{S}_{t}= S_{t}) = 1$ for all $t\geq0$.     So, Doob's regularity theorem implies that when working with martingales or submartingales we can assume that we work on its right continuous modification, and hence  Doob's optional stopping theorems  and maximum inequality apply to this modification.

 \subsection{$^{\spadesuit}$Local martingales} 
 \label{sec:localmartingale}
   
 A notable distinction between martingale theory in continuous time and martingale theory in discrete time is that in continuous time there exist processes  that  are not martingales,  even though they are locally driftless.   Such,  processes   are called {\bf local martingales}, and just as martingales they play an important role  in the theory of stochastic processes in continuous time.  
 
{
The formal definition for a local martingale goes as follows: 
 \begin{leftbar}
   We say that a process $L_t$ is a {\bf local martingale} if there exists a sequence of nondecreasing stopping times $\T_n$  with $n\in \mathbb{N}$  such that \cite{ito1965transformation}
   \begin{itemize}
   \item with probability one $\lim_{n\rightarrow \infty}\T_n= \infty$;
   \item the stopped process $L(t\wedge \T_n)$ is a uniform integrable martingale for each $n$.  
   \end{itemize}   
 \end{leftbar}
}
{A martingale is a local martingale, since we can set $\T_n = n$.  We speak of a {\bf strict local martingale} if a stochastic process   is a local martingale but not a martingale \cite{elworthy1999importance}.     }
In discrete time,  local martingales are  martingales, see Theorem VII.1 in \cite{Shir}, and hence {\bf strict local martingales}  are a distinct feature of continuous-time processes.

%\begin{leftbar}
%For local martingales $L_t$ there exists a \textbf{random-time transformation}
%\begin{equation*}
%   t\to  \tau(X_{[0,t]}),
%\end{equation*}
%such that  $L_{\tau}$ is a martingale.   A random time $\tau(X_{[0,t]})$ is, for any trajectory of $X$, an increasing process in $t$, which provides a unique mapping between $L_t$ and $L_{\tau}$.
%\end{leftbar}
%In  Appendix~\ref{app:locDef} we provide a formal definition of a local martingale in terms of stopping times, which is equivalent to the random time picture.   
  
  {
 One way to realise the sequence of stopping times $\T_n$ is through a random time transformation.    A {\bf random time} $\tau(X_{[0,t]})$ is a nonnegative and increasing process in $t$, and it can be used to define a sequence of stopping times by 
\begin{equation}
 \T_n  =  {\rm inf}\left\{ t\geq 0: \tau(\Xot) = n\right\}. 
 \end{equation}
     This yields the following alternative characterisation of local martingales.  
\begin{leftbar}
For local martingales $L_t$ there exists a \textbf{random-time transformation}
\begin{equation*}
   t\to  \tau(X_{[0,t]}),
\end{equation*}
such that  $L_{\tau}$ is a martingale.  
\end{leftbar}}

\subsubsection*{{It\^{o}-integrals and random time transformations}}
The importance of local martingales follows from the fact that It\^{o} integrals of the form Eq.~(\ref{eq:Ito}), copied here for convenience
\begin{equation*}
     I_t =  \int^{t}_{0} Z_s dB_s ,
\end{equation*}
are local martingales.  Indeed,  
 It\^{o} integrals exist for  integrands $D_s$ that obey
\begin{equation}
\mathcal{P}\left(\int^t_0Z^2_{s}ds<\infty\right) = 1,  \label{eq:PD}
\end{equation}
which is a weaker condition than  Eq.~(\ref{eq:AverageD}), that {for}  convenience we copy here as well,
\begin{equation*}
    \int^{t}_0 \langle Z^2_s \rangle {\rm d}s <\infty.
\end{equation*}
   While the latter condition implies that $I_t$ is a martingale, the previous condition Eq.~(\ref{eq:PD}) implies that $I_t$ is a local martingale, see Ref.~\cite{oksendal2003stochastic}.     Indeed,  consider a general  It\^{o} integral 
\begin{eqnarray}
\frac{dI_t}{dt}= Z_t\frac{ dB_t}{dt},  \label{eq:XBLocal}
 \end{eqnarray} 
 with $ Z_t = Z(I_{[0,t]},t) \geq 0$ and $B_t$ a Brownian motion.  Define the random time 
 \begin{equation}
     \frac{d \tau_t}{dt} = Z^2_t ,
 \end{equation}
 with time change rate $Z^2_t$.    It then holds that \cite{oksendal2003stochastic}
  \begin{equation}
    \frac{ dI_{\zeta_\tau} }{d\tau}   =     \frac{d\tilde{I}_{\tau} }{d\tau} =     \frac{ dB_{\tau}}{d \tau}  , \label{eq:randomI}
 \end{equation}
 with $\tau\in \mathbb{R}^+$ the time parameter, and  where
   \begin{equation}
   \zeta_\tau = {\rm inf}\left\{s\geq 0: \tau_s(I_{[0,s]}) \geq \tau \right\}
 \end{equation}
is the functional inverse of $\tau_t(I_{[0,t]})$.  {Note that according to Eqs.~(\ref{eq:XBLocal}-\ref{eq:randomI}) a rescaling of the form $Z_tdB_t=dB_\tau$ requires  that  $d\tau = Z^2_t dt$,   which  follows from the fundamental property $\langle B^2_t\rangle = t$ of the Brownian motion.}   In physics notation, we drop the tilde, writing  $\tilde{I}_{\tau} = I_{\tau}$ and   understanding that this is $I$ expressed in the  time $\tau$.    Hence, according to Eq.~(\ref{eq:randomI}),  $I_{\tau}$ is a Brownian motion and thus a martingale, and therefore $I_t$ is a local martingale.   

\subsubsection*{Sufficient conditions for martingality of a local martingale}
 
{     
 We discuss here a few criteria to determine whether a local martingale is a martingale.     If  the local martingale $L_t$ is bounded, i.e.,  $\langle {\rm sup}_{s\leq t} |L_s| \rangle < \infty$, then it will be martingale (see Theorem 51 in  chapter I page 38 of \cite{protter2005stochastic}).   Another criterion uses the quadratic variation   (Corollary 3 of Theorem 27 in chapter II of \cite{protter2005stochastic}).  
 \begin{leftbar}
 \begin{theorem}[\textbf{Condition for a local martingale to be a  martingale}] \label{theorem:CondMart} 
$\\$ A local martingale $L_t$ is a martingale with $\langle L^2_t\rangle<\infty$ for all $t\geq 0$ if and only if  $\langle [L, L]_t\rangle<\infty$ for all $t\geq 0$.
Moreover, it holds that    
 \begin{eqnarray}\
 \langle L^2_t\rangle  =  \langle [L, L ]_t\rangle. \label{eq:LContMArt}
  \end{eqnarray}
 \end{theorem} 
 \end{leftbar}
 The formula (\ref{eq:LContMArt}) is called the It\^{o} isometry.   Theorem \ref{theorem:CondMart} implies that the It\^{o} isometry is a fundamental property of square integrable martingales.  
 Finally,  if $M$ is a nonnegative, local martingale with $\langle M_0\rangle < \infty$, then $M$ is a supermartingale (Lemma 14.3 in section IV.14 of \cite{williams1979diffusions}).   This clarifies why in   in the panel (b) of Figure~\ref{fig:mart} the mean value $\langle X_t\rangle$ is a decreasing function. 
}

\subsubsection*{Example of a local martingale}   We  consider an example of  a strict local martingale, i.e., a local martingale that is not a martingale.   Consider the
 It\^{o} unidimensional stochastic differential equation \cite{kotani2006condition,cherstvy2013anomalous}: 
 \begin{eqnarray}
\dot{I}_t=  I_t^{k} \dot{B}_t,  \label{eq:XBx}
 \end{eqnarray}
 with $I_0 = 1$,  $k$  a real number, and $B_t$ a Brownian motion as before.  
 
\begin{figure}[H]
{\includegraphics[width=0.45\textwidth]{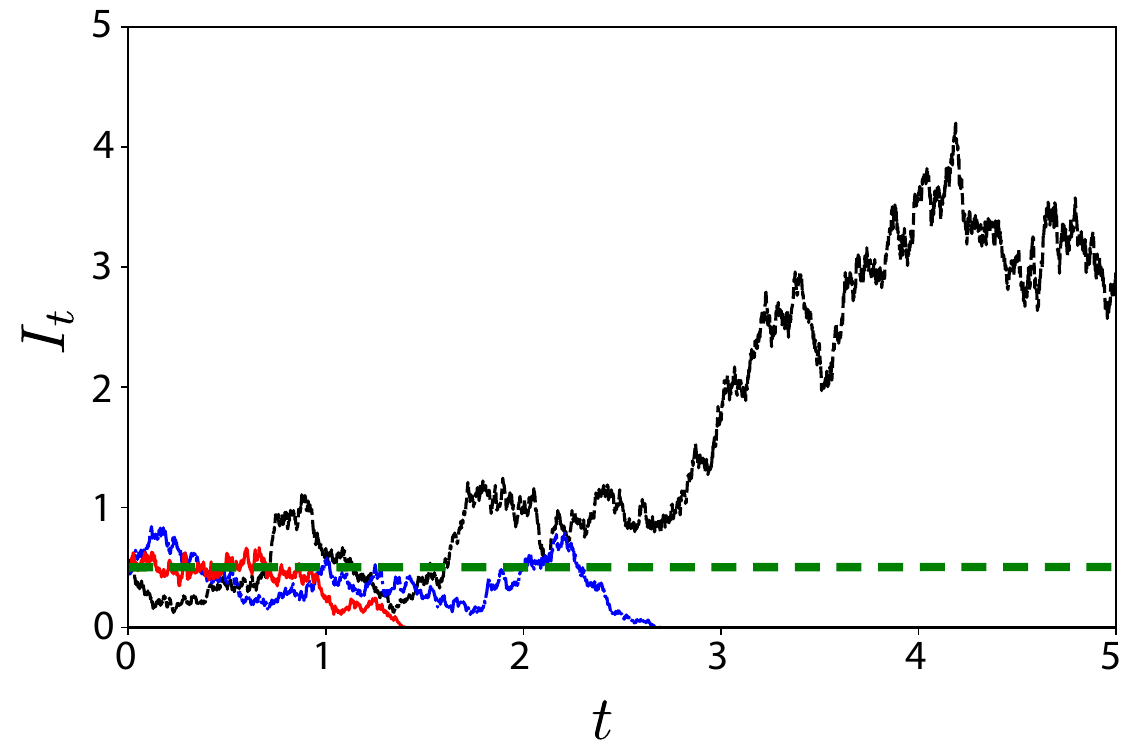}}
{\includegraphics[width=0.45\textwidth]{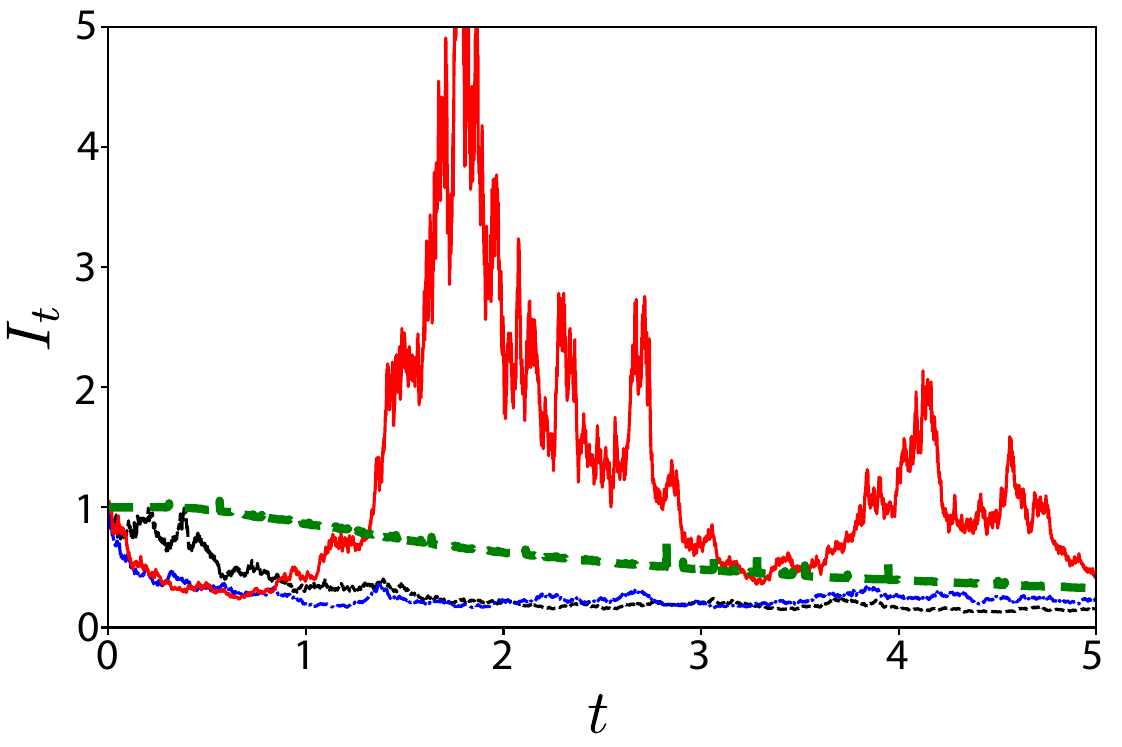}}
\caption{Illustration of  a martingale (left) and a strict local martingale (right).     We show three realisations of the process (\ref{eq:XBx})  for ${k}=0.5$  (left) and ${k}=1.5$ (right).   The  dotted green curve is an estimate of the average $\langle I_t \rangle$ based on an empirical average over~$10^6$ realisations of the process.    If   ${k}=0.5$, then $\langle I_t\rangle = 1$ and the process is driftless, whereas for ${k}=1.5$ the mean value    $\langle I_t\rangle$ decreases as a function of $t$.  \label{fig:mart}}
\end{figure}

 The physical picture is as follows: the process  $I_t$ is nonnegative and it has an  absorbing state at $I_t = 0$.    If $k>1$, then the diffusion constant gets  small  enough for $I_t\rightarrow 0$, such that $I_t$ gets trapped near the origin. As a consequence, $\langle I_t \rangle$ decreases as a function of $t$ and the It\^{o} integral $I_t$ is not a martingale.  On the other hand, when $k< 1$, then the diffusion constant does not decay fast enough for $X_t\rightarrow 0$ and  the process reaches the origin in a finite time.    In other words,  if $\T_0 = {\rm inf}\left\{t>0: X_t = 0\right\}$ then $\mathcal{P}(\T_0<\infty)=1$.   In this case, the process $X_{t\wedge \T_0}$ is a martingale as shown in Ref.~\cite{kotani2006condition} and illustrated  in Figure~\ref{fig:mart}.

      \subsection{Doob-Meyer decomposition}\label{sec:doobmeyer}
  Local martingales appear in the decomposition of a process into a martingale and a predictable process, which extends the Doob decomposition theorem, given by Theorem~\ref{doobD},  to processes in continuous time (Theorem 16 in chapter III on page 116 in \cite{protter2005stochastic}).   In  continuous time,  a stochastic process $A_t$ is {\bf predictable} if   $\langle A_t| X_{[0,t-dt]}\rangle = A_{t}+O(dt) $. 

 \begin{leftbar}
 \begin{theorem}[\textbf{Doob-Meyer Decomposition}]$\\$  \label{doobM}
 Let $Y_t$ be a right-continuous stochastic process function of the set of trajectories $X_{[0,t]}$, and integrable (i.e., $\langle  |Y_t| \rangle<\infty$ for all $t$).
 Then it can be uniquely decomposed as
 \begin{eqnarray}
   Y_{t} = Y_{0} + L_{t} + \underbrace{\int_{0}^{t} v_{s}ds}_{\displaystyle  A_{t}},
  \end{eqnarray}
  where we have introduced the conditional velocity
 \[
v_s=\lim_{h\rightarrow0^{+}}\left\langle \left.\frac{Y_{s+h}-Y_{s}}{h}\right|X_{[0,s]}\right\rangle 
\]
   The predictable process $A_t$ is  called  compensator and $L_t$ is a local martingale with respect to the underlying process $X_t$.% As in discrete time, 
   %a predictable process  is a   c\`{a}gl\`{a}d function applied on the set of trajectories $X_{[0,t]}$.    
  \end{theorem}
  \end{leftbar}
   % We call a process $A(t)$ increasing if with probability one $A(t)\geq A(s)$ for all $t\geq s$ and $A(0) = 0$.
  
  We now give some remarks about Doob-Meyer decomposition theorem.
  \begin{itemize}
      \item If $Y_t$ is a submartingale (supermartingale) then  $v_{s}\geq 0$ ($v_{s}\leq 0$) and then the compensator $A_t$ is increasing (decreasing).
       \item   The compensator of the square $X_t^2$ of a stochastic process is denoted by  $\langle X_t,X_t\rangle$ and called the  \textbf{predictable quadratic variation} or sharp bracket of $X$, see Ref.~\cite{protter2005stochastic}.   For  continuous processes, the predictable quadratic variation equals the quadratic variation defined  in Eq.~(\ref{eq:quadrVar}), but for processes  with jumps these are in general different.   Take for example the counting process $N_t$ of example Eq.~(\ref{eq:countingM}).  In this case, $[N_t,N_t] = N_t$, whereas $\langle N_t,N_t\rangle = \lambda_t$.   On the other hand, for the Brownian motion, $[B_t,B_t] = \langle B_t,B_t\rangle =  t $.    
\item      Theorem \ref{th:3} directly gives   the Doob-Meyer decomposition for  a real-valued bounded function $f_{t}\left(X_{t}\right)$  evaluated on  a Markovian process $X_{t}$, viz.,
 \begin{eqnarray}
  f_{t}(X_t)=  f(X_0)  + M_t+ \underbrace{\int_{0}^{t}ds\left(\partial_{s}f_{s}+\mathcal{L}_{s}f_{s}\right)(X_{s})}_{\displaystyle  A_t}, \label{eq:funcmarkovProcess}
  \end{eqnarray}
  where $M_t$ is  Dynkin's additive martingale, as defined in (\ref{eq:markovProtc}),  and  $\mathcal{L}_{s}$ is the generator of  $X_{t}$.

  \end{itemize}

\subsection{Continuous martingales}  

We consider the case of continuous martingales, i.e., martingales  with trajectories that  are continuous functions of time.  The main result we discuss here is the martingale representation theorem, which states that  for square integrable, continuous martingales the  integrator $dM_s$ in the It\^{o} integral can be assumed to be a Brownian motion.

As discussed before, an It\^{o} integral $ I_t =  \int^{t}_{0} Z_s dB_s $, as defined in Eq.~(\ref{eq:Ito}), with an integrand $Z_t$ that obeys Eq.~(\ref{eq:AverageD}), i.e. $\int^{t}_0 \langle Z^2_s \rangle {\rm d}s <\infty$, is a martingale.   In addition, it is square integrable.   Indeed,  from It\^{o}'s formula, see Appendix~\ref{app:Ito-lemma}, it follows that
\begin{equation}
    I^2_t  = 2\int^{t}_0 I_s Z_s dB_s + \int^{t}_0 Z^2_s ds
\end{equation}
and since  $\langle \int^{t}_0 I_s Z_s dB_s \rangle = 0$, 
\begin{equation}
\langle I^2_t \rangle  = \Big\langle \int^{t}_0 Z^2_s ds \Big\rangle ,
\end{equation}
which is finite, as assumed with Eq.~(\ref{eq:AverageD}).

Remarkably, the converse is also true, i.e., a square integrable martingale with respect to the Brownian motion $B_{[0,t]}$ is an It\^{o} integral.   This constitutes the martingale representation theorem (Theorem 4.3.4 in \cite{oksendal2003stochastic}).  
\begin{leftbar}
\begin{theorem}[\textbf{Martingale representation theorem}]
$\\$ Suppose $M_t$ is a martingale relative to $B_t$ and suppose that $\langle M^2_t\rangle<\infty$ for all $t\geq0$.   Then there exists a unique $Z_t$ evaluated on the trajectories $X_{[0,t]}$  that satisfies $\int^{t}_0 \langle Z^2_s \rangle {\rm d}s <\infty$ and that satisfies with probability one
\begin{eqnarray}
M_t = \langle M_0 \rangle + \int^t_0 Z_s dB_s
 \end{eqnarray} 
for all $t\geq 0$.
 \end{theorem}
 \end{leftbar} 
 
 As an illustrative example, consider the martingale $B_t^2-t$, see Eq.~(\ref{trieste2}), which can be expressed as an It\^{o} integral as follows
 \begin{equation}
    M_t = B^2_t -t = 2\int^t_0B_s  dB_s,
 \end{equation}
 where the second equation follows from applying It\^{o}'s lemma, see Eq.~\eqref{eq:itoFormula0}.

\subsection{$^{\spadesuit}$Stochastic exponential}\label{sec:stochExp}
 As we will see in the next chapter, the exponentiated, negative, fluctuating, entropy production of a nonequilibrium stationary process is a stochastic exponential. For this reason, we discuss here stochastic exponentials   in more detail.

Let $X_t\in \mathbb{R}^d$ be a possibly multidimensional c\`{a}dl\`{a}g process, i.e., a process with right-continuous   trajectories ($X_{t_+}=X_t$) that have left limits everywhere ($X_{t^{-}}$ exists), and let $Y_t(X_{[0,t]})\in \mathbb{R}$  be a stochastic process defined on  $X_t$. 
The stochastic (Dol\'{e}ans-Dade) exponential ~\cite{Doleans1970quelques}  associated with  $Y$   is  the solution of the stochastic differential  equation~\cite{protter2005stochastic}
 \begin{eqnarray}
\dot{\mathcal{E}}_t(Y) =  \mathcal{E}_{t-}(Y) \dot{Y}_t ,   \label{eq:stochExp}
 \end{eqnarray}
 where $\mathcal{E}_{t-}  = \lim_{\epsilon\rightarrow 0^+}\mathcal{E}_{t-\epsilon}$ and with $\mathcal{E}_{0}=1$.

 The stochastic exponential is specified by the process $Y_t$ and therefore we denote it by $\mathcal{E}_t(Y)$; sometimes we drop $Y$ because it is clear which process is meant.   We remark that  the notation $ \mathcal{E}_t(Y)$  is done in analogy with exponentials, yet the process $ \mathcal{E}_t(Y)$ in Eq.~\eqref{eq:stochExp} is a functional of the trajectory $Y_{[0,t]}$.   For the particular case of $Y_t=B_t$ we recover, using Eq.~\eqref{eq:stochExp}, the stochastic exponential associated with the Wiener process, whose solution is given by Eq.~\eqref{eq:BExp} with $z=1$, i.e.
 \begin{eqnarray}
\mathcal{E}_t(B) = \exp\left( B_{t} - \frac{t}{2} \right). \label{eq:BExp22}
\end{eqnarray}
  Note that interpreting Eq.~(\ref{eq:stochExp}) in Stratonovich, we would obtain the solution $\exp(Y_t-Y_0)$.   However, the stochastic exponential use this equation in the  It\^{o} interpretation, leading to a different stochastic process.  

If $Y_t = L_t$, a local martingale, then also $\mathcal{E}(L)$ is a local martingale, and hence the stochastic exponential inherits the local martingale property.   In addition, if $\mathcal{E}_t(L)>0$, then it is a positive supermartingale~\cite{protter2005stochastic, revuz2013continuous}.    If $\mathcal{E}_t(L)>0$, then  a necessary and sufficient condition for the martingality of a stochastic exponential is  that
\begin{equation}
    \langle \mathcal{E}_t(L) \rangle = 1
    \label{eq:iftDoleans}
\end{equation}
holds for all $t$, which is reminiscent of the integral fluctuation relation, see below. Equation~\eqref{eq:iftDoleans}  follows from the fact that $\mathcal{E}_t(L) $ is a supermartingale with constant expectation, see the discussion around Eq.~(\ref{eq:constCond}). {In the present case, for which $\mathcal{E}_t(L)>0$ and (\ref{eq:iftDoleans}) holds}, {we can define the path probability 
\begin{equation}
\mQ(X_{[0,t]}) \equiv   \mathcal{E}_t({L}) \mP(X_{[0,t]})
\end{equation}
}
so that 
  \begin{eqnarray}
   \mathcal{E}_t({L}) = R_t =  \frac{\mQ(X_{[0,t]})}{\mP(X_{[0,t]})}.
 \end{eqnarray}
 Hence, not all stochastic exponentials are Radon-Nikodym derivative processes, but if $\mathcal{E}_t$ is a positive, martingale, then it is.

On the other hand, unlike for the local-martingale property, the stochastic exponential does not inherit the martingale property.  Indeed, if $Y_t=M_t$, a martingale process, then it is not guaranteed that $\mathcal{E}_t(M)$ is a martingale.   Instead, one needs to verify some additional conditions that we discuss below.

Let us consider a few examples of stochastic exponentials: 

\subsubsection{Stochastic exponential of a differentiable function}
If $I_t=f_t$, with $f_t\in\mathbb{R}$ a differentiable function evaluated on $t$, then we obtain the differential equation 
 \begin{eqnarray}
\dot{\mathcal{E}}_t(f) =  \mathcal{E}_t (f) \dot{f}_t \label{eq:trivial}
 \end{eqnarray}
 with solution 
 \begin{equation}
     \mathcal{E}_t(f)= \exp(f_{t}-f_0).  
 \end{equation} 
 Notice that this is because It\^{o} and Stratonovich calculus are the same for differentiable functions.   
 
\subsubsection{Stochastic exponential of a continuous process}
Let $X$ be a possibly multidimensional process, and let $Y_t(X_{[0,t]})\in\mathbb{R}$ be a continuous c\`{a}dl\`{a}g process.   Equation~(\ref{eq:stochExp}) then reads
\begin{equation}
    \dot{\mathcal{E}_t} = \mathcal{E}_t \dot{Y}_t
\end{equation}
and is solved by 
 \begin{eqnarray}
 \mathcal{E}_t(Y)= 
 \exp\left(Y_t - Y_0 - \frac{1}{2}[Y,Y]_t\right), \label{eq:expS2}
 \end{eqnarray} 
 where $[Y,Y]_t$ is the quadratic variation defined in Eq.~(\ref{eq:quadrVar})~\footnote{Indeed, if we apply It\^{o}'s formula to  $\exp(Z)$, see Appendix~\ref{app:Ito-lemma}, with $Z = Y_t - Y_0 - [Y,Y]_t/2$, and use that $[Y,[Y,Y]] = 0$  and $[[Y,Y],[Y,Y]] = 0$, we obtain  Eq.~(\ref{eq:stochExp}).  
 Note that the correction term inside the exponential can be understood from the passage of Eq.~\eqref{eq:stochExp} from  the It\^{o} convention to the Stratonovich convention (see Appendix~\ref{app:Ito-lemma} on stochastic integrals).}.

As an example of stochastic exponential of a continuous process, consider the case of Eq.~(\ref{eq:expst2}), copied here for convenience, 
\begin{equation}
\frac{d}{dt}\underbrace{\exp(-S_t)}_{\displaystyle\mathcal{E}_t(Y)} = - \underbrace{\exp(-S_t)}_{\displaystyle\mathcal{E}_t(Y)} \sqrt{2 D_t }\dot{B}_t.   \label{eq:expst2nc}
\end{equation}
In this case, $Y_t = -\int^t_0 ds \sqrt{2D_s}\dot{B}_s$  and the  quadratic variation 
\begin{equation}
    [Y,Y]_t = 2\int^{t}_0 D_s ds ,
\end{equation} 
so that 
\begin{equation}
\mathcal{E}_t(Y)  = \exp\left(Y_t-Y_0-\int^{t}_0 D_s ds \right) =  \exp(-S_t).
\end{equation}

%{(RC : Sentence told before no ? )} {\bf [IN: yes, but we did not say the converse.   Maybe we should indeed mention the converse before too.]} 

We now give some remarks about the martingale structure of the stochastic exponential of continuous stochastic processes.
\begin{itemize}
    \item If $Y_t=L_t$ is a local martingale,  then $\mathcal{E}(L)$ is a local martingale, and the converse is also true, i.e., a strictly positive, continuous, local martingale takes the form of stochastic exponential $\mathcal{E}_t(L)$, see Ref.~\cite{revuz2013continuous}.
    \item  If $Y_t=M_t$ is a continuous  martingale,  then $\mathcal{E}_t(M)$ is a martingale when 
  Novikov's condition  \cite{novikov1973identity}, 
  \begin{eqnarray}
 \left\langle  \exp\left(\frac{1}{2} [M,M]_t\right)\right\rangle<\infty, \label{eq:novikov}
 \end{eqnarray}
holds for all $t\geq 0$.    
Notice that the Novikov condition is  a sufficient, and not a necessary condition for martingality.    However, this condition is often not very practical as we will see in the next chapter on thermodynamics.  
\item Another necessary condition for martingality is the Kazamaki condition~\cite{kazamaki1977problem}, which states that if 
$\exp(L_t/2)$
is a submartingale, then $\mathcal{E}_t(L) $ is a martingale.    These conditions have been refined, see for example references \cite{Kramkov, cherny2000criteria}.     
\item  {See Ref.~\cite{protter2008no} for a generalisation of the stochastic exponential  $\mathcal{E}_t(Y)$ to the case of processes with jumps.}
\end{itemize}

%\subsubsection{Stochastic exponential of a cÃ dlÃ g process}  When $I$  may have jumps $\Delta I_t = I_t-I_{t-}\neq 0$, then~\cite{protter2008no}
%%{(Raph : I doubt about $[I,I]_t$, I imagine that it must be here just the continuous part of thus quadratic variation, that you note  $[I,I]_c$ below. Moreover, this formula is not used in the review (could be eventually in section 5.4.2) : so this is just cultural !  )}{[\bf IN:  the idea was to use it for the exponential of -S in markov jump processes?]} 
%   \begin{eqnarray}
% \mathcal{E}_t(I) = \exp\left(I_t - I_0 - \frac{1}{2}[I,I]^c_t\right)\prod_{s\in \mathcal{I}(t)}(1+\Delta I_s)\exp(-\Delta I_s), \label{eq:expS3}
% \end{eqnarray} 
% where $[I,I]^c_t$ is the continuous part of the quadratic variation $[I,I]_t$, as defined in Eq.~(\ref{eq:quadrVar}).
% Hence, if $\Delta I_t>-1$, then $ \mathcal{E}_t(I)$ is positive.  If $I_t=M_t$ is a martingale and $\Delta M_t>-1$, then Novikov's condition has been generalised as follows \cite{protter2008no}
%   \begin{eqnarray}
% \left\langle  \exp\left(\frac{1}{2} \langle M^c,M^c\rangle_t + \langle M^d,M^d\rangle_t\right)\right\rangle<\infty, \quad \forall t\geq 0, \label{eq:novikov2}
% \end{eqnarray}
% where $\langle M^c,M^c\rangle_t$ and $\langle M^d,M^d\rangle_t$ are  the predictable quadratic variations, as defined in Sec.~\ref{sec:doobmeyer},  of the continuous and discontinuous parts of $M$, respectively, such that $M=M^c+M^d$.  If Eq.(\ref{eq:novikov2}) holds, then $\mathcal{E}$ is again a path probability ratio.  
%
%

%\chapter[Martingales in stochastic thermodynamics I

\chapter{Martingales in stochastic thermodynamics I: Introduction}
\label{sec:martST1} 

%\chapter{\textcolor{blue}{Appetizer to Stochastic Thermodynamics}}
\epigraph{\textit{Voudriez-vous bien passer vos jours\\
A faire le Sardanapale,\\
Et servir une martingale ?}  \vspace{0.5cm}\\
(Would you like to spend your days\\
To do the Sardanapale,\\
And serve a martingale?)\\
Paul Scarron, Le Virgile travesti, Ch.~IV (1648).}

%{\bf [IN: Outline for this section?  "Introductory Example" should be "Introductory Examples" ?, since we have multiple examples. ]}

Since the origins of thermodynamics in the  XIX Century, physicists have been intrigued by the implications of the second law of thermodynamics  at the mesoscopic level.  One of the first references to thermodynamics at the mesoscopic scale appeared in Tait's {\em Sketch of Thermodynamics}  (1878), on which  J.~C.~Maxwell commented  "{\it a finite number of molecules [...] are still and every now and then still deviating very considerably from the theoretical mean of the whole system [they belong to]. [...] Hence the second law of thermodynamics is continually being violated, and that to a considerable extent, in any sufficiently small group of molecules belonging to a real body}"~\cite{tait1877sketch,maxwell1878tait}.  

The pioneering thoughts of Tait and Maxwell illustrate the puzzle of  formulating a second law of thermodynamics for mesoscopic systems.   This puzzle has, to a large extent, been resolved in the past decades  with proper definitions of heat and entropy production based on the theory of stochastic processes.    According to stochastic thermodynamics, entropy production can be transiently negative, but is on average positive.   Moreover, the fluctuations of negative entropy production are constrained by fluctuation relations.    
  
  Several of  the standard results of stochastic thermodynamics can be understood and improved with  martingale theory.  
In the present chapter, we provide an introduction to martingale theory in stochastic thermodynamics. After briefly reviewing key definitions and results in stochastic thermodynamics,  we show  how martingales naturally appear  in the theory of stochastic thermodynamics.  In particular, with two  examples of stochastic processes, namely,   one-dimensional overdamped Langevin processes and Markov jump processes, we show that for stationary processes  the exponentiated negative entropy production is  a martingale, which is the central result in martingale theory for stochastic thermodynamics.      Through the study of  two simple examples, the present chapter  sets the stage for the next  three chapters that discuss the theory in a more general setup (Chapter~\ref{ch:thermo2}) and  provide a  detailed analysis of the implications of martingale theory (Chapter~\ref{sec:martST2}) and  (Chapter~\ref{sec:martST3}).

This chapter is structured as follows: In Sec.~\ref{sec:41}, we introduce the setup of an overdamped, one-dimensional, isothermal, Langevin process, and subsequently we review the basic thermodynamics results for this setup.   In Sec.~\ref{sec:42}, we show for this setup that if the process is stationary, then  the exponential of the negative entropy production is a martingale.   % In particular,  we discuss the martingale fluctuation relations, the martingale version of the second law of thermodynamics, and the implications of martingale theory for fluctuations of entropy production.   
Subsequently, in Sec.~\ref{sec:43}, we review thermodynamics for Markov jump processes, and in Sec.~\ref{sec:44} we discuss  the thermodynamics of Markov jump processes with martingale theory.

%\section{%Primer on Stochastic Thermodynamics
\section{Introduction: Langevin equation and thermodynamics}\label{sec:41}

Before embarking on a journey through thermodynamics with martingales, we  derive the "standard" first and  second laws of thermodynamics for nonequilibrium isothermal processes described by a one dimensional, overdamped, isothermal, Langevin equation.   Notice  that since the focus of this paper is on martingales, and since there  exist already several textbooks and review papers on stochastic thermodynamics,   we review  here  the essentials of stochastic thermodynamics, referring the interested reader to the   Refs.~\cite{LNP, seifert2012stochastic, peliti2021stochastic, chetrite2008fluctuation} for further details.

\begin{figure}[h!]
    \centering
    \includegraphics[width=11cm]{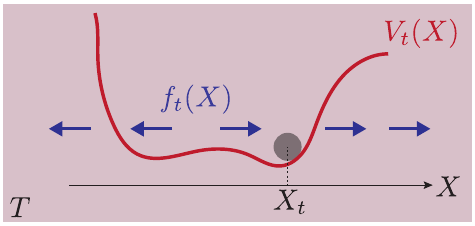}
    \caption{Illustration of the paradigmatic model   discussed in Sec.~\ref{sec:41}. A Brownian particle (gray circle) confined in a one dimensional  potential that may be time-dependent ($V_t(X)$, red line) is subject to an external force that may depend on time and space ($f_t(X)$, blue arrows). The position $X_t$ of the particle at time $t$ evolves according to 
    Eq.~\eqref{eq:lang2}. In this model, the particle fluctuates moving along the potential  and under the action of the external force field---the blue arrows illustrate the direction of the external force  and the length of the arrows its magnitude (note that the external force $f_t(x)$ is in general different from minus the instantaneous  value of the slope of the  potential $-\partial_x V_t(x)$).}
    \label{fig:STillus}
\end{figure}

\subsection{System setup}

Consider a  particle with mass $m$ that moves  with homogeneous mobility $\mu$ (or equivalently, friction coefficient $\gamma = 1/\mu$)   in a homogeneous thermal bath in equilibrium and at a constant temperature $T$, as illustrated in Fig.~\ref{fig:STillus}.   The particle is subject to a potential $V_t(x)\equiv V(x,\lambda_t)$ whose shape is controlled by a time-dependent deterministic protocol $\lambda_t$. Moreover, a non-conservative force $f_t(x)$ (e.g. solenoidal) is exerted on the particle.   The dynamics of the particle is described by the underdamped Langevin 
equation 
\begin{equation}
\begin{cases}
 \dot{X}_{t}=P_{t}/m,\\
 \dot{P}_{t}=-\displaystyle\frac{\gamma}{m}P_{t}-\left(\partial_{x}V_{t}\right)(X_{t})+f_{t}(X_{t})+\sqrt{2T\gamma}\dot{B}_t ,
\end{cases}
\label{ILEK2}
\end{equation}
where $B_t$ is a  Brownian noise, 
 where $P_t$ is the momentum of the particle at time $t$, and where $X_t$ is its position.
The first order equations~\eqref{ILEK2} can be written  equivalently as the   one-dimensional second order equation 
\begin{equation}
    m \ddot{X}_t = -\gamma \dot{X}_t  - (\partial_x V_t)(X_t) + f_t(X_t) + \sqrt{2 T\gamma}\dot{B}_t.
    \label{eq:lang1}
\end{equation}

For reasons of simplicity, in stochastic thermodynamics it is customary to consider  the overdamped limit, which we introduce in the following.
\begin{leftbar} 
In the overdamped limit, $m \mu \ll 1$, the position $X_t$ of the particle is described by the 
  \textbf{overdamped, isothermal Langevin equation} 
\begin{equation}
     \dot{X}_t = - \mu(\partial_x V_t)(X_t) +  \mu f_t(X_t) + \sqrt{2\mu  T }\, 
    \dot{B}_t,
    \label{eq:lang2}
\end{equation}
where we have used the notation $-(\partial_x V_t)(X_t)= -\left.\partial_{x} V_t(x) \right|_{x=X_t}$ for the value of the conservative force evaluated at $X_t$. 
 \end{leftbar}
 Notice that Eq.~(\ref{eq:lang2}) is    the   one-dimensional version  ($d=1$) of  Eq.~(\ref{eq:LE}) with  a homogeneous diffusion constant $D=\mu T$ determined by    Einstein's relation Eq.~(\ref{ER}). 
 % In  Appendix~\ref{app:underd} we provide a derivation of Eq.~(\ref{eq:lang2}) starting from the (underdamped) Langevin equation {(RC : NO, it is not this. What is written now in ~\ref{app:underd} is just the way to pass from Eq.~(\ref{eq:LE}) to underdamped Langevin equation.)}{\bf [IN: not clear to me how to change.   Lets discuss. ]}  .
% {In  Appendix~\ref{app:underd}, }
  Despite its simplicity, the Langevin Eq.~\eqref{eq:lang2} contains all the minimal ingredients of stochastic thermodynamics, namely fluctuations (thermal noise), energy (potential), and nonequilibrium forces (a time dependent potential and external forces).

\subsection{First law of  thermodynamics}  
\label{sec:1LLang}

We follow the conventional route in thermodynamics~\cite{tait1877sketch,maxwell1878tait}: we first define the work done on the system, and consequently we obtain the heat from the first law of thermodynamics.   

The work  done on the system in the time interval $[t,t+dt]$ consists of two contributions, namely, the work due to a  changing potential $(\partial_t V_t)(X_t,t)$ and the work  due to a nonconservative force $f_t$. 
%\begin{equation}
%(\partial_t V)(X_t,t)=\dot{\lambda}_t(\parti%al_\lambda V)(X_t,\lambda_t),  
%\end{equation}
%due to 
%a
%changing potential $V_t$   
Adding the two contributions, we obtain  that the power exerted on the system in $[t,t+dt]$ is \cite{Sekimoto1997, Sekimoto1998}
\begin{equation}
    \dot{W}_t \equiv (\partial_t V_t)(X_t) + f_t(X_t)\circ \dot{X}_t ,
    \label{eq:ws1}
\end{equation} 
{where $\circ$ denotes  the Stratonovich product (see Sec.~\ref{sec:stochasticcalculus} for a reminder on stochastic calculus).}
%which is the sum of
%\begin{equation}
%(\partial_t V)(X_t,t)=\dot{\lambda}_t(\partial_\lambda V)(X_t,\lambda_t),   % 
%\end{equation}
%due to the  protocol $\lambda_t$ changing the shape of the potential, and the power of the work $f_t$ done by a non-conservative force on the system. 
\begin{leftbar}
Integrating over time, we find the  \textbf{stochastic work} $W_t =\int_0^t \dot{W}_s ds$ done on the system along a stochastic trajectory $\Xot$, which using Eq.~\eqref{eq:ws1} reads
\begin{equation}
    W_t = \int_0^t \Big[ (\partial_{s} V_s)(X_{s})\text{d}s + f_s(X_{s})\circ dX_{s}\Big].
    \label{W}
\end{equation}
%which is also a stochastic  quantity (an adapted functional $W_t=W[X_{[0,t]}]$).  {[\bf IN: do we need the adapted part?]}
\end{leftbar}
Note that in Eqs.~(\ref{eq:ws1}-\ref{W}) we have used a  Stratonovich integral to define the work  done by a non-conservative force  on the system, and not an It\^{o} integral, 
%We have used  a Stratonovich integral because   it changes sign under time reversal, 
and this will prove to be  important for developing a thermodynamically consistent picture.   
 
Given the work $W_t$, we use the first law of thermodynamics to obtain an explicit expression for the heat.
\begin{leftbar}
 The \textbf{first law of stochastic thermodynamics} reads~\cite{Sekimoto1997, Sekimoto1998}
 \begin{equation}\label{eq:1LST}
    Q_t + W_t=V_{t}(X_{t})-V_{0}(X_{0}),
 \end{equation}
 which we assume to hold along any trajectory $X_{[0,t]}$ traced by a nonequilibrium system described by the isothermal Langevin equation~\eqref{eq:lang2}.
  %and holds for nonequilibrium systems described by the isothermal Langevin equation~\eqref{eq:lang2}.
  %when evaluated along any trajectory $X_{[0,t]}$ traced by a nonequilibrium system described by the isothermal Langevin equation~\eqref{eq:lang2}.
 \end{leftbar}
  The first law of thermodynamics, Eq.~(\ref{eq:1LST}), defines  the heat $Q_t$.    In rate form, Eq.~(\ref{eq:1LST}) reads
\begin{equation}
    \dot{Q}_t+\dot{W}_t  = \dot{V}_t = (\partial_t V_t)(X_t) +(\partial_x V_t)(X_t)\circ \dot{X}_t.
    \label{eq:fl1}
\end{equation}
Substituting  Eq.~(\ref{eq:ws1}) in Eq.~(\ref{eq:fl1}) we find 
\begin{equation}
 \dot{Q}_t =  -F_t(X_t)\circ \dot{X}_t, \label{eq:qs2}
 \end{equation}
 where  the total force
  \begin{equation}
    F_t(X_t)= - (\partial_x  V_t)(X_t) + f_t(X_t)
    \label{eq:ftot}
\end{equation}
 contains, in general, a conservative (first) and a non-conservative (second) term; this  
is the one-dimensional version of the more general expression Eq.~(\ref{F}).  Note that   the heat absorbed per unit of time in $[t,t+dt]$,  Eq.~\eqref{eq:qs2}, can also be expressed by 
\begin{equation}
    \dot{Q}_t = (-\gamma \dot{X}_t+\sqrt{2\gamma T}\dot{B}_t)\circ \dot{X}_t,
\end{equation} which was the original expression for the stochastic heat in overdamped Langevin systems obtained by Sekimoto~\cite{Sekimoto1997}.

%{\bf [start with work]}
% Following Sekimoto's papers~\cite{Sekimoto1997,Sekimoto1998}, we define the {\em stochastic} heat $Q_t\equiv Q[X_{[0,t]}]$ associated with the overdamped isothermal Langevin equation~\eqref{eq:lang2}.    
 \begin{leftbar}
Integrating $\dot{Q}_s$ over the interval $s\in [0,t]$, we obtain the    \textbf{stochastic heat} $Q_t=\int_0^t \dot{Q}_s ds $ absorbed by the system  along a stochastic trajectory $\Xot$~\cite{Sekimoto1997,Sekimoto1998},
\begin{equation}
    Q_t  =  -\int_0^t F_{s}(X_{s})\circ dX_{s} .\label{eq:qs3}
\end{equation}
\end{leftbar}
 Note that the Stratonovich rule implies that for  time-homogenous total forces, $F_{s}(x)=F(x)$,   $Q_t$ changes sign under time reversal.

  The formalism presented here has been extended  to underdamped Langevin systems for which the kinetic energy change leads to an additional term in the stochastic heat, see  e.g., Refs.~\cite{kurchan1998fluctuation, LNP,chetrite2008fluctuation}. 

\subsection{Second law of stochastic thermodynamics}  
\label{sec:2LLang}

Consider the Fokker-Planck equation
\begin{equation}
 \partial_{t}\rho_{t}(x)=- \partial_x  J_{t,\rho}(x) 
\label{eq:fpe1d}
\end{equation}
for the instantaneous density $\rho_{t}(x)=\langle\delta(X_t-x)\rangle $, which is the one-dimensional version of Eq.~\eqref{ce}.  According to Eq. ~\eqref{curr},   the hydrodynamic current is given by 
\begin{equation}
J_{t,\rho}(x)= \mu F_t(x) \rho_{t}(x)-\mu  T \partial_{x}\rho_{t}(x).
\label{eq:JFP}
\end{equation}

\begin{leftbar} Given a state $X_t$,   Shannon's instantaneous information   content  is given by  \cite{cover1999elements, seifert2005entropy}
\begin{equation}
    S^{\rm sys}_t \equiv - \ln \rho_t(X_t),
    \label{eq:defSSysonetime}
\end{equation}
which we identify as the  \textbf{nonequilibrium system entropy} for the system  in state $X_t$ at time $t$. The (nonequilibrium) system entropy change associated with $\Xot$ is thus given by
\begin{equation}
     \Delta S^{\rm sys}_t\equiv S^{\rm sys}_t - S^{\rm sys}_0 = \ln \frac{\rho_0(X_0)}{\rho_t(X_t)} .
\end{equation}
\end{leftbar}

Now, we review the notion of stochastic environmental entropy change, as commonly used in the stochastic thermodynamics of isothermal systems.
\begin{leftbar} Since  the environment is in a state of thermal equilibrium at temperature  $T$, the {\bf entropy change of the environment} is given by  Clausius' statement 
\begin{equation}
    S^{\rm env}_t = -\frac{Q_t}{T},
    \label{eq:Senvdefinition}
\end{equation}
where we recall that $Q_t$ is the stochastic heat given by Eq.~\eqref{eq:qs3}. Equation~\eqref{eq:Senvdefinition} thus provides the definition for the stochastic environmental entropy change along a trajectory of an isothermal, overdamped, Langevin equation.
\end{leftbar}

To obtain a balance equation for entropy, we determine the rate of  change of the nonequilibrium system entropy.  An explicit calculation yields
\begin{eqnarray}
    \dot{S}^{\rm sys}_t &=& \frac{d\left(-\ln \left(\rho_{t}\left(X_t\right)\right)\right)}{dt} \label{eq:us1}\\
    &=&  - \frac{\left(\partial_t \rho_{t}\right)\left(X_t\right)}{\rho_{t}\left(X_t\right)} - \frac{\left(\partial_x \rho_{t}\right)\left(X_t\right)}{\rho_{t}\left(X_t\right)}\circ \dot{X}_t\label{eq:us22}\\ 
    &=& \underbrace{- \frac{\left(\partial_t \rho_{t}\right)\left(X_t\right)}{\rho_{t}\left(X_t\right)}+ \frac{J_{t,\rho}(X_t)}{\mu T \rho_{t}\left(X_t\right)}\circ \dot{X}_t}_{\displaystyle \dot{S}_{t}^{\rm tot}} - \underbrace{\frac{F_t(X_t)}{T}\circ \dot{X}_t}_{\displaystyle \dot{S}_{t}^{\rm env}}. \label{eq:us33}
\end{eqnarray}
The steps we have used in Eqs.~(\ref{eq:us1}-\ref{eq:us33}) are the following: in Eq.~\eqref{eq:us1}, we have used the definition of stochastic system entropy ~\eqref{eq:defSSysonetime}. In Eq.~\eqref{eq:us22}, we have used Stratonovich  rules of calculus, which are formally identical to those of standard calculus. In Eq.~\eqref{eq:us33} we have used the definition of the probability current~\eqref{eq:JFP}. Lastly, in Eq.~\eqref{eq:us33},  we  have identified the second term as the change of the environmental entropy   $\dot{S}_t^{\rm env}=-\dot{Q}_t/T$, taking into account the expression~\eqref{eq:qs2} for the stochastic heat  $\dot{Q}_t= -F(X_t,t)\circ \dot{X}_t$.    

The  first two terms in the right-hand side of Eq.~\eqref{eq:us33} are changes in the system's entropy that do not involve environmental entropy changes, thus we identify them as the stochastic entropy production rate in $[t,t+dt]$
\begin{equation}
   \dot{S}^{\rm tot}_t = - \frac{\left(\partial_t \rho_{t}\right)\left(X_t\right)}{\rho_{t}\left(X_t\right)} + \frac{J_{t,\rho}(X_t)}{\mu T \rho_{t}\left(X_t\right)}\circ \dot{X}_t.
    \label{eq:StotStrato}
\end{equation} 
Note that the definition~\eqref{eq:StotStrato} is consistent with Prigogine's balance equation~\cite{kondepudi2014modern}
\begin{equation}
\dot{S}^{\rm tot}_{t} \equiv \dot{S}^{\rm sys}_{t}  + \dot{S}^{\rm env}_{t} ,  \label{eq:balanceS}
\end{equation}
with $\dot{S}^{\rm sys}_{t}$ given by Eq.~\eqref{eq:defSSysonetime}  and $\dot{S}^{\rm env}_{t}$ given by Eq.~\eqref{eq:Senvdefinition}. 
\begin{leftbar}
The {\bf stochastic entropy production} associated with a trajectory $\Xot$ of an overdamped Langevin equation~\eqref{eq:lang2} equals  the sum of the system entropy change plus the environmental entropy change,
\begin{equation}
    S^{\rm tot}_t = \Delta S^{\rm sys}_t -\frac{Q_t}{T} .
    \label{eq:Stotlangevin0}
\end{equation}
Integrating Eq.~\eqref{eq:StotStrato} over time, we get the explicit expression \cite{seifert2005entropy}
\begin{equation}
    S^{\rm tot}_t =  \int_0^t \left[-\frac{\left(\partial_s \rho_{s}\right)\left(X_s\right)}{\rho_{s}\left(X_s\right)} + \frac{J_{s,\rho}(X_s)}{\mu T \rho_{s}\left(X_s\right)}\circ \dot{X}_s\right] ds.
    \label{eq:Stotlangevin}
\end{equation}
\end{leftbar}
Note that the stochastic entropy production $S^{\rm tot}_t$ is a stochastic process that thus   fluctuates in time, and as we show below it  can  take negative values. On the other hand, a second law is recovered for the average  of $S^{\rm tot}_t$. {Here and in the following,  we will use the interpretation $\langle\dot{Z}_t\rangle\equiv (d/dt)\langle Z_t\rangle$ for all functionals $Z_t=Z[X_{[0,t]}]$.}

\begin{leftbar}
Indeed, averaging Eq.~(\ref{eq:StotStrato}) over many realisations of the process, we  find that the average rate of entropy production is non-negative 
\begin{equation}
    \langle \dot{S}^{\rm tot}_t\rangle   = \frac{1}{\mu T}\int_{\X}   \frac{\left(J_{t,\rho}(x)\right)^2 }{\rho_{t}(x)}dx\geq 0.
    \label{footsoon}
    \end{equation}
 Thus, we call Eq.~(\ref{footsoon})  the  {\bf second law of thermodynamics} for overdamped Langevin equations.   Integrating over time and using $S^{\rm tot}_0 =0$ we obtain
   \begin{equation}
       \langle S^{\rm tot}_t\rangle \geq 0. \label{eq:secondlaw}
   \end{equation}   
   {See Sec.~\ref{sec:martlanggent} for a detailed derivation of Eq.~\eqref{footsoon} in a more general setting beyond the unidimensional case. }
  \end{leftbar}

If we call the system together with its environment the {\bf universe}, then   the second law of thermodynamics  states that on average the total entropy of the universe increases.  
  
To derive the  second law of thermodynamics, given by  Eq.~(\ref{footsoon}), for overdamped Langevin equations, we  convert the Stratonovich integral in Eq.~(\ref{eq:StotStrato}) into an It\^{o} integral and use  the Langevin equation~\eqref{eq:lang2} for $\dot{X}_t$, yielding  (see Appendix~\ref{app:derStot})
\begin{equation}
\dot{S}^{\rm tot}_t
=-2\frac{\left(\partial_{t}\rho{}_{t}\right)(X_t)}{\rho{}_{t}(X_t)}+\underbrace{\frac{1}{\mu T}\left(\frac{J_{t,\rho}(X_t)}{\rho_{t}(X_t)}\right)^2}_{\displaystyle v^S_t(X_t)}+\underbrace{\left(\sqrt{\frac{2}{\mu T}}\frac{J_{t,\rho}(X_t)}{\rho_{t}(X_t)}\right)}_{\displaystyle\sqrt{2v^S_t(X_t)}}\dot{B}_{t},
\label{forGirs-1-1-1aa}
\end{equation}
where $v^S_t$ is the time-dependent {\bf entropic drift}~\cite{Pigolotti:2017}, which we discuss further in Sec.~\ref{sec:ItoApproach}.
Averaging Eq.~\eqref{forGirs-1-1-1aa} over the noise, the first and  the third term in \eqref{forGirs-1-1-1aa} vanish.   Indeed, the first term has zero average due to conservation of probability and the third term because it is a martingale.   The average of the second term yields precisely the right-hand side in Eq.~(\ref{footsoon}).    In Sec.~\ref{sec:martlanggent}, we generalise Eq.~(\ref{forGirs-1-1-1aa})  to the case of $d>1$    dimensions.

 \subsection{{Stratonovich and Ito formulations: recap}}
 
 {We provide here for readers' ease a short recap on the formulation of the first and second laws of thermodynamics in Stratonovich and Ito formulations. To this aim, we collect results from the previous subsections~\ref{sec:1LLang} and~\ref{sec:2LLang} and provide the stochastic rates of heat, work, energy and entropy production in $[t,t+dt]$ associated with the overdamped Langevin dynamics~\eqref{eq:lang2}.}

 {\textbf{Stratonovich formulation.} The work and heat exchanges read
\begin{eqnarray}
    \dot{W}_t &=& (\partial_t V_t)(X_t) + f_t(X_t)\circ \dot{X}_t ,
    \label{eq:wr1}\\
    \dot{Q}_t  &=& (\partial_x V_t)(X_t)\circ \dot{X}_t - f_t(X_t)\circ \dot{X}_t,
    \label{eq:hr1}
\end{eqnarray} 
which leads to the first law
\begin{equation}
    \dot{V}_t  = (\partial_t V_t)(X_t) + (\partial_x V_t)(X_t)\circ \dot{X}_t.
    \label{eq:er1}
\end{equation}
Note that Eq.~\eqref{eq:er1} could be retrieved from standard rules of calculus (chain rule for differentiation) that apply in the Stratonovich convention.   The rate of stochastic entropy production reads
\begin{equation}
   \dot{S}^{\rm tot}_t = - \frac{\left(\partial_t \rho_{t}\right)\left(X_t\right)}{\rho_{t}\left(X_t\right)} + \left[\frac{J_{t,\rho}}{\mu T \rho_{t}}\right] \left(X_t\right)\circ \dot{X}_t,
    \label{eq:stotr1}
\end{equation}
which leads to the second law at the average level $\langle \dot{S}^{\rm tot}_t\rangle \geq 0$.}

{\textbf{Ito formulation.} The work and heat exchanges can be retrieved by applying  to Eqs.~\eqref{eq:wr1} and~\eqref{eq:hr1} the rules of conversion between Stratonovich and Ito products (Theorem~\ref{th:0})
\begin{eqnarray}
    \dot{W}_t &=& [ \partial_t V_t +  \mu T (\partial_x f_t)] (X_t) + f_t(X_t) \dot{X}_t   ,
    \label{eq:wr2}\\
    \dot{Q}_t  &=& \mu T [ \partial_x^2 V_t -   (\partial_x f_t)] (X_t)  + [\partial_x V_t- f_t](X_t)\dot{X}_t,
    \label{eq:hr2}
\end{eqnarray} 
which leads to the first law
\begin{equation}
    \dot{V}_t  = (\partial_t V_t)(X_t) + (\partial_x V_t)(X_t) \dot{X}_t + \mu T (\partial_x^2 V_t)(X_t) ,
    \label{eq:er2}
\end{equation}
which  can be retrieved directly applying Ito rules of calculus, i.e. Ito's lemma [see Eq.~\eqref{eq:itoFormula0}].   On the other hand, the rate of stochastic entropy production reads
\begin{equation}
   \dot{S}^{\rm tot}_t = \left[- \frac{\left(\partial_t \rho_{t}\right)}{\rho_{t}} + \partial_x \frac{J_{t,\rho}}{\rho_{t}}\right] (X_t)+ \left[\frac{J_{t,\rho}}{\mu T \rho_{t}}\right] \left(X_t\right) \dot{X}_t,
    \label{eq:stotr2}
\end{equation}
which follows from applying to Eq.~\eqref{eq:stotr1} Theorem~\ref{th:0} for conversion of Stratonovich to Ito product. Next, replacing $\dot{X}_t$ in Eq.~\eqref{eq:stotr2} by the Langevin dynamics~\eqref{eq:LE2}, one finds 
\begin{equation}
   \dot{S}^{\rm tot}_t = \left[-2\frac{\left(\partial_t \rho_{t}\right)}{\rho_{t}} + v^S_t\right ] (X_t)+ \left[
   \sqrt{2v^S_t}\right](X_t) \dot{B}_t,
    \label{eq:stotr3}
\end{equation}
which reveals the martingale structure of $\exp(-S^{\rm tot}_t)$ in time-homogeneous stationary states, as we will show in the next Sec.~\ref{sec:42}.}

{Taken together, the results in this subsection illustrate the fact that Stratonovich convention provides a more simple mathematical formulation of the first law of thermodynamics, whereas the Ito convention is more suitable to discuss the second law. Generalizations of Eqs.~(\ref{eq:wr1}-\ref{eq:stotr3})  to $d-$dimensional overdamped Langevin dynamics can be found in Ref.~\cite{Pigolotti:2017}.}

\section[Martingales in stationary Langevin processes]{%Primer on 
Martingale theory for  stationary 1D isothermal Langevin processes}\label{sec:42}

A central result of  martingale theory for stochastic thermodynamics is that in   nonequilibrium stationary processes the exponentiated negative entropy production is  a martingale.     Therefore, we first derive this result, and subsequently,  we discuss  some interesting implications of the martingality of the   exponentiated negative entropy production.   Notably,  we discuss  here   % martingale fluctuation relations, the martingale version of the second law of thermodynamics, and 
some of the universal fluctuation properties of entropy production that can be derived from martingale theory.   A more extensive overview of the implications of martingale theory for thermodynamics is presented  in  Chapters 6 to 9, which includes  the martingale fluctuation relations and the martingale version of the second law of thermodynamics.

For reasons of clarity, we restrict ourselves to the simplest case of one-dimensional, stationary, overdamped, Langevin processes.    Nevertheless, martingale theory for thermodynamics is general, and applies also to  nonstationary, underdamped, or multidimensional Langevin processes, see e.g., Refs.~\cite{Chetrite:2011,Pigolotti:2017, chetrite2019martingale, neri2020second, manzano2021thermodynamics} or Chapter~\ref{ch:thermo2}.   Therefore, we encourage the reader, based on the derivations below,  to derive the corresponding results for, e.g., the multidimensional, underdamped, or nonstationary cases (this is  fun!).  

\subsection{Stationary overdamped Langevin processes}

A  Langevin process is {\bf stationary} when the initial distribution obeys
\begin{equation}
     \rho_t(x)= \rho_{\rm st}(x),
     \label{eq:st}
\end{equation}
for all $t\geq 0$ and $x\in \mathcal{X}$, 
where $\rho_{\rm st}$ is the stationary probability distribution solving Eq.~(\ref{eq:FPI}).   
Analogously, the stationary current is defined as the hydrodynamic current~\eqref{eq:JFP} associated with the  stationary distribution~\eqref{eq:st}, i.e.
\begin{equation}
J_{t,\rm st}(x)\equiv J_{t,\rho_{\rm st}}(x) =  \mu F_t(x) \rho_{\rm st}(x)-\mu  T \partial_{x}\rho_{\rm st}(x).
\label{eq:JFP2}
\end{equation}
Note that the stationary distribution solves $\partial_x J_{t,\rm st}(x)=0$, see Eq.~\eqref{eq:fpe1d}. 

%\begin{leftbar}
We say that the  Langevin equation \eqref{eq:lang2} is  {\bf time homogeneous} when  the  conditions
\begin{equation}
f_t = f \quad  {\rm and} \quad V_t=V,   \label{eq:statCond}
\end{equation}
are satisfied, and this is  a necessary condition for stationarity when the mobility matrix is independent of time; notice that this is not  the case with time-dependent mobilities.   When (\ref{eq:statCond}) holds, then also the total force is time independent, 
\begin{equation}
     F(x) = F_t(x)=(\partial_x V)(x) +   f(x).
\end{equation}

\begin{leftbar}
Throughout this section we assume {\bf stationarity and time-homogeneity}, in other words, we assume that $X_t$ obeys the Langevin equation~\eqref{eq:lang2} 
\begin{equation}
     \dot{X}_t = - \mu F(X_t) + \sqrt{2 T \mu }    \dot{B}_t,
    \label{eq:langTHNESS} 
\end{equation}
and  $\rho_0(x)=\rho_{\rm st}(x)$. Following Eq.~\eqref{eq:JFP2}, the stationary current is time independent and reads
\begin{equation}
J_{\rm st}(x)=  \mu F(x) \rho_{\rm st}(x)-\mu  T \partial_{x}\rho_{\rm st}(x).
\label{eq:JFP2st}
\end{equation}
\end{leftbar}

\subsection{Martingality of the exponentiated negative entropy production}  \label{sec:martingaleExpS}
We show that for stationary processes  $X_{t}$, described by Eq.~\eqref{eq:lang2} with the stationarity condition given by Eq.~(\ref{eq:st}), the   exponentiated negative entropy production $\emStot$ is a martingale.   To this aim, we use three distinct, but equivalent, approaches, namely, we show that $\emStot$ is (a)  an   It\^{o} integral of the form Eq.~\eqref{eq:Ito}; (b) a Radon-Nikodym derivative process (or path-probability ratio) of the form Eq.~(\ref{eq:radon}); and (c)  a Dynkin's martingale of the form Eq.~\eqref{eq:markovProtc}. {Note that the latter approach (Dynkin's) is new to our knowledge and thus first shown here}. While initially we will not bother too much with the distinction between  local martingales and martingales, we will come back on this point at the end of the section.

\subsubsection{ It\^{o}-integral approach}\label{sec:ItoApproach}
As discussed in Sec.~\ref{sec:examplescont}, It\^{o} integrals of the form Eq.~(\ref{eq:Ito}) that satisfy Eq.~(\ref{eq:AverageD})  are martingales.    Here, we show that $\exp(-S^{\rm tot}_t)$ is an It\^{o} integral.

\begin{leftbar} For  time-homogeneous stationary processes for which $\partial_{t}\rho_{t}(x)=0$ for all $x$,  the It\^{o} stochastic differential equation for $S^{\rm tot}_t$, given by~\eqref{forGirs-1-1-1aa}, simplifies into the compact form
\begin{equation}
      \dot{S}^{\rm tot}_t =    v^{S}(X_t)  +\sqrt{2 v^S(X_t)}\dot{B}_t, \label{eq:entropy}
 \end{equation}
 where  $v^S(X_t)$ is the so-called {\em entropic drift}~\cite{Pigolotti:2017} defined by 
 \begin{eqnarray}
     v^S (X_t) \equiv \frac{1}{\mu T} \left( \frac{J_{\rm st}(X_t)}{\rho_{\rm st}(X_t)}   \right)^2 \geq 0,
     \label{eq:vs}
 \end{eqnarray}
 and 
 where the noise $\dot{B}_t$ is the same noise as  in the Langevin  Eq.~\eqref {eq:lang2} for the dynamics of the particle. 
\end{leftbar} 
 
  Since $v^{S}_t\geq 0$,  it follows readily from Eq.~(\ref{eq:entropy}) that $S^{\rm tot}_t$ is a submartingale.    Note   that according to Eq.~(\ref{eq:entropy}) the drift and diffusion coefficients of $S^{\rm tot}_t$   are identical, which is reminiscent of the  Einstein relation Eq.~(\ref{ER}).  
  
The equality of the drift and diffusion coefficients of $S^{\rm tot}_t$ determines  the martingality of $\exp(-S^{\rm tot}_t)$.    
  Indeed, applying It\^{o}'s formula, see Eq.~(\ref{eq:itoFormula}) in Appendix~\ref{app:ItoFormula}, to the variable change $S^{\rm tot}_t \to \exp(-S^{\rm tot}_t) $, and using Eq.~\eqref{eq:entropy}, we obtain
 \begin{equation}
\frac{d\exp(-S^{\rm tot}_t)}{dt}=
-\sqrt{2 v^S(X_t)}\, \emStot \dot{B}_t, 
\label{eq:gbm}
\end{equation}
and hence $\exp(-S^{\rm tot}_t)$ is an It\^{o} integral;  notice the formal analogy between Eqs.~(\ref{eq:gbm}) and (\ref{eq:expst2nc}). In addition,  Eq.~(\ref{eq:gbm}) shows that $\exp(-S^{\rm tot}_t)$  is the stochastic exponential $\mathcal{E}_t(M)$ of the martingale
\begin{equation}
    M_t=-\int^{t}_0ds \sqrt{2v^S(X_s)} \dot{B}_s;
\end{equation}  the latter process is a martingale according to Eq.~(\ref{eq:AverageD})  as    $\langle v^S(X)\rangle = \langle \dot{S}^{\rm tot}\rangle <\infty$.   

\subsubsection{Path-probability-ratio approach }

We show that $\exp(-S^{\rm tot}_t)$ takes the form of a path-probability ratio by   identifying a suitable measure  $\mathcal{Q}$ for which $\exp(-S^{\rm tot}_t)$  can be written as a  Radon-Nikodym derivative process of the form  Eq.~(\ref{eq:radon})~\cite{maes1999fluctuation, maes2009selection, chetrite2008fluctuation, Chetrite:2011,   neri2017statistics,cates2022stochastic}.   
\begin{leftbar}
To this purpose, we introduce the time-reversal map  $\Theta_{t}$ that acts on the trajectories  $x_{\left[0,t\right]}$ through
\begin{equation}
\left[\Theta_{t}\left(x_{\left[0,t\right]}\right)\right]_{s}\equiv   x_{t-s} \quad {\rm for} \quad s\leq t. \label{eq:timeR}
\end{equation}
Subsequently, we show one of the central results in stochastic thermodynamics, namely
\begin{equation}
   S^{\rm tot}_{t}=\ln \frac{\mathcal{P}\left(X_{[0,t]}\right)}{\left(\mathcal{P}\circ \Theta_{t}\right)\left(X_{\left[0,t\right]}\right)},
   \label{eq:meanttoshow0}
\end{equation} 
or equivalently,
\begin{equation}
    \exp(-S^{\rm tot}_{t}) = \frac{\left(\mathcal{P}\circ \Theta_{t}\right)\left(X_{\left[0,t\right]}\right)}{\mathcal{P}\left(X_{[0,t]}\right)}. \label{eq:meanttoshow}
\end{equation} 
\end{leftbar}
For time-homogeneous stationary processes, the measure $ \mathcal{P}\circ \Theta_{t}$ appearing in the numerator of Eq.~\eqref{eq:meanttoshow} is  independent of $t$, 
see 
Refs.~\cite{nagasawa1964time,chung2006markov,chetrite2018peregrinations}, and hence the measure $\mQ$ in Eq.~(\ref{eq:radon}) is in this case $ \mathcal{P}\circ \Theta_{t}$.  This can be understood heuristically as follows. The map $\Theta_t$ is a time-reversal map that mirrors trajectories around the reflection  point $t/2$.  Since by assumption $\mP$ is a stationary measure, the location of the reflection point does not alter the statistics determined by  $\mP$.    We come back to this point at the end of the derivation.

\subsubsection*{Proof of relation~\eqref{eq:meanttoshow0}}
Let us now prove the relation \eqref{eq:meanttoshow0}.  Using the Onsager-Machlup path-integral approach, see Eqs.~(\ref{eq:onsagerMachlupApproach1})-(\ref{eq:onsagerMachlupApproach3}) and Eqs.~(\ref{eq:additiveA}), we can write explicit expressions for the conditional path probabilities \footnote{The $\dot{X}_{s}$ in these relations must be interpreted in Stratonovich convention.} , viz.,
\begin{equation}
  \mathcal{P}(X_{[0,t]}|X_0) = \frac{1}{\mathcal{N}} \exp\displaystyle \left(-\frac{1}{4\mu T}\int_0^t \left\{ \left[ \dot{X}_{s} -\mu F(X_{s})\right]^2  +  \frac{\mu}{2} (\partial_x F) (X_{s}) \right\} ds \right), \label{eq:pfpath}
\end{equation}   
and analogously, 
\begin{equation}
    \mathcal{P}(\Theta_t(\Xot)|X_t)
    = \frac{1}{\mathcal{N}}\exp\displaystyle \left(-\frac{1}{4\mu T}\int_0^t \left\{ \left[ -\dot{X}_{s} - \mu F(X_{s})\right]^2  +  \frac{\mu}{2} (\partial_x F) (X_{s}) \right\} ds \right).\label{eq:pbpath}
\end{equation}   
Taking the ratio of Eqs.~\eqref{eq:pfpath} and~\eqref{eq:pbpath},  
we obtain the so-called  local detailed balance condition,   
\begin{equation}
\displaystyle \frac{ \mP(\Theta_t(\Xot)|X_t)  }{\mathcal{P}(X_{[0,t]}|X_0) }  =  \exp\displaystyle \left(-\frac{1}{T}\int_0^t  F(X_{s}) \circ \dot{X}_{s} \, \text{d}s \right)=\exp\left(\frac{Q_t}{T} \right)=\exp\left(-S_t^{\rm env} \right),
\label{dimanche}
\end{equation}
that relates the stochastic heat $Q_t$ to the path probabilities.  
Lastly, multiplying Eq.~(\ref{dimanche}) by
$\exp(S^{\rm sys}_0-S^{\rm sys}_t) = \rho_{\rm st}(X_t)/\rho_{\rm st}(X_0)$ (see ~\eqref{eq:defSSysonetime}) we obtain 
\begin{equation}
\displaystyle 
 \frac{ \mP(\Theta_t(X_{[0,t]}))}{\mathcal{P}(X_{[0,t]})  } =
\frac{ \mP(\Theta_t(\Xot)|X_t)\rho_{\rm st}(X_t)  }{\mathcal{P}(X_{[0,t]}|X_0)\rho_{\rm st}(X_0)}  = \exp(-S^{\rm env}_t - \Delta S^{\rm sys}_t  ) = \emStot
\label{eq:Nirvana}
\end{equation}
which is Eq.~(\ref{eq:meanttoshow}) that we were meant to show. 

As promised, we now show that $\mathcal{P}[\Theta_t(X_{[0,t]})] = \mathcal{Q}[X_{[0,t]}]$, and hence there is no explicit time-dependence on $t$.   For this, we  show that the Lagrangian of $\mathcal{P}[\Theta_t(X_{[0,t]})]$ contains no explicit time dependency on $t$ ---see Eq.~ \eqref{eq:LagrangianFormula} for the definition of a Lagrangian.  Equation \eqref{eq:pbpath} can be rewritten as 
\begin{align*}
\mathcal{P}\left(\Theta_{t}\left(X_{\left[0,t\right]}\right)\right) & =\rho_{\rm st}\left(X_{t}\right)\frac{1}{\mathcal{N}}\exp\left(-\frac{1}{4\mu T}\int_{0}^{t}\left\{ \left(\dot{X}_{s}+\mu F\left(X_{s}\right)\right)^{2}+\frac{\mu}{2}\left(\partial_{x}F\right)\left(X_{s}\right)\right\} ds\right)\\
 & =\rho_{\rm st}\left(X_{0}\right)\frac{1}{\mathcal{N}}\exp\left(-\int_{0}^{t}\left[\begin{array}{c}
\frac{1}{4\mu T}\left\{ \left(\dot{X}_{s}+\mu F\left(X_{s}\right)\right)^{2}+\frac{\mu}{2}\left(\partial_{x}F\right)\left(X_{s}\right)\right\} \\
-\partial_{x}\left(\ln\rho_{\rm st}\right)\left(X_{s}\right)\dot{X}_{s}
\end{array}\right]ds\right).
\end{align*}
Hence, the Lagrangian transforms under reversal as 
\begin{equation}
\left(\Theta_{t}\mathcal{L}\right)\left[X_{s},\dot{X}_{s}\right]=\frac{1}{4\mu T}\left(\left(\dot{X}_{s}+\mu F\left(X_{s}\right)\right)^{2}+\frac{\mu}{2}\left(\partial_{x}F\right)\left(X_{s}\right)\right)-\partial_{x}\left(\ln\rho_{st}\right)\left(X_{s}\right)\dot{X}_{s}.
\end{equation}
The absence of an explicit $t-$dependence in the right hand side of the last relation shows that the measure $\mathcal{P}\circ \Theta_{t}$
is not explicitly dependent on $t$, as claimed before. This allows us to conclude that  $\exp(-S^{\rm tot}_{t})$ is a martingale.  

 In  Sec.~\ref{eq:martsigmafunct},  we  give an alternative proof of the martingality of   $\left(\mathcal{P}\circ \Theta_{t}\right) (\Xot)/\mP(\Xot)$ in  stationary processes.  In addition, in Sec.~\ref{sec:6p1},   we extend  the path probability ratio formula (\ref{eq:meanttoshow})  to  the non-stationary and/or time-inhomogeneous set up.  %We refer the reader to Sec.~\ref{sec:6p1}   for this.   
In this non-stationary and/or time-inhomogeneous set up,   the explicit time  dependency of the measure $\mathcal{Q}^{(t)}$ in the numerator prevents  us from proving   that $\mathcal{Q}^{(t)}(\Xot)/\mP(\Xot)$ is a martingale, as done for discrete time in \eqref{eq:Der}; the latter is developed in  Sec.~\ref{eq:martsigmafunct}.

We end this section with a comment on the second law of thermodynamics. 

\begin{leftbar}  The  second law of thermodynamics is recovered when averaging the stochastic entropy production over the probability $ \mathcal{P}(X_{[0,t]})$, as this yields   the Kullback-Leibler divergence between the forward and reverse path probabilities~\cite{parrondo2009entropy}:
\begin{equation}\label{eq:SKLD}
\langle S^{\rm tot}_t\rangle= \int \mathcal{D}x_{[0,t]}  \mathcal{P}(x_{[0,t]})    \ln \displaystyle \frac{  \mathcal{P}(x_{[0,t]}) }{ \mP(\Theta_t(x_{[0,t]})) } = D_{\rm KL}[  \mathcal{P}(x_{[0,t]}) ||  \mP(\Theta_t(x_{[0,t]})) ]\geq 0.
\end{equation}
\end{leftbar}

\subsubsection{$^{\spadesuit}$Dynkin's martingale approach}
According to Theorem~\ref{Th:markov}  and Eq.~(\ref{eq:harmtc}),    harmonic functions of the generator of a Markov process define martingales.    We show here that  $\exp(-s)$ is a harmonic function of the corresponding generator.   {This provides a third  derivation of the martingale property of $\exp(-S^{\rm tot}_t)$, which to the best of our knowledge has not appeared before in the literature.}

Consider the  two-dimensional joint process 
$\mathscr{X}_t= (S^{\rm tot}_{t}, X_{t})^{\rm T}$
which  according to Eqs.~(\ref{eq:lang2}) and (\ref{eq:entropy}) solves the 
stochastic differential equations
\begin{equation}
\begin{cases}
 \dot{S}^{\rm tot}_t =    v^{S}(X_t)  +\sqrt{2 v^S(X_t)}\dot{B}_t,\\
\dot{X}_t =  \mu F(X_t) +    \sqrt{2\mu  T }  \dot{B}_t,
\end{cases}
\label{eq:dimancheAout}
\end{equation}
 with common noise $\dot{B}_t$
The Markovian generator associated with the two-dimensional diffusion process $\mathscr{X}_t$ given by Eqs.~\eqref{eq:dimancheAout} is  [see Eq.~(\ref{eq:gpdL})]
\begin{equation}
\mathcal{L} = v^{S}(x)\partial_s  + \mu F(x)\partial_x 
+v^{S}(x) \partial^2_s +
\mu T \partial^2_x  
+ 2 \sqrt{T \mu v^S(x)}   \partial_x\partial_s .
\label{eq:dimaout-gen}
\end{equation} 
We readily verify that 
\begin{equation}
\mathcal{L}\left[ \exp(-s)\right](s,x)= 0,
 \end{equation}
and hence $\exp(-s)$ is  a harmonic function of the generator $\mathcal{L}  $, implying, according to  Theorem~\ref{Th:markov} and Eq.~(\ref{eq:harmtc}), that $\emStot$ is  a martingale.     
We also find that 
\begin{equation}
\mathcal{L} [s](s,x)= v^S(x) \geq 0,
\end{equation}
and thus $s$ is a subharmonic function of  the generator, which implies that   $S^{\rm tot}_t$ is a submartingale~\cite{bremaud2013markov}.

Now, we write the two-dimensional stochastic differential equation~\eqref{eq:dimancheAout} in the Langevin form~\eqref{eq:LE} associated with the joint process 
$\mathscr{X}_t= (S^{\rm tot}_{t}, X_{t})^{\rm T}$,
\begin{eqnarray}
\dot{\mathscr{X}}_{t} = -\left(D\nabla \mathscr{V}\right)\left(\mathscr{X}_{t}\right)  
+ \left(\nabla  \cdot D\right)   \left(\mathscr{X}_{t}\right) 
+ \sigma (\mathscr{X}_t)  \dot{B_t},  \label{eq:Langx}
\end{eqnarray}
where  $\nabla$ is  in this case the gradient in $(x,s)$-space with components $\nabla_1 = \partial_s$ and $\nabla_{2} = \partial_x$. In Eq.~\eqref{eq:Langx} we have also introduced
\begin{equation}
\sigma (s,x) = \left(\begin{array}{c}
\sqrt{2 v^S(x)}
\\
 \sqrt{2 T \mu }
\end{array}\right), 
\end{equation}
 the generalized diffusion matrix 
\begin{eqnarray}
D(s,x)= \frac{\sigma (x,s) \sigma^{\dagger} (x,s)}{2}= \left(\begin{array}{cc} v^{\rm S}&\sqrt{T\mu v^{S}(x)}\\\sqrt{T\mu v^{S}(x)}   & T\mu \end{array}\right), 
\end{eqnarray} 
and the generalized time homogeneous potential 
\begin{eqnarray}
\mathscr{V}(s,x) =  -s - \ln  \rho_{\rm st}(x). 
\label{eq: Dalida}
\end{eqnarray}

We remark that  %with respect to \eqref{eq:LE} the equation  
Eq.~\eqref{eq:Langx} has a mobility matrix equal to the diffusion matrix, which is  reminiscent of Einstein's relation. The form of Eq.~\eqref{eq:Langx} readily implies that the generalized Boltzmann distribution
\begin{equation}
  \rho_{\rm st}(s,x) =   \exp(-\mathscr{V}(s,x))=\rho_{\rm st}(x)\exp(s)
  \label{eq:Sunday-Monday}
\end{equation}
is the      invariant measure.
Note that this measure is not normalizable, which follows from the fact that the generalized potential $\mathscr{V}(s,x)$ given by Eq.~\eqref{eq: Dalida}  is not confining. Physically, the latter statement means  that the $S^{\rm tot}_t$ is extensive in time. Note also that the factorisation property, revealed by Eq.~\eqref{eq:Sunday-Monday}, suggests an asymptotic independence between $X_t$ and $S^{\rm tot}_{t}$.

\subsubsection{$^{\spadesuit}$Martingale or strict local martingale?}\label{sec:spadeMart}

Is the exponentiated negative entropy production a martingale ($\langle \emStot\rangle = 1$) or a  strict local martingale ($\langle \emStot\rangle < 1$)?

Formally, Eq.~(\ref{eq:gbm}) implies that $\exp(-S^{\rm tot}_t)$ is a local martingale, and to prove martingality we need to show that Eq.~(\ref{eq:AverageD}) holds.
   Alternatively, according to Eq.~(\ref{eq:gbm}), $\exp(-S^{\rm tot}_t)$ is the stochastic exponential %{(RC:suppress the differential notation)}
\begin{equation}
  \exp(-S^{\rm tot}_t) =   \mathcal{E}\left(M_t \right),
\end{equation} 
as defined in Eq.~(\ref{eq:stochExp}),
of the martingale 
\begin{equation}
    M_t = -\int^{t}_0  \: \sqrt{2 v^S(X_u)}\dot{B}_u du.
\end{equation}   Hence 
%{(RC : Decide what notation for exponential)}
$\emStot$ is a martingale when  Novikov's condition Eq.~(\ref{eq:novikov}) holds, which here reads
\begin{equation}
 \left\langle  \exp\left( \int^{t}_0 v^S(X_s) ds \right)\right\rangle<\infty \label{eq:novikovExp}
\end{equation}
for all $t\geq 0$.  

In the Radon-Nikodym derivative approach, we also need  Novikov's condition Eq.~(\ref{eq:novikovExp}) to guarantee that  $\emStot$ is a martingale.   Indeed, the Onsager-Machlup path integral method, widely used in physics \cite{seifert2012stochastic}, assumes that  $\mathcal{P}\circ\Theta_t$ is absolutely continuous with respect to $\mathcal{P}$.   However, there is no guarantee that this is actually the case, and we need an additional condition, such as the Novikov condition\footnote{The Dynkin  martingale   approach does not provide a rigorous proof of martingality neither, as $\exp(-s)$ is not a bounded function, which is required to show martingality, see Theorem~\ref{Th:markov}.} to demonstrate this.

 Note that  Novikov's condition is a mathematical requirement for martingality, but currently we are not aware of physical examples for which $\emStot$ is a  local martingale but not martingales.

  \subsubsection{On the non-submartingality of the environmental entropy change}
          \label{NMEC}
         In general, the stochastic heat and  environmental entropy change are not martingales. In particular, for time-homogeneous stationary states, we obtain from  Eqs.~(\ref{eq:Senvdefinition}),~(\ref{eq:qs2}),~(\ref{eq:lang2}) and~(\ref{eq:JFP2}) the following stochastic differential equation  for the  environmental entropy change: 
          \begin{equation}
              \dot{S}^{\rm env}_t = v^{E}(X_t) + \sqrt{2v^{E}(X_t)}\circ \dot{B}_t, \qquad {\rm where}\quad  v^E (X_t) \equiv \frac{\mu F^2(X_t)}{T}\geq 0. \label{eq:enventropy22} 
          \end{equation}
Note that in  Eq.~\eqref{eq:enventropy22} $B_t$ is the same  noise that enters  in the Langevin equation for $X_t$~(\ref{eq:lang2}), and that the equation should be interpreted in the Stratonovich sense.  On the other hand, using It\^{o}'s convention, we get 
\begin{equation}
              \dot{S}^{\rm env}_t = \mu \left(\partial_x F\right) (X_t)+  v^{E}(X_t) + \sqrt{2v^{E}(X_t)} \dot{B}_t. \label{eq:enventropy222} 
          \end{equation} 
This implies that  for a generic $F$ it does not hold, in general, that  $S^{\rm env}_t$ has  positive drift, even though $v^E$ is non-negative. 
Nevertheless, if  $ \partial_x F(x) + F^2(x)/T \geq 0$  holds for all $x$, then  $\dot{S}^{\rm env}_t$ is a submartingale.  This is the case, among others,  when $\partial_x F(x)=0$, such that  $F(x)$ is homogeneous and independent of $x$ (see e.g.~the example in Ch.~\ref{Ring}). In such a case, $v^{E}(X_t)=v^E$ is independent of $X_t$ and Eq.~\eqref{eq:enventropy222} is equivalent to $ \dot{S}^{\rm env}_t = v^{E} + \sqrt{2v^{E}} \dot{B}_t$,  similar to Eq.~\eqref{eq:entropy22} for $\Stot$ in time-homogeneous stationary processes.

 \subsubsection{Non-stationary processes}
          
       We consider the dynamics of $S^{\rm tot}_t$ for non-stationary and/or non {time-}homogeneous processes $X$. The It\^{o} stochastic differential equation for $\Stot$~\eqref{forGirs-1-1-1aa} reads then
    \begin{equation}
      \dot{S}^{\rm tot}_t =  -2\left(\partial_t \ln \rho_t\right)(X_t) +   v_t^{S}(X_t)  +\sqrt{2 v_t^S(X_t)}\dot{B}_t, \label{eq:entropy223}
    \end{equation}
    which is the Doob-Meyer decomposition of $S^{\rm tot}_t$ (see Theorem~\ref{doobM}).  We recall readers the definition of time-dependent entropic drift $v_t^S(X_t)$ given in Eq.~\eqref{forGirs-1-1-1aa}.
    Since for nonstationary processes  $\partial_t  \rho_t\neq 0$, the first term in Eq.~\eqref{eq:entropy223} does not vanish and can be negative, which implies that the predictable process in the Doob-Meyer decomposition is not increasing, and as a consequence $S^{\rm tot}_t$ is not a submartingale.   %cannot write $\Stot$ as the sum of a predictable increasing process plus a martingale, as in   Eq.~\eqref{eq:entropyDM}.    
    In addition, the drift and diffusion constants in Eq.~(\ref{eq:entropy223}) are not equal as in Eq.~(\ref{eq:entropy22}) for stationary processes, and as a consequence the  statistical properties (e.g. global infimum) described above  are not universal for non-stationary overdamped Langevin processes.   In Chapter~\ref{ch:thermo2} we elaborate further on stochastic thermodynamics in nonstationary processes,  and in particular we discuss  thermodynamics martingale processes for this case.  
    
    Notice that the fact that a process is not stationary does not prevent that there exist other thermodynamic quantities whose negative exponential is an It\^{o}  integral. Indeed, Refs.~\cite{chun2019universal,chetrite2019martingale} showed that the so-called housekeeping entropy production  obeys an equation analogous to Eq.~\eqref{eq:gbm} (and is thus an It\^{o}  integral) for any Markovian process that may be non-stationary. We refer the readers to Eq.~\eqref{eq:Sehk} and Refs.~\cite{oono1998steady,hatano2001steady,van2010three} for further details on the concept of housekeeping (also called adiabatic~\cite{van2010three}) entropy production.

\subsection{Universal properties for the fluctuations of the  stochastic entropy production}\label{sec:randomtime}
The (local) martingale property of  $\exp(-S^{\rm tot}_t)$ together with the continuity of the process $S^{\rm tot}_t$ as a function of time implies that several fluctuation properties of $S^{\rm tot}_t$ are universal.  Here, following  Ref.~\cite{Pigolotti:2017},  we derive the universal properties of $S^{\rm tot}_t$ directly from the evolution Eq.~\eqref{eq:entropy} for entropy production, while in the next chapter we use Doob's theorems, as reviewed in Chapter~\ref{ch:3}, to derive these results.   

\subsubsection{Entropic random-time change}

\begin{figure}[h]
    \centering
    \includegraphics[width=9cm]{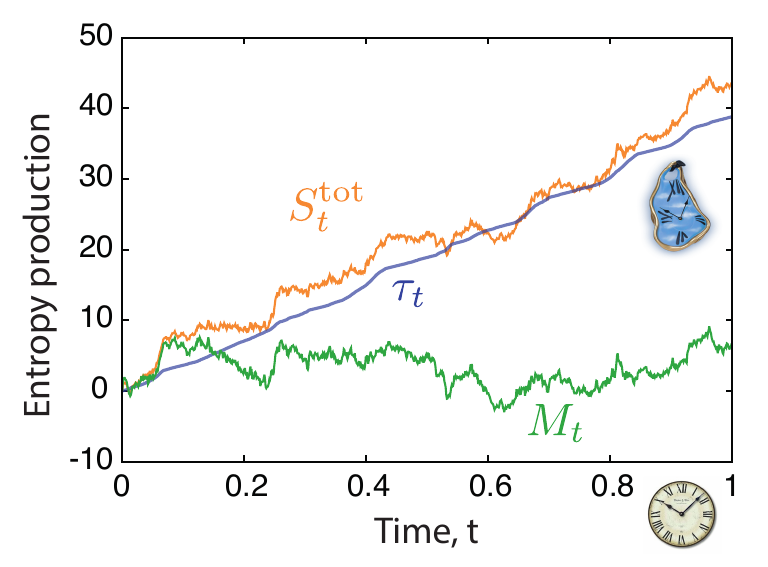}
    \caption{Illustration of the Doob-Meyer decomposition of entropy production,  Eq.~(\ref{eq:entropyDM}).  In time-homogeneous nonequilibrium stationary states, the stochastic entropy production $S^{\rm tot}_t$ (orange line) can be decomposed as the sum of the entropic time $\tau_t$ (blue line) plus a martingale $M_t$ (green line). The white clock in the $x$-axis illustrates the regular passage of time $t$  whereas the blue {\em Dalinean} clock illustrates the irregular passage of entropic time $\tau$  which depends on the states visited by the system.  Figure adapted from Ref.~\cite{Pigolotti:2017}. }
    \label{fig:ET}
\end{figure}

Our starting point is the It\^{o} stochastic differential equation for $\Stot$ in   time-homogeneous stationary states, see Eqs.~(\ref{eq:entropy}-\ref{eq:vs}) and copied here for convenience,
\begin{equation}
      \dot{S}^{\rm tot}_t =    v^{S}(X_t)  +\sqrt{2 v^S(X_t)}\dot{B}_t, \qquad  {\rm with} \qquad v^S (X_t) \equiv \frac{1}{\mu T} \left( \frac{J_{\rm st}(X_t)}{\rho_{\rm st}(X_t)}   \right)^2. \label{eq:entropy22}
 \end{equation}
Now, consider  the following time reparametrization
\begin{equation}
    dt \to d\tau_{t} (X_t) \equiv v^S (X_t) \,dt ,
    \label{eq:rtc1}
\end{equation}
such that $d\tau_t$  quantifies the expected  entropy production   in $[t,t+dt]$ given that the system was at state $X_t$ at time $t$. This is an example of a {\em random-time} transformation (see Sec.~\ref{sec:localmartingale} and also Sec.~8.5 in~\cite{oksendal2003stochastic}) of a stochastic process, in which a "clock" ticks faster (slower) whenever the system passes by a state of large (small) local entropy production. Following a single realization of duration $t$ its associated entropic random time is given by
\begin{eqnarray}
     \tau_t  = \int_0^t d\tau_s (X_s)  =  \int_0^t v^{S}(X_{s})\text{d}s,
     \end{eqnarray}
     which highlights the fact that $\tau_t$ is a functional of the trajectory $\Xot$.  Because $\langle  \dot{S}_{\rm s}^{\rm tot} |  X_{[0,t]} \rangle = v^S (X_t)$, see Eq.~\eqref{eq:entropy22}, the entropic time can be interpreted as the expected entropy production given that the system has traced a specific trajectory $\Xot$. 
     Integrating Eq.~\eqref{eq:entropy22} over time, we get
\begin{equation}
    S_t^{\rm tot} =  \tau_t + M_t,
    \label{eq:entropyDM}
\end{equation}
where   $M_t=\int_0^t \sqrt{2 v^S(X_{s})}\dot{B}_sds$ is a martingale %that can take any value 
and $\tau_t\geq 0$ is a monotonously nondecreasing  process, as $v^S(x)\geq 0$ for all $t$ and $x$.  In  martingale theory, this decomposition of entropy production (a submartingale) in the sum of the entropic time (a predictable process) and a noise process (martingale) is known as   the  {\em Doob-Meyer decomposition}, see Theorem~\ref{doobM} in Ch.~\ref{ch:3}. Applying the entropic random-time change given by Eq.~\eqref{eq:rtc1} to Eq.~\eqref{eq:entropyDM},  we get
     \begin{equation}
             \dot{S}^{\rm tot}_\tau = 
  1 + \sqrt{2}\,\dot{B}^{\rm S}_{\tau},  \label{eq:stot_tau}
     \end{equation}
     where  $\dot{B}^{\rm S}_{\tau}$  is a Gaussian white noise with $\langle \dot{B}^{\rm S}_{\tau} \rangle=0$ and $\langle \dot{B}^S_\tau \dot{B}^S_{\tau'}\rangle=\delta (\tau-\tau')$; note that  here the dot stands for the derivative with respect to $\tau$.

     \subsubsection{Universal properties in stationary states}
     Equation~\eqref{eq:stot_tau} reveals that, for any Langevin model described by Eq.~(\ref{eq:lang2}), $S_{\rm tot}$ obeys a drift-diffusion equation  with both drift  and diffusion coefficient equal to one   when measuring time in units of $\tau$.  This means that any statistical property of $S_{\rm tot}$ that is independent of $\tau$ is universal in this class of models. For example, even though the distribution  of $ \rho_{S^{\rm tot}_t} (s)$ at a fixed time $t$ is model dependent, the distribution of $S^{\rm tot}_\tau$ evaluated at entropic times is universal and given by
     \begin{equation}
         \rho_{S^{\rm tot}_\tau} (s) = \frac{\exp \left(-(s-\tau)^2/4\tau\right)}{\sqrt{4\pi\tau}}.
         \end{equation}

\begin{figure}[h]
    \centering
    \includegraphics[width=\textwidth]{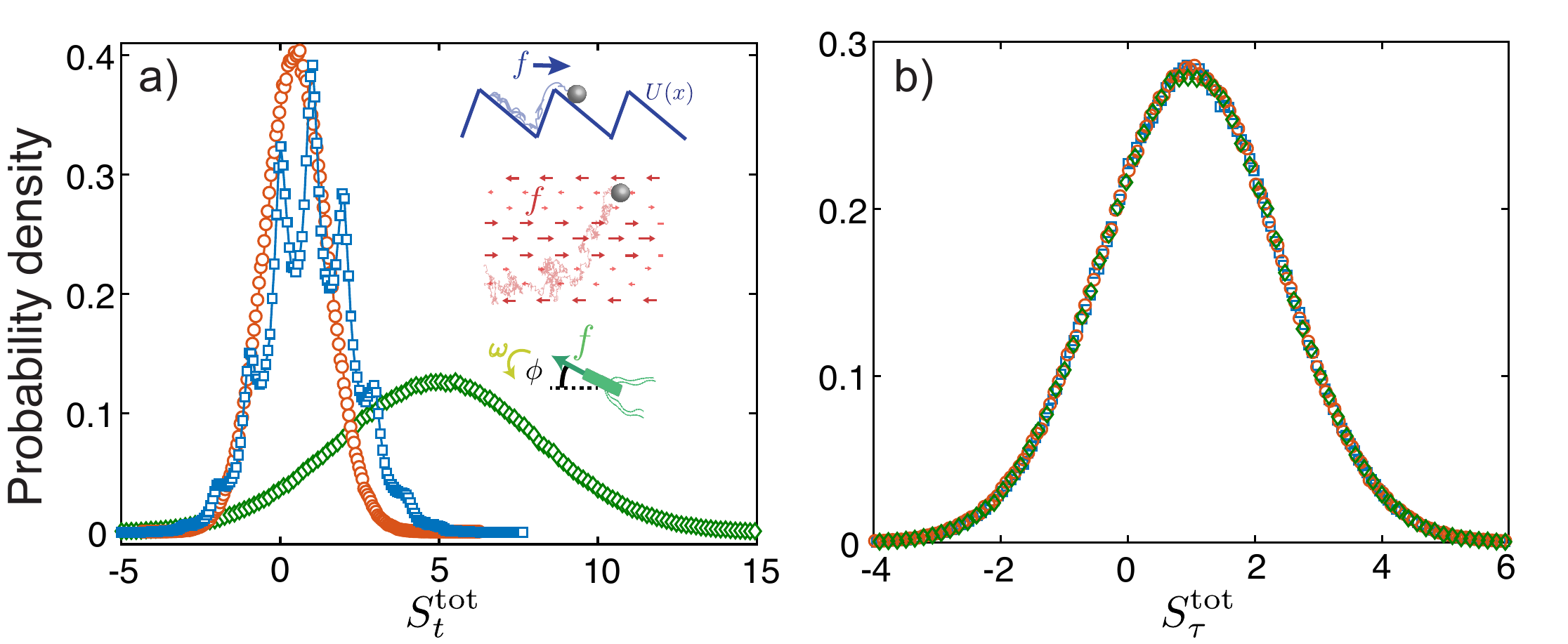}
    \caption{Universality of entropy production at entropic  times. Distributions of stochastic entropy production obtained from numerical simulations at fixed time $t=1$ (a) and at fixed entropic time $\tau=1$ (b).  The three different symbols are obtained from numerical simulations of the three models sketched in the caption in (a), see Fig.~\ref{fig:3l} for further details and  Ref.~\cite{Pigolotti:2017} for details and parameter values of the simulations. }
    \label{fig:ET2}
\end{figure}

The universality of entropy production revealed here extends to multidimensional overdamped Langevin systems, for which $\Stot$ also obeys a It\^{o} stochastic differential equation of the form~\eqref{eq:entropy} with a  entropic drift that is generalized to $d>1$ dimensions ---see Eqs.~\eqref{forGirs-1-1-1} and~\eqref{forGirs-1-1-1-8}. 
      We illustrate this universality principle in Fig.~\ref{fig:ET2} where we plot the distributions of stochastic entropy production for a driven colloidal particle ($d=1$-dimensional Langevin equation), a 2D diffusion in a space-dependent velocity field ($d=2$), and an active Brownian chiral swimmer ($d=3$). 
         
         \begin{leftbar}
               Furthermore, Eq.~\eqref{eq:stot_tau} reveals that any statistical property of $S^{\rm tot}_t$ that is independent of time contractions and dilations falls in the universality class of the standard one-dimensional drift diffusion process with unit  drift  and diffusion constant. For example, the global infimum of entropy production, defined as the minimum value that $S^{\rm tot}$ can take at any time, i.e.
         \begin{equation}
             S^{\rm inf}=\inf_{t\geq 0}S^{\rm tot}_t,
             \label{eq:minuit}
         \end{equation}
         is a  universal property for overdamped Langevin systems. This is because the value of $S^{\rm inf}$ associated with a given trajectory is independent of when it occurs, and thus on the value of $\tau$. As a result, its probability distribution can be found from that of the minimum of the 1D drift diffusion process,
         \begin{equation}
         \rho_{S^{\rm inf}} (s) = \exp (s), \quad {\rm with } \quad s\in \mathbb{R}^-,
         \label{eq:pinf1}
         \end{equation}
         %[{\bf IN: compare with Eq.~(\ref{eq:entropyNeg})}]
         i.e., it is an exponential distribution  with mean $\langle S^{\rm inf}\rangle = -1$.  One may also consider  the finite-time entropy-production infimum
         \begin{equation}
             S^{\rm inf}_t=\inf_{0\leq t'\leq t}S^{\rm tot}_t,
         \end{equation}
         that is, the minimum value that entropy production takes over a finite time interval $[0,t]$. The random variable $S^{\rm inf}_t\leq 0$ is always larger than its long-time limit $S^{\rm inf}_t\geq S^{\rm inf}$, which together with~\eqref{eq:pinf1} implies for Langevin systems the so-called infimum law
         \begin{eqnarray}
              \langle S^{\rm inf}_t\rangle \geq -1.
              \label{eq:inflaw}
         \end{eqnarray}%
        %In words, the average of the finite-time infimum of entropy production cannot be below minus the Boltzmann constant. 
         %We illustrate Eq.~\eqref{eq:inflaw} in Fig. [FIG] with numerical simulations  of a colloidal particle moving in a tilted, one-dimensional, potential.
         As shown below in Sec.~\ref{sec:infimaentropy}, the {\bf infimum law}~\eqref{eq:inflaw} extends for a broader class of  nonequilibrium stationary processes. %For overdamped Langevin models, the  universal limiting distribution for the global infimum of $S^{\rm tot}$ is illustrated in Fig. [FIG]. 
         \end{leftbar}
         
\begin{figure}[h]
    \centering
    \includegraphics[width=0.9\textwidth]{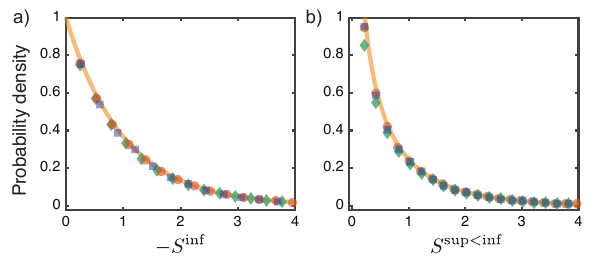}
    \caption{Universal properties of entropy production at entropic  times in time-homogeneous stationary states. Distributions of minus the infimum $-S^{\rm inf}$ (a) and supremum before the infimum  $S^{{\rm sup}<{\rm inf}}$ (b) associated with the stochastic entropy production $\Stot$ obtained for the model examples sketched in the caption Fig.~\ref{fig:ET2}a. The different symbols correspond to results from numerical simulations done for each model:  particle in a periodic potential  (blue squares),  particle in a 2D force field (red circles), active Brownian chiral swimmer (green diamonds). The orange lines are given by the analytical distributions   obtained from the drift-diffusion process with unit drift and diffusivity $\dot{X}_t = 1 + \sqrt{2}\dot{B}_t$:  $\rho_{S^{\rm inf}} (s)= \exp (s), \quad {\rm with } \quad s\in \mathbb{R}^-$ (a), and $\rho_{S^{{\rm sup}<{\rm inf}}} (s) = 2\exp (s) \text{acoth}(2\exp(s)-1)-1, \quad {\rm with } \quad s\in \mathbb{R}^+$ (b). 
    See Ref.~\cite{Pigolotti:2017} for further details. }
    \label{fig:ET3}
\end{figure}

         Other universal properties that can be identified from the entropy-production random-time change are the following [see Fig.~3 in~\cite{Pigolotti:2017}]:
         \begin{itemize}
             \item The maximum value that entropy production attains before reaching its global infimum;
             \item The number of crossings that entropy production crosses from $-s_0$ to $s_0$ with $s_0>0$ a positive real number;
             \item The number of "record breaking" events before reaching the global supremum/infimum.
         \end{itemize}
         Notably, one  can  identify an {\em infinite}  number of universal properties from the random-time stochastic differential equation for entropy production. 
         Moreover, the distribution of such universal quantities can be retrieved from the one-dimensional drift-diffusion process with both drift velocity  and diffusion coefficient equal to one, such as the  the distribution of the global infimum given by Eq.~(\ref{eq:pinf1}). See Fig.~\ref{fig:ET3} for two examples of such universal properties. %for such process is an exponential with mean $-1$, i.e., $\rho_{S^{\rm inf}}(s) = \exp(s)$.
         On the other hand, statistical properties that depend on the measurement of time, e.g. the first-passage time to reach a positive threshold, are not necessarily universal, and thus their distribution depends, in general, on the model details.

\section{Thermodynamics for isothermal Markov jump processes}\label{sec:43}

As a second example, we revisit the thermodynamics of isothermal Markov jump processes $X_t\in\mathcal{X}$, as defined in  Sec.~\ref{eq:jumpprocconta}, for which   $\mathcal{X}$ is a discrete phase space.  We assume that the transition rates satisfy the local detailed balance  condition given by Eq.~\eqref{eq:EDBx}, copied here for convenience \begin{equation}
        \frac{\omega_t(x,y)}{\omega_t(y,x)}=\exp\left(\frac{-(V_t(y)-V_t(x)) + f_{t}\:r(x,y)+ \sum^{m}_{
    a=1}\mu^{(a)}\:n_{a}(x,y)}{T}\right). 
        \label{eq:EDBx2}
\end{equation}
First, we derive the  first and second law of thermodynamics within this setup, see  Refs.~\cite{LNP, seifert2012stochastic, peliti2021stochastic} for more details, and then we revise martingale theory   for the thermodynamics of Markov jump processes.

\subsection{First law of stochastic thermodynamics} 
We define work at the level of a single trajectory, $X_{[0,t]}$, and subsequently use the first law of thermodynamics to obtain an expression for the heat.  

Recall that for Markov jump processes, trajectories are piecewise constant functions of the form Eq.~(\ref{eq:defsXMJP}).
%We assume that there is a time dependent potential $V_t(x)$  that represents a thermodynamic potential evaluated at $x\in \mathcal{X}$, and which we will refer to as the (free) energy of the system.  This potential is manipulated by an external agent, which clarifies its explicit dependence on time.    
The work done by an external agent on the system is  %{(RC : This must be written for \eqref{eq:EDBx2} )} 
\begin{equation}
    W_t=\int_{0}^{t}  ds\left( \partial_s V_{s}\right)(X_{s})  +  \sum_{j=1}^{N_t} 
   f_{\T_j}\: r(X_{\T^-_{j}},X_{\T^+_{j}})
    \label{Wsj}
\end{equation}
where the first term represents the energy change of the system due to a protocol that changes the shape of the potential  $V_t$, and the second term represents the work done on the system   by the nonconservative force $f_t$.      For example,  $f_t$ could be an external {mechanical} force and $r(x,y)$ the distance travelled  by the system in   the jump from $x$ to $y$.   

The first law of  thermodynamics  reads
 \begin{equation}\label{eq:1LST2}
     Q_t + W_t=V_{t}(X_{t})-V_{0}(X_{0}),
 \end{equation}
which holds at the level of individual trajectories $X_{[0,t]}$. 

Using  Eq.~\eqref{Wsj} in Eq.~(\ref{eq:1LST2}), we obtain  the heat
%Following Ref.~\cite{LNP}, we define heat as the amount of energy exchanged between the system and the microscopic degrees of freedom in the environment.   Mathematically, interactions between the system and the environment are modelled through the  jumps  in the Markov jump process.   Therefore the total heat absorbed by the system is  
%\sum^{N_t}_{j=1} \delta Q_{\T_{j}}(X_{\T_j}, X_{\T_{j+1}})
\begin{equation}
Q_t =   \sum_{j=1}^{N_t}    \left( V_{\T_j}(X_{\T^+_j})-V_{\T_j}(X_{\T^-_j})\right) - \sum_{j=1}^{N_t} 
   f_{\T_j}\: r(X_{\T^-_{j}},X_{\T^+_{j}}).  \label{eq:hj3}
\end{equation}   
Notice that the heat can be expressed in terms of the individual contributions 
\begin{equation}
    \Delta Q_{\T_{j}}(X_{\T_j}, X_{\T_{j+1}}) =  V_{\T_j}(X_{\T^+_j})-V_{\T_j}(X_{\T^-_j}) - 
   f_{\T_j}\: r(X_{\T^-_{j}},X_{\T^+_{j+1}})
\end{equation}
for each jump in the trajectory $X_{[0,t]}$.

\subsection{Second law of stochastic thermodynamics}
 The derivation is analogous to the one presented for Langevin processes in Sec.~\ref{eq:defSSysonetime}, and hence we will follow it closely here.     
 
 We define the system entropy as in Eq.~(\ref{eq:defSSysonetime}), viz.,
\begin{eqnarray}
S^{\rm sys}_t-S^{\rm sys}_0 &=& -\ln \left(\frac{\rho_{t}\left(X_t\right)}{\rho_{0}\left(X_0\right)}\right) \label{eq:Ms1}\\
    &=&  - \int_{0}^{t} ds \left(\partial_s \left( \ln \rho_{s}\right)\right)\left(X_s\right) -\sum_{i=1}^{N_t}  \ln \left( \frac{\rho_{\T_i}(X_{\T^+_i})}{\rho_{\T_i}(X_{\T^-_i})}\right) .
  \label{eq:Ms2}
  %  &=& \underbrace{- \frac{\left(\partial_t \rho_{t}\right)\left(X_t\right)}{\rho_{t}\left(X_t\right)}+ \frac{j_{t,\rho}(X_t)}{\mu  \rho_{t}\left(X_t\right)}\circ \dot{X}_t}_{\displaystyle \dot{S}_{t}^{\rm tot}} - \underbrace{\frac{F_t(X_t)}{T}\circ \dot{X}}_{\displaystyle \dot{S}_{t}^{\rm env}}, \label{eq:Ms3}
\end{eqnarray}  

Subsequently, we use the fact that the environment consists of a thermal reservoir at temperature $T$ plus  $n$  particle reservoirs with chemical potentials $\mu^{(a)}$, and hence the environment entropy change according to standard thermodynamics is  \cite{callen1998thermodynamics}
\begin{equation}
    S^{\rm env}_t =  \frac{-Q_t+ \sum^{N_j}_{j=1}\sum^n_{a=1}\mu^{(a)}n_{a}(X_{\T^-_{j}},X_{\T^+_{j}})}{T}. 
\end{equation}
Substituting  the heat, given by Eq.~(\ref{eq:hj3}), in the above equation, we obtain 
\begin{eqnarray}
    S^{\rm env}_t &=& \sum_{j=1}^{N_t} \frac{- \left( V_{\T_j}(X_{\T^+_j})-V_{\T_j}(X_{\T^-_j})\right)- 
   f_{\T_j}\: r(X_{\T^-_{j}},X_{\T^+_{j}})+ \sum^n_{a=1}\mu^{(a)}n_{a}(X_{\T^-_{j}},X_{\T^+_{j}})}{T}  
   \nonumber\\ 
   &=& \sum_{j=1}^{N_t} {\ln} \left( \frac{\omega_{\T_j}(X_{\T^-_{j}},X_{\T^+_{j}})}{\omega_{\T_j}(X_{\T^+_{j}},X_{\T^-_{j}})}\right), \label{eq:SEnvJump}
\end{eqnarray}
where the last line follows from the local detailed balance formula Eq.~(\ref{eq:EDBx2}).    Lastly,  
%local detailed balance formula Eq.~(\ref{eq:SEnvLoc}), {(RC: NO, we must use Clausius as in Langevin !, it will give the same result for the considered class of Isothermal process but the philosophy is better)} to express the environment entropy as 
%\begin{equation}
 %   \dot{S}^{\rm env}_t = \sum_{(x,y)\in\mathcal{G}_t} \ln  \left(\frac{\omega_{t}(y,x)}{\omega_{t}(x,y)}\right) \delta_{X(t),y}\delta_{X(t-),x}. 
%\end{equation} 
adding Eqs.~(\ref{eq:Ms2}) and (\ref{eq:SEnvJump}), and using the balance Eq.~(\ref{eq:balanceS}), we find 
\begin{equation}
    S^{\rm tot}_t =  -\int_{0}^{t}ds(\partial_{s}\ln\rho_{s})(X_{s}) - \sum^{N_t}_{j=1} \ln \left( \frac{\rho_{\T_j}(X_{\T^+_j})\omega_{\T_j}(X_{\T^+_j},X_{\T^-_j})}{\rho_{\T_j}(X_{\T^-_j})\omega_{\T_j}(X_{\T^-_j},X_{\T^+_j})} \right) . \label{eq:defStot} 
\end{equation}
Taking the ensemble average of the above equation, we obtain 
\begin{equation}
    \langle \dot{S}^{\rm tot}_t \rangle =   \sum_{(x,y)\in\mathcal{X}^2}\rho_t(x)\omega_{t}(x,y) \ln \left( \frac{\rho_t(x)\omega_{t}(x,y)}{\rho_t(y)\omega_{t}(y,x)}\right)\geq0, \label{eq:SDotResult}
\end{equation}
which is the second law of thermodynamics for Markov jump processes.
{To pass from Eq.~\eqref{eq:defStot}  to Eq.~\eqref{eq:SDotResult} we proceeded as follows. 
The  averge of the first term in Eq.~\eqref{eq:defStot} vanishes because of conservation of probabiliy
\begin{eqnarray}
\left\langle \int_{0}^{t}ds\left(\partial_{s}\ln\rho_{s}\right)\left(X_{s}\right)\right\rangle &=&\int_{0}^{t}ds\int_{\mathcal{X}}dx\rho_{s}(x)\left(\partial_{s}\ln\rho_{s}\right)\left(x\right)\nonumber\\
&=&\int_{0}^{t}ds\int_{\mathcal{X}}dx\left(\partial_{s}\rho_{s}\right)\left(x\right)\nonumber\\
&=&\int_{0}^{t}ds\partial_{s}\int_{\mathcal{X}}dx\rho_{s}\left(x\right)=0.\label{eq:proofzerossys}
\end{eqnarray}
On the other hand, using the definition of transition rates one gets that the average of the second term in the right-hand side of Eq.~\eqref{eq:defStot}  yields the right-hand side of Eq.~\eqref{eq:SDotResult}.}
Moreover, we derive the inequality in Eq.~(\ref{eq:SDotResult}) in Appendix~\ref{appendix:5NewSub}.   We have used the convention $\ln 0/0=0$.   
%{\bf [IN:  I cant prove the above inequality, but I can prove the equation below.    Why do we keep insisting on the nonstationary case?   ]}
  For stationary processes, the average rate of entropy production and the second law simplify into 
\begin{equation}
    \langle \dot{S}^{\rm tot}_t \rangle =   \sum_{(x,y)\in\mathcal{X}^2}\rho_{\rm st}(x)\omega(x,y) \ln \left( \frac{\rho_{\rm st}(x)\omega(x,y)}{\rho_{\rm st}(y)\omega(y,x)}  \right)\geq 0. \label{eq:SDotResultx}
\end{equation}

\section{Martingale theory for stationary  Markov jump processes} \label{sec:44}

We show that  $\exp(-S^{\rm tot}_{t})$ is a martingale within the context of stationary Markov jump processes.   %Consequently, the martingale fluctuation relations and second law of thermodynamics, as discussed in Secs.~\ref{sec:martFr} and \ref{sec:martSecond}, also apply for stationary Markov jump processes.   
However,  we show that   universal properties  that apply to Langevin processes do not apply to the Markov jump processes, as the latter  are not continuous.   In this section, we assume that the dynamics is time-homogeneous ($\omega_t(x,y) = \omega(x,y)$ for all $t$)  and   stationary ($\rho_0(x) = \rho_{\rm st}(x)$ and $\partial_t\rho_t(x)=0$).     For stationary Markov jump processes, \begin{equation}
    \sum_{x\in\mathcal{X}}\rho_{\rm st}(x)\omega(x,y) =   \sum_{y\in\mathcal{X}}\rho_{\rm st}(y)\omega(y,x). \label{eq:statcond}
\end{equation}

%\subsection{{Stationary process}}
%{(RC : I inverse the 2 next notions because same before :  it is strange, but a process can be stationary without being time homogeneous.)}

%\begin{leftbar}
%A  Markov jump process is {\bf stationary} when 
%\begin{equation}
%   \rho_0(x) = \rho_{\rm st}(x).
%\end{equation}
%\end{leftbar}
%{\color{red} ER: Repetition, why a box?? You are true, we can just write that it is as for diffusion. }
%{We remark that this definition is distinct in general from the notion of time-homogeneous jump process. }
%\begin{leftbar} We say that the Markov jump process is {\bf time homogeneous} when  
%\begin{equation}
%    \omega_t(x,y) = \omega(x,y).
%\end{equation}
%\end{leftbar}
%{In the following of this section, we suppose stationarity and time-homogeneity.}

\subsection{ The martingality of the exponentiated negative entropy production}

We show,  using three approaches, that the exponentiated negative entropy production is a martingale.

\subsubsection{$^{\spadesuit}$Dynkin's martingale approach}
%{I would send this  to a new section just after 7.1.5.1 call ''Martingale structure of stochastic entropy production for Jump process '' which is missing with respect to Langevin case (7.1.5.4) and in fact was present in a very preliliminary version that I supress myself (because it is proven in a general case in 7.2...but now that it is written here...). Same that before, we don't need 3 proofs...this break the philosophy that this chapter is readable for pedestrian. }

We show that $\exp(-s)$ is a harmonic function of the generator of the join process $\mathscr{X}(t)= (S_{\rm tot}(t), X(t))^{\rm T}$, and hence a martingale, and consequently according to the Theorem~\ref{Th:markov} it is a martingale.      Moreover, we show that $s$ is a subharmonic function of this generator, and thus a submartingale.  

From the Eqs.~\eqref{eq:defStot} and (\ref{eq:genps}), we find the following expression for the generator of the joint process that acts on functions $\phi(x,s)$ as 
\begin{eqnarray}
\mathcal{L}\left[\phi\right] (s,x) =\int_{\mathbb{R}}\,d\tilde{s}  \sum_{y\in\mathcal{X}}  \omega(x,y)\left(\phi(\tilde{s},y)-\phi(s,x)\right) \delta\left(\tilde{s}-s-\ln \left( \frac{\rho_{\rm st}(x)\omega(x,y)}{\rho_{\rm st}(y)\omega(y,x)}\right)\right), \nonumber\\
\end{eqnarray}
where $\delta$ is the Dirac delta distribution.   

The generator acting on $\exp(-s)$ gives , 
\begin{eqnarray}
\mathcal{L}\left[\exp(-s)\right]  &=& \exp(-s)\sum_{y\in\mathcal{X}}\, \omega(x,y)\left(\frac{\rho_{\rm st}(y)\omega(y,x)}{\rho_{\rm st}(x)\omega(x,y)}-1\right) \nonumber\\ 
&=& \frac{\exp(-s)}{\rho_{\rm st}(x)}\sum_{y\in\mathcal{X}}\, \left(\rho_{\rm st}(y)\omega(y,x)-\rho_{\rm st}(x)\omega(x,y)\right)  = 0, \label{eq:harmExpS}
\end{eqnarray} 
where in the last step we have used the   stationarity condition Eq.~(\ref{eq:statcond}).   Equation~(\ref{eq:harmExpS}) states that for stationary Markov jump processes   
 $\exp(-s)$ is a harmonic function of the generator $\mathcal{L}$, and hence $\exp(-S^{\rm tot}_t)$ is a martingale.    Also, 
\begin{eqnarray}
\mathcal{L}\left[s\right]  =   \sum_{y\in\mathcal{X}} \, \omega(x,y) \ln\left( \frac{\rho_{\rm st}(x)\omega(x,y)}{\rho_{\rm st}(y)\omega(y,x)}\right)  \geq 0,
\end{eqnarray}
where the last inequality follows from the stationarity condition Eq.~(\ref{eq:statcond})
and $s$ is thus a subharmonic function.  
%{\bf [IN: is the latter true?  Probably it is, but can you check]} {Note sure! For diffusion, it is true that $Lf=0\Rightarrow L\left(\ln f\right)\leq0$ }
Indeed, the positivity comes from writing
\begin{eqnarray}
\mathcal{L}\left[s\right]  = \frac{1}{\rho_{\rm st}(x)}   \sum_{y\in\mathcal{X}} \, \rho_{\rm st}(x)\omega(x,y) \ln\left( \frac{\rho_{\rm st}(x)\omega(x,y)}{\rho_{\rm st}(y)\omega(y,x)}\right)
\end{eqnarray}
and  the elementary convexity relation $a\ln (a/b)-a+b \geq 0$ for all $a\neq b \geq 0$.   Indeed, we can identify $a(x,y)=\rho_{\rm st}(x)\omega(x,y)$ and $b(x,y)=\rho_{\rm st}(y)\omega(y,x)$, and  the stationarity condition Eq.~(\ref{eq:statcond})  implies
$\sum_{y\in\mathcal{X}} \left(a(x,y)-b(x,y)\right)=0 $.

\subsubsection{Path-probability-ratio approach}\label{sec:radNikoMarkovJump}
We use the path-probability-ratio approach to show that $\exp(-S^{\rm tot}_t)$ is a martingale.    The rationale goes as follows: (i) we demonstrate that Eq.~(\ref{eq:meanttoshow})  also holds for Markov jump processes; (ii) we show that    $\mathcal{P}\circ \Theta_{t} \equiv \mathcal{Q}$ does not depend explicitly on  $t$; (iii) the martingality of  $\exp(-S^{\rm tot}_t)$ is concluded following the derivation Eq.~(\ref{eq:Der}) that holds for all ratios of the form Eq.~(\ref{eq:radon}).

First, we show that Eq.~(\ref{eq:meanttoshow})  also holds for Markov jump processes.  To this aim we use the Onsager-Machlup approach.     Assuming that $\mathcal{P}$ and $\mathcal{P}\circ\Theta$ are  mutually absolutely continuous, we can use the action $\mathcal{A}(X_{[0,t]})$ given by  Eq.~(\ref{eq:RFormCont}).   The corresponding action of the time-reversed process is %{(RC : wrong for time inhomogeneous process ! for treat this,  we must use the tilte as in our previous version. It a pity for me, but we must just write here for time homogeneous case)}
\begin{equation}
\mathcal{A}[\Theta_{t}(X_{[0,t]})] = -\ln \left(    \rho_{\rm st}(X_t)\right)  -\sum^{N_t}_{i=1} \ln \left( \omega(X_{\T^+_i},X_{\T^-_i})\right)+ \int^{t}_0 ds \lambda(X_s).\label{eq:RFormContTR}
\end{equation} 
Taking the ratio 
\begin{eqnarray}
\frac{(\mathcal{P}\circ\Theta_{t})[X_{[0,t]}]}{\mathcal{P}[X_{[0,t]}]}     &=& 
\exp\left(-\mathcal{A}[\Theta(X_{[0,t]})]+\mathcal{A}[X_{[0,t]}]\right)  \nonumber\\ 
&=& 
\exp\left(
\ln\left( \frac{\rho_{st}(X_t)}{\rho_{st}(X_0)}\right) 
+ \sum^{N_t}_{i=1} \ln 
\left( 
\frac{\omega(X_{\T^+_i},X_{\T^-_i})}{\omega(X_{\T^-_i},X_{\T^+_i})}
\right)
\right)=
\emStot. \nonumber\\ 
\label{IZ}
\end{eqnarray}

Second, in Appendix~\ref{app:indepMarkov} we show that  $\mathcal{P}\circ\Theta_t$  is not $t$ explicitly dependant.  

Finally, the martingality of $\exp(-S^{\rm tot}_t)$ is concluded from the derivation in  Eq.~(\ref{eq:Der}).

 In  Sec.~\ref{eq:martsigmafunct},  we  give an alternative proof that  $\left(\mathcal{P}\circ \Theta_{t}\right) (\Xot)/\mP(\Xot)$ is a  martingale in the stationary setup.

%We prove this in Sec.~\ref{eq:martsigmafunct}.  {\bf [IN: does Eq.~(\ref{IZ}) not prove that $\mathcal{P}\circ\Theta_t$ is independent of $t$?  We have no explicit time-dependence in the action!]}

\subsubsection{$^{\spadesuit}$It\^{o}'s integral approach}\label{sec:ItoIntegralAppraochSec}

%\textcolor{red}{(RC : I strongly doubt that this will be in the review on this form. If we want this, then we must define in chap 3 integration with respect to a Martingale with jump....I would just put the 2 next approaches, anyway all is proven in chap 7 for general process ! )}

 The exponentiated negative entropy production, $\exp\left(-S^{\rm tot}_{t}\right)$, is  a stochastic exponential $\mathcal{E}_t(M)$ of a martingale $M$, just as was the case for Langevin processes, see  Sec.~\ref{sec:ItoApproach}.   Indeed, as we show in  Appendix~\ref{sec:itoMarkovJumpApp} that  the stochastic exponential solves the Eq.~(\ref{eq:stochExp}), i.e., 
\begin{equation}
   \frac{d \exp(-S^{\rm tot}_t)}{dt} =  \exp(-S^{\rm tot}_{t-}) \dot{M}_t   \label{eq:stochExponJump}
\end{equation}
 with $M_t$  the martingale 
\begin{equation}
M_t =    \sum_{x,y\in \mathcal{X}^2} \left(\frac{\rho_{\rm st}(y) \omega(y,x)}{\rho_{\rm st}(x) \omega(x,y)}-1\right) (N_{t}(x,y) -  \tau_t(x) \omega(x,y)) , \label{eq:MtMarkovJumpStoch}
\end{equation}
and 
where we have used  $N_{t}(x,y)$
for the total number of times $X$ has jumped from $x$ to $y$  in the interval $[0,t]$, and  $\tau_t(x)$ for the total amount of time the process $X$ has spent  in the state $x$ in the interval $[0,t]$, see Eqs.~(\ref{eq:NtDef}) and (\ref{eq:tauDef}) for definitions.      Note that $M_t$ is a martingale as it is the sum of martingales of the form Eq.~(\ref{eq:countingM}), which can be derived with Dynkin's martingales, see   Eq.~(\ref{eq:rateDepDynkin1}).

\subsubsection{$^{\spadesuit}$Novikov's condition for Markov jump processes}\label{sec:novikovMJ}
  Just as was the case for Sec.~\ref{sec:martingaleExpS}, the  three approaches presented above demonstrate that $\emStot$ is a local martingale, and to confirm martingality we need to consider Novikov's condition.    Using   Novikov's condition for the stochastic exponential of a jump process,  we derive in  Appendix~\ref{app:novikov} the condition  
 \begin{eqnarray}
 \Bigg\langle  \exp\left(\sum_{x,y\in\mathcal{X}} \left(\frac{\rho_{\rm st}(y) \omega(y,x)}{\rho_{\rm st}(x) \omega(x,y)}-1\right)^2 \omega(x,y)\tau_t(x)\right)\Bigg\rangle<\infty, \quad \forall t\geq 0 .   \label{eq:novikovMJ}
 \end{eqnarray}

\begin{leftbar}
\subsubsection*{Three shades of  martingality}
We conclude that there  are three (equivalent) ways of representing the martingality of $\emStot$  in time-homogeneous nonequilibrium stationary processes:
\begin{itemize}
    \item 
The exponentiated negative entropy production is the stochastic exponential of a martingale $M_t$, 
\begin{equation}
\frac{d\exp\left(-S^{\rm tot}_t\right)}{dt} = \exp\left(-S^{\rm tot}_{t^-}\right) \dot{M}_t.  
\end{equation}
    \item  The  function $\exp(-s)$ is a harmonic function of the  generator $\mathcal{L}$ of the joint process $(S^{\rm tot}_t,X_{t})$, 
    \begin{equation}
        \mathcal{L}[\exp(-s)] = 0.
    \end{equation}
    \item The exponentiated negative entropy production is a path probability ratio 
    \begin{equation}
        \exp\left(-S^{\rm tot}_t\right)= \frac{(\mP\circ\Theta_t)[X_{[0,t]}]}{\mP(X_{[0,t]})},
    \end{equation}
    where $\mP \circ \Theta_t$ has no explicit dependency on time $t$. 
\end{itemize}
\end{leftbar}

\subsection{Non-universal properties for the fluctuations of the  stochastic entropy production} \label{sec:absence}

Unlike for Langevin processes where we showed in Sec.~\ref{sec:randomtime} that a random-time change renders the fluctuations of $\Stot$ universal, such property is not inherited by Markov-jump processes, even when their continuum limit is a Langevin process. 
However, as we show in  Chapter~\ref{sec:martST2}, for processes with jumps there exist  universal  bounds on the fluctuation properties of entropy production, i.e., bounds that are  valid for all time-homogeneous stationary processes.  Here, we anticipate  and illustrate some of these results    on a  paradigmatic model of a discrete process, namely, a biased random walk, and in the  Chapter~\ref{sec:martST2} we review results in a generic setup.  

Let us consider the paradigmatic example of a biased random walk given by a continuous-time Markov jump process in one dimension, with periodic boundary conditions. We also assume a homogeneous bias, i.e., transitions from site $x$ to $x+1$ occurring at a space-independent rate  $\omega(x,x+1)=\omega_+$, and transitions in the opposite direction to  
at a  space-independent rate $\omega(x,x+1)=\omega_-$. Following Sec.~\ref{motors}, we introduce an "affinity" bias parameter $A$  through the local detailed balance condition
\begin{equation}
\frac{\omega_+}{\omega_-}=\exp (A),
\label{eq:affinitymotors2}
\end{equation} 
and a kinetic rate $\nu=\sqrt{\omega_+\omega_-}$, such that $\omega_\pm = \nu \exp(\pm A/2)$, see also e.g., Ref.~\cite{baiesi2018life}. For  the case of molecular motors, $A$ can be related to the hydrolysis free energy of ATP hydrolyzation, the work done by an external force, and the temperature of the environment, see Eq.~\eqref{eq:affmotor} in    Sec.~\ref{motors}. The homogeneous bias together with the periodic boundary conditions induces a homogeneous stationary density, which implies that $\Delta S^{\rm sys}_t = -\ln[ \rho_{\rm st}(X_t)/\rho_{\rm st}(X_0)]=0$, and thus
\begin{equation}
    \Stot = S^{\rm env}_t = \ln\left( \left( \frac{\omega_+}{\omega_-}\right)^{X_t - X_0}\right).
    \label{eq:stotmotorper}
\end{equation}
In Eq.~\eqref{eq:stotmotorper} we have used Eq.~\eqref{eq:SEnvJump} for the environmental entropy change of a Markov-jump process and the fact that $X_t-X_0$ equals to the net number of jumps in the positive direction up to time $t$.  Using Eq.~\eqref{eq:affinitymotors2} and~\eqref{eq:stotmotorper} yields the martingale
\begin{equation}
    \emStot = \exp[-A (X_t-X_0)].
\end{equation}

The martingality of $\exp(-S^{\rm tot}_t)$ implies integral fluctuation relations at stopping times. 
Let us consider the stopping time
\begin{equation}
    \T = \inf \{ t\geq 0 \;\;\vert \;\; (X_t-X_0)\notin (-x_-,x_+)  \},
    \label{eq:stoptimemot}
\end{equation}
i.e., the first escape time of the position (relative to its initial value) from the interval $(-x_-,x_+)$ with $x_-$ and $x_+$ two finite positive integers. For the stopping time~\eqref{eq:stoptimemot}, we have that $ \langle |\exp(-S^{\rm tot}_\T)|\rangle<\exp[-A\min(x_-,x_+)]<\infty$ and $\mP(T<\infty)<1$  ($\T$ is finite), and we can thus readily apply Doob's optional stopping Theorem  ~\ref{eq:doobStop1}
\begin{equation}
\langle \exp(-S^{\rm tot}_{\T})\rangle=\langle \exp(-S^{\rm tot}_{0})\rangle=1,
\label{eq:Doobnis}
\end{equation}
which follows from $S^{\rm tot}_{0}=0$.
Furthermore, we can unfold the average at the stopping time~\eqref{eq:stoptimemot}  as
\begin{equation}
\langle \exp(-S^{\rm tot}_{\T})\rangle=P_+(x_+,x_-)\exp(-Ax_+)+P_-(x_+,x_-)\exp(Ax_-),
\label{eq:Doobnis2}
\end{equation}
where   $P_+(x_+,x_-)$ and $P_-(x_+,x_-)$ denote the absorption probabilities at $x_+$ and $-x_-$, respectively. 
Using  $P_+(x_+,x_-)+P_-(x_+,x_-)=1$ together with Eqs.~(\ref{eq:Doobnis}-\ref{eq:Doobnis2}), we find
\begin{equation}
P_-(x_+,x_-)=\frac{1-\exp(-Ax_+)}{\exp(Ax_-)-\exp(-Ax_+)}  .
\end{equation}
For $x_+\rightarrow \infty$ we have the absorption probability at position $-x<0$
\begin{equation}
P_-(x)=\exp(-Ax)  ,
\label{eq:absprobuseful}
\end{equation} 
which gives the probability that the relative position with respect to the initial value $X_t-X_0$ ever reaches the value $-x$. When $\omega_+>\omega_-$, $A>0$ and thus $P_-(x)<1$, i.e., if the drift is positive, the probability to ever reach a negative threshold is smaller or equal than one. 
Similarly, for $x_-\rightarrow \infty$ and absorption at position $x>0$, we have $P_+(x)=1$ for $A>0$, i.e. the walker reaches with probability one a single absorbing positive boundary when the drift is positive.

\begin{figure}[h]
    \centering
    \includegraphics[width=\textwidth]{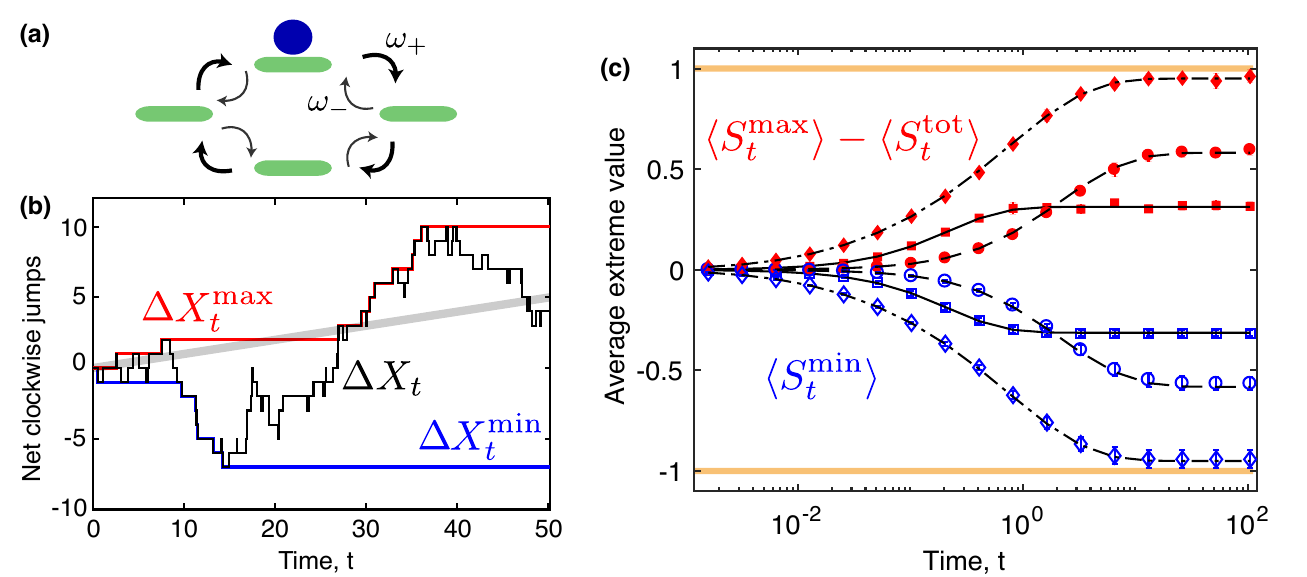}
    \caption{(a) Illustration of the minimal stochastic model of molecular motor motion, given by a continuous-time biased random walk in a discrete lattice with periodic boundary conditions. The transition rates are given by $\omega_+=\omega(x,x+1)= \nu \exp(A/2)$ and $\omega_-=\omega(x,x-1)=\nu\exp(-A/2)$,  for forward (clockwise) and backward (counterclockwise) stepping respectively. (b) Net number of clockwise jumps  as a function of time in an example trajectory of the model ($X_t - X_0$, black line), together with the average value  over many realizations $\langle \Delta  X_t\rangle$ (thick black line). The finite-time maximum $\Delta X^{\rm max}_t=\max_{s\in[0,t]} (X_s-X_0)$ and minimum of the trajectory $\Delta X^{\rm min}_t=\min_{s\in[0,t]} (X_s-X_0)$ are displayed with red and blue lines, respectively. (c)  Averages of the finite-time minimum $\langle S^{\rm min}_t\rangle$ (blue symbols) and $\langle S^{\rm max}_t\rangle-\langle S^{\rm tot}_t\rangle$ with $S^{\rm max}_t$ the finite time maximum of entropy production  (red symbols)  as a function of time $t$. Different symbols are obtained for different degrees of nonequilibrium: $A=1,\nu=0.5$ (squares), $A=2,\nu=2$ (circles), and $A=1,\nu=100$ (diamonds). The horizontal lines are set to $\pm 1$.
    In (c), symbols are obtained from numerical simulations and the lines are obtained from analytical calculations, see~\cite{guillet2020extreme} for further details. }
    \label{fig:motors}
\end{figure}

In what follows, we  assume $\omega_+>\omega_-$ i.e. $A>0$ (positive average velocity) without loss of generality. The analytical expression~\eqref{eq:absprobuseful} for the absorption probability in a negative boundary  can be used to obtain the statistics of extremal values
of position
\begin{eqnarray}
\Delta X^{\rm min}_t&\equiv &\min_{s\in[0,t]}(X_s-X_0), \\ 
\Delta X^{\rm max}_t&\equiv&\max_{s\in[0,t]}(X_s-X_0)  ,
 \end{eqnarray}
 as well as, of entropy production 
 $S^{\rm min}_t= A\Delta X^{\rm min}_t$ and $S^{\rm max}_t=  A \Delta X^{\rm max}_t$. We first consider
 the long time limit $\Delta X^{\rm min}\equiv\lim_{t\rightarrow\infty}\Delta X^{\rm min}_t$. The probability that the
 global minimum is at $-x<0$ is 
 \begin{equation}
 \rho_{\Delta X^{\rm min}}(-x)=P_-(x)-P_-(x+1),
 \end{equation}
 and therefore
 \begin{equation}
 \rho_{\Delta X^{\rm min}}(-x)=\rho_{S^{\rm min}}(-Ax)=\exp(-Ax)(1-\exp(-A)) .
 \end{equation}
 The averages of minima of position and entropy production are then given by
 \begin{eqnarray}
\langle S_{\rm min}\rangle &=& \frac{- A}{\exp A-1} \label{eq:infj},
 \end{eqnarray}
and
 \begin{eqnarray}
 \langle X_{\rm min}\rangle &=& \frac{-1}{\exp A-1} \label{eq:infjX}
 .
 \end{eqnarray}
The global minimum of entropy production Eq. (\ref{eq:infj}) therefore satisfies the infimum law $\langle S_{\rm min}\rangle\geq-1$.
Note however that in the case of continuous processes the infimum law at infinite time imposes precisely
$\langle S_{\rm min}\rangle=-1$, which is, as we have derived here, not obeyed for the biased random walk. Instead for the model discussed here, the average global infimum of entropy production is {\em not universal} as it depends on the model parameter $A$, see Eq.~\eqref{eq:infj}. 

The limit of a continuous process is reached when taking the diffusion limit
where the Peclet number ${\rm Pe}=v/D=2 \tanh(A)$ is small, ${\rm Pe}\ll 1$, where
$v=(\omega_+-\omega_-)=2\nu\sinh(A/2)$
and $D=(\omega_+ +\omega_-)/2$ is the
effective diffusion coefficient. This diffusion limit therefore 
corresponds to the regime of small $A$. In this limit Eq. (\ref{eq:infj}) approaches indeed the infimum law of entropy production for continuous stochastic processes $\langle S_{\rm min}\rangle=-1$. Interestingly, in this
limit the velocity $v$ is small and the motor close to stall. However the fluctuations become large for
small $A$ which is reflected in a divergence of the
average minimum $ \langle X_{\rm min}\rangle$ according to Eq.~(\ref{eq:infjX}). Numerical and analytical illustrations of the non-universal feature of the global infimum of entropy production are provided in Fig.~\ref{fig:motors} (see Ref.~\cite{guillet2020extreme}).

 Lastly, we would like to point to an interesting symmetry between the extrema of 
 entropy production $S_t^{\rm min}$ and $S_t^{\rm max}$ during the time interval $[0,t]$ during which the
 entropy production changes from $S^{\rm tot}_0=0$ to $S^{\rm tot}_t$. Indeed, the 
 reduction of entropy $S^{\rm tot}_0-S_t^{\rm min}\geq 0$ between start and minimum 
 obeys the same statistics as the reduction of entropy $S^{\rm max}_t-S^{\rm tot}_t\geq 0$.
 This follows from considering the time reversed process with trajectories $\tilde X_{[0,t]}=\{X_{t-u}\}_{u=0}^t$
 with path distribution ${\cal Q}(\Theta_t X_{[0,t]})={\cal P}(X_{[0,t]})$. This statistics of $\tilde X$
 is generated by the same hopping process but with rates $\omega_+$ and $\omega_-$ exchanged 
 or equivalently with $A\rightarrow -A$.
 The entropy production of the time reversed process therefore is
 $\tilde S^{\rm tot}_u =  A(X_t-\tilde X_u)$, where $\tilde X_u = X_{t-u}$.
 Note that  extrema of $X_t$ and its time reverse are the same,
$\tilde X^{\rm max}_t=X^{\rm max}_t$ and $\tilde X^{\rm min}_t=X^{\rm min}_t$. 
 The extrema of entropy production of the time reversed process are therefore
$\tilde S^{\rm min}_t=S^{\rm tot}_t-S^{\rm max}_t$ and $\tilde S^{\rm max}_t=S^{\rm tot}_t-S^{\rm min}_t$. 
Because ${\cal Q}(\tilde X_{[0,t]})={\cal P}(X_{[0,t]})$, the statistics of $\{\tilde X_u-X_t\}_{u=0}^t$ and 
$\{X_u-X_0\}_{u=0}^t$ are the same. Therefore the statistics of $\tilde S^{\rm tot}_s$ and $S^{\rm tot}_s$ are also the same
as well as those of minima $\tilde S^{\rm min}_t$, $S^{\rm min}_t$ and those of maxima 
$\tilde S^{\rm max}_t$, $S^{\rm max}_t$.
 As a consequence the distributions of minima and maxima of entropy production obey the symmetry relation
 \begin{equation}
 \rho_{S_0^{\rm tot}-S^{\rm min}_t}(s)=\rho_{S^{\rm max}_t-S^{\rm tot}_t}(s) \label{eq:symmaxmin} ,
 \end{equation}
 i.e. the reduction of entropy from time $t=0$ until the minimum values $S^{\rm min}_t$ 
 has the same statistics as the reduction of entropy
 from the maximum value $S^{\rm max}_t$ until it reaches $S^{\rm tot}_t$ at time $t$.
 A special case of the general statement (\ref{eq:symmaxmin}) is that the averages are the same,
 \begin{equation}
 \langle S^{\rm max}_t -\langle S^{\rm tot}_t\rangle\rangle = - \langle S^{\rm min}_t\rangle  .
 \end{equation}
 The definitions of $\tilde S^{\rm tot}_t$ and $S^{\rm tot}_t$ further imply that $\exp(-\tilde S^{\rm tot}_t)$
 and $\exp(S^{\rm tot}_t)$ are both martingales with respect to the distribution 
 ${\cal Q}$, while  $\exp(\tilde S^{\rm tot}_t)$
 and $\exp(-S^{\rm tot}_t)$ are both martingales with respect to ${\cal P}$.

%\chapter[Martingales in stochastic thermodynamics II]
\chapter{Martingales in stochastic thermodynamics II: Formal foundations}

\label{ch:thermo2}

\epigraph{\textit{As far as I see, all a priori statement in Physics have their origin in symmetry}. Herman Weyl 1952.}

After the works of the founding fathers of thermodynamics, among others, Clausius, Maxwell, and Boltzmann, the concept of entropy has become the cornerstone of the second law of thermodynamics.  Entropy is a source of continuous discussion with a common theme:
there does not exist a unique fully satisfactory notion of entropy and    the different definitions of entropy introduced in the literature are interesting for different  applications/perspectives~\cite{grad1961many,penrose2005foundations,wehrl1978general,mackey1989dynamic,wehrl1991information,balian2004entropy,maes2009selection}.  In the present chapter, we review different notions of entropy as they have been used   in stochastic thermodynamics, and discuss their relation with martingale theory.  

 The present chapter builds further on Ch.~\ref{sec:martST1}, where  we have developed martingale theory for stationary processes in two simple examples, namely, the one-dimensional overdamped Langevin process and Markov jump  processes.   The aim of the present chapter is  to extend martingale theory in thermodynamics for general processes that may be nonstationary.   To this aim, we use path probability ratios, which provide a versatile tool to construct martingales in stochastic thermodynamics, and which will correspond to different notions of entropy.

This chapter is organized into three sections.  In the first Sec.~\ref{sec:6p1}, we  introduce the entropic functionals, which are a generic classes of functionals defined through  path probability ratios. Furthermore, we provide examples of  entropic functionals that play a central role in stochastic thermodynamics, such as, the entropy production, work, heat, and we illustrate these on specific models, such as,  Langevin processes and jump processes.     In   the second Sec.~\ref{sec:6p2},  we derive rigorously the martingale structure for the functionals introduced in Sec.~\ref{sec:6p1}. In the last Sec.~\ref{sec:623}, we introduce the generalised entropic functionals and discuss their relevance for stochastic  thermodynamics and martingale theory.

To develop formal foundations in this  chapter, unless specified otherwise, the physical process $X_t$ with associated path probability $\mP$ is a {\em generic stochastic process}, which can be both in discrete or continuous time, and is not necessarily stationary and/or Markovian.

\section{Stochastic entropic functionals and fluctuation relations} 
\label{sec:6p1}

\subsection{Notation and preliminaries}
\label{sec:st}

We review the notation that we use for the  path probability of a trajectory $x_{[0,t]}$  of a process $X_t$
%in a time window $[0,t]$
for  discrete time and space, even though  we apply it throughout this section  in continuous time and space.    We denote the path probability to observe a  trajectory  $x_{[0,t]}$ in the observation time window $[0,t]$ by 
\begin{equation}
\mP (x_{[0,t]})\equiv \mP(X_0=x_0, X_{1}=x_{1},\dots, X_{t-1}=x_{t-1},  X_{t}=x_t),
\label{eq:P_definition}
\end{equation}
%{\bf [IN: I find this confusing.   Why is there a dependence on $x_{[0,t]}$ and not $x_{[r,s]}$?   I can write $\mP_{[r,s]} (x_{[r,s]})$ and it is the same, why do we then need $x_{[0,t]}$? ]} {(The answer is on the discussion after \eqref{GEF}, this notation is usueful just starting there. In a sens I agree with you, but I don't find a better solution.) }
%{[\bf IN: if it is $x_{[0,t]}$, shouldn't it then read $X_r = x_0$?]}
where  $ t\in\mathbb{N}$ are natural numbers and $x_s\in \mathcal{X}$ for all values of $s\in [0,t]$.  An analogous definition  can be formulated for   continuous time and/or space, which is nota bene the typical  setup for physics. 

%We use $[r,s]=[0,t]$, unless specified otherwise. Whenever $[r,s]=[0,t]$, 
%we  often use the shorthand notation $\mP(x_{[0,t]}) =  \mP_{[0,t]} (x_{[0,t]})$.  
%Otherwise when $[r,s]\subset[0,t]$ we  use $ \mP_{[r,s]}$. 
Since~$\mP$ denotes the path probability of the physical process $X$ of interest, we  often use the simplified notation 
\begin{equation}
 \langle \;\cdot\; \rangle  \equiv  \langle  \;  \cdot\;  \rangle_{\mP}   
\end{equation}
for averages over the physical path probability $\mathcal{P}$.
%Let  $X_{[0,t]}=(X_0,\dots,X_t)$  be the stochastic trajectory of a physical process. 
In this chapter, an important quantity  is    the path probability $\mP$ evaluated on the stochastic process $X_{[0,t]}$, which we denote by  $\mP(X_{[0,t]})$. We emphasize that $\mP(X_{[0,t]})$ is itself  a stochastic process.   

We often consider a second stochastic process~$\mQ(X_{[0,t]})$,  which is the  path probability $\mQ$ evaluated on the same stochastic trajectory $X_{[0,t]}$.   The path    probability $\mQ$ may correspond to  another physical process,  called  the \textbf{auxiliary process}.
%but evaluated on the same stochastic trajectory $X_{[0,t]}$.
Note that in general $\mP(X_{[0,t]})\neq \mQ(X_{[0,t]})$. 
%The object $\mP(X_{[0,t]})$ means that path probabilities of the type~\eqref{eq:P_definition} are evaluated on a given physical trajectory $X_{[0,t]}$.
Throughout this chapter, we  assume that the path probabilities~$\mP$ and $\mQ$ are  mutually {\em absolutely continuous}, which in the discrete case means that for all trajectories $x_{[0,t]}$ for which $\mQ(x_{[0,t]})=0$,  also  $\mP(x_{[0,t]})=0$, and vice versa. We also assume the {\em microreversibility}, i.e.,  $\mP$ and $\mQ$ are mutually  absolutely continuous when $\mQ$ is evaluated on a time-reversed trajectory.

 The time reversed trajectory denoted by   $\Theta_{t}\!\left(x_{\left[0,t\right]}\right)$ 
    is   the time-reversed path of $x_{[0,t]}$  whose value at time $s\leq t$ is given by
    \begin{equation}
    \left[\Theta_{t}\left(x_{\left[0,t\right]}\right)\right]_{s}\equiv
      x_{t-s}.
    \end{equation}
   In general, the time reversed trajectory could also include spatial involution of $\mathcal{X}$, i.e.   
    \begin{equation}
    \left[\Theta_{t}\left(x_{\left[0,t\right]}\right)\right]_{s}\equiv
      x^{\star}_{t-s},
      \label{RG}
    \end{equation}
    where the involution $x^\star$ has the property  $\left(x^{\star}\right)^{\star}=x$. In particular, for general Kramers-Einstein-Smoluchowski equation  \eqref{eq:lang1} $X_t$ may contain both position and momenta variables, and the time reversal operation involves a change of sign of all the momentum degrees of freedom.  However, for simplicity, we do not consider momentum-like degrees of freedom   in this chapter,  and we refer the reader to  Refs.~\cite{chetrite2008fluctuation} and~\cite{chetrite2018peregrinations} for further analyses.
%As we will show later, time-reversed trajectories appear in functionals  that relate to entropy production. 

We also consider families of path probabilities denoted by
\begin{equation}
\mP^{(u)} (x_{[0,t]}) \equiv \mP^{(u)}\big(X_0=x_0, X_{1}=x_{1},\dots, X_{t-1}=x_{t-1},  X_{t}=x_t\big),
\label{eq:P_definition}
\end{equation}
where $u,t\geq 0$, and analogously for $\mQ^{(u)}$.  

\begin{figure}[h]
\centering
\includegraphics[width=0.9\textwidth]{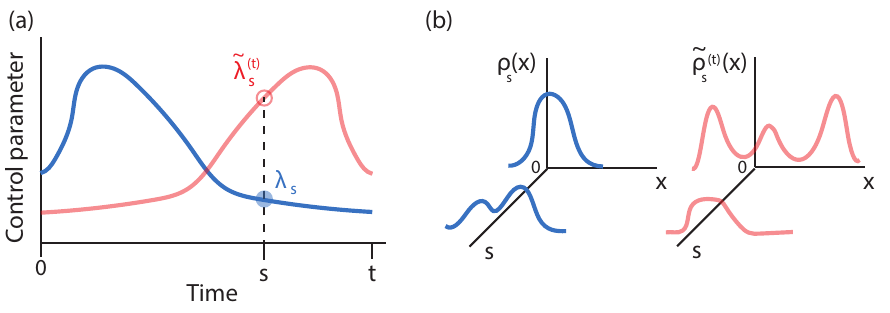}
\caption{ Panel (a):   Illustration of a physical process through the evolution of a control parameter $\lambda$ as a function of time (blue line) and of the backward, time-reversed protocol associated with the forward physical process (red line). We highlight the value of the forward (blue filled circle) and backward (red open circle) protocol at a time $s$ smaller than the time $t$ with respect which the time reversal is applied. Note that for $s\leq t$, the reversed protocol is defined as $\tilde{\lambda}_s=\lambda_{t-s}$, i.e.  in general $\tilde{\lambda}_s\neq \lambda_{s}$. Panel (b) Left: Illustration of the time evolution of an initial probability density in the forward process. Right: Illustration of the evolution of an arbitrary probability density in the backward process. }
\label{fig:timereversed}
\end{figure}

Lastly, let us discuss an important choice of $\mQ^{(t)}$  that appears in the Markovian context.  
In the Markovian context, when $\mP$ is the path probability of a process $X$ with Markovian generator $\mathcal{L}_t$, the most important choice for $\mQ^{(t)}$ corresponds with the {\bf time-reversed protocol}, which we denote by $\tilde{\mP}^{(t)}$ (see Fig.~\ref{fig:timereversed} for an illustration). 
In this case, for each  fixed $t$ the  auxiliary process is the  Markov process with time-reversed Markovian generator
\begin{equation}
\tilde{\mathcal{L}}^{(t)}_s \equiv \mathcal{L}_{t-s},
\label{Ltilde}
\end{equation}
for all $0\leq s \leq t$, and with  a given  {\em arbitrary}  initial density 
\begin{equation}
  \tilde{\rho}^{(t)}_0\equiv  \rho_0^{\tilde{\mP}^{(t)}}.  
\end{equation}
Analogously, we denote the  instantaneous density of $X$  associated with $\tilde{\mP}^{(t)}$  by 
\begin{equation}
\tilde{\rho}^{(t)}_s\equiv  \rho_s^{\tilde{\mP}^{(t)}}, 
\label{eq:tiltpho}
\end{equation}
for all $0\leq s \leq t$. {Note that this is not the time-reversal process that appears often in probabilistic literature~\cite{nagasawa1964time,chung2006markov,chetrite2018peregrinations} in which, differently to as in Fig.~\ref{fig:timereversed}, the instantaneous density is the time reversal of the original. }
%of $\tilde{\mP}$  is {\em arbitrary}
%, and we denote the path probability of this Markov process also by  $\mQ^{(t)}=\tilde{\mP}^{(t)}$.

\textbf{Meet the entropic functionals.} As shown in Ch.~\ref{sec:martST1}, key quantities in stochastic thermodynamics are expressed as {\em functionals} that take a specific value when evaluated over  stochastic trajectories $\Xot$. These {\em entropic} functionals (e.g. stochastic entropy production, stochastic environmental entropy change, etc.)  take the form of path-probability ratios.  In this chapter, we present some of the most relevant entropic functionals in stochastic thermodynamics, and discuss their martingale properties from both a mathematical and physical viewpoint. To guide the reader in this journey through the almanac of probability ratios, we provide here a quick summary of the entropic functionals that we  define later in this chapter: 
\begin{itemize}
    \item The $\Lambda$-stochastic entropic functionals involve the statistics of the physical process $\mP(\Xot)$ and  that of an arbitrary auxiliary process $\mQ^{(t)}(\Xot)$, both  evaluated  over the trajectories of $X$. A physical example of a $\Lambda$-stochastic entropic functional is the housekeeping entropy production $S^{\rm hk}_t$.
    
    \item The $\Sigma$-stochastic entropic functionals involve the statistics of the physical process $\mP(\Xot)$ evaluated over $\Xot$,  and that of an arbitrary auxiliary process $\mQ^{(t)}(\Theta_t(\Xot))$ evaluated over the time-reversal $\Theta_t(\Xot)$ of $\Xot$.  Two physical examples of  $\Sigma$-stochastic entropic functionals are the $\Sigma^{\rm tot}$-stochastic entropic functional, which is a stochastic process proportional to the fluctuating work dissipated in an isothermal system, and the $\mQ$-stochastic entropy production, which we discuss in the next bullet point. 
    %{\bf [IN: last sentence not clear?]}
    
    \item The $\mQ$-stochastic entropy production is a   $\Sigma$-stochastic  entropic functional for which  the initial distribution of auxiliary process $\mQ^{(t)}$ equals the instantaneous density $\rho_t$ of the  process  $\Xot$.
   Physical examples of the $\mQ$-stochastic entropy production are the total stochastic entropy production  $S^{\rm tot}_t$ and the excess stochastic entropy production $S^{\rm ex}_t$.
    %{\bf [IN: is total stoch entropy different from stoch entropy discussed in last bullet point?  If yes, this is not clear.]}
    
    \item The generalized $\Sigma$-stochastic entropic functionals have an analogous structure to the  $\Sigma$-stochastic entropic functionals, except that they involve probability ratios over arbitrary intervals   $[r,s]\subseteq [0,t]$. As we show in chap.~\ref{TreeSL}, the generalized $\Sigma$-stochastic entropic functionals yield a plethora of different formulations of the second law, some of which are well-known, and others that we derive in this {Treatise} for the first time.
\end{itemize}

{

\subsection{Definitions of $\Sigma$- and $\Lambda$-stochastic entropic functionals}
 Key quantities in stochastic thermodynamic quantities, such as, work, heat, entropy, and  energy, are %introduced in Ch.~\ref{sec:martST1}
%are  % 
formally functionals of stochastic trajectories. 
Here we introduce the  $\Sigma$-stochastic   and $\Lambda$-stochastic entropic functionals  as two classes of functionals that involve two (different) path probabilities, generalising  the formulae \eqref{dimanche}, \eqref{eq:Nirvana} and \eqref{IZ} of the previous chapter.  

\begin{leftbar}
We define the  \textbf{$\Sigma$-stochastic entropic functionals} and  the \textbf{$\Lambda$-stochastic entropic functionals}, both  associated with a generic stochastic process $X_t$, by 
\begin{eqnarray}
\Sigma_t^{\mathcal{P},\mathcal{Q}}\equiv&  \Sigma_t^{\mathcal{P},\mathcal{Q}} \left(X_{\left[0,t\right]}\right) \equiv& \ln\left[\frac{\mathcal{P}\left(X_{\left[0,t\right]}\right)}{\mathcal{Q}^{(t)}\left(\Theta _{t}\!\left(X_{\left[0,t\right]}\right)\right)}\right],\label{eq:actionfunc}
\end{eqnarray}
and
\begin{eqnarray}
\Lambda_t^{\mathcal{P},\mathcal{Q}}\equiv&\Lambda^{\mathcal{P},\mathcal{Q}}\left(X_{\left[0,t\right]}\right) \equiv& \ln\left[\frac{\mathcal{P}\left(X_{\left[0,t\right]}\right)}{\mathcal{Q}^{(t)}\left(X_{\left[0,t\right]}\right)}\right],\label{eq:lambdafunct}
\end{eqnarray}
%{\bf [IN: as we said before, $\mathcal{P}_{[0,t]} =\mathcal{P}$, and same for $\mQ$.  Why are we then using the complicated notation if there is a simpler one? ]}
where $\mathcal{P}$ is the path probability describing the statistics of the physical process of interest, $X_t$, and $\mathcal{Q}^{(t)}$ is a sequence of path probabilities describing the statistics of auxiliary processes. 
%{\bf [IN: comments should come outside the gray box, as otherwise it is too big (gray boxes should be as small as possible.)]}
\end{leftbar}

Now, we discuss  a few key properties related to   $\Sigma$-stochastic entropic  and $\Lambda$-stochastic entropic functionals:
\begin{itemize}

\item The $\Sigma$-stochastic entropic functional is also known as the  action functional, see Ref.~\cite{lebowitz1999gallavotti,maes1999fluctuation}.
\item The  role of the  time  index $t$  in the superscript of $\mQ^{(t)}$ is different from the one that appears in the subscript of  $X_{[0,t]}$. 
%family of path probabilities  $\mQ^{(t)}(X_{[0,t]})$ in Eqs.~\eqref{eq:actionfunc} and \eqref{eq:lambdafunct}.
Indeed, the $t$ in the subindex $[0,t]$ of $X_{[0,t]}$ determines the time window over which the path probability $\mQ$ is evaluated; $\mQ(X_{[0,t]})$ is obtained through marginalisation of $\mQ(X_{[0,\infty)})$.  On the other hand,  the superindex~$(t)$  in  $\mQ^{(t)}$ indicates a supplementary  dependency on time  that represents a sequence of path probabilities.      Note that the supplementary dependency on $t$ is not related to  nonstationarity    or time-inhomogeneity of the process $X$, as both $\mP$ and $\mQ^{(t)}$ for fixed $t$ can represent time-inhomogeneous processes.  
In stochastic thermodynamics,  the supplementary dependence on $t$ in  $\mQ^{(t)}$   originates from  reversing the direction of time relative to time $t$.  For example,  in Markov processes   $\mQ^{(t)}$  represents often a time-reversed Markov process determined by  a   reversed protocol Eq.~\eqref{Ltilde} $\tilde{\mathcal{L}}^{(t)}_s=\mathcal{L}_{t-s}$, which depends on the time-reversal reflection point $t$. 
%We come back below on this dual dependance in $t$ of $\mQ^{(t)}_{[0,t]}$.   
For a first reading of this chapter, we advice to focus on the particular  case of $\mQ^{(t)}=\mQ$. 
Note  that it is unnatural to  consider the analogous case   $\mathcal{P}^{(t)}$,  because $\mathcal{P}$ is the path probability of the   physical process $X$, and hence there is   no reason to have an additional dependency on $t$.  
%Note that this last dependence is not mandatory, for example the process \eqref{DG} or  for the time-reversed protocol in the case of a time homogeneous generator $\mathcal{L}_t=\mathcal{L}$. 

\item The mathematical  properties  of $\Sigma_t^{\mathcal{P},\mathcal{Q}}$ and $\Lambda_t^{\mathcal{P},\mathcal{Q}}$ are  similar (see below). Moreover,  $\Sigma_t^{\mathcal{P},\mathcal{Q}}$ functionals can be written as  $\Lambda^{\mathcal{P},\mathcal{Q'}}_{t}$ functionals (and reciprocally) by using a suitable choice for the path probability $\mathcal{Q}'$, which is called time reversal of $\mathcal{Q}$ in the probability theory literature, see Refs.~\cite{nagasawa1964time,chung2006markov,chetrite2018peregrinations}.

Therefore, it is natural to ask why there is a need  to introduce  the two  entropic functionals $\Sigma$ and $\Lambda$?  The answer is {\em blowin' in the wind} of martingales:  
as we   show in Sec.~\ref{sec:6p2},   $\exp(-\Lambda_t^{\mathcal{P},\mathcal{Q}})$  can be a martingale with respect to $\mP$ even if $\mathcal{Q}$ is non stationary,  whereas $\exp(-\Sigma_t^{\mathcal{P},\mathcal{Q}})$ requires in general a stationary $\mathcal{Q}$ to be a martingale.
%the Martingale property will be strongly affected by this passage from $\Sigma_t^{\mathcal{P},\mathcal{Q}}$ to  $\Lambda^{\mathcal{P},\mathcal{Q'}}_{[0,t]}$.
An intuitive idea behind this result is that the sequence $\mathcal{Q'}^{(t)}$ that satisfies $\Sigma_t^{\mathcal{P},\mathcal{Q}}=\Lambda^{\mathcal{P},\mathcal{Q'}^{(t)}}_{t}$  depends in general explicitly  on $t$, even when $\mathcal{Q}$ is a non stationary path probability without  explicit $t$-dependence. %{(except in the case where $\mathcal{Q}$  is a stationary path probability)}. 
%{\bf [IN: I dont understand the intuitive idea Shouldn't it be the other way around:$\Sigma_t^{\mathcal{P},\mathcal{Q}'}=\Lambda^{\mathcal{P},\mathcal{Q}}_{t}$?   ]}

%{\bf [IN: is 7.4  a particular case of 7.5?   Please clarify.]} {(RC : Yes ! Look the sentence p 139 after the grey box)}

%\end{leftbar}

\item 
Let us illustrate the difference between the two $t$-dependencies in $\mQ^{(t)}(X_{[0,t]})$ on the example of a Langevin process\footnote{See  also the footnote  in Sec.~\ref{sec:Examples}  for an example in the  discrete-time setup.}. For a sequence  of path probabilities   $\mQ^{(t)}$, the  corresponding sequence $\mathscr{L}_{s}^{(t)}$ of Lagrangians reads, see Eq.~(\ref{eq:additiveA})  
\begin{eqnarray}
\mathscr{L}_{s}^{(t)}(X_{s}, \dot{X}_s) &\equiv&  
\frac{1}{4}  \left( \dot{X}_{s}-\bmu^{(t)}_{s} (X_s) F^{(t)}_{s}(X_s)  \right) \left(\mathbf{D}^{(t)}_{s}\right)^{-1}  \left( \dot{X}_{s}-\bmu^{(t)}_{s} (X_s) F^{(t)}_{s}(X_s)  \right)
\nonumber\\ 
&& +  \frac{1}{2} \nabla\cdot \left( \bmu^{(t)}_{s} F^{(t)}_{s} \right)(X_s).
 \label{eq:LagrangianFormula2}
\end{eqnarray}
{We recall readers Eq.~\eqref{action} for the definition of Lagrangians $L$ in this context. } 
 The corresponding actions defining $\mQ^{(t)}$ are, see Eqs.~  (\ref{eq:onsagerMachlupApproach1}-\ref{eq:onsagerMachlupApproach2})  
\begin{equation}
    \mathcal{A}^{(t)}(X_{[0,t]}) ={ -\ln ( \rho_0(X_0)) + \int^t_0 ds L^{(t)}_s(X_s,\dot{X}_s)}.
\end{equation} 
Notice that the explicit $t$-dependency  of $\mQ^{(t)}$, denoted by the superscript in $\mQ^{(t)}$, is due to the second $t$-dependency in the mobility matrix $\bmu^{(t)}_{s}$, diffusion matrix $\mathbf{D}^{(t)}_{s}$, and total force $F^{(t)}_{s}$.  Nevertheless, for each fixed value of $t$, the Lagrangians $\mathscr{L}_{s}^{(t)}$ describe  time-inhomogeneous Langevin processes, as $\bmu^{(t)}_{s}$, $\mathbf{D}^{(t)}_{s}$, and $F^{(t)}_{s}$ depend explicitly on $s$.

\item A key feature of  $\Sigma$-stochastic entropic functionals \eqref{eq:actionfunc} (resp.,  $\Lambda$-stochastic entropic functionals)   are the  duality relations
\begin{eqnarray}
\Sigma_t^{\mathcal{P},\mathcal{Q}} \left( \Theta_{t} (X_{[0,t]})\right) &=&-\Sigma^{\mathcal{Q},\mathcal{P}}_{t}  (\Xot)\label{eq:duality1}
\end{eqnarray}
and
\begin{eqnarray}
\Lambda^{\mathcal{P},\mathcal{Q}} \left(X_{[0,t]}\right) &=&-\Lambda^{\mathcal{Q},\mathcal{P}} (\Xot).
\label{eq:duality2}
\end{eqnarray}
In words, $\Sigma$  changes sign under  the simultaneous reversal of  time  and the exchange of the measures $P\leftrightarrow Q$, whereas $\Lambda$ changes sign under exchange of  the  measures  $P\leftrightarrow Q$.
\item  Unlike the entropy production of the macroscopic second law of thermodynamics, see e.g.~Refs.~\cite{de2013non, kondepudi2014modern}, both the   $\Sigma$-stochastic entropic functional $\Sigma_{t}^{\mP,\mQ}$ and the  $\Lambda$-stochastic entropic functional $\Lambda_{t}^{\mP,\mQ}$ can take negative values. 
However,   average values of  entropic functionals are positive (see below).

\item The existence of  $\Sigma_{t}^{\mathcal{P},\mathcal{Q}}$ and  $\Lambda_t^{\mathcal{P},\mathcal{Q}}$  requires that $\mathcal{P}$ is {\em absolutely continuous} with respect to $\mathcal{Q}^{(t)}\Theta_{t}$ and $\mathcal{Q}^{(t)}$, respectively (see Chapter~\ref{ch:2a} for a discussion of the continuous case).   For example,  in discrete space $\Sigma_{t}^{\mathcal{P},\mathcal{Q}}$ and  $\Lambda_t^{\mathcal{P},\mathcal{Q}}$  are well defined  if  for all trajectories $X_{[0,t]}$ for which $\mathcal{Q}^{(t)}\left(\Theta_{t}\left(X_{[0,t]}\right)\right)=0$  or $\mathcal{Q}^{(t)}\left(X_{[0,t]}\right)=0$  it holds that  also  $\mP_{[0,t]}(X_{[0,t]})=0$.  These  conditions  ensure that the  $\Sigma$-stochastic entropic and the  $\Lambda$-stochastic entropic functionals,  respectively,  do not diverge when evaluated along a stochastic trajectory.

\item In Sec.~\ref{sec:623}, we will  introduce the \textbf{generalized $\Sigma$-stochastic entropic functionals}, which will  provide martingales  that lead to refinements of the second law of thermodynamics in chapter .~\ref{TreeSL}.

 \end{itemize}

\begin{leftbar}
The average values with respect to $\mathcal{P}$ of both the $\Sigma$-stochastic entropic and $\Lambda$-stochastic entropic functionals   are Kullback-Leibler divergences, viz.,
\begin{eqnarray}
\big\langle \Sigma_t^{\mathcal{P},\mathcal{Q}} \big\rangle &=&D_{\rm KL}\left[\mathcal{P}\left(\Xot \right)||\mathcal{Q}^{(t)}\left(\Theta_{t}\left(\Xot\right)\right)\right]\label{eq:AER-12}
\end{eqnarray}
and 
\begin{eqnarray}
\big\langle \Lambda_t^{\mathcal{P},\mathcal{Q}} \big\rangle &=&D_{\rm KL}\left[\mathcal{P}\left(\Xot \right)||\mathcal{Q}^{(t)}\left(\Xot\right)\right].
\label{eq:AER-1}
\end{eqnarray}
 As $\mP$,  $\mQ$ and $\mathcal{Q}^{(t)}\Theta_{t}$  are normalized path probabilities, the Kullback-Leibler divergences in the right-hand sides of Eqs.~(\ref{eq:AER-12}-\ref{eq:AER-1}) are greater or equal than zero, which imply the \textbf{"second laws"} 
\begin{equation}
\big\langle \Sigma_t^{\mathcal{P},\mathcal{Q}} \big\rangle\geq 0, \qquad {\rm and} \qquad \big\langle \Lambda_t^{\mathcal{P},\mathcal{Q}} \big\rangle\geq 0.
\label{poslun}
\end{equation}
\end{leftbar}

\subsection{Fluctuation relations for stochastic entropic functionals}
\label{FREF}

Fluctuation relations follow readily from 
 the  definitions Eqs.~\eqref{eq:actionfunc} and \eqref{eq:lambdafunct}, as can be seen from  the following central equations. 
\begin{leftbar}
 The
\textbf{"mother" fluctuation relations} for an arbitrary  functional $Z\left(\Xot\right)$ read~\cite{chetrite2008fluctuation}
\begin{eqnarray}
\Big\langle\, Z \left(\Theta_t \!\left(\Xot\right)\right)\, \Big\rangle_{\mQ^{(t)}}&=&\Big\langle \exp\left(-\Sigma_{t}^{\mP,\mQ}\right) Z \left(\Xot\right)\Big\rangle \label{eq:motherfluctrelation2}
\end{eqnarray}
and 
\begin{eqnarray}
\Big\langle\, Z \left(\!\Xot\right)\,\Big\rangle_{\mQ^{(t)}}&=&\Big\langle \exp\left(-\Lambda_{t}^{\mP,\mQ}\right) Z \left(\Xot\right)\Big\rangle. \label{eq:motherfluctrelation}
\end{eqnarray}
Notably, Eqs.~\eqref{eq:motherfluctrelation2}-\eqref{eq:motherfluctrelation} hold for any functional $Z$ and any pair  $\mathcal{P}$ and $\mathcal{Q}^{(t)}$ of absolutely continuous path probabilities.  
\end{leftbar}

Setting $Z(\Xot) = \delta (\Sigma^{\mP,\mQ}_t - \sigma) $ and $Z(\Xot) = \delta (\Lambda^{\mP,\mQ}_t - \lambda) $  in the first and second lines of  Eqs.~\eqref{eq:motherfluctrelation}, respectively, and using the duality relations \eqref{eq:duality1}-\eqref{eq:duality2}, we obtain the following \textbf{generalized Crooks fluctuation relations}~\cite{crooks1999entropy}
\begin{equation}
\Big\langle\, \delta (\Sigma^{\mQ,\mP}_{t} + \sigma) \Big\rangle_{\mQ^{(t)}}=\exp(-\sigma) \Big\langle\, \delta (\Sigma_t^{\mP,\mQ} - \sigma) \Big\rangle,
\label{eq:crooks}
\end{equation}
and
\begin{equation}
\Big\langle\, \delta (\Lambda^{\mP,\mQ}_{t} - \lambda) \Big\rangle_{\mQ^{(t)}}=\exp(-\lambda) \Big\langle\, \delta (\Lambda^{\mP,\mQ}_{t} - \lambda) \Big\rangle.
\label{eq:lambda13s}
\end{equation}
{Using Eq.~\eqref{eq:duality2}, one can rewrite Eq.~\eqref{eq:lambda13s} as
\begin{equation}
\Big\langle\, \delta (\Lambda^{\mQ,\mP}_{t} + \lambda) \Big\rangle_{\mQ^{(t)}}=\exp(-\lambda) \Big\langle\, \delta (\Lambda^{\mP,\mQ}_{t} - \lambda) \Big\rangle.
\label{eq:lambda13s2}
\end{equation}}
{Equations~\eqref{eq:crooks} and~\eqref{eq:lambda13s2} can also be written as}
\begin{equation}
\frac{\rho^{\mP}_{\Sigma_t^{\mP,\mQ}}(\sigma)}{\rho^{\mQ^{(t)}}_{\Sigma_t^{\mQ,\mP}}(-\sigma)} = \exp(\sigma ), \quad {\rm and} \quad 
\frac{\rho^{\mP}_{\Lambda_t^{\mP,\mQ}}(\lambda)}{\rho^{\mQ^{(t)}}_{\Lambda_t^{\mQ,\mP}}(-\lambda)} = \exp(\lambda).
\label{eq:crooks2}
\end{equation}
In the first relation of Eq.~(\ref{eq:crooks2}), the numerator (denominator) denotes  the  probability density %{\bf [IN:  did we not decide to use different notation for probability density?]} 
of $\Sigma_t^{\mP,\mQ}$ ( $\Sigma_t^{\mQ,\mP}$) under the probability law $\mP$ ($\mQ^{(t)}$), and analogously for the $\Lambda$-stochastic entropic functional in the second equation.
For the choice $Z(\Xot)=1$, Eqs.~\eqref{eq:motherfluctrelation} become the \textbf{generalized integral fluctuation relations} given by
\begin{equation}
\left\langle \exp\left(-\Sigma_t^{\mP,\mQ}\right)\right\rangle=1,\qquad {\rm and} \qquad \left\langle \exp\left(-\Lambda^{\mP,\mQ}_{t}\right)\right\rangle=1.%_{\mP}=1.
\label{eq:jarz0}
\end{equation}
Note that the generalised integral fluctuation relations hold for any (normalised) path probability   $\mQ^{(t)}$ that is  absolutely continuous with respect to $\mP$.

In the following, by considering specific  choices for the path probability  $\mQ^{(t)}$ of the auxiliary process, we  discuss examples of  $\Sigma$-stochastic entropic functionals and  $\Lambda$-stochastic entropic functionals that are relevant for physics. 

\subsection{$\mQ$-stochastic entropy production}
\label{sec:Qstot}

%\subsection{Stochastic Entropy production as peculiar \textbf{$\Sigma$-stochastic entropic functional} and associated stochastic environmental entropy flow}

We review the   $\mQ$-stochastic entropy production, which is a $\Sigma$-stochastic entropic functional  for a specific choice of $\mathcal{Q}^{(t)}$  that is widely used in stochastic thermodynamics, see e.g.~Refs.~~\cite{maes1999fluctuation,chetrite2008fluctuation,seifert2012stochastic,harris2007fluctuation,chernyak2006path}.  
In particular, we assume that 
\begin{equation}
    \rho^{\mQ^{(t)}}_0(x) \equiv  \langle  \delta(X_0-x)\rangle_{\mQ^{(t)}} = \langle  \delta(X_t-x)\rangle \equiv \rho_t(x), \label{eq:condQ}
\end{equation}
where $\rho_t(x)$ is the probability density of $X_t$ under its native dynamics,   determined by $\mP$.

\begin{leftbar} When specializing the  $\Sigma$-stochastic entropic functional ~\eqref{eq:actionfunc} to  $\mQ^{(t)}$ that satisfy Eq.~(\ref{eq:condQ}),  we obtain the so-called \textbf{$\mQ$-stochastic entropy production}, which we denote by $S_t^{\mP,\mQ}$, i.e.,    \cite{maes1999fluctuation,chetrite2008fluctuation,seifert2012stochastic,harris2007fluctuation,chernyak2006path}
 %{\bf [ IN: box for definition of $\mathcal{Q}$-stochastic entropy production?  This would be more clear in a box.]}
\begin{equation}
S_t^{\mP,\mQ} \equiv \Sigma_{t}^{\mathcal{P},\mathcal{Q}}=\ln\left(\frac{\mathcal{P}\left(X_{\left[0,t\right]}\right)}{\mathcal{Q}^{(t)}\left(\Theta _{t}\!\left(X_{\left[0,t\right]}\right)\right)}\right), \quad t\in \mathbb{R}^+.
\label{eq:Qstot}
\end{equation} 
\end{leftbar}
% Using  Bayes' theorem and definition~\eqref{eq:defSSysonetime} for the  stochastic system entropy,   we can split the  $\mQ$-stochastic entropy production into system and environment contributions. 
%We call \textbf{\color{blue}$\mQ$-}stochastic entropy production} during $[0,t]$, {\color{blue} and  note $S^{\mathcal{P},\mathcal{Q}}_{t}$, the particular case of Eq.~\eqref{eq:actionfunc} for which the initial density of the path measure $\mQ$ is {\color{blue} the instantaneous density at time $t$ of the path measure $\mP$, i.e.   $\rho^{\mQ}_{0}=\rho_t$. 

Using Bayes' law, 
the $\mQ$-stochastic entropy production~\eqref{eq:Qstot} can  be split into two parts, namely, a   system  entropy change $\Delta S^{\rm sys}_t$ and an environmental  $\mQ$-stochastic entropy change $S^{\rm env, \mathcal{P},\mathcal{Q}}_t$, viz., %during the time interval $[0,t]$

\begin{equation}
S^{\mathcal{P},\mathcal{Q}}_{t}=\underbrace{\ln \left( \frac{\rho_0(X_0)}{\rho_t(X_t)}\right)}_{\displaystyle \Delta S^{\rm sys}_t}+ \underbrace{\ln \displaystyle 
\left(
\frac{  \mathcal{P}\left(X_{[0,t]}|X_0\right) }
{\mathcal{Q}^{(t)}\left(\Theta_{t}\!\left(X_{\left[0,t\right]}\right)|X_t\right) }\right)}_{\displaystyle  \equiv S^{\rm env, \mathcal{P},\mathcal{Q}}_t},
 \label{eq:decom1}
\end{equation}
where for consistency with  Eq.~\eqref{dimanche}  we  have omitted the superscript $\mP$ in the system entropy $\rho^{\mP}_t$, and where the conditioning in  the numerator and the denominator of the environment entropy change is on the respective initial state.   More generally,  we have for a   $\Sigma$-stochastic entropic functional 
 %{\bf [IN: Equation below is not related to Q entropy production, so should not appear here but in section 6.1.2.]}
\begin{equation}
\Sigma^{\mQ,\mP}_{t}=\ln \left( \frac{\rho_0(X_0)}{\rho^{\mQ^{(t)}}_{0}(X_t)}\right)+ \underbrace{\ln \displaystyle 
\left(
\frac{  \mathcal{P}\left(X_{[0,t]}|X_0\right) }
{\mathcal{Q}^{(t)}\left(\Theta_{t}\!\left(X_{\left[0,t\right]}\right)|X_t\right) }\right)}_{\displaystyle  \equiv S^{\rm env, \mathcal{P},\mathcal{Q}}_t}, 
 \label{eq:decom}
\end{equation} 
where the conditioning in the numerator and the denominator is  again initial conditioning.

%{ER: text missing}

The decomposition~\eqref{eq:decom1} is one of the cornerstones of stochastic thermodynamics; it is the fluctuating version of the second law for open systems 
%{[\bf: IN: do we need open systems?]} {(as you want, you don't like ?) }
\begin{equation}
 \la S^{\mathcal{P},\mathcal{Q}}_{t}\ra=\la\Delta S^{\rm sys}_t\ra + \la S^{\rm env, \mathcal{P},\mathcal{Q}}_t\ra\geq 0.
 \label{Nis}
\end{equation}
which was introduced for a specific choice of $\mathcal{Q}$ by Prigogine et al. in the 1950s~\cite{prigogine1947etude,nicolis1977self}.  Of course, we should keep in mind that  the appropriate choice of $\mQ^{(t)}$ leading to an environment entropy change  $S^{{\rm env},\mP,\mQ}_t$ with physical content depends on the physical context.   
 
The choice of the initial density $\rho_0^{\mQ^{(t)}}=\rho_t$ in~\eqref{eq:Qstot} is not arbitrary. In particular,    this choice of $\rho_0^{\mQ^{(t)}}$  minimizes the average value of  $\Sigma_t^{\mP,\mQ}$.  Indeed, taking the average of the difference between Eqs.~\eqref{eq:decom} and~\eqref{eq:actionfunc} we obtain
%{\bf [IN: srangely enough, I cant find (6.16)?]}
\begin{equation}
    \langle \Sigma_{t}^{\mP,\mQ}\rangle - \langle  S_{t}^{\mP,\mQ}\rangle = \int dx\, \rho_t(x) \ln \frac{\rho_t(x)}{\rho_0^{\mQ^{(t)}}(x)} = D_{\rm KL}[\rho_t(x) ||\rho_0^{\mQ^{(t)}}(x) ] \geq 0.
\end{equation}
%where here we have assumed that $\Sigma_{t}^{\mP,\mQ}$ and $S_{t}^{\mP,\mQ}$  are associated with physical processes  the same "bulk" dynamics for $\mP$ and $\mQ$ but  and different initial density for the auxiliary process.  
%{\bf [IN: we first say that it minimises the average of $\Sigma^{\mP,\mQ}$, but in fact it minimises the average of the diference $\Sigma^{\mP,\mQ}-S^{\mP,\mQ}$ {(but we told more or less this just 3 line ago))}.  {\bf [IN: ah right!, please lets discuss]} Also, can't we just say directly that it minimises (6.19)?  ]} {(But 6.19 has many other dependance that initial density. )}
This result justifies the name "$\mQ$-entropy production",  as the  $\Sigma-$entropic functional contains an additional cost resulting from the initial density of the auxiliary process, while for the $\mQ$-entropy production the  cost from the initial state vanishes on average, and hence the average "$\mQ$-entropy production" is determined by the   dynamics described by $\mQ$. 
%particular case of $\Sigma-$entropic functional with $\rho_0^{\mQ^{(t)}}=\rho_t$, because using the average of the generic $\Sigma-$entropic functional involves an additional cost resulting from the choice of the initial density of the auxiliary process, and not from the dynamics itself. 
%{\bf [IN: I dont understand the last sentence]} % ({and not from the dynamics}).  

In the following, we  show that for  specific  choices of  $\mathcal{Q}$,  the 
$\Sigma$-entropic functional $\Sigma^{\mathcal{P},\mathcal{Q}}_{[0,t]}$,  the $\mathcal{Q}-$entropy production $S^{\mathcal{P},\mathcal{Q}}_{t}$, and  the environmental $\mathcal{Q}-$stochastic
entropy change $S^{\rm env, \mathcal{P},\mathcal{Q}}_{t}$, identifies with usual quantity which are   commonly introduce in stochastic thermodynamics.      We refer to Refs.~\cite{gallavotti1995dynamical,chetrite2008fluctuation,chetrite2009fluctuation,Chetrite:2011,baiesi2015inflow} for other interesting choices of $\mQ$, such as, those leading to universal fluctuations relations for phase-space contraction and/or multiplicative fluctuation relations for the finite-time Lyapunov exponents.

%\textcolor{blue}{
\subsection{Total  $\Sigma$-stochastic entropic functionals and stochastic entropy production for  Markovian processes }

%\subsubsection{{\color{blue} Total  $\Sigma$-entropic functional, total environmental flow and total entropy production}}

%\subsubsection{ Definition and decomposition} 

We define the  total  $\Sigma$-stochastic entropic functional $\Sigma_{t}^{\rm tot}$ as the  $\Sigma$-stochastic entropic functional, given by \eqref{eq:actionfunc}, specialized to the following choices of $\mP$ and $\mQ$:
\begin{itemize}
    \item  The statistics $\mP$ of the physical process $X$ are generated by a generic, Markovian, non-equilibrium process with  Markov generator $\mathcal{L}$. 
    \item  The satistics  $\mQ^{(t)}$ of the  auxiliary process are determined by the time reversed Markov process defined in \eqref{Ltilde}. 
     %e{\mathcal{L}}^{(t)}_s \equiv \mathcal{L}_{t-s}
%\label{Ltilde}
%\end{equation}
%and an {\em arbitrary} initial density $\rho_0^{\mQ^{(t)}}$, and we denote the path probability of this Markov process also by  $\mQ^{(t)}=\tilde{\mP}^{(t)}$.
    %$\mQ^{(t)} =\tilde{\mP}^{(t)}$.  {\bf [IN: is this the first time ever we inroduce the $\tilde{\mP}$?  Is $\tilde{P}$ always this in the review?  Just wondering whether we need to put this somewhere else and more clearly...  ]} {(RC : This is now the first time, before it's arrive in 5) }The path probability  of the auxiliary process is that associated with a prescribed "naive" time-reversed generator given by 
   % \begin{equation}
%\tilde{\mathcal{L}}^{(t)}_s \equiv \mathcal{L}_{t-s},
%\label{Ltilde}
%\end{equation}
 %{The super-index $t$ on $\tilde{\mP}^{(t)}$ and $\tilde{\mathcal{L}}^{(t)}_s$ indicate an explicit dependence on the final time $t$.  }
\end{itemize}
%$\mathcal{Q}=\tilde{\mathcal{P}}$, 
\begin{leftbar}
The {\bf total $\Sigma-$stochastic entropic} functional \eqref{eq:actionfunc} is defined by   
%{\bf [IN: box for definition of  total $\Sigma-$entropic functional?  This would be more clear in a box. ]}
\begin{equation}
    \Sigma^{\rm tot}_{t} \equiv  \Sigma_t^{\mP,\tilde\mP ^{(t)}} = \ln \left(\frac{\mP(\Xot)}{\tilde\mP^{(t)}(\Theta_t(\Xot))}\right).
    \label{AFS2}
\end{equation}
%Here, $\tilde\mP$ is the
\end{leftbar}
Following analogous steps as in Sec.~\ref{sec:Qstot}, we can split $\Sigma^{\rm tot}_t$ into a system and an environment entropy changes  during the time interval $[0,t]$, see also Eq.~\eqref{eq:decom},
\begin{equation}
    \Sigma^{\rm tot}_t  =\ln\left(\frac{\rho_0(X_0)}{\tilde{\rho}^{(t)}_0(X_t)}\right) +\underbrace{\ln \displaystyle \left(\frac{  \mathcal{P}\left(X_{[0,t]}|X_0\right) }
{\mathcal{\tilde\mP}^{(t)}\left(\Theta_{t}\!\left(X_{\left[0,t\right]}\right)|X_t\right) }\right)}_{\displaystyle \equiv S_{t}^{\rm env}}.
\label{Senvtot}
\end{equation}
The second term in the right hand side of~\eqref{Senvtot}  is the so-called \textbf{stochastic environmental entropy flow}, 
%{\bf [IN:  we called it before environment entropy change?  Now we speak of environmental entropy flow?]} {(yes, must be change in chap 5, we choose '' environmental entropy flow '' before.)} 
which has a similar structure as the environmental entropy change   $S^{\rm env}_{t}$ given by\eqref{dimanche}.   The first term in~\eqref{Senvtot} is a generalized system entropy change, which involves the initial density of the physical process and the  probability density  $\tilde{\rho}_{0}^{(t)}(X_t)$, which is the initial density of the auxiliary process evaluated at final state of the trajectory $\Xot$.
%\begin{equation} 
%S^{env,tot}_{t} \equiv  S^{\rm env, %\mathcal{P},\mathcal{\tilde\mP}}_{t}=\ln \displaystyle 
%\left(
%\frac{  \mathcal{P}\left(X_{[0,t]}|X_0\right) }
%{\mathcal{\tilde\mP}\left(R_{t}\!\left(X_{\left[0,t\right]}\right)|X_t\right) }
%\right). 
%\end{equation}
%which is the  formulae \textcolor{blue}{given by the two extremal side of}   ~\eqref{eq:defSenv}. 

\begin{leftbar}
%Let us now consider the case in which, for the auxiliary process, in addition to fixing the driving as $\tilde{\mathcal{L}}_s = \mathcal{L}_{t-s}$ we also set the initial  distribution    to be the distribution of $\mP$ at time $t$, i.e. $\rho_0^{\mQ^{(t)}}=\rho_t$. 
If  $\tilde{\rho}_{0}^{(t)}=\rho_t$  in Eq.~(\ref{Senvtot}), then  $\Sigma^{\rm tot}_t$ is also called  the stochastic entropy production, denoted by $S^{\rm tot}_t$, i.e., 
%{\bf [IN: better to define stochastic entropy production in a box. ]}
\begin{equation}
    S^{\rm tot}_t  \equiv\underbrace{\ln\left(\frac{\rho_0(X_0)}{\rho_{t}(X_t)}\right)}_{\displaystyle \Delta S^{{\rm sys}}_t} +\underbrace{\ln \displaystyle \left(\frac{  \mathcal{P}\left(X_{[0,t]}|X_0\right) }
{\mathcal{\tilde \mP}^{(t)}\left(\Theta_{t}\!\left(X_{\left[0,t\right]}\right)|X_t\right) }\right)}_{\displaystyle S^{\rm env}_{t}}.
\label{Stotcopied}
\end{equation}
\end{leftbar}

In Chapter~\ref{sec:martST1} we have studied $S^{\rm tot}$ for one-dimensional Langevin processes and Markov jump processes.   
In the following Secs.~\ref{sec:markovjumpsigma} and~\ref{sec:langsigma}, we  provide for  illustrative purposes explicit expressions of    $\Sigma^{\rm tot}_t$ and $S^{\rm tot}_t$  for  Markov jump processes  and  diffusion processes in arbitrary dimensions. 

%\begin{equation}
%    \langle \Sigma_t \rangle 
%\end{equation}

%{ Finally, the associated $\tilde\mP$-stochastic entropy production 
%$ S^{\rm tot}_{t} \equiv  S^{ %\mathcal{P},\mathcal{\tilde\mP}}_{t} $ is call \textbf{total stochastic entropy production}. We refund the  formulae given by the two extremal side of~\eqref{eq:defStot}.  \vspace{1cm}
%I dont understand this sentence so I would remove.}

%The decomposition\eqref{eq:decom} becomes then for entropy production 
%\begin{equation}
%S^{\rm tot}_{t}=\Delta S^{sys}_{t} + S^{\rm env, tot }
% \label{eq:decom3}
%\end{equation}
%}
%{\color{blue} We note that} for \textbf{total stochastic entropy functional} \Sigma^{tot}_{t}$ the generalized Crooks relation~\eqref{eq:crooks2} gives
%\begin{equation}
%\frac{\mP (\Sigma^{tot} _t=\sigma)}{\tilde\mP (\tilde\Sigma^{tot} %_t=-\sigma)} = \exp(\sigma)\quad,
%\label{eq:crooks3}
%\end{equation}
%with $\tilde\Sigma^{tot}_{t} \equiv\Sigma_{t}^{\tilde\mP,\mathcal{P}}$. In this case,  Eq.~\eqref{eq:crooks3} implies the integral fluctuation relation~$\langle \exp(-\Sigma^{tot} _t)\rangle=1$.  All the above results hold for any Markovian process.

%{\textcolor{blue}{
\subsubsection{Markov-jump processes }
\label{sec:markovjumpsigma}

For a Markov jump process defined  by time-dependent, transition  rates $\omega_t (x, y)$ for all $x,y\in\mathcal{X}$ (see section \eqref{eq:jumpprocconta}),  the total  $\Sigma$-stochastic entropic functional~\eqref{Senvtot}  
%{\bf [IN: didn't we call this differently??, such as "total $\Sigma-$entropic functional". ]} 
is given by 
\begin{equation}
\Sigma^{\rm tot}_{t} =\ln\left(\frac{\rho_0(X_0)}{\tilde{\rho}^{(t)}_0(X_t)}\right)+{\displaystyle\sum_{j=1}^{N_t}\ln\left[\frac{\omega_{\T_j}(X_{\T_j^{-}},X_{\T_j^{+}})}{\omega_{\T_j}(X_{\T_j^{+}},X_{\T_j^{-}})}\right]},
\label{eq:WLSPS}
\end{equation}
where the $\T_j$  in the right-hand side are the times when $X$ jumps between different states, with    $0\leq\T_1\leq \T_2\leq \cdots \T_{N_t}\leq t$, and  
 $N_t$ is the total number of jumps in the trajectory $\Xot$.   Note that it is also 
 possible to write  analogous explicit expressions for the general markovian $\mQ$ in $\Sigma$-stochastic entropic functionals, given by   \eqref{eq:actionfunc},  and the $\Lambda$-stochastic entropic functionals, given by \eqref{eq:lambdafunct}, associated to such pure jump processes.   
 %{\bf [IN: Really? \eqref{eq:actionfunc} and \eqref{eq:lambdafunct} are defined for general $\mQ$.   I do not see how an "analogous" explicit expression can be written for general $\mQ$.   ]} \textcolor{red}{(RC : too supress after reading..  By exemple, we can show that for Markovian general $\mQ$ (we can add the word ''Markovian '' after ''general'', i do)  %\begin{equation}
%\Sigma^{\mQ,\mP}_{t}=\ln \left( \frac{\rho_0(X_0)}{\tilde{\rho}_0(X_t)}\right)+ \sum_{j=1}^{N_{t}}\ln\left[\frac{W_{\mathcal{T}_{j}}(X_{\mathcal{T}_{j}^{-}},X_{\mathcal{T}_{j}^{+}})}{W_{t-\mathcal{T}_{j}}^{Q}(X_{\mathcal{T}_{j}^{+}},X_{\mathcal{T}_{j}^{-}})}\right]+\int_{0}^{t}ds\int_{\mathcal{X}}dy\left(W_{t-s}^{Q}(X_{s},y)-W_{s}(X_{s},y)\right)
%\end{equation} )}
 
 Note  that $\Sigma^{\rm tot}_{t}$ in~\eqref{eq:WLSPS}  exists if the so-called {\em microreversibilty} condition holds, viz.,  for all $x,y\in \mathcal{X}$, $\omega_t (x, y)>0$  implies $\omega_t (y, x)>0$; these conditions are  equivalent to the assumed absolute continuity between $\mathcal{P}$ and
${\mathcal{\tilde\mP}\Theta_{t}}$.

For a microreversible Markov jump process, the  total stochastic entropy production $S^{\rm tot}_t$ is given by 
\begin{equation}
S^{\rm tot}_t =\underbrace{\ln\left(\frac{\rho_0(X_0)}{\rho_{t}(X_t)}\right)}_{\displaystyle \Delta S^{{\rm sys}}_t}+\underbrace{\displaystyle\sum_{j=1}^{N_t}\ln\left[\frac{\omega_{\T_j}(X_{\T_j^{-}},X_{\T_j^{+}})}{\omega_{\T_j}(X_{\T_j^{+}},X_{\T_j^{-}})}\right]}_{\displaystyle S^{\rm env}_t},
\label{eq:WLSPSJP}
\end{equation}
where we decomposed $S^{\rm tot}_t$ in terms of  system entropy change $\Delta S^{\rm sys}_t$ and the environment entropy change  $S^{\rm env}_t$.   Rewriting the first term in~Eq.~\eqref{eq:WLSPSJP}, $S^{\rm tot}_t$      can  be expressed  in its alternative form  Eq.~(\ref{eq:defStot})
\begin{equation}
S^{\rm tot}_t = -\int_{0}^{t}ds(\partial_{s}\ln\rho_{s})(X_{s})+\displaystyle\sum_{j=1}^{N_t}\ln\left[\frac{\rho_{\T_j}(X_{\T_j^-})\omega_{\T_j}(X_{\T_j^{-}},X_{\T_j^{+}})}{\rho_{\T_j}(X_{\tau_j^+})\omega_{\T_j}(X_{\T_j^{+}},X_{\T_j^{-}})}\right],
\label{fatigue}
\end{equation}

For  a system in equilibrium, $S^{\rm tot}_t=0$, as
 $\rho_{s}=\rho_{st}$ is independent of time  and the  detailed balance relation  $\rho_{\rm st}(x)\omega_{s}(x,y)=\rho_{\rm st}(y)\omega_{s}(y,x)$ holds for all $x,y\in \mathcal{X}$ and $s\geq 0$.  On the other hand, for a  nonequilibrium system  the average total entropy production  reads (see also  Eq.~(\ref{eq:SDotResult}) {and Eq.~\eqref{eq:proofzerossys}})
\begin{eqnarray}
\langle S^{\rm tot}_t  \rangle &=&\int_{0}^{t}ds \int_{\mathcal{X}}dx\int_{\mathcal{X}}dy \;\rho_{s}(x)\omega_{s} (x,y)\ln\left[{\frac{\rho_{s}(x)\omega_{s}(x,y)}{\rho_{s}(y)\omega_{s}(y,x)}}\right]\\
&=&{\frac{1}{2}} \int_{0}^{t}ds \int_{\mathcal{X}}dx\int_{\mathcal{X}}dy\; J_{s,\rho}(x,y) \ln\left[{\frac{\rho_{s}(x)\omega_{s}(x,y)}{\rho_{s}(y)\omega_{s}(y,x)}}\right], \label{eq:schnack}
\end{eqnarray} 
where in the second equality we have used the definition of the instantaneous probability current  $J_{s,\rho}(x,y)=\rho_{s}(x)\omega_{s}(x,y)-\rho_{s}(y)\omega_{s}(y,x)$, with $s\in [0,t]$.  Equation~\eqref{eq:schnack} is the celebrated Schnakenberg formula for the entropy production of Markovian systems~\cite{schnakenberg1976network}, which was derived two decades before the origins of stochastic thermodynamics.

\subsubsection{Multidimensional overdamped Langevin processes} 
\label{sec:langsigma}
We consider a multidimensional Langevin process described by      Eq.~\eqref{eq:LE}, and which  we rewrite here for convenience, 
\begin{equation}
\dot{X}_{t}= (\bmu_{t}F_{t})(X_{t})+\left(\nabla\, \mathbf{D}_{t}\right)(X_{t})+\sqrt{2\mathbf{D}_{t}(X_{t})} \dot{B}_{t}\;.\label{eq:LE2}
\end{equation}
Recall that  $F_t(x) = -\nabla V_t(x) + f_{t}(x)$ is a generic force which has a conservative part  $-\nabla V_t(x)$ and a non-conservative $f_t(x)$ part, and both contributions can depend explicitly on time, see Eq.~\eqref{F}.      

The total $\Sigma$-stochastic entropic functionals associated with trajectories generated by the overdamped Langevin equation~\eqref{eq:LE2}  are given by, see e.g.~\cite{lebowitz1999gallavotti,chetrite2008fluctuation},   
\begin{equation}
\Sigma^{\rm tot}_{t} 
=\ln\left(\frac{\rho_0(X_0)}{\tilde{\rho}^{(t)}_0(X_t)}\right)
+\underbrace{\int_{0}^{t} 
\Big(\left(\bmu_s F_{s}\right)\textbf{D}_{s}^{-1}\Big)(X_{s}) \circ \dot{X}_{s}ds}_{\displaystyle S^{\rm env}_t}.
\label{forGirs-1-1a}
\end{equation}
Here, $\mathbf{D}_t$ needs to be invertible,  which implies that this result does not hold for underdamped Langevin equations.   
Note that it is possible to write  analogous explicit expressions for the general $\Sigma$-stochastic entropic functional~\eqref{eq:actionfunc} associated to markovian $\mQ$  and $\Lambda$-stochastic entropic functional~\eqref{eq:lambdafunct}  associated to  multidimensional Langevin equations, see e.g. Ref.~\cite{chetrite2008fluctuation}.

The total stochastic entropy production of a multidimensional Langevin process is given by \begin{equation}
S^{\rm tot}_t 
=
\underbrace{\ln\left(\frac{\rho_0(X_0)}{\rho_{t}(X_t)}\right)}_{\displaystyle \Delta S^{{\rm sys}}_t}
+\underbrace{\int_{0}^{t} 
\Big(\left(\bmu_s F_{s}\right)\textbf{D}_{s}^{-1}\Big)(X_{s})\circ \dot{X}_{s}ds} _{\displaystyle S^{\rm env}_t}.
\label{forGirs-1-1-E}
\end{equation}
Using the definition of the probability current~Eq.~\eqref{curr} 
\begin{equation}
\ensuremath{
J_{t,\rho}(x)\equiv \left(\left(\bmu_t F_{t}\right)\rho_{t}\right)(x)
%\left(\bmu_{t}(x) F_{t}(x)\right)\rho_{t}(x)
-\ensuremath{\left(\textbf{D}_{t}\nabla\rho_{t}\right)(x)},
}
\end{equation} 
and using the Stratonovich (i.e. standard)  rules of calculus,  
we can rewrite Eq.~\eqref{forGirs-1-1-E}  as 
\begin{equation}
S^{\rm tot}_t 
=
-\int_{0}^{t}
\left(\partial_{s}\ln\rho_{s}\right)(X_{s})ds+\int_{0}^{t} (J_{s,\rho}\left(\rho_{s}\textbf{D}_{s}\right)^{-1})(X_{s})\circ \dot{X}_{s}ds, 
\label{eq:forGirs-1-1}
\end{equation}
which generalizes Eq.~\eqref{eq:StotStrato} to the multidimensional case. {To pass from Eq.~\eqref{forGirs-1-1-E} to Eq.~\eqref{eq:forGirs-1-1} we used the relation
\begin{equation}
\ln\left(\frac{\rho_{0}\left(X_{0}\right)}{\rho_{t}\left(X_{t}\right)}\right)=-\int_{0}^{t}d\left(\ln\left(\rho_{s}\left(X_{s}\right)\right)\right)=-\int_{0}^{t}\left(\left(\partial_{s}\ln\rho_{s}\right)\left(X_{s}\right)ds+\left(\nabla\ln\rho_{s}\right)\left(X_{s}\right)\circ\dot{X}_{s}ds\right).
\end{equation}}

For equilibrium processes,  the total stochastic entropy production vanishes, even at the stochastic level. This is because equilibrium dynamics satisfy  $\rho_{t}=\rho_{\rm st}$ and  the "detailed balance" condition  $J_{s,\rho_{\rm st}}=0$, and hence the two terms in~\eqref{eq:forGirs-1-1} vanish.
 On the other hand, for nonequilibrium processes  $S^{\rm tot}_t $ can take  any value (positive or negative), yet its average is positive.  
Indeed,   the average total stochastic entropy production  is a quadratic  form of the probability current, viz.,
\begin{equation}
\langle S^{\rm tot}_t  \rangle =\int_{0}^{t}ds\int_{\mathcal{X}} dx\, \left( J_{s,\rho}\left(\rho_{s}\textbf{D}_{s}\right)^{-1}J_{s,\rho}\right)(x)\geq 0,
\label{eq:stotavgj}
\end{equation} 
and the positivity  follows from $\left(J_{s,\rho}\left(\rho_{s}\textbf{D}_{s}\right)^{-1}J_{s,\rho})\right)(x)\geq 0 $ for all values of $x\in \mathcal{X}$; we refer to  Sec.~\ref{sec:martlanggent} for a derivation of Eq.~\eqref{eq:stotavgj}.
 %Then, this relation is the proof that the relation\eqref{footsoon} of the previous chapter is still true in multidimensional context and with space-time inhomogeneous mobility, i.e for  \textbf{ Langevin equation} \eqref{eq:LE}, at least if the diffusion matrix $\textbf{D}_t$ is invertible.

\subsubsection{Overdamped  isothermal Langevin equation}
\label{sec:setupJarzover}
Consider now the Langevin dynamics described by Eq.~\eqref{eq:LE2} with the  Einstein relation \eqref{ER}   fulfilled,  i.e., $\textbf{D}_{t}(x)= T\bmu_t(x)$, with $\bmu_t(x)$  a symmetric  mobility matrix. We call this the {\em overdamped  isothermal Langevin equation}.  
%, i.e. for overdamped Isothermal Langevin equation~\eqref{eq:LEO} with the constraint
For overdamped  isothermal Langevin processes, the stochastic environmental entropy change \eqref{forGirs-1-1-E}   is given by 
 \begin{equation}
S^{\rm env}_{t} = \frac{1}{T} \int_{0}^{t}  F_{s}(X_{s}) \circ\dot{X}_{s}ds = -\frac{Q_t}{T},
\label{eq:6p33}
\end{equation}
where in the second equality we have used the relation~\eqref{eq:qs3} for the stochastic heat absorbed by the system. 
Equation~\eqref{eq:6p33} shows that the Clausius relation between environmental entropy change and heat  also holds for isothermal  multidimensional Langevin system.  % and with inhomogeneous mobility, i.e, for overdamped isothermal Langevin processes given by Eq.~\eqref{eq:LE2} and the additional constraint $\textbf{D}_{t}(x)= T\bmu_t(x)$. 

Now, we explcit $\Sigma^{\rm tot}_t$ in an important physical example.   Suppose that the potential $V$ is determined   by a  deterministic protocol $\lambda_s$ ($s\in [0,t]$), and that the system is initially described by an equilibrium ensemble with initial density
\begin{equation}
    \rho_{0}(x)=\exp\left(-\frac{(V_{0}(x)-G_{0}^{\rm eq})}{T}\right),
    \label{eq:canonincalinitial}
\end{equation}
where 
\begin{equation}
G_{t}^{\rm eq}\equiv -T\ln\left(\int_{\mathcal{X}}dx\exp\left(-\frac{V_{t}(x)}{T}\right)\right), 
\label{Feq}
\end {equation}
is the equilibrium free energy at time $t\geq 0$. 
Note that if  an external force is present, $\rho_0(x)$ is not a steady state,  even if the potential is constant.
Consider now as auxiliary reference process with initial density equal to  
\begin{equation}
    \rho_{0}^{\mQ}(x)=\exp\left(-\frac{(V_{t}(x)-G_{t}^{\rm eq})}{T}\right),
      \label{eq:canonincalinitialbackward}
\end{equation}
 which coincides with the stationary equilibrium distribution that the system may have if  the driving is stopped at time $t$ (i.e. for $s\geq t$ we have $f_{s}=0$ and  $V_{s}=V_{t}$). 
Moreover, we assume that the driving of the auxiliary process is the "time-reversal" $\lambda^{\mQ}_s=\lambda_{t-s}$ for $s\in [0,t]$. The associated  $\Sigma^{\rm tot}$-entropic functional given by Eq.~ \eqref{forGirs-1-1a}  reads 
\begin{align}
\Sigma_{t}^{\rm tot} & =\frac{V_{t}(X_{t})-V_{0}(X_{\text{0}})+G_{0}^{\rm eq}-G_{t}^{\rm eq}+\int_{0}^{t}F_{s}\left(X_{s}\right)\circ \dot{X}_{s}ds}{T},\nonumber\\
 & =\frac{G_{0}^{\rm eq}-G_{t}^{\rm eq}+\int_{0}^{t}\left(f_{s}\left(X_{s}\right)\circ \dot{X}_{s}ds+\left(\partial_{s}V_{s}\right)\left(X_{s}\right)ds\right)}{T},
 \label{jeudi}
\end{align}
where in the second equality we used the Stratonovich (i.e. standard) rules of calculus and Eq.~\eqref{F} for the total force $F_s(X_s)= -\partial_x V_s(X_s) + f_s(X_s)$ for this particular dynamics. %{Comment about external force which is absence at time 0.} {\color{red}No! the external force is present at time 0. } 
As shown below, Eq.~\eqref{jeudi} together with the martingale properties of $\Sigma^{\rm tot}_t$  allows us to derive the celebrated Jarzynski's equality~\cite{jarzynski1997nonequilibrium} and Crooks' fluctuation relation~\cite{crooks1999entropy} involving the fluctuating work done and the equilibrium free energy changes in driven overdamped isothermal systems.

\begin{leftbar}
For isothermal overdamped Langevin systems that are driven away  by a time-dependent deterministic protocol from an initial, thermal state \eqref{eq:canonincalinitial},  Eq.~\eqref{jeudi} relates  $\Sigma^{\rm tot}_t$  to  the fluctuating work done on the system and to the equilibrium free energy change in the interval $[0,t]$, viz.,
\begin{equation}
\Sigma_{t}^{\rm tot} =\frac{ W_{t}- (G_{t}^{\rm eq}-G_{0}^{\rm eq})}{T}.
\label{W-w2}
\end{equation}
Equation~\eqref{W-w2}  follows from identifying the integral in the right-hand side of Eq.~\eqref{jeudi} as the stochastic work exerted on the system, see Eq.~\eqref{W} for the expression of the stochastic work $W_t$ for the one-dimensional case.
Here, we have also used that $G_{t}^{\rm eq}$ is the \textbf{equilibrium free energy} defined in \eqref{Feq}. 
Specializing the integral fluctuation relation for $\Sigma-$entropic functionals~\eqref{eq:jarz0} to the choice Eq.~\eqref{W-w2}, we obtain  \textbf{Jarzynki's equality}~\cite{jarzynski1997nonequilibrium} 
\begin{equation}
\left\langle \exp\left(-\frac{W_{t}}{T}\right)\right\rangle =\exp\left(-\frac{(G_{t}^{\rm eq}-G_{0}^{\rm eq})}{T}\right). 
\label{eq:jarzynskiequality}
\end{equation}
We remark that  the average in the left-hand side in Eq.~\eqref{eq:jarzynskiequality}  is done over all trajectories starting from  the initial canonical distribution given by Eq.~\eqref{eq:canonincalinitial}. 
%Similarly,  the relation~\eqref{eq:crooks2} for $\Sigma-$entropic functionals allows to derive \textbf{Crooks' fluctuation relation}~\cite{crooks1999entropy}
%{\color{blue}
%\begin{equation}
%\frac{\mP (W_t=w)}{\widetilde\mP (W_t=-w)} = \exp\left(\frac{w-(G_{t}^{\rm eq}-G_{0}^{\rm eq})}{T}\right),
%\label{eq:crooksW}
%\end{equation}}
%where $\rho_t(W_t)$ and  $\tilde\rho_t(W_t)$ are respectively the probability density for the stochastic work  in the forward and reverse process at time $t$ before the beginning of these processes, with the backward process starting from the canonical state given by Eq.~\eqref{eq:canonincalinitialbackward}. 
\end{leftbar}

 We note that the relation~\eqref{W-w2} can  also be derived from the expression~\eqref{eq:WLSPS} for   $\Sigma^{\rm tot}$ associated with  isothermal Markov-jump processes~\eqref{eq:EDBx}.  Moreover, for  general underdamped isothermal Langevin systems \eqref{eq:lang1}, the stochastic work exerted on the system on the time interval $\left[0,t\right]$ can still be related  to a $\Sigma$-entropic functional, see e.g. relations (7.16-7.17) in~\cite{chetrite2008fluctuation}. 

\subsubsection{Martingale structure of  the stochastic entropy production for  Langevin processes}
\label{sec:martlanggent}

Now, we study in more detail the martingale structure of $\exp(-S^{\rm tot}_t)$ for   multidimensional Langevin equations.  
%{\bf [IN: not sure that i understand this sentence?]}

The explicit expression for $S^{\rm tot}_t$,  given by Eq.~\eqref{eq:forGirs-1-1}  in the Stratonovich form, can be rewritten as follows  in the It\^{o} form,  %using the  relation~\eqref{I-S}
\begin{eqnarray}
S^{\rm tot}_t &=&
-\int_{0}^{t}ds
\left(\partial_{s}\ln\rho_{s}\right)(X_{s})+\int_{0}^{t} (J_{s,\rho}\left(\rho_{s}\textbf{D}_{s}\right)^{-1})(X_{s}) \dot{X}_{s}ds\nonumber\\
&+&\int_{0}^{t} ds\big[\textbf{D}_{s}\nabla\left(J_{s,\rho}\left(\rho{}_{s}\textbf{D}_{s}\right)^{-1}\right)\big](X_{s}).
\label{eq:forGirs-1-1-I}
\end{eqnarray}
{The conversion of $S^{\rm tot}$ from Stratonovich [Eq.~\eqref{eq:forGirs-1-1}]  to It$\hat{\rm o}$ [Eq.~\eqref{eq:forGirs-1-1-I}] follows from Eq.~\eqref{eq:StratotoItoLang}, copied here for convenience
\begin{align*}
 \int_{0}^{t}g_s(X_{s})\circ\dot{X}_{s}ds =\int_{0}^{t}g_s(X_{s})\dot{X}_{s}ds+\int_{0}^{t}\textbf{D}_{s}(X_{s})\left[\left(\nabla g_s\right)(X_{s})\right]ds,
\end{align*}
which is valid  for any  function $g_t(x)$ that is smooth on $t$ and $x$. }
%S^{\rm tot}_t  = \int_{0}^{t} \left(-\left(\partial_{s}\ln\rho_{s}\right)+\textbf{D}_{s}\nabla\left(j_{s,\rho}\left(\rho{}_{s}\textbf{D}_{s}\right)^{-1}\right)\right)(X_{s})ds+ (j_{s,\rho}\left(\rho_{s}\textbf{D}_{s}\right)^{-1})(X_{s}) dX_{s}\right).
Plugging  the Langevin equation~\eqref{eq:LE2} in Eq.~\eqref{eq:forGirs-1-1-I}, we obtain 
\begin{eqnarray}
S^{\rm tot}_t & 
=&
\int_{0}^{t}
\left[
-\left(\partial_{s}\ln\rho_{s}\right)
+\textbf{D}_{s}\nabla\left(J_{s,\rho}\left(\rho{}_{s}\textbf{D}_{s}\right)^{-1}\right)+
J_{s,\rho}\left(\rho_{s}\textbf{D}_{s}\right)^{-1}\bmu_{s}F_{s}
\right](X_{s}) ds \nonumber\\
&&+\int_{0}^{t}
\left[J_{s,\rho}\left(\rho_{s}\textbf{D}_{s}\right)^{-1}\left(\nabla\, \mathbf{D}_{s}\right)
\right](X_{s}) ds
+\int_{0}^{t}
\left[ J_{s,\rho}\left(\rho_{s}\textbf{D}_{s}\right)^{-1}\sqrt{2\textbf{D}_{s}}\right] (X_{s}) \dot{B}_{s}
ds.\nonumber
\end{eqnarray}
Expanding the second term of the first line, and simplifying some terms, we find 
\begin{eqnarray}
S^{\rm tot}_t & 
=&
\int_{0}^{t}
\left[
\left(
-\left(\partial_{s}\ln\rho_{s}\right)+ \frac{\nabla J_{s,\rho}}{\rho{}_{s}}
-\frac{J_{s,\rho}\nabla\rho{}_{s}}{\rho{}_{s}^2} +
J_{s,\rho}\left(\rho_{s}\textbf{D}_{s}\right)^{-1}\bmu_{s}F_{s}
\right)(X_{s})\right] ds \nonumber\\
&& +\int_{0}^{t}
\left[ (J_{s,\rho}\left(\rho_{s}\textbf{D}_{s}\right)^{-1}\sqrt{2\textbf{D}_{s}})(X_{s}) \dot{B}_{s}
\right] 
ds.
\label{eq:forGirs-1-1-details}
\end{eqnarray}
Lastly, using the Fokker-Planck equation~\eqref{ce}, and the definition of the probability current~\eqref{currL},   we get 
\begin{equation}
S^{\rm tot}_t 
=
\int_{0}^{t}\left[\left(
-2\left(\partial_{s}\left(\ln\rho{}_{s}\right)\right)+{\frac{J_{s,\rho}\textbf{D}_{s}^{-1}J_{s,\rho}}{\left(\rho{}_{s}\right)^{2}}}\right)(X_{s})
+J_{s,\rho}\left(X_{s}\right)\left(\rho_{s}\textbf{D}_{s}\right)^{-1}\left(X_{s}\right)\sqrt{2\textbf{D}_{s}\left(X_{s}\right)}\dot{B}_{s}
\right]ds. 
\label{foot}
\end{equation}
%{\bf [IN: equation does not fit the line]}
 Taking the average of Eq.~\eqref{foot} over the Brownian noise yields the second law~Eq.~\eqref{eq:stotavgj}.   In addition, Eq.~\eqref{foot} together with the rules of  It\^{o}  calculus allows us to  uncover  the martingale structure of $\exp(-S^{\rm tot}_t)$. 
\begin{leftbar}
\textbf{It\^{o} stochastic differential equation for stochastic entropy production in multidimensional Langevin processes and martingality.} Deriving Eq.~\eqref{foot} with respect to time we get
\begin{equation}
\dot{S}^{\rm tot}_t=\left[
-2\partial_{t}\ln\rho{}_{t}+\frac{J_{t,\rho}\textbf{D}_{t}^{-1}J_{t,\rho}}{\left(\rho{}_{t}\right)^{2}}\right](X_{t})+\left(\sqrt{2}\frac{J_{t,\rho}}{\rho_{t}}\textbf{D}_{t}^{-1/2}\right)\left(X_{t}\right)\dot{B}_{t}.
\label{forGirs-1-1-1}
\end{equation}
 We can define a new scalar white noise $\dot{B}_{t}^{S}$, with zero mean $\langle \dot{B}_{t}^{S}\rangle =0$ and autocorrelation $\langle \dot{B}_{t}^{S}\dot{B}_{s}^{S} \rangle= \delta(t-s)$,  such that Eq.~\eqref{forGirs-1-1-1}  takes the form 
\begin{equation}
\dot{S}^{\rm tot}_t=
-2\left(\partial_{t}\ln\rho{}_{t}\right)(X_{t})+\underbrace{\left(\frac{J_{t,\rho}\textbf{D}_{t}^{-1}J_{t,\rho}}{\left(\rho{}_{t}\right)^{2}}\right)(X_{t})}_{ \equiv \displaystyle v^S_t(X_t)}+
\underbrace{\sqrt{\left(2
\frac{J_{t,\rho}\textbf{D}_{t}^{-1}J_{t,\rho}}{\left(\rho{}_{t}\right)^{2}}
\right)(X_t)}}_{\equiv \displaystyle \sqrt{2v^S_t(X_t)}
}
\dot{B}^{S}_{t}.
%{\frac{d\left(\exp(-S^{\rm tot}_t)\right)}{dt}}=-2\exp(-S^{\rm tot}_t)\left(\partial_{t}\ln\rho_{t}\right)(X_{t})-\exp(-S^{\rm tot}_t) \sqrt{\left(\frac{J_{t,\rho}}{\rho_t}\textbf{D}_{t}^{-1}\frac{J_{t,\rho}}{\rho_t}\right)(X_t)}\,\dot{B}_{t}^{S}
\label{forGirs-1-1-1-8}
\end{equation}
{Averaging over many realizations, we get
\begin{equation}
\frac{d}{dt}\langle {S}^{\rm tot}_t  \rangle =\int_{\mathcal{X}} dx\, \left( J_{s,\rho}\left(\rho_{s}\textbf{D}_{s}\right)^{-1}J_{s,\rho}\right)(x)\geq 0,
\label{eq:stotavgj1}
\end{equation} 
which is equivalent to Eq.~\eqref{eq:stotavgj}. }
\end{leftbar}
Equation~\eqref{forGirs-1-1-1-8} has  an analogous structure  to the unidimensional case Eq.~\eqref{eq:entropy}, but with an entropic drift 
\begin{equation}
    v^S_t(x)=\left(\frac{J_{t,\rho}\textbf{D}_{t}^{-1}J_{t,\rho}}{\rho{}_{t}^{2}}\right)(x).
\end{equation}
Applying the multidimensional It\^{o} formula, see Appendix~\eqref{app:ItoFormula2}, to the change of variable $S^{\rm tot}_t\to \exp(-S^{\rm tot}_t)$,  we obtain from \eqref{forGirs-1-1-1} the stochastic differential equation
\begin{equation}
\hspace{-0.5cm}\frac{d\exp(-S^{\rm tot}_t)}{dt}
=-2\exp(-S^{\rm tot}_t)\left(\partial_{t}\ln\rho{}_{t}\right)(X_{t})-\exp(-S^{\rm tot}_t) \left(\sqrt{2}\frac{J_{t,\rho}}{\rho_{t}}\textbf{D}_{t}^{-1/2}\right)\left(X_{t}\right)\dot{B}_{t}           . 
\label{forGirs-1-1-1-E}
\end{equation}
Note that this is not a closed set of  stochastic differential equations because it is not autonomous in $\exp(-S^{\rm tot}_t)$, but the joint process $(X_{t},\exp(-S^{\rm tot}_t))$ admits a closed set of stochastic differential equations. Equations~\eqref{forGirs-1-1-1-8} and~\eqref{forGirs-1-1-1-E} extend the Eqs.~\eqref{eq:entropy} and~\eqref{eq:gbm}  to the multidimensional context and with space-time inhomogeneous mobility.

\begin{leftbar}
Because of the presence of a non-vanishing drift in Eq.~\eqref{forGirs-1-1-1-E},  $\exp(-S^{\rm tot}_t)$ \textbf{is not a martingale} in general.   Instead,  $\exp(-S^{\rm tot}_t)$ is a martingale  \textbf{if and only if} $\partial_{t}\rho_{t}=0$, i.e.,   in a stationary state (which can be an equilibrium state %$J_{\rho_{inv}}=0$ 
or   nonequilibrium steady state).  %$\nabla J_{\rho_{inv}}=0$ but $J_{\rho_{inv}} \neq 0$). 
Thus for time-homogeneous  nonequilibrium stationary processes, the exponentiated, negative, total entropy production is an exponential martingale and the martingale fluctuation relation 
\begin{equation}
\langle\, \exp(-S^{\rm tot}_{t}) \,|\, X_{[0,s]}\, \rangle = \exp(-S^{\rm tot}_{s})\quad
\label{eq:mart_Stot}
\end{equation}
holds, 
which  implies  a  conditional second law (submartingale property) for the total entropy production in steady state
\begin{equation}
\langle\, S^{\rm tot}_t  \,|\, X_{[0,s]} \,\rangle \geq  S_{s}^{\rm tot} \quad, 
\label{eq:submart_Stot}
\end{equation}
for any  $0 \leq s\leq t$.
\end{leftbar}
%{\bf [IN:  Why having full derivations in gray boxes?   Can't we restrict gray boxes to important results/definitions?]}{ER: yes here we exaggerated}

 In the forthcoming Sec.~\ref{sec:6p2}, we will come back to this martingale properties in more fundamental way, and for a more  generic setup that includes also other entropic functionals and  jump processes.
%{\color{magenta}ADD $\exp(-S^{\rm tot}_t)$ is a subharmonic function $f(x,u)=u$ of the joint process ($X,S^{\rm tot}$) .....I  am not sure to add this because it is more complicated prove that add almost nothing w.r.t what I wrote here.  }

\subsection{$^{\spadesuit}$Excess and housekeeping entropy production for Markovian processes}\label{sec:Pex}

Now,  we  review the notions of excess and housekeeping entropy production as introduced by Oono and Paniconi~\cite{oono1998steady} and further explored in Refs.~\cite{hatano2001steady, speck2005integral,chetrite2008fluctuation,esposito2010three,van2010three,harris2007fluctuation} in the context of fluctuation relations within stochastic thermodynamics. We choose here to keep the original terminology used by Oono and Paniconi~\cite{oono1998steady} for isothermal Markovian processes despite the setup that we consider is  more general. We also note that a popular alternative terminology was introduced by Esposito and Van den Broeck~\cite{esposito2010three,van2010three}, where they substitute the word "excess" by "non-adiabatic" and the word "housekeeping" by  "adiabatic" in the context of non-isothermal environments.
 
\begin{leftbar}

For a  nonequilibrium Markovian stochastic process $X_t$ with arbitrary Markovian generator $\mathcal{L}_t$,  the  fluctuating total entropy production $S^{\rm tot}_t$ given by Eq.~\eqref{Stotcopied}  can be decomposed as the sum of two terms:
 \begin{equation}
 S^{\rm  tot}_t = S^{\rm  ex}_t + S^{\rm  hk}_t,
 \label{OOP}
 \end{equation}
where $S^{\rm ex}_{t}$  and $S^{\rm hk}_t$ are respectively the so-called excess stochastic entropy production and housekeeping stochastic entropy production  which are defined below.  
\end{leftbar}
\begin{leftbar}
 The \textbf{excess stochastic entropy production} $S^{\rm ex}_{t}$ is a  $\mQ$-stochastic entropy production  of the form~\eqref{eq:Qstot} specialized to the  choice  $\mathcal{Q}^{(t)}=\mathcal{P}^{\rm ex,(t) }$ (see~Ref.~\cite{chetrite2018peregrinations} for details),
\begin{eqnarray}
S^{\rm ex}_{t} &\equiv& S_{t}^{\mathcal{P},\mathcal{P}^{\rm ex,(t)}}=\ln\left(\frac{\mathcal{P}\left(X_{\left[0,t\right]}\right)}{\mathcal{P}^{\rm ex,(t)}\left(\Theta _{t}\!\left(X_{\left[0,t\right]}\right)\right)}\right)\nonumber \\
&=&-\ln\left(\frac{\rho_{t}}{\pi_t} (X_t)\right)+\ln\left(\frac{{\rho}_0}{\pi_0} (X_0)\right)-\int_0^t\text{d}s \,(\partial_s \ln\pi_s)  (X_s).
\label{eq:excessentropy}
\end{eqnarray}
Here, the  path probability     $\mathcal{P}^{\rm ex,(t)}$ is  associated with the dynamics generated by "dual" time-reversed generator~\cite{chetrite2018peregrinations} given for all $0\leq s \leq t$ by 
\begin{equation}
\mathcal{L}^{\rm ex,(t)}_{s} \equiv \pi_{t-s}^{-1} \circ \mathcal{L}_{t-s}^{\dagger} \circ\pi_{t-s},
\label{AFS}
\end{equation}
where $\circ$ denotes  here the composition operator, and  $\pi_{t}$ is the so-called \textbf{accompanying density}~\cite{hanggi1982stochastic} which obeys
\begin{equation}
\mathcal{L}_{t}^{\dagger}\pi_{t}=0,\label{eq:acc}
\end{equation}
for all values of time $t\geq 0$.  
\end{leftbar}
Note that $\pi_{t}$ would be the stationary density of the process if the external parameters are constant and equal to those at time $t$.  We give below further remarks and clarifications about the accompanying density, which is not equal to the instantaneous density of the process generating $X_t$. 
 To  further clarify the notation in~\eqref{AFS}, we note that $\mathcal{L}^{\rm ex,(t)}_{s}$ acts on a function $f(x)$  as follows: 
\begin{equation}
(\mathcal{L}^{\rm ex,(t)}_s f)(x)= \frac{(\mathcal{L}_{t-s}^{\dagger} (\pi_{t-s} f) )(x)}{\pi_{t-s}(x)}.
\label{Lex}
\end{equation}

%{\color{blue}Finally, note that for time homogeneous dynamics ($\pi_t=\rho_{st}$) and stationary dynamics  ($\rho_{t}=\rho_{0}=\rho_{st}$) then the excess stochastic entropy production \eqref{eq:excessentropy} vanish.}
%{\color{red} Don't we say this below?} {\color{magenta} true !}. 
\begin{leftbar}
From~\eqref{OOP}, the \textbf{housekeeping stochastic entropy production}  $S^{\rm hk}_t=S^{\rm tot}_t-S^{\rm ex}_t$, is defined as the difference between the total and excess stochastic entropy production, which, after some cumbersome algebra given in Sec.~10.5.2 in~\cite{chetrite2018peregrinations},  can be written for generic Markov processes in the form of a $\Lambda-$entropic functional
%initially expressed as the difference thanks to   between $S^{\rm  tot}$ and $S^{\rm  ex}$,  can then be put on the form (citer Habilitation Chetrite Paragraph 10.5.2) 
 \begin{equation}
 S^{\rm hk}_t = \Lambda_{t}^{\mathcal{P},\mP^{\rm hk}}= \ln \left(\frac{\mP\left(\Xot\right)}{\mP^{\rm hk}\left(\Xot\right)}\right). 
 \label{eq:Sehk}
 \end{equation}
Here,  $\mP^{\rm hk}$ is the path  probability associated  with the same initial density $\rho_0$ and with  the "dual"  Markovian generator given  for all $ s \geq 0$ by  
 \begin{equation}
\mathcal{L}^{\rm hk}_s \equiv \pi_{s}^{-1} \circ \mathcal{L}_{s}^{\dagger} \circ \pi_{s}.
\label{DG}
 \end{equation}
 %\end{itemize}
%A key feature of the \textbf{housekeeping entropy production} contrary to any $\mQ$-stochastic entropy production define in \eqref{eq:Qstot},  is an  $\Lambda$-entropic functional. 
 \end{leftbar}

% {\bf [IN: gray box too big, can we only keep the essential definitions, or cut it into small boxes?   ]} 

Let us now give some important remarks concerning the definition of housekeeping and excess entropy production.
 \begin{itemize}
 
 \item    The fact that, $S^{\rm tot}_t-S^{\rm ex}_t$,  a difference of two $\Sigma-$stochastic entropic functionals, can be expressed as  $S^{\rm hk}_t$,    a $\Lambda-$entropic functional, is a special property that does not hold  in general for arbitrary $\Sigma$-stochastic functionals. 
 
 \item A key insight often overlooked in the literature is that the accompanying density  $\pi_{t}$ given by the solution of Eq.~\eqref{eq:acc} is {\em not} in general a solution of the  Fokker-Planck equation~(\ref{eq:FP}) associated with the dynamics of the process $X_t$, i.e. in general
\begin{equation}
\partial_t \pi_t \neq \mathcal{L}^{\dagger}_t\pi_t,    
\end{equation}
%because $\mathcal{L}^{\dagger}_t\pi_t=0$
%because the left hand side of this equation don't vanish for $\pi_{t}$, otherwise the right hand side vanish. 
On the other hand, $\pi_{t}$ satisfies  $\mathcal{L}_{t}^{\dagger}\pi_{t}=0$ at all times $t$, i.e. it coincides with the stationary density of a process on which  Markov generator would be frozen for at its value at time $t$. In other words,  $\pi_{t}$ is the instantaneous density of the process if and only if the dynamics is either stationary or quasistatic at all times.

\item The average value of the excess entropy production~\eqref{eq:excessentropy}   reads      
\begin{equation}
\langle   S^{\rm ex}_t\rangle = D_{\rm KL} \left[ \rho_0 || \pi_0 \right]-D_{\rm KL} \left[ \rho_t || \pi_t \right] - \int_0^t \text{d}s\, \int_{\mathcal{X}} \text{d}x\,\rho_s(x) (\partial_s\ln\pi_s) (x)\quad.
\label{Sexmean}
\end{equation}
 Choosing $\rho_0=\pi_0$, the positivity of $\langle S^{\rm ex}_t \rangle$ given in (\ref{poslun}),  allows one to derive the result by  Vaikuntanathan and Jarzynski~\cite{vaikuntanathan2009dissipation} $
- \int_0^t \text{d}s\, \int_{\mathcal{X}} \text{d}x\,\rho_s(x) (\partial_s\ln\pi_s) (x) \geq D_{\rm KL} \left[ \rho_t || \pi_t \right]$. 

Moreover, the formulae \eqref{Sexmean} can also be written ~\cite{chetrite2018peregrinations}
\begin{equation} \langle S^{\rm ex}_t\rangle = \int_0^t \text{d}s\, \int_{\mathcal{X}} \text{d}x\,\  \left(\partial_{s}{\rho}_s(x)\right) \left( \ln\frac{\pi_s}{\rho_s}\right)(x).
\label{AFSVendredi}
\end{equation}
Equation~\eqref{AFSVendredi} implies that $\langle S^{\rm ex}_t\rangle$ is close to zero for adiabatic processes, i.e. when $\rho_s\simeq \pi_s$.

\item {\em Physical interpretation of houskeeping and excess entropy production}. Note that if instantaneous detailed balance holds, i.e. $\pi_{s} \circ \mathcal{L}_s \equiv  \mathcal{L}_{s}^{\dagger}\circ \pi_{s}$, then $\mathcal{L}^{\rm hk}_s=\mathcal{L}_s$ and  $S^{\rm hk}_t=0$, see Eq.~\eqref{eq:Sehk}, yielding  $S_t^{\rm tot}=S_t^{\rm ex}$.   This clarifies the adjective "housekeeping"  from the fact that it corresponds to the entropy production  that results from the violation of instantaneous detailed balance, even if the process is stationary. On the other hand, the excess entropy production vanishes on average [see Eq.~\eqref{AFSVendredi}] for stationary processes and otherwise it is non-zero, even when instantaneous detailed balance holds.  

Two important paradigmatic examples are the following: (i)  a nonequilibrium stationary state ($\rho_0=\rho_t=\rho_{\rm st}$) with time-independent driving, one has $\pi_t=\rho_{st}$, which implies $S^{\rm tot}_t=S^{\rm hk}_t$; (ii)  a  non stationary relaxation with instantaneous detailed balance with respect to $\pi$, i.e.  $\pi_{s} \circ \mathcal{L}_s \equiv  \mathcal{L}_{s}^{\dagger}\circ \pi_{s}$,  of a system from an arbitrary initial distribution to a final state, for which one gets $S^{\rm tot}_t=S^{\rm ex}_t$. For most nonequilibrium process however, $S^{\rm hk}_t$ and   $S^{\rm ex}_t$ may both be nonzero fluctuating quantities. 
 \item %Contrary to the case of the total $\Sigma$-stochastic entropic functional  \eqref{AFS2} considered in the previous paragraph, 
 The expression~\eqref{eq:excessentropy} for $S^{\rm ex}_t$   and \eqref{eq:Sehk} for $S^{\rm hk}_t$  are generic for  Markovian processes,  without the need to restrict to pure Jump or  diffusion processes, e.g. it holds also for Markovian stochastic equation with Gaussian and Poissonian white noise. %This is not the case of the the housekeeping entropy production  \eqref{eq:Sehk} for who explicit expression depends of the class of Markovian process. 
 We provide in Appendix~\ref{appendix:6}  alternative explicit expressions for the  excess~\eqref{eq:excessentropy} and housekeeping~\eqref{eq:Sehk} stochastic entropy production when process are restricted to Markov-jump and to  multidimensional Langevin processes.

 %\end{itemize} 

\item Because $S^{\rm ex}_t$ and $S^{\rm hk}_t$ are examples of $\Sigma-$stochastic entropic and $\Lambda-$stochastic entropic functionals respectively, they obey mother fluctuation theorems~\eqref{eq:motherfluctrelation}, which imply Crooks-like~\eqref{eq:crooks2} and Jarzynski-like~\eqref{eq:jarz0} fluctuation relations for both quantities. The latter are given by
\begin{equation}
\langle \exp\left(-S_t^{\rm ex}\right)\rangle=1,\qquad \langle \exp\left(-S_t^{\rm hk}\right)\rangle=1,%_{\mP}=1.
\label{eq:jarz02}
\end{equation}
where the first equality is often known as the Hatano-Sasa relation~\cite{hatano2001steady} (see also~\cite{crooks2000path,hummer2001free}  for previous derivations of similar results)  and the second equality as the integral fluctuation relation for the housekeeping entropy production, which for the case of one-dimensional Langevin equations with additive noise is known as the Speck-Seifert relation~\cite{speck2005integral}.
A corollary of these fluctuation relations is the second laws 
\begin{equation}
    \langle S^{\rm ex}_t\rangle\geq 0,\qquad \langle S^{\rm hk}_t\rangle\geq 0,
    \label{eq:jarz03}
\end{equation}
which hold for arbitrary nonequilibrium processes. The inequality $\langle S^{\rm ex}_t\rangle\geq 0$  has been found to be crucial to define the efficiency of active-matter heat engines~\cite{datta2021second}.  Finally, the relation \eqref{eq:jarz03} together with  the Oono-Paniconi decomposition~\eqref{OOP}  implies  the "refinement" of the second law for $S_t^{\rm tot}$ : 
\begin{equation}
\langle S_t^{\rm tot}\rangle\geq \sup\left(\langle S_t^{\rm ex}\rangle,\langle S_t^{\rm hk}\rangle  \right)\geq0\quad, 
\label{RHSL}
\end{equation} 
which is the main result of this theory.  {Note that other approaches to the Oono-Paniconi decompositions are available even for quantum systems, where e.g. the positivity of the adiabatic entropy is not guaranteed at discrete times~\cite{manzano2018quantum}.}

%\item Due  to the identification of $S^{\rm hk}_t$ and $S^{\rm ex}_t$ as   $\Lambda$-Entropic and $\Sigma$-Entropic functionals,   fluctuation theorems for these quantities follow almost straightforwardly.
%For example, the integral fluctuation theorem~\eqref{eq:jarz0} leads to the Hatano-Sasa~\cite{hatano2001steady} relation~\eqref{eq:jarz02}
%which implies using Jensen's inequality the second law $\langle S_t^{\rm hk}\rangle \geq 0$, cf. Eq.~\eqref{poslun}. This second law together with  the Oono-Paniconi decomposition~\eqref{OOP} imply then the "refinement" of the second law~\eqref{poslun}
%\begin{equation}
%\langle S_t^{\rm tot}\rangle\geq \sup\left(\langle S_t^{\rm ex}\rangle,\langle S_t^{\rm hk}\rangle  \right)\geq0\quad.
%\label{RHSL}
%\end{equation} 
%where the two terms in the sup of the right-hand side  are positive.

 \end{itemize}

%Analogously, the associated \textbf{$\mathcal{P}^{\star}$-stochastic environmental entropy flow \eqref{Senv}} during $[0,t]$ is noted $S^{env,ex}_{t}$ and given by
%\begin{equation} 
%S^{env,ex}_{t}\equiv  S^{\rm env, \mathcal{P},\mathcal{P}^{\star}}_{t} = \ln\left[\pi_{t} (X_t)\right]-\ln\left[\pi_0 (X_0)\right]-\int_0^t\text{d}s \,[\partial_s \ln\left(\pi_s\right)](X_s).
%\label{Senvex}
%\end{equation}
%\end{leftbar}
 %Finally, the associated \textbf{{\color{blue}$\tilde\mP$-stochastic entropy production }} is noted $ S^{ex}_{t}$, call  \textbf{excess stochastic entropy production} and is given by
 %\begin{leftbar}
%\begin{equation}
%S^{ex}_{t} \equiv  S^{ \mathcal{P},\mathcal{P}^{\star}}_{t}=-\ln\left[\frac{\rho_{t}}{\pi_t} (X_t)\right]+\ln\left[\frac{{\rho}_0}{\pi_0} (X_0)\right]-\int_0^t\text{d}s \,[\partial_s \ln\left(\pi_s\right)]  (X_s)\quad.
%\label{eq:Exentropy}
%\end{equation}

%The associated mean entropy flow read \begin{equation} \langle S^{\rm ex}_t\rangle = \int_0^t \text{d}s\, \int_{\Omega} \text{d}x\,\ \left(\partial_{s}{\rho}_s\right) (x)  \ln\left[\pi_s} (x)\right]\quad.
%\label{eq:flowexc}
%\end{equation}
%\end{itemize}
%\end{leftbar}

%\subsection{$\Lambda$-Entropic functionals and  Oono-Paniconi decomposition}

 %The rationale behind the adjective "total" entropy flow comes from this decomposition. The ''house'' is probably a metaphor for steady state. 
 %From Eqs.~\eqref{eq:totalentropy} and~\eqref{eq:excessentropy} we can prove,  using the property $S^{\rm env, hk}_t =  S^{\rm env,tot}_t - S^{\rm env, ex}_t  =    S^{\rm tot}_t  -S^{\rm ex}_t $, that

%\end{leftbar}

\section{Martingale structure of entropic functionals }
\label{sec:6p2}

In this Section,   we %provide a roadmap to this framework by  
identify  martingales  with respect to a physical stochastic process $X_t$ that play an important role in stochastic  thermodynamics.  The martingales that we identify are   exponentials of specific examples of  $\Sigma$-stochastic entropic and $\Lambda-$stochastic entropic functionals, as  introduced in Sec.~\ref{sec:6p1}, multiplied by minus one.   %Note that in general the exponentiated negative $\Sigma$-stochastic entropic and $\Lambda-$stochastic entropic functionals are not martingales. 

For simplicity, we  consider in the proofs of this section that time is discrete, so that~$t \in \mathbb{N}$.     In this case, $\mP(X_{[0,t]})$ and $\mQ^{(t)}(X_{[0,t]})$    are normalised path probabilities.     Nevertheless, the results obtained below are also valid for   the continuous-time setup, which is the usual setup of  stochastic thermodynamics.  

%{\color{blue} To me this section is MUCH BETTER written than 6.1 and one could start the chapter from 6.2 without 6.1. Now, I get the impression 6.1  should be reduced considerably to make this chapter readable, or slaughtered in Ch4, appendix, and a little bit in Ch6.}

%{\color{blue} Another way, we start with 6.2. and whenever a new entropic functional appears we give the definition. It seems more natural to start with the "EASY" ones to go to MEDIUM and HARD later, instead of starting with $\Sigma^{P,Q}$. Because we need (6.63) to define (6.3)!}

 %\subsection{Martingales from path probability ratios without time reversal ($\Lambda$-entropic functional) }
  %\subsection{Exponentiated, negative, $\Lambda$-stochastic entropic functionals  are martingales   if the   reference measure   satisfies  $\mQ^{(t)}=\mQ$  }
  \subsection{When are  exponentiated, negative $\Lambda$-stochastic entropic functionals   exponential martingales? } 
%\textcolor{red}{Change this title ? because it is a little lying }  
  
%{\bf [IN:Simpler title:$\Lambda$-entropic functionals are martingales ]}{Good idea !}

%We begin by considering a generic nonequilibrium  stochastic process describing e.g. the dynamics of a mesoscopic system under stationary or non-stationary driving. We denote by $X_t$ the state of the system (e.g. its position and momentum) at time $t$.  

%Initially, we assume  in this section for simplicity  that   $\mQ^{(t)}=\mQ$, and at the end of this section we relax this assumption. 
\begin{leftbar}
Assume that the path probability $\mQ$  has no  supplemental $t$ dependence, i.e. $\mQ^{(t)}=\mQ$.  It holds then that   the $\Lambda$-stochastic entropic functional
\begin{equation}\Lambda_t^{\mathcal{P},\mathcal{Q}}= \ln [\mathcal{P}\left(X_{\left[0,t\right]}\right) /\mathcal{Q}\left(X_{\left[0,t\right]}\right) ]
\end{equation}
is a {\bf submartingale} and the process $\exp(-\Lambda_{t}^{\mP,\mQ})$ is a {\bf martingale}, both with respect to $\Xot$. In particular, for all $t\geq s\geq 0$ it holds that
\begin{equation}
\Big\langle \exp\left(-\Lambda_{t}^{\mP,\mQ}\right)\Big | X_{[0,s]} \Big\rangle = \exp\left(-\Lambda_{s}^{\mP,\mQ}\right),
\label{eq:mart1index}
\end{equation}
and 
\begin{equation}
\langle\, \Lambda_{t}^{\mP,\mQ} \,|\, X_{[0,s]} \,\rangle \geq  \Lambda_{s}^{\mP,\mQ}.\label{eq:submart1index} \end{equation}
%which holds for any trajectory $X_{[0,s]}$ drawn with normalized probability $\mP_t$ {\bf [IN:  I dont think that is required.  This is conditioning so the probability measure from which $X_{[0,s]}$  is drawn is irrelevant!]}. 
%The averages in Eqs.~\eqref{eq:mart1index} and \eqref{eq:submart1index}  are with  respect to   the probability $\mP$ of the physical process generating the trajectories $X_{[0,t]}$. 
\end{leftbar}
Hence, all   $\Lambda$-stochastic entropic functionals of the form~\eqref{eq:lambdafunct} with the additional condition  $\mQ^{(t)}=\mQ$ increase conditionally with respect to time.
 
For  $\mQ^{(t)}=\mQ$,  the martingale property in~\eqref{eq:mart1index} follows from  a derivation similar to the one presented in  Eq.~(\ref{eq:Der}) of Chapter~\ref{ch:2a}, viz., 
\begin{eqnarray}
\Big\langle \exp\left(-\Lambda_{t}^{\mP,\mQ}\right)\Big |X_{[0,s]} \Big\rangle &=& 
\int \mathcal{D} x_{[s+1,t]}\frac{\mQ(X_{[0,s]}, x_{[s+1,t]})}{\mP(X_{[0,s]}, x_{[s+1,t]})}\mP(x_{[s+1,t]}|X_{[0,s]}) \label{eq:mart1index_p1} \\
&=& \int \mathcal{D} x_{[s+1,t]}\frac{\mQ(X_{[0,s]}, x_{[s+1,t]})}{\mP(X_{[0,s]}, x_{[s+1,t]})}\frac{\mP(X_{[0,s]}, x_{[s+1,t]})}{\mP(X_{[0,s]})}  \label{eq:mart1index_p2} \\
&=& \frac{\int \mathcal{D} x_{[s+1,t]}\mQ(X_{[0,s]}, x_{[s+1,t]})}{\mP(X_{[0,s]})}  \label{eq:mart1index_p3} \\
&=& \frac{\mQ(X_{[0,s]})}{\mP(X_{[0,s]})}=\exp\left(-\Lambda_{s}^{\mP,\mQ}\right) . \label{eq:mart1index_p4} 
\end{eqnarray} 
%{\bf [IN:  There is no theorem, proof, so why square?] }
In Eq.~\eqref{eq:mart1index_p1} we have used the definition~\eqref{eq:lambdafunct} of the $\Lambda$-entropic functional; in  Eq.~\eqref{eq:mart1index_p2} we have used Bayes' theorem; in Eq.~\eqref{eq:mart1index_p3} we have used  the fact that $\mP(X_{[0,s]})$  is independent of $X_{[s+1,t]}$; and in Eq.~\eqref{eq:mart1index_p4} we have marginalised $\mQ$. The marginalization step from Eq.~\eqref{eq:mart1index_p3} to Eq.~\eqref{eq:mart1index_p4} is crucial for the proof of martingality, which in this case follows  immediately from the fact that   $\mQ$ is a path probability, i.e.,
\begin{eqnarray}
\int \mathcal{D} x_{[s+1,t]}\mQ(X_{[0,s]}, x_{[s+1,t]}) &=& \int \text{d}x_{s+1} \cdots \int \text{d}x_t\, \mQ(X_0,\dots ,X_s, x_{s+1},\dots , x_{t}) \nonumber \\
&=& \mQ(X_0,\dots ,X_s) = \mQ(X_{[0,s]})\, . 
\end{eqnarray}
%{\bf [IN: do we need definition symbol in the last step?]}

For path probabilities with supplementary $t$-dependence, denoted by $\mQ^{(t)}$,  martingality requires the marginalisation property (see last step of previous proof): 
\begin{equation}
\int \mathcal{D} x_{[s+1,t]}\mQ^{(t)}(x_{[0,t]})=\mQ^{(s)}(x_{[0,s]}). 
\end{equation}
Relevant examples of (sequences) of path probabilities $\mQ^{(t)}$ that contain a   supplementary $t$-dependence are:  $\mQ=\mP^{\rm ex,(t)}$  with $\mP^{\rm ex,(t)}$  the Markovian path probability associated with the generator Eq.~\eqref{AFS}, and $\mQ=\tilde{\mP}^{(t)}$ with $\tilde{\mP}^{(t)}$ the Markovian path probability associated with the generator Eq.~\eqref{Ltilde}.   Moreover, as we  show in the next paragraph, when the path probability $\mQ$ involves  time-reversal maps $\Theta_t$, then $\mQ$ has a  supplementary $t$-dependence.
%{\bf [IN: what do you mean?  Can you put equation? ]}
%This is because then $\mQ$ has a   supplementary $t$-dependence,  written as $\mQ^{(t)}$.   

Moreover, for any functional $\Lambda_{t}$ that obeys  the following two conditions  it holds that $\exp(-\Lambda_{t})$ is a  martingale: (i) the $\Lambda_{t}$ functional is  additive in time, i.e.,   $\Lambda_{t}=\Lambda_{s}+\Lambda_{[s,t]}$ for any $0\leq s\leq t$ ; and (ii) the $\Lambda_{t}$ functional  obeys the Jarzynski-like equality  $\langle \exp(-\Lambda_{[s,t]}) | X_{[0,s]} \rangle  =1$  for any $t\geq s \geq 0$~\footnote{Here, $\Lambda_{[s,t]}$ should depend  on $X_{[s,t]}$ only. }.
 These two conditions imply that $\langle \exp(-\Lambda_{t}) | X_{[0,s]} \rangle = \exp(-\Lambda_{s})\langle \exp(-\Lambda_{[s,t]}) | X_{[0,s]} \rangle = \exp(-\Lambda_{s})$, and hence $\exp(-\Lambda_{t})$ is a martingale. 
 Note that the additive structure  $\Lambda_{t}=\Lambda_{s}+\Lambda_{[s,t]}$ is not a generic property for  $\Lambda-$stochastic entropic functionals.    However, the additive structure is fulfilled  by  $\Lambda$-stochastic functionals of the form Eq.~(\ref{eq:lambdafunct}) for which both $\mP$ and $\mQ$ are  by path probabilities of  Markovian processes.  As shown below, conditions (i) and (ii) are sufficient but not necessary conditions for  $\exp(-\Lambda_{t})$ to be an exponential martingale.

\subsubsection*{Example: housekeeping entropy production  of a  Markovian processes} 
%{\bf [IN: can we remove 7.2.1.1 ?]}{(Why ? It is a place with physics)}

The  housekeeping entropy production  $S^{\rm hk}_t$, as defined by Eq.~\eqref{eq:Sehk},  is an example of a $\Lambda$-stochastic entropy functional that results from the choice $\mQ=\mP^{\rm hk}$, where $ \mP^{\rm hk}$ the Markovian path probability associated with the "dual"  $t$-independent generator,  defined by Eq.~\eqref{DG}.
\begin{leftbar}
Therefore, Eq.~\eqref{eq:mart1index} implies that $\exp(-S^{\rm hk}_t)$ is a martingale, i.e., 
\begin{equation}
\langle\, \exp(-S^{\rm hk}_t ) \,|\, X_{[0,s]}\, \rangle = \exp(-S^{\rm hk}_s ),
\label{eq:mart_Shk}
\end{equation}
which holds for  any $t\geq s\geq 0$ and any $X_{[0,s]} $. 
\end{leftbar}

\begin{leftbar}
Applying Jensen's inequality to Eq.~\eqref{eq:mart_Shk} we obtain a  {\bf conditional second law} for the {\bf housekeeping entropy production}, viz., 
\begin{equation}
\langle\, S^{\rm hk}_t  \,|\, X_{[0,s]} \,\rangle \geq  S^{\rm hk}_s,
\label{eq:submart_Shk}
\end{equation} for any $t\geq s\geq 0$.  
In other words, $S^{\rm hk}_t$ is a submartingale, and the  housekeeping entropy production is \textit{conditionally increasing} with time.
\end{leftbar}
Specializing Eq.~\eqref{eq:mart_Shk} to $s=0$  and taking the average over the initial state,   we obtain as a corollary the integral fluctuation relation
\begin{eqnarray}
    \langle\, \exp(-S^{\rm hk}_t )  \rangle &=&   \int_{\mathcal{X}}dx_0 \,\rho_0(x_0)\langle\, \exp(-S^{\rm hk}_t ) \,|\, X_0=x_0\, \rangle\nonumber  \\
    &=&   \int_{\mathcal{X}}dx_0 \,\rho_0(x_0) \left\langle \exp(-S^{\rm hk}_0 ) \right\rangle\nonumber\\ 
    &=&  1.  \label{eq:derivFTHSKP}
\end{eqnarray}
The second equality in Eq.~(\ref{eq:derivFTHSKP}) comes from the martingale condition  Eq.~\eqref{eq:mart_Shk}, and the third equality comes from  Eq.~\eqref{eq:mart_Shk} for $t=0$ and  $ S^{\rm hk}_0=0$. 
Similarly, using the submartingale condition~Eq.~\eqref{eq:submart_Shk}, the second-law like inequality $\langle\, S^{\rm hk}_t  \,\rangle  \geq  0$ follows,  which in fact holds for any initial density $\rho_0$.  Further details about the martingale structure of the exponentiated negative housekeeping entropy production can be found in Refs.~\cite{chetrite2019martingale,chun2019universal}.

%\subsection{Martingales from path probability ratios with time reversal ($\Sigma$-entropic functional)}
%\subsection{$^{\spadesuit}$  Exponentiated, negative, $\Sigma$-stochastic entropic functionals are martingales if the reference measure satisfies $\mQ^{(t)}=\mQ^{\rm st}$}
\subsection{$^{\spadesuit}$When are  exponentiated, negative, $\Sigma$-stochastic entropic functionals   exponential martingales?}
\label{eq:martsigmafunct}

%{\bf [IN:Simpler title:$\Sigma$-entropic functionals are in general  not martingales ]}{Good idea !}

%\begin{leftbar}
%\begin{equation}
%\Sigma_{t}^{\mP,\mQ}\equiv\ln\!\left[\frac{\mP_{[0,t]}(X_{[0,t]})}{\mQ_{[0,t]}\!\left(\Theta_{t} X_{[0,t]}\right)}\right]\quad.

%\label{eq:functional422}
%\end{equation}
%Here, the reference measure  $\mQ_{[0,t]}\left(\Theta_{t} X_{\left[0,t\right]}\right)$ is the probability to observe the time-reversed trajectory $\Theta_{t} X_{\left[0,t\right]}$ in another physical process (with path probability $\mQ$),  and $\Theta$ denotes time reversal, with  $(\Theta_{t} (X_{\left[0,t\right]})_s = X_{t-s}$ the value of the time-reversed trajectory at any time  $0 \leq s\leq t$.   
%\end{leftbar}

Contrarily to $\Lambda$-stochastic functionals, it holds that $\exp(-\Sigma^{\mP,\mQ}_t)$ is in general not a martingale even when the path probability $\mQ$  has no  supplemental $t$ dependence.     Indeed, following similar steps as for the $\Lambda$-stochastic entropic functional in the previous section,   we find that the  martingale condition is, in general, not fulfilled:
% Note that in the case of e.g. total entropic functionals~\eqref{AFS2} and of the excess entropy production~\eqref{eq:excessentropy}, the path probability $\mQ$ depends crucially on the time $t$ at which the  time-reversal operation is applied. This fact is present in our formalism. But in fact, to lighten the notation we did not make explicit until now in our notation that e.g. the generators  $\tilde{\mathcal{L}}_s$ (equation \eqref{Ltilde}) and $\mathcal{L}_s^{\rm ex}$ (equation \eqref{AFS}) for $s\leq t$ involve time reversals with respect to time $t$.  This fact will now play an capital role to investigate the Martingale property of $\Sigma^{\mP,\mQ}_t$ with respect to $t$.
\begin{eqnarray}
\hspace{-1cm}\left\langle \exp\left(-\Sigma_{t}^{\mP,\mQ}\right)\Big | X_{[0,s]} \right\rangle &=& 
\int \mathcal{D}x_{[s+1,t]}\frac{\mQ_{[0,t]}^{(t)}(\Theta_{t} (X_{[0,s]},x_{[s+1,t]}))}{\mP_{[0,t]}(X_{[0,s]},x_{[s+1,t]})}\mP(x_{[s+1,t]}|X_{[0,s]}) \label{eq:mart2index_p1} \\
&=& \int \mathcal{D}x_{[s+1,t]}\frac{\mQ_{[0,t]}^{(t)}(\Theta_{t} (X_{[0,s]},x_{[s+1,t]}))}{\mP_{[0,t]}(X_{[0,s]},x_{[s+1,t]})}\frac{\mP_{[0,t]}(X_{[0,s]},x_{[s+1,t]})}{\mP_{[0,s]}(X_{[0,s]})}  \label{eq:mart2index_p2} \\
&=& \frac{\int \mathcal{D}x_{[s+1,t]}\mQ_{[0,t]}^{(t)}(\Theta_{t} (X_{[0,s]},x_{[s+1,t]}))}{\mP_{[0,s]}(X_{[0,s]})}  \label{eq:mart3index_p3} \\
&\neq & \frac{\mQ^{(s)}_{[0,s]}(\Theta_s X_{[0,s]})}{\mP_{[0,s]}(X_{[0,s]})}=\exp\left(-\Sigma_{s}^{\mP,\mQ^{(s)}}\right) . \label{eq:mart2index_p4} 
\end{eqnarray}
Here, the key step is the inequality~\eqref{eq:mart2index_p4}, which can be written  more explicitly as 
\begin{eqnarray}
\int \mathcal{D}x_{[s+1,t]}\mQ_{[0,t]}^{(t)}(\Theta_{t} (X_{[0,s]},x_{[s+1,t]})) &=& \int \text{d}x_{s+1} \cdots \int \text{d}x_t\, \mQ^{(t)}_{[0,t]}(x_t,\dots ,x_{s+1}, X_{s},\dots , X_{0})\nonumber \\
&=& \mQ^{(t)}_{[t-s,t]}(X_s,\dots ,X_0)\\
&=& \mQ^{(t)}_{[t-s,t]}(\Theta_s X_{[0,s]}) \neq \mQ^{(s)}_{[0,s]}(\Theta_s X_{[0,s]}).%\\
%&\neq&  \mQ^{(s)}_{[0,s]}(\Theta_s X_{[0,s]})\equiv \mQ^{(s)}(\Theta_s X_{[0,s]}) .  
\end{eqnarray}  

Note that for bookkeeping purposes, we have used the subindices $[0,t]$, $[0,s]$, and $[t-s,t]$ to denote marginalised path probabilities of $\mP$ and $\mQ^{(t)}$.  For example, $\mP_{[0,t]}\left(x_{(0,t)}\right)$ denotes the marginal of $\mP\left(x_{[0,\infty]}\right)$ for which all variables $x_{[t,\infty]}$ have been integrated out.   Analogously, $\mQ^{(t)}_{[0,s]}\left(x_{[0,s]}\right)$ denote the marignal of $\mQ^{(t)}\left(x_{[0,\infty]}\right)$ for which all variables $x_{[s,\infty]}$ have been integrated out, and so forth.    

Hence, Eqs.~(\ref{eq:mart2index_p1}-\ref{eq:mart2index_p4}) imply that   for general driven nonequilibrium processes 
\begin{equation}
\Big\langle \exp\left(-\Sigma_{t}^{\mP,\mQ}\right)\Big | X_{[0,s]} \Big\rangle =     \exp\left(-\Sigma_{s}^{\mP,\mQ}\right) \frac{\mQ^{(t)}_{[t-s,t]}(\Theta_s X_{[0,s]})}{\mQ^{(s)}_{[0,s]}(\Theta_s X_{[0,s]})} ,
\label{april}
\end{equation}
 any $0\leq s\leq t$.   
 \begin{leftbar}
 Consequently, in general, $\exp\left(-\Sigma_{t}^{\mP,\mQ}\right)$ are {\bf not} martingales, i.e.
\begin{equation}
    \Big\langle \exp\left(-\Sigma_{t}^{\mP,\mQ}\right)
    \Big | X_{[0,s]} \Big\rangle \neq \exp\left(-\Sigma_{s}^{\mP,\mQ}\right).
\end{equation} 
\end{leftbar}

In special cases, the equality  
\begin{equation}
    \mQ^{(t)}_{[t-s,t]}(\Theta_s X_{[0,s]})=\mQ_{[0,s]}^{(s)}(\Theta_{s} X_{[0,s]}) \label{eq:eqSigmaMart}
    \end{equation} 
    required for the martingality of $\exp(-\Sigma^{\mP,\mQ}_t)$, holds.    In particular, Eq.~(\ref{eq:eqSigmaMart})  holds when the following conditions are met: (i) $\mQ^{(t)}$ is independent of $(t)$, i.e., $\mQ^{(t)}=\mQ$; (ii) $\mQ$ is a stationary measure; and (iii) $\mQ$ is time homogeneous, i.e., $\mQ^{(t)}=\mQ^{\rm st}$.   If conditions (i)-(iii) hold, then   $\exp(-\Sigma^{\mP,\mQ}_t)$ is a martingale.    A notable example is  the process $\exp(-S^{\rm tot}_t)$, where $S^{\rm tot}_t$ is the entropy production of  a time-homogeneous, stationary process $X$, as discussed in  Sec.~\ref{sec:martlanggent} (see also below for details).

\begin{leftbar}
Hence, $\exp(-\Sigma^{\mP,\mQ})$ is a {\bf martingale} when  $\mQ^{(t)}=\mQ^{\rm st}$ is  a
$t-$independent, stationary, and  time homogeneous path probability.   In this case, $\mQ^{(t)}_{[t-s,t]}= \mQ^{st}_{[0,s]}$, and the martingale property of $\exp(-\Sigma^{\mP,\mQ}))$ is restored, viz.,    
\begin{equation}
\langle\, \exp(-\Sigma_{t}^{\mP,\mQ^{\rm st}}) \,|\, X_{[0,s]}\, \rangle = \exp(-\Sigma_{s}^{\mP, \mQ^{\rm st}} ),
\label{eq:mart_Stot1}
\end{equation}
for all  $0 \leq s\leq t$.    
\end{leftbar}

\begin{leftbar}
Using Jensen's inequality on Eq.~(\ref{eq:mart_Stot1}), we find  that 
\begin{equation}
 \quad\langle\, \Sigma_{t}^{\mP,\mQ^{\rm st}} \,|\, X_{[0,s]} \,\rangle \geq  \Sigma_{s}^{\mP,\mQ^{\rm st}} ,
 \end{equation}
and hence for   $t-$independent, stationary, and  time homogeneous path probabilities $\mQ^{(t)}=\mQ^{\rm st}$, the process  $\Sigma_{t}^{\mP,\mQ^{\rm st}}$ is a {\bf submartingale}.
\end{leftbar}

Note that  that the martingale property~\eqref{eq:mart_Stot1} does not require that $\mP$ is  stationary and/or Markovian.

If $\mP$ and/or $\mQ$  are nonnormalized, then Eq.~\eqref{eq:mart_Stot1} does not hold due to breaking of marginalization property.   A notable example is  the environmental $\mathcal{Q}$-stochastic entropy change $S^{\rm env, \mathcal{P},\mathcal{Q}}_t$, as defined in Eq.~\eqref{eq:decom}, 
%---which includes  the classic stochastic environmental entropy flow $S_{t}^{\rm env}$ define in 
%\eqref{Senvtot}---, 
for which $\exp(-S^{\rm env, \mathcal{P},\mathcal{Q}}_t)$ is  not  a martingale  (see also Sec.~\ref{NMEC}). 
 This in spite of the fact that, according to  the decomposition  \eqref{eq:decom1}, $S^{\rm env, \mathcal{P},\mathcal{Q}}_t$  is a $\Sigma$-stochastic entropic functional when  $\rho_0(x)=\rho^{\mQ}_{0}(x)=1$.   However, in this case, $\mP$ and $\mQ$ are not normalized, and therefore  $\exp(-S^{\rm env, \mathcal{P},\mathcal{Q}}_t)$ is not a martingale. 

We further discuss two  examples of  $\Sigma$-stochastic entropic functionals that are important for stochastic thermodynamics:
\begin{itemize}
\item For Markovian processes the condition $\mQ^{(t)}=\mQ^{\rm st}$ is  equivalent to the three conditions
 \begin{enumerate}
 \item The family $\mQ^{(t)}$ has no supplementary dependence on the  final time $t$, i.e., $\mQ^{(t)}=\mQ$ for a certain path probability $\mQ$. 
 \item   In addition to Condition 1, the Markovian generator of $\mQ$ is time homogeneous. 
 
 \item  In addition to Condition 1, the initial density of $\mQ$ is the associated stationary density, i.e.,  $\rho^{\mQ}_0=\rho^{\mQ}_{st}=\rho^{\mQ}_t$ for all $t\geq0$. 
  \end{enumerate}
 In one side, conditions (1), (2) and (3) together,   are sufficient conditions for the martingale property~\eqref{eq:mart_Stot1}.   But from another side, in Sec.~\ref{sec:martlanggent}, we have shown that   for stationary, multidimensional Langevin processes $\exp(-S^{\rm tot}_t)$ is a martingale, even when condition (2) does not hold~\footnote{An example of Markov process where we have condition (3) without condition (2) is a general Isothermal Langevin equation : ~\eqref{eq:LE} with Einstein relation \eqref{ER},  without external force $f_t=0$, generic  time-homogeneous potential $V_t=V$, and with the  mobility matrix $\bmu_{t}$  having an explicit time dependence. For this example, the  stationary density is the Gibbs density  $\rho_{\rm st}\sim \exp(-H(x)/T)$,  and if moreover $\rho_{0}=\rho_{st}$, we have (3) without (2).}.   Hence,  Conditions (1-3) are sufficient but not necessary.    An another interesting example is the excess entropy $S^{\rm ex}_{t}$, as defined in~\eqref{eq:excessentropy}, which is also a $\Sigma$-stochastic entropic functional.    In this case, $\exp(-S^{\rm ex}_{t})$ is not a martingale, and $S^{\rm ex}_{t}$ does not satisfy any of the conditions 1, 2 and 3, except in the trivial case where     $S^{\rm ex}_{t}=0$ for all $t$.
 
 %On the other hand, we can note that the  $S^{\rm ex}_{t}$~\eqref{eq:excessentropy} (which is also a $\Sigma$-stochastic entropic functional)  never satisfies conditions  1, 2 and 3,  except in trivial  case $S^{\rm ex}_{t}=0$ for all $t$.  %Because the definition of the dual generator  \eqref{Lex} show that $\mQ={\mP^{ex}}$ is stationary if : (i) $\pi_t=\rho_{st}$, (ii) $\mP$ is time homogeneous and (iii) $\rho_0^{\mQ}=\rho_{t}=\rho_{st}$.  Then, theses relation imply that $\mP$ is stationary and then the excess stochastic entropy production \eqref{eq:excessentropy} vanish.

 \item   In the case where  the path measure $\mQ$ satisfies  the Conditions 1 and 2 of the previous item, and not the Condition 3., i.e.,  when $\mQ$ represents a time-homogeneous system that relaxes to its stationary state, then the bulk term in the ratio $\mQ^{(t)}_{[t-s,t]}/\mQ^{(s)}_{[0,s]}$ cancels out, and the Eq.~\eqref{april}   takes the form 
\begin{equation}
\Big\langle \exp\left(-\Sigma_{t}^{\mP,\mQ}\right)\Big | X_{[0,s]} \Big\rangle =     \exp\left(-\Sigma_{s}^{\mP,\mQ}\right) \frac{\rho_{t-s}^{\mathcal{Q}}}{\rho_{0}^{\mathcal{Q}}} (X_{s}).  
\label{april2}
\end{equation}
In this case it is possible to ''martingalize''the Eq.~\eqref{april2}   by   eliminating the border term as follows, 
 \begin{equation}
\langle\, \exp(-\Sigma_{t}^{\mP,\mQ} -\alpha_t^{\mQ,(t)}) \,|\, X_{[0,s]}\, \rangle =\exp(-\Sigma_{s}^{\mP,\mQ} -\alpha_s^{\mQ, (t)}),
 \label{april10}
 \end{equation}
 for all  $0\leq s \leq t$, and
 where 
 \begin{equation}
     \alpha_s^{\mQ,(t)}=\ln \left(\frac{\rho_{0}^{\mathcal{Q}}(X_{s})}{\rho_{t-s}^{\mathcal{Q}}(X_{s})}\right).
 \end{equation}

%   {??  Note that $  \alpha_t^{\mQ,(t)}=0$.}
%Then, in the set-up, $\exp(-\Sigma_{s}^{\mP,\mQ} -\alpha_s^{\mQ,(t)})$ is a Martingale in $s$. 
\end{itemize}

\subsubsection*{Example of total entropy production  for Markovian processes}
As shown in Sec.~\ref{sec:martlanggent}, for stationary processes  $X_t$ the exponential $\exp\left(-S_{t}^{\rm tot}\right)$ of the total stochastic entropy production   $S_{t}^{\rm tot}$  is a martingale.  
Otherwise,  if $X_t$ (and thus $\mP$) is a non-stationary process,  then Eq.~\eqref{april} for   $\mQ^{(t)} =\tilde{\mP}^{(t)}$, where  $\tilde{\mP}^{(t)}$ is the path probability    associated with a protocol that has been reversed at time $t$,  yields
\begin{equation}
\Big\langle \exp\left(-S_{t}^{\rm tot}\right)\Big | X_{[0,s]} \Big\rangle =     \exp\left(-S_{s}^{\rm tot}\right) \frac{\tilde{\mP}^{(t)}_{[t-s,t]}(\Theta_s X_{[0,s]})}{\tilde{\mP}^{(s)}_{[0,s]}(\Theta_s X_{[0,s]})} . 
\label{april1000}
\end{equation}
Simplifying the ratio $\tilde{\mP}^{(t)}_{[t-s,t]}/\tilde{\mP}^{(s)}_{[0,s]}$ in~Eq.~\eqref{april1000},  we obtain the relation
%\begin{equation}
%\Big\langle \exp\left(-\Sigma_{t}^{tot}\right)\Big | X_{[0,s]} \Big\rangle =     \exp\left(-\Sigma_{s}^{tot}\right) \frac{\rho_{t-s}^{\tilde{\mP}^{(t)}}}{\rho_{0}^{\tilde{\mP}^{(s)}}} (X_{s}).  
%\label{april222}
%\end{equation}
%Now, in the particular case of \textbf{total entropy production} $S_{t}^{\rm tot}$, the relation \eqref{april222}   take with $\rho_{0}^{\tilde{\mP}^{(s)}}}=\rho_{s}$,  the form
\begin{equation}
\Big\langle \exp\left(-S_{t}^{\rm tot}\right)\Big | X_{[0,s]} \Big\rangle =     \exp\left(-S_{s}^{\rm tot}\right) \frac{\rho^{\tilde{\mP}^{(t)}}_{t-s}(X_{s})}{\rho_{s}(X_{s})} ,
\label{april22}
\end{equation}
where $\rho^{\tilde{\mP}^{(t)}}_{t-s}= \tilde{\rho}^{(t)}_{t-s}$ is the instantaneous density at time $t-s$ resulting from the evolution of the initial density  $\rho^{\tilde{\mP}^{(t)}}_{0}=\tilde{\rho}^{(t)}_{0}= \rho_{t}$ by the dynamics with the protocol that has been time-reversed at time $t$.
 This comes from the fact that in this case the initial density of $\tilde{\mP}^{(t)}_{[t-s,t]}$ in \eqref{april1000}  is $\rho^{\tilde{\mP}^{(t)}}_{t-s}= \tilde{\rho}^{(t)}_{t-s}$ and the initial density of $\tilde{\mP}^{(s)}_{[0,s]}$ in \eqref{april1000}  is $\rho^{\tilde{\mP}^{(s)}}_{0}=\tilde{\rho}^{(s)}_{0}=\rho_{s}$, see Fig.~\ref{fig:timereverseddelta} for an illustration. 

\begin{figure}[h]
\centering
\includegraphics[width=0.8\textwidth]{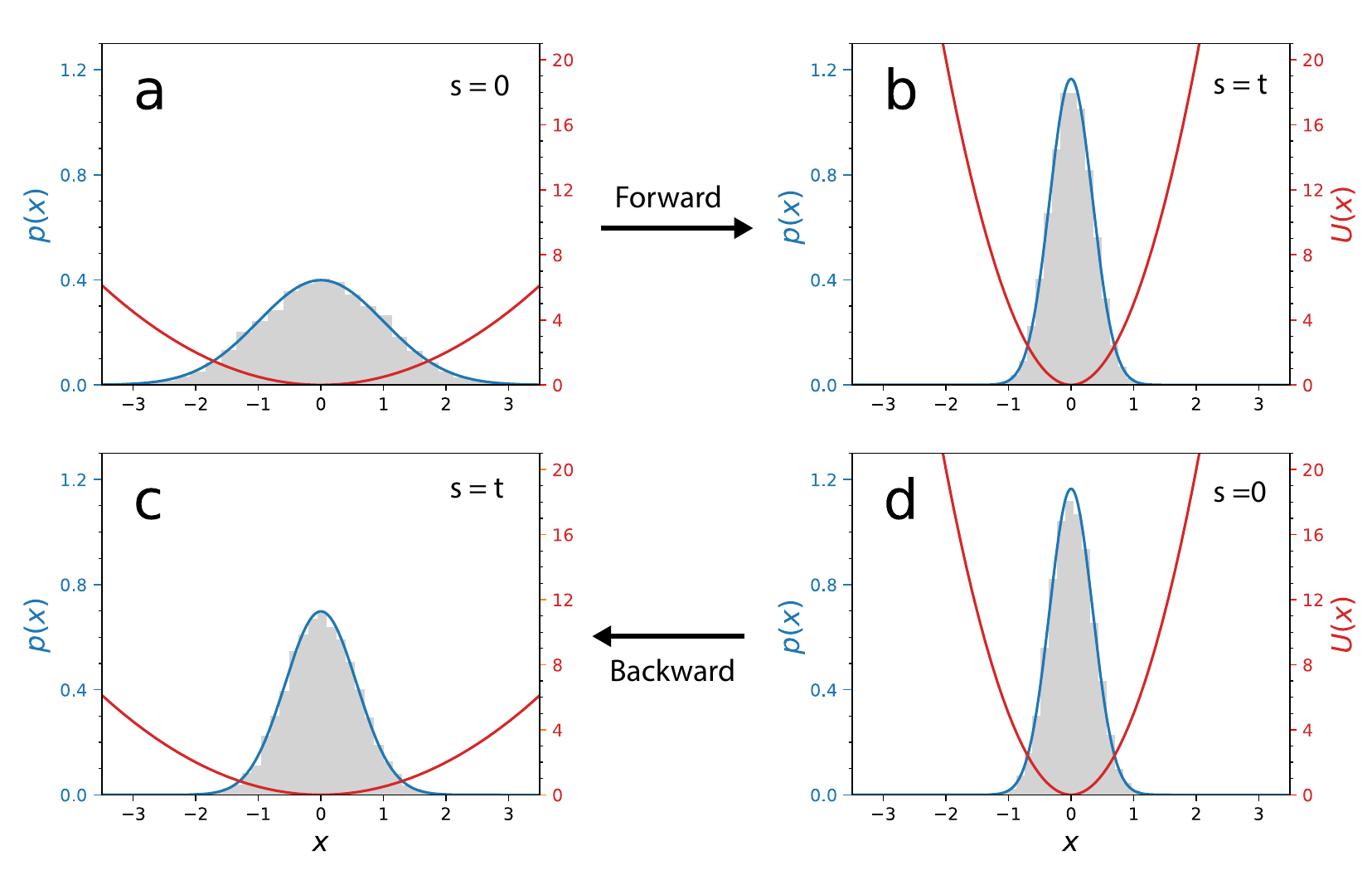}
\caption{ Distributions $\rho(x)$ (blue lines) of the position of a particle $x$ in a time dependent harmonic potential $U(X)$ (red lines), obtained from samples at different times $s=0,t$ (see legends) during a forward (a$\to$b) and a backward (d$\to$c) process, with the latter initialized with the final distribution of the forward process.  The results are obtained for a system described by the Langevin equation $\dot{X}_s = -\mu \kappa_t X_s + \sqrt{2D}\dot{B}_s$, with $\kappa_s = \kappa_0 + rs$ in the forward process, and $\tilde{\kappa}_s= \kappa_{t-s}$ in the backward process. The gray bars are obtained from numerical simulations and the blue line from analytical calculations. 
Values of the parameters: $\mu=10$, $\kappa_0 = 1$, $r=9/t$,    $t=0.05$, $D=1$, simulation time step $5\times 10^{-5}$, $10^4$ realizations. Figure courtesy of Tarek Tohme. }
\label{fig:timereverseddelta}
\end{figure}

%\begin{equation}
%\langle\, \exp(-\Sigma_{t}^{tot} -\alpha_t^{tot,(t)}) \,|\, X_{[0,s]}\, \rangle =\exp(-\Sigma_{s}^{tot} -\alpha_s^{tot,(t)}),
% \label{april1000}
 %\end{equation}
  %with  (Note that $ \alpha_{t}^{(t)}=0$.) 
 %\begin{equation}
  %   \alpha_s^{tot,(t)}=\left(
   %  \ln \frac{\rho_{0}^{\tilde{\mP}^{(s)}}}{\rho_{t-s}^{\tilde{\mP}^{(t)}}}\right)(X_{s}).
    % \label{deltatot}
 %\end{equation}
%And  it holds that 
Equation~\eqref{april22} implies that we  can ''martingalize'' $\exp(-S^{\rm tot}_t)$ in generic nonequilibrium Markovian processes, as we discuss now.

\begin{leftbar}
Indeed, for all $0\leq s\leq t$ it holds that
\begin{equation}
\langle\, \exp(-S_{t}^{\rm tot} -\delta_{t}^{(t)}) \,|\, X_{[0,s]}\, \rangle =\exp(-S_{s}^{\rm tot} -\delta_{s}^{(t)}),
 \label{april100}
 \end{equation}
 with 
 \begin{equation}
     \delta_{s}^{(t)}=
     \ln \left( \frac{\rho_{s}(X_s)}{\tilde{\rho}^{(t)}_{t-s}(X_s)}\right);
     \label{deltatot}
 \end{equation}
notice that $ \delta_{t}^{(t)}=0$.    
\end{leftbar}

The relations~(\ref{april22}-\ref{deltatot}) are extension in general set-up of the martingale integral fluctuation relation~Eq.~\eqref{eq:mart_Stot}.
%Then, in conclusion  $\exp(-S_{s}^{tot} -\delta_{s}^{(t)})$  are Martingale in $s$.
The term $ \delta _s^{(t)}$ is the so-called {\em stochastic distinguishability} between conjugate times in the forward and backward process, and $ \delta _s^{(t)}$ vanishes for (possibly nonequilibrium) stationary states --for which $\rho_s$ and $\tilde{\rho}^{t}_{s}$ are independent on time-- where one recovers the martingale condition~\eqref{eq:mart_Stot}.  For non-stationary states, one has in general $\rho_{s}(x)\neq \tilde{\rho}^{(t)}_{t-s}(x)$ (see e.g. Figs.~\ref{fig:timereverseddelta}(a,d)), and $ \delta _s^{(t)}$ fluctuates in time $s$. 
 See also Ref.~\cite{Chetrite:2011} for the appearance of the stochastic distinguishability, but for the generalized $\Sigma$-stochastic entropic functional introduced in the next section.

Note  that by the tower property of condition expectations  { [see Eq.~\eqref{eq:Tower1}]},  Eq.~\eqref{april22} implies for all $0\leq u\leq s\leq t$ that
 \begin{eqnarray}
 \langle\, \exp(-S_{s}^{\rm tot} -\delta_{s}^{(t)}) \,|\, X_{[0,u]}\, \rangle &=&  \langle\, \langle\, \exp(-S_{t}^{\rm tot} -\delta_{t}^{(t)}) \,|\, X_{[0,s]}\, \rangle \,|\, X_{[0,u]}\, \rangle\nonumber\\
 &=& \langle\, \exp(-S_{t}^{\rm tot} -\delta_{t}^{(t)}) \,|\, X_{[0,u]}\, \rangle\nonumber\\
 &=& \exp(-S_{u}^{\rm tot} -\delta_{u}^{(t)}).
 \label{april100000}
 \end{eqnarray}
Thus we conclude that   $\exp(-S_{s}^{\rm tot} -\delta_{s}^{(t)})$  are Martingales.

Applying Doob's optional stopping theorem (Theorem~\ref{Doob2})  to a stopping time $\mathcal{T}$ with  $\mathcal{T}\leq t$,  we obtain \eqref{april100}  (see  Ref.~\cite{manzano2021thermodynamics} for the original proof)
\begin{equation}
\langle\, \exp(-S_{\mathcal{T}}^{\rm tot} -\delta_{\mathcal{T}}^{(t)})  \rangle =\langle\, \exp(-S_{0}^{\rm tot} -\delta_{0}^{(t)})  \rangle = \int dx \rho_0(x)  \left[\frac{\tilde{\rho}^{(t)}_{t}(x)}{\rho_0 (x)}\right]=\int dx   \tilde{\rho}^{(t)}_{t}(x)=1 
 \label{april1final}. 
 \end{equation}
The third equality uses  that $S^{\rm tot}_0=0$. The last equality is the normalisation of $\tilde{\rho}^{(t)}_{t}(x)$.  
%Finally, by Jenssen inequality on  \eqref{april1final}, we obtain the gambling second law \eqref{eq:secondlaw_stop_nonstat00}.

\section{$^{\spadesuit}$ Generalized $\Sigma$-stochastic entropic functional }\label{sec:623}
%{\bf [IN: This should be 6.3 and not 6.2.3. as it defines somethin completely new, the generalised $\Sigma$-entropic functional]}

As shown in the previous section, a  $\Sigma_{t}^{\mP,\mQ}$ functional may obey an integral fluctuation relation $\langle \exp(-\Sigma_t^{\mP,\mQ})\rangle=1$, even though  $\exp(-\Sigma_t^{\mP,\mQ})$ is not a martingale.  This follows from the "mother" fluctuation relation  Eq.~\eqref{eq:jarz0}; a notable example is when $\Sigma^{\mP,\mQ}_t = S^{\rm ex}_t$, the excess entropy production.   
 To rationalize this fact, and find the lost martingale behind this integral fluctuation relation,   we  introduce in this section the {\it generalized} $\Sigma$-stochastic entropic functionals introduced in Ref.~\cite{Chetrite:2011}.   With these functionals we can disentangle the connection between  integral fluctuation relation and  the    martingality of  a stochastic process.  
 
 \subsection{Definition of  generalized  $\Sigma$-stochastic entropic functionals}
 
Just as was the case for $\Sigma$-stochastic entropic functionals,   {\it generalized}  $\Sigma$-stochastic entropic functionals involve  two path probabilities, viz., the path  probability $\mP$ evaluated on the trajectory $\Xot$, and a second  $\mQ$ evaluated on the time-reversed trajectory $\Theta_t (\Xot)$. The difference between $\Sigma$-stochastic entropic functionals and {\it generalized}  $\Sigma$-stochastic  functionals lies in the fact that  generalised $\Sigma$-stochastic entropic functionals are evaluated over \textbf{subset} intervals $[r,s]\subseteq [0,t]$, as  described below.
 \begin{leftbar}
The \textbf{generalized  $\Sigma$-stochastic entropic functionals} are functions defined on the paths $X_{[r,s]}$ associate with subsets  $[r,s]\subseteq[0,t]$ of the  time interval  $[0,t]$, which is the   time interval to which the time reversal operation  $\Theta_t$ applies.    The generalized  $\Sigma$-stochastic entropic functionals are defined by
\begin{equation}
\Sigma_{[r,s];t}^{\mP,\mQ}\equiv \Sigma_{[r,s];t}^{\mP,\mQ}\left(X_{[r,s]}\right)\equiv\ln\!\left[\frac{\mP_{[r,s]}(X_{[0,t]})}{\mQ^{(t)}_{[t-s,t-r]}\!\left(\Theta_{t} X_{[0,t]}\right)}\right],
\label{GEF}
\end{equation}
with $0\leq r \leq s\leq t$, and where   $\mP_{[r,s]}(X_{[0,t]})$ is the marginal  of $\mP_{[0,t]}(X_{[0,t]})$ defined on the time-window $[r,s]$, and hence $\mP_{[r,s]}(X_{[0,t]})$ depends only   on $X_{[r,s]}$; for discrete time and space, we can write 
    \begin{equation}
\mP_{[r,s]} (x_{[0,t]}) \equiv \mP(X_r=x_r, X_{r+1}=x_{r+1},\dots, X_{s-1}=x_{s-1},  X_{s}=x_s).
\label{eq:PGaf}
\end{equation}
%  So, $\mP_{[r,s]} (x_{[0,t]})$  depends only on  $X_{[r,s]}$ and %could also be denoted more naturally as $\mP_{[r,s]}(X_{[r,s]})$.  
Analogously,  $\mQ^{(t)}_{[t-s,t-r]}\left(\Theta_{t} X_{[0,t]}\right)$  is the marginal of $\mQ^{(t)}_{[0,t]}\left(\Theta_{t} X_{[0,t]}\right)$  on  the time-window $[t-s,t-r]$, and also only depends on $X_{[r,s]}$; for  discrete time and space,
\begin{equation}
\mQ^{(t)}_{[t-s,t-r]} (\Theta_t x_{[0,t]}) \equiv \mQ^{(t)}(X_{t-s}=x_s, X_{t-s+1}=x_{s-1},\dots, X_{t-r-1}=x_{r-1},  X_{t-r}=x_r).
\label{eq:QGaf}
\end{equation} 
%This confirm that it depends also just of  $X_{[r,s]}$; it could also be denoted $\mQ^{(t)}_{[t-s,t-r]}\!\left( X_{[s,r]}\right)$ where $X_{[s,r]}$ is the reversal of $X_{[r,s]}$. 
%{\bf [IN:  why do we introduce the notation $\mP_{[r,s]}(X_{[0,t]})$ and $\mQ^{(t)}_{[t-s,t-r]}(\Theta_t X_{[0,t]})$ when two steps later we say it is indeed confusing?]}
\end{leftbar}
Note that the  $\Sigma$-stochastic entropic functional, given by Eq.~\eqref{eq:lambdafunct}, is a generalised $\Sigma$-stochastic entropic functional of the form Eq.~\eqref{GEF} for the choice $r=0$ and $s=t$ : 
\begin{equation}
\Sigma_{[0,t];t}^{\mP,\mQ}= \Sigma_{t}^{\mP,\mQ}
\label{eq:equivalencefunctionals2}.
\end{equation}

   Also, when $Q^{(t)}=Q_{\rm st}$ is $t$-independent and stationary, then (see p.168 in~\cite{chetrite2018peregrinations}) 
\begin{equation}
\Sigma_{[0,s];t}^{\mP,\mQ_{\rm st}}= \Sigma_{s}^{\mP,\mQ_{\rm st}}
\label{eq:equivalencefunctionals}.
\end{equation}
for all $0 \leq s\leq t$.

%{\bf [IN: I think the point below can be said more shortly and clearly:  the generalised entropic functional is well defined as numerator and denominator  are both be  functions of  $X_{[r,s]}$] }
%We remark that the definition of the generalized $\Sigma-$entropic functional is that the path probabilities have to be defined carefully.  {\bf[ IN:  I dont understand the last sentence??  Isn't this a tautology?  ]}
%To give further insights, we illustrate the form of these path probabilities for the case of discrete time and space. The path probability in the numerator reads explicitly {[\bf IN: why suddenly now small $x$, while in the numerator of the definition we have big $X$?  I understand, but  maybe the reader not...]}and the path probability in the denominator

%Note that here, $x_i$ is the value of the random variable $X_i$. 
The choice of the time window $[t-s,t-r]$ for $\mQ^{(t)}$  leads to   path probabilities in the numerator and denominator of the generalised $\Sigma$-stochastic entropic functional, as given by Eqs.~\eqref{eq:PGaf} and~\eqref{eq:QGaf}, respectively, that are evaluated on the same part of the trajectory $x_{[0,t]}$. Indeed,  if instead we would have used
\begin{equation}
\mQ^{(t)}_{[r,s]} (\Theta_t x_{[0,t]}) = \mQ^{(t)}(X_{r}=x_{t-r+1}, X_{r+1}=x_{t-r},\dots, X_{s-1}=x_{t-s},  X_{s}=x_{t-s+1})\quad,
\end{equation}
then the denominator would not be compatible with 
Eq.~\eqref{eq:PGaf}. 
%{\bf [IN: not comparable is too vague]} {(''Absolutely continous'' is better ?  is also not exactly the good word because here the measure disappear)} 
%Therefore, looking at the interval $[r,s]$ in $\mP$ involves the  same snapshots of the original trajectory that are inspected  in the time-reversed trajectory by looking at the interval $[t-s,t-r]$ in $\mQ$.

Similar to the case of  $\Sigma$-stochastic entropic functionals in Chapter~\ref{sec:6p1}, it holds that: 
\begin{itemize}

\item The {\it generalized}  $\Sigma$-stochastic entropic functionals verify the  duality relation \cite{Chetrite:2011}
\begin{eqnarray}
\Sigma_{[r,s];t}^{\mP,\mQ} \left( \Theta_{t} (X_{[0,t]})\right) &=&-\Sigma_{[t-s,t-r];t}^{\mQ,\mP} (X_{[0,t]}), \label{eq:duality122}
\end{eqnarray}
for all  $0\leq r \leq s\leq t$.

\item The average values with respect to $\mathcal{P}$ of generalized $\Sigma-$stochastic entropic functionals are Kullback-Leibler divergences \cite{Chetrite:2011}, viz.,
\begin{eqnarray}
\big\langle \Sigma_{[r,s];t}^{\mP,\mQ}\big\rangle &=&D_{\rm KL}\left[\mP_{[r,s]}(X_{[0,t]})||\mQ^{(t)}_{[t-s,t-r]}\!\left(\Theta_{t} X_{[0,t]}\right)\right].\label{eq:AER-1222}
\end{eqnarray}
As both $\mP$ and $\mQ$ are normalized path probabilities, the Kullback-Leibler divergence in the right-hand side of Eqs.~(\ref{eq:AER-1222}) is greater or equal than zero, which implies the  \textbf{"second laws"} \cite{Chetrite:2011} 
\begin{equation}
\big\langle \Sigma_{[r,s];t}^{\mP,\mQ}\big\rangle\geq 0,
\label{poslun22}
\end{equation}
for all  $0\leq r \leq s\leq t$.
\end{itemize}

 \subsection{Fluctuation relation for  generalized $\Sigma$-stochastic entropic functionals}
Following similar steps as in Chapter \ref{sec:6p1} for $\Sigma$-stochastic entropic functionals, we derive     fluctuation relations for the generalised $\Sigma$-stochastic entropic functionals, as defined in Eq.\eqref{GEF}.
\begin{leftbar}
 The
\textbf{"mother" fluctuation relation}~\cite{Chetrite:2011} for  arbitrary  functionals $Z[X_{[r,s]}]$ reads
\begin{eqnarray}
\Big\langle\, Z [ \Theta_t \!\left(X_{[r,s]}\right)]\, \Big\rangle_{\mQ^{(t)}}&=&\Big\langle \exp\left(- \Sigma_{[r,s];t}^{\mP,\mQ}\right) Z \left[X_{[r,s]}\right]\Big\rangle, \label{eq:motherfluctrelation22}
\end{eqnarray}
 for all  $0\leq r \leq s\leq t$.
\end{leftbar}

Setting $Z[X_{[r,s]}] = \delta (\Sigma_{[r,s];t}^{\mP,\mQ} - \sigma) $ and using the duality relations \eqref{eq:duality122}, we obtain the  \textbf{generalized Crooks fluctuation relation}~\cite{Chetrite:2011}
%{\bf [IN: this seems like a new concept?   Functional that is a distribution?  Do we need to comment? ]} {(No, it is usual in physics, remember field theory.)} {\bf [IN: okay agreed, maybe a reminder that $\delta$ is a dirac distribution, it only appeared once or twice before?]}
%in the first and second line in Eqs.~\eqref{eq:motherfluctrelation}, respectively, we get  using  the duality relations~\eqref{eq:duality1}-\eqref{eq:duality2}, Eqs.~\eqref{eq:motherfluctrelation}  give the \textbf{generalized Crooks fluctuation relations}~\cite{crooks1999entropy} 
\begin{equation}
\Big\langle\, \delta (\Sigma_{[t-s,t-r];t}^{\mQ,\mP} + \sigma) \Big\rangle_{\mQ^{(t)}}=\exp(-\sigma) \Big\langle\, \delta (\Sigma_{[r,s];t}^{\mP,\mQ} - \sigma) \Big\rangle,
\label{eq:crooks22}
\end{equation} for all  $0\leq r \leq s\leq t$. 
This can also be expressed as
\begin{equation}
\frac{\rho^{\mP}_{\Sigma_{[r,s];t}^{\mP,\mQ}}(\sigma)}{\rho^{\mQ^{(t)}}_{\Sigma_{[t-s,t-r];t}^{\mQ,\mP}}(-\sigma)} = \exp(\sigma ), 
\label{eq:crooks222}
\end{equation} for all  $0\leq r \leq s\leq t$. 

With the choice $Z[X_{[r,s]}]=1$, Eq.~\eqref{eq:motherfluctrelation22} becomes the \textbf{generalized integral fluctuation theorems} given by
\begin{equation}
\left\langle \exp\left(-\Sigma_{[r,s];t}^{\mP,\mQ}\right)\right\rangle=1,
\label{eq:jarz0}
\end{equation} for all  $0\leq r \leq s\leq t$. 
Note that the generalised integral fluctuation relation holds for any (normalised) path probability   $\mQ^{(t)}$ that is  absolutely continuous with respect to $\mP$.

\subsection{Exponentiated, negative, generalized $\Sigma$-stochastic entropic functional are  martingales}

\begin{leftbar}
Exponentiated, negative, generalized $\Sigma-$stochastic entropic functionals $\exp\left(-\Sigma_{[r,s];t}^{\mP,\mQ}\right)$ with  $[r,s]\subseteq [0,t]$ are  \textbf{martingales} with respect to the final time $s$ when $r$ and $t$ are fixed. 
 %{\bf [IN:  What do we mean by martingale with respect to a time?  Martingale is defined with respect to a process not a time]} {(RC : For a two time indexed process $X_{n,m}$, yo must precise with respect to what time you will write the Martingality. But what i mean precicely is in the next equation)} 
Indeed,  it holds that 
\begin{equation}
\Big\langle \exp\left(-\Sigma_{[r,s'];t}^{\mP,\mQ} \right)\Big | X_{[r,s]} \Big\rangle =\exp\left(-\Sigma_{[r,s];t}^{\mP,\mQ}\right), \label{eq:backmart4index_p4a} 
\end{equation}
for  all $0\leq r \leq s\leq s'\leq t$.
Applying Jensen's inequality to Eq.~\eqref{eq:backmart4index_p4a}   we get that $\Sigma_{[r,s];t}^{\mP,\mQ}$ are \textbf{submartingales} with respect to the final time $s$ when $r$ and $t$ are fixed. More precisely, 
\begin{equation}
\langle\, \Sigma_{[r,s'];t}^{\mP,\mQ}  \,|\, X_{[r,s]} \,\rangle \geq  \Sigma_{[r,s];t}^{\mP,\mQ},
\label{eq:forwardsubmartsigmag} 
\end{equation}
for all $0\leq r \leq s\leq s'\leq t$.
%Note that in Eqs () the conditional expectation is done over trajectories which have a {\em future} constraint, as $[s,s']$ comes after $[r,s']$.
\end{leftbar}
Now, we derive Eq.~\eqref{eq:backmart4index_p4a}. 
%This generalization permits to restore the exponential martingale structure as follows.
For all $0\leq r \leq s\leq s'\leq t$, it holds that 
%{ER: I changed this to the new notation but it needs more work}
\begin{eqnarray}
\hspace{-1cm}\lefteqn{\Big\langle \exp\left(-\Sigma_{[r,s'];t}^{\mP,\mQ}\right)\Big | X_{[r,s]} \Big\rangle }&& \nonumber\\
&=& 
\int \mathcal{D}x_{[s+1,s']}\frac{\mQ^{(t)}_{\left[t-s',t-r\right]}(\Theta_{t} \left(X_{[0,s]},x_{[s+1,s']},X_{[s',t]}\right))}{\mP_{\left[r,s'\right]}(X_{[0,s]},x_{[s+1,s']},X_{[s',t]})}\mP_{\left[r,s'\right]}(x_{[s+1,s']}|X_{[r,s]}) 
\label{eq:mart4index_p1} 
\\
&=& 
\int \mathcal{D}x_{[s+1,s']}
\frac{\mQ^{(t)}_{\left[t-s',t-r\right]}(\Theta_{t} \left(X_{[0,s]},x_{[s+1,s']},X_{[s',t]} \right))}
{\mP_{\left[r,s'\right]}(X_{[0,s]},x_{[s+1,s']},X_{[s',t]})}\frac{\mP_{\left[r,s'\right]}(X_{[0,s]},x_{[s+1,s']},X_{[s',t])})}{\mP_{\left[r,s'\right]}(X_{[r,s]})}  \nonumber\\ \label{eq:mart4index_p2} 
\\
&=& \frac{\int \mathcal{D}x_{[s+1,s']}Q^{(t)}_{\left[t-s',t-r\right]}(\Theta_{t} \left(X_{[0,s]},x_{[s+1,s']},X_{[s',t]} \right))}{\mP_{\left[r,s\right]}(X_{[0,t]})}  
\label{eq:mart4index_p3} 
\\
&=& \frac{\mQ^{(t)}_{\left[t-s,t-r\right]}(\Theta_{t} X_{[0,t]})}{\mP_{\left[r,s\right]}(X_{[0,t]})}=\exp\left(-\Sigma_{[r,s];t}^{\mP,\mQ}\right).  
\label{eq:mart4index_p4} 
\end{eqnarray}
The relation \eqref{eq:mart4index_p1}  follows from the fact that  the left-hand side of  Eqs.~(\ref{eq:PGaf}) and~(\ref{eq:QGaf})  are independent of $x_{[0,r-1]}$ and $x_{[s+1,t]}$. We  also use this property to obtain the denominator of the last term of Eq.~\eqref{eq:mart4index_p2}. Then, to obtain  Eq.~\eqref{eq:mart4index_p3}, we use the marginalisation $\mP_{\left[r,s'\right]}(X_{[r,s]})=\mP_{\left[r,s\right]}(X_{[r,s]})$ for all $0\leq s\leq s'$,  and the previous  independence property  to obtain  $\mP_{\left[r,s'\right]}(X_{[r,s]})=\mP_{\left[r,s\right]}(X_{[r,s]})=\mP_{\left[r,s\right]}(X_{[0,t]})$. Finally,
the first equality in~\eqref{eq:mart4index_p4} follows from the integration of~\eqref{eq:QGaf} which yields
\begin{eqnarray}
\int \mathcal{D}x_{[s+1,s']}Q^{(t)}_{\left[t-s',t-r\right]}(\Theta_{t} x_{[0,t]}) &=&  \mQ^{(t)}(X_{t-s}=x_{s}, X_{t-s+1}=x_{s-1},\dots, X_{t-r}=x_r)  \nonumber\\ 
&\equiv& \mQ^{(t)}_{\left[t-s,t-r\right]}(\Theta_{t} x_{[0,t]}).
\end{eqnarray}

It may appear surprising that the quantity $\exp(-\Sigma_{[r,s];t}^{\mP,\mQ^{(t)}})$, which is a martingale with respect to the final time $s$, contains as a particular case (\eqref{eq:equivalencefunctionals2}) the exponentials of $\Sigma$-stochastic entropic functionals $\exp(-\Sigma_{t}^{\mP,\mQ^{(t)}})$,  that are  not  martingales. This comes from the fact that by choosing $r=0, s'=t$ the forward martingale property \eqref{eq:backmart4index_p4a}   becomes, for all $0\leq s\leq t$,
\begin{equation}
\Big\langle \exp\left(-\Sigma_{t}^{\mP,\mQ}\right)\Big |X_{[0,s]} \Big\rangle=
\exp\left(-\Sigma_{[0,s];t}^{\mP,\mQ}\right)\neq  \exp\left(-\Sigma_{s}^{\mP,\mQ}\right),
\end{equation}
except for $t$ independent and  stationary $\mQ$, when we have the relation~\eqref{eq:equivalencefunctionals}.

%At the beginning of this Review we have defined  the   martingale property forwards in time (i.e. conditioned on past observations).  However in probability theory it is also common to discuss {\em backward martingales}, which are defined with respect to conditioning  on trajectories in the future. 
Moreover, in \cite{Chetrite:2011, chetrite2018peregrinations}, it is shown that the generalized $\Sigma-$stochastic entropic functionals $\Sigma_{[r,s];t}^{\mP,\mQ}$ do not only have an exponential martingale structure as a function of the final time $s$  when conditioning  over the past, but they also have a backward martingale structure as a function of the initial time $r$ when conditioning on the future.
\begin{leftbar}
 The exponentiated, negative, generalized $\Sigma-$stochastic entropic functionals $\exp\left(-\Sigma_{[r,s];t}^{\mP,\mQ}\right)$ with $[r,s]\subseteq [0,t]$ are \textbf{backward martingales} with respect to the initial time $r$ when $s$ and $t$ are fixed.   Indeed,  it holds that~\cite{chetrite2018peregrinations}
\begin{equation}
\Big\langle \exp\left(-\Sigma_{[r,s];t}^{\mP,\mQ}\right)\Big | X_{[r',s]} \Big\rangle =\exp\left(-\Sigma_{[r',s];t}^{\mP,\mQ}\right), \label{eq:backmart4index_p4} 
\end{equation}
for all  $0\leq r \leq r'\leq s\leq t$.
Applying Jensen's inequality to Eq.~\eqref{eq:backmart4index_p4}, we find that $\Sigma_{[r,s];t}^{\mP,\mQ}$ are \textbf{backward submartingales} with respect to the initial time $r$ when $s$ and $t$ are fixed.  In particular, 
\begin{equation}
\langle\, \Sigma_{[r,s];t}^{\mP,\mQ}  \,|\, X_{[r',s]} \,\rangle \geq  \Sigma_{[r',s];t}^{\mP,\mQ},
\label{eq:superbackmart4index_p4} 
\end{equation}
for all  $0\leq r \leq r'\leq s\leq t$.
%This means that $ \Sigma_{[r,s]}^{\mP,\mQ}$ is conditionally increasing with respect to $s$ and conditionally decreasing with respect to $r$.
\end{leftbar}
Note that in Eqs.~(\ref{eq:backmart4index_p4}-\ref{eq:superbackmart4index_p4})  the conditional expectation is done over trajectories which have a {\em future} constraint, as {$[r',s]$ comes after $[r,r']$}.  In other words, the generalized $\Sigma-$stochastic entropic functionals  \textbf{conditionally increase backwards in time} when looking at the initial time of the scanned interval $[r,s]$. As we will show below in Sec.~\ref{sec:historical}, the backward martingale  structure of $\exp\left(-\Sigma_{[r,s];t}^{\mP,\mQ}\right)$  is instrumental to recover some traditional formulations of the second law of thermodynamics and derive also new universal principles.

Taken all together, we conclude that the generalized $\Sigma-$entropic functionals on $[r,s]\subseteq[0,t]$ have a "two-faced" martingale structure. They are {\em forward submartingales} with respect to the final time $s$ and {\em backward submartingales} with respect to the initial time $r$. In other words,
 $ \Sigma_{[r,s]}^{\mP,\mQ}$  conditionally increases with respect to $s$ and conditionally decreases with respect to $r$.

\subsection{Generalized $\Sigma$-stochastic entropic functional  for Markovian processes }
\label{nischapter}

%\begin{leftbar}
We discuss   generalized $\Sigma$-stochastic entropic functionals for Markovian processes.    The Markov property implies: 
\begin{itemize}\item  First, a decomposition of the  generalized $\Sigma$-stochastic entropic functional in terms of  the environmental  $\mQ$-stochastic entropy change, as defined in~\eqref{eq:decom1}, and a boundary term : 
\begin{equation}
\Sigma_{\left[0,s\right],t}^{\mathcal{P},\mathcal{Q}}=\ln\left(\frac{\rho_{0}\left(X_{0}\right)}{\rho_{t-s}^{Q^{(t)}}\left(X_{s}\right)}\right)+
S_{s}^{\rm{env},\mathcal{P},\widehat{\mQ}^{(t,s)}},
\label{sep2}
\end{equation}
for  all $0 \leq s\leq t$.  In this relation, the environment entropy change $S_{s}^{\rm{env},\mathcal{P},\widehat{\mQ}^{(t,s)}}$ is ~\eqref{eq:decom1}
\begin{equation}
S_{s}^{{\rm env},\mathcal{P},\widehat{\mQ}^{(t,s)}}= \ln\left(\frac{\mathcal{P}_{\left[0,s\right]}\left(\left.X_{\left[0,s\right]}\right|X_{0}\right)}{\left[\widehat{\mQ}^{(t,s)}\right]_{\left[0,s\right]}\left(\left.\Theta_{s}X_{\left[\text{0},s\right]}\right|X_{s}\right)}\right),
\end{equation}
with the path probability $\widehat{\mQ}^{(t,s)}$ is defined by iterating the   reversed protocol, see Eq.~\eqref{Ltilde}, twice, viz,
\begin{equation}
\widehat{\mQ}^{(t,s)}\equiv\widetilde{\widetilde{\mQ^{(t)}}^{(t)}}^{(s)},
\label{foot2}
\end{equation}
{where we recall that in  Sec.~\ref{sec:6p1} we defined the measure $\widetilde{Q}^{(t)}$ as  time-reversed protocol of the path measure $Q$ with respect to the reference time $t$.}
This apparently-complicated object $\widehat{\mQ}^{(t,s)}$ is in fact the path probability of a   Markovian process with  generator
\begin{equation}
\left(\mathcal{L}^{\widehat{\mQ}^{(t,s)}}\right)_{u}=\left(\mathcal{L}^{\left(\widetilde{Q}^{(t)}\right)}\right)_{s-u}=\mathcal{L}_{t-s+u}, 
\label{eq:nov}
\end{equation} for all $0\leq u\leq s \leq t$.  
In other words,  the  iteration of two reversed protocols is just a time translation. 

The relation \eqref{sep2} follows from the equality
\begin{equation}
\mathcal{Q}^{(t)}_{\left[t-s,t\right]}\left(\Theta_{t}X_{\left[0,t\right]}\right) =\rho_{t-s}^{Q^{(t)}}\left(X_{s}\right)\left[
\widehat{\mQ}^{(t,s)}\right]_{\left[0,s\right]}\left(\left.\Theta_{s}X_{\left[\text{0},s\right]}\right|X_{s}\right), 
\label{Nis-Nis}
\end{equation}  which holds for all $0\leq r \leq s\leq t$. We advice  readers to prove the  relation~\eqref{Nis-Nis} for Langevin systems with additive noise by using the Lagrangian given by Eq.~\eqref{eq:LagrangianFormula}.

\item Second, the factorisation of the path probability   resulting from Markov property, permits to obtain  for all $0\leq r \leq s\leq t$ the decomposition formulae of generalized $\Sigma$-stochastic entropic functional\footnote{For all $0\leq r \leq s\leq t$, we have the  decomposition of the Markovian path probabilities 
\begin{equation}
\mathcal{P}_{\left[0,s\right]}\left(X_{\left[0,t\right]}\right)=\frac{\mathcal{P}_{\left[0,r\right]}\left(X_{\left[0,t\right]}\right)\mathcal{P}_{\left[r,s\right]}\left(X_{\left[0,t\right]}\right)}{\rho_{r}\left(X_{r}\right)},
\end{equation} and
\begin{equation}
\mathcal{Q}_{\left[t-s,t\right]}^{(t)}\left(\Theta_{t}X_{\left[0,t\right]}\right)=\frac{\mathcal{Q}_{\left[t-s,t-r\right]}^{(t)}\left(\Theta_{t}X_{\left[0,t\right]}\right)\mathcal{Q}_{\left[t-r,t\right]}^{(t)}\left(\Theta_{t}X_{\left[0,t\right]}\right)}{\rho_{t-r}^{Q^{(t)}}\left(X_{r}\right)}.
\end{equation}}:  
\begin{equation}
\Sigma_{\left[r,s\right],t}^{\mathcal{P},\mathcal{Q}}
  =\ln\left(\frac{\rho_{r}\left(X_{r}\right)}{\rho_{t-r}^{Q^{(t)}}\left(X_{r}\right)}\right)+\Sigma_{\left[0,s\right],t}^{\mathcal{P},\mathcal{Q}}-\Sigma_{\left[0,r\right],t}^{\mathcal{P},\mathcal{Q}}.
  \label{sep}
\end{equation}

\end{itemize}
\begin{leftbar}
Combining Eq.~\eqref{sep2} and Eq.~\eqref{sep}, we obtain the general formulae 
\begin{equation}
\ensuremath{\Sigma_{\left[r,s\right],t}^{\mathcal{P},\mathcal{Q}}}=\ensuremath{\ln\left(\frac{\rho_{r}\left(X_{r}\right)}{\rho_{t-s}^{\mQ^{(t)}}\left(X_{s}\right)}\right)}+\ensuremath{S_{s}^{env,\mathcal{\mP},\widehat{\mQ}^{(t,s)}}}-\ensuremath{S_{r}^{env,\mathcal{\mP},\widehat{\mQ}^{(t,r)}}}. 
\label{beau}
\end{equation} 
Moreover,  using the decomposition~\eqref{eq:decom1} of the $\mQ$-stochastic entropy production, the relation~\eqref{beau} can also be written   as a $\mQ$-stochastic entropy production 
\begin{equation}
\ensuremath{\Sigma_{\left[r,s\right],t}^{\mathcal{P},\mathcal{Q}}}=\ensuremath{\ln\left(\frac{\rho_{s}\left(X_{s}\right)}{\rho_{t-s}^{\mQ^{(t)}}\left(X_{s}\right)}\right)}+\ensuremath{S_{s}^{\mathcal{\mP},\widehat{\mQ}^{(t,s)}}}-\ensuremath{S_{r}^{\mathcal{\mP},\widehat{\mQ}^{(t,r)}}}. 
\label{beauE}
\end{equation} 
Equations~\eqref{beau} and~\eqref{beauE} provide a interpretation of the generalized $\Sigma$-stochastic entropic functional  for Markovian processes. Moreover, the forward and backward martingale property of $\Sigma_{[r,s],t}^{\mathcal{P},\mathcal{Q}}$ proven in this chapter, implies that the right hand side of~\eqref{beau} and~\eqref{beauE} have the same martingale structure.
\end{leftbar}

\begin{leftbar}
Specializing  the relation \eqref{foot}  to the particular case $\mQ^{(t)} =\widetilde{\mP}^{(t)}$ gives that $\widehat{\mQ}^{(t,s)}=\widetilde{\mP}^{(s)}$ and $\widehat{\mQ}^{(t,r)}=\widetilde{\mP}^{(r)}$, and then Eq.~\eqref{beau} yields
\begin{equation}
\ensuremath{\Sigma_{\left[r,s\right],t}^{\mathcal{\mP},\widetilde{\mP}^{(t)}}=
\underbrace{
\ensuremath{\ln\left(\frac{\rho_{r}\left(X_{r}\right)}{\tilde{\rho}^{(t)}_{t-s}\left(X_{s}\right)}}\right)}}_{\displaystyle\equiv \alpha_{r,s}^{(t)}}+\ensuremath{S_{s}^{\rm env}}-\ensuremath{S_{r}^{\rm env}},
\label{beau2}
\end{equation} 
   for all  $0\leq r \leq s\leq t$, where $S_{s}^{env}$ is the environment entropy change defined in Eq.~\eqref{Senvtot}, and where 
\begin{equation}
\alpha_{r,s}^{(t)}\equiv \ln\left(\frac{\rho_{r}\left(X_{r}\right)}{\tilde{\rho}^{(t)}_{t-s}\left(X_{s}\right)}\right).
\label{Nis-fat}
\end{equation}
The  martingale property of $\exp(-\ensuremath{\Sigma_{\left[r,s\right],t}^{\mathcal{\mP},\widetilde{\mP}^{(t)}}})$ allows us to retrieve the theory of Ref.~\cite{neri2020second} (see  Ch.~\ref{sec:Jarzysnk}) within the general context of generalized $\Sigma$-stochastic entropic functionals.  The generalized integral fluctuation relations ~\eqref{eq:jarz0} read here
\begin{equation}
\left\langle \exp\left(- \ensuremath{S_{s}^{\rm env}}+\ensuremath{S_{r}^{\rm env}}-\alpha_{s,r}^{(t)}\right)\right\rangle=1,
\label{eq:jarz022}
\end{equation} for all  $0\leq r \leq s\leq t$.
\end{leftbar}

\begin{leftbar}
Lastly, by using the decomposition \eqref{Stotcopied} of total entropy production, the relation \eqref{beau2} can also be written   for all  $0\leq r \leq s\leq t$.
\begin{equation}
\Sigma_{[r,s];t}^{\mP,\widetilde{\mP}^{(t)}}
=\underbrace{\ln\left(\frac{\rho_s(X_s)}{\tilde{\rho}^{(t)}_{t-s}(X_s)}\right)}_{\delta_{s}^{(t)}}
+
\displaystyle S^{\rm tot}_s-S^{\rm tot}_r,  
\label{forGirs-1-1aa11}
\end{equation}
where the stochastic distinguishability $\delta_{s}^{(t)}$ was  define in relation \eqref{deltatot}. 
This time, the induced  martingality property of the right hands sides  allows  to retrieve  the results~\eqref{april100} and~\eqref{deltatot} (see also  Ref.~\cite{manzano2021thermodynamics}). Moreover, the generalized integral fluctuation theorems~\eqref{eq:jarz0} become here 
\begin{equation}
\left\langle \exp\left(- S^{\rm tot}_s+S^{\rm tot}_r- \delta_{s}^{(t)}\right)\right\rangle=1,
\label{eq:jarz022}
\end{equation} for all  $0\leq r \leq s\leq t$.
\end{leftbar}

%\item In another hand, by choosing the particular case $\mQ^{(t)} =\mP^{\rm ex,(t)}}$ where $\mP^{\rm ex,(t)}$ is the markovian path measure associated to the generator \eqref{AFS},  the relation \eqref{foot} give that $\widehat{\mQ}^{(t,s)}}=\widetilde{\mP^{\rm hk}}^{(s)}$   and  $\widehat{\mQ}^{(t,r)}}=\widetilde{\mP^{\rm hk}}^{(r)}$, and then \eqref{beauE} give 
%\begin{equation}
%\ensuremath{\Sigma_{\left[r,s\right],t}^{\mathcal{P},\widetilde{\mP}^{\rm ex,(t)}}=\ensuremath{\ln\left(\frac{\rho_{s}\left(X_{s}\right)}{\rho_{t-s}^{\mQ^{}}\left(X_{s}\right)}\right)}+\ensuremath{S_{s}^{\rm ex}}-\ensuremath{S_{r}^{\rm ex}}}. 
\label{beauEE}
%\end{equation} 
%The induced  martingality property of the right hand side  permits then to refind the result \eqref{april100},\eqref{deltatot} (which come from  Ref.~\cite{manzano2021thermodynamics}).  

To give examples, the relation~\eqref{beau2} for the case of multidimensional Langevin process described by      Eq.~\eqref{eq:LE} reads
\begin{equation}
\ensuremath{\Sigma_{\left[r,s\right],t}^{\mathcal{\mP},\widetilde{\mP}^{(t)}}=\ensuremath{\underbrace{
\ensuremath{\ln\left(\frac{\rho_{r}\left(X_{r}\right)}{\tilde{\rho}^{(t)}_{t-s}\left(X_{s}\right)}\right)}}_{ \displaystyle\alpha_{r,s}^{(t)}}}+\ensuremath{\underbrace{
\int_{r}^{s} 
\Big(\left(\bmu_u F_{u}\right)\textbf{D}_{u}^{-1}\Big)(X_{u}) \circ \dot{X}_{u}du}_{\displaystyle S^{\rm env}_s-S^{\rm env}_r} }}.
\label{beau3}
\end{equation}
In Sec.~\ref{sec:Jarzysnk}, we will give another proof of the martingale property of $\exp(-\ensuremath{\Sigma_{\left[r,s\right],t}^{\mathcal{\mP},\widetilde{\mP}^{(t)}})$ in this setup. For general jump processes,  the relation~\eqref{beau2} becomes 
\begin{equation}
\Sigma_{[r,s];t}^{\mP,\widetilde{\mP}^{(t)}}
= \ensuremath{\underbrace{
\ensuremath{\ln\left(\frac{\rho_{r}\left(X_{r}\right)}{\tilde{\rho}^{(t)}_{t-s}(X_{s})}}\right)}}_{ \displaystyle\alpha_{r,s}^{(t)}}}
+
\underbrace{
\displaystyle\sum_{j \vert  r \leq \T_j \leq s  }\ln\left(\frac{\omega_{\T_j}(X_{\T_j^{-}},X_{\T_j^{+}})}{\omega_{\T_j}(X_{\T_j^{+}},X_{\T_j^{-}})}\right)\quad}_{\displaystyle S^{\rm env}_s-S^{\rm env}_r}.
\label{forGirs-1-1aaa}
\end{equation}

%\chapter[Martingales in stochastic thermodynamics III]
\chapter{Martingales in stochastic thermodynamics III: %Implications for physics of 
Stationary states}

\label{sec:martST2}

\epigraph{\textit{As far as we know today, there is no automatic, permanently effective perpetual motion machine, in spite of the molecular fluctuations, but such a device might, perhaps, function regularly if it were appropriately operated by intelligent beings..}\\ Smoluchowski, Vortr\"{a}ge \"{u}ber die kinetische Theorie der Materie u. Elektrizitat, (1914, p.89).}

In this Chapter, we show how several  classical results of stochastic thermodynamics can be significantly improved  with  martingale theory.   In particular, we derive more general versions of the  second law of thermodynamics and   fluctuation relations.  Moreover,    using the powerful technology of martingales, as discussed in Chapters~\ref{ch:2a}, \ref{ch:2b}, and \ref{ch:3}, we exactly describe  certain fluctuation properties  of entropy production, notably, for  their infima, first-passage times, and splitting probabilities.   Lastly, we discuss how these results can be used to (apparently) overcome  classical thermodynamic limits  by cleverly exploiting the fluctuations in a stochastic process.  

\section{Setup: nonequilibrium stationary states}
 Throughout this Chapter, we focus on time-homogeneous, stationary processes.   Figure~\ref{fig:NESSmodel} depicts two paradigmatic examples of  such processes.     Figure~\ref{fig:NESSmodel}(a)  shows a Brownian particle that  moves in a periodic potential under the action of a constant, non-conservative force.    The non-conservative force  induces a net current along the ring, which results in a net dissipation of heat to the environment.  {Since the process is stationary}, we assume that the initial distribution of the system is given by  its nonequilibrium, stationary distribution.
\begin{figure}[H]
    \centering
    \includegraphics[width=\textwidth]{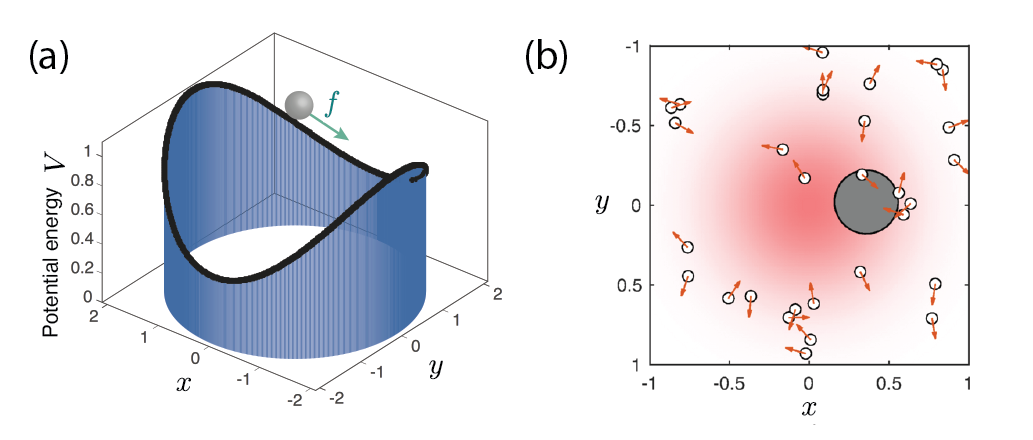} \label{fig:NESSmodel}
    \caption{Panel (a): Illustration of a  paradigmatic, single-particle  model for a nonequilibrium, time-homogeneous, stationary process. A Brownian particle (gray sphere)   moves on top of a "rollercoaster" potential under the action of an external, constant, force $f$. Figure adapted from Ref.~\cite{neri2017statistics}.  For the special case of a flat potential, this model is equivalent to a driven particle on a ring, as illustrated in Fig.~\ref{fig:ring}. Panel (b): Illustration of a many-particle model for a  nonequilibrium, time-homogeneous, stationary process. An overdamped Brownian particle  (gray sphere) immersed in a fluid  with periodic boundary conditions is trapped within a potential (red). The particle interacts with an ensemble of $N>1$ overdamped "active" Brownian particles (white circles) that are self-propelled in randomly-varying directions (red arrows); see Ref.~\cite{datta2021second} for a study of thermodynamics in active matter systems.   }
\end{figure}
Further examples of physical systems belonging to this class are, e.g., systems described by multidimensional Langevin equations and stationary Markov-jump processes.  See Fig.~\ref{fig:NESSmodel}(b) for a many-particle example relevant in the study of active matter systems~\cite{datta2021second}, and {the inset in Fig.~\ref{fig:ET2}(a)} for sketches of some other multidimensional overdamped Langevin models.

The general philosophy of the Chapter {goes} follows:  we assume   from the get-go that $X$ is a stationary, stochastic process for which  the exponentiated, negative, total entropy production  during $[0,t]$   takes the form \eqref{Stotcopied}
\begin{equation}
    \emStot= \frac{(\mathcal{P}\circ\Theta_t)(X_{[0,t]})}{\mathcal{P}(X_{[0,t]})} . \label{eq:EspSM}
\end{equation} 
Consequently,  as shown in Chapters~\ref{sec:martST1} and \ref{ch:thermo2},  $  \emStot$ is  a martingale (because of stationarity), which is a  fundamental fact in nonequilibrium thermodynamics.   Subsequently, we derive various results based on the martingality of $  \emStot$.

Notice  that  Eq.~(\ref{eq:EspSM}) could also describe the thermodynamics of   active matter systems, as long as $X$ describes the trajectories of  all degrees of freedom that are driven out of equilibrium (in the example of Panel(b) in Fig.~\ref{fig:NESSmodel} this involves the dynamics of both the gray and white spheres).  

We start this Chapter with Sec.~\ref{sec:conv} that summarises    results in  conventional stochastic thermodynamics, and which forms a useful point of reference  for the more general  results that follow from  martingale theory and are derived  in the later sections of this Chapter.  
 Subsequently, in Sec.~\ref{sec:marSecondFr}, following Refs.~\cite{Chetrite:2011,neri2017statistics, chetrite2019martingale,chetrite2018peregrinations, neri2019integral}, we review some of the central results from martingale theory for thermodynamics, namely, the martingale versions of the fluctuation relations and  the ensuing  versions of the second law of thermodynamics.     In Sec.~\ref{sec:stop7}, we  review  results on splitting probabilities, and the statistics of first-passage times, and extreme values of entropy production, taken mainly from Refs.~\cite{neri2017statistics, neri2019integral}.   Next  we  review  thermodynamic bounds  on first-passage times of dissipative currents, taken from Refs.~\cite{gingrich2017fundamental, roldan2015decision, neri2022universal, neri2022estimating, neri2022extreme}.       The last Sec.~\ref{sec:overcoming} discusses an application, namely how to overcome classical limits on thermodynamic processes by stopping a stochastic process at a cleverly chosen moment~\cite{neri2019integral}.

\section{Conventional  fluctuation relations}\label{sec:conv}

Fluctuation relations are mathematical relations that constrain the statistics of stochastic thermodynamic quantities.  These results were introduced in the 1990s and are also referred to as fluctuation theorems, see~Refs.~\cite{seifert2012stochastic,Sekimoto1997,peliti2021stochastic,gallavotti1995dynamical,lebowitz1999gallavotti,kurchan1998fluctuation,van2010three} for some classical references.  Here we review some celebrated fluctuation relations that are generic for time-homogeneous, nonequilibrium, stationary states.

The detailed fluctuation relation, 
\begin{equation}\label{eq:dft2x}
  \frac{\rho_{S^{\rm tot}_t}(s)}{{\rho}_{S^{\rm tot}_t}(-s)}=\exp(s).  
\end{equation}
states that in a stationary process  the probability density of the stochastic entropy production evaluated at $S^{\rm tot}_t=s>0$  is exponentially larger than the probability density evaluated at $S^{\rm tot}=-s<0$; note that this is a special case of  Eq.~\eqref{eq:crooks2} valid for the total, stochastic, entropy production of a nonequilibrium stationary state.   
   The mathematical derivation of fluctuation relations can be found in Ch.~\ref{ch:thermo2} of this {Treatise},   see Eq.~\eqref{eq:crooks2} with \eqref{AFS2} and \eqref{Stotcopied}.

From~Eq.~\eqref{eq:dft2x}   follows the integral fluctuation relation 
\begin{equation}\label{eq:iftxx}
    \langle\exp(-S^{\rm tot}_t) \rangle = \int_{-\infty}^{\infty} ds \rho_{S^{\rm tot}_t}(s)\exp(-s) = \int_{-\infty}^{\infty} ds \,\rho_{S^{\rm tot}_t}(-s) = 1.
\end{equation} 
Applying Jensen's inequality 
\begin{equation}
\langle \exp(-X) \rangle \geq \exp(-\langle X\rangle)  \label{eq:jensen}
\end{equation}
to $X=S^{\rm tot}_t$, 
we obtain the  second law of stochastic thermodynamics
\begin{equation}
    \langle S^{\rm tot}_t\rangle \geq 0, \label{eq:secondlawxx}
\end{equation} 
which is illustrated in Fig.~\ref{fig:MT_1x}.   For stationary systems, the stronger version 
\begin{equation}
    \langle 
    \dot{S}^{\rm tot}_t\rangle \geq 0, \label{eq:secondlawxxx}
\end{equation} 
of the second law holds, because  $\langle 
    \dot{S}^{\rm tot}_t\rangle= \langle 
    \dot{S}^{\rm tot}_0\rangle=\langle 
    {S}^{\rm tot}_t\rangle/t\geq 0$. 

\begin{figure}[h]
    \centering
    \includegraphics[width=0.8\textwidth]{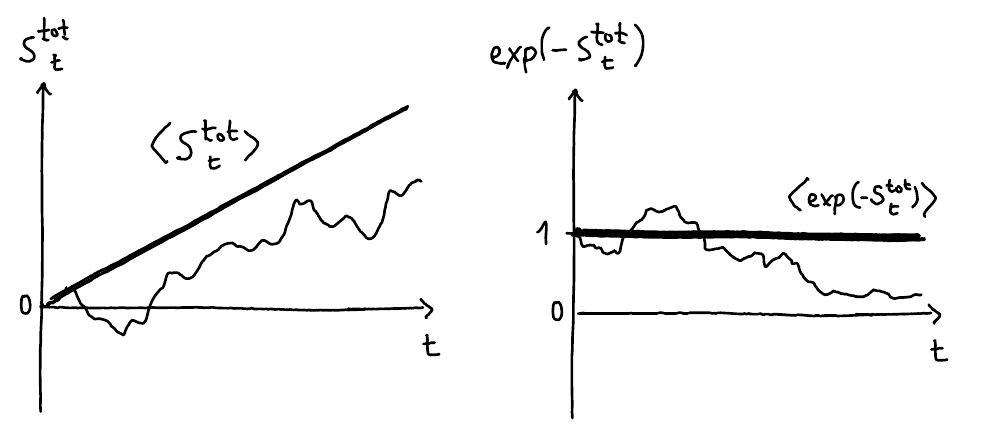}
    \caption{Illustration of the (classic) second law of thermodynamics $\langle S^{\rm tot}_t\rangle \geq0$ and the (classic) integral fluctuation relation $\langle \exp(-S^{\rm tot}_t)\rangle=1$. We sketch a single trajectory  of the stochastic entropy production  $S^{\rm tot}_t$ (left,  thin  lines) and of its negative exponential  $\exp(-S^{\rm tot}_t)$ (right,  thin line) in nonequilibrium stationary states. The thick lines in both panels illustrate the  values of $S^{\rm tot}_t$ (left panel) and $\exp(-S^{\rm tot}_t)$ (right panel) averaged over many different realizations.} %Stochastic entropy production is a submartingale with positive drift, whereas its negative exponential is martingale and has no drift. }
    \label{fig:MT_1x}
\end{figure}

Another interesting consequence of the integral fluctuation relation  is that negative fluctuations of entropy must exist in  nonequilibrium processes.   Applying   Markov's inequality Eq.~(\ref{eq:PA}) to $A = \emStot$ and using the integral fluctuation relation, we obtain the constraint \cite{seifert2012stochastic}
\begin{equation}
    \mathcal{P}\left(S^{\rm tot}_t \leq -s\right) \leq \exp(-s) , \quad {\rm for}  \quad s\geq 0, \label{eq:entropyNegx} 
\end{equation} 
on negative fluctuations of entropy production.

In what follows, we  use martingales to significantly extend these classical results from stochastic thermodynamics, i.e., the second law of thermodynamics Eq.~(\ref{eq:secondlawxx}), the integral fluctuation relation Eq.~(\ref{eq:iftxx}), and the bound on negative fluctuations of  entropy Eq.~(\ref{eq:entropyNegx}).

\section[Martingale fluctuation relations and   second laws]{Martingale fluctuation relations and martingale versions of the second law}\label{sec:marSecondFr}
\subsection{Martingale  integral fluctuation relations}\label{sec:martFr}

We derive extensions for the integral fluctuation relation Eq.~(\ref{eq:iftxx}) that follow from martingale theory~\cite{neri2019integral}.  

\begin{leftbar}
As the exponentiated, negative,  entropy production is a martingale, the relation \eqref{eq:mart_Stot} implies that~\cite{Chetrite:2011, neri2017statistics},
\begin{equation}
\langle\, \exp(-S^{\rm tot}_t ) \,|\, X_{[0,s]}\, \rangle = \exp(-S^{\rm tot}_s ),
\label{eq:mart_Stot4}
\end{equation}
which is known as the {\bf martingale  integral fluctuation relation}.  
\end{leftbar}
We provide an illustation of the martingale  integral fluctuation relation~\eqref{eq:mart_Stot4} in Fig.~\ref{fig:illusIFTM}.
\begin{figure}[H]
    \centering
    \includegraphics[width=\textwidth]{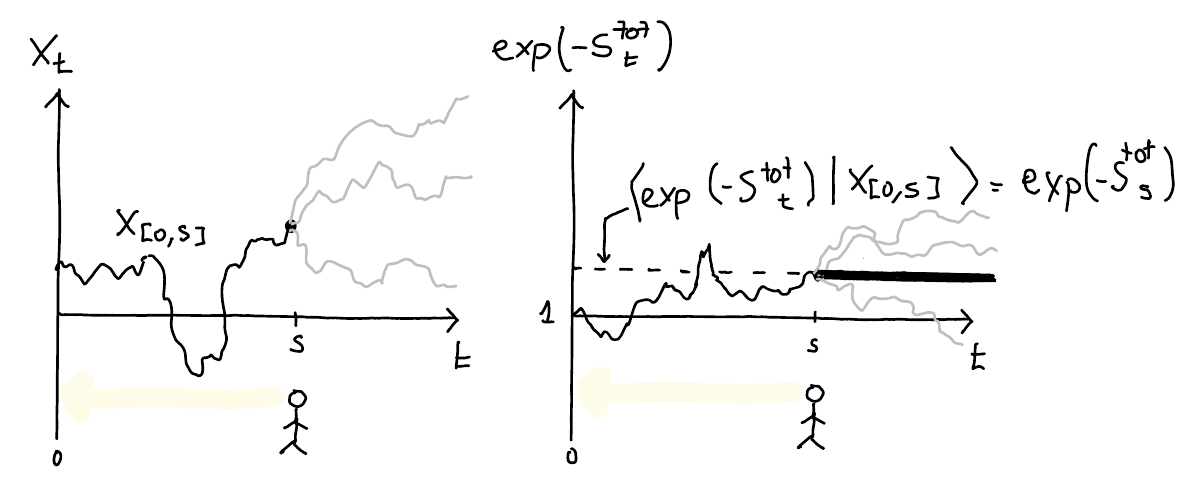}
    \caption{Sketch of the martingale integral fluctuation relation given by Eq.~\eqref{eq:mart_Stot4}. Left: An observer tracks the evolution of a process up to time $s$, recording a stochastic trajectory $X_{[0,s]}$ (black line). The evolution of the process at later times, given $X_{[0,s]}$, is stochastic and can have different outcomes (gray lines). Right: Given $X_{[0,s]}$, the value of  the exponentiated negative entropy production is known up to time $s$. The martingale condition~\eqref{eq:mart_Stot4} implies that  future average values of $\exp(-S^{\rm tot}_t)$ for $t\geq s$, given $X_{[0,s]}$, remain constant and equal to the value $\exp(-S^{\rm tot}_s)$ (black thick horizontal line). }
    \label{fig:illusIFTM}
\end{figure}

According to Theorem~\ref{Theorem:Equiv}, the martingale integral fluctuation relation~\eqref{eq:mart_Stot4} is equivalent to the following  integral fluctuation relation at {\em stopping times}.   
\begin{leftbar} 
Applying Doob's optional stopping theorems (see Sec.~\ref{sec:stop}) to $\emStot$, we obtain the  {\bf integral fluctuation relations at stopping times}
\begin{equation}
\langle\, \exp(-S^{\rm tot}_\mathcal{T} ) \rangle = 1, \label{eq:SStop}
\end{equation}
which holds when either the stopping time $\mathcal{T}$ is bounded, or $\mathcal{T}$ is with probability one finite and $S^{\rm tot}_t$ is bounded for all $t<\mathcal{T}$.    
\end{leftbar} 
The integral fluctuation relations at stopping times reveal a new level of universality, as they hold for  stopping times satisfying one of the following two conditions:
 \begin{itemize}
     \item $\mathcal{T}\in [0,t_{0}]$ for a fixed time $t_0\in \mathbb{R}^+$;
     \item $\mathcal{P}(\mathcal{T}<\infty)=1$ and $|S^{\rm tot}_{t}|<c$ for all $t\in [0,\mathcal{T}]$.
 \end{itemize}
Later in this Chapter, we determine the statistics of extreme values of entropy production and the splitting probabilities of entropy production  by specializing the integral fluctuation relation at stopping times~\eqref{eq:SStop} to specific classes of stopping times.   But, first we use in the next section the  martingale  fluctuation relations to derive  martingale versions of the second law of thermodynamics.

\subsection{Martingale versions of the second law of thermodynamics}\label{sec:martSecond}  

Although the second law of thermodynamics Eq.~(\ref{eq:secondlawxxx}) implies that on average the entropy of the universe increases,   this result is not entirely satisfactory.   Indeed, since for mesoscopic systems negative fluctuations of entropy production exist, as implied by Eq.~(\ref{eq:iftxx}),  it is not excluded  that   an intelligent being, say a demon, can anticipate when entropy decreases, and this question has   puzzled physicists, see e.g.~Refs.~\cite{szilard1964decrease, leff2002maxwell}.      However, the following two martingale versions of the second law of thermodynamics state that negative fluctuations  of entropy cannot be anticipated.  

\begin{leftbar}
Since $S^{\rm tot}_t$ is a submartingale, the relation \eqref{eq:submart_Stot} implies the {\bf  conditional strong second law} of thermodynamics, i.e., 
\begin{equation}
\langle\, S^{\rm tot}_t  \,|\, X_{[0,s]}\, \rangle \geq  S^{\rm tot}_s .
\label{eq:mart_Stot5}
\end{equation}
\end{leftbar}
Taking the average over $X_{[0,s]}$ in Eq.~(\ref{eq:mart_Stot5}) we readily obtain the "classical" second law of stochastic thermodynamics given by Eq.~(\ref{eq:secondlawxx}).

\begin{leftbar}
Applying Jensen's inequality (\ref{eq:jensen}) for $X=S^{\rm tot}_{\T}$  to Eq.~(\ref{eq:SStop}), we obtain the {\bf second law of thermodynamics at stopping times}, viz.,
\begin{equation}
 \langle S^{\rm tot}_\mathcal{T}  \rangle \geq  0. \label{eq:secondlawStop}
\end{equation} 
\end{leftbar}
Note that the martingale version of the second law, Eq.~(\ref{eq:mart_Stot5}), implies the second law Eq.~(\ref{eq:secondlawxx}) and is a significantly stronger result.       Even though in stochastic processes  negative fluctuations of entropy production exist, according to the martingale second law, Eq.~(\ref{eq:mart_Stot5}), an observer cannot anticipate those so-called transient "violations" of the second law based on the  past history $X_{[0,s]}$ of the process!    {Hence, Eq.~(\ref{eq:mart_Stot5}) is a stochastic version of the second law of thermodynamics, in the same way that (\ref{eq:1LST}) is a stochastic version of the first law of thermodynamics. }

The second law of thermodynamics at stopping times provides a different, but equivalent, perspective: an observer cannot reduce entropy by stopping the processes at a cleverly chosen moment. 
 
\begin{figure}[H]
    \centering
    \includegraphics[width=0.75\textwidth]{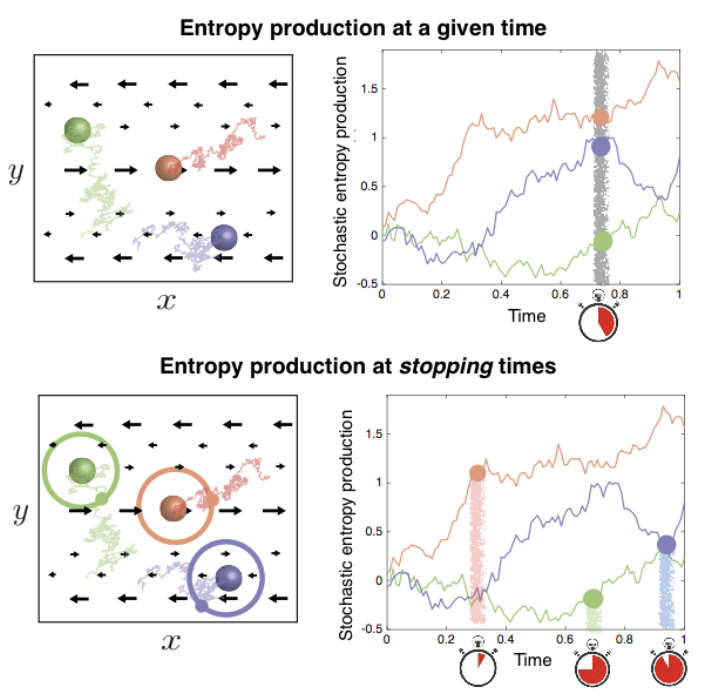}
    \caption{Illustration of the implications of the second law of thermodynamics at fixed times and stopping times. Top left: illustration of three stochastic trajectories drawn from a nonequilibrium stationary state (force field in black arrows). Top right: stochastic entropy production associated with the three trajectories. Their value at a fixed time (clock) is, according to the second law $\langle S^{\rm tot}_t\rangle\geq 0$, on average positive. Bottom left: illustration of three trajectories that are "stopped" when they cross a circle of a given radius centred at their initial position (rings). Bottom right: the  stochastic entropy production along the stopped trajectories. We highlight in filled circles the values that $S^{\rm tot}_{\mathcal{T}}$ takes when the particles cross the circle. Their average over many trajectories obeys the second law at stopping times $\langle S^{\rm tot}_{\mathcal{T}}\rangle\geq 0$. }
    \label{fig:MT_2}
\end{figure}

\label{sec:iFTSLST}

We illustrate the second law at stopping times~\eqref{eq:secondlawStop}   in Fig.~\ref{fig:MT_2} for the example of non-interacting colloidal particles moving  in a two dimensional fluid under the influence of a force field. In this example, the stopping time is  the first exit time of a particle from a circle centred at the initial position of the particles and with a fixed positive radius.

\section{Statistics of stopping times and extreme values}\label{sec:stop7}

We review several results on stopping times and extreme values in stationary processes.  

\subsection{Splitting  probabilities for   entropy production}\label{sec:abssurvent}
In the present Section, the stopping time $\mathcal{T}$ determines the stopping problem
\begin{equation}
    \mathcal{T} \equiv  \left\{t\geq 0: S^{\rm tot}_t \notin (-s_-, s_+) \right\}, 
    \label{eq:stsplit}
\end{equation}
where $s_-,s_+\geq 0$,
and we denote the corresponding  splitting probabilities  by
\begin{equation}
P_+(s_+,s_-) \equiv \mathcal{P}\left(S^{\rm tot}_\mathcal{T}\geq s_+\right) \quad {\rm and} \quad P_-(s_+,s_-) \equiv \mathcal{P}\left(S^{\rm tot}_\mathcal{T}\leq - s_-\right).  
\label{eq:p+p-def}
\end{equation}
The stopping problem Eq.~(\ref{eq:stsplit}) is illustrated in   Fig.~\ref{fig:pppmilus}.    Following~\cite{neri2017statistics, neri2019integral}, we derive now explicit expressions for $P_+$ and $P_-$.
\begin{figure}[H]
    \centering
    \includegraphics[width=0.45\textwidth]{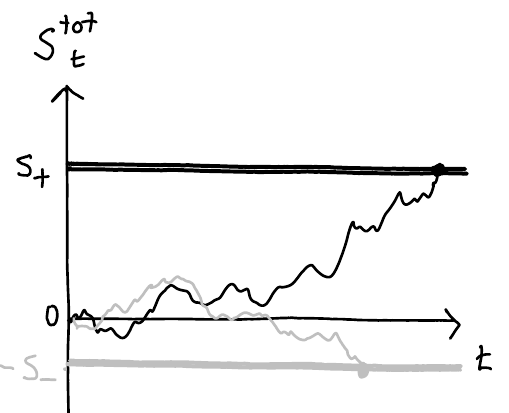}
    \caption{Illustration of two  trajectories of stochastic entropy production escaping from the interval $(-s_-, s_+)$ through the positive threshold (black line) and through the negative threshold (gray line).}
    \label{fig:pppmilus}
\end{figure}

For nonequilibrium stationary states, $S^{\rm tot}_t$ grows indefinitely, and hence 
\begin{equation}
P_+(s_+,s_-) + P_-(s_+,s_-) = 1.  \label{PpPm}
\end{equation}
Moreover, using the integral fluctuation relation at stopping times, Eq.~(\ref{eq:SStop}), on the stopping time~\eqref{eq:stsplit} we obtain 
\begin{equation}
P_+(s_+,s_-) \langle \exp(-S^{\rm tot}_\T) \rangle_+ + P_-(s_+,s_-) \langle \exp(-S^{\rm tot}_\T) \rangle_- = 1.  \label{PpPm2}
\end{equation}
Here, we have introduced the conditional averages
\begin{equation}
    \langle \;\;\cdot\;\;  \rangle_+  = \langle \;\;\cdot\;\; | S^{\rm tot}_\T \geq s_+  \rangle_+,\qquad {\rm and} \qquad \langle \;\;\cdot\;\;  \rangle  = \langle \;\;\cdot\;\; | S^{\rm tot}_\T \leq -s_-  \rangle. 
\end{equation}
Solving the set of Eqs.~(\ref{PpPm}-\ref{PpPm2}), we obtain the solution 
\begin{equation}
P_+(s_+,s_-)  = \frac{\langle \exp(-S^{\rm tot}_{\T})\rangle_{-}-1}{\langle \exp(-S^{\rm tot}_{\T})\rangle_{-}-\langle \exp(-S^{\rm tot}_{\T})\rangle_+},\label{eq:pprob3}
\end{equation}
and 
\begin{equation}
 P_-(s_+,s_-)  =  \frac{1-\langle \exp(-S^{\rm tot}_{\T})\rangle_+}{\langle \exp(-S^{\rm tot}_{\T})\rangle_{-}-\langle \exp(-S^{\rm tot}_{\T})\rangle_+}.\label{eq:pprob4}
\end{equation}

\begin{leftbar}
For time-homogeneous,  stationary states with    $S^{\rm tot}_{\mathcal{T}} \in \left\{-s_-, s_+\right\}$, which includes diffusion processes for which  $S^{\rm tot}_{t}$ is continuous in $t$,  we obtain the following {\bf universal} expressions for the {\bf splitting probabilities of entropy production}, 
\begin{eqnarray}
P_+(s_+,s_-) &=& \frac{\exp(s_-)-1}{\exp(s_-)-\exp(-s_+)}, \label{eq:pprob2}
\end{eqnarray}
and
\begin{eqnarray}
P_-(s_+,s_-)  &=& \frac{1-\exp(-s_+)}{\exp(s_-)-\exp(-s_+)}.\label{eq:pprob2n}
\label{eq:pprob}
\end{eqnarray} 
\end{leftbar}

Remarkably, the splitting probabilities are  independent of the finite-time moments of  entropy production, such as, the rate of entropy production; the universality of splitting probabilities can also be understood with the  random time transformation  discussed in Sec.~\ref{sec:randomtime}.

Note that the {\bf splitting-probability  fluctuation relation} 
\begin{eqnarray}
\frac{P_-(s_+,s_-)}{P_+(s_-,s_+)} = \exp(-s_-),
\end{eqnarray}
holds, which is reminiscent of the detailed fluctuation relation Eq.~(\ref{eq:dft2x}). 
However,  contrarily to the detailed fluctuation relation, the splitting probability fluctuation relation compares the splitting probabilities $P_-$ and $P_+$ in two different stopping problems,  except when $s_-=s_+=s$ in which case we obtain
\begin{eqnarray}
\frac{P_-(s,s)}{P_+(s,s)} = \exp(-s).
\label{eq:FTpassagesym}
\end{eqnarray}

For processes with jumps, we do not obtain universal expressions for $P_-$ and $P_+$, in correspondence with the results in Sec.~\ref{sec:absence}.   Nevertheless, we can derive universal bounds on $P_-$ and $P_+$.   
\begin{leftbar}
Using that $\langle \exp(-S^{\rm tot}_{\T})\rangle_+ \leq \exp(-s_+)$  and $\langle \exp(-S^{\rm tot}_{\T})\rangle_- \geq \exp(s_-)$, we obtain from  Eqs.~(\ref{eq:pprob3}) and (\ref{eq:pprob4}) the following  {\bf universal bounds on splitting probabilities} 
\begin{eqnarray}
 P_+(s_+,s_-) &\geq & 1- \frac{1}{\exp(s_- ) - \exp(-s_+ )}, \end{eqnarray}
and
\begin{eqnarray}
 P_-(s_+,s_-) &\leq &  \frac{1}{\exp(s_- ) - \exp(-s_+ )},
\label{eq:pprob4x}
\end{eqnarray}
which hold for time-homogeneous, stationary states.
\end{leftbar}

Another interesting quantity is the  survival probability $P_{\rm surv}(\tau)$ for $S^{\rm tot}$  to stay below a positive  threshold $s_+>0$ in a finite time $\tau\geq 0$. This survival probability  can be tackled  by specializing the integral fluctuation theorem at stopping times for $T_{\rm surv} = \T\wedge \tau = \text{min}(\T,\tau)$, with $\T$ the first-passage time to reach the threshold $s_+$ with $s_-\gg 1$. Following analogous steps as for $\T$, we get
\begin{equation}
    P_{\rm surv}(\tau) = \frac{\langle \exp(-S^{\rm tot}_{\T})\rangle-1}{\langle \exp(-S^{\rm tot}_{\T})\rangle-\langle \exp(-S^{\rm tot}_{\tau})\rangle_{\rm surv}},
\end{equation}
where $\langle \cdot \rangle_{\rm surv}$ denotes an average over all the trajectories that did not cross the threshold in the finite-time interval of duration $\tau$.

\subsection{Extreme-value statistics of  entropy production}
\label{sec:infimaentropy}

The results obtained in Sec.~\ref{sec:abssurvent} for the splitting probabilities can be used to determine the extreme-value statistics of entropy production.

The global infimum of  entropy production, defined by 
\begin{equation}
    S^{\rm inf} \equiv\inf_{t\geq 0} S^{\rm tot}_t,
\end{equation}
is the largest lower bound  of entropy production  along a  trajectory.
Because $S^{\rm tot}_0=0$,  $S^{\rm inf} $ can only take {nonpositive} values, i.e.,  $ S^{\rm inf}  \leq 0$. We  use martingale theory to determine the  statistical properties of the global infimum of entropy production.

First we tackle the cumulative distribution of $S^{\rm inf}$. The probability $\mP(S^{\rm inf} \leq -s)$ that the infimum $S^{\rm inf}$ is smaller or equal than $-s$, with $s>0$, equals the probability that entropy production crosses  at  any time $t\geq 0$ an absorbing boundary located at  $-s$. 
\begin{leftbar}
Taking the limit $s_-\to s$ and $s_+\to \infty$ in the right-hand side of Eq.~\eqref{eq:pprob4x},  gives the {\bf  universal  bound }
\begin{equation}
    P(S^{\rm inf} \leq -s)   \leq \exp(-s), \quad {\rm for}  \quad s\geq 0,
    \label{eq:Sinfglobaldist2}
\end{equation}
on  {\bf extreme negative fluctuations of entropy production}.  
\end{leftbar} 

The bound Eq.~(\ref{eq:Sinfglobaldist2}) from martingale theory should be compared with the weaker bound Eq.~(\ref{eq:entropyNegx}) from "classical" stochastic thermodynamics~\cite{seifert2012stochastic}.  
In this regard, note that Eq.~(\ref{eq:Sinfglobaldist2}) is a stronger result as $S^{\rm inf} \leq S_t$ for all values of $t$.   Moreover, for stationary diffusion  processes the equality in the bound Eq.~(\ref{eq:Sinfglobaldist2}) is attained, and the bound is thus as good as it gets.  Indeed,  taking the limit $s_-\to s$ and $s_+\to \infty$ of the right-hand side in Eq.~\eqref{eq:pprob}, we obtain 
\begin{equation}
    P(S^{\rm inf} \leq -s)  = \exp(-s).
    \label{eq:Sinfglobaldist}
\end{equation}
Eq.~(\ref{eq:Sinfglobaldist}) implies that the entropy-production global infimum in a continuous stochastic process  follows an exponential distribution with mean equal to $-1$, i.e.,
\begin{equation}
    \rho_{S^{\rm inf}}(s)   = \exp(s), \quad \forall s\leq 0.
\end{equation} 

From the bound Eq.~(\ref{eq:Sinfglobaldist2}) we obtain a second-law-like relation on the infimum of entropy production that was coined the  infimum law in Ref.~\cite{neri2017statistics}.
\begin{leftbar}
The inequality Eq.~(\ref{eq:Sinfglobaldist2}) implies the  {\bf infimum law} 
\begin{equation}
    \langle S^{\rm inf}\rangle \geq -1,  \label{eq:infimum}
\end{equation}
where the equality is attained for driven diffusion processes (i.e. when $X_t$ is a continuous process in $t$).  
\end{leftbar}

\begin{figure}[H]
    \centering
    \includegraphics[width=0.99\textwidth]{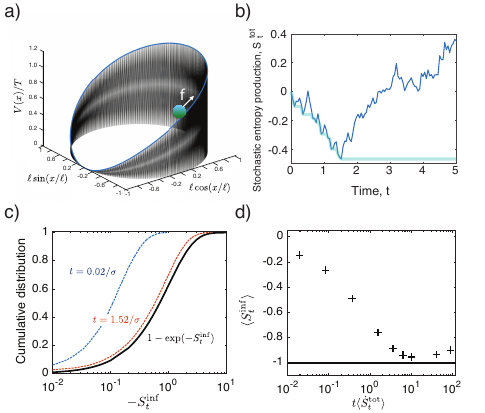}
    \caption{Illustration of the universal bounds on the statistics of extreme values of  entropy production in  the one-dimensional Langevin process described by Eq.~(\ref{eq:lang2}) with periodic boundary conditions, constant external force $f$, and potential $V(x) = T \ln \left(\cos(x)+2\right)$.   Panel a) Graphical illustration of the model. Panel  b) {Example trajectory of the stochastic entropy production $S^{\rm tot}_t$  associated with a stochastic trajectory of the particle position (blue thin line). The finite time infimum $S^{\rm inf}_t$ of entropy production associated with the  stochastic trajectory  of $S^{\rm tot}_t$ is shown in thick cyan line}. Panel  c) Cumulative distribution of the entropy production finite-time infimum $S^{\rm inf}_t$ for different values of $t$ (in units of the rate of entropy produciton $\langle\dot{S}^{\rm tot}\rangle$) obtained from simulations, see legend. The dashed lines are obtained from numerical simulations of the model shown in~(a) and compared with the  right-hand side of the universal bound from the infimum law (black thick line), Eq.~\eqref{eq:Sinfglobaldist3}. Panel  d) Entropy production finite-time infimum $S^{\rm inf}_t$ averaged over many simulations, as a function of time $t$.  See Ref.~\cite{neri2017statistics} for further details.   }
    \label{fig:infimacdf}
\end{figure}

The bound  Eqs.~\eqref{eq:Sinfglobaldist2} on extreme negative fluctuations of entropy production and the infimum law~\eqref{eq:infimum} are illustrated in  Fig.~\ref{fig:infimacdf} for the one-dimensional Langevin process of Eq.~(\ref{eq:lang2}) with periodic boundary conditions, potential $V(x) = T \ln \left(\cos(x)+2\right)$, and constant external force $f_t = f$.  

\begin{leftbar}
The universal bounds for the statistics of the global infimum of $\Stot$ serve to tackle the statistics of the finite-time infimum of entropy production (also called {\em running minimum} in the random-walk literature~\cite{benichou2016temporal}), defined as
\begin{equation}
    S^{\rm inf}_t \equiv\inf_{s\in [0,t]} S^{\rm tot}_s,
    \label{eq:sinftdef}
\end{equation}
and illustrated in Fig.~\ref{fig:finitetimeinf}. Because  the finite-time infimum is greater or equal than the global infimum $S^{\rm inf}_t \geq = \lim_{t\to \infty}S^{\rm inf}_t \equiv S^{\rm inf}$, Eqs.~\eqref{eq:Sinfglobaldist2} and~\eqref{eq:infimum} imply respectively the universal bounds
\begin{equation}
    P(S^{\rm inf}_t \leq -s)   \leq \exp(-s), \quad {\rm for}  \quad s\geq 0,
    \label{eq:Sinfglobaldist3}
\end{equation}
and
\begin{equation}
    \langle S^{\rm inf}_t\rangle \geq -1.  \label{eq:infimum2}
\end{equation}
\begin{figure}[H]
    \centering
    \includegraphics[width=0.45\textwidth]{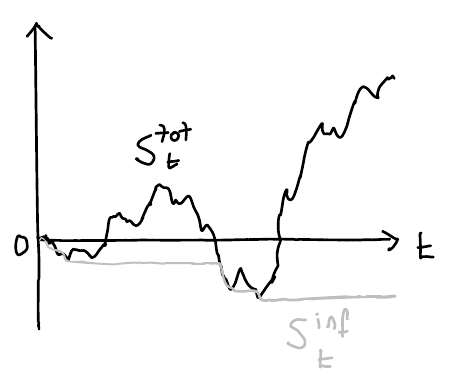}
    \caption{Illustration of a  trajectory of stochastic entropy production $\Stot$ (black line) and its finite-time infimum given by Eq.~\eqref{eq:sinftdef}.}
    \label{fig:finitetimeinf}
\end{figure}
\end{leftbar}

 Experimental tests  of Eqs.~\eqref{eq:Sinfglobaldist3} and the infimum law~\eqref{eq:infimum2} have been reported in  electronic double dots~\cite{singh2019extreme}  and in a Brownian motor immersed in a granular gas~\cite{cheng2020infima}. 
 In recent papers \cite{neri2022extreme, polettini2022phenomenological}, the present arguments for the extreme values of entropy production have been extended to the case of arbitrary edge currents in  a Markov jump process, and it was proven that the statistics of extreme values of a generic edge current are described by a geometric distribution characterised by an effective affinity.    Moreover in Ref.~\cite{manzano2021survival}, bounds tighter for than the infimum law have been derived using Doob's $L^p$ inequalities~\cite{williams1991probability}, and applied to bound the survival statistics of the work in steady-state heat engines.
 
 \subsubsection*{Negative fluctuations of entropy production  on   a ring}
{With an illustrative example we show that  infima of entropy  production are more effective in probing  negative fluctuations of entropy production and testing fluctuation relations than classical results based on fixed time observables.}     For this, let us consider the unidimensional drift-diffusion process on a ring introduced in Eq.~\eqref{eq:le1d}, i.e., 
 \begin{equation}
    \dot{X}_t = \mu f + \sqrt{2\mu T}\dot{B}_t\label{eq:Driftdiffusion} \end{equation}
  This is in fact a particular example  of  Eq.~(\ref{eq:lang2}) for a conservative force that is  homogeneous in time and space, i.e. $f_{t}(x)=f$.   The entropy production solves Eq.~(\ref{eq:entropy}), i.e.,
  \begin{equation}
      \dot{S}^{\rm tot}_t = v^S + \sqrt{2v^S}\dot{B}_t,
  \end{equation}
    with the homogeneous entropic drift given by
    \begin{equation}
        v^{S} = \frac{\mu f^2 }{ T}.
    \end{equation}
Thus for this example, $S^{\rm tot}_t$ is a drift-diffusion process with  the distribution 
\begin{equation}
    \rho_{S^{\rm tot}_t}(s) = \frac{1}{\sqrt{4\pi v^S}} \exp\left(-\frac{(s-v^St)^2}{4v^S}\right)
\end{equation}
and with the  cumulative distribution 
\begin{equation}
    \mathcal{P}\left(S^{\rm tot}_t \leq -s \right) = \frac{1}{2}\left(1+{\rm erf}\left(\frac{s-v^{S}t}{2\sqrt{v^S}}\right)\right), \label{eq:PTStotPlot}
\end{equation}
where ${\rm erf}(x) = (2/\sqrt{\pi})\int^x_0\exp(-y^2)dy$ is the error function.   On the other hand, the cumulative distribution of the entropy-production infimum is given by 
\begin{equation}
\mathcal{P}\left(S^{\rm inf}_t \leq -s \right) =  \frac{1}{2}\left(1+{\rm erf}\left(\frac{s-v^St}{2\sqrt{v^St}}\right)\right)  + \frac{\exp(s)}{2}\left(1-{\rm erf}\left(\frac{-s-v^St}{2\sqrt{v^St}}\right)\right),\label{eq:PTStotPlot2} 
\end{equation}
which follows from the exact expression for the infimum distribution of a 1D drift diffusion process, see Ref.~\cite{neri2017statistics}.

Figure~\ref{fig:FTBound} shows the  two cumulative distributions $   \mathcal{P}\left(S^{\rm tot}_t \leq -s \right)$ and  $\mathcal{P}\left(S^{\rm inf}_t \leq -s \right)$  and compares them with the exponential bounds Eqs.~(\ref{eq:entropyNegx}) and (\ref{eq:Sinfglobaldist3}), respectively.    From  Fig.~\ref{fig:FTBound} it is apparent that   Eq.~(\ref{eq:entropyNegx}) is a loose bound for all values of $t$, while Eq.~(\ref{eq:Sinfglobaldist3}) is tight in the limit of large $t$, as predicted by martingale theory.   Also, note that the quality of the bound (\ref{eq:entropyNegx})  worsens as a function of $t$.

\begin{figure}[h]
    \centering
    \includegraphics[width=0.5\textwidth]{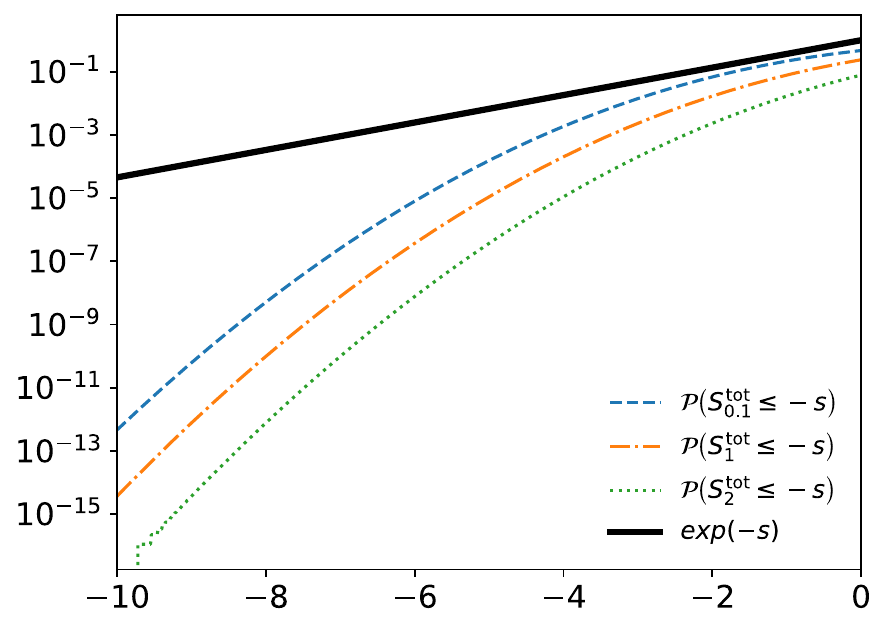}
     \put(-110,-10){$-s$} 
    \includegraphics[width=0.5\textwidth]{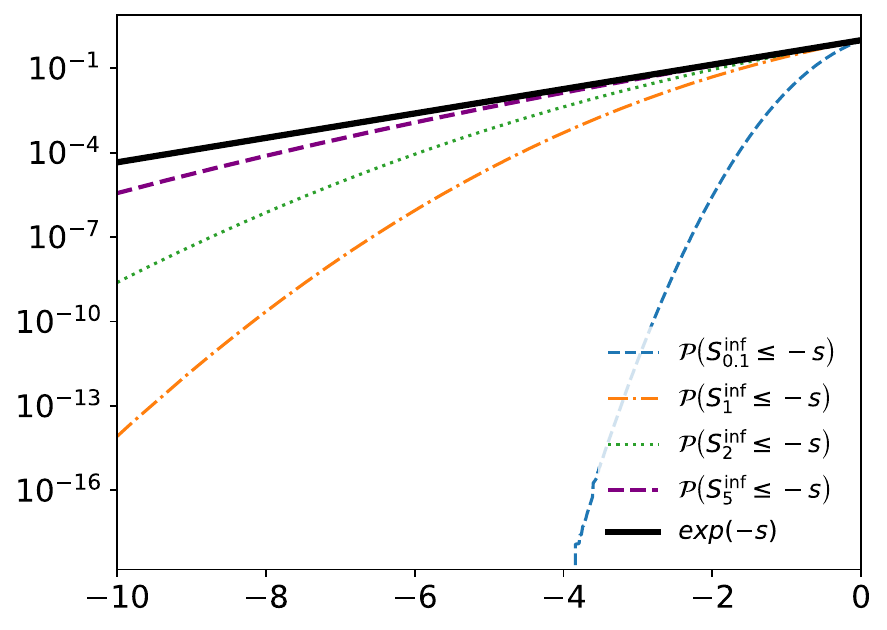}
          \put(-110,-10){$-s$} 
    \caption{Comparison between the upper bounds on negative fluctuations of entropy production from classical stochastic thermodynamics Eq.~(\ref{eq:entropyNegx})  (left) and from martingale stochastic thermodynamics Eq.~(\ref{eq:Sinfglobaldist2}) (right) on  the example of the drift-diffusion process on a ring described by Eq.~(\ref{eq:Driftdiffusion}). Left: plot of $\mathcal{P}\left(S^{\rm tot}_t \leq -s \right)$ from Eq.~(\ref{eq:PTStotPlot}) as a function of $-s<0$ for $v^S=1$ and given values of $t$, and comparison with the upper bound $\exp(-s)$.  Right:  plot of $\mathcal{P}\left(S^{\rm inf}_t \leq -s \right)$ from Eq.~(\ref{eq:PTStotPlot2}) as a function of $-s<0$ for the same values of $v^S$ and $t$ as in the left panel, and comparison with the upper bound $\exp(-s)$ from  thermodynamics with martingales, see Eq.~\eqref{eq:Sinfglobaldist3}.  }
    \label{fig:FTBound}
\end{figure}

\subsection{First-passage-time fluctuation relation for Langevin processes}
We review the first-passage-time fluctuation relations for    entropy production in  stationary Langevin processes~\cite{neri2017statistics}.  This fluctuation relation considers the statistics of the first-passage time Eq.~(\ref{eq:stsplit}) for symmetric thresholds $s_+=s_-=s$.   %, i.e., the first time entropy production goes below the  negative  threshold $-s_-=-s$ or above the positive  threshold $s_+=s$, with $s>0$.   

To state the first-passage-time fluctuation relation, we  define the following stopping times for entropy production  (see Fig.~\ref{fig:FPTFTaa} for an illustration)
\begin{itemize}
    \item $\T_+$  is the first time when  $S^{\rm tot}_t$  reaches the positive threshold $s>0$, given that $S^{\rm tot}_t$  did not pass  below $-s_- = -s$ at earlier times $t'< t$; if $S^{\rm tot}_t$  escapes first through the negative threshold, then  we set $\T_+=\infty$.
    \item $\T_-$  is the first time when  $S^{\rm tot}_t$  reaches the negative  $-s<0$, given that $S^{\rm tot}_t$ did not go above $s>0$ at earlier times $t'<t$; if $S^{\rm tot}_t$ first escapes through the positive threshold, then $\T_-=\infty$.
\end{itemize} 
\begin{figure}[h]
    \centering
    \includegraphics[width=0.5\textwidth]{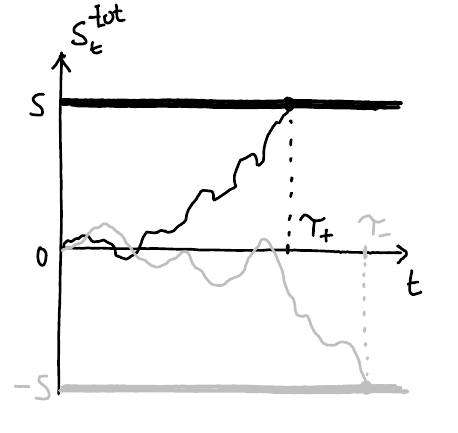}
    \caption{Illustration of the stopping times $\T_+$ and $\T_-$ for entropy production to first escape the interval $(-s,s)$ from its positive and negative boundaries, respectively.}
    \label{fig:FPTFTaa}
\end{figure}
Remarkably, the  cumulative probabilities for $\T_+$ and $\T_-$ obey the relation
\begin{equation}
    \frac{\mP(\T_+\leq t)}{\mP(\T_-\leq t)}=\exp(s),
    \label{eq:condft}
\end{equation}
which holds for all $t\geq 0$ and for all $s\geq 0$.

We sketch a proof of Eq.~\eqref{eq:condft}:
\begin{eqnarray}
\mP(\T_+\leq t) &=& \langle\, \theta(t-\T_+)\,\rangle \\
&=& \int \mathcal{D}\xot \mP(\xot) \theta(t-\T_+[\xot]) \\
&=& \int \mathcal{D}\xot \exp( S^{\rm tot}(\xot) )\mP(\Theta_t\xot)\theta(t-\T_+(\xot)) \\
\label{metalica}&=& \int \mathcal{D}\xot\exp(-S^{\rm tot}(\Theta_t\xot))\mP(\Theta_t\xot) \theta(t-\T_-(\Theta_t\xot)) \\ \label{maiden}
&=& \int \mathcal{D}\xot\exp(-S^{\rm tot}(\xot))\mP(\xot) \theta(t-\T_-(\xot))\\\label{Pinkfloyd}
&=& \langle \emStot|\T_- \leq t \rangle \mP(\T_-\leq t) \\ \label{Supertramp}
&=&  \langle\exp(-S^{\rm tot}_{\T_-}) \rangle\mP(\T_-\leq t) \\ \label{Bowie}
&=& \exp(s)\mP(\T_-\leq t) .
\end{eqnarray}
We provide details on the most involved steps in the derivation shown above.   In Eq.~(\ref{maiden}) we have used that $S^{\rm tot}(x_{[0,t]}) = -S^{\rm tot}(\Theta_t(x_{[0,t]}))$ and  $\T_+(x_{[0,t]})= \T_-(\Theta_t(x_{[0,t]}))$.   In Eq.~\eqref{maiden}, we have used the fact that the Jacobian of the transformation $x_{[0,t]}\rightarrow  \Theta_t x_{[0,t]}$ is one. In Eq.~\eqref{Supertramp}, we  have used the martingality of $\emStot$ and  Eq.~(\ref{eq:OTcomplicated}) of Doob's optional stopping Theorem~\ref{Doob1x} for $\T_1=\T_-$, $\T_2 =t$, and for the uniformly integrable martingales  $\exp(-S^{\rm tot}_{t'\wedge t})$ that are defined at   fixed values of  $t\geq 0$ and for  $t'\in[0,t]$.     Lastly, in  Eq.~\eqref{Bowie} we have used the fact that $X_t$ is a diffusion process.       We have also used  here $\theta$ for the Heaviside theta function, not to be confused with the time-reversal operator $\Theta_t$.

\begin{leftbar}
Equation~(\ref{eq:condft}) implies that the first-passage  densities  obey the \textbf{first-passage-time fluctuation relation} for  the  stochastic entropy production in nonequilibrium stationary processes~\cite{neri2017statistics,bauer2014affinity}, viz., 
\begin{equation}
    \frac{\rho_{\T_+}(t) }{\rho_{\T_-}(t) } = \exp(s).  
    \label{eq:fptft}
\end{equation}
Note that the $s$-dependency on the left hand side of Eq.~(\ref{eq:fptft}) is hidden in the boundary conditions of the stopping times $\T_{\pm}$. 
\end{leftbar}
Notice that $\rho_{\T_+}(t)\text{d}t$ and $\rho_{\T_-}(t)\text{d}t$ are  defined  by
$\rho_{\T_\pm}(t)\equiv \mP(\T_\pm \in [t,t+dt])/dt$, but we remark that these are unnormalized densities because the splitting probabilities obey $\mP(\T_{\pm}\leq \infty)=P_{\pm}(s,s)<1$. This motivates us to define the conditional (normalized) densities 
\begin{equation}
    \hat{\rho}_{\T_\pm}(t)\equiv \frac{\mP(\T_\pm \in [t,t+dt]|\T_\pm<\infty)}{dt} = \frac{\rho_{\T_\pm }(t)}{P_\pm (s,s)}.
    \end{equation}

\begin{leftbar}
From Eq.~\eqref{eq:fptft}, and the fluctuation relation for the splitting probabilities~$P_+(s,s)/P_-(s,s) = \exp(s)$ (see  Eq.~\eqref{eq:FTpassagesym}), we find that
\begin{equation}
    \hat{\rho}_{\T_+}(t) = \hat{\rho}_{\T_-}(t).
    \label{eq:FPTsymmetry}
\end{equation}
In words, Eq.~\eqref{eq:FPTsymmetry} states that it takes the same amount of time to increase entropy production by   $s$  as  it takes to reduce entropy production by  $-s$. 
\end{leftbar}
Relations analogous to the remarkable symmetry given by Eq.~\eqref{eq:FPTsymmetry} have been derived in the context of Haldane equalities in enzyme kinetics (see e.g. Ref.~\cite{qian2006generalized}) and for first-passage-time dualities in diffusion processes (see e.g. Ref.~\cite{krapivsky2018first}).  Figure~\ref{fig:FPTFT} shows a numerical test  for the symmetry relation~\eqref{eq:FPTsymmetry}  for first-passage times and  the fluctuation relation~\eqref{eq:FTpassagesym}  for splitting probabilities.  
\begin{figure}[H]
    \centering
    \includegraphics[width=\textwidth]{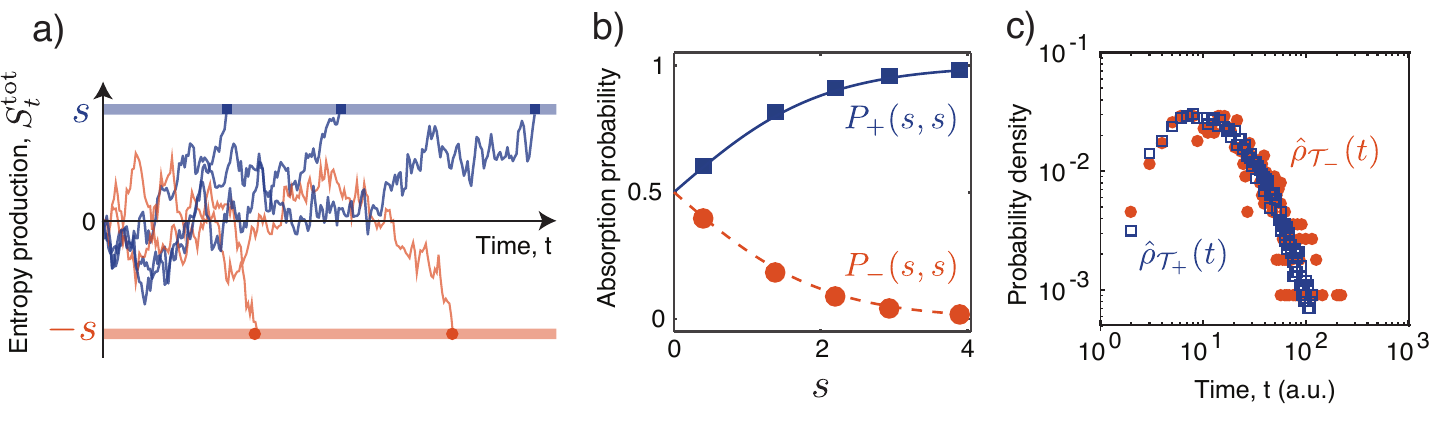}
    \caption{The escsape problem for entropy production out of a symmetric interval $(-s,s)$.     Panel (a): Example trajectories of $\Stot$ as a function of time. We highlight in blue three trajectories that first reach $s$ before $-s$  (at stochastic times $\T_+$) , and in red   two trajectories that first reach $-s$ before $s$  (at stochastic times $\T_-$).  Panel (b): Splitting probabilities $P_+(s,s)$ that $\Stot$  goes above $s$ before it  (possibly)  passes below $-s$ (blue squares), and $P_-(s,s)$ for $\Stot$ to first pass below $-s$ before it goes above $s$ (red circles).    Markers are obtained from numerical simulations, and the lines denote the analytical expressions~\eqref{eq:pprob2} [$P_+(s,s)$, blue solid line] and~\eqref{eq:pprob2n}  [$P_+(s,s)$, red dashed line] for $s_+=s_-=s$. (c) Conditional normalized densities for $\T_+$ (blue open squares)  and for $\T_-$ (red filled circles) obtained from numerical simulations.  Results illustrate the symmetry relation given by Eq.~\eqref{eq:FPTsymmetry}.    All  numerical results are obtained for the model sketched in Fig.~\ref{fig:infimacdf}a (see Ref.~\cite{neri2017statistics} for further details).  }
    \label{fig:FPTFT}
\end{figure}

%\subsection{{First-passage-time fluctuation relation for Jump process : Example of Jump process by Franck}}

\subsection{Trade-offs between speed, uncertainty, and dissipation}
\label{sec:TURs}

A  recurrent theme in  nonequilibrium thermodynamics is that processes far from thermal equilibrium are governed by a trade-off between  speed, uncertainty, and dissipation.   Indeed, concrete examples of this  thermodynamic trade-off have been found  in kinetic proof reading \cite{murugan2012speed, sartori2015thermodynamics, mallory2019trade}, sensory adaptation \cite{lan2012energy}, and microscopic heat engines \cite{pietzonka2018universal}.   Even though  speed and uncertainty are quantified differently in these examples, they  are suggestive  of  universal inequalities describing a  trade-off between speed, uncertainty, and dissipation in nonequilibrium systems.    

%\begin{figure}[h!]
%    \centering
%    \includegraphics[width=0.6\textwidth]{pres.pdf}
%    \put(-320,100){dissipation}    %\put(-320,200){speed}
%    \put(-320,100){uncertainty}
%    \caption{  Illustration of the trade-off between speed, uncertainty, and dissipation.  The idea is to determine a surface in a three-dimensional space that separates nonpermissible processes below the surface with processes that are permissible above the surface.   Figure taken from \cite{neri2022universal}.}
%    \label{fig:tradeoff}
%\end{figure}
In recent years, universal  inequalities expressing trade-offs in generic,  nonequilibrium, stationary states have been derived for Markov jump processes and overdamped Langevin processes.      We revisit here two inequalities  based on first-passage times, namely,  the  speed-uncertainty-dissipation trade-off relation   \cite{roldan2015decision, neri2022universal, neri2022estimating} and the  thermodynamic uncertainty relation \cite{gingrich2017fundamental}.

As discussed in Sec.~\ref{sec:stop}, martingale theory provides a powerful set of tools to study processes at stopping times, and we will use this here to study  nonequilibrium trade-off relations involving first-passage times.  In particular,  we use martingale theory to show  that the speed-uncertainty-dissipation trade-off relation is optimal in a  specific sense that we discuss below, and we also use martingale theory to evaluate the trade-off relations in a simple example of a nonequilibrium process.

\subsubsection{Setup: empirical current and stopping time}  
Let $J_t$ be an empirical integrated current in a stochastic process $X_t$ that is either a stationary Markov jump process   or an overdamped Langevin process, and assume without loss of generality that $\langle J_t\rangle>0$.   In a Markov jump process, an empirical current    takes the form 
\begin{equation}
    J_t =  \sum_{(x,y)} c(x, y) J_t(x,y),
\end{equation}
where $J_t(x,y) = N_t(x,y) - N_t(y,x)$ is the difference between the number of jumps $N_t(x,y)$ from $x$ to $y$ minus the number of jumps $N_t(y,x)$ from $y$ to $x$ counted  in the time-interval~$[0,t]$ (see definition in Eq.~(\ref{eq:NtDef})), and $c(x, y) \in \mathbb{R}$ quantifies the "resource" transported when the process jumps from $x$ to $y$.   The stochastic entropy production, defined by Eq.~\eqref{eq:WLSPSJP}, takes here the form 
\begin{equation}
S^{\rm tot}_t = \frac{1}{2}\sum_{x,y} \ln \left( \frac{\rho_{\rm st}(x)\omega(x,y)}{\rho_{\rm st}(y)\omega(y,x)}\right) J_t(x,y),
\end{equation}
and is a particular example of empirical current, where we identify   $c(x, y)$  in this case as the total entropy change in a jump. 
An analogous formalism  applies to Langevin processes in which  case  empirical currents are  Stratonovich integrals, viz.,  
\begin{equation}
    J_t = \int^t_0 c(X_s) \circ dX_s.  
\end{equation} 
In what follows, and throughout this Sec.~\ref{sec:TURs}, we rely on   the first-passage time 
\begin{equation}
    \T \equiv {\rm inf}\left\{t\geq 0 : J_t\notin (-\ell_-,\ell_+)\right\} \label{eq:TJDef}
\end{equation} 
for the current $J_t$ to exit the open interval defined by the    thresholds $\ell_-,\ell_+>0$, and we  consider the limit   $\ell_{\rm min}\gg 1$, where  $\ell_{\rm min} = {\rm min}\left\{\ell_-,\ell_+\right\}$.   

\subsubsection{Trade-off relations based on first-passage times}
We review  thermodynamic, trade-off relations between speed, uncertainty and dissipation that are based on first-passage processes.   

%{\color{red} ER: to be fixed by Izaak. No mention to martingale here. See comment 5 in Ch7 by the Referee. I copy it here: "Section 7.4.4. the presentation of the trade-off relations is sketchy and more discussion would be required. In particular, the important result (7.56) is stated rather abruptly, without proof. Besides, the relation of (7.56) with martingales is not clear. Up to that point, things were self-contained in the review, but this result (7.56) seems to pop up from nowhere."}

%{\color{blue} IN:    To me this seems perfect.   I don't know what there is to fix.  I think this is more about explaining to Referee then about changing the main text.}

\begin{leftbar}
The {\bf nonequilibrium, thermodynamical trade-off} relations  we consider take  the form  
\begin{equation}
a \:\epsilon_{\rm unc} \langle \dot{S}^{\rm tot}_t\rangle \langle \T\rangle(1+o_{\ell_{\rm min}}(1)) \geq  1, \label{eq:tradeoff}
\end{equation}
 where $a\in \mathbb{R}^+$ is a constant; where $\langle \dot{S}^{\rm tot}_t\rangle$ is the average entropy production rate that quantifies {\bf dissipation};
 where $\langle \T\rangle$ is the  mean first-passage time  that quantifies {\bf speed}; and where $\epsilon_{\rm unc}$ is a dimensionless observable that  quantifies  {\bf uncertainty} in the process.    Later when considering specific examples of such trade-off relations we define $a$ and $\epsilon_{\rm unc}$.   
 \end{leftbar}

 The factor $1+o_{\ell_{\rm min}}(1)$ represents an arbitrary function that converges to zero when $\ell_{\rm min} \gg 1$ and implies that Eq.~(\ref{eq:tradeoff}) is an asymptotic relation that holds in the limit of large  values of the first-passage thresholds $\ell_+$ and $\ell_-$.   Dissipation  $\langle \dot{S}^{\rm tot}_{t}\rangle$ is 
given by  \eqref{eq:schnack} for Markov jump processes and by \eqref{eq:stotavgj} for Langevin processes. 
 The trade-off relation Eq.~(\ref{eq:tradeoff}) states that processes that are fast, have a small amount of  fluctuations, and dissipate little, are physically nonpermissible, see Panel (a) of Fig.~\ref{fig:tradeoff2} for an illustration.  
 
 Below we review two examples of  trade-off relations that take the form of Eq.~(\ref{eq:tradeoff}), but differ in  the way that uncertainty  $\epsilon_{\rm unc}$ is quantified.   
 \begin{leftbar}
 The first relation we discuss is the {\bf speed-uncertainty-dissipation trade-off relation}~ \cite{roldan2015decision, neri2022universal,neri2022estimating}, which is the inequality Eq.~(\ref{eq:tradeoff}) for
 \begin{equation}
     \epsilon_{\rm unc} \equiv \frac{1}{|\ln P_-|}, \quad {\rm and} \quad a \equiv\frac{\ell_-}{\ell_+}, 
 \end{equation}
 where 
 \begin{equation}
     P_- \equiv  \mathcal{P}\left[J(\T)\leq -\ell_-\right], 
\end{equation}
is the probability that $J_t$ leaves for the first time the interval $(-\ell_-, \ell_+)$ through the lower threshold. 
\end{leftbar}
The measure $\epsilon_{\rm unc}\in [0,1/|\ln 2|]$ takes the value $\epsilon_{\rm unc} = 0$ for processes without fluctuations ($P_-=0$) and takes the value $\epsilon_{\rm unc} = 1/|\ln 2|$ for processes with a large amount of fluctuations ($P_-=1/2$).

\begin{leftbar}
The second relation we consider is the {\bf thermodynamic uncertainty relation}~\cite{gingrich2017fundamental}, which is the inequality Eq.~(\ref{eq:tradeoff}) for    
 \begin{equation}
     \epsilon_{\rm unc} \equiv \frac{\langle \T^2\rangle - \langle \T\rangle^2}{\langle \T\rangle^2} \quad {\rm and} \quad a \equiv \frac{1}{2}.  
 \end{equation}
 \end{leftbar}
In the thermodynamic uncertainty relation,  uncertainty is determined by the variance of the first-passage time; an equivalent uncertainty relation  holds at fixed times \cite{barato2015thermodynamic,gingrich2016dissipation, pietzonka2016universal}. 
 It should be emphasized that  both the  speed-uncertainty-dissipation trade-off relation  and the   thermodynamic uncertainty relation  are  generically valid for nonequilibrium stationary states of Markov jump processes and Langevin processes.

\begin{figure}[h!]
    \centering
        {\includegraphics[width=0.45\textwidth]{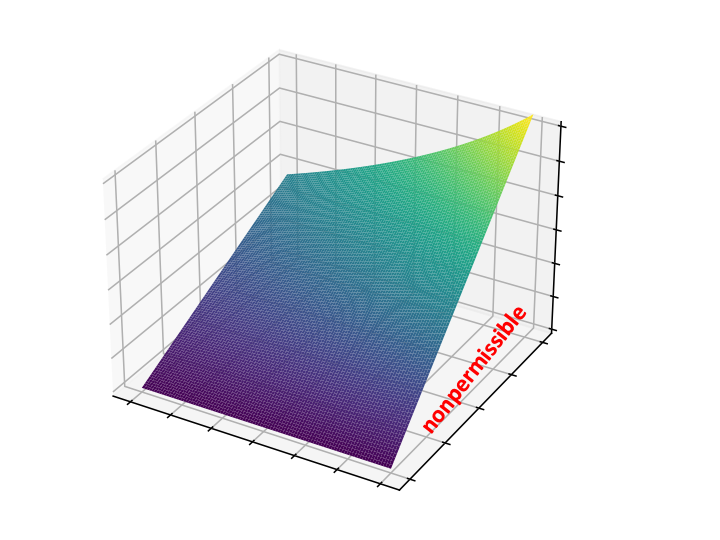}
             \put(-200,120){(a)} 
     \put(-40,100){$\langle\dot{S}^{\rm tot}_t\rangle$}
        \put(-155,20){$\epsilon_{\rm unc}$} 
 \put(-60,30){$\langle \T\rangle$} }
 {   \includegraphics[width=0.45\textwidth]{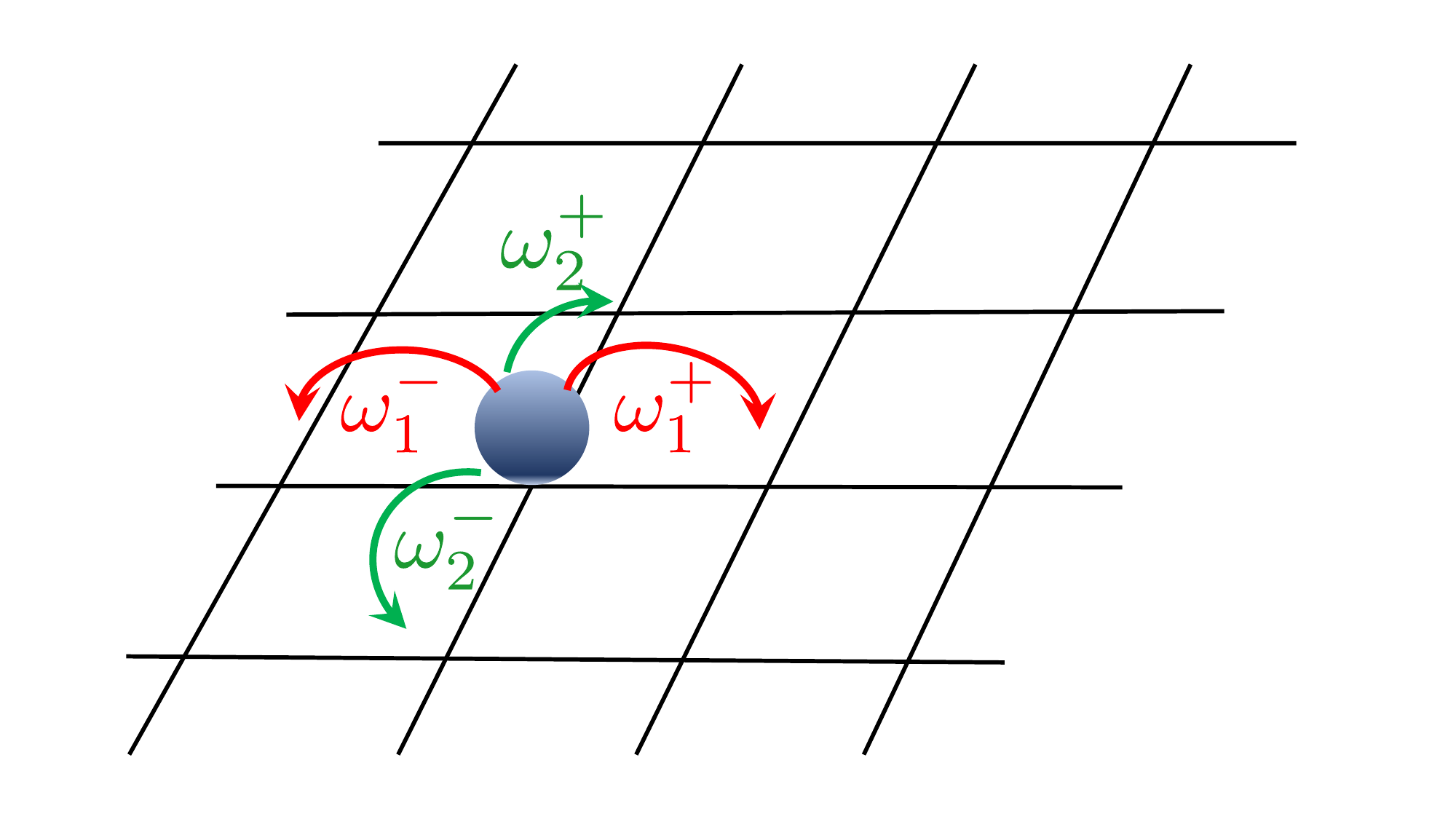}
         \put(-200,120){(b)}} 
     { \includegraphics[width=0.4\textwidth]{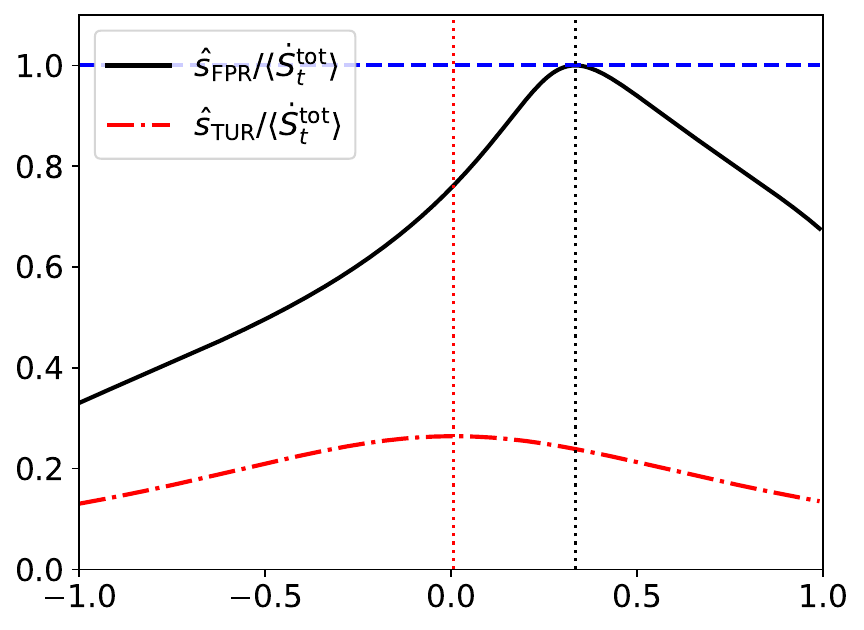}
         \put(-200,120){(c)} 
        \put(-92,-10){$\Delta$} }
       { \includegraphics[width=0.4\textwidth]{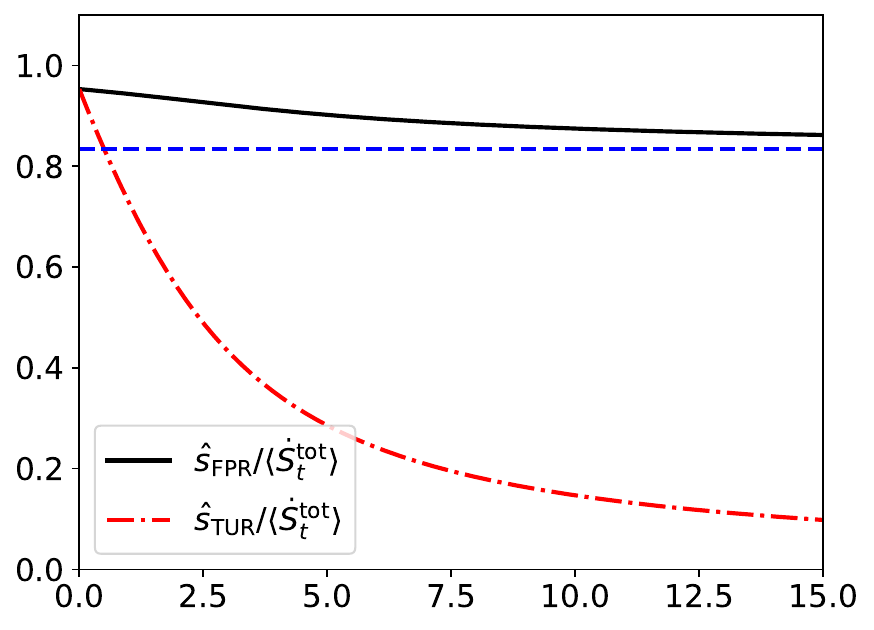}
             \put(-195,120){(d)} 
 \put(-110,-10){$\langle\dot{S}^{\rm tot}_t\rangle/\omega_{\rm total}$} }
    \caption{Trade-off between speed, uncertainty and dissipation.     Panel (a): Illustration of the trade-off relation  Eq.~(\ref{eq:tradeoff}).    Processes under the plotted surface are nonpermissible.   These processes are fast, fluctuate little, and dissipate little.   Panel (b):  Illustration of the random walk model on a two-dimensional lattice, as defined in Sec.~\ref{sec:simple2D}. Panel (c): Plot of the ratios $\hat{s}_{\rm FPR}/\langle \dot{S}^{\rm tot}_t\rangle$ and $\hat{s}_{\rm TUR}/\langle \dot{S}^{\rm tot}_t\rangle$ for the model illustrated in Panel (b) as a function of the parameter $\Delta$ that defines the current, see Eq.~(\ref{eq:JDef}).  Parameters $\omega^+_1 = \exp(2.5)/[4\cosh(2.5)]$,  $\omega^-_1 = \exp(-2.5)/[4\cosh(2.5)]$, $\omega^+_2 = \exp(5)/[4\cosh(5)]$,  and $\omega^-_2 = \exp(-5)/[4\cosh(5)]$.  Vertical dashed lines denote the locations of the maxima of $\hat{s}_{\rm FPR}$ and $\hat{s}_{\rm TUR}$, the former corresponding with Eq.~(\ref{eq:DeltaKM2}).  Panel (d):  Similar plot as in Panel (c), but now the ratios are plot as a function of $\langle \dot{S}^{\rm tot}_t\rangle/\omega_{\rm total}$, where $\omega_{\rm total} = \omega^-_1 + \omega^+_1 + \omega^-_2 + \omega^+_2$.   To this aim, the model parameters are set to  $\omega^+_1 = \exp(\nu/2)/[4\cosh(\nu/2)]$,  $\omega^-_1 = \exp(-\nu/2)/[4\cosh(\nu/2)]$, $\omega^+_2 = \exp(\nu)/[4\cosh(\nu)]$,  and $\omega^-_2 = \exp(-\nu)/[4\cosh(\nu)]$ and $\nu$ is varied.     All figures are taken from Refs.~\cite{neri2022universal} and  \cite{neri2022estimating}. }
    \label{fig:tradeoff2}
\end{figure}
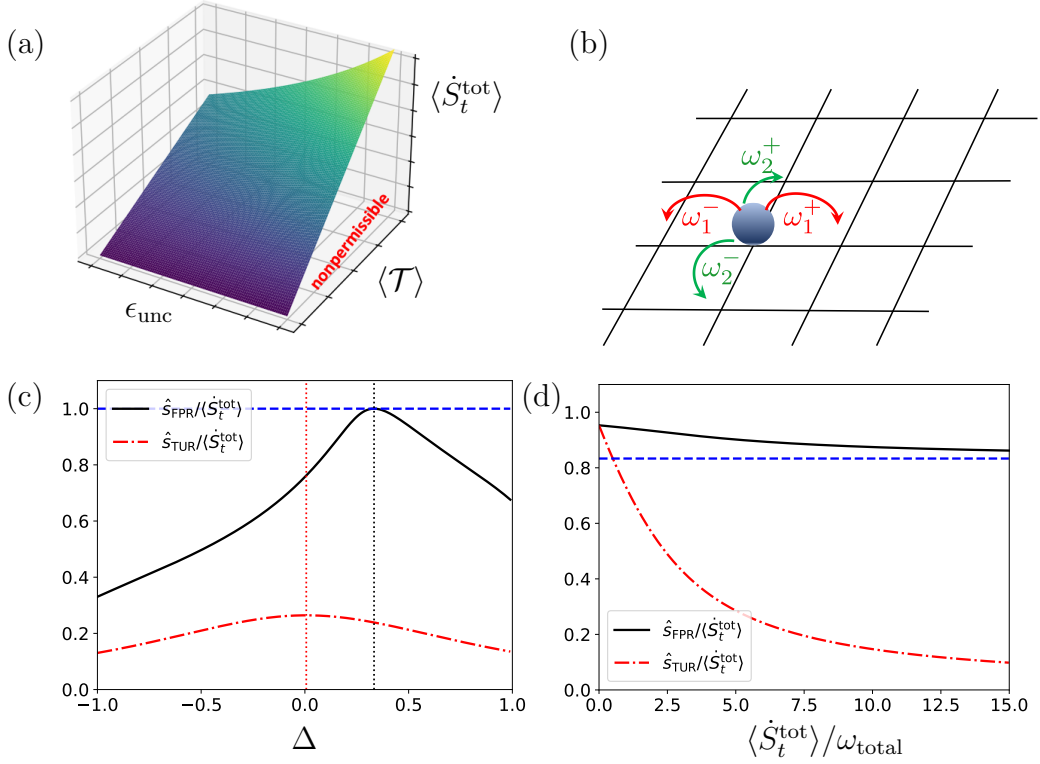
\subsubsection{Comparing the quality of different trade-off relations}

%{\color{red}Page 172: Could you give some details (or references) for the proofs of (7.61)
%and (7.62)? It is just written ‘we show that...’ without extra explanations}. 

%{\color{blue} IN:  the proof appears just below in equations (7.63)-(7.65) and for 7.63 we refer to 7.18 where this is shown.   It is not clear to me how to make this more clear, but I am open to suggestions.          }

To compare the quality  of the two trade-off relations, we  evaluate the  following estimates 
\begin{equation}
    \hat{s}_{\rm FPR} \equiv  \frac{\ell_+}{\ell_-} \frac{|\ln P_-|}{ \langle \T\rangle}  \leq \langle \dot{S}^{\rm tot}_t\rangle   \quad {\rm and} \quad \hat{s}_{\rm TUR} \equiv  2 \frac{\langle \T\rangle}{ \langle \T^2\rangle - \langle \T\rangle^2}  \leq \langle \dot{S}^{\rm tot}_t\rangle  \label{eq:defSFPR}
\end{equation}
of dissipation based on first-passage times.
 The ratios $\hat{s}_{\rm FPR}/\langle \dot{S}^{\rm tot}_t\rangle$ and $\hat{s}_{\rm TUR}/\langle \dot{S}^{\rm tot}_t\rangle$ determine the fraction of the average rate of dissipation $\langle \dot{S}^{\rm tot}_t\rangle$ captured by the estimators of dissipation $\hat{s}_{\rm FPR}$  and $\hat{s}_{\rm TUR}$ based on the trade-off relation between speed, uncertainty and dissipation or the thermodynamic uncertainty relation, respectively.   The closer the ratios $\hat{s}_{\rm FPR}/\langle \dot{S}^{\rm tot}_t\rangle$ and $\hat{s}_{\rm FPR}/\langle \dot{S}^{\rm tot}_t\rangle$ are to one, the tighter are the inequalities in Eq.~(\ref{eq:defSFPR}), and hence the better is the quality of the trade-off relation. 

%The thermodynamic uncertainty relation  quantifies uncertainty  with the ratio $\langle \T\rangle^2/\langle \T^2\rangle - \langle \T\rangle^2$ of the first-passage time, leading to the inequality \cite{gingrich2017fundamental}
%\begin{equation}
%   \hat{s}_{\rm TUR} \equiv   \frac{2\langle \T\rangle }{\langle \T^2 \rangle - \langle \T\rangle^2}  \geq \dot{s} (1+o_{\ell_{\rm min}}(1)), \label{eq:TUR}
%\end{equation}
%where  $o_{\ell_{\rm min}}(1)$ is the little-o notation denoting a function that converges to zero when  $\ell_{\rm min}\gg 1$.    On the other hand, the speed-uncertainty-dissipation trade-off relation quantifies uncertainty with the splitting probability
%\begin{equation}
%     p_- \equiv \log \mathcal{P}\left[J(\T)\leq -\ell_-\right], 
%\end{equation}
% to hit the negative threshold first, leading to the %inequality 
%\begin{equation}
% \hat{s}_{\rm FPR} \equiv  \frac{\ell_+}{\ell_-}  \frac{|\log p_-|}{\langle \T\rangle}  \geq \dot{s} (1+o_{\ell_{\rm min}}(1)). \label{eq:tightIneq}
%\end{equation}

Using martingale methods, we  show that for currents that are proportional to the entropy production, viz., 
\begin{equation}
    J_t=cS^{\rm tot}_t,
\end{equation}
where $c$ is a constant,  it holds that  
\begin{equation}
\hat{s}_{\rm FPR} = \langle \dot{S}^{\rm tot}_t\rangle, 
\end{equation}
and hence the  speed-uncertainty-dissipation trade-off relation is optimal in this case.   
 Indeed,   Eq.~(\ref{eq:pprob4})  in the limit of $\ell_-,\ell_+\gg 1$ implies 
\begin{equation}
    P_- = \exp(-\ell_-/c (1+o_{\ell_{\rm min}}(1))). \label{eq:pMAsympt}
\end{equation}
In addition, since in this case ~\cite{neri2022universal},
\begin{equation}
    \langle \T\rangle = \frac{\ell_+}{c\langle S^{\rm tot}_t\rangle }(1+o_{\ell_{\rm min}}(1)),   \label{eq:meanFPTLarge}
\end{equation}
we obtain from Eqs.~(\ref{eq:pMAsympt}) and (\ref{eq:meanFPTLarge}) the  equality
\begin{equation}
    \hat{s}_{\rm FPR} = \langle \dot{S}^{\rm tot}_t\rangle
\end{equation}
for currents  that are proportional to $S^{\rm tot}_t$.

\subsubsection{Comparing $\hat{s}_{\rm FPR}$ with $\hat{s}_{\rm TUR}$ in a simple example of a nonequilibrium process}\label{sec:simple2D}
Let us now compare $\hat{s}_{\rm FPR}$ with $\hat{s}_{\rm TUR}$ for the  general case of currents $J_t$ that are not necessarily proportional to $S^{\rm tot}_t$ in a simple model of a nonequilibrium process $X$, as  done in  Ref.~\cite{neri2022estimating}.

We consider the process $X=(X^{(1)}_t,X^{(2)}_t)$ describing the position of a  particle that jumps on a two-dimensional lattice at rates $\omega^+_{1}$, $\omega^-_1$, $\omega^+_2$, and $\omega^-_2$, for which we assume that  $\omega^+_{1}>\omega^{-}_1$ and $\omega^+_2>\omega^-_2$, see Panel (b) of Fig.~\ref{fig:tradeoff2} for an illustration.   In this example, empirical currents take the form 
\begin{equation}
    J_t = (1-\Delta)X^{(1)}_t + (1+\Delta)X^{(2)}_t, \label{eq:JDef}
\end{equation}
and when 
\begin{equation}
    \Delta = \frac{\ln (\omega^+_2/\omega^{-}_2) - \ln (\omega^{+}_1/\omega^{-}_1) }{\ln (\omega^{+}_2/\omega^{-}_2) + \ln (\omega^{+}_1/\omega^{-}_1) }, \label{eq:DeltaKM2}
\end{equation}
the current $J_t$ is proportional to $S^{\rm tot}_t$.

The mean rate of dissipation is, from definition Eq.~(\ref{eq:SDotResultx}), given by 
\begin{equation}
   \langle \dot{S}^{\rm tot}_t\rangle = (\omega^+_1-\omega^-_1) \ln \frac{\omega^+_1}{\omega^-_1} + (\omega^+_2-\omega^-_2)  \ln \frac{\omega^+_2}{\omega^-_2} \label{eq:sdotAverageHere} .
\end{equation}
{Note that here we have applied Eq.~(\ref{eq:SDotResultx}) and assumed periodic boundary conditions in the two-dimensional lattice which leads to a homogeneous steady-state density, i.e. $\rho_{\rm st}(x)$ to be independent of $x$.}

To determine $\hat{s}_{\rm FPR}$ and $\hat{s}_{\rm TUR}$, we use   in  Appendix~\ref{sec:MFPT}
 martingales and   the technology of  Doob's    optional stopping theorems, as discussed in  Sec.~\ref{sec:stop}, to determine  an explicit expression  for the  splitting probability $P_-$, the mean first-passage time $\langle \T\rangle$, and the variance $\langle \T^2\rangle -\langle \T\rangle^2$, yielding 
\begin{equation}
    \hat{s}_{\rm FPR} =  |z^\ast|\left((1-\Delta)(\omega^+_1-\omega^-_1) + (1+\Delta)(\omega^+_2-\omega^-_2)\right)(1+o_{\ell_{\rm min}}(1)),  \label{eq:SFPR}
 \end{equation} 
 where $z^\ast$ is the nonzero solution to  
 \begin{eqnarray}
  0&=&  [1-\exp(z^\ast (1-\Delta))]\omega^{+}_1  + [1-\exp(-z^\ast  (1-\Delta))]\omega^{-}_1\nonumber\\
  &+&  [1-\exp(z^\ast  (1+\Delta))]\omega^{+}_2 + [1-\exp(-z^\ast  (1+\Delta))]\omega^{-}_2, \label{eq:fDefx}
 \end{eqnarray}
and 
\begin{equation}
    \hat{s}_{\rm TUR} = 2 \frac{[(1-\Delta)(\omega^+_1-\omega^-_1) +(1+\Delta)^2(\omega^+_2-\omega^-_2)]^2}{(1-\Delta)^2(\omega^+_1+\omega^-_1) +(1+\Delta)^2(\omega^+_2+\omega^-_2) }. \label{eq:STUR}
\end{equation}

In Panel (c) of Fig.~\ref{fig:tradeoff2}, we use the Eqs.~(\ref{eq:sdotAverageHere}), (\ref{eq:SFPR}) and (\ref{eq:STUR}), to  plot $\hat{s}_{\rm FPR}/\langle \dot{S}_{\rm tot}\rangle$ and $\hat{s}_{TUR}/\langle \dot{S}_{\rm tot}\rangle$ as a function of $\Delta$.    Observe that for $\Delta$ given by Eq.~(\ref{eq:DeltaKM2}), as indicated by the vertical dotted line in Fig.~\ref{fig:tradeoff2}, the inequality for $\hat{s}_{\rm FPR}$ is tight, as predicted by martingale theory.   In addition, for all values of $\Delta$ it holds that $\langle \dot{S}^{\rm tot}_t\rangle \geq \hat{s}_{\rm FPR}\geq \hat{s}_{\rm TUR}$, and hence  $\hat{s}_{\rm FPR}$ is in this example a better estimator of dissipation.

In Panel (d) of Fig.~\ref{fig:tradeoff2}, we plot  $\hat{s}_{\rm FPR}$ and $\hat{s}_{\rm TUR}$,  as a function of $\langle \dot{S}_{\rm tot}\rangle$.   This figure   reveals that $\hat{s}_{\rm FPR} = \hat{s}_{\rm TUR}$ near equilibrium ($\langle \dot{S}_{\rm tot}\rangle\approx 0$), whereas in the opposing nonequilibrium limit it holds that $\hat{s}_{\rm TUR}/\langle \dot{S}^{\rm tot}_t\rangle \rightarrow 0$,   whereas $\hat{s}_{\rm FPR}/\langle \dot{S}^{\rm tot}_t\rangle$ converges to a finite nonzero  value   for increasing values of $\langle \dot{S}^{\rm tot}_t\rangle$, which is indicated by the blue dashed line in the figure.    Hence, far from equilibrium $\hat{s}_{\rm TUR}$ captures a negligible fraction of the dissipation, while $\hat{s}_{\rm FPR}$ captures a finite fraction of the dissipation.

% In addition, we show that 
%\begin{equation}
%\langle \T\rangle = 
%\end{equation}
%and  the variance  
%\begin{equation}
%\langle \T^2\rangle - \langle \T\rangle^2 = .
%\end{equation} 
%This provides 
%\begin{equation}
%    q_{\rm FPR} = 
%\end{equation}
%and 
%\begin{equation}
%    q_{\rm TUR} = 
%\end{equation}
%both of which are plotted in Fig.~\ref{}. 

\section[Overcoming classical limits with stopping times]{Overcoming classical thermodynamic limits by stopping at a clever moment}\label{sec:overcoming}

The second law of thermodynamics at stopping times, given by Eq.~(\ref{eq:secondlawStop}), states that it is not possible to reduce entropy by stopping at  a clever moment.     This law applies to the total entropy production $S^{\rm tot}_t$ and implies that a demon cannot reduce entropy, not even when it is infinitely smart and has complete knowledge of the past.    

However, there exist observables $Y_t(X_{[0,t]})$   that obey a classic second law of thermodynamics, in the sense that 
\begin{equation}
    \langle Y_t\rangle \geq0,
\end{equation}
but do not obey a second law at stopping times, in the sense that 
\begin{equation}
    \langle Y_\T\rangle \ngeq0,
\end{equation}
i.e. its average at stopping times is {\em not necessarily} greater or equal than zero.
A notable example of such an observable is the  %negative 
heat dissipated $-Q_t$, as defined in Eq.~(\ref{eq:qs2}), for stationary,  isothermal, overdamped, unidimensional  Langevin processes given by Eq.~(\ref{eq:lang2}).  In this case, the second law of thermodynamics implies that the heat decreases on average, but  nevertheless, a demon can use stopping times $\T$ to overcome this classical thermodynamic limit.  More generally, for generic stationary   processes the environment entropy change $S^{\rm env}_t$, defined in~\eqref{Stotcopied}, 
 obeys, in one hand\footnote{This follows from  stationarity which gives  $\langle \Delta S^{\rm sys}_t\rangle = 0 $, and the fact that the second law~\eqref{Nis} holds for all normalized $\mQ$.} 
\begin{equation}
    \langle S^{\rm env}_t\rangle \geq 0,  \label{eq:classicallimit}
\end{equation}
even though  $S^{\rm env}_t$ is not a submartingale.
On the other hand,
\begin{equation}
    \langle S^{\rm env}_\T\rangle \geq -\langle \Delta S^{\rm sys}_\T \rangle \ngeq 0, 
    \label{eq:Senvnosecondlaw}
\end{equation}
i.e. the average environmental entropy change at stopping times is not necessarily greater or equal than zero.
Equation~\eqref{eq:Senvnosecondlaw} implies that a demon can overcome the classical limit Eq.~(\ref{eq:classicallimit})  by  stopping a process at a cleverly chosen moment $\T$, as  anticipated by Maxwell, see e.g~\cite{leff2002maxwell}.    Note that the operation of such a  demon relies crucially on  (i) the possibility to stop a process at a  random  time; and (ii) the fact that we ignore changes in the entropy of the demon itself.    

In what follows, we discuss two examples of cases for which a demon can use stopping times to overcome the  classical limit Eq.~(\ref{eq:classicallimit}) in isothermal (Sec.~\ref{sec:isothdemon}) and non-isothermal (Sec.~\ref{sec:gyrdemon})  conditions.

\subsection{Heat extraction from stopping at a cleverly chosen moment}
\label{sec:isothdemon}

\begin{figure}[h]
    \centering
      \includegraphics[width=0.45\textwidth]{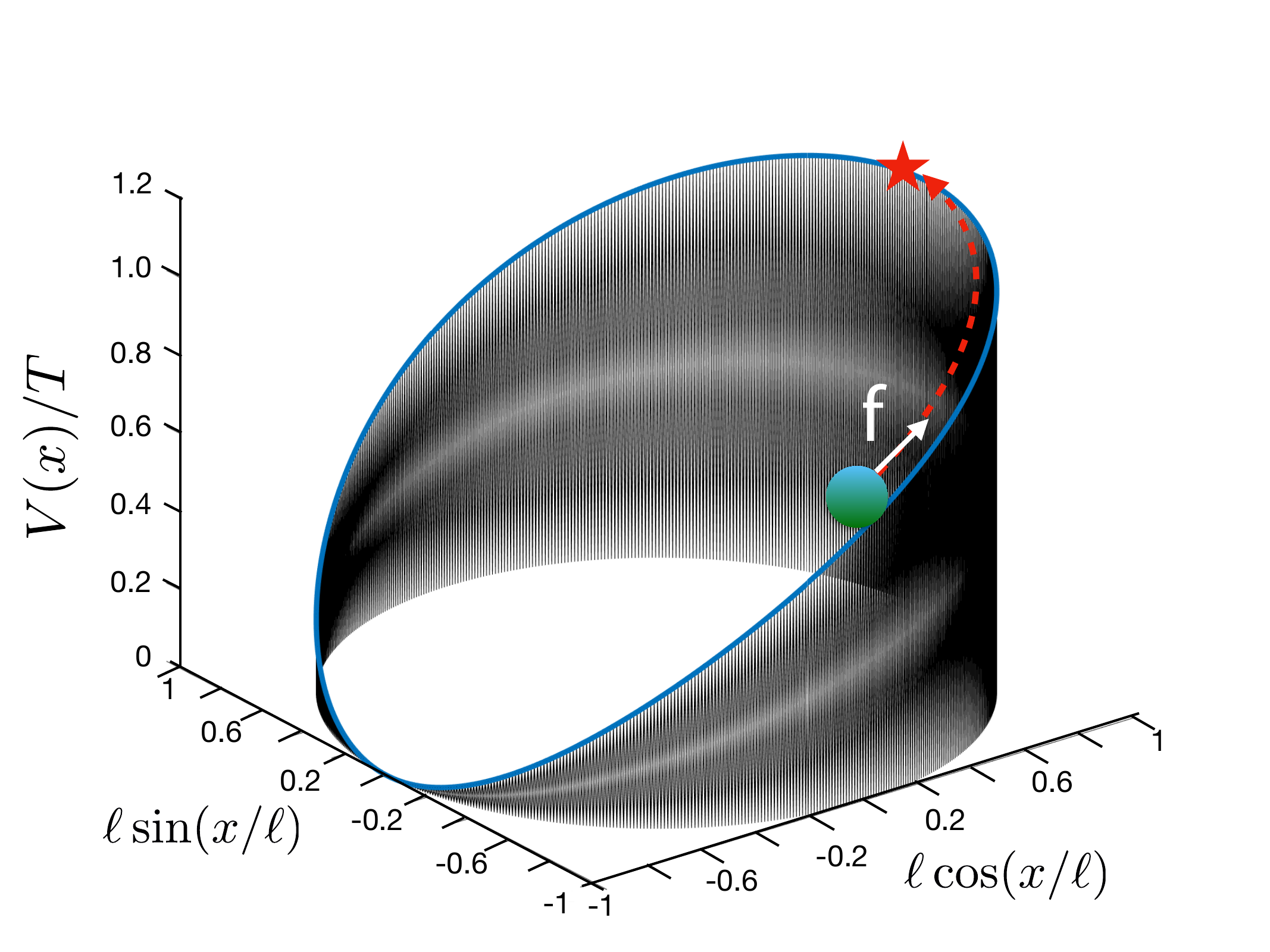}
    \includegraphics[width=0.5\textwidth]{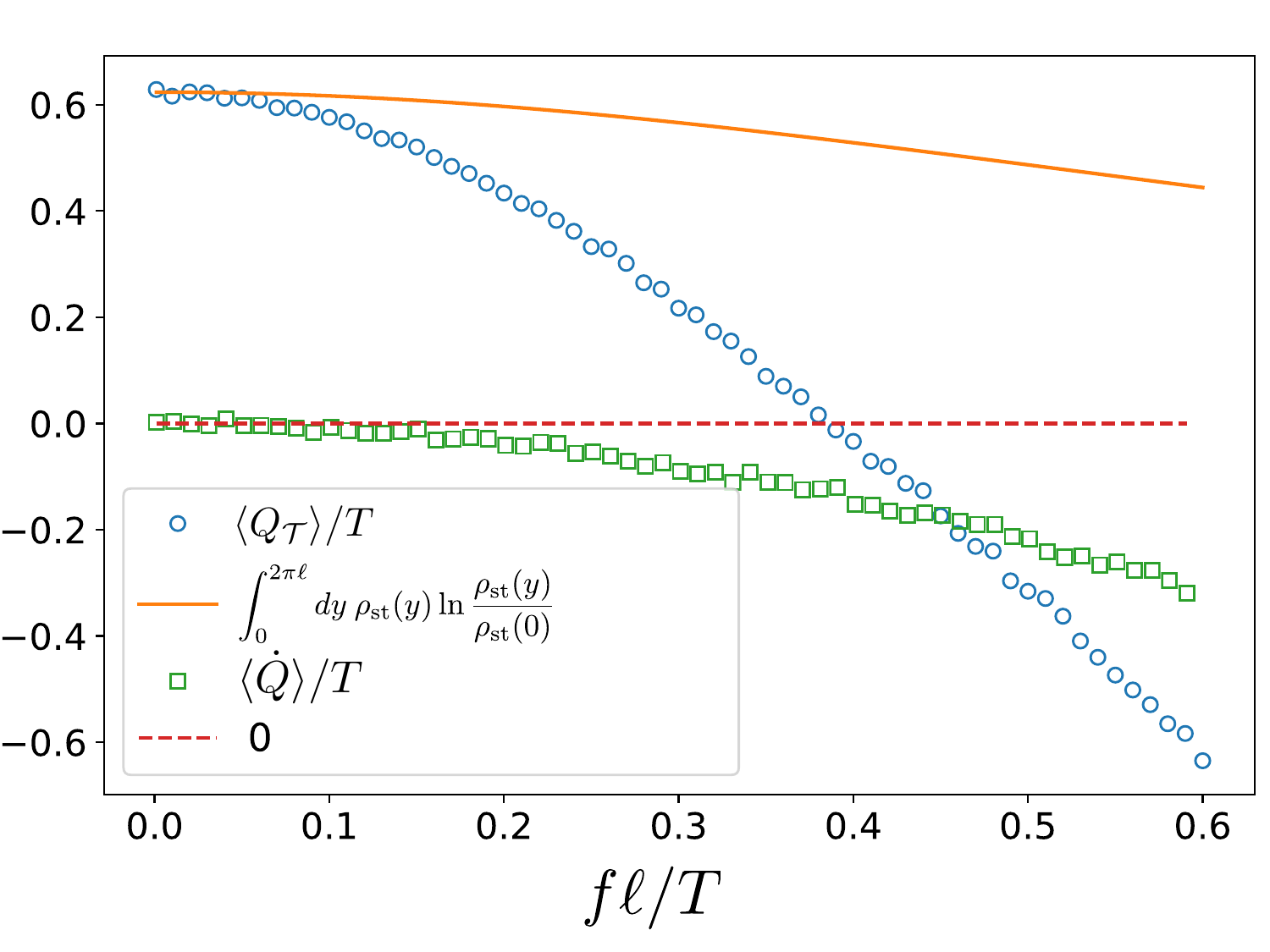}
     \caption{Heat extraction by a colloidal particle from an  reservoir at temperature $T$.     Left Panel: illustration of the model used in the right panel, viz.,  a colloidal particle on a ring.   The position of the particle is described by Eq.~(\ref{eq:lang2}) with potential  $V(x) =  T\ln(\cos (x/\ell)+2)$, and constant force $f$.    Right Panel: the average heat $\langle Q_\T\rangle$ at the stopping time $\T$, defined in Eq.~(\ref{eq:stopUsed}) and illustrated in the Left Panel by a star,  and the average heat rate $\langle \dot{Q}\rangle$, both plotted as a function of the forcing $f\ell/T$. {Here, we estimate the average heat rate as $\langle \dot{Q}\rangle \simeq \langle Q_t\rangle/t $ using empirical averages and~$t$ sufficiently large, which leads to $\langle \dot{Q}\rangle\leq 0$ by virtue of $\langle Q_t\rangle = -T\langle S^{\rm env}_t \rangle  \leq 0$, see Eq.~\eqref{eq:classicallimit}.  %The model considered is shown  in the Left Panel.
     Simulation results are in agreement with the classical second law $\langle Q_t\rangle\leq 0$ [Eq.~\eqref{eq:SecondQt}] and the second law at stopping times Eq.~\eqref{eq:SLSTI}, specialized for this example as $\langle Q_\T\rangle \leq T\int^{2\pi}_0 dy \rho_{\rm st}(y)\ln [\rho_{\rm st}(y)/\rho_{\rm st}(0)]$ [Eq.~\eqref{eq:SLSTI3}]. } %     compared with the upper bounds   [right-hand side of Eq.~(\ref{eq:SLSTI}) multiplied by $-1/T$]  and $0$ [right-hand side of Eq.~(\ref{eq:SecondQt})] for $\langle Q_\T\rangle$ and $\langle \dot{Q}\rangle$, respectively.   
     In the simulations the parameters were set to $T=1$, $\mu=1$, and $\ell=1$.    
     Figures taken from Ref.~\cite{neri2019integral}. }
    \label{fig:th}
\end{figure}
%\subsection{Stationary case}
Let $X$ be the position of a colloidal particle described by the one-dimensional Langevin process Eq.~(\ref{eq:lang2}). 
As already anticipated in the introduction of this section,  the negative heat $-Q_t$ obeys the classical second law
\begin{equation}
    -\langle Q_t\rangle \geq 0 ,  \label{eq:SecondQt}
\end{equation}
which follows from inserting Eq.~(\ref{eq:Stotlangevin0}) into Eq.~(\ref{eq:secondlaw}) and using that for   stationary systems $\langle \Delta  S^{\rm sys}_t\rangle = 0$;  $S^{\rm sys}_t$ is the system entropy as defined in Eq.~(\ref{eq:defSSysonetime}).  
On the other hand, from the  second law at stopping times  Eq.~(\ref{eq:secondlawStop}) it follows that 
\begin{equation}
-    \la Q_\T \ra \geq   - T\la  S^{\rm sys}_\T\ra +  T\la S^{\rm sys}_0\ra .
    \label{eq:SLSTI}
\end{equation} 
Since the right-hand side of Equation~\eqref{eq:SLSTI} can be negative, the negative heat does not satisfy a second law at stopping times, and  a colloidal particle can in principle absorb heat from a thermal reservoir by stopping at  a time $\T$ defined by a suitable prescribed criterion.  

In Fig.~\ref{fig:th}, we illustrate  heat extraction  for   a colloidal particle that moves in a nonconstant potential on a ring under the influence of a  nonconservative force.   The position of the colloidal particle is described by   Eq.~(\ref{eq:lang2}) with periodic boundary conditions,  a constant, nonconservative force $f$,  and  a  potential $V(x) = T \ln(\cos(x/\ell)+2)$, as considered before {in Fig.~\ref{fig:infimacdf}}.      The stopping criterion we implement is shown in the left panel of Fig.~\ref{fig:th}:   we stop the process as soon as the colloidal particle reaches the peak of the potential located at $x=0$, i.e., 
\begin{equation}
    \T = {\rm inf}\left\{t\geq 0: X_t = 0\right\}.  \label{eq:stopUsed} 
\end{equation}
As shown in the right panel of Fig.~\ref{fig:th}, as long as $f$ is small enough,  the system extracts on average heat from the thermal reservoir at the stopping time $\T$, i.e., $\langle Q_\T\rangle \geq 0$ {(see blue squares in the right panel in Fig.~\ref{fig:th})}, and the amount of heat that can be extracted is upper bounded by the   second law at stopping times given by Eq.~(\ref{eq:SLSTI}).
{More precisely, in this example the system's dynamics is initialized in the stationary state $\rho_{\rm st}(x)$ whereas $\rho_{\rm st}(X_\T)=\rho_{\rm st}(0)$. Thus the average system entropy change up to the stopping time~\eqref{eq:stopUsed} reads $\la  S^{\rm sys}_{\T}\ra  -\la S^{\rm sys}_0 \ra =\int^{2\pi}_0 dy \rho_{\rm st}(y)\ln \rho_{\rm st}(y)/\rho_{\rm st}(0)$.  As a result, the second law at stopping times~\eqref{eq:SLSTI}, copied here for convenience
\begin{equation*}
   -   \frac{ \la Q_\T \ra}{T} +[\la  S^{\rm sys}_\T\ra -  \la S^{\rm sys}_0\ra]  \geq  0  ,  
\end{equation*}
is specialized for this example as an upper bound for the averaged absorbed heat up to the stopping time $\T$ given by Eq.~\eqref{eq:stopUsed}, i.e.
\begin{equation}
  \frac{ \la Q_\T \ra}{T} \leq \int^{2\pi}_0 dy \rho_{\rm st}(y)\ln \frac{\rho_{\rm st}(y)}{\rho_{\rm st}(0)}   .
    \label{eq:SLSTI3}
\end{equation} 
We provide a numerical verification of the inequality~\eqref{eq:SLSTI3} in the right panel of  Fig.~\ref{fig:th}, which  shows  that such second law at stopping times is tight when the system is near equilibrium, i.e., when $f\approx 0$. }

\subsection{Super Carnot efficiency at stopping times}
\label{sec:gyrdemon}

{\em Steady-state heat engines} are thermal machines that are permanently in contact with  two thermal reservoirs, one at hotter    $T_{\rm h}$ and another at a colder $T_{\rm c}\leq T_{\rm h}$ temperature. After a transient, the engine achieves an average stationary heat flow from the hot to the cold reservoir that can be used to extract power. A key example of a steady-state heat engine is {\em Feynman's ratchet} where a ratchet and a pawl are immersed in two gas containers held at different temperatures. As shown earlier~\cite{leighton1965feynman,parrondo1996criticism}, the nonequilibrium constraint  $T_{\rm c}\neq T_{\rm h}$ results in a net extraction of work which can be used e.g. to lift a weight against the gravitational pull.

It is well known  the key role of fluctuations in determining the thermodynamic performance of steady-state heat engines~\cite{hanggi2009artificial,seifert2012stochastic}.  However, only very recently  thermodynamic insights of such machines at stopping times have been unveiled with the help of martingales~\cite{neri2019integral,manzano2021survival}.   For example, an important question is what is the average heat transfer  between two "main events" corresponding to two consecutive passages  in the teeth of Feynman's ratchet wheel?

The average thermodynamic fluxes in steady-state heat engines over a fixed time interval $[0,t]$  obey $\langle W_t\rangle\leq 0$ (work extraction), $\langle Q_{\rm h}\rangle\geq 0$ (absorption of heat from the hot bath), and $\langle Q_{\rm c}\rangle\leq 0$ (dissipation of heat in the cold bath). The first law of thermodynamics implies
\begin{equation}
    \langle \dot{W} \rangle + \langle \dot{Q}_{\rm c}\rangle+ \langle \dot{Q}_{\rm h}\rangle=0
    \label{eq:1lsshe}
\end{equation}
 and the second law  for  steady-state heat engines
 \begin{equation}
     \langle \dot{S}^{\rm tot}_t\rangle = -\langle \dot{Q}_{\rm c}\rangle/T_{\rm c}- \langle \dot{Q}_{\rm h}\rangle/T_{\rm h}\geq 0,
     \label{eq:2lsshe}
 \end{equation}
which follows from stationarity. Combining Eq.~\eqref{eq:1lsshe} and~\eqref{eq:2lsshe}, one finds that the long-time efficiency (defined analogously as in classical heat engines) of the engine is always smaller or equal than Carnot efficiency, i.e.
\begin{equation}
    \eta = \frac{-\langle \dot{W} \rangle}{\langle \dot{Q}_{\rm h}\rangle}\leq \underbrace{1-\frac{T_{\rm c}}{T_{\rm h}}}_{\displaystyle\equiv \eta_C}.
    \label{eq:standardeff}
\end{equation}

We now ask the question: what are the implications of the second law of thermodynamics at stopping times~\eqref{eq:secondlawStop} concerning the efficiency achieved by a steady-state heat engine cleverly stopped at a stochastic time $\T$? To this aim, we consider the {\em stopping-time efficiency} $\eta_{\T}$   associated with the stopping time $\T$ as 
%enable work extraction $\langle W_t\rangle\leq 0 $, w
\begin{eqnarray}
\eta_{\T} \equiv \frac{-\langle W_\T\rangle/\langle\T \rangle }{\langle Q_{\rm h,\T}\rangle/\langle\T \rangle } = \frac{-\langle W_\T\rangle}{\langle Q_{\rm h,\T}\rangle},
\label{eq:stoppingtimeeff}
\end{eqnarray}
where $-\langle W_\T\rangle$  and $\langle Q_{\rm h,\T}\rangle$ are respectively the average work extracted and the average heat absorbed from the hot bath in the time interval $[0,\T]$. In general, trajectories $X_{[0,\T]}$ are not cyclic, i.e.  $\rho_{\T}\neq \rho_0 $.  This implies that the first law averaged over many  trajectories $X_{[0,\T]}$ stopped at a stochastic time $\T$ reads 
\begin{equation}\label{eq:1lavstop}
    \langle {W}_\T \rangle + \langle {Q}_{{\rm c},\T}\rangle+ \langle {Q}_{{\rm h},\T}\rangle=\langle \Delta V_\T \rangle ,
\end{equation}
Here, $\Delta V_\T= V(X_\T)-V(X_0)$ is the energy change in $[0,\T]$, which one cannot simply neglect with respect to the average heat and work done up to the stopping time--as in the traditional first law~\eqref{eq:1lsshe}. Similarly, the second law of thermodynamics at stopping times~\eqref{eq:secondlawStop} reads in this case 
\begin{equation}\label{eq:2lavstop}
    \langle S^{\rm tot}_\T \rangle = \langle \Delta S^{\rm sys}_\T \rangle -\langle Q_{{\rm c},\T}\rangle/T_{\rm c}- \langle Q_{{\rm h},\T}\rangle/T_{\rm h}\geq 0,
\end{equation}
with $\Delta S^{\rm sys}_\T$ the system entropy change in $[0,\T]$. 
The second law~\eqref{eq:2lavstop} reveals something interesting, namely,  the stopping time carries an addtional system entropy term with respect to the traditional second law~\eqref{eq:2lsshe}.
We also note that $Y_t = -Q_c/T_{\rm c} - Q_h/T_{\rm h}$ satisfies a second law of thermodynamics at fixed times but not at stopping times, and this is a key property that allows to overcome classical limits.
In particular, combining  Eqs.~\eqref{eq:stoppingtimeeff},~\eqref{eq:1lavstop} and~\eqref{eq:2lavstop} we obtain
 \begin{eqnarray} 
\eta_{T} \    \leq    \eta_{C} -  \frac{\langle \Delta G^{\rm ne}_{{\rm c},\T}\rangle}{\langle Q_{{\rm h},\T}\rangle }, \label{eq:carnotStoppxxxx}
  \end{eqnarray} 
with 
\begin{equation}
G^{\rm ne}_{\rm c,\T}= V(X_\T)\ -T_{\rm c} S^{\rm sys}_{\T},
\end{equation}
is the  nonequilibrium free energy of the system at stopping times with respect to the cold thermal bath.    Notably,   the second term in the right-hand side of (\ref{eq:carnotStoppxxxx}) may be positive for specific "clever" choices of stopping times.  Therefore, the second law of thermodynamics at stopping times does not prevent stopping-time efficiencies  $\eta_\T$
to surpass the  Carnot efficiency.
%of steady-state  heat engines at stopping times can be greater than Carnot efficiency. %This is because using a stopping time the system is, in general, no longer stationary or cyclic.
\begin{figure}[h!]
    \centering
    \includegraphics[width=1\textwidth]{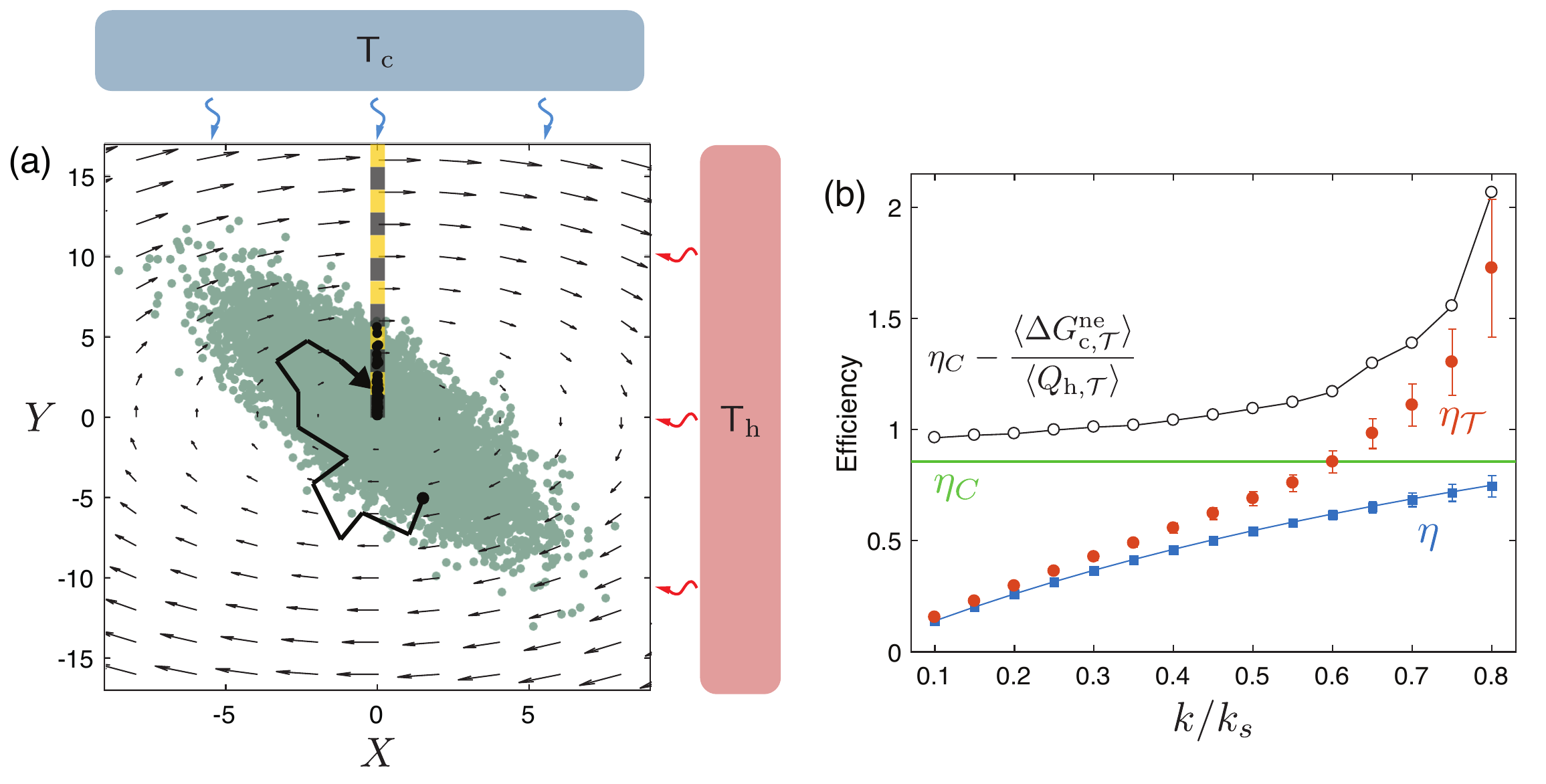}
    \caption{Efficiency of a gyrator that stops at a cleverly chosen time. (a) Illustration of the model: an overdamped Brownian particle moves in one dimensions, $X$ and $Y$, trapped in an elliptical potential~\eqref{eq:potgyrator},  under the action of an external, non-conservative force field~\eqref{eq:forcesgyrator} (black arrows) and subject to thermal fluctuations of different temperatures $T_{\rm h}$ and $T_{\rm c}<T_{\rm h}$ along the $X$ and $Y$ coordinates, respectively. The thick black arrow illustrates a single trajectory stopped at $\mathcal{T}$ given by Eq.~\eqref{eq:Tme} being the first time at which the "barrier" (thick line with black and orange stripes) is crossed by crossing from the second to the first quadrant. The green filled circles show the initial state of $100$ independent realizations drawn from the stationary state, whereas the black filled circles are their value at the stopping time $\mathcal{T}$. (b)   Efficiencies as a function of the stiffness $k$ of the non-conservative force (in units of the "stalling" stiffness $k_{\rm s} \equiv c \eta_{ C}/(2-\eta_{C})$ at which the net current vanishes), viz., the
    long-time  efficiency $\eta$ (Eq.~\eqref{eq:standardeff}, blue squares, simulations; blue solid line, theory), the stopping-time  efficiency $\eta_\T$ (Eq.~\eqref{eq:stoppingtimeeff}, red circles), and  the upper bound to the stopping time efficiency dictated by the second law of thermodynamics at stopping times (right-hand side in Eq.~\eqref{eq:carnotStoppxxxx}, black open circles). 
    The horizontal green  line is set at Carnot efficiency $\eta_C=1-(T_{\rm c}/T_{\rm h})=7$. Parameter values:  $\mu=1$, $u_1=1$, $u_2=1.2$, $T_{\rm c}=1$, $T_{\rm h}=7$,  $c=0.9$,  $10^4$ independent realizations, and simulation time step $\Delta t=10^{-3}$, see~\cite{neri2019integral} for further details.
    }
    %at fixed times $\eta_T =-\langle W(T)\rangle/\langle Q_{\rm h}(T)\rangle$  as a function of $k$ for fixed $T= 5$ (blue squares) and  efficiencies $\eta_{T_{\rm me}} =-\langle W(T_{\rm me})\rangle/\langle Q_{\rm h}(T_{\rm me})\rangle$ at the main event  $T=T_{\rm me}$ defined by Eq.~(\ref{eq:Tme}) (red circles), both  obtained from numerical simulations.  The blue line is the theoretical value of the  efficiency  at large times given by Eq.~(\ref{eq:etaTh}),  and the black circles correspond to  the bound~(\ref{eq:carnotStoppxxxx}); the black line is a guide to the eye. The horizontal black dashed line is set at Carnot efficiency. Values of the parameters are $\mu=1$, $u_1=1$, $u_2=1.2$, $\mathsf{T}_{\rm c}=1$, $\mathsf{T}_{\rm h}=7$, and $c=0.9$; the averages  are estimated from $10^4$ independent realisations initially sampled  from the stationary state. Numerical simulations are performed with the   Heun's numerical integration scheme with a time step $\Delta t=10^{-3}$ \cite{sekimoto2010stochastic}. Error bars  denote the  standard errors of the empirical mean. }
    \label{fig:gyr}
\end{figure}

For illustrational purposes, we borrow  from Ref.~\cite{neri2019integral} the illustration of the bound~\eqref{eq:carnotStoppxxxx} applied
a paradigmatic model of a steady-state engine, namely the Brownian gyrator which was introduced in Ref.~\cite{filliger2007brownian} and realized experimentally in~\cite{argun2017experimental}, see also Refs.~\cite{pietzonka2018universal,cerasoli2018asymmetry,manikandan2019efficiency} for theoretical insights. The model is described by a two-dimensional Langevin equation describing e.g. the  motion of an overdamped Brownian particle in an  elliptical confining potential   that is subject to two nonequilibrium constraints: (i) two thermal baths at temperatures $T_{\rm h}$ and $T_{\rm c}<T_{\rm h}$ each acting only along the $x$  and $y$ axes, respectively; and (ii) an external torque generated by external, non-conservative forces. See Fig.~\ref{fig:gyr}a for an illustration of the Brownian gyrator. The equations of motion of the model read (cf. Eq.~\eqref{eq:diff})
\begin{equation}
{\left(\begin{array}{c}
\dot{X}_{t}\\
\dot{Y}_{t}
\end{array}\right)}=-\mu\left(\begin{array}{c}
\partial_{x}V\left(X_{t},Y_{t}\right)\\
\partial_{y}V\left(X_{t},Y_{t}\right)
\end{array}\right)+\mu\left(\begin{array}{c}
f_{x}\left(X_{t},Y_{t}\right)\\
f_{y}\left(X_{t},Y_{t}\right)
\end{array}\right)+\left(\begin{array}{cc}
\sqrt{2 \mu T_{h}} & 0\\
\text{0} & \sqrt{2 \mu T_{c}}
\end{array}\right)\left(\begin{array}{c}
\dot{B}_{x}\\
\dot{B}_{y}
\end{array}\right).
\label{eq:gyrator}
\end{equation}
   %\begin{eqnarray}
  %\dot{X}_t&=& -\mu \partial_x V(X_t,Y_t) + \mu   f_x (X_t,Y_t)+  \sqrt{2T_{\rm h}} \dot{B}_1, \label{eq:gyrator1} \\
  %\dot{Y}_t&=& -\mu \partial_y V(X_t,Y_t)  + \mu   f_y (X_t,Y_t)+  \sqrt{2T_{\rm c}} \dot{B}_2. \label{eq:gyrator2}
   % \end{eqnarray} 
    In Eq.~(\ref{eq:gyrator}), the potential    \begin{equation}
    V(x,y)=\frac{1}{2}\left( u_1 x^2 + u_2 y^2 + c xy  \right),
    \label{eq:potgyrator}
    \end{equation}
    with $u_1,u_2>0$, and $0<c< \sqrt{u_1 u_2}$ (see Ref.~\cite{pietzonka2018universal}). Furthermore,   the two components of the external non-conservative force are 
    \begin{equation}
    f_x (x,y) = k y,\quad f_2 (x,y) = -kx, \label{eq:forcesgyrator}
    \end{equation}
    and $B_1$ and $B_2$ are two independent Wiener processes, 

%is the average work exerted on the system in the time interval $[0,T]$, and $\langle Q_{\rm h}(T)\rangle$ is the heat absorbed by the system from the hot reservoir at temperature $\mathsf{T}_{\rm h}$ in the time interval $[0,T]$. For example, if $T$ is the time that a steady-state heat engine takes to reach a  region  $\Gamma$ in phase space, $\eta_T$ equals to the ratio of the average work extracted and the heat absorbed from the hot reservoir along trajectories that reach, at stochastic times, the region $\Gamma$. 

 Amongst the infinite possible choices of stopping strategies,  Ref.~\cite{neri2019integral} considered the stopping time of first occurrence of the "main event" 
\begin{eqnarray}
\T= {\rm inf}\left\{t>0:  \varphi_{t^-} > \pi/2 , \; \varphi_{t^+} \leq  \pi/2   \right\},
\label{eq:Tme}
\end{eqnarray} 
where we assume that at $t=0$ the system is initialized in its stationary state (see green circles in Fig.~\ref{fig:gyr}a). In Eq.~\eqref{eq:Tme} the  variable $\varphi_t = \tan^{-1} (Y_t/X_t)\in [-\pi , \pi]$ is the phase associated with the state $(X_t,Y_t)$, thus $\T$ corresponds to the first crossing from the second quadrant to the first quadrant.
{The distribution at stopping times $\rho_{X_{\T},Y_{\T}}$ is concentrated near the positive $y$ axis (black circles in Fig.~\ref{fig:gyr}a), and is less broad than the initial distribution (green circles in Fig.~\ref{fig:gyr}a). Thus, the system entropy change in $[0,\T]$, $\Delta S^{\rm sys}_{\T}=S^{\rm sys}_\T - S^{\rm sys}_0$, is often negative for this example and this choice of stopping time. This result, together with the fact that   the system energy change in~$[0,\T]$, $\Delta U_{\T}=U_\T - U_0$, is often smaller than minus the system entropy change times the temperature $-T \Delta S^{\rm sys}_{\T}$, leads to positive free energy changes $\langle \Delta G^{\rm ne}_{{\rm c},_{\T}}\rangle$, which  opens up the possibility for stopping time efficiencies above the Carnot limit, see Eq.~\eqref{eq:carnotStoppxxxx}.  
Readers are referred to Ref.~\cite{neri2019integral} for details on the calculations of the free energy change at stopping times. }
 We show in Fig.~\ref{fig:gyr}b with results obtained from numerical simulations, that the stopping-time efficiency associated with the stopping time $\T$ (red circles), which satisfies the bound~\eqref{eq:carnotStoppxxxx}, can surpass the Carnot efficiency near equilibrium, a result that is inaccessible by stopping trajectories at a fixed time (blue squares). %     As said before, this is because using the stopping time $\T$ stopping time the stopped trajectories are no longer stationary or cyclic.

%\chapter[Martingales in stochastic thermodynamics IV]
\chapter{Martingales in stochastic thermodynamics IV: %Implications for physics of 
Non-stationary processes}
\label{sec:martST3}

\epigraph{\textit{La martingale est introuvable comme l'\^ame.} \\ (The martingale is as elusive as the soul.)\\Alexandre Dumas, La Femme au collier de velours, Ch.~XVIII (1850).}

We  use martingales to  further  extend classical results in stochastic thermodynamics, but this time for nonstationary processes.

To this purpose, we use  the generalized  $\Sigma$-stochastic entropic functionals, as defined in Eq.~\eqref{GEF}, for the special case of     $r=0$ and  $\mQ^{(t)}=\widetilde{\mP}^{(t)}$, where $\widetilde{\mP}^{(t)}$ is a sequence of probability measures associated with the time-reversed protocol.  For simplicity, we denote here  such    generalized  $\Sigma$-stochastic entropic functionals by $\hat{\Sigma}_s$. 
Note that in Sec.~\ref{sec:623}  we have shown  that  $\hat{\Sigma}_s$ can be decomposed  in terms of a stochastic environmental entropy flow, given by Eq.~\eqref{Senvtot}, or equivalently, in terms of a stochastic total entropy production, given by Eq.~\eqref{Stotcopied},  viz., 
\begin{equation}
\hat{\Sigma}_s
=
\ensuremath{\ln\left(\frac{\rho_{0}\left(X_{0}\right)}{\tilde{\rho}^{(t)}_{t-s}\left(X_{s}\right)}\right)}+\ensuremath{S_{s}^{\rm env}}=\ln\left(\frac{\rho_s(X_s)}{\tilde{\rho}^{(t)}_{t-s}(X_s)}\right)
+
\displaystyle S^{\rm tot}_s,
\label{forGirs-1-1aa1}
\end{equation}
which holds for $0\leq s\leq t$.
Here,  $\tilde{\rho}^{(t)}$ is the  instantaneous density associated with the time-reversed protocol for a specified  initial distribution $\tilde{\rho}^{(t)}_0=\rho_{t}$ (see Eq.~\eqref{eq:tiltpho} for its definition).

Importantly, as shown in Sec.~\ref{sec:623},  the process   $\exp(-\hat{\Sigma}_s) $ is  a  martingale with respect to $X_{[0,s]}$.  In this Chapter, we use  the martingale property of $\exp(-\hat{\Sigma}_s) $ to  derive fluctuation relations at stopping times for nonstationary nonequilibrium processes.

We initiate this Chapter with  Sec.~\ref{sec:81} that reviews     Jarzynski's equality.   Subsequently,   following  Refs.~\cite{neri2020second, Chetrite:2011},  in Sec.~\ref{sec:Jarzysnk} we extend  Jarzysnki's equality to an equality that applies at stopping times, and discuss applications of this result.   In Sec.~\ref{sec:GamblingDem}, we review another extension of Jarzynski's equality for non-stationary processes, which is then  used to design {\em gambling demons} that can extract on average more work than the free energy difference at the stopping time.

\section{Jarzynski's equality}\label{sec:81}
 Jarzynski's celebrated equality, introduced in Ref.~\cite{jarzynski1997nonequilibrium}, provides an equality between the statistics of the stochastic work done on the system and the (deterministic) equilibrium free energy change between the initial and the final states of a nonequilibrium protocol. 
As reviewed in Sec.~\ref{sec:setupJarzover} (see Eq.~\eqref{eq:jarzynskiequality}), Jarzynski's equality is given by
\begin{equation}
\left\langle \exp\left(-\frac{W_t-\Delta G^{\rm eq}}{T}\right)\right \rangle = 1,  \label{eq:jarz}
\end{equation}
where the average $\langle  \cdot\rangle$ is taken over the trajectories $X_{[0,t]}$ of a mesoscopic process that is initially at time $s=0$ in an equilibrium state,
and is for $s\in[0,t]$ driven away from equilibrium by an external protocol (after which it can be asumed to relax again to an equilibrium state).    The $\Delta G^{\rm eq}$ in Eq.~(\ref{eq:jarz}) denotes the free energy difference between the final and initial state.  
The Eq.~(\ref{eq:jarz})  was   first derived for  a Hamiltonian system in ~\cite{jarzynski1997nonequilibrium}, {and  was later extended with a Master equation approach to stochastic processes, including, Langevin processes in Refs.~\cite{rahav2013nonequilibrium, chernyak2006path, jarzynski2004classical}};  in Sec.~\ref{sec:setupJarzover} we have {rederived} the Jarzynski equality for overdamped isothermal  Langevin processes.     The Jarzynski equality implies the second law 
\begin{equation}
\langle W_t\rangle \geq \Delta G^{\rm eq},  \label{eq:JarzSecond}
\end{equation}
which states that the work done on a system must on average be larger than the free energy difference between the final and the initial state.

\section{Jarzynski equality at stopping times} \label{sec:Jarzysnk}
Events in mesoscopic systems can  happen at random times.   Hence  in order to address questions of the sort "{\it how much work is needed on average for a particle to escape a metastable state?}" or "{\it how much work is required to stretch a polymer to a certain predefined fixed length?}", we  need a  formulation of the second law  of thermodynamics that holds at random times~\cite{neri2020second}.    

We further detail the latter example, which serves as a canonical example in this Section.  
Consider a polymer with one end  attached to an anchor  fixed at position $x=0$, and the second  (dangling) end attached by a spring to a molecular motor positioned at $\lambda_0 = \lambda_{\rm i}$, as shown in the upper panel of Fig.~\ref{fig:polymer}.     At time $t=0$ the motor starts moving forwards.   Our event of interest is the binding of the   second endpoint of the polymer to an anchor located at $x=\ell$, as shown in the bottom panel of Fig.~\ref{fig:polymer}.     How much work does the motor perform on average on the polymer to complete this event of interest, and what is the corresponding second law of thermodynamics?  

Since the polymer is a mesoscopic system the position of its end point is a stochastic process, and therefore the time $\T$ when the event of interest happens  is a random variable.   Consequently,  the classical second law  of thermodynamics, Eq.~(\ref{eq:JarzSecond}), does not apply.   Instead, following Refs.~\cite{neri2020second, Chetrite:2011} we  present a generalisation of the second law of thermodynamics  that applies at random times.

\subsection{System setup}

For simplicity, we focus here on the  one-dimensional Langevin process 
\begin{equation}
    \dot{X}_s = -\mu \partial_x V(X;\lambda_s) + \sqrt{2 \mu T} \: \dot{B}_s, \label{eq:noneqLang}
\end{equation} 
where $s\geq 0$ is the time index, and 
\begin{equation}
    \lambda_s \equiv \left\{\begin{array}{ccc} \lambda_{\rm i} &{\rm if}& s\leq 0, \\ \lambda_{s} &{\rm if}& s\in [0,\tau], \\ 
    \lambda_{\rm f} &{\rm if}& s\geq  \tau, 
    \end{array}\right.
\end{equation} 
denotes a protocol that runs over a time interval   $s\in [0,\tau]$ of finite duration $\tau$; notice that we use here $s$ as a time index instead of $t$ in order to have a notation consistent with Sec.~\ref{sec:623} on generalised  entropic functionals, as it will turn out that the central quantity of interest is a  generalised $\Sigma$-stochastic entropic functional.
%; this is the same notational convention as used in Sec.~\ref{sec:623} of Chapter~\ref{ch:thermo2}, but different form Ref.~\cite{neri2020second}. 
Equation~(\ref{eq:noneqLang}) equals  Eq.~(\ref{eq:lang2}) in the absence of a nonconservative force $f_s = 0$ and for $ V_s(x) = V(x;\lambda_s)$.   We assume that the initial state 
\begin{equation}
    \rho_0(x) = \rho^{\rm eq}(x;\lambda_{\rm i})
    \label{nis-2}
\end{equation}
where 
\begin{equation}
 \rho^{\rm eq}(x;\lambda) = \exp\left(-\frac{V(x;\lambda)+G^{\rm eq}(\lambda)}{T}\right) , \quad \forall x\in \mathcal{X},  \label{eq:boltz}
\end{equation}
is the Boltzmann distribution, and 
\begin{equation}
    G^{\rm eq}(\lambda) =T \ln \left( \int_{\mathcal{X}} dx  \exp\left(-\frac{V(x;\lambda)}{T}\right)\right)
    \label{nis33}
\end{equation}
is the equilibrium free energy for a given value of the parameter $\lambda$.    
\begin{figure}[H]
    \centering
    \includegraphics[width=0.55\textwidth]{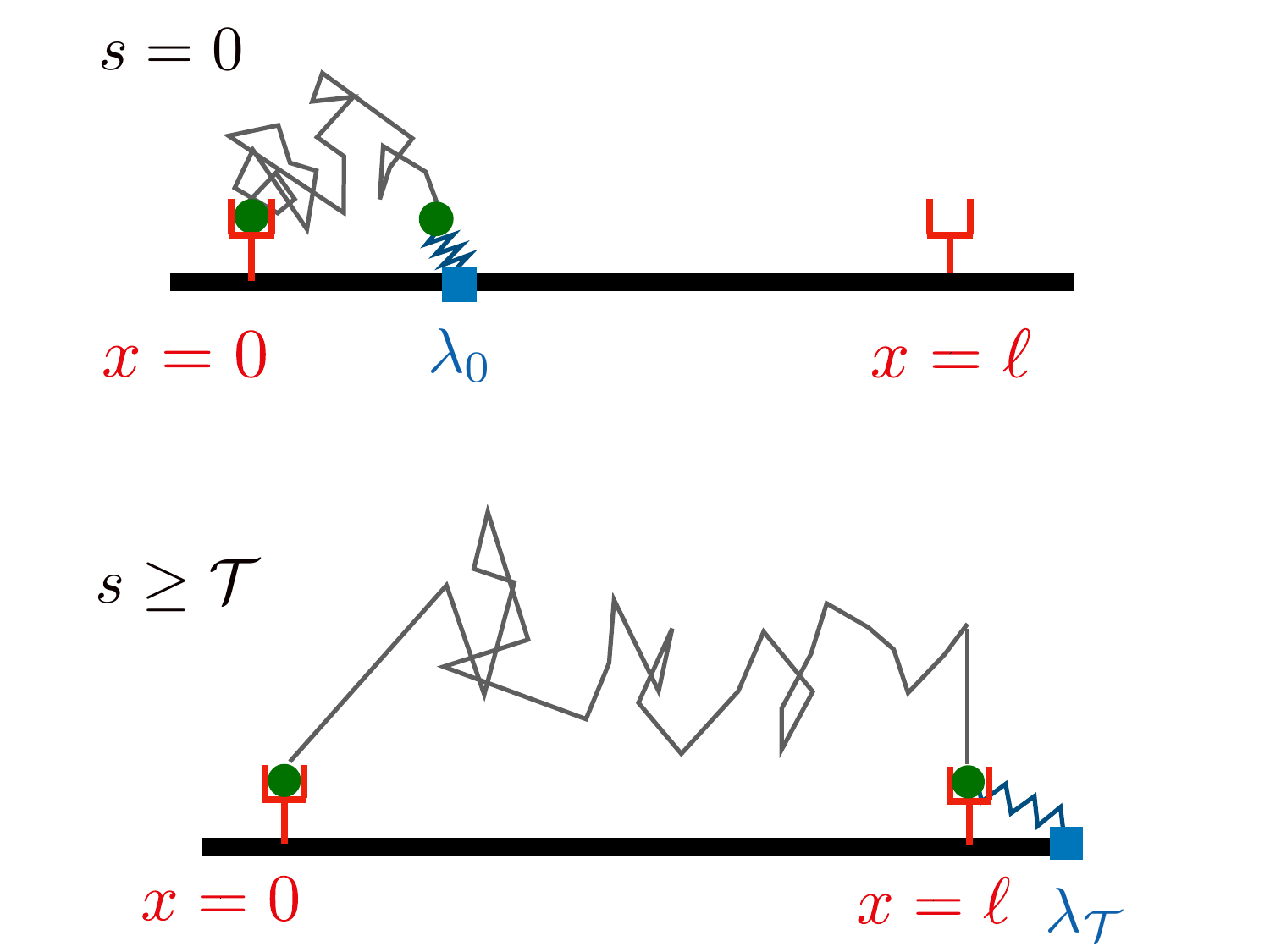}
    \caption{How much work is required to  stretch a polymer to a certain fixed length $\ell$?    A polymer (gray zigzag line) has one   end point (green circle) bound to an anchor point  (red symbol)  located at $x=0$, and has another end point    coupled to a spring representing a molecular motor located initially located at    $\lambda_0$  (blue zigzag line).    At times $s>0$ the motor  moves forwards and when  the dangling end point of the polymer reaches the anchor point located at $x=\ell$, it binds to it.   Since the dangling end point of the polymer is fluctuating, also the time $\T$ of arrival at $x=\ell$ is random, and hence for this event the second law of thermodynamics at stopping times Eq.~(\ref{eq:WTStop}) applies.       Figure taken from~\cite{neri2020second}.}
    \label{fig:polymer}
\end{figure}

\subsection{Martingale associated with $X$}\label{sec:martAsX}  
We identify a martingale, which we denote by $\exp(-\hat{\Sigma}_s)$, associated with the process $X_s$.     

%The key object of interest here is the  submartingale 
Consider the process 
\begin{equation}
\hat{\Sigma}_s = -\frac{Q_s}{T} + \ln \left(  \rho^{\rm eq}(X_0;\lambda_{\rm i})\right) -\ln \left(\tilde{\rho}^{(\tau)}_{\tau-s}(X_s)\right),  \label{eq:Ss}
\end{equation}
where
\begin{equation}
    \dot{Q}_s =  \partial_xV(X;\lambda_s)\circ \dot{X}_{s} \label{eq:heat83}
\end{equation} 
is the rate of heat absorbed by the system, as defined in   Eq.~(\ref{eq:qs3}) or \eqref{eq:6p33}, and  
where  $\tilde{\rho}^{(\tau)}_{\tau-s}$  is the solution to the Fokker-Planck equation
 \begin{equation}
    \partial_{s}\tilde{\rho}^{(\tau)}_{s}+\partial_x  \tilde{J}_{s,\tilde{\rho}^{(\tau)}}=0\label{eq:FPR1}
\end{equation} 
with the probability current 
\begin{equation}
\tilde{J}_{s,\tilde{\rho}^{(\tau)}}  = -\mu \partial_x V(x;\tilde{\lambda}_s) \tilde{\rho}^{(\tau)}_s(x)-\mu  T \partial_{x}\tilde{\rho}^{(\tau)}_{s}(x),   \label{eq:FPR2}
\end{equation} 
 with the time-reversed protocol
\begin{equation}
    \tilde{\lambda}^{(\tau)}_s \equiv \left\{\begin{array}{ccc} \lambda_{\rm f} &{\rm if}& s\leq 0, \\ \lambda_{\tau-s} &{\rm if}& s\in [0,\tau], \\ 
    \lambda_{\rm i} &{\rm if}& s\geq  \tau,   
    \end{array}\right. \label{eq:tildelambda} 
    \end{equation}  
    and with the initial state 
    \begin{equation}
\tilde{\rho}^{(\tau)}_0(x) = \rho^{\rm eq}(x;\lambda_{\rm f}).    
\label{Nis88}
\end{equation} 
  The constant term $\ln \left( \rho_{\rm eq}(X_0;\lambda_i)\right)$ in the expression (\ref{eq:Ss}) of $\hat{\Sigma}_s$ assures that 
    \begin{equation}
        \langle \exp(-\hat{\Sigma}_0)\rangle = \int_{\mathcal{X}} dx \tilde{\rho}^{(\tau)}_{\tau}(x) = 1. 
    \end{equation}

As suggested by the notation, the process $\hat{\Sigma}_s$ given by Eq.~\eqref{eq:Ss} is a particular case of the generalized $\Sigma$-stochastic entropic functional \eqref{forGirs-1-1aa1} for  $t=\tau$,  for $\tilde{\rho}^{(t)}_0$ given by Eq.~\eqref{nis-2}, and for  $S^{\rm env}$ given by Claussius' statement~Eq.~\eqref{eq:6p33}.   

%$\tilde{\rho}_{\tau-s} \equiv \rho_{\tau-s}^{\widetilde{\mP}^{(\tau)}$ is the solution at time $\tau-s$ to the Fokker-Planck equation with time-reversed protocol at time $\tau$  that starts in the equilibiurm state $\rho_{\rm eq}(x;\lambda_{\rm f})$ associated with the final state $\lambda_f$ of the protocol, viz., 
%\begin{equation}
%\tilde{\rho}_0(x) = \rho_{\rm eq}(x;\lambda_{\rm f}).    
%\label{Nis88}
%\end{equation} 
%As will become soon evident, the martingality of $\exp(-\hat{\Sigma}_s)$ requires that
%The Fokker-Planck equation associated with %$\tilde{\rho}$ is then given by 
% \begin{equation}
%    \partial_{s}\tilde{\rho}_{s}+\partial_x  \tilde{J}_{s,\tilde{\rho}}=0. \label{eq:FPR1}
%\end{equation} 
%with a probability current 
%\begin{equation}
%\tilde{J}_{s,\tilde{\rho}}  = -\mu \partial_x V(x;\tilde{\lambda}_s) \tilde{\rho}_s(x)-\mu  T \partial_{x}\tilde{\rho}_{s}(x),   \label{eq:FPR2}
%\end{equation} 
%and with a time-reversed protocol
%\begin{equation}
%    \tilde{\lambda}_s \equiv \left\{\begin{array}{ccc} \lambda_{\rm f} &{\rm if}& s\leq 0, \\ \lambda_{\tau-s} &{\rm if}& s\in [0,\tau], \\ 
 %   \lambda_{\rm i} &{\rm if}& s\geq  \tau.   
 %   \end{array}\right. \label{eq:tildelambda} 
%    \end{equation}  

In Appendix~\ref{app:chapt8}, we use the  It\^{o} integral approach from  Sec.~\ref{sec:ItoApproach} to derive a  compact It\^{o} stochastic differential equation for $\hat{\Sigma}_s$, viz., 
\begin{eqnarray}
\frac{d}{ds}\hat{\Sigma}_s = v^S_s(X_s)  + \sqrt{2 v^S_s(X_s)} \dot{B}_s, \label{eq:StVSNonStat}
\end{eqnarray}
where 
 \begin{equation}
    v^S_s= \frac{1}{\mu T} \left(\frac{\tilde{J}_{\tau-s,\tilde{\rho}}(X_s)}{  \tilde{\rho}^{(\tau)}_{\tau-s}\left(X_s\right)}\right)^2. \label{eq:vSs}
\end{equation} 
Note that Eq.~(\ref{eq:StVSNonStat}) has the same form as Eq.~(\ref{eq:entropy}), albeit with  an entropic drift $v^S_s$ that exhibits an explicit dependence on time $s$.

Applying It\^{o}'s formula (see Appendix~\ref{app:ItoFormula}) to the variable transformation $\hat{\Sigma}_s \to \exp(-\hat{\Sigma}_s) $ and using Eq.~\eqref{eq:StVSNonStat}, we obtain
 \begin{equation}
\frac{d\exp(-\hat{\Sigma}_s)}{ds}=
-\sqrt{2 v^S_s(X_s)}\, \exp(-\hat{\Sigma}_s) \dot{B}_s, 
\label{eq:gbm2}
\end{equation}
and hence $\exp(-\hat{\Sigma}_s)$ is an It\^{o} integral.   Hence,  according to Eq.~(\ref{eq:gbm2})  $\exp(-\hat{\Sigma}_s)$ is the stochastic exponential $\mathcal{E}_s(M)$ of the martingale
\begin{equation}
    M_s=\int^{s}_0du \sqrt{2v^S_u(X_u)} \dot{B}_u.
\end{equation}
Hence, we have "rediscovered"   (see previous  Sec.~\ref{sec:ItoApproach} and Sec.~\ref{sec:spadeMart}) in an explicit way that $\exp(-\hat{\Sigma}_s)$ is a martingale  provided  Novikov's condition is satisfied, which we assume to be the case in what follows.   Moreover, since $v^S_s\geq 0$, it holds that $\hat{\Sigma}_s$ is a submartingale, and it satisfies a conditional strong second law 
\begin{equation}
\langle \hat{\Sigma}_s |X_{[0,u]}\rangle \geq  \hat{\Sigma}_u     \label{eq:CSSLChp8}
\end{equation}
for all $0 \leq u \leq s$.

Note that  this example  has the appealing property that the origin of time reversal $t=\tau$  is immaterial.   Indeed, the same process $\hat{\Sigma}_s$ is obtained for all $t \geq \tau$, as we show in  Appendix~\ref{app:chapt82}.     This is because the initial state is given by Eq.~\eqref{Nis88} and the protocol has finite duration.     

%Alternatively, the  martingality of $\exp(-\hat{\Sigma}_s)$ can be understood from the fact that  $\hat{\Sigma}_s$ is a generalised entropic functional  of the form Eq.~(\ref{beau2}), viz., 
%\begin{equation}
 %   \hat{\Sigma}_s = \Sigma_{\left[0,s\right],\tau}^{\mathcal{\mP},\widetilde{\mP}^{(\tau)}}(X_{[0,s]}) = \ln \frac{\mathcal{\mP}\left(X_{[0,s]}\right)}{\tilde{\mathcal{\mP}}^{(\tau)}\left(\Theta_\tau\left(X_{[0,s]}\right)\right)}, \label{eq:timeRevSigma}
%\end{equation}
%where $\widetilde{\mP}$  is the measure describing the statistics in the  time-reversed protocol  $\tilde{\lambda}_s$ and with initial distribution $\rho_{\rm eq}(x;\lambda_f)$ [see the Fokker-Planck  equation~(\ref{eq:FPR1}-\ref{eq:FPR2})].     Since generalised entropic functional  are martingales, also $\exp(-\hat{\Sigma}_s)$ is a martingale. 

\subsubsection*{$^{\spadesuit}$Note on uniform integrability}
An important distinction between the process $\emStot$ for stationary $X$, as defined in Chapter~\ref{sec:martST1}, and the process $\exp(-\hat{\Sigma}_s)$ defined in \eqref{forGirs-1-1aa1}-(\ref{eq:Ss}) for nonstationary $X$, is that $\exp(-\hat{\Sigma}_s)$ is (in general) a uniformly integrable for $s\in \mathbb{R}^+$, while   $\emStot$ is not uniformly integrable for $t\in \mathbb{R}^+$.  This can be understood as follows.   

Both $\emStot$  and $\exp(-\hat{\Sigma}_s)$ are bounded from below, and hence according to the  martingale convergence theorem,  Theorem \ref{convergenceTheorem},  limits
\begin{equation}
    \exp(-S^{\rm tot}_\infty) = \lim_{t\rightarrow \infty}  \emStot
\end{equation}
and 
\begin{equation}
    \exp(-\hat{\Sigma}_\infty) = \lim_{s\rightarrow \infty}  \exp(-\hat{\Sigma}_s)
\end{equation}
exist.   According to condition Eq.~(\ref{eq:UICond}),  if in addition $\langle  \exp(-S^{\rm tot}_\infty) \rangle = 1$ and $\langle  \exp(-\hat{\Sigma}_\infty) \rangle = 1$, then  $\exp(-S^{\rm tot}_t)$ and $ \exp(-\hat{\Sigma}_s)$ are, respectively,  uniformly integrable processes.  

However, for stationary processes 
\begin{equation}
0 = \langle \exp(-S^{\rm tot}_\infty) \rangle  \neq    \langle  \emStot \rangle  = 1,
\end{equation}
and hence $\exp(-S^{\rm tot}_t)$ is not uniformly integrable. 
This is because with probability one  $\lim_{t\rightarrow  \infty}S^{\rm tot}_t=+\infty$.  

On the other hand, 
\begin{equation}
\langle \exp(-\hat{\Sigma}_\infty) \rangle  =    \langle  \exp(-\hat{\Sigma}_s) \rangle  = 1,
\end{equation}
as with probability one $\lim_{s\rightarrow  \infty}\hat{\Sigma}_s\in \mathbb{R}^+$.

\subsection{Derivation of the Jarzynski equality at stopping times }\label{sec:JarzM}
To obtain a Jarzynski equality at stopping times, we  rewrite the process $\hat{\Sigma}_s$ in terms of the stochastic work $W_t$  done on the system  and the  equilibrium free energy $G^{\rm eq}(\lambda_s)$, given by Eq.~\eqref{nis33}.   Using the first law of thermodynamics, Eq.~(\ref{eq:1LST}) and the Boltzmann distribution, Eq.~(\ref{eq:boltz}), we obtain
\begin{equation}
\hat{\Sigma}_s=  \frac{W_s-\Delta G^{\rm eq}(\lambda_s)}{T} -\pi_s,
\end{equation}
where the {\em equilibrium} free energy difference between the final and initial states reads \eqref{eq:Ss}
\begin{equation}
\Delta G^{\rm eq}(\lambda_s) = G^{\rm eq}(\lambda_s) - G^{\rm eq}(\lambda_{\rm i}),
\end{equation}
and
where the remainder term
\begin{equation}
    \pi_s \equiv  \ln \left( \frac{\tilde{\rho}^{(\tau)}_{\tau-s}(X_s)}{\rho^{\rm eq}(X_s;\lambda_s)}\right).
\end{equation}

\begin{leftbar}
Since for finite $\tau$ the process $\exp(-\hat{\Sigma}_s)$ is a uniformly integrable martingale, see note on uniform integrability in Sec.~\ref{sec:martAsX}, Doob's optional stopping theorem, Theorem~\ref{Doob1x}, applies,  yielding the  {\bf Jarzynski equality at stopping times} \cite{neri2020second}, i.e., 
\begin{equation}
\Bigg\langle \exp\left(-\frac{W_\T-\Delta G^{\rm eq}(\lambda_\T) }{T}+ \pi_\T\right)\Bigg \rangle = 1.  \label{eq:JarzStop}
\end{equation}
\end{leftbar}

The Eq.~(\ref{eq:JarzStop}) is reminiscent of  Jarzynski's equality Eq.~(\ref{eq:jarz}), except for the presence of the remainder term $\pi_\T$ that includes  the nontrivial contributions to $\hat{\Sigma}_s$ due to the fact that we stopped the process $X$ at a random time $\T$.  Nevertheless, it is justified to call Eq.~(\ref{eq:JarzStop})     a Jarzynski equality at stopping times as in several limiting cases it holds that  $\pi_\T=0$ yielding the good-looking equality 
\begin{equation}
\Bigg\langle \exp\left(-\frac{W_\T-\Delta G^{\rm eq}(\lambda_\T) }{T}\right)\Bigg \rangle = 1,  \label{eq:JarzStop2}
\end{equation}
which is Eq.~(\ref{eq:jarz}) for $t\rightarrow \T$.

Equation (\ref{eq:JarzStop2}) applies in the 
following limiting cases for which it holds that $\pi_\T=0$:  
\begin{enumerate}[label=(\roman*)]
    \item $\T=\tau$:  indeed, in this case $\tilde{\rho}^{(\tau)}_{0}(x) = \rho^{\rm eq}(x;\lambda_{\rm f})$ and thus $\pi_\tau = 0$.   In this case, Eq.~(\ref{eq:JarzStop}) is identical to the Jarzynski equality Eq.~(\ref{eq:jarz}) as  $\T=\tau$.   
    \item  the stopping time $\T$ is larger or equal than $\tau$:  indeed,  $\tilde{\rho}^{(\tau)}_{\tau-\T}(x) = \rho^{\rm eq}(x;\lambda_{\rm f})$ for $\T>\tau$, and thus $\pi_s=0$ for $s>\tau$.
    \item the driving $\lambda_s$ is quasi-static: in this case, $\tilde{\rho}^{(\tau)}_{\tau-s}(x) = \rho^{\rm eq}(x;\lambda_s)$ for all $s$,  such that $\pi_s  = 0$.
    \item the protocol is quenched (i.e., $\lambda_s = \lambda_{\rm f}$ for  $s>0$) and the stopping time is with probability one greater than zero (i.e., $\mP\left(\T>0\right)=1$): this is a special case of (iii). 
\end{enumerate}

We derive now a second law of thermodynamics at stopping times based on the Jarzynski equality at stopping times. 
 
 \begin{leftbar}
Jensen's inequality Eq.~(\ref{eq:jensen}) applied to $X=S_s$, together with Jarzynski's equality at stopping times, Eq.~(\ref{eq:JarzStop}), yields the {\bf second law of thermodynamics at stopping times}  \cite{neri2020second}
\begin{equation}
\langle W_\T \rangle - \langle \Delta G^{\rm eq}(\lambda_{\T}) \rangle + T \langle \pi_\T \rangle  \geq 0 .  \label{eq:WTStop}
\end{equation}
\end{leftbar} 

Although here, for reasons of  simplicity we have derived Eqs.~(\ref{eq:JarzStop}) and (\ref{eq:WTStop}) for one dimensional, overdamped Langevin processes, these relations are generally valid for multidimensional  overdamped Langevin processes and Markov jump processes, see Ref.~\cite{neri2020second}.

Note that for the special cases where $\pi_\T=0$, as discussed below Eq.~(\ref{eq:JarzStop2}), we obtain the appealing bound 
\begin{equation}
\langle W_\T \rangle \geq  \langle \Delta G^{\rm eq}(\lambda_{\T}) \rangle .  \label{eq:WTStopxx}
\end{equation}
  In other words, the average amount of work we need to perform on a system in order for a certain event of interest to happen, as determined by the stopping time $\T$, must be greater or equal  than the average increase in free energy.  This second law of thermodynamics holds  for quenched protocols for which the event happens with probability one at nonzero times, and for quasistatic protocols.  
  
  Although the  remainder term $\pi_\T$  in Eq.~(\ref{eq:WTStop}) spoils in general the more practical inequality Eq.~(\ref{eq:WTStopxx}), the remainder is at the origin of interesting phenomena, such as, events in which on average  an agent increases  the  free energy of a system more than the work it does on it.   
  
  In what follows, we illustrate the second law of thermodynamics Eq.~(\ref{eq:WTStop}), as well as Eq.~(\ref{eq:WTStopxx}), on the canonical example of the polymer in Fig.~\ref{fig:polymer}, and we discuss the role of the remainder term $\pi_\T$.

\begin{figure}[H]
    \centering
    \includegraphics[width=0.55\textwidth]{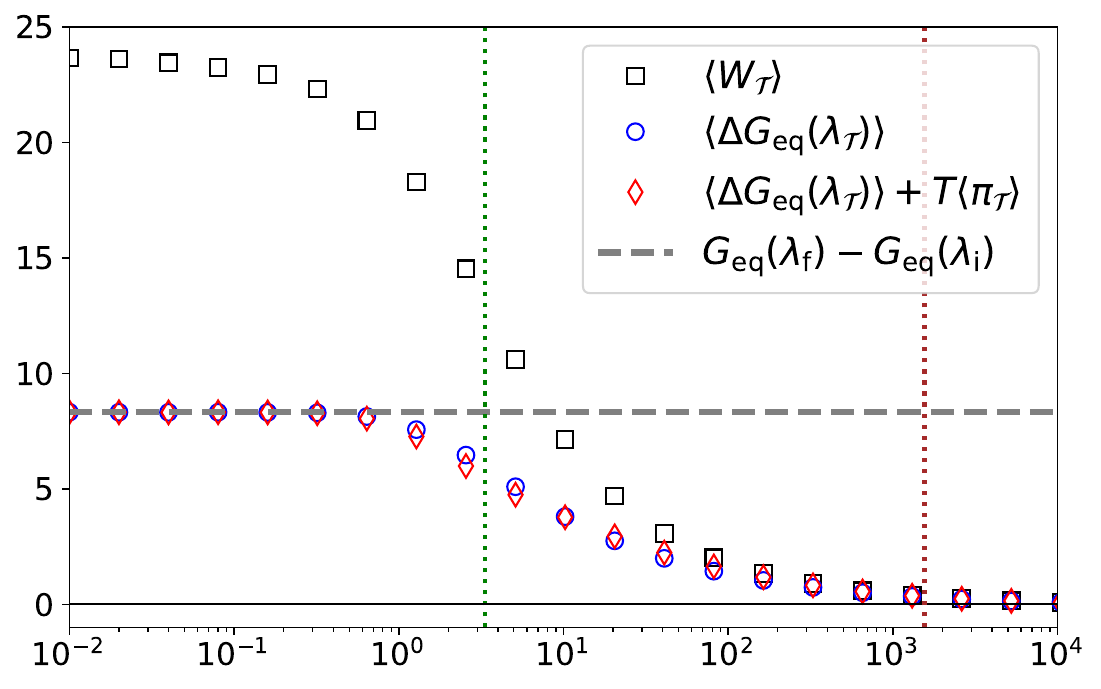} \put(-130,-10){$\tau_{\rm prot}$} 
    \caption{Simulation results demonstrating the second law of thermodynamics at stopping times, Eq.~(\ref{eq:WTStop}), for the model defined in Sec.~\ref{sec:JarzM} and illustrated in Fig.~\ref{fig:polymer}.    The parameters used in simulations are $\ell=2.2$, $\mu=0.1$, $T=\kappa_{\rm p} = 1$, $\kappa_{\rm m}=2$, $\lambda_{\rm i} = 0.2$, $\lambda_{\rm f}=5$, and $\tau=10^6$.  The vertical dotted lines denote the relaxation time $\tau_{\rm rel}=10/3$ of the polymer towards equilibrium and the mean first-passage time $\tau_{\rm fp} = 1560$ for the polymer to reach the dangling end point from the initial point in the absence of a driving protocol.   The black solid line indicates zero and is a guide to the eye. 
     Figures are taken from Ref.~\cite{neri2020second}.}
    \label{fig:polymer2}
\end{figure}

\subsection{Canonical example illustrating the second law of thermodynamics}
We illustrate the second law at stopping times, Eq.~(\ref{eq:WTStop}), on the example of Fig.~\ref{fig:polymer2}.   

We assume that the position $X$ of the dangling end point is well described by Eq.~(\ref{eq:noneqLang}) with the thermodynamic potential
\begin{eqnarray}
V(x;\lambda_s)  =   \frac{\kappa_{\rm p}}{2} \: x^2 + \frac{\kappa_{\rm m}}{2} \left(x-\lambda_s\right)^2, \label{eq:potR}
\end{eqnarray}
which is  the sum of the    potential $ \kappa_{\rm p} x^2/2$  of a polymer with one  of its end points anchored  to the substrate at $x=0$, and the  potential
$\kappa_{\rm m} \left(x-\lambda_s\right)^2/2$, of the spring that connects the dangling end point of the polymer to the molecular motor with its centre of mass located at  $\lambda_s$.     At time $s=0$ this  motor-polymer  system  is in thermal equilibrium with its surroundings, and at time $s>0$ the motor starts moving forwards.   The dynamics of the center of mass of the molecular motor  is described by
\begin{eqnarray}
\lambda_s = \lambda_{\rm i}  + (\lambda_{\rm f}-\lambda_{\rm i})\frac{1-\exp\left(-s/\tau_{\rm prot}\right)}{1-\exp\left(-t/\tau_{\rm prot}\right)}, \quad s\in[0,t],  \label{eq:prot}
\end{eqnarray}   
where $\tau_{\rm prot}>0$ is the time scale determining the protocol speed.    
The polymer relaxes over a  time scale    $\tau_{\rm rel} = 1/(\mu(\kappa_{\rm m}+\kappa_{\rm p}))$.     If $\tau_{\rm prot} \ll \tau_{\rm rel}$, then the molecular motor quenches the polymer, whereas if $\tau_{\rm prot} \gg \tau_{\rm rel}$, then the motor stretches the polymer in a quasi-static manner.  

We determine the average work $\langle W_\T\rangle$ that the motor performs on the polymer in order to bring the second end point of the polymer to the location  $X(t) = \ell$.    Hence, the stopping time is defined by 
\begin{equation}
    \T = {\rm inf}\left\{s\geq 0: X_s = \ell\right\}. 
\end{equation}
Simulation results in  Fig.~\ref{fig:polymer2} show numerically that the second law of thermodynamics at stopping times, Eq.~(\ref{eq:WTStop}), holds.  We observe two regimes, viz., the quenched regime for $\tau_{\rm prot}<\tau_{\rm rel}$, in which case the dissipated work $\langle W_{\T}\rangle-\langle \Delta G^{\rm eq}(\lambda_{\T})\rangle$ is large, and the opposing quasi-static limit of $\tau_{\rm prot}>\tau_{\rm fp}$, for which  $\langle W_{\T}\rangle-\langle \Delta G^{\rm eq}(\lambda_{\T})\rangle\approx 0$.  Another relevant time scale for this problem is the mean first-passage time $\tau_{\rm fp}$ that $X$ needs to reach $X=\ell$ when $\lambda_{\rm f}=\lambda_{\rm i}$.   If  $\tau_{\rm prot}>\tau_{\rm fp}$, then $\langle W_{\T}\rangle\approx 0$.

An interesting feature of the second law, which becomes evident  from    Fig.~\ref{fig:polymer2}, is that  $\langle \pi_\T\rangle \approx 0$, and hence the appealing bound Eq.~(\ref{eq:WTStopxx}) ensues.     The approximation $\langle \pi_\T\rangle \approx 0$  follows from the fact that $\langle \pi_\T\rangle = 0$ in the two limiting cases $\tau_{\rm prot}\gg\tau_{\rm rel}$ and  $\tau_{\rm prot}\ll\tau_{\rm rel}$, for which the as  protocol is quasistatic and quenched, respectively.   As discussed below Eq.~(\ref{eq:JarzStop2}), in these two limiting cases $\pi_\T=0$.   In the intermediate regime $\pi_\T\neq 0$, but simulation results in Fig.~\ref{fig:polymer2} show that nevertheless $\langle \pi_T\rangle \approx 0$.   

Taken together,  it often holds that $\langle \pi_\T\rangle\approx 0$ and hence the practical inequality Eq.~(\ref{eq:WTStopxx}) applies.  This inequality  states that also at random times on average the free energy of  a system cannot increase more than the average work done on it, in accordance with the classical result  Eq.~(\ref{eq:JarzSecond}).

\subsection{Overcoming classical limits by stopping at a clever moment: $\langle W_{\T}\rangle\leq \langle \Delta G^{\rm eq}(\lambda_\T)\rangle $}  

As discussed in Sec.~\ref{sec:overcoming},  it is possible to (apparently) overcome classical limits by stopping a process at a clever moment.     We consider now this question from the perspective of a nonstationary process, which is significantly more subtle than the stationary case.  

It is the remainder term $\pi_\T$ in the second law Eq.~(\ref{eq:WTStop}) that describes the possibility to increase on average the free energy of a system more than the work put into it.   To achieve this, we need a large enough positive value of $\langle\pi_\T\rangle$ as  the dissipated work is lower bounded by $-T\langle\pi_\T\rangle$, viz.,
\begin{equation}
    \langle W_\T \rangle - \langle \Delta G^{\rm eq}(\lambda_{\T}) \rangle \geq - T \langle \pi_\T \rangle  .
\end{equation}
In order to have $\langle W_\T\rangle$ small enough we need a large enough value of $\langle \pi_\T \rangle$, which as discussed in the previous section  can be attained when  $\mP(\T=0)>0$.

We illustrate this in Fig.~\ref{fig:polymer3} for the same model for $X$ as considered in Fig.~\ref{fig:polymer2}, i.e., the Langevin Eq.~(\ref{eq:noneqLang}) with potential Eq.~(\ref{eq:potR}).  A notable difference is that the stopping event is defined by 
\begin{equation}
    \T = {\rm min}\left\{s\geq 0: X_s\leq \ell\right\} 
\end{equation}
so that {$\mP(\T=0)>0$}.    The numerical results in Fig.~\ref{fig:polymer3} show that there exists a region at intermediate protocol speeds $\tau_{\rm prot}$ for which $\langle W_\T\rangle < \langle \Delta G^{\rm eq}(\lambda_{\T}) \rangle$, demonstrating that  the classical limit $\langle W_t\rangle >  \Delta G^{\rm eq}$ can be overcome by stopping a process at a cleverly chosen moment.

\begin{figure}[H]
    \centering
    \includegraphics[width=0.55\textwidth]{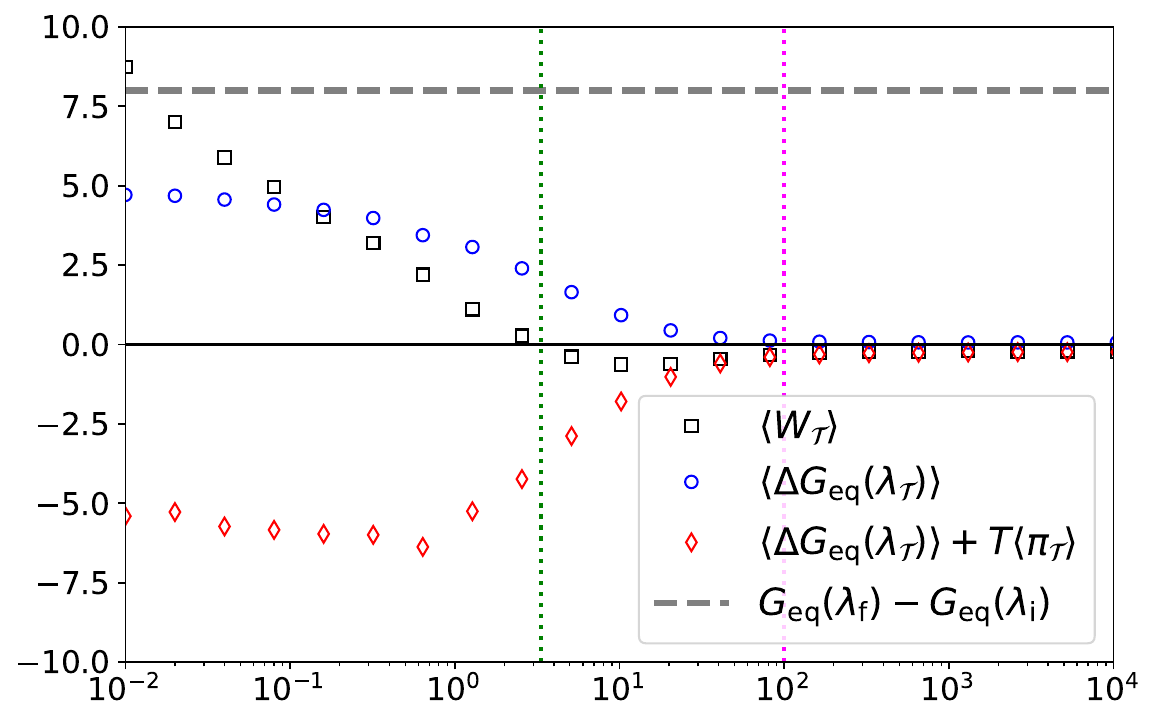}
    \put(-130,-10){$\tau_{\rm prot}$}
    \caption{Overcoming classical limits by stopping at a clever moment, viz. $\langle W_\T\rangle \leq \langle \Delta G^{\rm eq}(\lambda_\T)\rangle$.   The model used in the one of Fig.~\ref{fig:polymer} and  defined in Sec.~\ref{sec:JarzM}.    The parameters used in simulations are $\ell=0.2$, $\mu=0.1$, $T = 10$, $\kappa_{\rm p}=1$ $\kappa_{\rm m}=2$, $\lambda_{\rm i} = 1$, $\lambda_{\rm f}=5$, and $\tau=50$.   The stopping time used is     $\T = {\rm min}\left\{s\geq 0: X_s\leq \ell\right\}$.    The black solid line indicates zero and is a guide to the eye. 
     Figures are taken from \cite{neri2020second}.}
    \label{fig:polymer3}
\end{figure}

\section{Second law at stopping times and gambling demons}\label{sec:GamblingDem}

%Sections~\ref{sec:5p1} and~\ref{sec:5p2} revealed that $S^{\rm tot}_t$ has an exponential martingale structure in Langevin and Markov-jump stationary processes, and exploited some consequences of this finding. When dealing  with Markovian non-stationary processes, it is convenient to develop a generic approach, treating Langevin and Markov-jump processes in a unified framework in which the statistics of the process of interest is given through a path probability $\mP$ that satisfies some mild assumptions given below. 

The objects of interest in this section will be 
the generalized stochastic entropic functional given by Eq.~\eqref{forGirs-1-1aa1}:
{
\begin{equation}
\hat{\Sigma}_s =
\underbrace{\ln\left(\frac{\rho_s(X_s)}{\tilde{\rho}^{(t)}_{t-s}(X_s)}\right)}_{\displaystyle\delta_{s}^{(t)}}
+
\displaystyle S^{\rm tot}_{s}
\label{forGirs-1-1aa111},
\end{equation}}
where the first term  
is denoted as the {\em stochastic distinguishability}   between conjugate times in the forward and backward process~\cite{manzano2021thermodynamics}; it is 
 given by Eq.~\eqref{deltatot}, copied here for convenience
\begin{equation}
     \delta_{s}^{(t)}=
     \ln \left(\frac{\rho_{s}(X_s)}{\tilde{\rho}^{(t)}_{t-s}(X_s)}\right).
     \label{deltatot00}
 \end{equation}
We recall that here the stochastic total entropy production $ S^{\rm tot}_{s}$ is given by    \eqref{AFS2}-\eqref{Stotcopied}  
\begin{equation}
S^ {\rm tot}_{s}=\ln\left[\frac{\mathcal{P}\left(X_{\left[0,s\right]}\right)}{\widetilde{\mathcal{P}}^{(s)}\left(\Theta_{s}\!\left(X_{\left[0,t\right]}\right)\right)}\right],
\label{eq:actionfunc_0}
\end{equation}
and we will consider $0\leq s\leq t$ throughout this section.
Here, the path probabilities $\mP$ and $\tilde\mP^{(t)}$ are defined as follows:
%{\color{magenta}I write more precise things of the blue part below in chapter 6.2.2.1 :  the good place for this. Here it must stay just a litte presentation and the relations \eqref{eq:IFT_stop_Stot_nonstat}, \eqref{eq:secondlaw_stop_nonstat}, \eqref{eq:aprilfatigue} which are applications.}  
\begin{itemize}
    \item Forward process is a nonequilibrium Markovian process with initial state drawn from $\rho_0(x)$ and  driven through a deterministic protocol $\lambda_u$ to a final state with distribution $\rho_{t}(x)$. In the forward process, a given trajectory $x_{\left[0,t\right]}$  is produced with probability $\mathcal{P}(x_{\left[0,t\right]})$. 
    
    \item Auxiliary backward process starts from state drawn from the final distribution of the forward process $\tilde{\rho}^{(t)}_0 (x) = \rho_t(x)$. It is driven by a protocol that is the time-reversal mirror of the forward protocol $\tilde{\lambda}_{s} = \lambda_{t-s}$. in the backward auxiliary process, a given trajectory $x_{\left[0,t\right]}$  is produced with probability $\widetilde{\mathcal{P}}^{(t)}\left(x_{\left[0,t\right]}\right)$. See Fig.~\ref{fig:timereversedJohn} for an illustration of a forward and a backward process.
\end{itemize}

\begin{figure}[h]
\centering
\includegraphics[width=0.9\textwidth]{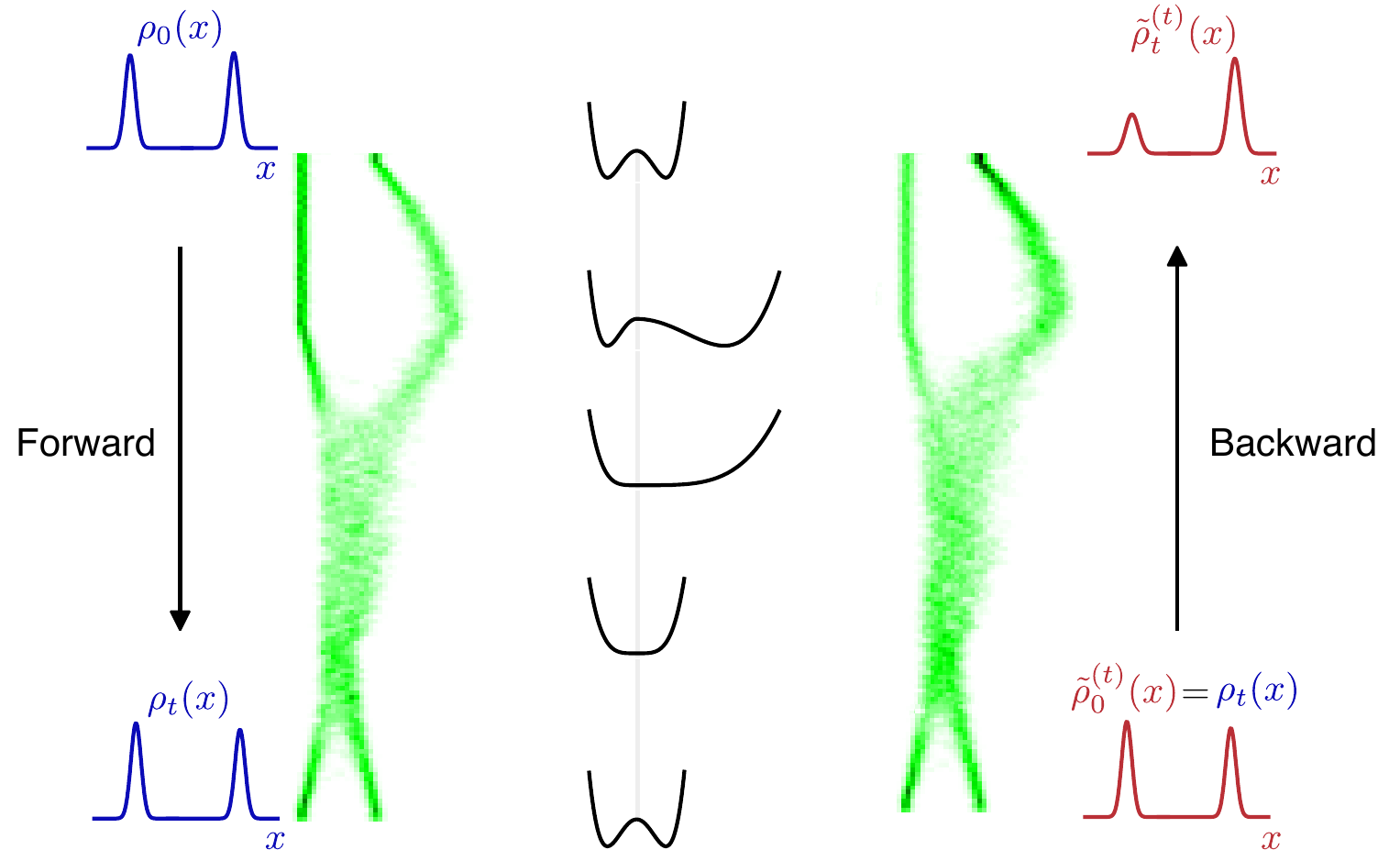}
\caption{Illustration of a forward  and a backward  process used in Sec.~\ref{sec:GamblingDem}, where a Brownian particle immersed in a fluid is externally-controlled with a feedback trap creating a time-dependent potential (black line in the central column). In the forward (backward) process, the particle is initially drawn from the distribution $\rho_0(x)$ ($\rho^{(t)}_0(x)$) and the potential evolves as in the central column from top to bottom (bottom to top), reaching a final state characterized by the distribution $\rho_t(x)$ ($\rho^{(t)}_t(x)$). The green dots illustrate histograms of the particle position taken during the evolution of the forward (left) and backward (right) processes, revealing the time-reversal asymmetry in the statistics of $X$, i.e.  in general 
$\rho_{s}(x)\neq \tilde{\rho}^{(t)}_{t-s}(x)$  for $s\in(0,t]$, see Eq.~\eqref{deltatot00}. Figure adapted from~\cite{gavrilov2016arbitrarily} with permission.
} %Note that even if the initial distribution of the backward process $\rho^{(t)}_0(x)=\rho_t(x)$ equals to the final distribution of the forward process, one has at  times  $0<s\leq t$  $\rho^{(t)}_s(x)\neq\rho_{t-s}(x)$.
\label{fig:timereversedJohn}
\end{figure}

%{\color{magenta} Problem :what is $tau$ below  ....if $0\leq t \leq \tau$ as claim below, this means that it is not the total entropy production but a generalized entropic functional and then the Martingale relation which permits to arrive to the relation \eqref{eq:IFT_stop_Stot_nonstat}, \eqref{eq:secondlaw_stop_nonstat}, \eqref{eq:aprilfatigue}   must be the relaton 153 and 146 of two refreshing view...Otherwise, if in fact $ \tau=t$,  it is below that there are some problem...\eqref{eq:mart_Stot_nonstat} is good but (5.73) and  \eqref{eq:mart_Stot_nonstat2} must be little changed, and \eqref{eq:secondlaw_nonstat2} is wrong. See section 6.2.2.1.  }

 In the following we make use of the mathematical power of the martingales to extract knowledge about entropy production at stopping times  for Markovian processes that are in general non-stationary. First, we report  recents result (see  Refs.~\cite{manzano2021thermodynamics,neri2020second}) that revealed that the stochastic total entropy production $S^{\rm tot}_t$ is not an exponential  martingale in generic non-stationary nonequilibrium processes.
 
 \begin{leftbar} 
 For generic non-stationary Markovian processes, the stochastic entropy production $S^ {\rm tot}_{t}$ given by Eq.~\eqref{eq:actionfunc_0} \textbf{is not an exponential martingale}, i.e. in general $\langle\, \exp(-S_{t}^{\rm tot} ) \,|\, X_{[0,s]}\, \rangle \neq \exp(-S_{s}^{\rm tot})$. However, as we saw in two different ways ---in equation \eqref{april100} in Sec.~\ref{eq:martsigmafunct} and in relation~\eqref{forGirs-1-1aa11} in Sec.~\eqref{nischapter}---  it is possible to ''martingalize'' $S^{\rm tot}_t$ in non-stationary nonequilibrium processes, i.e. find a process related to $S^{\rm tot}_t$  that is an  exponential martingale. In particular, it follows that for generic (even non-stationary) nonequilbrium processes, 
for $0\leq u\leq s\leq t$, it holds that \eqref{eq:backmart4index_p4} 
 \begin{equation}
\langle\, \exp(-S_{s}^{\rm tot} -\delta_{s}^{(t)}) \,|\, X_{[0,u]}\, \rangle =  \exp(-S_{u}^{\rm tot} -\delta_{u}^{(t)}).
 \label{april10000}
 \end{equation} 
 %with 
 %\begin{equation}
  %   \delta_{s}^{(t)} \equiv
   %  \ln \left[\frac{\rho_{s}(X_s)}{\tilde\rho_{t-s}(X_s)}\right].
    % \label{deltatot00}
 %\end{equation}
  Note that $\delta_{t}^{(t)}=0$ for all $t$ and that the superindex in $ \delta _s^{(t)}$ denotes the time with respect one does the time-reversal operation, $t$. The stochastic distinguishability vanishes at all times  for (possibly nonequilibrium) stationary states ---for which $\rho_s$ and $\rho_{s}^{\tilde{\mP}^{(t)}}$ are independent on time $s$. For non-stationary processes, $\delta_{s}^{(t)}$ fluctuates and can in principle take any value.
 \end{leftbar} 

%A detailed and rigorous mathematical proof of Eq.~\eqref{april10000} is given below in Sec.~\ref{eq:martsigmafunct}.
%{RC : we want now underline a tricky point : for $0\leq u\leq s\leq t$, we have 
%\begin{equation}
%\langle\, \exp(-S_{s}^{\rm tot} -\delta_{s}^{(t)}) \,|\, X_{[0,u]}\, \rangle \neq \exp(-S_{u}^{\rm tot} -\delta_{u}^{(t)}),
 %\label{april100000}
 %\end{equation}
 %So the family $\exp(-S_{s}^{\rm tot} -\delta_{s}^{(t)})$ is not a Martingale at fixed $t$. Moreover, we have also for $0\leq s\leq t$ 
%\begin{equation}
%\langle\, \exp(-S_{t}^{\rm tot} -\delta_{t}^{(t)})  \,|\, X_{[0,s]} \rangle = \langle\,\exp(-S_{t}^{\rm tot})  \,|\, X_{[0,s]}\, \rangle \neq \exp(-S_{s}^{\rm tot}) = \exp(-S_{s}^{\rm tot} -\delta_{s}^{(s)}),
 %\label{april1000000}
 %\end{equation} 
 %and so the family $\exp(-S_{s}^{\rm tot} -\delta_{s}^{(s)})$ is  no more a Martingale. So we have the Martingale ''like'' relation   \eqref{april10000} without a clear Martingale quantity !  }

Applying Jensen's inequality to the ''martingale property''~\eqref{april10000}, we obtain that for any $0\leq s \leq t$ we have the sub-Martingale relation
\begin{equation}
\langle\,   S^{\rm tot}_t+ \delta_{t}^{(t)}  \,|\, X_{[0,s]}\, \rangle \geq   S^{\rm tot}_s+ \delta_{s}^{(t)}.
\label{eq:submart_Stot_nonstat22}
\end{equation} 

Specializing the ''submartingale'' condition~\eqref{eq:submart_Stot_nonstat22} to $s=0$,  noting that $S^{\rm tot}_0=0=\delta_{t}^{(t)}$, and averaging with respect to $X_0$, we get the refined second law for non-stationary Markovian  processes \eqref{poslun22}
\begin{equation}
\langle\, S^{\rm tot}_t\,\rangle \geq  \langle\, \delta_{0}^{(t)}\,\rangle ,
\label{eq:secondlaw_nonstat2222}
\end{equation}
where  \eqref{deltatot00}
\begin{equation}
   \langle\, \delta_{0}^{(t)}\,\rangle = \int_{\mathcal{X}} dx \rho_0(x) \ln \left[ \frac{\rho_0 (x)}{\tilde{\rho}^{(t)}_{t}(x)} \right] \geq 0,
\end{equation}
is the Kullback-Leibler divergence between the distribution $\rho_0$ and $\tilde{\rho}^{(t)}_{t}$. It is equal to zero for $t=0$ and it is positive otherwise. We will generalize this second law in section~\ref{sec:6321} within the context of deterministic refinements of the second law.

The fact that for any $0\leq s \leq t$,  $S_{s}^{\rm tot} +\delta_{s}^{(t)}$ is an exponential martingale  has other important consequences for stochastic thermodynamics, which can be found applying Doob's optional stopping theorems. Similarly to the integral fluctuation theorem~\eqref{eq:SStop} at stopping times for stationary processes, $\langle \exp(-S^{\rm tot}_\T)\rangle = 1$, for non-stationary processes one can show (see Sec.~\ref{eq:martsigmafunct})  that an integral fluctuation theorem holds.
\begin{leftbar}
\textbf{Integral fluctuation relation at stopping times} for driven Markovian processes that may not be stationary. 
For a stopping time $\mathcal{T}\leq t$, 
\begin{equation}
\langle\, \exp(-(S^{\rm tot}_\mathcal{T}+\delta^{(t)}_\mathcal{T}) )\, \rangle = 1,
\label{eq:IFT_stop_Stot_nonstat00}
\end{equation}
see  Eq.~\eqref{april1final}  and mathematical derivation in Sec.~\ref{eq:martsigmafunct}. Note that here, it is crucial to note that the quantity $\delta^{(t)}_\T=\ln \left(\rho_{s}(X_s)/\tilde\rho^{(t)}_{t-s}(X_s)\right)\vert_{s=\T}$ results from evaluating the instantaneous densities $\rho_s(X_s)$ and $\tilde\rho^{(t)}_{t-s}(X_s)$ at (stochastic) stopping times  $s=\T$ that  are extracted from the forward process. For stationary processes $\delta^{(t)}_\T=0$,  and thus one recovers $\langle \exp(-S^{\rm tot}_\T)\rangle = 1$, see Eq.~\eqref{eq:SStop}.
\end{leftbar}

The fact that the stochastic distinguishability can in principle take any value at stopping times  has implications regarding the extension of the second law for $S^{\rm tot}_t$ in generic Markovian nonequilibrium processes, as we show below, in terms of the so-called second law at stopping times.
\begin{leftbar}
\textbf{Second law at stopping times} for driven Markovian processes that may not be stationary. Applying Jensen's inequality to \eqref{eq:IFT_stop_Stot_nonstat00}, we find that for any stopping time $\T\leq t$, one has
\begin{equation} 
\langle\, S^{\rm tot}_{\mathcal{T}}\,\rangle \geq - \langle\, \delta^{(t)}_{\mathcal{T}}\,\rangle ,
\label{eq:secondlaw_stop_nonstat00}
\end{equation}
where 
\begin{equation}
    \langle\, \delta^{(t)}_{\mathcal{T}} \rangle = \int_{0}^{t} ds \int_{\mathcal{X}} dx \;\rho_{X_{\mathcal{T}},\mathcal{T}}(x,s) \ln  \left[\frac{\rho_{s}(x)}{\tilde{\rho}^{(t)}_{t-s}(x)}\right] .
    \label{eq:aprilfatigue00}
\end{equation}
Here $\rho_{X_{\mathcal{T}},\mathcal{T}}$ is the joint probability density
for the stopping time to take the value~$\T=s$ and for  the system to be at state $X_\T=x$ when the stopping condition happens. On the other hand, the densities $\rho_{s}(x)$ and $\tilde{\rho}^{(t)}_{t-s}(x)$ denote the instantaneous density of the forward and backward process evaluated at  times $s$ and $t-s$, respectively.

\end{leftbar}

Note that, using Bayes' formula, we have in Eq.~\eqref{eq:aprilfatigue00} that $\rho_{X_{\T},\mathcal{T}}(x,s)=  \rho_{\mathcal{T}} (s) \rho_{X_{\mathcal{T}}\vert\mathcal{T}}(x\vert s)$, however in general $\rho_{X_{\mathcal{T}}\vert\mathcal{T}}(x\vert s) \neq \rho_{s}(x)$. This highlights the fact that the right-hand side of Eq.~\eqref{eq:aprilfatigue00} is not a Kullback-Leibler divergence, hence it is not obvious the sign of the term $\langle\, \delta^{(t)}_{\mathcal{T}}\,\rangle$.
In the following we present a physical example of a system in which using stopping strategies one can find negative average stochastic entropy production at stopping times, i.e. $\langle\, S^{\rm tot}_\T\,\rangle <0$, a feature that is not  forbidden by the second law at stopping times~\eqref{eq:secondlaw_stop_nonstat00}.

\subsubsection*{Experimental implementation with single electron transistors}
%{(RC : I dont check notation...I will do with Edgar)}

We now discuss a recent application of the second law at stopping times given by Eq.~\eqref{eq:secondlaw_stop_nonstat00} in the context of information demons, see Ref.~\cite{manzano2021thermodynamics} for details.   
 Maxwell's demon thought experiment is considered the cornerstone of information thermodynamics. Such a "demon" is able to e.g. induce a net heat flow from a cold to a hot reservoir by using information acquired from the bath molecules in a clever way. In Maxwell's original proposal, an external controller ("demon")  is allowed to open and close a tiny gate separating two gas containers that are held at different temperatures. Such demon acts at stochastic times, it opens the gate only when a particle get sufficiently close to the gate. Moreover, it applies a feedback protocol, as it opens the gate only to particles coming from the cold bath than are colder than the average, and to particles coming from the hot reservoir that are hotter than the average. This way, the demon applies feedback control on the entire system by changing the concentration of particles in each of the baths, which results in a net heat flow from the cold to the hot bath, in an apparent violation of the second law. Such conundrum have been thoroughly studied within the framework of information thermodynamics~\cite{parrondo2015thermodynamics}, which established the minimal energetic costs and the entropy production associated with measurement and feedback, which led to the derivation of second laws in the presence of information processing.

 \begin{figure}[h]
    \centering
    \includegraphics[width=\textwidth]{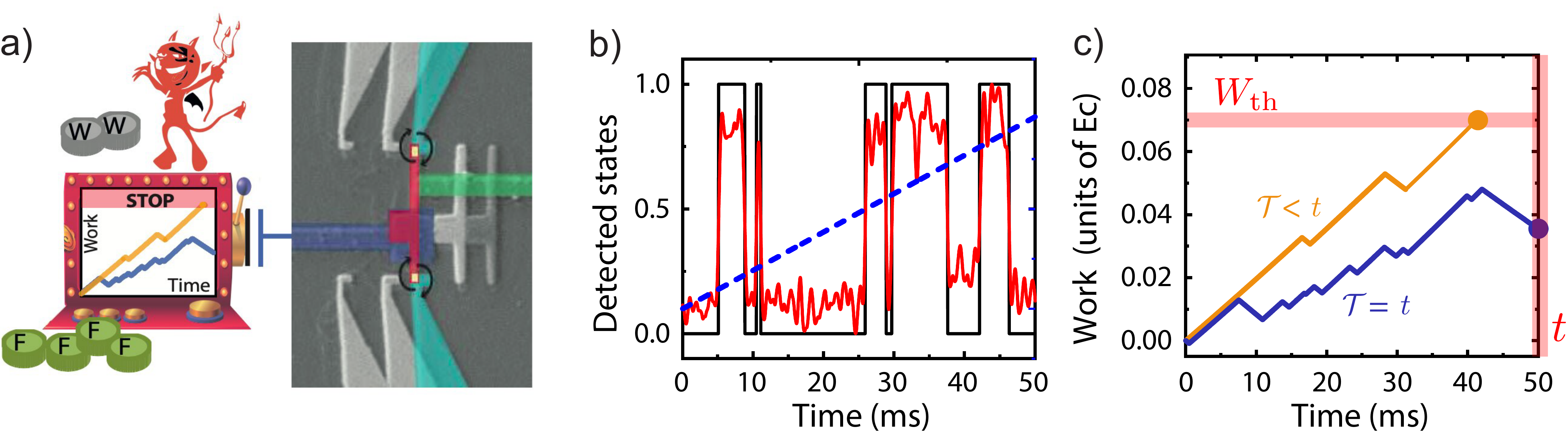}
    \caption{Experimental realization of a Gambling demon. a) Sketch of the gambling demon and experimental setup. An external controller ("demon") monitors the fluctuations of a mesoscopic system that is driven out of equilibrium by a deterministic protocol of a prefixed duration $t$.  The demon gambles with the information retrieved from the system, by stopping its evolution when a specific criterion is first met. In this case, it stops the external driving if the work done on the system exceeds a threshold value (orange line), or in the contrary at time $t$ (blue line). As a result of this procedure, the demon expects to extract on average more free energy (gold coins) than the work invested (silver coins), an outcome that is inacessible without using gambling strategies (i.e. stopping the dynamics always at time $t$). Such idea was realized in~\cite{manzano2021thermodynamics} with an electronic system in which individual electrons can tunnel (black arrows) into a metallic island (red) whose voltage is controlled  in time. b) The experimental value of the detected state of the island (red line) is digitized (black line) and used for gambling. The blue line shows the expected value of the state of the electron averaged over many trajectories in the absence of gambling. c) Experimental values of the work done on the electron until the stopping event of the gambling protocol takes place in two example trajectories that stop at $\T<t$ (orange) and at $\T=t$.   Figure adapted from Ref.~\cite{manzano2021thermodynamics}.}
    \label{fig:GD1}
\end{figure}

 We now ask the question: what is the entropy production associated with a demon that is only able to stop the dynamics of a physical process at stochastic times using suitable {\em gambling} strategies?  Such scenario may result from considering a Maxwell-like demon that is able to terminate a process at a random time (open/close a gate) but does not apply feedback control after taking such action. We exemplify this question on an experiment in which an isothermal system at temperature $T$ is driven out of equilibrium through a time-dependent protocol of a fixed finite duration~$t$. By varying this protocol, the potential of the system is switched from $V_{0}(x)$ to $V_t(x)$.  When averaging over many repetitions of the same protocol, the second law of thermodynamics implies that 
 \begin{equation}
     \langle W_t  \rangle  - \langle\Delta G^{\rm ne}_t\rangle \geq 0  ,
     \label{eq:secondlawnoneq}
 \end{equation}
 where $ \Delta G^{\rm ne}_t = G^{\rm ne}_t-G^{\rm ne}_0$ is the nonequilibrium free energy difference between the final and initial states of the system\footnote{The nonequilibrium free energy is formally defined as $G^{\rm ne}_t = V_t - T S^{\rm sys}_t$, see Eq.~\eqref{eq:noneqfree} in Ch.~\ref{TreeSL}, with $E_t$ and $S^{\rm sys}_t$ the (stochastic) energy and nonequilibrium entropy of the system at time $t$. We will provide a proof of the second law~\eqref{eq:secondlawnoneq}  in Ch.~\ref{TreeSL}, see Eq.~\eqref{eq:refin11}.}.

 A relevant question in this context is the following.
 Can one find a suitable stopping strategy ---in particular a bounded  stopping time $\T\leq t$--- that results on an average work extracted that is above the  free energy difference $\langle \Delta G^{\rm ne}_\T\rangle$ averaged over all stopped trajectories? Note that here, $\langle \Delta G^{\rm ne}_\T\rangle = G^{\rm ne}_\T - G^{\rm ne}_0$ is calculated between the state at the stopping time and the initial state, therefore it involves trajectories of {\em stochastic} duration~$[0,\T]$. From the second law at stopping times~\eqref{eq:secondlaw_stop_nonstat00} and noting that \cite{esposito2011second} $S^{\rm tot}_s = (W_s- \Delta G^{\rm ne}_s)/T$  for isothermal systems, one has
 \begin{equation}
     \langle W_\T  \rangle - \langle \Delta G^{\rm ne}_\T\rangle \geq -T \langle \delta^{(\tau)}_{\mathcal{T}}\rangle  ,
 \label{eq:2Lgambling}
 \end{equation}
 where the stochastic distinguishability term $\langle \delta^{(t)}_{\mathcal{T}}\rangle $ is given as in Eq.~\eqref{eq:aprilfatigue00}. Equation~\eqref{eq:2Lgambling} opens the possibility for average work extraction beyond the nonequilibrium free energy change using stopping times.

In Ref.~\cite{manzano2021thermodynamics}, a gambling demon was proposed theoretically and realized  with a single-electron transistor (SET) experimental setup. Briefly, the dynamics of an electron hopping in an out of  metallic island was tracked in time. The energy of the island was externally controlled through a deterministic protocol that was repeated many times to extract sufficient statistics. The stochastic dynamics of the electron resembles that of a two level system with states $0$ and $1$ and time-dependent transition rates. Under the assumption of local detailed balance the transition rates between the two states obey
\begin{equation}
    \omega(0,1)/\omega(1,0) = \exp(-\Delta V /T)
\end{equation}
where $\Delta V$ is the energy difference between the two levels at time $s\in[0,t]$.
A useful choice of gambling strategy is given by the family of stopping times 
\begin{equation}
    \T  = \min ( \T_{\rm wth} ,t),
    \label{eq:stopcond}
\end{equation}
where $\T_{\rm wth}$ is the first passage time of the work done on the system to reach a predefined threshold value $W_{\rm th}\geq 0$. For the two-level model system considered here, the work done up to time $s\leq t $ reads 
\begin{equation}
    W_s = V_s(X_s)-V_0(X_0)     -\sum_{j=1}^{N_s}    \left[V_{\T_j}(X_{\T_j^+}) -  V_{\T_j}(X_{\T_j^-})\right],
    %\sum_{\big\{s=0 \, \big\vert X_{s^-}\neq  X_{s^+}\big\}}^{t} \big[ E_{s^+}(X_{s^+})-E_{s^-}(X_{s^-})\big,
    \label{WsjGD}
\end{equation}
which follows from Eq.~\eqref{Wsj}. We recall here that the second term in~\eqref{WsjGD} is the heat absorbed by the system, which involves the energy change of the system at the  $j-$th jump between states $X_{\T_j^-}\to X_{\T_j^+} $, and that $N_t$ is the total number of jumps in the trajectory~$\Xot$.  
We recognize in the right hand side of~\eqref{WsjGD} the first term as the energy change and the second term as the heat absorbed by the system up to time~$t$. Note also that here $X_t\in \{ 0,1 \}$ for all $t$ and time is assumed to be continuous. The gambling strategy resulting from executing the stopping condition~\eqref{eq:stopcond} is such that it satisfies $\T\leq t$, as required by the second law at stopping times~\eqref{eq:2Lgambling}. It is important to remark that other strategies involving stopping times would also satisfy the same constraint. The strategy defined by~\eqref{eq:stopcond} is such that the work at the end of the gambling protocol $W_\T $ is a random variable which takes the value 
\begin{equation}
W_\T = \begin{cases}
  W_{\rm th} &  \text{ if}\quad \T<t \\
  W_{t}\leq W_{\rm th} & \text{ if}\quad \T=t
\end{cases}.
\end{equation}
Because $W_{t}$ is a random variable, $W_\T$ is also a random variable whose distribution depends crucially on the threshold value $W_{\rm th}$. 

Experimental results in Ref.~\cite{manzano2021thermodynamics} explored the fluctuations of $W_\T$, with $\T$ defined by Eq.~\eqref{eq:stopcond}, for different values of the work threshold $W_{\rm th}$, see Fig.~\ref{fig:GD2}.
\begin{figure}[h!]
    \centering
    \includegraphics[width=0.9\textwidth]{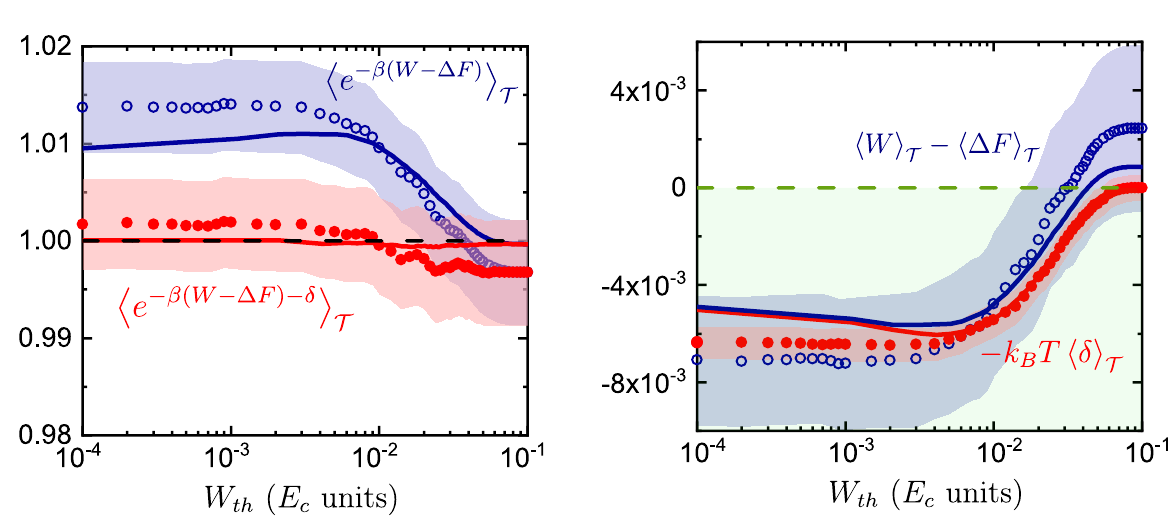}
    \caption{Experimental verification of the integral fluctuation theorem at stopping times~\eqref{eq:IFT_stop_Stot_nonstat00_isothermal} (left) and of the second law at stopping time~\eqref{eq:2Lgambling} (right) for the gambling demon setup: experimental values (circles) and theoretical predictions (lines). In both panels we plot the results obtained as a function of the work threshold value $W_{\rm th}$  used in the stopping rule given by Eq.~\eqref{eq:stopcond}. Here $E_c=109\mu$eV is the charging energy of the island. 
    For large threshold values, Jarzynski's equality (top) and the standard second law (bottom) are recovered, as expected.
    See Ref.~\cite{manzano2021thermodynamics} for details.  }
    \label{fig:GD2}
\end{figure}
Figure~\ref{fig:GD2}a shows that the fluctuations of the work done up to the stopping time $\T$ defined by Eq.~\eqref{eq:stopcond} does not satisfy Jarzynski's equality, i.e.
\begin{equation}
\langle\, \exp(-(W_\T - \Delta G^{\rm ne}_\T)/T)  \, \rangle \neq 1.
\label{eq:violationChrisstoppingtime}
\end{equation}
Notably, one recovers $\langle\, \exp(-(W_\T - \Delta G^{\rm ne}_\T)/T)  \, \rangle = 1$ for the case of $W_{\rm th}$ large, which corresponds to the case in which no gambling is executed at all, and all trajectories have the same duration  $\T = t$, as  in Jarzynski's  setup. The experimental results are however in excellent agreement, see Fig.~\ref{fig:GD2}a, for all threshold values $W_{\rm th}$ with  the integral fluctuation relation at stopping times
\begin{equation}
\langle\, \exp(-(W_\T - \Delta G^{\rm ne}_\T)/T)  \exp(-\delta^{(t)}_\mathcal{T} )\, \rangle = 1,
\label{eq:IFT_stop_Stot_nonstat00_isothermal}
\end{equation}
which is a special case of Eq.~\eqref{eq:IFT_stop_Stot_nonstat00} for isothermal systems. Consistent with Eq.~\eqref{eq:IFT_stop_Stot_nonstat00_isothermal}, the average work done on the system by gambling along trajectories of stochastic duration $\T$ obeys the second law at stopping times  $\langle W_\T  \rangle - \langle \Delta G^{\rm ne}_\T\rangle \geq -T \langle \delta^{(t)}_{\mathcal{T}}\rangle$, see Eq.~\eqref{eq:2Lgambling} which follows from applying Jensen's inequality to~\eqref{eq:IFT_stop_Stot_nonstat00_isothermal}. For the experimental conditions used in~\cite{manzano2021thermodynamics}, the term $\langle \delta^{(t)}_{\mathcal{T}}\rangle$
 was positive for all the choices of the work threshold $W_{\rm th}$, see Figure~\ref{fig:GD2}b (red circles).  Moreover, the second law at stopping times~\eqref{eq:2Lgambling} provides a tight bound in this system, which leads to values of work extraction at stopping times beyond the free energy change along the stopped trajectories, i.e.   $\langle W_\T  \rangle \leq  \langle \Delta G^{\rm ne}_\T\rangle$, a result that is forbidden by the standard second law, i.e. without using gambling or feedback control. Moreover as it was shown in~\cite{manzano2021thermodynamics} that the extent at which the "traditional" second law is violated, measured by how negative can  $\langle W_\T  \rangle - \langle \Delta G^{\rm ne}_\T\rangle$ be, depends on the degree of time-asymmetry induced by  the external protocol, which can be rationalized as follows. When the system is driven slowly (fast), the statistics of the  forward and backward protocols are similar (fast) at stopping times, which makes the stochastic distinguishability term to be small (large).
 
%%%%%%%
%\include{chap7}

%[Martingales in stochastic thermodynamics V]
\chapter{Martingales in stochastic thermodynamics V: The "tree" of  second laws}

\label{TreeSL}

\epigraph{\textit{H\"anggi's Law: The more trivial your research, the more people will read it and agree. You write a nontrivial paper and you likely will be the only one who will remember it.} \\ Arthur Bloch, Murphy's Law: Book three (1985).}

%\section{The "tree" of  Second Laws for Markovian processes}

\begin{figure}[h!]
\centering
\includegraphics[width=0.7\textwidth]{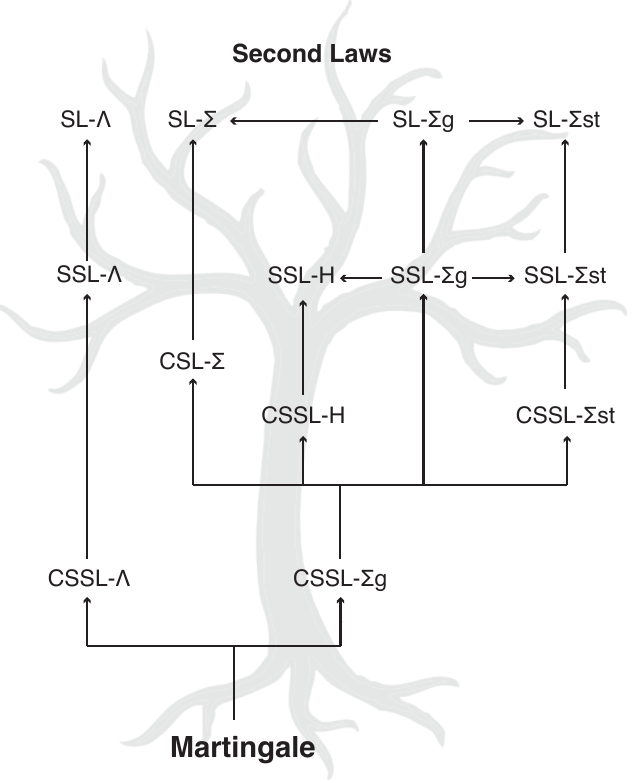}
\caption{ A "tree" of second laws emerge with its root at the martingale property of the entropic functionals  introduced in Sec.~\ref{sec:6p1},~\ref{sec:6p2} and \ref{sec:623} . The second laws are arranged in a hierarchical structure with the arrows denoting  which laws follow as specific examples of more general results.  
The different acronyms stand for different formulations of the second law introduced below. See Sec.~\ref{sec:CSSL} for the definitions of the conditional strong second laws (CSSL).  See Sec.~\ref{sec:CSL} for the definitions of the conditional second laws (CSL). See Sec.~\ref{sec:SSLs} for strong second laws (SSL).  See Sec.~\ref{sec:SL} for second laws (SL). }
\label{fig:tree}
\end{figure}

This Chapter provides different formulations of the second law of thermodynamics descending from the martingale properties unveiled in Ch.~\ref{ch:thermo2}. As fruits of the martingale theory of stochastic thermodynamics, we derive a plethora of second-law-like inequalities from the submartingale conditions of chapter Ch.~\ref{ch:thermo2}, from which the second laws Eqs.~(\ref{eq:mart_Stot5}), (\ref{eq:CSSLChp8}) and (\ref{eq:submart_Stot_nonstat22}) from Chapters ~\ref{sec:martST2} and \ref{sec:martST3} are specific examples.     

The "classical" second law of thermodynamics that appears in stochastic thermodynamics takes the form 
\begin{equation}
 \langle\, Z(\Xot)\;  \rangle \geq  0,   
 \end{equation}  
where $Z$ is a functional evaluated over stochastic trajectories $\Xot$.    Instead, martingale theory provides  second laws  involving   conditional expectations 
\begin{equation}
\langle\, Z(X_{[r,u]})\; | X_{[s,t]} \, \rangle  \geq Z(X_{[s,t]}), 
\end{equation}
 for all  $0\leq r\leq s\leq t\leq u$.   Hence, with martingale theory we can address how knowledge about a system's trajectory affects the second law of thermodynamics.

Figure~\ref{fig:tree} illustrates the "tree"-like  hierarchy of the different formulations of the Second Law of Thermodynamics  that we derive from the martingales   of Sec.~\ref{sec:6p2} -\ref{sec:623}.     The different formulations of the second law depend on the amount of  knowledge we have available about a system's trajectory.   The   versions of the second law of thermodynamics that appear at the bottom of the tree assume that the observer has detailed knowledge available about  the system's trajectory, while the observer's knowledge decreases when ascending the tree leading to weaker version of the second law of thermodynamics.

In this Chapter, we assume for simplicity that $X_t$ is  a     Markov process in discrete or continuous time, even though most of the  results  can also be formulated for generic stochastic processes.

\section{Conditional Strong Second
Laws  (CSSL)}
\label{sec:CSSL}

The martingale properties for entropic functionals discussed in Chapter~\ref{ch:thermo2} can be interpreted as conditional strong second laws, which constrain the average of entropic functionals in a future time $t$ conditioned on the fact that the system traces  a specific  trajectory $X_{[0,s]}$  up to a previous time $t\leq s$. 

\subsection{Conditional Strong Second
Law for $\Lambda$-stochastic entropic functionals (CSSL-$\Lambda$)}

The submartingale  condition~\eqref{eq:mart1index}  for $\Lambda$-stochastic entropic functionals  states that 
\begin{equation}
\left\langle \left. \Lambda_{t}^{\mathcal{P},\mathcal{Q}}\right|X_{\left[0,s\right]}\right\rangle \geq\Lambda_{s}^{\mathcal{P},\mathcal{Q}},\label{eq:CSSLst}
\end{equation}
 for all  $0\leq s\leq t$.  
In other words, it is not possible to anticipate a decrease in $\Lambda_{s}^{\mathcal{P},\mathcal{Q}}$ based on knowledge of the past trajectory $X_{[0,s]}$.   
%$\Lambda_{t}$ is \textit{conditionally increasing} with final time $t$. 
A physical  example  of a conditional strong second law, is 
\begin{equation}
    \langle S^{\rm hk}_t|X_{[0,s]}\rangle \geq S^{\rm hk}_s,
\end{equation}
where  $S^{\rm hk}_t$  is the 
the housekeeping entropy production, as defined in Eq.~\eqref{eq:Sehk}.

\subsection{Conditional Strong
Second Law for  $\Sigma$-stochastic entropic functionals when $\mQ=\mQ^{\rm st}$ (CSSL-$\Sigma_{st}$)}
\label{CSSLst1000}

As we have discussed in Chapter \ref{sec:6p2}, $\Sigma$-stochastic entropic functionals are  in general  not submartingales, unless the reference path probability $\mQ$ is time independent,   stationary, and time homogeneous, i.e.,    $\mathcal{Q}=\mathcal{Q}^{\rm st}$.      In this case, the  
the submartingale condition~\eqref{eq:mart_Stot1} reads
\begin{equation}
\left\langle \left. \Sigma_{t}^{\mathcal{P},\mathcal{Q}^{\rm st}}\right|X_{\left[0,s\right]}\right\rangle \geq\Sigma_{s}^{\mathcal{P},\mathcal{Q}^{\rm st}}.
\label{eq:CSSLst10}
\end{equation} 
for all  $0\leq s\leq t$,    
which means that it is not possible to anticipate a decrease in  $\Sigma_{t}^{\mathcal{P},\mathcal{Q}^{\rm st}}$ based on the knowledge of the past trajectory $X_{[0,s]}$.  
If moreover $\mathcal{P}$ (or $X_t$)  is a stationary Markov process, i.e., $\mathcal{P}=\mathcal{P}^{\rm st}$,  and it holds that $\Sigma_{t}^{\mathcal{P}^{\rm st},\mathcal{P}^{\rm st}} = S^{\rm tot}_t$, with $S^{\rm tot}_t$ the 
 total  entropy production  given by Eq.~\eqref{Stotcopied}, then   the  conditional strong second law Eq.~(\ref{eq:CSSLst10}) for $S^{\rm tot}_t$ reads  
\begin{equation}
\left\langle  S^{\rm tot}_t\big|X_{[0,s]}\right\rangle \geq S^{\rm tot}_s.
\label{eq:CSSLstEP}
\end{equation} 
Notice that (\ref{eq:CSSLstEP}) is   Eq.~(\ref{eq:mart_Stot5}) in Chapter~\ref{sec:martST2}.

 On the other hand, if  $\mathcal{P}$ (or $X_t$)  is nonstationary, then the total entropic functional  $\Sigma_{t}^{\rm tot}$ and the total stochastic entropy production  $S_{t}^{\rm tot}$ do not  satisfy   conditional strong second laws.  The same reasoning applies to the  excess stochastic entropy production $S^{\rm ex}_{t}$ given by Eq.~\eqref{eq:excessentropy}.

%\item Finally,  as a favorite example of physical system, we consider  Isothermal system such that the relation  \eqref{W-w}   is valid and give the relation between the total $\Sigma$-stochastic entropic Functional and the fluctuating work 
% \begin{equation}
%W_{t}=T\left(\varSigma_{t}^{tot}+F_{t}^{eq}-F_{0}^{eq}\right).
%\label{W-w3}
%\end{equation}
%Here $T$ is the Bath temperature.  This is the case by example of \textbf{Overdamped Isothermal Langevin equation \eqref{eq:LEO}} as we shown in Section 6.1 by choosing 
 
% \begin{equation}
%\mathcal{Q}=\widetilde{\mathcal{P}}\textrm{ and }\begin{cases}
%\rho_{0}(x)=\exp\left(-\frac{H_{0}(x)-F_{0}^{eq}}{T}\right)\\
%\rho_{0}^{\mathcal{Q}}(x)=\exp\left(-\frac{H_{t}(x)-F_{t}^{eq}}{T}\right)
%\end{cases}.
%\label{Choice}
%\end{equation}

\subsection{Conditional Strong Second
Law for the generalized $\Sigma$-stochastic entropic functional (CSSL-$\Sigma_{g}$)}

Generalized $\Sigma$-stochastic entropic functionals $\Sigma_{\left[r,s\right],t}^{\mathcal{P},\mathcal{Q}}$ with  $[r,s]\subseteq[0,t]$ are forward submartingales with respect to $s$ when $r$ and $t$ are fixed    (see Eq.~\eqref{eq:forwardsubmartsigmag}) and  backward submartingales with respect to $r$ when $s$ and $t$ are fixed (see Eq.~\eqref{eq:superbackmart4index_p4}).    We unify these two statements by formulating a conditional strong second law.  
\begin{leftbar}
The generalized $\Sigma$-stochastic entropic functional $\Sigma_{\left[r,s\right],t}^{\mathcal{P},\mathcal{Q}}$ with $[r,s]\subseteq[0,t]$, as defined in Eq.~\eqref{GEF}, obeys the following  {\bf conditional strong second law (CSSL-$\Sigma_{g}$)} 
\begin{equation}
\left\langle \left.\Sigma_{\left[r',s'\right],t}^{\mathcal{P},\mathcal{Q}}\right|X_{\left[r,s\right]}\right\rangle \geq\Sigma_{\left[r,s\right],t}^{\mathcal{P},\mathcal{Q}}, 
\label{eq:CSSLG}
\end{equation}
for all $0\leq r'\leq r\leq s\leq s'\leq t$.
In words, $\Sigma_{\left[r,s\right],t}^{\mathcal{P},\mathcal{Q}}$  \textit{conditionally increases} with respect to the final time $s$ and \textit{conditionally decreases} with respect to the initial time $r$.
\end{leftbar}
The CSSL-$\Sigma_{g}$ given by Eq.~\eqref{eq:CSSLG} implies all the conditional strong second laws presented in Sec.~\ref{CSSLst1000}, and is the root of many of the most well-known formulations of the second law of thermodynamics, see Fig.~\ref{fig:1}; for example, it  implies the second laws  Eqs.\eqref{eq:CSSLChp8} and \eqref{eq:submart_Stot_nonstat22}.

 The conditional strong second law  \eqref{eq:CSSLG}   together with the relation \eqref{beau} proved in Ch.~\ref{ch:thermo2}, gives for arbitrary Markovian process the following relation. For all $0\leq r'\leq r\leq s\leq s'\leq t$ it holds that
\begin{eqnarray}
&&\left\langle \left. \left( \ln\left(\frac{\rho_{r'}\left(X_{r'}\right)}{\rho_{t-s'}^{\mQ^{(t)}}\left(X_{s'}\right)}\right)+S_{s'}^{env,\mathcal{\mP},\widehat{\mQ}^{(t,s')}}-S_{r'}^{env,\mathcal{\mP},\widehat{\mQ}^{(t,r')}} \right)
\right|X_{\left[r,s\right]}\right\rangle \geq\nonumber\\
&&\hspace{2cm}\ln\left(\frac{\rho_{r}\left(X_{r}\right)}{\rho_{t-s}^{\mQ^{(t)}}\left(X_{s}\right)}\right)+S_{s}^{env,\mathcal{\mP},\widehat{\mQ}^{(t,s)}}-S_{r}^{env,\mathcal{\mP},\widehat{\mQ}^{(t,r)}}, 
\label{eq:CSSLG22}
\end{eqnarray}
in terms of  the environmental  $\widehat{\mQ}$-stochastic entropy change \eqref{eq:decom1}. Here $\widehat{\mQ}^{(t,.)}$ is the path probability of a   Markovian process with  generator given in 
\eqref{eq:nov}. In the same way, with the relation \eqref{beauE} proved in chapter \eqref{ch:thermo2}, we obtain  for arbitrarily Markovian process the  {\bf conditional strong second law} for all $0\leq r'\leq r\leq s\leq s'\leq t$  
\begin{eqnarray}
&&\left\langle \left. \left( 
\ln\left(\frac{\rho_{s'}\left(X_{s'}\right)}{\rho_{t-s'}^{\mQ^{(t)}}\left(X_{s'}\right)}\right)+S_{s'}^{\mathcal{\mP},\widehat{\mQ}^{(t,s')}}-S_{r'}^{\mathcal{\mP},\widehat{\mQ}^{(t,r')}}
\right)
\right|X_{\left[r,s\right]}\right\rangle \geq\nonumber\\
&&\hspace{2cm}\ln\left(\frac{\rho_{s}\left(X_{s}\right)}{\rho_{t-s}^{\mQ^{(t)}}\left(X_{s}\right)}\right)+S_{s}^{\mathcal{\mP},\widehat{\mQ}^{(t,s)}}-S_{r}^{\mathcal{\mP},\widehat{\mQ}^{(t,r)}}, 
\label{eq:CSSLG222}
\end{eqnarray}
in term of the $\widehat{\mQ}$-stochastic entropy production. 
\begin{leftbar}
As a special case of \eqref{eq:CSSLG22}, the relation \eqref{beau2} of Ch.~\ref{ch:thermo2}  allows to derive the  following {\bf conditional strong second law} for all $0\leq r'\leq r\leq s\leq s'\leq t$ :  
\begin{equation}
\left\langle 
\underbrace{
\ln\left(
\frac{\rho_{r'}\left(X_{r'}\right)}{\tilde{\rho}^{(t)}_{t-s'}\left(X_{s'}\right)}\right)}_{\displaystyle \alpha_{r',s'}^{(t)}}+S_{s'}^{\rm env}-S_{r'}^{\rm env}
\Big|X_{\left[r,s\right]}\right\rangle \geq
\underbrace{
\ln\left(
\frac{\rho_{r}\left(X_{r}\right)}{\tilde{\rho}^{(t)}_{t-s}\left(X_{s}\right)}\right)}_{\displaystyle \alpha_{r,s}^{(t)}}+S_{s}^{\rm env}-S_{r}^{\rm env}, 
\label{eq:CSSLG222222}
\end{equation}
which generalizes Eq.~ \eqref{eq:CSSLChp8} and where $S_{s}^{\rm env}$ is the environment entropy change defined in Eq.~\eqref{Senvtot}.  Furthermore, using the decomposition \eqref{Stotcopied} of total entropy production, we obtain the following {\bf conditional strong second law } for generic Markovian process and for all $0\leq r'\leq r\leq s\leq s'\leq t$ : 
\begin{equation}
\left\langle 
\underbrace{\ln\frac{\rho_{s'}(X_{s'})}{\tilde{\rho}^{(t)}_{t-s'}(X_{s'})}}_{\displaystyle\delta_{s'}^{(t)}}
+
\displaystyle S^{\rm tot}_{s'}-S^{\rm tot}_{r'}
\Big|X_{\left[r,s\right]}\right\rangle \geq
\underbrace{\ln\left(\frac{\rho_s(X_s)}{\tilde{\rho}^{(t)}_{t-s}(X_s)}\right)}_{\displaystyle\delta_{s}^{(t)}}
+
\displaystyle S^{\rm tot}_s-S^{\rm tot}_{r}, 
\label{eq:CSSLG222222222}
\end{equation}
which generalizes Eq.~\eqref{eq:submart_Stot_nonstat22}. This relation extends the  conditional strong second law~\eqref{eq:CSSLstEP} to the nonstationary setup. 
\end{leftbar}

%}

\subsection{Conditional version of the Historical Second
Law (CSSL-$H$) for Markovian processes}
\label{sec:historical}

From the CSSL-$\Sigma_{g}$ given by Eq.~\eqref{eq:CSSLG}, it is possible to derive many well-known formulations of the second law of thermodynamics.%, including some "historical" versions for $X_t$ Markovian such as the one we now consider. 

Let us consider the following $\Sigma$-stochastic entropic functional  that only depends on the state $X_r$ at the initial time $r$ of the interval of interest $[r,s]$, viz., 
%we have that for all $0\leq r\leq s\leq t$ : 
\begin{equation}
\Sigma_{\left[r,s\right],t}^{\mathcal{P},\mathcal{P}^{\rm h, (t)}}=\ln\left(\frac{\rho{}_{r}}{\rho'_{r}}\left(X_{r}\right)\right).
\label{eq:historicalsigmag}
\end{equation}
Here, $\rho'_{r}$ represents the  instantaneous density of a Markov process that has the same generator $\mathcal{L}$ as the process  $X_t$, but with an initial density  $\rho'_{0}$ that may be different from $\rho_0$, the probability density of $X_0$ under its native measure $\mP$.   

The  path probability $\mQ=\mathcal{P}^{ \rm h, (t)}$  that determines the generalised $\Sigma$-stochastic entropic functional in  Eq.~\eqref{eq:historicalsigmag}  has a similar structure to the excess path probability $\mathcal{P}^{ {\rm ex}, (t)}$, as defined in Sec.~\ref{sec:Pex}. In particular, $\mathcal{P}^{ \rm h, (t)}$ is the path probability of  a process with  initial density $\rho_{0}^{\mathcal{P}^{ \rm h, (t)}}=\rho'_{t}$ and with a Markovian generator that is given by the generalized Doob's h-transform   
\begin{equation}
\mathcal{L}_{s}^{h, (t)}\equiv\left(\rho'_{t-s}\right)^{-1}\circ\mathcal{L}_{t-s}^{\dagger}\circ\rho'_{t-s}-\left(\rho'_{t-s}\right)^{-1}\left(\mathcal{L}_{t-s}^{\dagger}\rho'_{t-s}\right),
\label{eq:genDoobhistoric}
\end{equation}
{which holds for  $s\leq t$, and $\circ$  denotes here the composition operator. See Ref.~\cite{chetrite2015nonequilibrium} for additional information about continuous-time Doob's $h$-transform.}   
Note that, if we replace in Eq.~\eqref{eq:genDoobhistoric} the density  $\rho'_t$ by the accompanying density $\pi_t$, as defined in Eq.~(\ref{eq:acc}), then  we get the Markovian generator associated with the "excess" dynamics $\mP^{\rm ex, (t)}$, see Eq.~\eqref{AFS}~\footnote{ See p. 174-175 in~\cite{chetrite2018peregrinations} for a detailed proof of Eq.~\eqref{eq:historicalsigmag}}. 
\begin{leftbar}
Specializing  the  backward submartingale relation~(\ref{eq:CSSLG})  to the choice~\eqref{eq:historicalsigmag} yields the following 
{\bf conditional version} of the  {\bf "historical" second
law (CSSL-$H$)}, viz.,
\begin{equation}
\left\langle \left.\ln \left( \frac{\rho{}_{r}\left(X_{r}\right)}{\rho'_{r}\left(X_{r}\right)}\right)\right|X_{\left[s,t\right]}\right\rangle \geq\ln \left( \frac{\rho{}_{s}\left(X_{s}\right)}{\rho'_{s}\left(X_{s}\right)}\right)\,
\label{eq:CSLH}
\end{equation}
for all $0\leq r\leq s\leq t$.  
Because $r\leq s$, Eq.~\eqref{eq:CSLH}  implies that  $\ln \frac{\rho{}_{r}\left(X_{r}\right)}{\rho'_{r}\left(X_{r}\right)}$ is conditionally increasing  in the reverse flow of time. 

%{We have \textbf{$\textbf{CSSL-$\Sigma_{g}$}\Rightarrow \textbf{ CSSL-H}.$} The name ''Conditional version of the Historical Strong Second Law'' will make sens by averaging \eqref{eq:CSLH} in the paragraph \ref{SSL-H}.  [TO Me THIS IS TRIVIAL AS WE ARE SAYING THIS IS A PARTICULAR CASE] } {\color{red}OK WITH YOU}
\end{leftbar}
%{ Then, the relation \eqref{eq:backmart4index_p4}  means that $\frac{\rho'_ r}{\rho_{r}}\left(X_{r}\right)$ is an Backward martingale, i.e. for all $0\leq r\leq s\leq t$ : 
%\begin{equation}
%\left\langle \left.\frac{\rho'_{r}}{\rho_{r,}}(X_{r})\right|X_{\left[s,t\right]}\right\rangle =\frac{\rho'_{s}}{\rho_{s}}(X_{s}).\label{eq:CSSLH-1}
%\end{equation}
%Do we need this? I think not.}
%{\color{red}We can suppress, but then is will be maybe more mysterious. As you want}
%This relation can also be shown with elementary direct calculation given in footnote (32) p 175 of the Habilitation. 

Now, we consider two examples  for which   the  conditional version of the historical strong second law CSSL-$H$ is particularly beautiful. 
\begin{itemize}

\item {\it "Canonical" setup}:  Let us consider $X_t$ a process  which starts from an arbitrary initial distribution $\rho_0(X_0)$ and  has a  stationary density  given by the  Gibbs canonical distribution $ \rho_{\rm st}(x)=\exp(-(H(x)-G^{\rm eq})/T)$, with the equilibrium free energy~$G^{\rm eq}=\int_{\mathcal{X}}dx \exp(-H(x)/T)$.
Such dynamics, starting from a  non-Gibbsian initial distribution,  is sometimes called {\em  relaxation process}. 
 This is the case for example of isothermal Langevin processes (Langevin equation \eqref{eq:LE} with Einstein relation \eqref{ER} with time-independent potential  and no external forces.  
 For such relaxation dynamics, we have 
\begin{eqnarray}
\ln\left(\frac{\rho_{t}(X_{t})}{\rho_{\rm st}(X_{t})}\right) & = & \ln\left(\rho_{t}(X_{t})\right)+\frac{H(X_{t})-G^{\rm eq}}{T},\label{eq:marre}
\end{eqnarray}
where in the right hand side, we recognize the \textbf{nonequilibrium 
free energy} which is defined as
\begin{equation}
G_{t}^{\rm ne}= H(X_{t})+T\ln\left(\rho_{t}(X_{t})\right),
\label{eq:noneqfree}
\end{equation}
which is a fluctuating quantity whose ensemble average is given by ~\cite{gaveau1997general,esposito2011second,parrondo2015thermodynamics}
\begin{equation}
    \left\langle G_{t}^{\rm ne}\right\rangle =\langle H(X_{t})\rangle +T\langle \ln\rho_{t}(X_{t})\rangle= \left\langle H(X_{t})\right\rangle -T\langle S_{t}^{\rm sys}\rangle  .
    \label{eq:noneqfreeavg}
\end{equation}

For the choice   $ \rho'_{t}=\rho_{\rm st}$, the CSSL-$H$~\eqref{eq:CSLH} with Eq.~\eqref{eq:marre} and using the definition~\eqref{eq:noneqfree}, we  derive a universal constraint for the expected value of the nonequilibrium free energy for such  relaxation processes.
\begin{leftbar}
Let $X_t$ represent a  process  that relaxes under isothermal conditions to  the stationary Gibbs canonical density~$\rho_{\rm st}(x)=\exp(-(H(x)-G^{\rm eq})/T) $, starting from an  arbitrary initial state $\rho_0(X_0)$.  In this case,  the  CSSL-$H$, given by Eq.~(\ref{eq:CSLH}), implies  that
\begin{equation}
\left\langle \left.G_{r}^{\rm ne}\right|X_{\left[s,t\right]}\right\rangle \geq G_{s}^{\rm ne},\label{eq:CSLH-22}
\end{equation}
for all $0\leq r\leq s\leq t$.   
Hence,  \textbf{nonequilibrium 
free energy}, given by Eq.~\eqref{eq:noneqfree}, of a  relaxation processes under isothermal conditions   is a \textbf{backward submartingale}.  
%In other words, the  nonequilibrium free energy conditionally
 %increases  in the reverse flow of time.  
\end{leftbar}
%Equation~\eqref{eq:CSLH-22} is a remarkable result. It tells us that, if one knows the values of the process in the future (at time $s\geq r$), it is possible to  "predict" (i.e. bound) the expected value of the nonequilibrium free energy in the past (at time $r$) even ignoring the outcomes of the process before time $s$.

\item {\it "Microcanonical" setup}:   Let us consider  $X_t$ a process which has a  homogeneous stationary  density $\rho_{\rm st}$  (e.g.  a driven Langevin process on a ring with constant force considered in Sec.~\ref{Ring}), the CSSL-$H$ given by Eq.~\eqref{eq:CSLH}  for the choice  $ \rho'_{t}=\rho_{\rm st}$,  gives that for all $0\leq r\leq s\leq t$, one has
\begin{equation}
\left\langle \left.\ln\left(\rho{}_{r}\left(X_{r}\right)\right)\right|X_{\left[s,t\right]}\right\rangle \geq\ln\left(\rho{}_{s}\left(X_{s}\right)\right).\label{eq:CSLH-1}
\end{equation}
The above equation can be formulated in terms of a constraint for the nonequilibrium system entropy $S^{\rm sys}_t = - \ln \rho_t(X_t)$, see Eq.~\eqref{eq:defSSysonetime}, as follows.
\begin{leftbar} 
For relaxation processes towards a homogeneous stationary state, the {\bf system entropy} is a {\bf backward supermartingale}, i.e., 
\begin{equation}
\left\langle \left.S_{r}^{\rm sys}\right|X_{\left[s,t\right]}\right\rangle \leq S_{s}^{\rm sys},\label{eq:CSLH-1-1}
\end{equation}
for all $0\leq r\leq s\leq t$.   Hence, in a  ''microcanonical'' setup, the   system entropy   conditionally decreases in the reverse flow of time.
\end{leftbar}
\end{itemize}
%\end{itemize} 

Note that  there exist two type of second laws, those that consider the expected value of an observable in the future given its past history, and those that consider the expected value of an observable in the past given its current history.   For example, the stochastic entropy production in a stationary process is a submartingale in the forward dynamics, implying  we cannot anticipate a decrease of entropy in the universe based on knowledge of the past's history of a system.   On the other hand, the nonequilibrium free energy in a relaxation process is a  submartingale in the backward dynamics, implying that  we expect free energy to have decreased in the past, irrespective of our knowledge of the system's trajectory
  Both laws imply that knowledge of a system's trajectory does not affect the second law, irrespective whether we look forwards or backwards in time.

%\subsection{CSSL-$\Sigma$ : Conditional Strong Second
%Law for $\Sigma$-stochastic entropic functional }

%The CSSL-$\Sigma_{g}$ (\ref{eq:CSSLG}) imply
%with the choice $r=0$ and $r'=t$ and %$\Sigma_{\left[0,t\right],t}^{\mathcal{P},\mathcal{Q}}=\Sigma_{\left[0,t\right]}^{\mathcal%{P},\mathcal{Q}}$, give that 
%that 
%\begin{leftbar}
%for all $0\leq s\leq s'\leq t$ : 
%\begin{equation}
%\left\langle \left.\Sigma_{\left[0,t\right]}^{\mathcal{P},\mathcal{Q}}\right|X_{\left[s,s'%\right]}\right\rangle \geq\Sigma_{\left[s,s'\right],t}^{\mathcal{P},\mathcal{Q}}
%\label{eq:CSLG-1}
%\end{equation}
%\end{leftbar}
%We note that this is valid even without stationnarity hypothesis for
%$\mathcal{Q}.$ But if $\mathcal{Q}=\mathcal{Q}_{st},$ the relation
%(6.85) premits to refind CSSL-$\Sigma_{st}$ \eqref{eq:CSSLst}) . 

\section{One-time Conditional Second Laws (CSL)}
\label{sec:CSL}

In the previous Section~\ref{sec:CSSL}, we  have introduced second-law-like inequalities for ensembles of trajectories satisfying  constraints that involve their values over a finite time window. Such conditional strong second laws  can be simplified when considering ensembles of trajectories $\Xot$ for which their value at a given time, e.g., $X_s$ for $s\leq t$ is constrained.  We call these relations one-time conditional second laws, which we abbreviate as CSL. 
\subsection{One-time Conditional Second Law for $\Sigma$-stochastic entropic functionals (CSL-$\Sigma$) and $\Lambda$-stochastic entropic functionals (CSL-$\Lambda$) }
\label{sec:6321}

From the definition of  generalized $\Sigma$-stochastic entropic functional over the subset interval $[r,s]\subseteq[0,t]$, see Eq.~\eqref{GEF},  we find 
\begin{equation}
    \Sigma_{\left[0,t\right],t}^{\mathcal{P},\mathcal{Q}}=\Sigma_{t}^{\mathcal{P},\mathcal{Q}},\qquad \Sigma_{\left[s,s\right],t}^{\mathcal{P},\mathcal{Q}}=\ln\left(\frac{\rho_{s}}{\rho_{t-s}^{\mathcal{Q}}}\left(X_{s}\right)\right).
\end{equation}
Then, the   Conditional Strong Second
Law for generalized  $\Sigma$-stochastic entropic functionals, i.e., the CSSL-$\Sigma_{g}$ given by Eq.~\eqref{eq:CSSLG},  implies a   one-time Conditional Second Law for the $\Sigma$-stochastic entropic functional (\textbf{CSL-$\Sigma$}), viz., for  $0\leq s\leq t$ and for an arbitrary auxiliary process $\mQ$,
\begin{equation}
\langle\Sigma_{t}^{\mP,\mQ} \,|\, X_s\, \rangle \geq \ln\left[\frac{\rho_{s}(X_s)}{\rho_{t-s}^{\mathcal{Q}}(X_s)}\right].
\label{eq:SL1}
\end{equation}
%We resume that we have  \textbf{CSSL-$\Sigma_{g}$}\Rightarrow \textbf{ CSL-$\Sigma$}.
Note that this result follows also from the choice $Z\left(\Xot\right)=\delta(X_s-x)$ in the mother fluctuation relation \eqref{eq:motherfluctrelation} and applying and Jensen's inequality.  We also note that the right-hand side of Eq.~\eqref{eq:SL1} can be negative.  Similarly, one can also prove  an analogous result, the  one-time conditional second law for  $\Lambda $-stochastic entropic functionals (\textbf{CSL-$\Lambda$}):
\begin{equation}
 \langle\Lambda_{t}^{\mP,\mQ} \,|\, X_s\, \rangle \geq \ln\left[\frac{\rho_{s}(X_s)}{\rho_{s}^{\mathcal{Q}}(X_s)}\right],
\label{eq:SL20}
\end{equation}
which holds for any $0\leq s\leq t$, and an arbitrary $\mQ$. 
  Averaging Eq.~\eqref{eq:SL1},~\eqref{eq:SL2} over all possible values of $X_s$,  we obtain for any $0\leq s\leq t$ the {\em deterministic refinements} of the second laws
\begin{eqnarray}
\langle\Sigma_{t}^{P,Q}\, \rangle& \geq & D\left[\rho_{s}(x)|| \rho_{t-s}^{\mathcal{Q}}(x)\right],\label{eq:SL2sigma}\end{eqnarray}
 and 
\begin{eqnarray}
 \langle\Lambda_{t}^{P,Q}\, \rangle &\geq& D\left[\rho_{s}(x)|| \rho_{s}^{\mathcal{Q}}(x)\right].
\label{eq:SL2}
\end{eqnarray}
 Notably, Eq.~\eqref{eq:SL2sigma}   extends the Kawai-Parrondo-Van~Den~Broeck relation derived in Ref.~\cite{Kawai2007} to arbitrary  nonequilibrium Markovian processes. 

%Recall that $D\left[\rho_{s}|| \rho_{t-s}^{\mathcal{Q}}\right]=\int_\mathcal{E} \text{d}x \rho_{s}(x) \ln [ \rho_{s}(x)/\rho_{t-s}^{\mathcal{Q}}(x)]$ is the KL divergence between the densities $\rho_{s}$ and $\rho_{t-s}^{\mathcal{Q}}$.
 
\subsection{One-time Conditional Second Law for isothermal Markovian systems }

For  $X_t$ an overdamped Markovian nonequilibrium process in isothermal conditions, we showed in Sec.~\ref{sec:langsigma} that  the total $\Sigma$-stochastic entropic functional  can be written in terms of  the fluctuating work  and the equilibrium free energy change, as $\Sigma_{t}^{\rm tot} = [W_{t}-(G_{t}^{\rm eq}-G_{0}^{\rm eq})]/T$, see Eq.~\eqref{W-w2}.
This result holds for driven isothermal systems  initially in thermal equilibrium, i.e. $\rho_{0}(x)=\exp\left(-(H_{0}(x)-G_{0}^{\rm eq})/T\right)$. As we showed in Sec.~\ref{sec:setupJarzover}, to obtain this simple relation between $\Sigma^{\rm tot}_t$ and $W_t$ one needs to choose as   auxiliary process that with initial density $\rho_{0}^{\mathcal{Q}}(x)=\exp\left(-(H_{t}(x)-G_{t}^{\rm eq})/T\right)$ with the "naive" time reversal of the Markov generator of the original process $\mathcal{L}^{\mQ}_s=\mathcal{L}_{t-s}$ ($s\leq t$). Applying the results from previous Sec.~\ref{sec:6321} to the functional $\Sigma_{t}^{\rm tot} = [W_{t}-(G_{t}^{\rm eq}-G_{0}^{\rm eq})]/T$ has important physical consequences that we explain below.
%\begin{equation}
%\mathcal{Q}=\widetilde{\mathcal{P}}\textrm{ and }\begin{cases}
%\rho_{0}(x)=\exp\left(-\frac{H_{0}(x)-G_{0}^{\rm eq}}{T}\right)\\
%\rho_{0}^{\mathcal{Q}}(x)=\exp\left(-\frac{H_{t}(x)-G_{t}^{\rm eq}}{T}\right)
%\end{cases}.
%\label{Choice1}
%\end{equation} 
\begin{enumerate} 
\item First, specializing Eq.~\eqref{eq:SL2sigma} to the choice~$\Sigma_{t}^{\rm tot} = [W_{t}-(G_{t}^{\rm eq}-G_{0}^{\rm eq})]/T$ and setting  $s=t$, one gets a refined  second law for the fluctuating work exerted on an isothermal system,
\begin{equation}
   \left\langle W_{t}\right\rangle \geq G_{t}^{\rm eq}-G_{0}^{\rm eq}+ T D_{\rm KL}\left[\rho_{t}(x)\bigg|\bigg| \exp\left(-\frac{H_{t}(x)-G_{t}^{\rm eq}}{T}\right)\right] = \left\langle G_{t}^{\rm ne}\right\rangle-\left\langle G_{0}^{\rm ne}\right\rangle. 
   \label{eq:refin11}
\end{equation}
The second equality in~\eqref{eq:refin11} follows from the definition~\eqref{eq:noneqfreeavg} for the average nonequilibrium free energy $\langle G^{\rm ne}_t\rangle= \langle H_t(X_t)\rangle  + T\langle \ln \rho_t(X_t)\rangle   $ and the fact that $\langle G^{\rm ne}_0\rangle=G^{\rm eq}_0$ since the system is initially in thermal equilibrium. This refinement of the second law was derived in~\cite{vaikuntanathan2009dissipation}, see also~\cite{parrondo2015thermodynamics}.

\item Second, specalizing Eq.~\eqref{eq:SL1} to the choice~$\Sigma_{t}^{\rm tot} = [W_{t}-(G_{t}^{\rm eq}-G_{0}^{\rm eq})]/T$, and setting $s=t$ equal to the final time, one retrieves the conditioned second law  
\begin{equation}
\left\langle \left.W_{t}\right|X_{t}\,\right\rangle +G_{0}^{\rm eq}\geq T\ln\left(\rho_{t}\left(X_{t}\right)\right)+H_{t}\left(X_{t}\right).
\label{eq:csl1tw}
\end{equation}
Then,  using Bayes theorem in~\eqref{eq:csl1tw} we obtain for any subset $\Omega\subseteq\mathcal{X}$ of the phase space $\mathcal{X}$,
\begin{eqnarray}
\left\langle  W_{t}\;\big|X_{t}\in \Omega \right\rangle +G_{0}^{\rm eq}&\geq& T\frac{\int_{\Omega}dx\rho_{t}\left(x\right)\ln\left(\rho_{t}(x)\right)}{\int_{\Omega}dx\rho_{t}\left(x\right)}+\frac{\int_{\Omega}dx\rho_{t}\left(x\right)H_{t}(x)}{\int_{\Omega}dx\rho_{t}\left(x\right)}\nonumber \\
&=&T\frac{\int_{\Omega}dx\rho_{t}\left(x\right)\ln\left(\frac{\displaystyle\rho_{t}(x)}{\displaystyle\exp\left(-H_{t}(x)/T\right)}\right)}{\int_{\Omega}dx\rho_{t}\left(x\right)}.
\label{2hdumat}
\end{eqnarray}
 Equation~\eqref{2hdumat} suggests introducing the {\em conformational free energy}~\cite{maragakis2008differential}
\begin{equation}
G_{t}^{\Omega}=-T\ln\left(\int_{\Omega}dx\exp\left(-\frac{H_{t}(x)}{T}\right)\right),
\end{equation}
which is in general different to the equilibrium free energy $G^{\rm eq}_{t}$ in which the integral is done over $\mathcal{X}$, see Eq.~\eqref{Feq}.  

\begin{leftbar}
Equation~\eqref{2hdumat} can be understood as a   conditional second law for the fluctuating work exerted along an arbitrary nonequilibrium process in isothermal conditions:
\begin{equation}
\left\langle \left.W_{t}\right|X_{t}\in \Omega\right\rangle \geq G_{t}^{\Omega}-G_{0}^{\rm eq}+ T\ln\left(\int_{\Omega}dx\,\rho_{t}\left(x \right)\right).
\label{eq:SB}
\end{equation}

Analogously,  plugging relation $\Sigma_{t}^{\rm tot} = [W_{t}-(G_{t}^{\rm eq}-G_{0}^{\rm eq})]/T$ in the equation \eqref{eq:SL1} but this time for $s=0$,  we get after some analogue algebra  the initial time  Condition Second Law for the fluctuating work exerted on an isothermal system 
\begin{equation}
\langle \, W_t \,|\, X_0\in  \Omega\, \rangle \geq   (G_t^{\rm eq} -  G_0^{\Omega}) -T \ln \left( \int_\Omega \text{d}x\, \rho_{t}^{\mathcal{Q}} (x)   \right)\quad.
\label{eq:SR}
\end{equation}
\end{leftbar}
Note that, because $\int_\Omega dx \rho_t(x) \leq \int_{\mathcal{X}}dx \rho_t(x) =1$, then the last term in  Eq.~\eqref{eq:SB} is negative, which implies that the average work done over trajectories that belong to the subset $X_t\in \Omega$ can be below  the conformational free energy change.

Proof of Eq.~\eqref{eq:SB}.  We have from  the relation \eqref{2hdumat} that 
\begin{equation}
\left\langle \left.W_{t}\right|X_{t}\in \Omega \right\rangle +G_{0}^{\rm eq}-G_{t}^{\Omega }\geq T\int_ {\Omega} dx\rho_{t,\Omega}\left(x\right)\ln\left(\frac{\rho_{t,\Omega}(x)\int_{\Omega}dy\rho_{t}\left(y\right)}{\exp\left(-\frac{H_{t}(x)-G_{t}^{\Omega}}{T}\right)}\right),
\label{eq:intermCSLFW}
\end{equation}
where 
\begin{equation}
    \rho_{t,\Omega}(x)=\frac{\displaystyle\rho_{t}\left(x\right)1_{\Omega}(x)}{\displaystyle\int_{\Omega}dx\rho_{t}\left(x\right)},
\end{equation} is the normalized density over the subset $\Omega$. Equation~\eqref{eq:intermCSLFW}
 can  also be written  as follows
\begin{eqnarray}
\left\langle \left.W_{t}\right|X_{t}\in \Omega\right\rangle +G_{0}^{\rm eq}-G_{t}^{\Omega}&\geq& T\int_ {\Omega} dx\rho_{t,\Omega}\left(x\right)\ln\left(\frac{\rho_{t,\Omega}(x)}{\exp\left(-\frac{H_{t}(x)-G_{t}^{\Omega}}{T}\right)}\right)+T\ln\left(\int_{\Omega}dy\rho_{t}\left(y\right)\right)\nonumber\\
&\geq& T\ln\left(\int_{\Omega}dx\rho_{t}\left(x\right)\right),
\end{eqnarray}
where in the second line we have used the fact that the first tem in the right hand side of the first line is positive because it is a Kullback-Leibler divergence. This concludes the proof of conditional second law for the fluctuating work~\eqref{eq:SB}.

Note that Eqs.~\eqref{eq:SB} and~\eqref{eq:SR}  were previously derived, respectively, in the context of the energetics of symmetry breaking and symmetry restoration  in Ref.~\cite{Roldan2014}. They provide generalization of Landauer's principle and a rationale for the energetics of Szilard's engine.  These relations, and related generalizations, have also been derived  in Refs.~\cite{junier2009recovery,alemany2012experimental,camunas2017experimental}, and  fruitfully applied to uncover thermodynamic properties of biopolymers in  single-molecule experiments.

\end{enumerate}

\section{Strong Second Laws  (SSL)}
\label{sec:SSLs}

The conditional strong second laws presented in Sec. \eqref{sec:CSSL} have as interesting colloraries the, so-called, strong second laws, which involve the rate of change of the average of $\Lambda$-stochastic, $\Sigma$-stochastic, and  generalized $\Sigma$-stochastic  entropic functionals. Moreover, it is also possible to recover a "historical" formulation of the second law, which we discuss below in Sec.~\ref{SSL-H}.

\subsection{Strong Second Law for $\Lambda$-stochastic  (SSL-$\Lambda$) and  $\Sigma$-stochastic  (SSL-$\Sigma_{\rm st}$) entropic functionals  of  stationary auxiliary process $\mQ = \mQ^{\rm st}$}

The conditional strong second law for entropic functionals CSSL-$\Lambda$, given by Eq.~\eqref{eq:CSSLst}, implies that the average of a  $\Lambda-$stochastic entropic functional   increases with time, i.e., 
\begin{equation}
\frac{d}{dt}\left\langle \Lambda_{t}^{\mathcal{P},\mathcal{Q}}\right\rangle \geq0,
\label{SSLst}
\end{equation} 
 for all $t\geq 0$. 
This motivates us to call $\Lambda_{t}^{\mathcal{P},\mathcal{Q}}$ a   Lyapunov function~\cite{baiesi2015inflow}, as it is, on average, an increasing function of time.   An example of   the strong second law~\eqref{SSLst} is when  $\Lambda_{t}^{\mathcal{P},\mathcal{Q}}$ is the housekeeping entropy production, see Eq.~\eqref{eq:submart_Shk}.

Similarly, the CSSL-$\Sigma_{\rm st}$ for $\Sigma$-stochastic entropic functionals with stationary auxiliary reference process implies a strong second law (SSL-$\Sigma_{\rm st}$), i.e.,  
\begin{equation}
\frac{d}{dt}\left\langle \Sigma_{t}^{\mathcal{P},\mathcal{Q}_{st}}\right\rangle \geq0,
\label{SSLst2}
\end{equation} 
 for all $t\geq 0$.
Equation~\eqref{SSLst} implies  that   functionals of the form $ \Sigma_{t}^{\mathcal{P},\mathcal{Q}_{st}}$ increase  on average in time. We note however that this result does not imply   the concavity in time, sometimes postulated for entropy production in classical thermodynamics~\cite{nicolis1977self}.

We  provide some remarks concerning the SSL-$\Lambda$~\eqref{SSLst} and the SSL-$\Sigma_{\rm st}$~\eqref{SSLst2}.
\begin{enumerate} 

\item Analogously as  what we have discussed in Sec. \ref{CSSLst1000}  for the CSSL-$\Sigma_{\rm st}$, if the process~$X_t$ is not stationary, then the total entropic functional  $\Sigma_{t}^{\rm tot}$ and the stochastic entropy production  $S_{t}^{\rm tot}$ do not necessarily obey  strong second laws. Analogously,  the excess stochastic entropy production $S^{\rm ex}_{t}$, given by Eq.~\eqref{eq:excessentropy},  does not satisfy, in general, a strong second law. 
%{This seems to contradict 6.120 which holds for generic $\mP$}

\item If the process $X_t$ is stationary, i.e.,  $\mathcal{P}=\mathcal{P}_{st}$, then the total $\Sigma$-stochastic entropic functional $\Sigma^{\rm tot}_{t}$ defined in~\eqref{AFS2} fulfills a strong second law. This includes the case of the total stochastic entropy production $S^{\rm tot}_t$~\eqref{Stotcopied}.  Lastly, for $X_t$ an overdamped isothermal stationary process, %system modelized by  stationary process, i.e.  $\mathcal{P}=\mathcal{P}_{st}$,  where the choice \eqref{Choice} give  the relation  \eqref{W-w30},
the associated second law \eqref{SSLst} is  the second law for fluctuating work exerted on isothermal system, viz.,~$
d\left\langle W_{t}\right\rangle /dt \geq0$.

%\item As a second preliminary conclusion, we have the hierarchy of second
%law 
%\textbf{$\textbf{CSSLst}\Rightarrow \textbf{SSLst}\Rightarrow \textbf{SLst}.$}
\end{enumerate} 

\subsection{Strong Second Laws for generalized $\Sigma$-stochastic entropic
functionals (SSL-$\Sigma_{g}$) }

Strong second laws also hold for the generalized $\Sigma$-stochastic entropic functionals in the subset interval $[r,s]\subseteq[0,t]$, for which we have shown that they fulfill two conditional strong second laws, one forward and another backwards in time, see Eq.~\eqref{eq:CSSLG}. This allows us to derive two strong second laws, one with respect to a decreasing   initial observation  time $r$, and another one with respect to an increasing final observation timed $s$.    
Indeed, averaging the CSSL-$\Sigma_{g}$ relation (\ref{eq:CSSLG}) implies  the following strong second laws for the  generalized $\Sigma$-stochastic entropic
functional (SSL-$\Sigma_{g}$),
\begin{equation}
\frac{\partial}{\partial s}\left\langle \Sigma_{\left[r,s\right],t}^{\mathcal{P},\mathcal{Q}}\right\rangle \geq0 
\end{equation}
and
\begin{equation}
\frac{\partial }{\partial r}\left\langle \Sigma_{\left[r,s\right],t}^{\mathcal{P},\mathcal{Q}}\right\rangle \leq 0,
\label{SSLg}
\end{equation}
where 
  $0\leq r\leq s\leq t$.
This result implies that  $\Sigma_{\left[r,s\right],t}^{\mathcal{P},\mathcal{Q}}$ is \textit{increasing} with time $s$ and decreasing with time $r$.
%We resume that we have  \textbf{CSSL-$\Sigma_{g}$}\Rightarrow \textbf{ SSL-$\Sigma_{g}$}
Note  that the SSL-$\Sigma$g does not imply that $\frac{\partial}{\partial t}\left\langle \Sigma_{t}^{\mathcal{P},\mathcal{Q}}\right\rangle \geq0$;  in general, the average $\Sigma$-stochastic entropic functional, given by  $\left\langle \Sigma_{t}^{\mathcal{P},\mathcal{Q}}\right\rangle= \left\langle \Sigma_{\left[0,t\right],t}^{\mathcal{P},\mathcal{Q}}\right\rangle  $, does  {\it not}  increase   monotonically a function of $t$~\footnote{Except for the case   $\mathcal{Q}=\mathcal{Q}_{st}$ where we have \eqref{SSLst}.}, as  setting $s=t$ in the relation~\eqref{SSLg} after taking  the derivative $\partial/\partial s$ is different from   setting  $s=t$ before taking  the derivative with respect to $s$.  %

\subsection{Historical Strong Second Law (SSL-H) for Markovian processes}
\label{SSL-H}
Below, we derive the  "historical" formulation of the strong second law as a  direct consequence  from the CSSL-$\Sigma_{g}$.   This is an important point, as it   reinforces the physical interest in the conditional strong second law.

%Averaging over  $X_{[s,t]}$  the relation  \eqref{eq:CSLH} ---which has conditional expectation for constrained trajectories $X_{[0,s]}$ up to time $s\leq t$---   we find for all $t\geq 0$ the contractive property of the Kullback-Leibler distance under the Markovian evolution.
\begin{leftbar}
Averaging  the Conditional version of the Historical Strong Second
Law~\eqref{eq:CSLH} over   $X_{[s,t]}$   (recall that $X_{[s,t]}$ is random in ~\eqref{eq:CSLH})  one retrieves the "historical" formulation of the second law (\textbf{SSL-H}) which is formulated as follows.  Let $\rho_t(x)$ be the instantaneous density at time $t$ of a generic stochastic process, and $\rho'_t(x)$ the density at the same time of a process which has the same dynamics but an arbitrary initial density $\rho'_0$ that may be different from the actual initial density of the process $\rho_0$.
For all $t \geq 0$  it follows that
\begin{equation}
\frac{d}{dt}D_{\rm KL}\left[\left.\rho_{t}(x)\right\Vert \rho'_{t}(x)\right]\leq0,
\label{HSL}
\end{equation}
with equality for the special case $\rho'_0(x)=\rho_0(x)$ which implies that $\rho_t(x)=\rho'_t(x)$ for all~$t$. 
Equation~\eqref{HSL} is considered by many authors ''the'' historical second law associated to a Markovian process in
many place in the literature, see the books and reviews~\cite{gardiner1985handbook,risken1996fokker,mackey1989dynamic,cover1999elements} and  also and the classic article~\cite{lebowitz1957irreversible}.
\end{leftbar}

We now provide some additional remarks about the SSL-H~\eqref{HSL}. 
\begin{enumerate}
%\item We can also find \eqref{HSL} from the second line of  \eqref{SSLg} by doing the choice  $\mathcal{Q}=\mathcal{Q}^{\ast}$ which permits to obtain (see )  for all $0\leq r\leq s\leq t$ :  $ \Sigma_{\left[r,s\right],t}^{\mathcal{P},\mathcal{Q}^{\ast}}=\ln\left(\frac{\rho{}_{r}}{\rho'_{r}}\left(X_{r}\right)\right). $   We resume that we have  \textbf{CSSL-H$\Rightarrow$ SSL-H} and \textbf{CSSL-$\Sigma_{g}\Rightarrow$} \textbf{SSL-$\Sigma_{g}\Rightarrow$ SSL-H}. 

\item Let us consider  the "microcanonical" relaxation setup introduced in Sec.~\ref{sec:historical},  i.e. a system with arbitrary initial distribution that relaxes towards a  homogeneous stationary distribution $\rho_{st}$. 
Averaring the CSSL-H~\eqref{eq:CSLH-1-1} over $X_{[s,t]}$, we obtain that the system entropy increases with time on average $d\left\langle S_{t}^{\rm sys}\right\rangle/dt \geq 0$.

\item Following an analogous procedure for the case of  "canonical" relaxations (i.e. a system such that its stationary density $\rho_{st}$ exist and is the Gibbs canonical
density) introduced in Sec.~\ref{sec:historical}, 
we obtain  $d\left\langle G_{t}^{\rm ne}\right\rangle/dt \leq 0$. 
\end{enumerate}

\section{Second Laws for entropic functionals (SL)}
\label{sec:SL}

To finalize our journey through the tree of second laws, Fig.~\ref{fig:tree}, we quote here second laws   that follow readily as corollaries   from  the   {\it strong} second laws   presented in Sec.~\ref{sec:SSLs}.

\subsection{Second Law for $\Sigma$-stochastic entropic functionals (SL-$\Sigma$) and  for $\Lambda-$ stochastic entropic functionals (SL-$\Lambda$)}

From the definitions of the $\Lambda-$stochastic  and $\Sigma$-stochastic entropic functionals, we have shown in Eq.~\eqref{poslun}   that
\begin{equation}
\langle\Lambda_{t}^{\mP,\mQ}\, \rangle   \geq 0, \qquad {\rm and} \qquad \langle\Sigma_{t}^{\mP,\mQ}\, \rangle   \geq 0,
\label{eq:SL3}
\end{equation}
for all $t \geq 0$,
which we refer to as the second law for $\Lambda-$stochastic entropic functionals (SL-$\Lambda$) and the second law for $\Sigma$-stochastic entropic functionals (SL-$\Sigma$), respectively.

%We  have 
%\textbf{\textbf{SSL-$\Lambda \Rightarrow$ SL-$\Lambda$}  and  \textbf{CSL-$\Sigma \Rightarrow$ SL-$\Sigma$}.
Consequently,  the SL-$\Sigma$  holds for all the  examples of $\Sigma$-stochastic entropic functionals introduced in Chapter~\ref{ch:thermo2}, inter alia,  $\Sigma^{\rm tot}_t$, $S^{\rm tot}_t$,  and $S^{\rm ex}_t$, and analogously, the SL-$\Lambda$ holds for all examples of  $\Lambda-$ stochastic entropic functionals considered, such as,  $S^{\rm hk}_t$. 

%\textbf{SL is valid for all stochastic $\mathcal{Q}$-Entropy production $S^{\mathcal{P},\mathcal{Q}}_{t}$ (total, excess,...) } which are particular case of $\Sigma$-stochastic entropic functional \eqref{eq:actionfunc}, but also for the Housekeeping entropy production which are particular case of $\Lambda$-Entropic functional.  

%\begin{enumerate}
%\item We can also see this relation as direct consequence of  $\textbf{CSL}$.

  %Note that the Oono-Paniconi refinement \eqref{RHSL} use the two present second law for conclude. 

%\item For  Isothermal system, plugging \eqref{W-w3}  in the associated left equation \eqref{eq:SL3}, give the second law  for the fluctuating work exerted on Isothermal system  
%\[
%\left\langle W_{t}\right\rangle \geq G_{t}^{eq}-G_{0}^{\rm eq}. 
%\]
%for the initial density $\rho_{0}(x)=\exp\left(-\frac{H_{0}(x)-F_{0}^{eq}}{T}\right)$
%\end{enumerate}

\subsection{Second Law for
generalized $\Sigma$-stochastic entropic functionals  (SL-$\Sigma_{g}$) }
We derive a second law for generalized $\Sigma$-stochastic entropic functionals  from the strong second law Eq.~(\ref{SSLg}).  

The definition of  $\Sigma^{\mP,\mQ}_{[r,s]}$,  given by Eq.~\eqref{GEF}, specialized to $s=r$, yields 
\begin{equation}
    \Sigma_{\left[r,r\right],t}^{\mathcal{P},\mathcal{Q}}=\ln\left(\frac{\rho_{r}}{\rho{}_{t-r}^{\mathcal{Q}}}\left(X_{s}\right)\right),
    \label{eq:sigmagrr}
\end{equation}
for  all $0\leq r\leq t$.    Using  Eq.~\eqref{eq:sigmagrr} in the   SSL-$\Sigma_{g}$ \eqref{SSLg}, we find that  
\begin{equation}
\left\langle \Sigma_{\left[r,s\right],t}^{\mathcal{P},\mathcal{Q}}\right\rangle \geq D_{\rm KL}\left[\left.\rho_{r}(x)\right\Vert \rho{}_{t-r}^{\mathcal{Q}}(x)\right],\label{eq:RWSL2}
\end{equation}
for all $0\leq r\leq s\leq t$.   As    the Kullback-Leibler divergence is nonnegative,  the second law
\begin{equation}
\left\langle \Sigma_{\left[r,s\right],t}^{\mathcal{P},\mathcal{Q}}\right\rangle \geq0, \label{eq:SLg}
\end{equation} 
for all $0\leq r\leq s\leq t$, ensues.   The second law Eq.~(\ref{eq:SLg})  holds for   generalized $\Sigma$-stochastic entropic functionals (SL-$\Sigma$), as given by Eq.~\eqref{poslun22}.    The  \eqref{eq:secondlaw_nonstat2222} is a specific example of the second law  Eq.~(\ref{eq:SLg}).
  
  This concludes our almanac of second laws derived from the martingale properties of entropic functionals.
\chapter{Martingales in progressive quenching} \label{sec:PQ}

\epigraph{We are from the very beginning illogical and thus unjust beings {\it and can recognize this}.\\   \vspace{0.1cm} F. Nietzsche, from ``Human, All Too {Human}''. }

In this chapter we meet with the martingale in physics in a different route from  the path probability ratio, which has been discussed in the previous two chapters.
We mostly use a discrete ``time'' variable.
We hope this chapter may provide with a new look at the martingale process in physics, and inspire the readers to explore its 
consequence in their domain of research.

\section{Introduction}\label{sec:PQintro}
 The conservation laws in physics are  in many cases  related to some form of invariance under symmetry operations. When a system has such a symmetry, the consequent conservation law imposes a ever-lasting memory of the initial condition.
The martingale property is a kind of {\it stochastic conservation} property. Unlike the sub- or super-martingale, the expectation of a random variable is kept constant once its value is observed at some point of time. Then the natural questions might be : 
(i) What form of memory is brought by the martingale property? 
and 
(ii) Is there any {invariance} behind its martingale property? 
Below we will give, through the study of the concrete model which we call Progressive Quenching (PQ), answers to these questions.

{As a part of} this review on the martingale in physics, this chapter brings two ingredients that might be of general interest for those who are entering this domain.
First we take the route to the martingale  through the so-called {\it tower rule} (or {\it tower property}) { [see Eq.~\eqref{eq:Tower1}]}, which is a  route distinct from the path probability ratio {mainly} discussed in the precedent chapters.
Secondly we show the case in which the martingale is found in the mean drift of the stochastic evolution of the principal process of interest. In the language of 
stochastic differential equations, $\dot{X}_t= G_t(X_t)+ \sqrt{2D} \dot{B}_t,$
it is $G_t(X_t)$ that is martingale. For such case we coin a word {\it hidden martingale} relative to the process~$X_t.$   

Below we will show that the martingale can be used for the inference of the past state and, moreover, for the prediction of the future probability distribution, beyond just some conditional expectations.
In Sec.~\ref{subsec:B} we introduce the notion of PQ process. Then in Sec.~\ref{subsec:C} we introduce the model we focus on, which is of discrete states and discrete time. We show the presence of martingale process behind the main stochastic process. In Sec.~\ref{subsec:D} we describe the consequences of the hidden martingale process, concerning the inference and the prediction. We conclude this Chapter in Sec.~\ref{subsec:E}.

%but the role of martingale in physics may have long been a vague question among the physicists 
   
\section{Progressive Quenching as a Neutral Operation}
\label{subsec:B}

We sometimes encounter the situations in which system's degrees of freedom become progressively fixed. 
When a molten material as a fluid system is pulled out as a string from a furnace and is quickly cooled down \cite{Damien2017}, the fluid degrees of freedom associated to fluid particles are progressively fixed (quenched).  See Fig.\ref{fig:Damien2017}.
%%%%%%%%%%%%%%%
\begin{figure}[h]
\centering
\includegraphics[width=5cm]{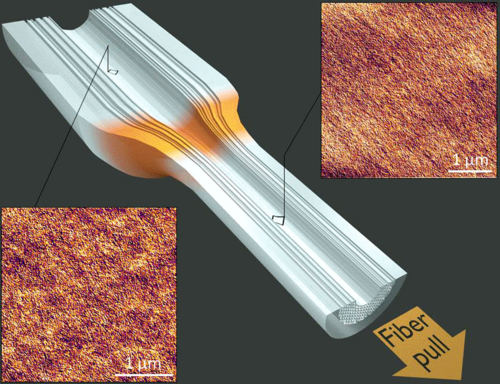}
\caption{Sketch of the drawing of hollow fibres. See \cite{Damien2017} for the details.
Figure taken from Fig.1 of \cite{Damien2017}.}
\label{fig:Damien2017}
\end{figure}
%%%%%%%%%%%%%%
The roughness exponents of the diffusion-type field, such as the surface undulation of the string, shows the modified and anisotropic exponents as compared with the equilibrium one \cite{phason-freezing-PhA}.
{Although the analogy is not close, we might also} consider the process of decision-making by a community, in which the members progressively make up her or his mind before the referendum.  In both examples, the already fixed part can influence the behavior of the part whose degrees of freedom are not yet fixed. 
We shall call these type of processes the progressive quenching, PQ.
It is largely unknown what generic aspects are in this type of problem.
In the non-equilibrium statistical mechanics viewpoint, the PQ should be categorised in such class that (1) the system's dynamics breaks the local detailed balance {(because the fixed part will never be unfixed afterwords)}, 
and that (2) the partition between the system and the external system is revised.
While the progress has been made a lot in understanding the repartition between the system and the bath since the last decade \cite{LNP}, the similar question for the system and the external system has been much less explored. %The hidden martingale was found by chance in the course of the exploitation of PQ \cite{PQ-KS-BV-pre2018}.

To have an intuition of PQ we first describe this process for a one-dimensional ferromagnetic Ising chain up to the second-nearest neighbour interaction \cite{PQ-chain}, whose energy $H$ can be written as 
\begin{equation}
- H= J_0\sum_{i=1}^{N-1} s_i s_{i+1}+J_1 \sum_{i=1}^{N-2} s_i s_{i+2}+h\sum_{i=1}^{N} s_i.
\end{equation} 
The protocol of PQ is described in {Fig.\ref{fig:schema1D}.} 
After an event of quenching (see below) is done, the unquenched part is re-equilibrated.
Then the polarity of a specified number of spins (one spin in the case of Fig.\ref{fig:schema1D}(a)) are fixed at their orientations that they took at the moment.
This is the quenching event.  The orientation of the newly fixed spins are, therefore, sampled from the equilibrium ensemble of the unquenched spins' configurations, but these spins are subject to the interactions with the {\it quenched spins} in addition to the interaction among the unquenched part.
We should note that this process is {\it not} quasi-static although {the unfixed spins are  completely re-equilibrated.} It is {in the sense that} the fixing of some spins implies to raise the barrier for the flipping of these spins so that the mean flipping interval exceeds the time-scale of observation/operation (see Chap.7.1 of \cite{LNP}).

If $J_1=0$ the system has only the nearest neighbour interaction and we can directly use the technique of the transfer matrix. For $J_1 > 0$ we can still use this technique by introducing the composite variable, $\xi_p\equiv \{s_{2p-1},s_{2p}\}.$ Using this technique {it was} found that, for all the four models of PQ shown in Fig.\ref{fig:schema1D}, the statistics of the finally quenched spins over the entire semi-infinite chain is identical to the equilibrium ensemble characterized by the temperature at which each spin has been quenched.
This result is somehow counterintuitive because the protocol of PQ is very far from equilibrium,  {breaking the local detailed-balance (LDB)} symmetry. A lesson that we might obtain from this solvable example is that PQ is a kind of {\it neutral} or non-invasive operation. For those unfixed spins which are just ahead of the quenching frontier, the fixation of the frontier spins is not ``sensed'' in the sense that the {\it statistics of}
their equilibrium average is not biased nor modified by this operation.
{In general the fixed part can cause the persistence in the process of unfixed part  through the coupling between fixed part and unfixed one.}
 %although the stochastic nature of the spin quenching makes vary its value upon the progress. This aspect leads to the (hidden) martingale of $m^{\rm (eq)}$ we will see in the next subsection.
%%%%%%%%%%%%%%%
\begin{figure}[h]
\centerline{%{schema3.pdf} {schema4.pdf} 
\hspace{0.5cm}
{\includegraphics[width=15cm]{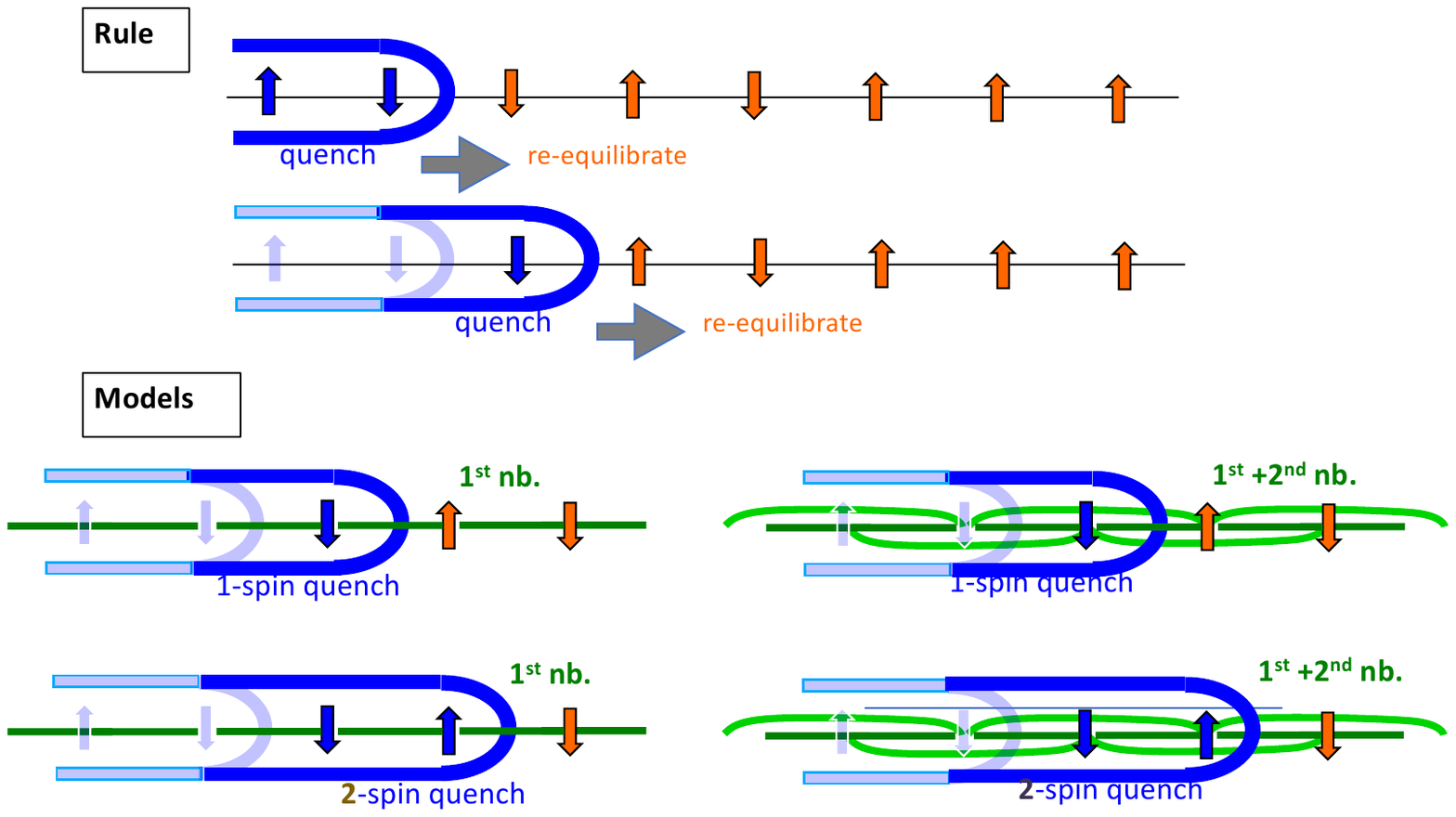}}  }
\caption{a) Elementary iterative step of progressive quenching applied to 1D Ising models.  After the unquenched part is re-equilibrated a specified number of spins are fixed at their orientation that they took at the moment 
From b) to e) present different systems and different quenching units  In b) and c) the spins interact with their own first nearest neighbours, while in d) and e) the spins interact also with their second nearest neighbours 
In b) and d) a single spin is quenched at a time, while in c) and e) a pair of spins are quenched at a time. (Figures are adopted from \cite{PQ-chain}.) }
\label{fig:schema1D}
\end{figure}
%%%%%%%%%%%%%%

\section{Globally Coupled Spin Model and Hidden Martingale\label{subsec:C}}
\subsection{Setup of model and protocol\label{sec:define}}
In Ref.~{\cite{PQ-KS-BV-pre2018}}  {the authors took}
the ferromagnetic Ising model on a complete network, that is, the model in which any one of the spins interacts with all the other spins with equal coupling constant, {}{$j_0/N_0,$ where $N_0$ is the total number of spins.} 
Each spin $s_k$ takes the value $\pm 1.$ 
We mean by the stage-${\mathsf{T}},$ or simply ${\mathsf{T}},$ the stage there are ${\mathsf{T}}$ {\it fixed} spins, see Fig.\ref{fig:model}{(a)} for illustration. 
%%%
The integer ${\mathsf{T}}$ act as a fictive time of discrete stochastic processes. In the present Chapter we avoid purposely the notation of usual time $t$  because ${\mathsf{T}}$ should be better understood as the parameter characterising the hybrid statistical ensemble consisting of the statistics of thermally fluctuating spins and that of fixed spins. 
%%%
At the stage-${\mathsf{T}}$, those $N=N_0-{\mathsf{T}}$ unfixed spins are subject under the field consisting of two parts, $h=(-\frac{j_0}{N_0}M)
+{h_{\rm ext}}$. 
The part $-\frac{j_0}{N_0}M$  is the ``molecular field'' due to the quenched magnetization $M=\sum_{k=1}^{{\mathsf{T}}}s_k,$ where we have relabelled the spins for our convenience. 
  The other part, ${h_{\rm ext}},$ is the genuine external field to perturb the process of PQ. The energy function then reads
\begin{equation}\label{eq:HTM}
\mathcal{H}_{{\mathsf{T}},M}=-\frac{j_0}{N_0} \sum_{{\mathsf{T}}+1\leq i<j\leq N_0} s_i s_j +\left(-\frac{j_0}{N_0}M +{h_{\rm ext}}\right)\sum_{i={\mathsf{T}}+1}^{N_0}s_i   .
\end{equation}

 The protocol of PQ is the cycle of re-equilibration of the unfixed spins and the fixation of a single spin at $\pm 1$ just in the state it took at the moment of fixation.
Because of the canonical equilibrium of unfixed spins, the probabilities for fixing in $\pm 1$ 
are, respectively, $ (1\pm m^{\rm (eq)}_{{\mathsf{T}},M})/2,$ see Fig.\ref{fig:model}(b), where $m^{\rm (eq)}_{{\mathsf{T}},M}$ is the canonical average of the unfixed spins with the probability weight $\exp(-\beta\mathcal{H}_{{\mathsf{T}},M}),$ where $\beta$ is the inverse of the temperature times the Boltzmann constant. 
If we are quenching the spin $s_{{\mathsf{T}}+1}$ having already quenched $\{s_1,\ldots,s_{\mathsf{T}}\}\equiv s_{[1,{\mathsf{T}}]},$
its conditional expectation, $\langle s_{{\mathsf{T}}+1}|s_{[1,{\mathsf{T}}]} \rangle,$ is  $m^{\rm (eq)}_{{\mathsf{T}},M_{\mathsf{T}}}$  where $M_{\mathsf{T}}=\sum_{i=1}^{\mathsf{T}} s_i.$ 
As we focus on the quenched magnetization, $M_{\mathsf{T}},$ this relation may be rather written as 
\begin{equation}\label{eq:PQforM}
M_{{\mathsf{T}}+1}=M_{\mathsf{T}}+s_{{\mathsf{T}}+1},\qquad \langle s_{{\mathsf{T}}+1}|  M_{[0,{\mathsf{T}}]} \rangle=m^{\rm (eq)}_{{\mathsf{T}},M_{\mathsf{T}}}.
\end{equation}

  Hereafter we shall use the energy unit so that $\beta=1.$
When we follow a process of PQ, the fixed magnetization, $M$, realises an {\it observable stochastic process}, if we regard ${\mathsf{T}}$ as the discrete time. We will denote this process by $M_{\mathsf{T}}.$
Besides, though it may be {\it hidden} behind $M_{\mathsf{T}},$ the equilibrium fixed spin, $m^{\rm (eq)}_{{\mathsf{T}},M_{\mathsf{T}}},$ also realises a stochastic process, which we will denote by $m_{\mathsf{T}}.$ 
Both processes, $M_{\mathsf{T}}$ and $m_{\mathsf{T}},$  will play  crucial roles in our analysis.
Fig.\ref{fig:model}(b) %{fig:intro-b} 
shows that the PQ is a Markovian stochastic process for $M_{\mathsf{T}}.$ 
When the coupling parameter $j_0$ is either too small (i.e. too high temperature) or 
the opposite (i.e. too low temperature) the process of PQ is trivial as shown in Fig.\ref{fig:model}(c), that is,  ${\null}{M}_{\mathsf{T}}$ undergoes either almost unbiased random walk or almost polarized, ${\null}{M}_{\mathsf{T}}\simeq \pm {\mathsf{T}},$ the polarity of which is determined during the first few stages, respectively. To explore the most non-trivial case, {$j_0$ will be chosen} at the ``critical'' point. 
Because of the finite size $N_0<\infty$ the true paramagnetic susceptibility $\chi$ is bounded as $\mathcal{O}(N_0)$. Therefore, the critical coupling, $j_{0,\rm crit},$ is determined as the best fit of  $\chi$ to the Curie's law; $\chi\sim (j_{0,\rm crit}-j_0)^{-1}.$ We found  $j_{0,\rm crit}\simeq {1.030}$ for $N_0=2^8.$ 

The stochastic process ${\null}{M}_{\mathsf{T}}$ starting with this critical coupling gives rise to the trajectories that are far from the unbiased random walks and look to follow more or less contour lines of $m^{\rm (eq)}_{{\mathsf{T}},M}$ as shown in Fig.\ref{fig:model}(d). %{fig:m-contour}. 
{This quasi-ballistic trajectory is a sort of persistent random walk.}
At the ensemble level, Fig.\ref{fig:model}(e) %{fig:evolution-00}
shows that the probability density of the mean {\it fixed spin,} $M_{\mathsf{T}}/{\mathsf{T}},$ evolves from a single peaked form to the double peaked one \cite{PQ-KS-BV-pre2018}. {Although one might suppose some spontaneous symmetry breaking mechanism behind the double peak, it is not the case because the effective coupling among the {\it unfixed} spins, $j_{\rm eff} = j_{0,\rm crit} (1-\frac{{\mathsf{T}}}{N_0})$ is below critical for ${\mathsf{T}}\ge 1$.}
%%%%%%%%%%%%%%%%
\begin{figure}[H]
\centering 
\subfigure(a)  % caption for subfigure 
{   \label{fig:intro-a}  \includegraphics[width=7.0cm]{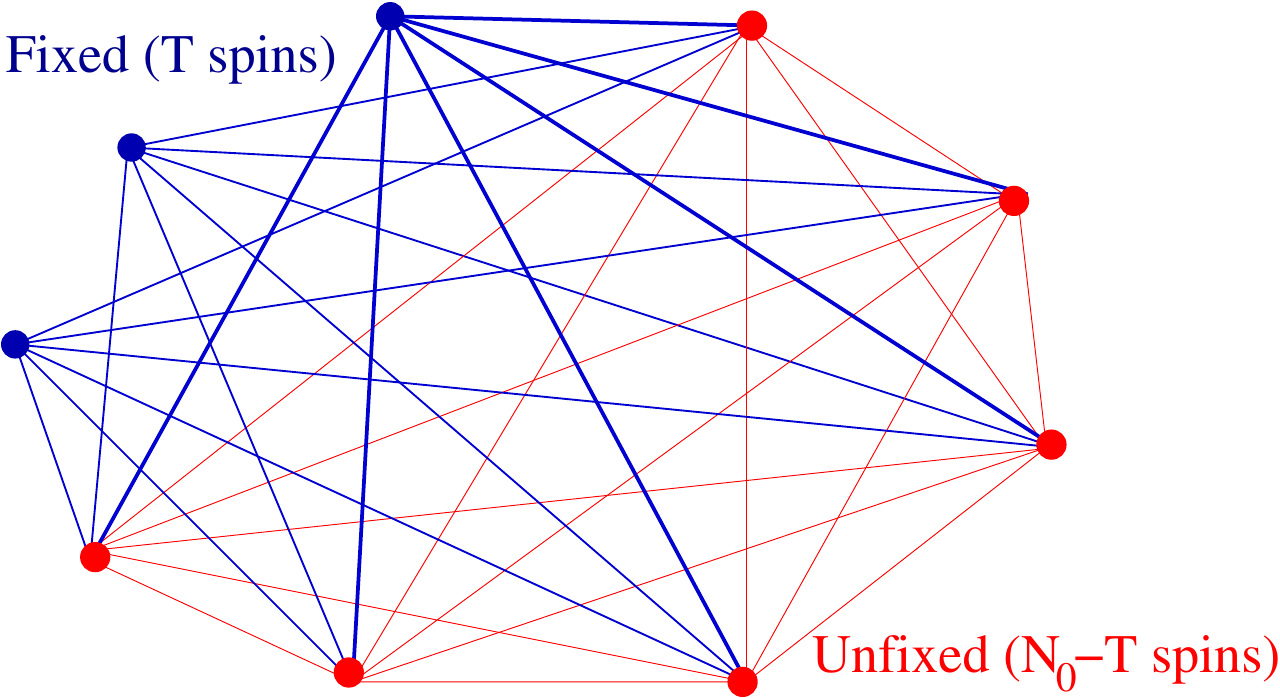}} 
\hspace{0.5cm}
\subfigure(b) % caption for subfigure 
{   \label{fig:intro-b}  \includegraphics[width=5.6cm]{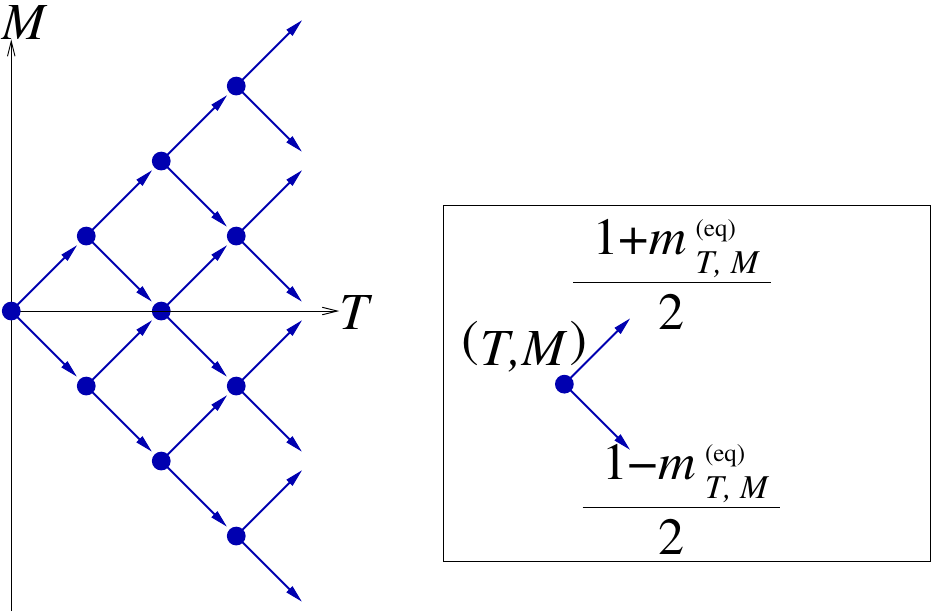}} 
\\
\subfigure(c)% caption for subfigure a
{\label{fig:off-critical-samples}    
\includegraphics[width=4.2cm]{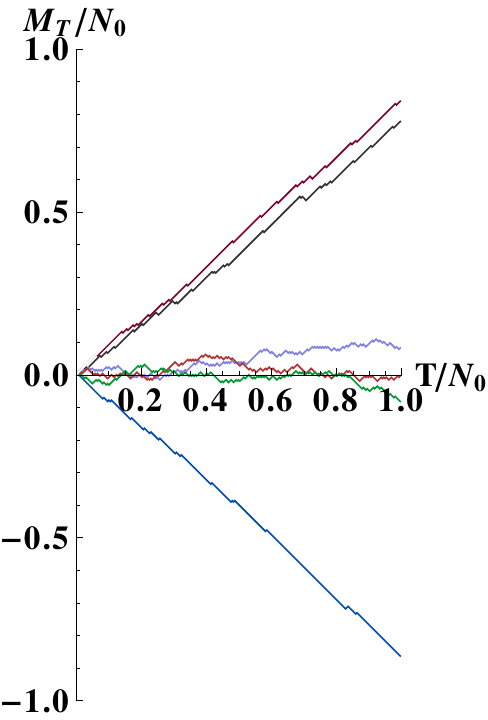}
}   
\hspace{5mm}
\subfigure(d)% caption for subfigure a
{\label{fig:m-contour}    
\includegraphics[width=7.7cm,angle=90.0]{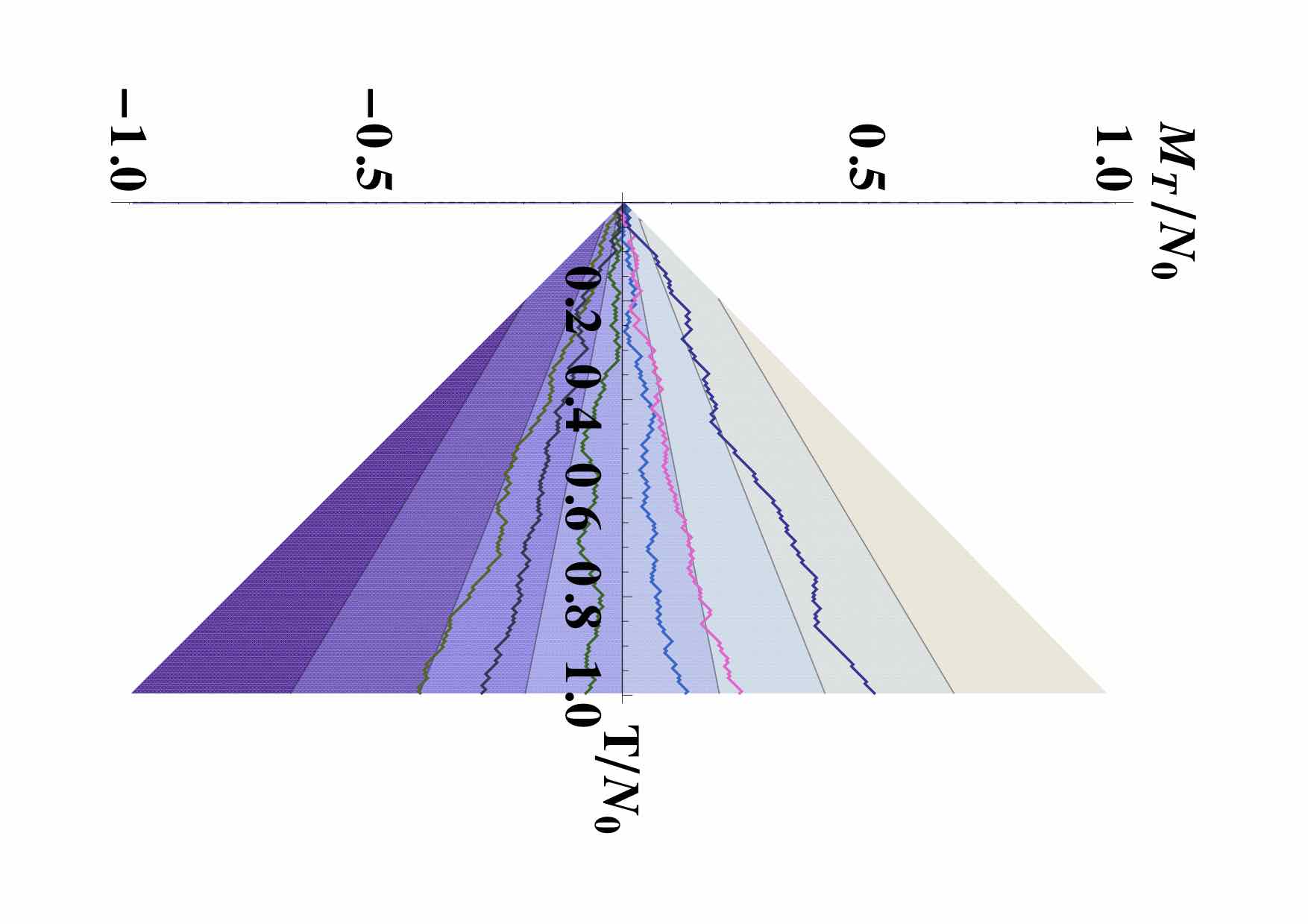}
} \\
\subfigure(e)% caption for subfigure a
{\label{fig:evolution-00}
\includegraphics[width=7.cm]{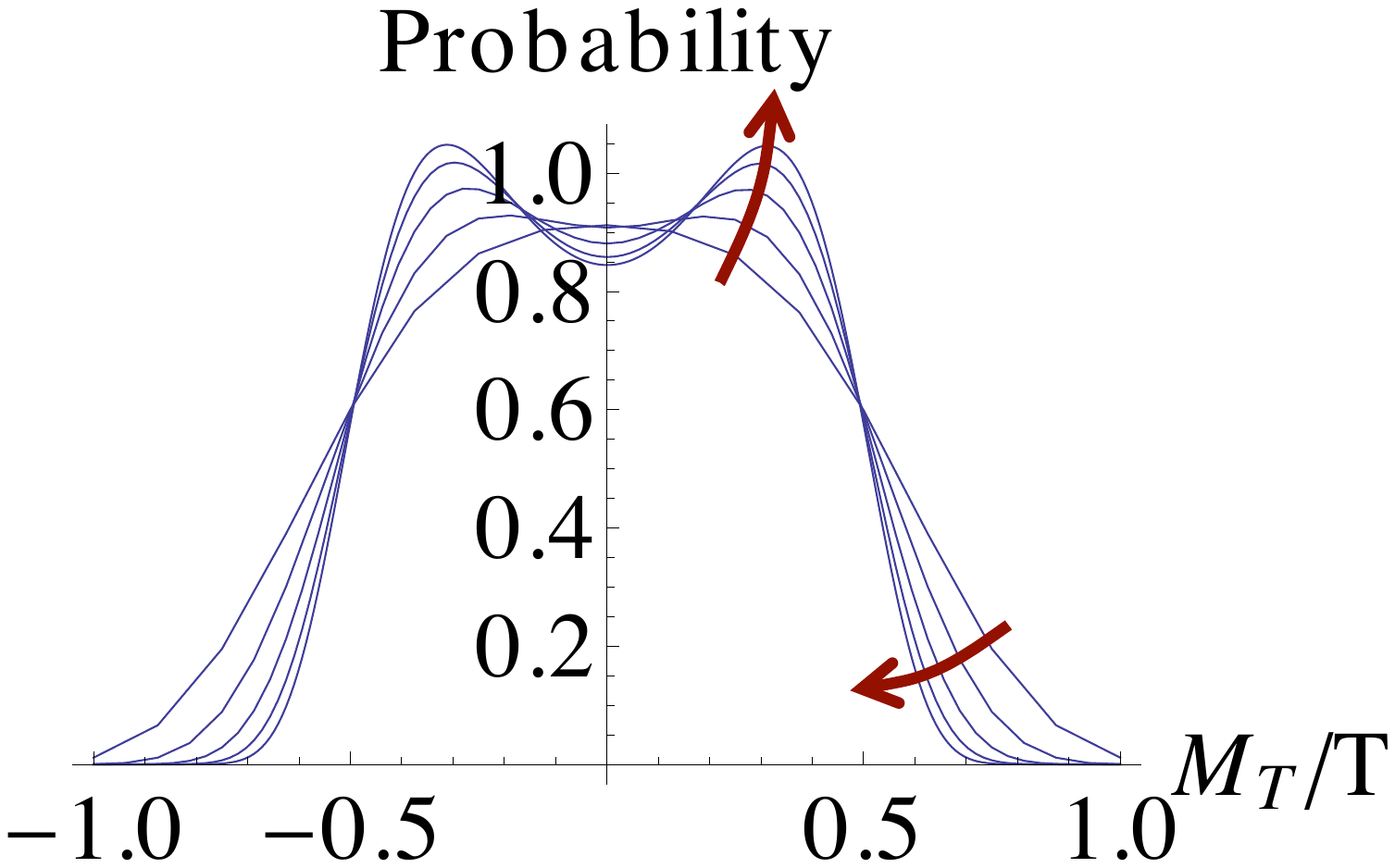}
}  
%%---%%%
\caption{(a) In the complete network of $N_0(=9)$ spins, ${\mathsf{T}}(=3)$ spins have been fixed and there remain $N_0-{\mathsf{T}}$ unfixed spins.
(b) PQ process of a complete spin network is a Markovian process on the 2D directed lattice coordinated by ${\mathsf{T}}$ and $M=\sum_{k=1}^{\mathsf{T}} s_k.$  Those lattice points which are not visited are masked.
(c) Three sample histories with $j_0=1.5$ (curves near the diagonals), and three others with $j_0=0$ (curves near the horizontal axis) are shown by different colors (brightness) for the system with the total size $N_0=256.$
(d) 
The six sample histories (curves of different colors (brightness)) with $j_0=j_{0,c}(\simeq {1.030}),$ the ``critical coupling''
with the size $N_0=256,$ are superposed 
on the contour plot of $m^{\rm (eq)}_{{\mathsf{T}},M}$ for the same $j_0$ 
(almost straight lines inside the triangle with gradient of color (brightness)).
 The value of $m^{\rm (eq)}_{{\mathsf{T}},M}$ is positive [negative], respectively, above [below] the horizontal axis.
%(a) and 
(e) Probability distributions of the mean {\it fixed} spin value, $M_{\mathsf{T}}/{\mathsf{T}}$,  at different stages,
${\mathsf{T}}=2^k$ with integers $k=4-8.$ The system size is $N_0=2^8=256.$
The initial conditions is %are (a) 
$M_0=0.$ % and (b) $M_1=1$, respectively. In both (a) and (b) the 
The increment of ${\mathsf{T}}$ is indicated by the thick red arrows. Figures adapted from \cite{CM-KS-2020} and \cite{PQ-KS-BV-pre2018}.
} 
\label{fig:model} % caption for the whole figure
\end{figure} 
%%%%%%%%%%%%%%%%%%%%%

\subsection{Hidden martingale process\label{sec:hidden}}
In \cite{PQ-KS-BV-pre2018} {it was found} that 
 the process ${\null}{M}_{\mathsf{T}}$ that shows apparently the long-term memory in 
Fig.\ref{fig:model}(d) can be characterized by the {\it hidden} martingale
of the stochastic process, $m_{\mathsf{T}}(\equiv m^{\rm (eq)}_{{\mathsf{T}},{\null}{M}_{\mathsf{T}}}).$ In mathematical term, {it can be shown} that, %for $N_0\gg 1$ and $N\equiv N_0-{\mathsf{T}}\sim N_0,$ 
\begin{equation} \label{eq:meqISmartingale}
\langle m_{{\mathsf{T}}+1} | M_{[0,{\mathsf{T}}]}\rangle=m_{{\mathsf{T}}}, %+
%E[m_{{\mathsf{T}}+1} | \mathcal{F}_{\mathsf{T}}]=m_{{\mathsf{T}}}, %+
\end{equation} 
where $\langle{X} |M_{[0,{\mathsf{T}}]}\rangle$ means to take the conditional expectation of ${X}$ under the given {\em sub-history}, $M_{[0,{\mathsf{T}}]},$ that is, under the specified data of $M_{\mathsf{T}}$ from ${\mathsf{T}}=0$ up to ${\mathsf{T}}$.  
In the present model of PQ, specifying the sub-history is equivalent to listing the values of the fixed spin up to the stage-${\mathsf{T}}$, i.e., $s_{[1,{\mathsf{T}}]},$ or $M_{[0,{\mathsf{T}}]}$ with $M_0=0$ is understood.
The value of  $m_{\mathsf{T}} $ on the right hand side is, therefore, known for a given $M_{[0,{\mathsf{T}}]}.$ 
{In the present} case the process $M_{\mathsf{T}}$ is Markovian and 
we could replace $M_{[0,{\mathsf{T}}]}$ by the information of the last stage, $M_{\mathsf{T}}.$

Eq.  (\ref{eq:meqISmartingale}) or, equivalently, 
$ \langle (m_{{\mathsf{T}}+1}-m_{{\mathsf{T}}})| M_{[0,{\mathsf{T}}]}\rangle= 0,$ 
  guides the evolution of the total fixed spin, ${\null}{M}_{{\mathsf{T}}+1}-{\null}{M}_{\mathsf{T}},$ which takes only the binary values, $\pm 1.$\footnote{ { Note that if we replaced $m_{{\mathsf{T}}+1}$ on the left hand side by ${\null}{s}_{{\mathsf{T}}+1}={\null}{M}_{{\mathsf{T}}+1}-{\null}{M}_{{\mathsf{T}}}$,  the equation  is nothing but the definition of our quenching protocol.} }
By inductively applying (\ref{eq:meqISmartingale}) we can show (see also \cite{OST-book})
\begin{equation} \label{eq:OST}
\langle m_{{\mathsf{T}}'}| M_{[0,{\mathsf{T}}]}\rangle=m_{\mathsf{T}},%+
%E[m_{{\mathsf{T}}'}| \mathcal{F}_{\mathsf{T}}]=m_{\mathsf{T}},%+
\quad  {\mathsf{T}}\le \forall {\mathsf{T}}' \le N_0.
\end{equation} 
%The property (\ref{eq:OST}) means that the hidden process $m_{\mathsf{T}}$ is a martingale process.
The relationship (\ref{eq:OST}) means that, as far as the average value is concerned, we need not integrate the discrete time master equation from ${\mathsf{T}}$ to ${\mathsf{T}}'$ to evaluate $m_{{\mathsf{T}}'}.$ 

In  \cite{PQ-KS-BV-pre2018}  (\ref{eq:meqISmartingale}) {has been derived}
up to a possible stochastic error of $\mathcal{O}({N_0}^{-2})$ {using}
the large $N_0$-expansion of the formula of quasi-canonical expectation of $m_{{\mathsf{T}}+1,M_{{\mathsf{T}}+1}}^{(\rm eq)}.$
More recently \cite{CM-KS-2021}, however, {it was noticed} that (\ref{eq:meqISmartingale}) holds precisely and from general principle of tower rule  { [see Eq.~\eqref{eq:Tower1}]}. The following argument follows the line of Appendix \ref{app:towerproperty}. Since those unquenched spins at the stage ${\mathsf{T}},$ i.e.  $\{s_{{\mathsf{T}}+1},\ldots,s_{N_0}\},$  
are all equivalent ({\it homogeneity}), we can replace $s_{{\mathsf{T}}+1}$ in the second part of (\ref{eq:PQforM}) by $s_{N_0},$ the last spin to be fixed. 
{\it If  $(m_{\mathsf{T}}=) \langle s_{N_0}|M_{[0,{\mathsf{T}}]} \rangle$ can be regarded as}  $\langle Z|X_{[0,{\mathsf{T}}]}\rangle$
 in Appendix \ref{app:towerproperty} with the mappings, $Z\mapsto s_{N_0}$ and 
 $X_t \mapsto M_t,$ then it follows the higher order tower rule   {[Eq.~\eqref{eq:Tower1}]}; {for $0< {\mathsf{T}}\le {\mathsf{T}}'\le N_0,$}
\begin{equation} \label{eq:towerPQ}
\langle \, \langle s_{N_0}| M_{[0,{\mathsf{T}}']}\rangle\, |\,  M_{[0,{\mathsf{T}}]} \rangle 
= \langle s_{N_0}|\,M_{[0,{\mathsf{T}}]} \rangle .
\end{equation}
This means (\ref{eq:OST}), i.e., $ \langle m_{{\mathsf{T}}'} |\,M_{[0,{\mathsf{T}}]} \rangle = m_{{\mathsf{T}}}.$
 We would stress that the hidden martingale property shown here holds irrespective of the initial coupling parameter $j_0,$ either near critical or not.

\section{Consequences of Hidden Martingale Process\label{subsec:D}}
The next step is to find the consequence of this (hidden) martingale property in the (principal)
stochastic process $\{M_{\mathsf{T}}\}.$ Noticing $M_{N_0}=M_{\mathsf{T}}+\sum_{j={\mathsf{T}}+1}^{N_0} s_j$
and $
\langle s_j |M_{[0,{\mathsf{T}}]}\rangle = \langle\, \langle s_j | M_{[0,j-1]} {}\rangle \,| M_{[0,{\mathsf{T}}]}\rangle =\langle m_{j-1}  \,| M_{[0,{\mathsf{T}}]} \rangle  = m_{\mathsf{T}}$ for $j>{\mathsf{T}},$
we have the {\it hidden maltingale formula} (discrete version):
\begin{equation} \label{eq:hidden-Mgl-formula}
\langle M_{N_0}| M_{[0,{\mathsf{T}}]}\rangle = M_{\mathsf{T}}+(N_0-{\mathsf{T}}) m_{{\mathsf{T}}} .%+\mathcal{O}(1).
\end{equation}
At the end of this section we will {discuss}  the continuum version of (\ref{eq:hidden-Mgl-formula}).

\subsection{Inference \label{sec:back}}
{Below are given the examples of}  the usage of the hidden martingale formula (\ref{eq:hidden-Mgl-formula})
 to infer the past stage, which will also provide with an elementary demonstration of this theorem. 

Suppose that the process starts by the stage-${\mathsf{T}}$ (${\mathsf{T}}>0$) with a fixed magnetization $M_{\mathsf{T}},$ 
and that we are given   $\langle M_{N_0}| M_{[0,{\mathsf{T}}]} \rangle$ 
 from a large ensemble of the final data $\{M_{N_0}\}.$ 
Now in (\ref{eq:hidden-Mgl-formula}) the value of the left hand side is known, while the 
right hand side is a function of unknown $M_{\mathsf{T}}$ with a given ${\mathsf{T}}.$ Therefore,
(\ref{eq:hidden-Mgl-formula}) is an (implicit) equation for $M_{\mathsf{T}}.$ In this manner,
we can infer $M_{\mathsf{T}}$ with the cost of calculation of $\sim N_0$ (for {a reliable} expectation of $M_{N_0}$)
instead of solving the master equation costing $\sim {N_0}^2.$
{It was} numerically verified that this scenario indeed works very well.

\subsection{Prediction \label{sec:forward}}
Another usage of the hidden martingale formula (\ref{eq:hidden-Mgl-formula}) is the prediction of the conditional expectation of the main stochastic process, $M_{{\mathsf{T}}'},$ 
for ${\mathsf{T}}'>{\mathsf{T}}.$
%especially when ${\mathsf{T}}'=\mathcal{O}(N_0)$ and ${\mathsf{T}}\ll N_0$ with $N_0\gg 1.$
In \cite{CM-KS-2020} the equivalent of (\ref{eq:hidden-Mgl-formula}) {was shown}:
\begin{equation}\label{eq:IntMg}
\left\langle \left.\frac{M_{{\mathsf{T}}'}-M_{\mathsf{T}}}{{\mathsf{T}}'-{\mathsf{T}}} \right| M_{[0,{\mathsf{T}}]} \right\rangle= m_{{\mathsf{T}}},
%E\left[\left.\frac{M_{{\mathsf{T}}'}-M_{\mathsf{T}}}{{\mathsf{T}}'-{\mathsf{T}}} \right|\mathcal{F}_{\mathsf{T}}\right]= m_{{\mathsf{T}}},
\quad 
{\mathsf{T}}'>{\mathsf{T}}.
\end{equation}
For example, in the case of $N_0={\mathsf{T}}'=100$ and ${\mathsf{T}}=5,$ we can predict $\langle
M_{100}| M_{[0,5]} \rangle$ to be $M_5+95\times m_{5}.$ 
%After submission of \cite{CM-KS-2020} we found a simpler derivation as described above.

\subsection{Prediction of probability distribution function
\label{sec:forward2}}
In {the present} model of PQ, the {hidden maltingale formula} (\ref{eq:hidden-Mgl-formula})
allows to predict
$M_{N_0}$ from the data of ${\mathsf{T}}$-th stage with ${\mathsf{T}}\ll N_0$ :
\begin{equation} \label{eq:optics}
{\null}{M}_{N_0}= M_{\mathsf{T}}+ (N_0-{\mathsf{T}}) m_{{\mathsf{T}}}+\mathcal{O}((N_0-{\mathsf{T}})^{\frac{1}{2}}),
\end{equation}
where we recall $m_{\mathsf{T}}\equiv m^{\rm (eq)}_{{\mathsf{T}},M_{\mathsf{T}}}$
and the last term, $\mathcal{O}((N_0-{\mathsf{T}})^{\frac{1}{2}}),$ represents the sum, $\sum_{k={\mathsf{T}}+1}^{N_0} (s_k-m_{\mathsf{T}}),$ consisting of the terms deemed to vanish individually upon the conditional average, $\langle\,\, | M_{[0,{\mathsf{T}}]}\rangle.$ 
The approximation (\ref{eq:optics}), which ignores the diffusive aspect of the process after the ${\mathsf{T}}$-th stage, may be called a {\it geometrical optics} approximation\footnote{{While the analogy is not close, let us regard the ensemble of the graphs $\{(t,M_t)\}_{T\le t\le N_0}$ representing the histories of the total magnetization on the $(t,M)$-plane as a light wave emitted from $(T,M_T)$ in
} {
the direction parallel to  $(1,m_T).$ In the {\it wave} optics, when the wavelength of the light is non-negligible against the aparture of the light source, the flux of light is broadened as it propagates while the location of the maximum intensity goes along the ``ray'', $M_t=M_T+(t-T)m_T$ for $T\le t\le N_0,$ according to the {\it geometrical} optics. Likewise, in the Progressive Quenching, the stochasticity causes diffusion of the trajectories around the mean history, $M_t=M_T+(t-T)m_T$ for $T\le t\le N_0.$ While the broadening of the light flux grows linearly with distance from the source, the trajectories of Progressive Quenching will diffuses like $\sim (t-T)^{1/2}$ for $T\le t\le N_0.$}}. 
{To know the probability distribution of $M_{\mathsf{T}}$ is a relatively easy task for ${\mathsf{T}}\ll N_0$ with the calculation cost of some power of ${\mathsf{T}}.$
The last formula (\ref{eq:optics}) then allows} to predict the final probability distribution of $M_{N_0}$ 
 \null{(for the numerical procedure, see  subsection~\ref{sec:optics}.)}
Fig.\ref{fig:fig3} demonstrates how it works well.
In the left part of the figure the PQ process is unbiased, where the distribution at ${\mathsf{T}}=2^4$ is symmetric and unimodal (inset) while the final one is bimodal (dense dotted curve). 
%We can predict the
The final distribution of $M_{N_0}$ {is predicted} by the piecewise linear curve with $2^4+1$ nodes.
In the right part of the figure the PQ process is unbiased except at the stage-$(2^4-1),$ 
when the infinite external field, $h_{\rm ext}=\infty,$ is applied to force ${\null}{s}_{2^4}$ is quenched 
to be $+1.$ The distribution at the stage-$2^4$ (inset) is almost equal to the unbiased case but 
suffers the shift by $\Delta M=+1.$ The subsequent unbiased PQ process leads then to the final bimodal but asymmetric distribution as shown by the dense dotted curve.
Also in this case the prescription described above (the piecewise linear curve with $2^4+1$ nodes) reproduces well the main feature of the full numerical result. 

Amazingly this method of hidden martingale can predict the binodal distribution in the 
far future ($N_0\gg {\mathsf{T}}$) given the data of unimodal distribution. Since  %``martingale drift,''
 $m^{\rm (eq)}_{{\mathsf{T}},M}$ is a monotonous function of $M$ (not shown), the results are 
far from trivial.
In case that $N_0$ and ${\mathsf{T}}$ constitutes the double hierarchy $1\ll {\mathsf{T}}\ll N_0,$ our methodology may serve as a reasonable tool of numerical asymptotic analysis.

\subsection{Numerical construction of the  distribution from (\ref{eq:optics})}
\label{sec:optics}
We notice that, in the absence of stochastic diffusion{, i.e. the term $\mathcal{O}((N_0-{\mathsf{T}})^{\frac{1}{2}})$ in (\ref{eq:optics}),}  the probability associated to any subset of the values of $M_{\mathsf{T}}$ at the stage ${\mathsf{T}}$ is directly conveyed to the corresponding subset of the values of $M_{N_0}$ at the finale stage, {somehow reminiscent of the Liouville's theorem that allows the probability to be carried along the Hamiltonian flow. }

Suppose that, at the stage ${\mathsf{T}}$, we have {an access to the probabilities}, $P^{({\mathsf{T}})}_i,$ of having  the fixed {magnetization} $M=-{\mathsf{T}}+2i \equiv \mu_i$ with $i=0,1,\ldots,{{\mathsf{T}}}.$
Also we prepare the data of $m^{\rm (eq)}_{{{\mathsf{T}}},M=\mu_i}$ with $i=0,1,\ldots,{{\mathsf{T}}}.$
{The object is to generate 
the normalized probability density, $p(M),$ with continuous variable $M\in [-N_0,N_0],$ of the final fixed magnetization through a piecewise linear approximation with ${\mathsf{T}}+1$ nodes.}
The assignment of the binning box may not be unique. Here {we follow 
the Appendix C of \cite{CM-KS-2020} to use} a simple trapezoidal rule to make Fig.\ref{fig:fig3}: 

%pre top
For the simplicity of notations, we introduce (see Eq.(\ref{eq:optics}))
\begin{eqnarray}
%\mu_i&=& {-{{\mathsf{T}}}+2i}
%\qquad
%  {\sigma_i}= m^{\rm (eq)}_{{{\mathsf{T}}},\mu_i} ,
%\qquad
 x_i= \mu_i+(N_0-{{\mathsf{T}}}) \,m^{\rm (eq)}_{{{\mathsf{T}}},\mu_i} , 
\end{eqnarray}
where $i=0,1,\ldots,{{\mathsf{T}}}.$
We will make up the final probability {\it density} $p(x)$ so that its normalization is
$\int_{x_0}^{x_{{{\mathsf{T}}}}}p(x)dx=1.$ We make a piecewise linear approximation of $p(x)$ whose
joint-points are $\{x_i,p(x_i)\}.$ The normalization condition then reads
{%\small
\begin{eqnarray}
1&=&\!\!\!\sum_{i=0}^{{{\mathsf{T}}}-1}\frac{p(x_i)+p(x_{i+1})}{2}(x_{i+1}-x_i)
\cr &=& p(x_0)\frac{x_{1}-x_0}{2}
+\!\!\!\sum_{i=1}^{{{\mathsf{T}}}-1}p(x_i)\frac{x_{i+1}-x_{i-1}}{2}
+p(x_{{{\mathsf{T}}}})\frac{x_{{{\mathsf{T}}}}-x_{{{\mathsf{T}}}-1}}{2}. 
\end{eqnarray}
}
Then we define $p(x_i)$ through
\begin{eqnarray}
p(x_0)\frac{x_{1}-x_0}{2}&=&P^{({{\mathsf{T}}})}_{0 },
\cr p(x_i)\frac{x_{i+1}-x_{i-1}}{2}&=& P^{({{\mathsf{T}}})}_{i } \qquad i=1,\ldots,{{\mathsf{T}}}-1
\cr p(x_{{{\mathsf{T}}}})\frac{x_{{{\mathsf{T}}}}-x_{{{\mathsf{T}}}-1}}{2}&=&P^{({{\mathsf{T}}})}_{ {\mathsf{T}}}
\end{eqnarray}
so that the ``ray'' of geometrical optics carries the probability from ${\mathsf{T}}={{\mathsf{T}}}$ to 
${\mathsf{T}}=N_0.$ (The uneven weight on both extremities is harmless because $P^{({\mathsf{T}})}_0$
and $P^{({\mathsf{T}})}_{{\mathsf{T}}}$ are very small.
The martingale prediction of the probability densities in Fig.\ref{fig:fig3} are thus made.
%pre end
Naturally, the prediction by hidden martingale give narrower distributions than the full numerical results because the former method ignores the diffusion, whose contribution would fatten the distributions by $\sim (256-16)^{\frac{1}{2}}\simeq 15.$ 

%%%%%%%%%%%%%%%%%%%%%%%%%%
%\vspace{-1cm}
%%%%%%%%%%%%%%%%%%%%%%%%%%%%%%%%%
\begin{figure}[h!!]
\centering
{\includegraphics[width=12cm]{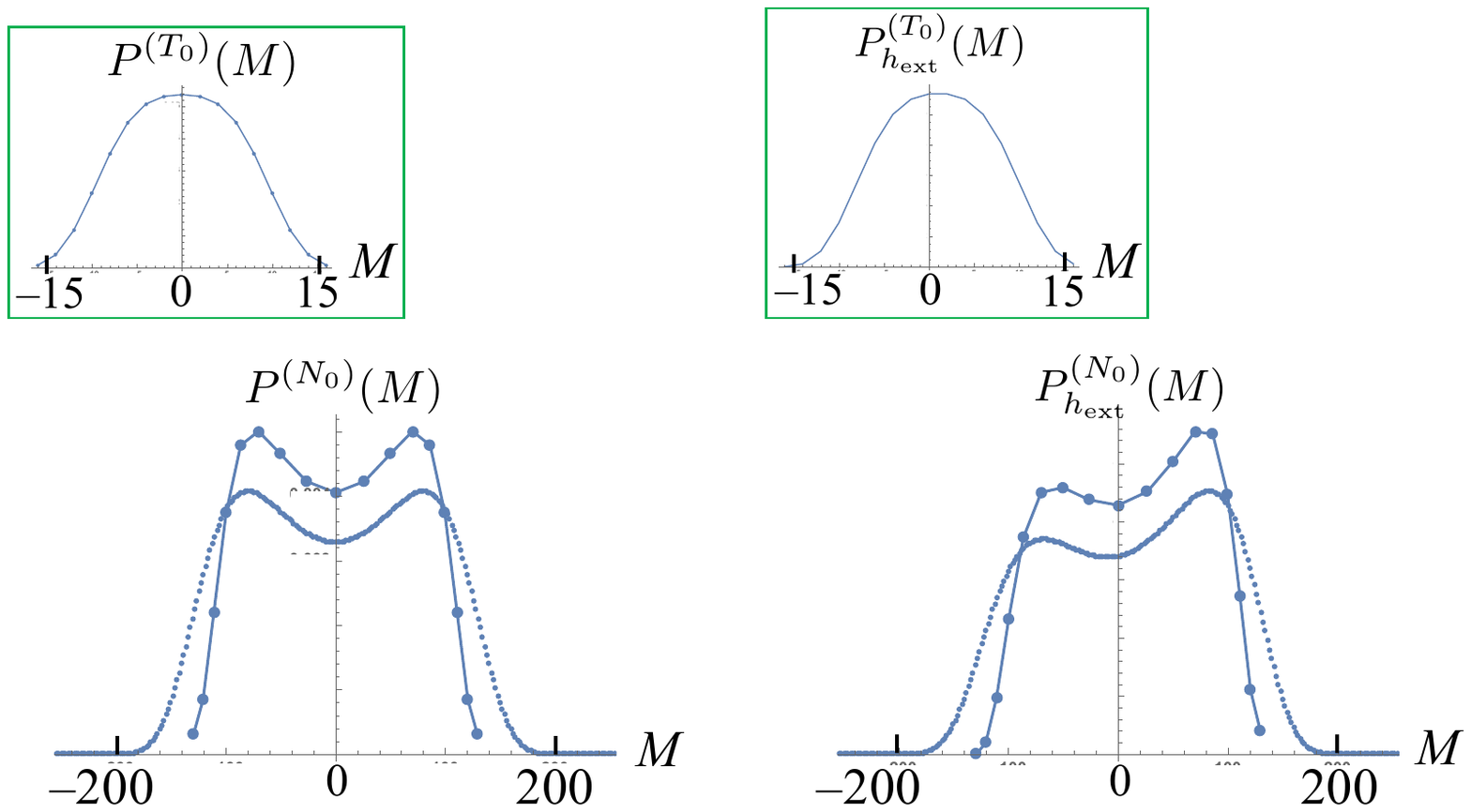}}
%\vspace{-1cm}
\caption{Comparison between the final distributions of $M_{N_0}$ predicted by the hidden martingale property (joined ${\mathsf{T}}+1$ dots) and those by full numerical solution (filled circles) with ${\mathsf{T}}=2^4$ and $N_0=2^8$ \cite{CM-KS-2020}.
See  text for  details.}
\label{fig:fig3} % caption for the whole figure
\end{figure}
%%%%%%

\mbox{}\\
\noindent{\bf Hidden martingale formula with continuous time  (Ref.~\cite{CM-KS-2020})}
 Before concluding the main results of this section, we give the  continuum version of (\ref{eq:hidden-Mgl-formula}).
Suppose that the stochastic process ${\null}{X}_t$ is generated by a hidden martingale system 
 through the stochastic differential equation (SDE),
\begin{equation}\label{eq:SDEx}
 {\dot{X}}_t=G_t(X_t)  + V_t(X_t) {\dot{B}}_t,
 \end{equation}
 where   ${\null}{B}_t$ is a Wiener (or martingale) process and $G_t(X_t) $ is the {\it martingale drift} satisfying
\begin{equation}\label{eq:mtgl-a}
 \langle G_t(X_t)
 | M_{[0,s]}\rangle=G_s(X_s), \qquad t\ge s. 
\end{equation}
Then 
\begin{equation} \label{eq:Img}
 \left\langle \left.\frac{{\null}{X}_{t}-X_s}{t-s} \right|M_{[0,s]} \right\rangle
% E\left[\left.\frac{{\null}{X}_{t}-X_s}{t-s} \right|\mathcal{F}_s\right]
 = G_s(X_s), \qquad t\ge s.
\end{equation}

\noindent {\it (Proof)} 
Taking the conditional expectation of Eq.(\ref{eq:SDEx}) with the condition $M_{[0,s]}$ we have for all $\tau\ge s,$
$$
{
\langle \dot{X}_{\tau}|
M_{[0,s]}
\rangle=\langle 
G_\tau(X_\tau)
| M_{[0,s]}\rangle
%E[d{\null}{X}_{\tau}|\mathcal{F}_s]=E[a(\tau,{\null}{X}_\tau)|\mathcal{F}_s]
=G_s(X_s) ,
}
$$
where (\ref{eq:mtgl-a}) has been used in the second equality.
By integrating the above equation with respect to $\tau$ from $s$ up to $t,$  we have
$\langle {\null}{X}_{t}-X_s| M_{[0,s]}\rangle
%E[{\null}{X}_{t}-X_s|\mathcal{F}_s]
=G_s(X_s)\,(t-s).$\, 
$\square$\\
In the {present model} of our PQ, we may approach our process to an SDE by
$dM_\tau:= M_{\tau+1}-M_{\tau}$ and $d\tau:=1$ for $\forall \tau\ge s.$ 
Then  a type of the Doob-Mayer decomposition, $d{\null}{M}_{\tau}=m_{\tau} d\tau+ ({\null}{s}_{\tau+d\tau}-m_{\tau}d\tau),$ gives what corresponds to (\ref{eq:SDEx}).

\noindent {\it (Remark)} 
While the generalization from discrete version is straightforward, there can be functional constraints on $G_t(X_t)$ in order for $G_t(X_t)$ to be martingale with respect to $X_{[0,s]}$. The practical application of the continuous version has not been tested yet. 
{Some analysis has been recently made for the case where $V_t(z)=1$ and 
$G_t(z)$ is independent of time \cite{KS-2023}.}

\section{Concluding discussion of this Chapter}\label{subsec:E}

When a martingale process is hidden behind the observed Markovian stochastic process, the former may bring long-lasting memory effects to the latter. 
% In the context specific to the progressive quenching (PQ), if the hidden martingale property might be attributed to the fact that the act of quenching is a neutral operation although non-quasistatic, the equilibration before each quenching step would not be an indispensable for this property. The exploration of new models are waited.
If we regard (\ref{eq:optics}) as a { geometrical optics}  approximation of the full evolution, there may be a route to reach this form through the Freidlin-Wentzell approach \cite{Freidlin-Wentzell}  under the constraint of hidden martingale. Further theoretical studies are also needed. 

{The authors of} \cite{CM-KS-2021} showed that, 
%together with the Ising and Markovian nature of PQ, 
 the hidden martingale property (\ref{eq:OST}) is equivalent to a {\it local invariance} of the path weights.
 %, 
This invariance may reflect {an} aspect of martingale as stochastic conservation although {such invariance is not found with any martingale other than the present PQ model. In the latter} case the local invariance implies a constrained canonical structure of the statistics of $M_{\mathsf{T}}$ \cite{CM-KS-2021}.
While the given quenched spins impose a permanent memory on the individual process, 
the neutral action of quenching allows to reflect the equilibrium statistics of the unquenched spins in the quenched ensemble. 
The constrained canonical structure makes compatible these two complementary aspects, {see also \cite{CM-KS-2023} }.

%In a wider context beyond the specific PQ model, the martingale based on the tower rule (\ref{eq:OST}) also applies to non-Markovian process $\{M_t\},$ and the stochastic process  need not be quasi-static as long as there exist a homogeneity of the form like $\langle s_{{\mathsf{T}}+1}|M_{[0,{\mathsf{T}}]}\rangle =\langle s_{N_0}|M_{[0,{\mathsf{T}}]}\rangle.$ We encourage that many explorations in a variety of domain will be done in future. 

%%%%%%
%\bibliographystyle{apsrev4-1.bst}  
 % \bibliography{bib-PQ} 

%%%%%%%
%\include{chap9-Simone}
\chapter{Martingales in population genetics}
\label{ch:popdyn}

%\epigraph{There exist only two kinds of modern mathematics books: one which you cannot read beyond the first page and one which you cannot read beyond the first sentence.\\ \vspace{0.1cm} Cheng Ning Yang. Physics Nobel prize 1957.}

%\epigraph{At present one may say that the mathematical theory of evolution is in a somewhat unfortunate position, too mathematical to interest most biologists, and not sufficiently mathematical to interest most mathematicians.\\ \vspace{0.1cm} J.B.S. Haldane, "Forty years of genetics" from "Background to Modern Science" (Cambridge University Press, 1938)} (Alternative quote)

\epigraph{It is remarkable, I think, that their behavior [of mutant frequencies] is calculable from the theory of stochastic processes, a theory which until recently has been regarded as too academic to have actual biological applications.\\  \vspace{0.1cm} Motoo Kimura, from ``The neutral theory of molecular evolution'', 1983 \cite{kimura1983neutral}. }

Individuals belonging to natural populations are characterized by a certain degree of genetic diversity. Population genetics studies the distribution of these genetic variants as effect of mutations, natural selection, stochasticity, and other evolutionary forces.  In particular, it is nowadays established that a large portion of mutations confer a negligible selective advantage (or disadvantage) to individuals carrying them. The fate of these mutations is therefore determined by pure chance without any deterministic selection. In population genetics, these mutations are called "neutral". The widespread occurrence and importance of neutral mutations was pointed out by Motoo Kimura \cite{kimura1983neutral}. Kimura's theory has encountered substantial resistance over the years -- partially due to the fact that, historically, evolution was implicitly thought to be a deterministic process. In contrast, Kimura's neutral theory is inherently stochastic. %In the introduction of his 1983 book \cite{kimura1983neutral}, Kimura writes ``It is remarkable, I think, that their behavior [of mutant frequencies] is calculable from the theory of stochastic processes, a theory which until recently has been regarded as too academic to have actual biological applications''.

The distinction between neutral, advantageous and deleterious mutations has become a cornerstone of modern population genetics. This concept provides us with a perfect example of the analogy between population genetics and non-equilibrium physical systems, and how martingales can be applied to population genetics.

\section{The Moran model}

To make our discussion more concrete, we introduce the Moran model of population genetics. The Moran model describes a population of $N$ individuals reproducing asexually. The total number of individuals $N$ is kept constant by resource availability, so that every time an individual dies another individual instantly reproduces.
A number of individuals $n$ in the population, with $0\le n\le N$, carry a given mutation. We call these individuals the ``mutants'' and the remaining $(N-n)$  ``wild-type individuals''.  For the time being, we assume the mutation to be neutral, i.e. mutants die and reproduce at the same rates as the wild type individuals. The number of mutants in the population evolves with rates
\begin{eqnarray}\label{eq:moran_neutral}
  n\rightarrow n+1 &\mbox{with}\ \mbox{rate} & \frac{n(N-n)}{N}    \nonumber\\
  n\rightarrow n-1 &\mbox{with}\ \mbox{rate} &\frac{n(N-n)}{N}
\end{eqnarray}
Eqs.~(\ref{eq:moran_neutral}) can be understood by thinking that the rate at which the number of mutants increase is proportional to the number $(n-N)$ of wild type individuals, times the probability $n/N$ that the dead individual is replaced by a copy of a mutant. Similar reasoning apply to the rate of decrease of $n$.  The master equation defined by the rates~(\ref{eq:moran_neutral}) is characterized by two absorbing states, $n=0$ and $n=N$. In the language of population genetics, if the absorbing state $n=N$ is reached we say that the mutation has "reached fixation". To understand the evolution of a population, it is important to compute the probability $P_+$ of this event. 
A short way of computing this probability is by noticing that $n_t$ is a martingale defined on a bounded interval, and therefore must satisfy Doob's optional stopping theorem. Calling $\tau$ the time at which one of the two absorbing states is reached, we obtain:
\begin{equation}\label{eq:moranfix}
  n_0=\langle n_\tau\rangle =0 \cdot P_- +N \cdot P_+ \rightarrow P_+=\frac{n_0}{N}.
\end{equation}
Therefore, in the neutral Moran model, the probability of a mutation to reach fixation is equal to its current fraction in the population. This is a basic yet fundamental result of neutral population genetics.

We now generalize the Moran model to a case in which the mutation possibly confers a selective advantage to individuals carrying it. We define a selective advantage $s$ as a relative increase in the reproduction rate. The transition rates of the model read
\begin{eqnarray}\label{eq:moran_select}
  n\rightarrow n+1 &\mbox{with}\ \mbox{rate} & (1+s) \frac{n(N-n) }{N}   \nonumber\\
  n\rightarrow n-1 &\mbox{with}\ \mbox{rate} & \frac{n(N-n)}{N}.
\end{eqnarray}
In the three cases $s>0$, $s=0$, and $s<0$ the process is a submartingale, martingale, and surmartingale, respectively.
In population genetics, if $s$ is negligible, the mutation is considered to be neutral; if $s$ is sufficiently large and positive the mutation is advantageous; and if $s$ is negative and sufficiently large in absolute value the mutation is deleterious. By analyzing the model, we will clarify what does it mean to be ``negligible'' and ``sufficiently large''. For simplicity, we study the model in the continuous approximation. Assuming $N$ to be large, the fraction $X=n/N$ of mutants satisfies the Langevin equation
\begin{equation}\label{eq:moran_langevin}
\dot{X}_t=sX_t(1-X_t)+\sqrt{\frac{2X_t(1-X_t)}{N}}\dot{B}_t.
\end{equation}
The Langevin equation~(\ref{eq:moran_langevin}) is interpreted in the It\^{o} sense and can be derived from the master equation by means of a Kramers-Moyal expansion, see e.g. \cite{gardiner1985handbook}. We truncated this expansion at the first order in $1/N$ and assumed $s$ to be order $1/N${, so that we neglected terms of order $s/N$.}

Also Eq.~(\ref{eq:moran_langevin}) is characterized by two absorbing states, in this case at $X=0$ and $X=1$. In this case, if $s\neq 0$ the process $X_t$ is not a martingale. However, performing a change variable to $Y_t=\exp(-sNX_t)$ by means of the It\^{o} formula we obtain
\begin{equation}\label{eq:moran_langevin2}
\dot{Y}_t=-sNY_t\sqrt{\frac{2X_t(1-X_t)}{N}}\dot{B}_t .
\end{equation} of $Y_t$ is governed by an It\^{o} stochastic differential equation without drift, the process $Y_t$ is a martingale. It is interesting to notice the analogy with stochastic thermodynamics, where entropy production is a submartingale whereas the exponential of minus the entropy production is a martingale. The range $X_t\in [0,1]$ corresponds to a range $Y_t\in[e^{-sN},1]$. We can therefore apply once more Doob's optional stopping theorem to the stopping time defined as the first time at which one of the two absorbing states is reached:
\begin{equation}\label{eq_select_doob1}
  Y_0= P_+  \exp\left(-sN\right)+ P_-  .
\end{equation}
Using that $P_++P_-=1$ and expressing the probabilities in terms of $X_0$, we find that the probability of fixation is
\begin{equation}\label{eq:kimuraform}
  P_+=\frac{1-\exp\left(-sNX_0\right)}{1-\exp\left(-sN\right)}.
\end{equation}
Equation~(\ref{eq:kimuraform}) is the celebrated Kimura's formula for the fixation probability of a mutation \cite{kimura1962probability}. It is analogous to the expression \eqref{eq:ppluspminusintro1} that we derived for the biased random walk. Equation (\ref{eq:kimuraform}) is singular for $s=0$. However, it correctly predicts the neutral result $p_1=X_0$ (see Eq.~(\ref{eq:moranfix})) in the limit $s\rightarrow 0$.

Importantly, Kimura's formula clarifies when a selective advantage is sufficiently large. Note that Kimura's formula depends on the parameters $s$ and $N$ only via the combination $sN$. It follows that mutation characterized by selective advantages $|s|\ll N^{-1}$ behave essentially as neutral. This fact has deep consequences for the evolution of natural populations. 

In population genetics, the model embodied in Eq.~\eqref{eq:moran_langevin} is used to describe the fate of mutations in real populations. However, the intensity of random fluctuations of mutation frequencies tend to be much larger than predicted by models such as Eq.~\eqref{eq:moran_langevin}. An explanation is that many simplifying assumptions underlying the Moran process do not hold in reality. One of the most important is the assumption of population size: it can be shown that, in populations of variable size, evolution is strongly affected by "bottlenecks", i.e. epochs in which the population size happened to be small \cite{gillespie2004population}.  To compensate for these effects, when using the Moran model to describe real populations, the parameter $N$ is taken as an effective parameter, called the ``effective population size''. For example, the effective population size estimated for humans from fluctuations of mutation frequencies is on the order of  $N=10^4$, whereas estimates for Escherichia Coli range between~$10^6$ to~$10^8$. In general, Equation~(\ref{eq:kimuraform}) reveals that mutations characterized by small selective advantages $s\ll 1/N$ do not significantly influence the fixation probability and therefore effectively behave as neutral. This fact implies that bacteria such as E. coli, characterized by a large effective population size, are much more sensitive to fitness differences than for example humans. For example, a mutation conferring a selective advantage $s=10^{-5}$ would be seen as neutral by a human population, but as strongly advantageous by most bacteria.

\begin{figure}
\centering 
  \includegraphics[width=10cm]{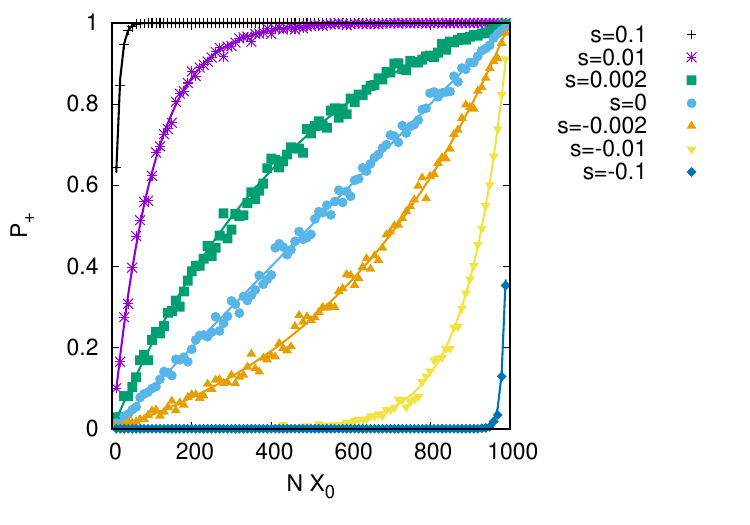}
  \caption{Kimura's fixation formula (Eq.(\ref{eq:kimuraform}) (lines) compared to simulations of the Moran process (Eq.(\ref{eq:moran_neutral})) with $N=1000$ (points). Each point is an average over $10^3$ Gillespie simulations of a master equation with rates (\ref{eq:moran_select}), where we set $\mu=1$ and $s$ as in the figure legend. }
  \label{fig:kimura}
\end{figure}

An alternative approach to study Eq.~(\ref{eq:moran_langevin2}) is to perform a random time change
\begin{equation}\label{eq:randomtime_genetics}
\D  \tau=\D t X_t(1-X_t)
\end{equation}
so that Eq.~(\ref{eq:moran_langevin2}) becomes
\begin{equation}\label{eq:randomtime_genetics2}
  \frac{\D}{\D \tau}X_\tau=s+\sqrt{\frac{2}{N}}\dot{B}'_\tau,
\end{equation}
where $\dot{B}'_\tau$ is also a white Gaussian noise. In terms of the random time, the population dynamics is described by a simple Langevin process with constant drift and diffusion terms.

\section{Duality and martingales}

So far we analyzed the Moran process using a diffusion approximation, which paved the way to an analysis using martingales. In the following we discuss another type of correspondence between discrete population models and Ito stochastic differential equations, based on the notion of duality, that does not rely on any approximation \cite{shiga1986stationary,doering2003interacting}. We consider a single population made up of a variable number $n$ of individuals. Each individual reproduce at rate $\gamma$ and die at rate $\chi(n-1)$, proportional to the number of other individuals due to competition for resources:
\begin{eqnarray}\label{birthcoagul}
  n\rightarrow n+1 & \quad \mbox{with}\ \mbox{rate}\quad& \gamma n\nonumber\\
  n\rightarrow n-1 & \quad \mbox{with}\ \mbox{rate}\quad& \chi n(n-1)
\end{eqnarray}
The corresponding master equation reads
% \begin{equation}\label{me_birthdeath}
% \partial_t \rho_t(n)= \sum_m M_{nm} \rho_t(m)
% \end{equation}
\begin{equation}\label{me_birthdeath}
\partial_t \rho_t(n)= \sum_m \omega(n,m) (\rho_t(m)-\rho_t(n))
\end{equation}
with the transition rates
% \begin{equation}\label{eq:transitionmatrix_moran}
%   M_{nm}=\gamma m (\delta_{n,m+1}-\delta_{nm})+ \chi m(m-1)(\delta_{n,m-1}-\delta_{nm}).
% \end{equation}
\begin{equation}\label{eq:transitionmatrix_moran}
  \omega(n,m)=\gamma m \delta_{n,m+1}+ \chi m(m-1)\delta_{n,m-1}.
\end{equation}
We now associate to the Master equation~(\ref{me_birthdeath}) a Langevin dynamics
% introduce U first so that it is not backward in time
\begin{equation}\label{eq:morandual}
\dot{Z}_t=-\gamma Z_t(1-Z_t)+\sqrt{2\chi Z_t(1-Z_t)}\dot{B}_t.
\end{equation}
{We note that Eq.~(\ref{eq:morandual}) has the same form of Eq.~(\ref{eq:moran_langevin}) if we perform the change of variable}
\begin{equation}
X_t=1-Z_t .
\end{equation}
{With this mapping, Eq.~(\ref{eq:morandual}) can be seen as a (truncated) Kramers-Moyal expansion of the particle model defined in Eq.~(\ref{eq:moran_select}), with selective advantage $\gamma$ and constant population size $\chi^{-1}$. }

{In this section, we shall instead relate Eq.~(\ref{eq:morandual}) with the Master equation~(\ref{me_birthdeath}), which does not conserve population size. This relation is very different in spirit to the one based on the Kramers-Moyal expansion and, in particular, does not rely on any approximation. The idea of this alternative approach is to combine the discrete process defined in Eq.~(\ref{birthcoagul}) with the continuous process described in Eq.~(\ref{eq:morandual}) to obtain a new process which is a martingale. To this aim,  }we consider the process $Z_t^m$, where $m$ is an arbitrary integer number. Applying the Ito formula~\eqref{eq:itoFormula0} yields
\begin{equation}\label{eq:zmm}
\dot{Z}^m_t=mZ_t^{m-1}\dot{Z}_t+\chi m(m-1)Z_t^{m-1}(1-Z_t).
\end{equation}
Substituting Eq.~(\ref{eq:morandual}) and Eq.~(\ref{eq:transitionmatrix_moran}) into Eq.~(\ref{eq:zmm}) we obtain 
% \begin{equation}\label{eq:zmm2}
% \dot{Z}_t^m=\sum_{n=0}^\infty M_{nm} Z_t^n+\sqrt{2\gamma} m Z_t^{m-1}\sqrt{Z_t(Z_t-1)}\dot{B}_t,
% \end{equation}
\begin{equation}\label{eq:zmm2}
\dot{Z}_t^m=\sum_{n=0}^\infty \omega(n,m) (Z_t^n-Z_t^m)+\sqrt{2\gamma} m Z_t^{m-1}\sqrt{Z_t(Z_t-1)}\dot{B}_t,
\end{equation}
%where the matrix $M_{mn}$ is the same appearing in Eq. (\ref{eq:transitionmatrix_moran}). 
We now introduce the quantity
\begin{equation}
  \mathcal{M}_t=\sum_{m=1}^\infty Z_t^m \rho_{T-t}(m),
\end{equation}
where $T$ is an arbitrary {(reference)} time. The process $\mathcal{M}_t$ combines a solution of the Langevin equation (\ref{eq:morandual}) with a backward solution $\rho_{T-t}(m)$ of the master equation (\ref{me_birthdeath}). Independently of the choice of the time $T$ and the initial conditions of the two processes, $\mathcal{M}_t$ is a martingale. We can in fact prove from Eqs. (\ref{me_birthdeath}) and (\ref{eq:zmm2}) that $\mathcal{M}_t$ is governed by an Ito process without drift:
\begin{eqnarray}
\dot{\mathcal{M}}_t&=& \sum_{m=1}^\infty \left(\dot{Z}^m_t \rho_{T-t}(m)
+Z^m_t \partial_t\rho_{T-t}(m)\right)\nonumber\\
&=& \sum_{m=1}^\infty \rho_{T-t}(m) \sqrt{2\gamma} m Z_t^{m-1}\sqrt{Z_t(Z_t-1)}\dot{B}_t.
% \left( \langle Z^n_t\rangle M_{nm} p_m(T-t)-\langle Z^m(t)\rangle M_{mn} p_n(T-t)   \right)=0 .
\end{eqnarray}
From its definition, the martingale $\mathcal{M}_t$ can be also expressed as
\begin{equation}
    \mathcal{M}_t=\langle {Z_t}^{N_{T-t}} \rangle_N,
\end{equation}
where with $\langle\dots\rangle_N$ we denote the expectation over {trajectories $N_{T-t}$ of the Master equation~(\ref{me_birthdeath}). We note that, while the continuous process $Z_t$ progresses forward in time $t$, the discrete process $N_{T-t}$ progresses backward in time.} The martingality of $\mathcal{M}_t$ implies for example that
\begin{equation}
  \langle Z_{t}^{N_0}\rangle_{Z,N} = \langle Z_0^{N_t}\rangle_{Z,N},
\end{equation}
where $\langle\dots\rangle_{Z,N}$ is the expectation over the forward continuous process and the backward discrete process.
We remark that this equality is valid for any $t$, and any choices of the initial conditions of the two processes. By appropriate choices of initial conditions, this relation can be exploited to derive useful properties of the two processes \cite{doering2003interacting}.

Interestingly, these techniques can be also applied to spatially extended populations. A prototypical stochastic model describing the dynamics of spatial populations is the stochastic Fisher-Kolmogorov equation
\begin{equation}\label{eq:stochastic_fisher}
\partial_tF_t(x)=sF_t(x)(1-F_t(x))+D\partial_x^2 F_t(x)+ \sqrt{\frac{2F_t(x)(1-F_t(x))}{N}}\xi_t(x).
\end{equation}
Using duality it can be shown that the probability of a small, localized population described by the Fisher-Kolmogorov equation to grow up to a large size is still governed by the formula (\ref{eq:kimuraform}) for the fixation probability of a well-mixed population \cite{doering2003interacting,pigolotti2013growth}.
%%%%%%%
%\include{chap10-shamik}
%\chapter 7
\chapter{Martingales in finance}
\label{chapter:finance}
\epigraph{\textit{October: This is one of the particularly dangerous months to invest in stocks. Other dangerous months are July, January, September, April, November, May, March, June, December, August and February.}   
\\ \vspace{0.5cm} Mark Twain, from ``Pudd'nhead Wilson'', 1894}

We give here an overview of the use of martingales in finance. Since the
theory of martingales had its early discussions in finance, it is no
wonder that a huge amount of literature exists on this subject. In the
treatment below, we do not
aim to be exhaustive or rigorous in any way, and our primary (and perhaps
only) motivation is
to introduce the basic terminologies of quantitative finance and
discuss how the theory of martingales arises naturally in this setting.
In the process, we hope to get the readers excited about the
field of quantitative finance. For further details, readers are directed to more specialized texts on the
subject, e.g., Refs.~\cite{Hull:2006,Wilmott:2006,Mantegna:2000,Bouchaud:2000}. A concise and self-contained review on
the topic, written from a physicist's point of view, is
Ref.~\cite{Vasconcelos:2009}. A reader aspiring to master all of stochastic
calculus required for a rigorous mathematical formulation of quantitative
finance may look up Refs.~\cite{Shreve:2000,Shiryaev:1999}.

%%%%%%%%%%%%%%%%%%%%%%%%%%%%%%%%%%%%%%%%%%%%%%%%%%%%%%%%%%%%%%%%%%%%%%%%%%%%%%%%%%%%%%%%%%%%%%%%%%%%%%%%%%%%%%%
\section{Riskless and risky financial assets: Bank deposits and stocks}
A riskless asset is one for which the return is fixed and guaranteed regardless of the market situation. A
prominent example is a bank deposit $\mathcal{B}_t$, with $t$ denoting time: an amount $\mathcal{B}_0$ deposited in a bank that offers a fixed interest
rate $r$  increases at a rate
\begin{equation}
\dot{\mathcal{B}_t}=r\mathcal{B}_t,
\label{eq:bank-return}
\end{equation} 
where the dot denotes derivative with respect to time.
The above evolution implies an
exponential growth in time, and yields a fixed
return with value $\mathcal{B}_t=\mathcal{B}_0 \exp(rt)$ at time $t$. Depending on $r$ and $\mathcal{B}_0$, although that does sound like a fortune, it could be possible that the depositor earns more through investments whose
worth is contingent on the evolution of the market. Such investments are
in general risky, since unlike riskless assets no fixed return is
guaranteed, but which when planned and managed well nevertheless offer
the investor the unique
opportunity to profit from market fluctuations. 
\begin{figure}[!ht]
\centering
\includegraphics[width=120mm]{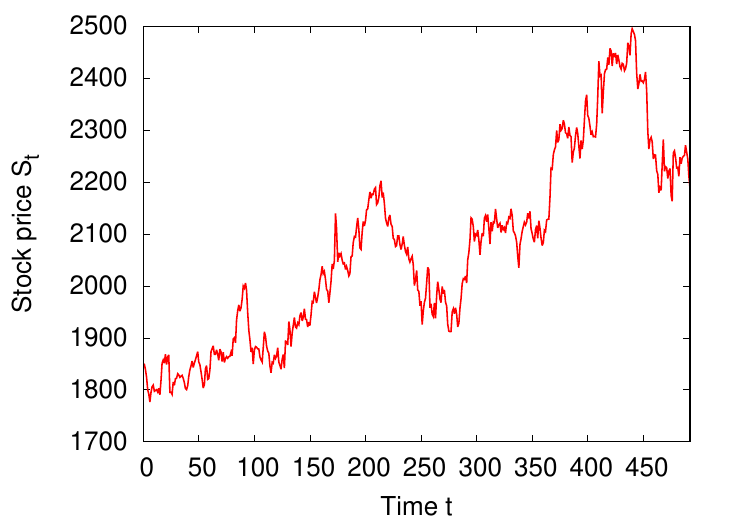}
\caption{Representative stock price fluctuations as a function of time. The figure depicts data on the closing price of HDFC Bank on a daily
basis from 19th September, 2017 to 18th September, 2019, i.e., over a period of 24 months. Source:
National Stock Exchange, India.}
\label{fig:stock-price-time}
\end{figure}

An example of risky
assets is what are called stocks or shares. A stock gives its holder the ownership
of a small part of the company issuing the stock. A company that
requires to raise its capital often does so by issuing stocks. By
selling many such stocks, the company is able to raise its
capital at typically lower costs than would have been possible if it were to
borrow money from banks, which would ask for high interests on the money
borrowed. It is evident that the stock price depends
on the overall worth of the company in the market\footnote{There are two types of market - primary and
secondary. When a company issues its shares, the process is  called Initial Public
Offering (IPO). Investors interested in buying the shares have to apply in order to procure the shares. In case there are
more applications than the number of shares issued, applicants are
chosen randomly.  Selected applicants buy shares directly from the
company. Stock exchanges have no part to play here. This is referred as
the primary market. After the above process is complete, the company
gets listed in the stock exchanges. Only after this can an investor
trade (buy or sell) the stock of the company in the exchanges from
another share holder. This is called the secondary market.},
which in turn depends
on how it has been performing in recent times, but also, interestingly, 
on how it is projected to perform in future. A small market fluctuation
due to, e.g., a Government decision, which is anticipated to affect the future
performance of the company, may lead to a change in the current price of
its stocks. All the aforementioned factors lead to stock prices behaving erratically in
time, an example of which is shown in Fig.~\ref{fig:stock-price-time}. In other words, the stock price $S_t$ is a random function of
time $t$; expressing its variation in time as
\begin{equation}
\dot{S}_t=R_tS_t,
\label{eq:stock-return}
\end{equation}
where $R_t$ is now the rate of return, which is itself a fluctuating quantity.   
In analogy with Eq.~(\ref{eq:bank-return}), we may expect the ``rate of
return" $R_t$, a random function of time, to have a part representing
the mean or expected rate of return and a part that varies randomly in time. The former part
may be deducible on the basis of the average of the company's past, present and 
projected future performance, and is thus a deterministic or a
predictable component, while all the uncertainty that got glossed over
in computing the average is included in the
random part. While there may be several ways to model the random part as a
function of time, one of the most popular and simple ones in the field
of quantitative finance is the
so-called Geometric Brownian Motion (GBM) model. In this model, the rate of 
change in the stock price  is 
\begin{equation}
\dot{S}_t=\left(\mu +\sigma \dot{B}_t\right) S_t,
\label{eq:GBM}
\end{equation}
where the constants $\mu$ and $\sigma$ represent respectively the
expected rate of return and the standard deviation of returns, also
called volatility, and where $B_t$
is the standard Brownian motion or a Wiener process, as defined in Sec.~\ref{sec:examplescont}. 
Volatility is a statistical measure of the dispersion of returns: the
higher the volatility, the riskier is the stock.

Equation~(\ref{eq:GBM}) is an example of a stochastic differential
equation (SDE), which may be solved subject to a given initial condition
$S_{t_0}=s_0$. In terms a new random variable $Z_t \equiv \ln S_t$, on
applying the It\^{o}'s formula, see Eq.~(\ref{eq:itoFormula}) in  Appendix~\ref{app:Ito-lemma}, and using Eq.~(\ref{eq:GBM}), we get 
\begin{equation}
\dot{Z}_t=\mu-\frac{\sigma^2}{2}+\sigma  \dot{B}_t,
\label{eq:GBM-Z}
\end{equation}
which on integration with respect to time gives 
\begin{equation}
Z_t=Z_{t_0}+\left(\mu-\sigma^2/2\right)(t-t_0)+\sigma\left(B_t-B_{t_0}\right),
\end{equation}
with $Z_{t_0}
= \ln s_0$; when expressed in terms of $S_t$, we get the following random
function of time for the stock price $S_t$:
\begin{equation}
S_t=s_0\exp\left(\left(\mu-\frac{\sigma^2}{2}\right)(t-t_0)+\sigma\left(B_t-B_{t_0}\right)\right).
\end{equation}
Equation~(\ref{eq:GBM-Z}) implies that $Z_t-Z_{t_0}$ is normally distributed with mean $\mu-\sigma^2/2)(t-t_0)$ and variance $\sigma^2(t-t_0)$, i.e., $Z_t-Z_{t_0}\sim \mathcal{N}((\mu-\sigma^2/2)(t-t_0),\sigma
\sqrt{t-t_0})$. It then follows that the probability density of the  stock price $S_t$ at time $t$,
subject to the initial condition $S_{t_0}=s_0$, is given by  the log-normal
distribution
\begin{equation}
\rho_{S_t}(s|S_{t_0}=s_0)=\frac{1}{s\sqrt{2\pi\sigma^2(t-t_0)}}\exp\left(-\frac{\left[\ln\left(\frac{s}{s_0}\right)-\left(\mu-\frac{\sigma^2}{2}\right)(t-t_0)\right]^2}{2\sigma^2
(t-t_0)}\right).
\label{eq:GBM-pdf}
\end{equation}

%%%%%%%%%%%%%%%%%%%%%%%%%%%%%%%%%%%%%%%%%%%%%%%%%%%%%%%%%%%%%%%%%%%%%%%%%%%%%%%%%%%%%%%%%%%%%%%%%%%%%%%%%%%%%%%
\section{Options and the Black-Scholes equation for option pricing}
\label{sec:BS}

Stocks are sold and bought (``traded") in organized stock exchanges,
such as the New York Stock Exchange, the NASDAQ Stock Market, etc. Every stock exchange devises an
index that is a representative of the daily average behavior of the
corresponding market. Different from stocks whose intrinsic values are
based directly on their market values and which therefore constitute primary
financial assets for the holder, there are financial instruments called
derivatives whose intrinsic values derive from the price of
some underlying primary assets. Derivatives are also referred to as contingent
claims as their values are contingent on that of
the underlying asset. One such basic derivative is what are called options, which we will deal with now.

An option is a contract between two parties to buy or sell in future an
underlying primary asset at an agreed price, regardless of the market
situation prevailing at the time the asset is bought or sold. The two sides of
the contract are called the buyer and the seller or the underwriter.
European options can be exercised only on the future date (the maturity
or expiration date) agreed in the
contract, while American options can be exercised at any point of time
until the expiration date. Here we will discuss only
European options. The two common types of European options are calls and puts. An European
call option gives the buyer the right, but not the obligation, to buy
the underlying asset (stock $S_t$) at the strike price $K$ specified in the
contract on the expiration date $\mathcal{T}$, regardless of the current price
(the spot price) $ S_\mathcal{T}$  of the asset. If the call buyer exercises his option, the seller is accordingly obliged
to sell the asset at price $K$. An European put option is
quite similar to the call option, excepting that it gives the buyer the
right, but not the obligation, to sell the underlying asset at price $K$
on date $\mathcal{T}$ regardless
of the spot price $S_\mathcal{T}$, and if
exercised, the seller is then obliged to buy the asset at price $K$.
Either way, due to the obligation to sell or buy
at a predetermined price on date $\mathcal{T}$ regardless of the spot price
$S_\mathcal{T}$, the seller may incur a loss, so that the buyer when entering into the option
contract must compensate somewhat by paying on-spot a certain amount called the option premium to
the seller. From the above, it is evident that investors buy calls or
sell puts (respectively, sell calls or buy puts) when
they anticipate that the price of the underlying asset will increase
(respectively, decrease) in
time.

%In Section~\ref{app:option}, we discuss in detail the concept of price of an option. 
Here, we discuss the concept of {\bf  price} of an option, from the point of
view of a potential buyer of a call option, which would help us fix our
ideas about option premium. If the spot price $S_t$ at any time $t$
exceeds the strike price $K$, it would make sense, in case it were
possible, to exercise the call
option, buy the asset from the
seller at price $K$ and sell it in the market at price $S_t$ (buy low
and sell high), thereby making a profit; we would then say that
the option has a positive intrinsic value given by the difference
$S_t-K$. If on the other hand one has
$S_t<K$, it is cheaper to buy in the market itself, and it would be
meaningless to exercise the call option; we would then say that the option has zero intrinsic value. This
leads us to define the intrinsic value of a call option at time $t$ to be the function ${\rm max}(S_t-K,0)$. Besides the
intrinsic value, the option would also have a time value that may be
understood thus. At any time $t<\mathcal{T}$, suppose that we have $S_t>K$. Now, since
there is still time left until expiration, there is a possibility that
in course of time until $\mathcal{T}$, $S_t$ will increase even further beyond $K$, which is
to say that the option has a certain positive time value. It is clear that the further 
$S_t$ is beyond $K$, higher is the probability that in the time until
$\mathcal{T}$, $S_t$ will increase even further beyond $K$, and so higher will be
the time value. How about the case $S_t<K?$ Again, since there is still
time until expiration, there is still a chance that $S_t$ will exceed
$K$: the lower $S_t$ is below $K$, of course, the smaller is this chance. All
these lead us to conclude that the time value of the option is a monotonically increasing
function of $S_t$. {Moreover, the further one is from expiration, the higher is the time value. This is because longer is the time until expiration, higher is the investor's expectation and consequently, higher is the probability that market fluctuations may cause $S_t$ to exceed $K$.}
The sum of the time value and the intrinsic
value gives the option price $C\equiv C(S_t,t;K, \mathcal{T})$, where we have shown
explicitly the factors on which the option price depends: the time at
which we value the option and the spot price of the underlying asset, as
well as the strike price $K$ and the expiration date $\mathcal{T}$. Since the time
value gets smaller as the expiration date gets closer, the call option
price as $t$ hits $\mathcal{T}$ is just the intrinsic value.
These points are shown schematically in Fig.~\ref{fig:call-put}(a). 

\begin{figure}[!ht]
\includegraphics[width=\textwidth]{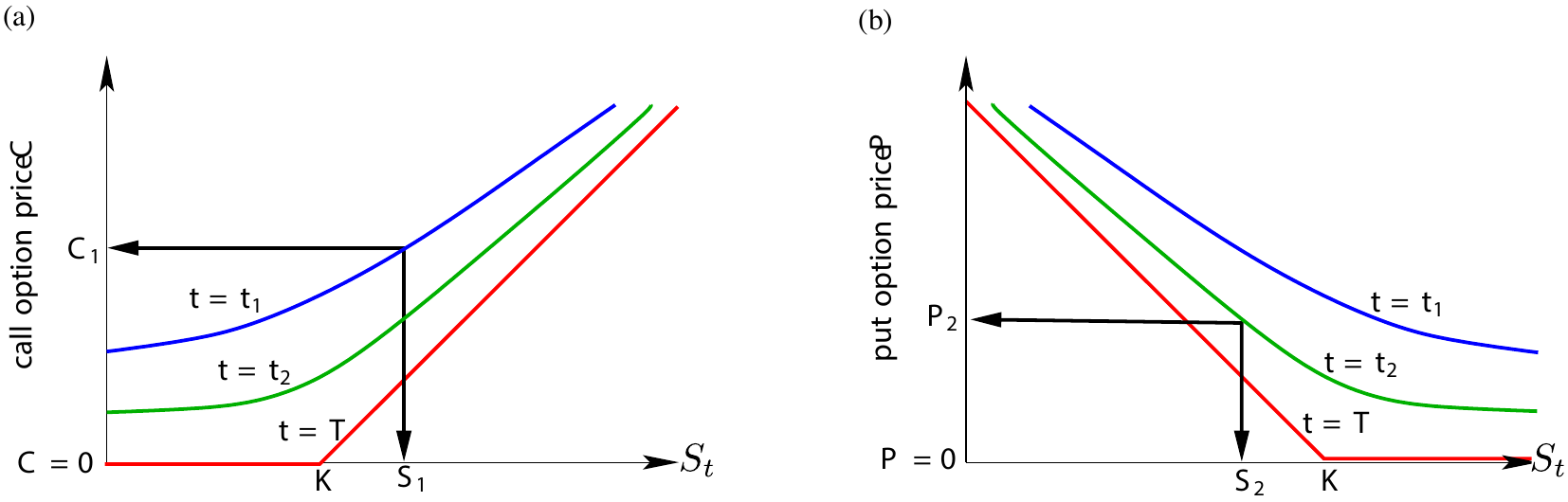}
\caption{Price of a European call and put option; here, we have $t_1 <
t_2 < \mathcal{T}$. The red line, which is
the limit $t \to \mathcal{T}$ of the $t< \mathcal{T}$-curves, is also
the intrinsic value of the option. Note that the origin of the $S_t$-axis
is certainly not zero: you cannot have a stock priced at zero!}
\label{fig:call-put}
\end{figure}

We now discuss the price of a put option, from the point of
view of a potential buyer. If the spot price $S_t$ at any time $t$
is below the strike price $K$, it would make sense, in case it were
possible, to exercise the put 
option, sell the asset to the
buyer at a higher price $K$, thereby making a profit; we would then say that
the option has a positive intrinsic value given by the difference
$K-S_t$. If on the other hand one has
$S_t>K$, it is better to sell in the market itself, and it would be
meaningless to exercise the put option; we would then say that the
option has zero intrinsic value. The intrinsic value of a put option at
time $t$ is then the function ${\rm max}(K-S_t,0)$. Coming to the time
value, suppose at any time $t<\mathcal{T}$, we have $S_t <K$. Now, since
there is still time left until expiration, there is a possibility that
in course of time until $\mathcal{T}$, $S_t$ will decrease even further below $K$,
which is tantamount to saying that the option has a certain positive
time value. The further 
$S_t$ is below $K$, higher is the probability that in the time until
$\mathcal{T}$, $S_t$ will decrease even further below $K$, and so higher will be
the time value. Summarizing, the time value of a put option is a monotonically decreasing
function of $S_t$, and the further one is from expiration, the
higher is the time value. The sum of the time value and the intrinsic
value gives the put option price $P\equiv P(S_t,t;K,\mathcal{T})$. Since the time
value gets smaller as the expiration date is approached, 
the put option price as $t$ hits $\mathcal{T}$ is made up entirely of the intrinsic value. 
The
discussed scenario is shown in Fig.~\ref{fig:call-put}(b). 

Now that we have discussed the scenario of a potential buyer of either a
call or a put option, let us now proceed to discuss the situation of one
who has already purchased the option. A buyer of a call option who has
purchased the option at time $t_1<\mathcal{T}$ when the spot price was $S_1 \equiv
S_{t_1}$ and the corresponding cost was $C_1\equiv C(S_1,t_1;K,\mathcal{T})$ had to pay as option premium the amount
$C_1$ to the seller. A payoff diagram summarizes the net worth of the
option from the point of view of the buyer, and is shown in
Fig.~\ref{fig:payoff}(a). On the expiration date, the payoff is $-C_1$
if $S_\mathcal{T}<K$ and is $S_\mathcal{T}-K-C_1$ if $S_\mathcal{T}>K$.
The payoff of a call option (buy) on maturity is thus given by
\begin{equation}
{\rm payoff}_{\rm call}^{\rm buy}={\rm max}(S_\mathcal{T}-K,0)-C_1.
\end{equation}
From the point of view of the seller, the call option payoff diagram is
evidently just the mirror image of that for the buyer
(Fig.~\ref{fig:payoff}(b)): the maximum
profit of the call option buyer is the maximum loss of the call option
seller, and vice versa. Moreover, the buyer has unlimited potential for
profit, and correspondingly, the seller has unlimited loss potential. 

\begin{figure}[!ht]
\centering
\includegraphics[width=\textwidth]{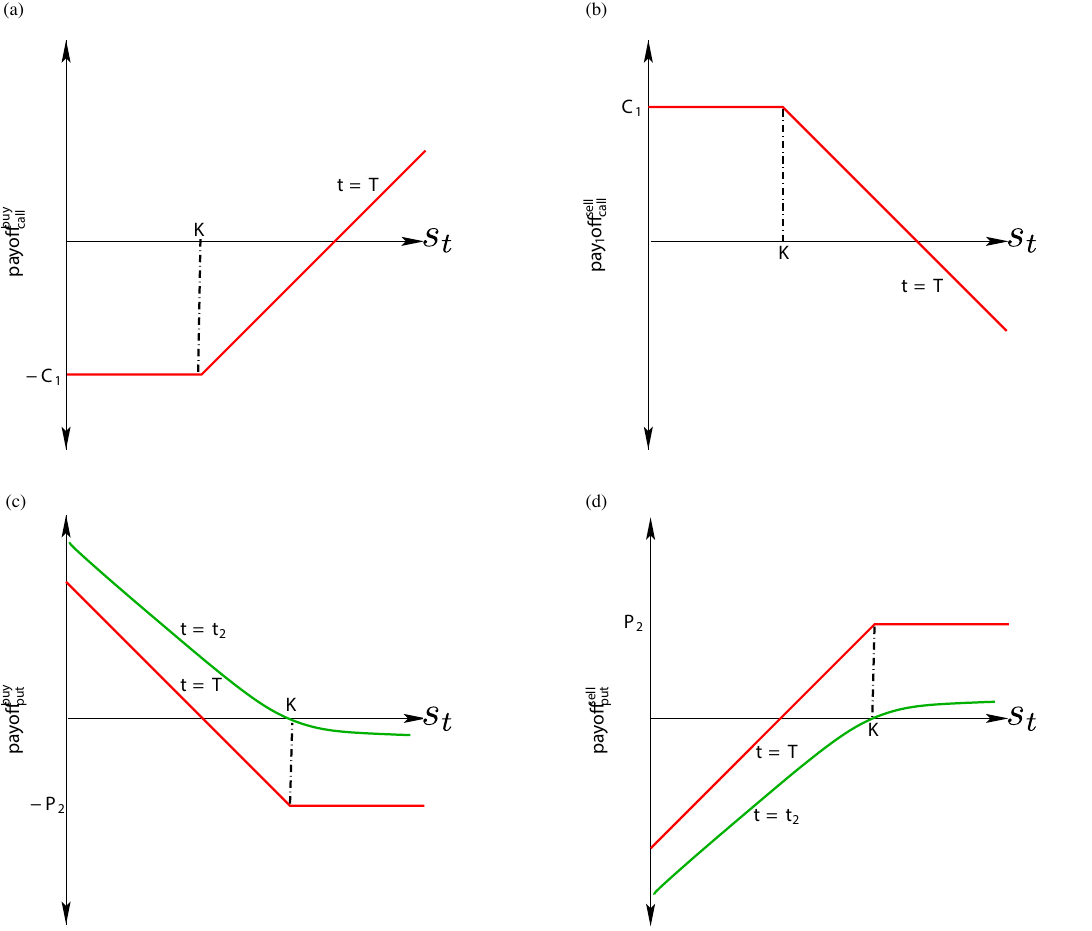}
\caption{Payoff diagrams for call and put options. The quantities $C_1$
and $P_2$ are defined in Fig.~\ref{fig:call-put}.}
\label{fig:payoff}
\end{figure}

Arguing as above, one may obtain the payoff diagram of a put option
(buy) on maturity. Thus, a buyer of a put option who has
purchased the option at time $t_2<\mathcal{T}$ when the spot price was $S_2 \equiv
S_{t_2}$ and the corresponding cost was $P_2\equiv P(S_2,t_2;K,\mathcal{T})$ has
his payoff given by
\begin{equation}
{\rm payoff}_{\rm call}^{\rm buy}={\rm max}(K-S_\mathcal{T},0)-P_2,
\end{equation}
while that from the point of view of the seller is the mirror image of that for the buyer
(Fig.~\ref{fig:payoff}(c),(d)). Comparing Figs.~\ref{fig:call-put}
and~\ref{fig:payoff}, we see that the payoff curve at time
$t<\mathcal{T}$ may be obtained from the corresponding cost curve in the former by shifting it (vertically down for the buyer and vertically up
for the seller) by an amount given by the cost at the corresponding
value of the stock (see Fig.~\ref{fig:payoff}(c),(d) for an
illustration).  
%In Section~\ref{app:option}, we discuss in detail the concept of price of an option. 

{From the above discussion,} it is evident that the buyer
and the seller of an option have different
expectations from the market, so that a central question as regards entering
into an option contract is: What should be the ``right" option price
$\mathcal{V}$
($=C$ for call and $=P$ for put) that 
would ensure that none of the two sides of an option contract have an a
priori advantage to make profit, for it is the magnitude of $\mathcal{V}$ that enters into the
payoff diagram for the buyer and the seller, see Fig.~\ref{fig:payoff}:
the buyer would not like a high $\mathcal{V}$ while the seller would very much
like a high $\mathcal{V}$. The value of $\mathcal{V}$ depends on the dynamics of
the underlying stock $S_t$. 
The question of finding the right $\mathcal{V}$, equivalent to finding a closed
form expression for $\mathcal{V}$ as a function of $K, \mathcal{T},S_t$ and time, has paramount
importance in option pricing. It was answered in the most remarkable way by economists Black and
Scholes \cite{Black:1973} and Merton \cite{Merton:1973}, a work that
earned Scholes and Merton (Black had died by then) the Nobel prize in
Economics in 1997. A crucial assumption behind deriving the model is
that of a market with ``no-arbitrage" opportunity, so we now digress to
discuss briefly what an arbitrage opportunity means.

An investor who does not wish to make any initial commitment of money
may still make money in the market in the following way. He may borrow a stock from someone
who has it, and sell it in the market. This
process of selling an asset that one does not own is called short
selling or shorting or taking a short position on the asset (buying the actual asset is what is
termed ``taking a long position on the asset"). The borrower has to eventually pay back the
lender (i.e., he has to ``close" his short position on the borrowed
asset), so what he may do
is to buy the same stock from the market on a later date, and
return it to the lender, and in the process, if the stock price
decreases, he makes a profit by this short selling; otherwise, he incurs a
loss. Thus, there is a risk involved in short selling, which may be
compensated thus: he chooses a company that is listed in two different
stock exchanges, say, exchange A and exchange B. Suppose he finds that at some
point in time, the last traded price $S^{(\rm A)}$ for selling a stock of the company
in A is higher than the last traded price $S^{(\rm B)}$ for buying a stock of the same
company in B. He may then with no initial
commitment short sell $N$ stocks in A and use the proceeds to close his
short position by buying
$N$ stocks in $B$, making in the process a riskless profit of
$N(S^{(\rm A)}-S^{(\rm B)})$. This process of making a riskless profit, with
no initial money at all, by entering simultaneously into transactions in
two or more markets is called an arbitrage opportunity or an arbitrage,
and those who exploit such opportunities are called arbitrageurs. A
minute's thought would reveal that such arbitrage opportunities cannot
last in the market for long, for selling the stock in A will decrease the price for selling a stock in A, while buying the stock in B
will increase the price to buy a stock in B. As a result of
these two competing tendencies, an equilibrium price for the stock in
both the exchanges will be reached in time and then the arbitrage
opportunity will no longer exist. The action of an arbitrageur is said to be
self-destroying in that it is destroying the action itself, but the
latter takes time, and in the process, the arbitrageur makes profit. An
efficient market is then one that satisfies the no-arbitrage condition, that
is, it does not allow anyone to make profit out of thin air (even if
some short-term profit may be possible, it would not allow for any
long-term profit). In common parlance, one says that there is no free
lunch possible in an efficient market. 

Besides arbitrageurs, there are hedgers in the market\footnote{In the market, there are also speculators who
unlike the hedgers like to take risks, by anticipating trends in the market
and exploiting them to make profit~\cite{Wilmott:2006}.}. A hedge is
defined to be an investment that protects one's finances from risks.
Hedgers may use derivatives to reduce the
risk in their portfolio in the following manner. Consider an investor
with a long position on a stock, for whom the risk is associated with
the possibility of the stock price going
down in time. In this case, a hedging strategy could be to buy a put
option on the stock, so that one would sell the stock only if the price
goes below a certain level, and can keep it with him if
the price goes up. In the former case, the proceeds from selling the
stock at a higher price (the strike price) than the spot price may
minimize or offset somewhat the risk associated with the long position, and
this comes at the price of the option premium that he paid in buying the
put option. We may think of the option as like an insurance in the
present scenario. 

With the above background, we now move
on to describe the Black-Scholes equation for option pricing.
Here, we will discuss the Black-Scholes equation in a simple setting, while
generalizations and a more detailed consideration may be found in, e.g.,
\cite{Wilmott:2006}. To derive the equation, assume within our simple
setting that there are two assets in the market: a bank deposit $\mathcal{B}_t$ and a
stock $S_t$, whose dynamics are given respectively by
Eqs.~(\ref{eq:bank-return}) and~(\ref{eq:GBM}). The quantity
$r$ in Eq.~(\ref{eq:bank-return}) is the interest rate offered by the bank to
a depositor, but is also the interest rate the bank charges on money borrowed
from the bank. Moreover, the market is
assumed to be free of arbitrage opportunities. For a more extensive list
of assumptions behind the Black-Scholes equation, the reader is directed to
Ref.~\cite{Wilmott:2006}.

Let us define a financial
portfolio as a combination of financial assets held by, e.g., individual
investors and/or managed by financial professionals. To derive the
Black-Scholes equation, consider a portfolio consisting of a long position on a European
call option and a short position on $\Delta$ stocks. The option has
strike price $K$ and expiration date $\mathcal{T}$ on the underlying stock $S_t$.
The question is what should be the value $C(S_t,t; K, \mathcal{T})$ of the option at time
$t$ subject to the boundary condition $C(S_\mathcal{T},\mathcal{T};K, \mathcal{T})={\rm
max}(S_\mathcal{T}-K,0)$. We take the
portfolio to be self-financing, that is, in course of time no money is
taken out of the portfolio and no additional money is put into it, so
that any change the portfolio value may undergo is due to change in
asset prices only. The value $V_t$ of the portfolio at time $t$ is
given by 
\begin{equation}
V_t=C(S_t,t)-\Delta~S_t,
\label{eq:V-1}
\end{equation}
where we have suppressed the dependence of $C$ on $K$ and $\mathcal{T}$, for ease
of notation. The minus sign on the right hand side of Eq.~(\ref{eq:V-1}) is a reminder of the
fact that we need to eventually close the short position on the stocks, and so we ``owe" the market an amount $\Delta ~S_t$. 
{As it will turn out, $\Delta$ will be a function of time: $\Delta=\Delta_t$. The portfolio is self-
financing, which implies the following. Let us specialize to discrete times. The value of
the portfolio at time $t = 0$ is
\begin{equation}
V_0 = C(S_0, 0) - \Delta_0S_0,
\end{equation}
which on its own will yield the value at the next time instant $t = 1$ as $C(S_1, 1) - \Delta_0 S_1$,
while our strategy being self-financing, the new portfolio at time $t = 1$ should be able to
be bought with the asset one has from the previous period, i.e.,
\begin{equation}
V_1 = C(S_1, 1) - \Delta_1 S_1 = C(S_1, 1) - \Delta_0 S_1,
\end{equation}
yielding
\begin{equation}
S_1(\Delta_1 - \Delta_0) = 0. 
\end{equation}
In continuous times, we thus have
\begin{equation}
S_t\mathrm{d}\Delta_t = 0.
\end{equation}
Using the above equation, we obtain from Eq.~(\ref{eq:V-1}) the rate of change in its value as
given by 
\begin{equation}
\dot{V}_t=\dot{C}-\Delta ~\dot{S}_t.
\label{eq:V-2}
\end{equation}} Equation~(\ref{eq:V-1}) represents what is known as a {\bf delta-hedging} 
portfolio. Delta hedging involves holding an option and shorting a
quantity $\Delta$ of the underlying. Its practical importance and
hedging implications will be discussed below.

Now, using Eq.~(\ref{eq:GBM}) and It\^{o}'s formula, see Appendix~\ref{app:ItoFormula}, we have
\begin{equation}
\dot{C}=\frac{\partial C}{\partial t}+\mu S_t \frac{\partial
C}{\partial S_t}+\frac{\sigma^2S_t^2}{2}\frac{\partial^2 C}{\partial
S_t^2}+\sigma S_t \frac{\partial C}{\partial S_t}\dot{B}_t,
\label{eq:dC}
\end{equation}
which when used in Eq.~(\ref{eq:V-2}) yields
\begin{equation}
\dot{V}_t=\frac{\partial C}{\partial t}+\mu S_t \frac{\partial
C}{\partial S_t}+\frac{\sigma^2S_t^2}{2}\frac{\partial^2 C}{\partial
S_t^2}-\mu \Delta~ S_t+\sigma S_t \left(\frac{\partial C}{\partial
S_t}-\Delta\right)\dot{B}_t.
\label{eq:V-3}
\end{equation}
%Note that the only stochasticity in the above equation that would make
%the portfolio risky in time is due to the last term on the right hand
%side. This risk can be eliminated by choosing $\Delta$ so as to satisfy
%\begin{equation}
%\Delta=\frac{\partial C}{\partial S_t}.
%\label{eq:Delta-choice}
%\end{equation}
%With the above choice, Eq.~(\ref{eq:V-3}) gives
%\begin{equation}
%\dot{V}_t=\frac{\partial C}{\partial t}+\frac{\sigma^2S_t^2}{2}\frac{\partial^2 C}{\partial
%S_t^2}.
%\label{eq:V-4}
%\end{equation}
{What the above equation gives is the value $V_t$ of the portfolio in the future on knowing its value at the current instant $t$ at which one knows with certainty the current stock price~$S_t$. Note that when one says that the stock price $S_t$ is a random function of time (see Eq.~(\ref{eq:GBM-pdf})), what one means is that although one knows the price at the current instant (one has to just visit a stock exchange), one cannot predict with certainty the price in the future. We will now show how, knowing $S_t$ and $V_t$ at the current instant $t$, the above equation allows to know with certainty the value of $\dot{V}_t$ and hence of $V_t$ in future. To this end, let us choose $\Delta$ in such a way that one gets rid of the term involving $\dot{B}_t$ on the right hand
side, namely, we choose $\Delta$ such that 
\begin{equation}
\Delta=\frac{\partial C}{\partial S_t}.
\label{eq:Delta-choice}
\end{equation}
With the above choice, Eq.~(\ref{eq:V-3}) gives
\begin{equation}
\dot{V}_t=\frac{\partial C}{\partial t}+\frac{\sigma^2S_t^2}{2}\frac{\partial^2 C}{\partial
S_t^2}.
\label{eq:V-4}
\end{equation}
It is now evident that knowing $S_t$ allows one to compute the right hand side and obtain with certainty the value of $\dot{V}_t$ (one has to evaluate the derivatives at $S_t$) and consequently, the value of $V_{t+\mathrm{d}t}$ (provided Eq.~(\ref{eq:Delta-choice}) remains valid during the interval $[t,t+\mathrm{d}t]$), and hence, there is no more randomness or stochasticity in the evolution of $V_t$. In other words,~$V_t$ has a deterministic evolution in time, and so the portfolio becomes {\bf risk free} for the choice given by Eq.~\eqref{eq:Delta-choice}, i.e. for $\Delta=\partial C/\partial S_t$.} 

Now that we have a risk-free portfolio and the market is by assumption
free of arbitrage opportunities, the portfolio would yield the same rate
of return as we would get if we had deposited an equivalent amount
of cash in a bank account, see Eq.~(\ref{eq:bank-return}). This may be
explained as follows: Suppose the rate of return $r_{\rm risk-free}$ from the risk-free
portfolio is different from $r$, and let us say that one has $r_{\rm risk-free}>r$.
Then, someone would borrow money from the bank (which would according to
our assumptions ask for an interest rate $r$ on the
lent amount), and would invest it in the
risk-free portfolio. He would then use the return from the portfolio to give back
the money he owes to the bank, and in the process, pocket the difference of the return from the
invested amount and the money given back to the bank, without making any
initial investment. The market being arbitrage-free would not allow for
such a possibility, and hence, we conclude that $r_{\rm risk-free}$
should equal $r$. Then, we may write 
\begin{equation}
\dot{V}_t=rV_t=r\left(C-\frac{\partial C}{\partial
S_t}S_t\right),
\label{eq:BS-r}
\end{equation}
where in obtaining the second equality we have used Eqs.~(\ref{eq:V-1})
and~(\ref{eq:Delta-choice}).
\begin{leftbar}
Comparing Eqs.~(\ref{eq:V-4}) and~(\ref{eq:BS-r}) yields the celebrated
\textbf{Black-Scholes equation}
\begin{equation}
\frac{\partial C}{\partial t}+\frac{\sigma^2
S_t^2}{2}\frac{\partial^2 C}{\partial S_t^2}+rS_t\frac{\partial C}{\partial
S_t}-rC=0,
\label{eq:BS}
\end{equation}
with the boundary condition
\begin{equation}
C(S_\mathcal{T}, \mathcal{T})={\rm max}(S_\mathcal{T}-K,0).
\label{eq:BS-boundary-condition}
\end{equation}
\end{leftbar}

As we have discussed earlier, hedging relates to reduction of risks in
one's financial portfolio. We saw above that choosing
$\Delta=\partial C/\partial S_t$, which corresponds to exploiting
correlation between the option and the stock making up the portfolio
(clearly, evolution of $C$ depends on the dynamics of $S_t$),
led to perfect elimination of risks in that the resulting portfolio has
completely deterministic evolution~(\ref{eq:V-4}). Such a strategy goes
by the name of delta hedging. Note that the quantity $\partial
C/\partial S_t$ continually changes in time. This implies that the
amount $\Delta$ of stocks that one needs to short to offset the risk
associated with the long position must change continually in time. Delta
hedging is thus an example of a dynamic hedging strategy. 

In obtaining the Black-Scholes equation~(\ref{eq:BS}), the only place where the
nature of the derivative enters into the derivation is through the
boundary condition~(\ref{eq:BS-boundary-condition}). Thus, it should be
possible to generalize the derivation for the price of an arbitrary
European option $F(S_t,t)$ with payoff $F(S_\mathcal{T}, \mathcal{T})=\Phi(S_\mathcal{T})$, where $\Phi(S)$ is a
known function. The price $F(S_t,t)$ should then follow the equation
\begin{equation}
\frac{\partial F}{\partial t}+\frac{\sigma^2
S_t^2}{2}\frac{\partial^2 F}{\partial S_t^2}+rS_t\frac{\partial F}{\partial
S_t}-rC=0,
\label{eq:BS-F}
\end{equation}
with the boundary condition
\begin{equation}
F(S_\mathcal{T}, \mathcal{T})=\Phi(S_\mathcal{T}).
\label{eq:BS-F-boundary-condition}
\end{equation}

%%%%%%%%%%%%%%%%%%%%%%%%%%%%%%%%%%%%%%%%%%%%%%%%%%%%%%%%%%%%%%%%%%%%%%%%%%%%%%%%%%%%%%%%%%%%%%%%%%%%%%%%%%%%%%%
%%%%%%%%%%%%%%%%%%%%%%%%%%%%%%%%%%%%%%%%%%%%%%%%%%%%%%%%%%%%%%%%%%%%%%%%%%%%%%%%%%%%%%%%%%%%%%%%%%%%%%%%%%%%%%%
\subsection*{Solution of the Black-Scholes equation~(\ref{eq:BS})}
\label{sec:BS-equation}
The treatment here follows the one given in
Refs.~\cite{Wilmott:2006,Vasconcelos:2009}. The Black-Scholes equation~(\ref{eq:BS}) may be solved by performing the following
transformation to a set of dimensionless variables that turns it into the heat equation or the Fokker Planck equation for a free Brownian particle, both well
known in physics:
\begin{eqnarray}
&&\tau \equiv \frac{\mathcal{T}-t}{2/\sigma^2},~~x\equiv
\ln(S_t/K),~~u(x,\tau)\equiv \exp(\alpha x + \beta^2\tau)\frac{C(S_t,t)}{K},\nonumber\\ &&\alpha \equiv
\frac{1}{2}\left(\frac{2r}{\sigma^2}-1\right),~~\beta=\frac{1}{2}\left(\frac{2r}{\sigma^2}+1\right).
\label{eq:BS-transformation}
\end{eqnarray}
In terms of transformed variables $u,x,\tau$, Eq.~(\ref{eq:BS})
reads~\cite{Vasconcelos:2009} (for details, see
Appendix~\ref{app:heat-derivation}):
\begin{equation}
\frac{\partial u}{\partial \tau}=\frac{\partial^2 u}{\partial x^2},
\label{eq:heat-equation}
\end{equation}
while the condition~(\ref{eq:BS-boundary-condition}) becomes an initial
condition thus:
Equation (\ref{eq:BS-boundary-condition}) gives the result $u(x,0)K/\exp(\alpha x)={\rm max}(S_t-K,0)$, that is,
\begin{equation}
u(x,0)={\rm max}(S_t\exp(\alpha x)/K-\exp(\alpha x),0)={\rm max}(\exp((\alpha+1)x)-\exp(\alpha x),0),
\end{equation}
where we have used the fact that $S_t/K=\exp(x)$. Next, using $\alpha+1=\beta$, we finally have the desired initial condition:
\begin{equation}
u(x,0)={\rm max}(\exp(\beta x)-\exp(\alpha x),0).
\label{eq:initial-condition}
\end{equation}

Now, the heat equation~(\ref{eq:heat-equation}) is solved as
\begin{equation}
u(x,\tau)=\int_{-\infty}^\infty {\rm d}x'~u(x',0)G(x,x'),
\label{eq:u-solution}
\end{equation}
in terms of the Green's function for the heat equation:
$G(x,x')=1/\sqrt{4\pi \tau}\exp(-(x-x')^2/(4\tau))$. 
Using the initial condition~(\ref{eq:initial-condition}) in the last
equation allows to write $u(x,\tau)$ as
\begin{equation}
u(x,\tau)=I(\beta)-I(\alpha);~~I(a)\equiv \frac{1}{\sqrt{4\pi
\tau}}\int_0^\infty {\rm
d}x'~\exp(ax'-(x-x')^2/(4\tau))=\exp(ax+a^2\tau)N(d_a),
\label{eq:u-solution-1}
\end{equation}
with
\begin{equation}
d_a\equiv \frac{x+2a\tau}{\sqrt{2\tau}},
\end{equation}
and $N(x)$ being the cumulative distribution for a Gaussian random
variable distributed as $\mathcal{N}(0,1)$:
\begin{equation}
N(x)=\frac{1}{\sqrt{2\pi}}\int_{-\infty}^x {\rm d}y~\exp(-y^2/2).
\end{equation}
Using Eq.~(\ref{eq:u-solution-1}) and reverting to the original variables
of $C,S_t,K$, etc by using Eq.~(\ref{eq:BS-transformation}) lead to the following result:
\begin{leftbar}
\textbf{The Black-Scholes formula for the price of a European call
option:}
\begin{eqnarray}
&&C(S_t,t)=S_tN(d_1)-K\exp(-r(\mathcal{T}-t))N(d_2);\nonumber \\&&d_1 = \frac{\ln
\left(S_t/K\right)+\left(r+\sigma^2/2\right)({\cal
T}-t)}{\sigma\sqrt{\mathcal{T}-t}},~~d_2 = \frac{\ln
\left(S_t/K\right)+\left(r-\sigma^2/2\right)({\cal
T}-t)}{\sigma\sqrt{\mathcal{T}-t}}. \nonumber \\ \label{eq:shamikscholes}
\end{eqnarray}
Equation~\eqref{eq:shamikscholes} is  known as the Black-Scholes formula for
option pricing: a closed expression to price an option in a market where
there are two assets, namely, a bank deposit $\mathcal{B}_t$ subject to interest
rate $r$ and a
stock $S_t$ with expiration $\mathcal{T}$ and strike price $K$, and with the dynamics of $\mathcal{B}_t$
and $S_t$ given respectively by Eqs.~(\ref{eq:bank-return})
and~(\ref{eq:GBM}).
\end{leftbar}

From the foregoing, it is easy to write down the solution to
Eq.~(\ref{eq:BS-F}) for an arbitrary European option. Defining
$u(x,\tau)\equiv \exp(\alpha x + \beta^2\tau)F(S_t,t)/K$, and following the
same line of argument as the one followed in
Eqs.~(\ref{eq:BS-transformation}) - (\ref{eq:u-solution}), one
obtains in analogy with Eq.~(\ref{eq:u-solution}) that
\begin{equation}
u(x,\tau)=\int_{-\infty}^\infty {\rm d}x'~\Phi(x')G(x,x'),
\end{equation}
which when expressed in terms of variables $F,S_t,K$, etc yields
\begin{equation}
F(S_t,t)=\frac{\exp(-r(\mathcal{T}-t))}{\sqrt{2\pi \sigma^2(\mathcal{T}-t)}}\int_0^\infty {\rm
d}S'~\frac{\Phi(S')}{S'}\exp\left[-\frac{\{\ln(S'/S_t)-(r-\sigma^2/2)({\cal
T}-t)\}^2}{2\sigma^2(\mathcal{T}-t)}\right]. 
\label{eq:F-solution}
\end{equation}
Equation~(\ref{eq:F-solution}) is the
Black-Scholes formula for an arbitrary European option with payoff
$F(S_\mathcal{T}, \mathcal{T})=\Phi(S_\mathcal{T})$, where $\Phi(S)$ is a
known function. For a European call option, $F(S_t,t)=C(S_t,t)$ and
$\Phi(S)={\rm max}(S-K,0)$, while for a European put option, $F(S_t,t)=P(S_t,t)$ and
$\Phi(S)={\rm max}(K-S,0)$, where $K$ is the strike price and $\mathcal{T}$ is the
expiration time.

%%%%%%%%%%%%%%%%%%%%%%%%%%%%%%%%%%%%%%%%%%%%%%%%%%%%%%%%%%%%%%%%%%%%%%%%%%%%%%%%%%%%%%%%%%%%%%%%%%%%%%%%%%%%%%%
%%%%%%%%%%%%%%%%%%%%%%%%%%%%%%%%%%%%%%%%%%%%%%%%%%%%%%%%%%%%%%%%%%%%%%%%%%%%%%%%%%%%%%%%%%%%%%%%%%%%%%%%%%%%%%%
\section{Efficient market and the martingale approach}
The so-called martingale approach to an efficient market offers an
alternative elegant way to arrive at the Black-Scholes equation. With
the preliminaries on elements of probability theory that may be found in {Appendix}~\ref{app:probability}, let us now describe the martingale approach to an
efficient market.

%%%%%%%%%%%%%%%%%%%%%%%%%%%%%%%%%%%%%%%%%%%%%%%%%%%%%%%%%%%%%%%%%%%%%%%%%%%%%%%%%%%%%%%%%%%%%%%%%%%%%%%%%%%%%%%
%%%%%%%%%%%%%%%%%%%%%%%%%%%%%%%%%%%%%%%%%%%%%%%%%%%%%%%%%%%%%%%%%%%%%%%%%%%%%%%%%%%%%%%%%%%%%%%%%%%%%%%%%%%%%%%

\subsection{Equivalent martingale measure and the Girsanov theorem}

{We have already seen that the standard
Brownian motion is a martingale.
We know that the probability density of the standard Brownian motion is a
Gaussian:
\begin{equation}
\rho^\mP_{B_t}(b)=\frac{1}{\sqrt{2\pi t}}\exp\left(-b^2/(2t)\right).
\end{equation}
Using the defining property of a martingale, we may
conclude that the motion with a drift, given by
\begin{equation}
        \widetilde{B}_t=at+B_t;~~0 \le t \le \mathcal{T},
\end{equation}
is not a martingale precisely because of the presence of the drift $a$.
The {\bf Girsanov theorem}~\cite{lip2013} states however that $\widetilde{B}_t$ becomes a
standard Brownian motion with respect to the probability measure $\mQ$
given by 
\begin{equation}
\rho^{\mQ}_{\widetilde{B}_t}(b)=M_{\widetilde{B}_t}(b)\rho^\mP_{\widetilde{B}_t}(b),
\label{eq:Girsanov}
\end{equation}
where $M_{\widetilde{B}_t}(b)$ is the stochastic process
\begin{equation}
M_{\widetilde{B}_t}(b)=\exp\left(-ab+\frac{a^2t}{2}\right).
\end{equation}
Indeed, using
\begin{equation}
\rho^\mP_{\widetilde{B}_t}(b)=\frac{1}{\sqrt{2\pi
t}}\exp\left(-\left(b-at\right)^2/(2t)\right),
\end{equation}
we see that
\begin{equation}
\rho^\mQ_{\widetilde{B}_t}(b)=\frac{1}{\sqrt{2\pi
t}}\exp\left(-b^2/(2t)\right),
\end{equation}
which indeed corresponds to the probability density of the standard Brownian motion.
Thus, with respect to measure $\mQ$, the process with drift,
$\widetilde{B}_t$, becomes a standard Brownian motion and is thus a
martingale. The measure $\mQ$ is
called the equivalent martingale measure.}

{Consider now the Geometric Brownian motion~(\ref{eq:GBM}), which we
rewrite below as
\begin{equation}
{\rm d}S_t=\sigma S_t\left(\frac{\mu}{\sigma}{\rm d}t+{\rm d}B_t\right)=\sigma
S_t{\rm d}\widetilde{B}_t;~~0 \le t \le \mathcal{T},
\label{eq:GBM-1}
\end{equation}
with
\begin{equation}
\widetilde{B}_t=\frac{\mu}{\sigma}t+B_t.
\end{equation}
The Girsanov theorem would make $\widetilde{B}_t$ a standard Brownian
motion with respect to the measure~(\ref{eq:Girsanov}) with $a=\mu/\sigma$, and then the SDE
(\ref{eq:GBM-1}) will have no drift so that the stochastic process
$S_t$ will be a martingale with respect to the measure $\mQ$.}

%%%%%%%%%%%%%%%%%%%%%%%%%%%%%%%%%%%%%%%%%%%%%%%%%%%%%%%%%%%%%%%%%%%%%%%%%%%%%%%%%%%%%%%%%%%%%%%%%%%%%%%%%%%%%%%
%%%%%%%%%%%%%%%%%%%%%%%%%%%%%%%%%%%%%%%%%%%%%%%%%%%%%%%%%%%%%%%%%%%%%%%%%%%%%%%%%%%%%%%%%%%%%%%%%%%%%%%%%%%%%%%
\subsection{The martingale approach to an efficient market}
{ With the above background, we now
 come to discuss about the main object of this section: the martingale
 approach to an arbitrage-free market. Consider a market comprising two assets $(\mathcal{B}_t,S_t)$, with $\mathcal{B}_t$ a risk-free asset (a bank deposit) and $S_t$ a risky asset such as a stock that is modelled as a stochastic process.  As
implied by Eq.~(\ref{eq:bank-return}), we have the growth law $\mathcal{B}_t=\mathcal{B}_0
\exp(rt)$, where $\mathcal{B}_0 \equiv \mathcal{B}_{t=0}$ is the initial amount deposited in
the bank.  
%We consider $\{S_t\}_{t \ge 0}$ to be a
%stochastic process on a probability space $(\Omega, \mathcal{F}, P)$ and
%adapted to a filtration $\{\mathcal{F}_t\}_{t \ge 0}$. 
Based on the
information available up to time $t_0$, the expected price of $S_t$ at a
later time $t > t_0$ is $\langle S_t |S_{[0,t_0]}\rangle$ with $S_{[0,t_0]}=\{S_s\}_{s \in [0,t_0]}$, so that
if the market is arbitrage-free, we now argue that the price at time $t_0$
should be $\langle S_t |S_{[0,t_0]}\rangle/\exp(r(t-t_0))$. For, if the stock is priced at time
$t_0$ at a value
$y<\langle S_t|S_{[0,t_0]}\rangle/\exp(r(t-t_0))$, then a buyer would take
advantage of the situation by borrowing an amount of money $y$ at time
$t_0$ to buy the asset and then selling at time $t$ to repay his debt of
$y\exp(r(t-t_0))$, thereby pocketing at time $t$ a positive profit of $\langle S_t
|S_{[0,t_0]}\rangle-y\exp(r(t-t_0))$. On the other hand, if the stock
is priced at $y>\langle S_t|S_{[0,t_0]}\rangle/\exp(r(t-t_0))$, then
a seller would take advantage of the situation by selling the stock at
time $t_0$ and lending an amount of money $y$ so that at time $t$, he
would receive an amount $y\exp(r(t-t_0))$ and would buy back the asset to make
a positive profit of $y\exp(r(t-t_0))-\langle S_t|S_{[0,t_0]}\rangle$. The market being arbitrage-free, it would not allow
for both these opportunities of making profit out of thin air, and
hence, the stock at time $t_0$ should be priced at $\langle S_t|S_{[0,t_0]}\rangle/\exp(r(t-t_0))$, which by definition is the actual price
$S_{t_0}$ at time $t_0$. Rewriting in terms of $\mathcal{B}(t)$, and recalling that
a bank deposit is risk-free, i.e., non-stochastic, we get 
\begin{equation}
\left\langle \frac{S_t}{\mathcal{B}_t}\Big|S_{[0,t_0]}\right\rangle_\mQ=\frac{S_{t_0}}{\mathcal{B}_{t_0}};~~t \ge t_0.
\label{eq:market-martingale}
\end{equation}
From the definition of a martingale, it then follows that the stochastic
process given by $\{S_t/\mathcal{B}_t\}_{t \ge 0}$ is a martingale. The ratio $S_t/\mathcal{B}_t$ is known as the
discounted price of the stock $S_t$. Note that
Eq.~(\ref{eq:market-martingale}) hold with the expectation calculated with respect to a suitable probability measure $\mQ$.}

{In the light of the foregoing, we now state the two fundamental
theorems of asset pricing.
\begin{leftbar}
\textbf{First Fundamental Theorem of asset pricing.}
{If in the
market there exists at least one probability
measure $\mQ$ such that the discounted price
$S_t/\mathcal{B}_t$ is a martingale with respect to the measure $\mQ$, that is,
\begin{equation}
\left\langle \frac{S_t}{\mathcal{B}_t}\Big|S_{[0,t_0]}\right\rangle_{\mQ}=\frac{S_{t_0}}{\mathcal{B}_{t_0}};~~t \ge t_0,
\label{eq:market-martingale-1}
\end{equation}
then the market does not admit arbitrage, or, in other words, the market is efficient.} 
In words, an efficient market is one for which it should not be possible
to make definite predictions about future price on the basis of the
information available today, so that the best
prediction that one can make for the expected future price discounted to the
present time is today's price itself. {One may ask when does the measure $\mQ$ exist? If the market is arbitrage-free, the measure $\mQ$ has to exist. For a mathematically-rigorous discussion of conditions for the existence of $\mQ$, beyond the scope of this review, the reader is referred to, e.g., Refs.~\cite{dalang,rydberg-finance}.} 
\end{leftbar}
In the above backdrop, we now turn to the Black-Scholes model of option
pricing. To this end, assume, as in Section~\ref{sec:BS}, that there are two assets in the market: a bank deposit $\mathcal{B}_t$ and a
stock $S_t$, whose dynamics are given respectively by
Eqs.~(\ref{eq:bank-return}) and~(\ref{eq:GBM}). Moreover, the market is
assumed to be free of arbitrage opportunities, which according to our
discussions above is to be regarded as an efficient market. Using the fact that in an efficient market, all financial assets are
martingales with respect to the measure $\mQ$, we may
now write for an arbitrary European option $F(S_t,t)$ with maturity
$\mathcal{T}$ and payoff function $\Phi(S)$ that
\begin{eqnarray}
\frac{F(S_t,t)}{\mathcal{B}_t}&=&\left\langle \frac{F(S_\mathcal{T},\mathcal{T})}{\mathcal{B}_\mathcal{T}}\Big|S_{[0,t]}\right\rangle_{\mQ} \nonumber \\
&=&
\left\langle
\frac{\Phi(S_\mathcal{T})}{\mathcal{B}_\mathcal{T}}\right\rangle_{\mQ; t,S_t},
\end{eqnarray}
where in obtaining the second line, we have used the fact that
$F(S_\mathcal{T}, \mathcal{T})=\Phi(S_\mathcal{T})$. In the second line, using
$\mathcal{B}_t=\mathcal{B}_0\exp(rt)$, and denoting by $\rho^{\mQ}_{S_t}(s|S_{t_0}=s_0)$ the probability density under the measure $\mQ$ of $S_t$ with initial value $S_{t_0}=S_0$, the
quantity $\langle \cdot \rangle_{\mQ;t.S}$ means the following: 
\begin{equation}
F(S_t,t)=\exp\left[-r(\mathcal{T}-t)\right]\int_0^\infty {\rm d}s'~\Phi(s')\rho^\mQ_{S_\mathcal{T}}(s'|S_t=S_t).
\label{eq:FSt}
\end{equation}
In order
to obtain $\mQ$, consider the stochastic process 
\begin{equation}
Z_t=\frac{S_t}{\mathcal{B}_t}=\exp(-rt)S_t,
\end{equation}
so that
\begin{equation}
{\rm d}Z_t=r\exp(-rt)S_t{\rm d}t+\exp(-rt){\rm d}S_t=\sigma Z_t{\rm
d}\widetilde{B}_t;~~\widetilde{B}_t=\frac{\mu-r}{\sigma}t+B_t,
\label{eq:Btwidetilde-1}
\end{equation}
where we have used Eq.~(\ref{eq:GBM}) to arrive at the second equality.
The latter when rewritten in terms of the process $\widetilde{B}_t$ reads 
\begin{equation}
{\rm d}S_t=rS_t{\rm d}t+\sigma S_t {\rm d}\widetilde{B}_t.
\label{eq:GBM-risk-free}
\end{equation}
From our previous discussion on the Girsanov theorem, we know that one
can construct a measure $\mQ$ with respect to which
the process $\widetilde{B}_t$ is a standard Brownian motion. Note that the measure $\mQ$ will be different for different $S_t$'s that would have in general different $\mu$ and different $\sigma$ that gets reflected in having correspondingly different $\widetilde{B}_t$'s, see Eq.~(\ref{eq:Btwidetilde-1}). for Then,
Eq.~(\ref{eq:GBM-risk-free}) has the same form as the Geometric Brownian
motion~(\ref{eq:GBM}), with mean rate of return given by $\mu=r$. The
latter fact, which implies that risky stocks guarantee the same mean
rate of return as the risk-free bank account, makes the pricing method based on the measure $\mQ$ sometimes referred to as
risk-neutral valuation. The measure $\mQ$ with respect to which
the discounted stock price $S_t/\mathcal{B}_t$ is a martingale is therefore said to be a
risk-neutral measure.}

{Using
Eq.~(\ref{eq:GBM-pdf}) and with the substitution $\mu=r,S_0=S_t$, we thus have
\begin{equation}
\rho^\mQ_{S_\mathcal{T}}(s'|S_t=s)=\frac{1}{s'\sqrt{2\sigma^2({\cal
T}-t)}}\exp\left[-\frac{\left\{\ln\left(s'/s\right)-\left(r-\sigma^2/2\right)({\cal
T}-t)\right\}^2}{2\sigma^2
(\mathcal{T}-t)}\right],
\end{equation}
which when used in Eq.~(\ref{eq:FSt}) yields
\begin{equation}
F(S_t,t)=\frac{e^{-r(\mathcal{T}-t)}}{\sqrt{2\pi\sigma^2(\mathcal{T}-t)}}\int_0^\infty {\rm
d}s'~\frac{\Phi(s')}{s'}\exp\left[-\frac{\left\{\ln\left(s'/S_t\right)-\left(r-\sigma^2/2\right)({\cal
T}-t)\right\}^2}{2\sigma^2
(\mathcal{T}-t)}\right],
\end{equation}
the same as Eq.~(\ref{eq:F-solution}).}

{In an efficient market, we know that at least one risk-neutral measure
$\mQ$ will exist. If a unique $\mQ$ exists, there is a unique arbitrage-free
price for every derivative, and the market is said to be complete. This brings us to the
second fundamental theorem of asset of pricing: 
\begin{leftbar}
\textbf{Second Fundamental Theorem of asset pricing.} An arbitrage-free
$(\mathcal{B}_t,S_t)$-market is complete if and only if the measure $\mQ$ is unique. 
\end{leftbar}
For rigorous mathematical proof and implications
of the two fundamental theorems of asset pricing, the reader is referred to
Ref.~\cite{Shiryaev:1999}.}

To conclude, we see in this brief overview on use of martingales in the
field of finance how an approach based on martingales allows to obtain
rather straightforwardly the solution of the Black-Scholes equation
without actually solving it using the rather nontrivial variable
transformation discussed in Section~\ref{sec:BS-equation}. The
martingality encodes the expectation that in an efficient market, all
relevant information is already reflected in the prices, so that the
best possible prediction for the expected future price would be today's price. The Nobel-winning Black-Scholes model for pricing an option contract with an underlying
martingale structure provided one of the earliest and remarkable mathematical
foundations to option-market activities around the world. The success of
the model led to an eventual boom in
options trading with people gaining confidence in engaging in such activities. The assumptions
behind the model have over the years been relaxed and generalized in
many directions, leading to a spectrum of models that are currently in
wide use in derivative pricing and risk management all over the world.

\chapter{Final remarks and discussion}
\label{ch:conclusion}
\epigraph{In the old days, you could type into our main computer ``Edit explain life" and you got the answer  ``Life is a supermartingale" \\ \vspace{0.5cm} {\em Obituary: Joseph Leonard Doob}, J.~L.~Snell, J. Appl. Prob. {\bf 42}, 247-256 (2005).}

\section*{Other revelations of martingales}

%Despite the broadness of contents that we aimed when writing this treatise,
There exist other fields in science where  martingales have found valuable applications. Here is a swift list of some of the miscellanous topics that we have not covered in this treatise.

  The main aim of {\bf decision theory} is to develop    algorithms that take fast and reliable decisions from the observations of a  noisy process.  Wald's sequential probability ratio test (SPRT) \cite{wald1992sequential,  wald1948optimum, tartakovsky2014sequential} is optimal amongst sequential hypothesis tests with a prescribed error probability when the observation process consists of a sequence of iid random variables, in the sense that it provides the minimum average time to decide   between two competing hypothesis.   In addition,  for a broad class of observations processes Wald's SPRT is  optimal  in the asymptotic limit of small prescribed error probabilities when neglecting subleading order terms~\cite{tartakovsky2014sequential}.   The   recent work~\cite{doerpinghaus2022optimal} shows that Wald's SPRT is optimal in an information theoretically sense for continuous observation processes, providing an information theoretical interpretation for the SPRT.   % More general, Wald's test achieves the minimal mean decision time for a given error probability for a broad class of stochastic processes describing the observed data \cite{wald1948optimum, tartakovsky2014sequential}.

The SPRT takes sequential observations from a stream of data coming from a stochastic process $X_t$,  and measures the weight of evidence through the log-likelihood ratio
\begin{equation}
    \Lambda_t = \log \frac{\mP(\Xot|H_1)}{\mP(\Xot|H_2)},
    \label{eq:LLRWald}
\end{equation}
where $\mP(\Xot|H_1)$ and $\mP(\Xot|H_2)$ are the path probabilities for the sequence $\Xot$ when the statistical hypothesis $H_1$ and $H_2$  are, respectively, true.    Wald's SPRT takes a decision when the log-likelihood ratio  leaves  the interval  $(-L_2,L_1)$ for the first time with $L_1>0$ and $L_2>0$ the decision thresholds, i.e.,
\begin{eqnarray}
\T = {\rm min}\left\{t\geq 0: \Lambda_t\leq -L_2 \ {\rm or} \  \Lambda_t\geq L_1 \right\}.  
  \end{eqnarray} 
When $\Lambda_{\T}\geq L_1$,  then the SPRT test decides  for $H_1$, whereas if  $\Lambda_{\T}\leq -L_2$, then  the SPRT decides   for   $H_2$.    The decision thresholds are set by the prescribed error probabilities, see~Refs.~\cite{wald1992sequential,  wald1948optimum, tartakovsky2014sequential}. 

Reference~\cite{roldan2015decision} uses Wald's SPRT to decide on  the direction of time's arrow from the observation of  a trajectory  drawn from a time-homogeneous  stationary process.  Interestingly, \cite{roldan2015decision} shows that  the mean decision  is related to the entropy production rate of the process. Moreover, as the log-likelihood ratio~\eqref{eq:LLRWald} has the form of a $\Lambda-$stochastic entropic functional (see Ch.~\ref{ch:thermo2}), it is possible to exploit the mathematical machinery of martingales to derive fluctuation relations for decision times \cite{Roldan-prX2017} and to develop quantitative criteria on how far from Wald's optimality is a decision maker \cite{dorpinghaus2018testing}.

A field in physics where martingales have found profound applications and we did not discuss in this {Treatise} are {\bf quantum measurements}. 
Briefly, in models of iterated  discrete
(continuous) time measurements, the collapse
of the system  at large times  can be rationalized in terms of the convergence theorem of submartingales (as Theorem~\ref{convergenceTheorem} in Ch.~\ref{ch:3} or more precisely version with almost sure convergence).
More precisely,  under a discrete iterated (resp. 
continuous) time measurement, the diagonal elements of the density matrix,
in a special basis called pointer basis and given by a non-demolition
hypothesis~\cite{wiseman2009quantum},   is a martingale. For a pure state, such martingale is given by 
 the modulus square of the projection of the ket in the pointer basis. This result has been shown for both continuous time~\cite{adler2001martingale} and discrete time~\cite{bauer2011convergence}.
 Let us now we give a smell of this formulation in a 
physical example. In continuous-time quantum measurements, the equation for the
evolution of the density matrix is called {\em quantum trajectory}~\cite{wiseman2009quantum}; it is given by a
matricial stochastic differential equation with Gaussian and/or Poissonian white noise.
 If the Hilbert space of the system is two-dimensional with orthonormal basis $\{\left|+\right\rangle , \left|-\right\rangle\}$~\footnote{Which will be here also the pointer basis.}, then the density matrix in this
basis is parametrized by 
$\rho_{t}=\left(\begin{array}{cc}
X_{t} & Z_{t}\\
\overline{Z_{t}} & 1-X_{t}
\end{array}\right)$ where $X_{t}\in\left[0,1\right]$ and $Z_t$ is the coherence of the density matrix. 
Then, the quantum trajectory with Gaussian noise which results from the continuous measurements of the operator \footnote{
This matrix is diagonal in the orthonormal basis $\left|\pm1\right\rangle $;  this is the meaning of the quantum non-demolition hypothesis here.}\
 $\frac{1}{4}\sigma_{z}=\frac{1}{4}\left(\begin{array}{cc}
1 & \text{0}\\
0 & -1
\end{array}\right)$, is given by the Ito stochastic differential equation for
$X_{t}$
\begin{equation}
\dot{X}_{t}=r X_{t}\left(1-X_{t}\right)\dot{B}_{t},
\label{sdecoll}
\end{equation}
 where $r$ is a parameter quantifying the rate of measurement. Figure~\ref{fig:quantummart} shows representative trajectories for  $X_t$ obtained from numerical simulations.
\begin{figure}[h]
\centering
\includegraphics[width=0.6\textwidth]{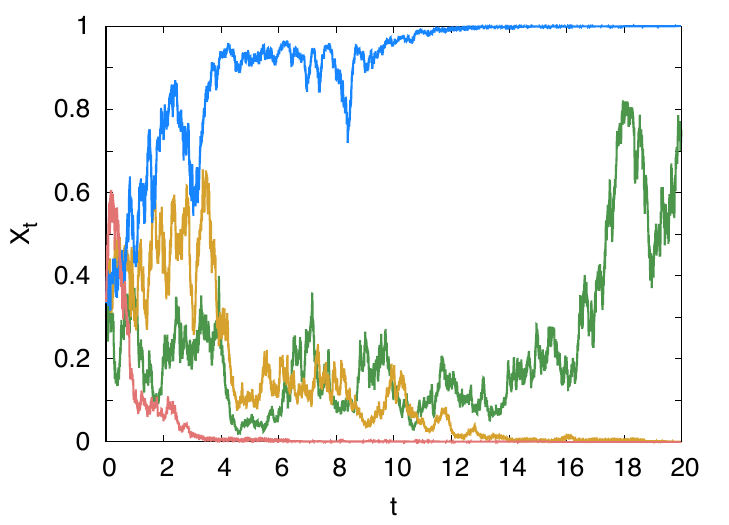}
\caption{Representative time series of the  process~\eqref{sdecoll}. Parameters:     $r =1$, $X_0=1/3$, $dt=10^{-3}$.}
\label{fig:quantummart}
\end{figure}

Then, from the absence of drift in this stochastic differential equation, $X_{t}$ is a bounded (local) Martingale, and converges to $X_{\infty}\in\left\{ 0,1\right\} $ by virtue of continuous-time version of  Theorem~\ref{convergenceTheorem} in Ch.~\ref{ch:3}. Moreover, the martingale property implies that $
X_{0}=\left\langle X_{\infty}\right\rangle =1\times P\left(X_{\infty}=1\right)+0\times P\left(X_{\infty}=0\right).
$
 We then find the Born law as a emergent property: 
\[
\begin{cases}
P\left(X_{\infty}=1\right)=X_{0}=\text{Tr}\left(P_{1}\rho_{0}\right)\\
P\left(X_{\infty}=0\right)=1-X_{0}=\text{Tr}\left(P_{-1}\rho_{0}\right)
\end{cases}\textrm{with}\quad \begin{cases}
P_{1}=\left|1\right\rangle \left\langle 1\right|\\
P_{-1}=\left|-1\right\rangle \left\langle -1\right|
\end{cases}.
\]
%\selectlanguage{french}%

Lastly, let us mention two more interesting applications of martingales in statistical physics.   In the context of {\bf spin glass theory}~\cite{sherrington1975solvable,mezard1984nature}, a full replica symmetry breaking theory for a spin glass on a Bethe lattice, which is one of the main open challenges in this research area,  has been formulated with the help of martingales \cite{concetti2018full, parisi2017marginally}.    In the {\bf theory of critical phenomena}, Cardy's formula for the crossing probability of a stochastic Loewner evolution, which in the case of percolation gives the probability  that there exist a percolating cluster, has been rederived and extended with martingales, see Ref.~\cite{bauer2003conformal}.  
{Martingales also  play a key role in the study of {\bf nonequilibrium properties of interacting particles}, see e.g. Spohn's treatise~\cite{spohn2012large}. In particular, martingale theory was applied to prove directly weak convergence of path probabilities, which go beyond convergence of moments as done in expansion techniques. Fruits of this approach,  explicit proofs for the Green-Kubo formula,   current statistics and  various hydrodynamic limits can be retrieved through elegant calculations, see also Refs.~\cite{de1986reaction,ferrari1988non,kipnis1998scaling}.}

%\section*{Recap of key results}
%\subsection*{How to identify a martingale?}
%{\bf [IN: discuss different approaches, including, Dynkin martingale, Ito integral, path probability ratio]}

%\subsection*{What to do with martingales?}
%{\bf [IN: discuss Doob's theorems and properties at stopping times and extreme values.] }

%\subsection*{Martingales in thermodynamics}
%{\bf [IN: discuss, among others, martingale versions of the second law of thermodynamics]}

\section*{Discussion}

This   {Treatise} highlights the use  of martingale theory in statistical physics, population dynamics and quantitative finance. Although  martingales have been used extensively in the  latter two research areas, its relevance and usefulness for statistical physics, notably stochastic thermodynamics, is a recent endeavour. %, emerging field exploited until now mainly in the context of stochastic thermodynamics.
Taken together, the  results and techniques reviewed here address {\em why  a statistical physicist should learn martingale theory.}  As we have shown, martingales are ubiquitous and their properties are fundamental in probability theory.  Therefore,   we think that martingale theory can be considered as relevant for statistical physics as, e.g.,  the theory of  Markov processes or large deviation theory.  Particularly interesting is the fact that once a martingale, submartingale or supermartingale has been identified, we can use theorems from martingale theory to unveil universal physical principles. 
For random walks, the "martingale" approach is particularly useful when dealing with first-passage properties and extreme-value statistics. As we have shown with several examples,  non-trivial extreme-value and first-passage-time calculations can be greatly simplified  upon using Doob's theorems for stopping times. This leads to another key concept for physics unveiled by this {Treatise}, viz., the {\it stopping time}.  We have thoroughly reviewed the concept of stopping times in the context of stochastic processes as generalized first-passage times. Furthermore, upon applying several well-known martingale theorems to physically-relevant stopping times, we have presented several "shortcuts" to calculations of, e.g., absorption probabilities,  first passage time statistics (mean, second moment, distributions), and finite-time statistics of extrema.

When dealing with the stochastic thermodynamics of small systems,
the martingale approach provides novel insights with respect to conventional fluctuation theorems developed in the 1990s and 2000s. On one hand, the martingale structure of thermodynamically-relevant probability ratios leads to a tree-like hierarchy of second laws of thermodynamics, among which only some of them were known previously  in the literature. Furthermore, applying mathematical properties of martingales to thermodynamics quantities unveils universal fluctuation relations for, e.g., stopping  times, extrema, and absorption probabilities of entropy production in stationary states.  Interestingly, for stationary processes  one can overcome classical limits for, e.g., the efficiency of thermal machines, by stopping the dynamics of a system upon a cleverly-chosen time. On the other hand, we have shown that extra care is required when  applying martingale concepts to  non-stationary processes, as the second laws at stopping times are in this context nontrivial generalizations of the traditional second laws  at fixed times. This   leads to the  so-called gambling opportunities, which allow an observer to extract more work from a system than given by  the  free energy difference between the initial and final state through several executions of a protocol stopped at a cleverly chosen strategy. {For future work, it will be interesting to relate the martingale bounds on work extraction to the performance of Szilard demons or engines~\cite{szilard1964decrease}.}

We expect that martingales will find use in statistical physics beyond the study of  fundamental principles in stochastic thermodynamics. In biophysics, recent work proposed that small living systems (e.g. cells) can take accurate rapid   decisions  in noisy environments through applying   threshold criteria (e.g. Wald's SPRT) to accumulated chemical species~\cite{siggia2013decisions,desponds2020mechanism}.  Similarly, in cognitive neuroscience it has been hypothesized that binary perceptual decisions taken by e.g. rhesus monkeys~\cite{gold2001neural,kira2015neural}  result from the accumulation of neural evidence in the brain and the implementation of log-likelihood-ratio threshold tests. The plethora of second laws for path-probability ratios discussed in this work and the trade-off relations between speed and accuracy may thus shed further light in understanding decision making of living systems from the sub-cellular to the whole organism level. Furthermore, stopping times form a versatile toolbox with  applications  in various research areas.   A notable example is  computer science~\cite{moore2011nature}, where the first  thermodynamics insights brought by e.g. Landauer and Bennett~\cite{landauer1961irreversibility,bennett1973logical} were rationalized by the field of information thermodynamics~\cite{parrondoreview}. The development of a comprehensive stochastic-thermodynamic framework of computation is however still in its infancy~\cite{wolpert2023stochastic}. Stopping-time statistics could be pushed forward in unveiling  novel generic thermodynamic laws that govern computational tasks executed by e.g. finite automata, Turing machines, and quantum computers.

% The field of mathematical finance,   boosted since 

%The universal nature of the thermodynamic laws developed with martingales could be exported to e.g. develop second laws for financial markets pushing forward recent developments on information-thermodynamic framework is still e.g. in developing second laws touzo2021information 

%universal laws that go beyond standard fluctuation theorems. Football club of second laws recoving well known results but highlighting new. Second laws at stopping times not trivial generalizations. Gambling demons extracting more work than free energy, efficiency at stopping times beyond Carnot.  Entropy production but more could be said.
%Future promising avenues: computation (automatas), second laws for finance, boomerang from Black and scholes, social dynamics (quenching), quantum martingales. Active matter. Decision theory and neuroscience.

%%%%%%%
 \renewcommand\pagenumbering[1]{}

\pagenumbering{arabic}
\chapter{Acknowledgements}

ER acknowledges support from ICTP, highlighting  the work of his entire research group, and the academic support of the QLS and the CMSP sections. He also thanks the following institutions for hospitality while writing this work: Universit\'e C\^{o}te d'Azur, DIPC---Donostia International Physics Centre,  ESPCI Paris, MISANU Belgrade and PMF Ni\v{s}. He is also grateful for fruitful scientific discussions on martingales to: Gonzalo Manzano, Rosario Fazio, the Pekola Lab, Rosemary Harris, Joachim Krug, Matteo Marsili, David Wolpert, Gulce Kardes, and Tarek Tohme. He thanks Pietro Luigi Muzzeddu, Debraj Das, Yonathan Sarmiento, John Bechhoefer,  Juan MR Parrondo,  Massimo Campostrini,  and Stefano Ruffo for feedback on the elaboration of this {Treatise}. 

RC is supported by the French National Research Agency through the projects QTraj (ANR-20-CE40-0024-01), RETENU (ANR-20-CE40-0005-01), and ESQuisses (ANR-20-CE47- 0014-01). 
He acknowledges his habilitation's committee for comments on preliminary versions of Ch.~\ref{ch:thermo2}  and Ch.~\ref{TreeSL}: Eric Akkermans, Giovanni Gallavotti, Giovanni Jona-Lasinio,  Senya Shlosman, Michel Bauer, Denis Bernard, Cedric Bernardin, Sergio Ciliberto, Bernard Derrida,  Krzysztof Gawedzki, Kirone Mallick, C\'ecile Monthus and R\'emi Rhodes. 
Finally, RC dedicates this {Treatise} to his master, Krzysztof Gawedzki, who left us in January 2022.

SG acknowledges support from the Science and Engineering Research Board
(SERB), India under SERB-MATRICS scheme Grant No. MTR/2019/000560, and SERB-CRG scheme Grant No. CRG/2020/000596. He also gratefully acknowledges the many clarifying and fruitful discussions
and constant guidance received from Debraj Das who literally ushered him
into the field of quantitative finance. {He also thanks ICTP Abdus Salam International Centre for Theoretical Physics, Trieste, Italy, for support under its Regular Associateship scheme.} He is grateful to Partha
Nag for help with the references, and thanks Debraj
Das, {Soumya Kanti Pal, Sayan Roy and C L Sriram for discussions and useful comments on the text. SG is particularly grateful to Rudra Pratap Jena for several insightful discussions regarding delta hedging.}

KS thanks  Charles Moslonka (Gulliver ESPCI-PSL) who contributed essentially to the main part of Ch.~\ref{sec:PQ}. KS and Charles Moslonka are grateful to ER and Guilhem Semerjian (ENS-PSL) for their constructive and valuable comments.

This research was supported in part by the International Centre for  Theoretical Sciences (ICTS) for the online program "Stochastic Thermodynamics: Recent Developments"  (code: ICTS/strd2022/06) and for the program "Workshop on Martingales in Finance and Physics" (ICTP) .

\appendix

\chapter[Appendix to Chapter 1]{Appendix to Chapter~\ref{ch:1}}\label{appendix:0} 

\section{Random walk between two absorbing boundaries}
Consider a random walker moving  in discrete time steps on  a one-dimensional lattice, with $X_t \in \left\{-L,-L+1,\ldots,L\right\}$ the sites of the lattice. 
   At every discrete time step, the walker hops to its right-neighbor site with probability $0 \le q \le 1$ and to its left-neighbour site with a complementary probability $1-q$.   The process terminates at the random time $\T$ when either $X_t=L$ or $X_t=-L$.         This is the classical gambler's ruin problem, as formulated, for example, in Feller's treatise on probability theory \cite{feller1957introduction}   Following Ref.~\cite{feller1957introduction}, we determine here the splitting probabilities and mean-first passage time of the gambler's ruin problem, see also~\cite{redner2001guide}.

\section{Splitting probabilities}  
We determine the probabilities $P_+(i)$ and $P_-(i)$  that the walker ends its excursion at $X_{\T}=L$ or $X_{\T}=-L$, respectively, given that the walker started its excursion  from site $X_0=i$.

The splitting probabilities satisfy the following recurrence equation
\begin{equation}
    P_-(i)=qP_-(i+1)+(1-q)P_-(i-1), \quad {\rm for}\;\;  i \in \left\{-L+1,-L+2,\ldots,L-1\right\},  \label{eq:PRec}
\end{equation} 
 with boundary conditions  $P_-(-L)=1$ and $P_-(L)=0$.      For $q\neq1/2$, the  Eqs.~(\ref{eq:PRec})   admit solutions of the form $P_-(i) = \alpha^i$.    Substitution in Eqs.~(\ref{eq:PRec}), gives 
 $\alpha =  q\alpha^2 + (1-q)$, which admits two solutions, $\alpha=1$ and $\alpha = (1-q)/q$.    Consequently, 
 \begin{equation}
     P_-(i) = \alpha_0 + \beta_0 \left(\frac{1-q}{q}\right)^{i}
 \end{equation}
where $\alpha_0$ and $\beta_0$ are determined by the boundary conditions.    Consequently, 
\begin{equation}
P_-(i) = \frac{\left(\frac{1-q}{q}\right)^{2L}-\left(\frac{1-q}{q}\right)^{i+L}}{\left(\frac{1-q}{q}\right)^{2L}-1}  \label{eq:ppluspminusintro3}
\end{equation}
and analogously, 
\begin{equation}
P_+(i) =  \frac{\left(\frac{1-q}{q}\right)^{i+L}-1}{\left(\frac{1-q}{q}\right)^{2L}-1}  .
\end{equation}
Note that since the solution to Eq.~(\ref{eq:PRec}) with boundary conditions $P_-(-L)=1$ and $P_-(L)=0$ is unique, these are the expressions for the splitting probabilities.  
 
 For $q=1/2$, we suggest a linear solution of the form 
 \begin{equation}
 P_-(i) = \alpha_0 + \beta_0 i 
 \end{equation}
leading to 
\begin{equation}
 P_-(i) =  1-\frac{i+L}{2L} .\label{eq:ppluspminusintro4}
\end{equation}  
The Eqs.~(\ref{eq:ppluspminusintro1}) and (\ref{eq:ppluspminusintro2}) in the main text are obtained by setting $i=0$ in Eqs.~(\ref{eq:ppluspminusintro3}-\ref{eq:ppluspminusintro4}).

\begin{figure}[H]
    \centering
    \includegraphics[width=0.7\textwidth]{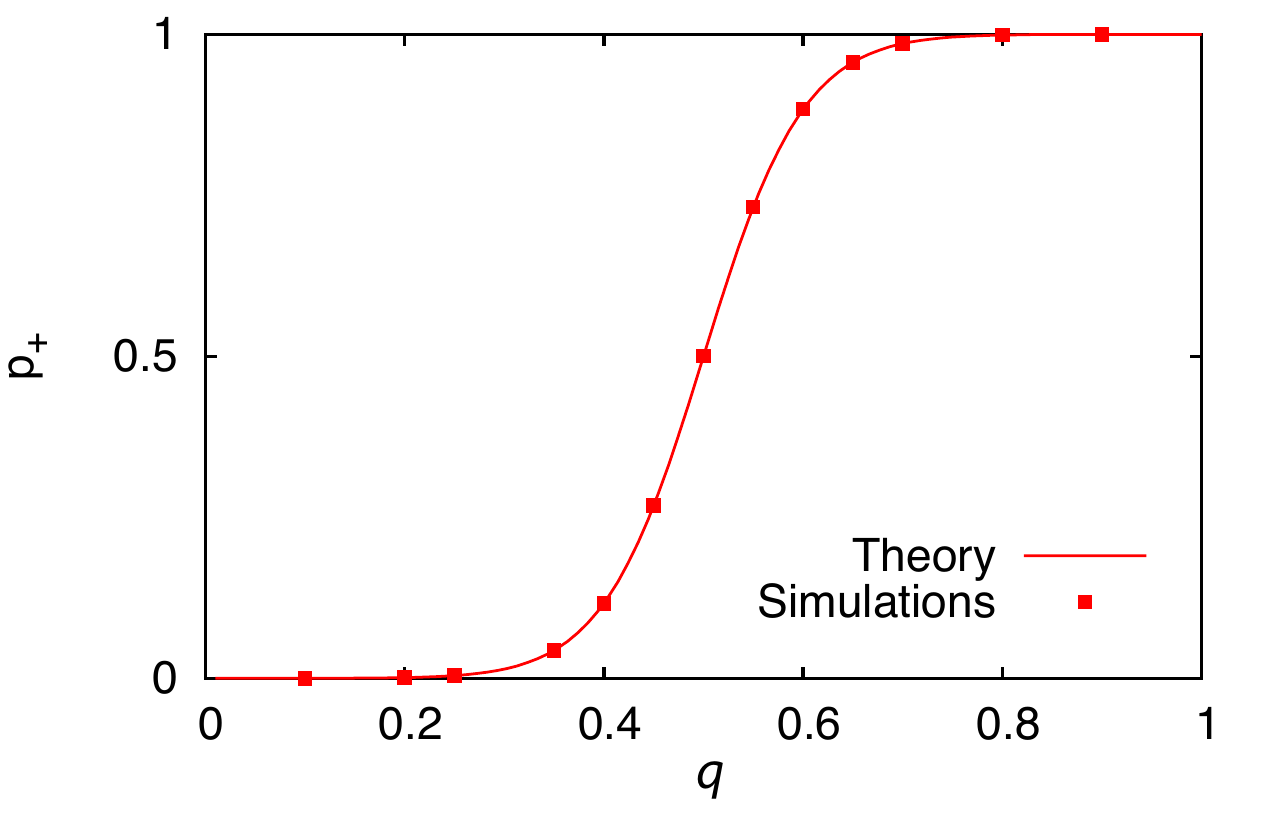}
    \caption{Comparing Eq.~(\ref{eq:ppluspminusintro1}) with results from simulations for $L=5$. The data involve sampling $10^6$ independent dynamical realizations.}
    \label{fig:compare-pplus}
\end{figure}

\section{Mean first-passage time}

We determine the  mean duration of the process,  $\tau_i = \langle \T| X_0=i \rangle$, which obey the recurrence relations 
\begin{equation}
\tau_i = q\tau_{i+1} + (1- q) \tau_{i-1} +1
\end{equation}
with boundary conditions 
\begin{equation}
\tau_{-L} = \tau_L = 0.
\end{equation}

For $q\neq 1/2$, the solution takes the form 
\begin{equation}
\tau_i = \frac{i+L}{1-2q} + \alpha_0 + \beta_0 \left(\frac{1-q}{q}\right)^{i}
\end{equation}
Using the boundary conditions, we find 
\begin{equation}
\tau_i =  \frac{L+i}{1-2q}  - \frac{2L}{1-2q}  \frac{1- \left(\frac{1-q}{q}\right)^{i+L}}{1- \left(\frac{1-q}{q}\right)^{2L}} \label{eq:TAvDoobIllu3}
\end{equation}

On the other hand, for $q=1/2$ the solution takes a quadratic form 
\begin{equation}
\tau_i = -i^2 + \alpha_0 + \beta_0 i,
\end{equation}
such that with boundary conditions 
\begin{equation}
\tau_i = (i+L)(L-i). \label{eq:TAvDoobIllu4}
\end{equation}

The Eqs.~(\ref{eq:TAvDoobIllu1}) and (\ref{eq:TAvDoobIllu2}) in the main text are obtained by setting $i=0$ in Eqs.~(\ref{eq:TAvDoobIllu3}) and (\ref{eq:TAvDoobIllu4}).

%%%%%%%
%\include{chap_app1}

\chapter[Appendix to Chapter 2]{Appendix to Chapter~\ref{ch:2a}}\label{appendix:A} 

%%%%%%%%%%%%%%%%%%%%%%%%%%%%%%%%%%%%%%%%%%%%%%%%%%%%%%%%%%%%%%%%%%%%%%%%%%%%%%%%%%%%%%%%%%%%%%%%%%%%%%%%%%%%%%%
\section{A primer on probability theory}
\label{app:probability}

Here, we provide a primer on probability
theory, emphasizing in particular the elements that may prove to be
both essential and useful in reading this review. For a more
extensive treatise on probability theory within the ambit of
quantitative finance, the reader is referred to
Ref.~\cite{Shreve:2000}. While a physicist's notion of probability and
measure may suffice to understand martingales, the rigorous mathematical
foundation of probability theory, a glimpse of which is provided below,
is absolutely necessary to comprehend scientific papers (regular
postings may be found on the
arXiv: \url{https://arxiv.org/list/q-fin/new}) and standard mathematical treatise on quantitative
finance, e.g., Ref.~\cite{Shreve:2000}

%%%%%%%%%%%%%%%%%%%%%%%%%%%%%%%%%%%%%%%%%%%%%%%%%%%%%%%%%%%%%%%%%%%%%%%%%%%%%%%%%%%%%%%%%%%%%%%%%%%%%%%%%%%%%%%
%%%%%%%%%%%%%%%%%%%%%%%%%%%%%%%%%%%%%%%%%%%%%%%%%%%%%%%%%%%%%%%%%%%%%%%%%%%%%%%%%%%%%%%%%%%%%%%%%%%%%%%%%%%%%%%
\subsection{Probability space and $\sigma$-algebra}
In discussing probability, one talks about a random experiment or a
random trial, namely, an experiment whose outcome is random, i.e., one gets in general a different outcome
every time the experiment is repeated under identical conditions. Let
$\Omega$ denote the sample space, i.e., the set of all
possible elementary outcomes $\omega$ of the random trial.
An event $A$ is a subset\footnote{In mathematics, a set $A$ is considered a subset of a set $B$, or, equivalently,
$B$ is a superset of $A$, if all elements of $A$ are also elements of
$B$.
} of
$\Omega$. The set of observable events is the collection ${\cal F}$ of subsets of
$\Omega$ (conventionally called the family of
subsets of $\Omega$) with the following properties:
\begin{enumerate}
\item $\emptyset \in {\cal F}~{\rm and}~\Omega \in {\cal F}$; Here,
$\emptyset$ is the empty set, denoting the event ``nothing
happens", while $\Omega$ denotes the event ``something happens."
\item $A \in {\cal F} \implies A^c \in {\cal F}$, where $A^c$ is the
complement of $A$ (if $A$ is an event, ``$A$ does not happen" is also an event).
\item $A_1,A_2,A_3, \ldots \in {\cal F} \implies \cup_i A_i
\in {\cal F}$ (if a sequence of events can occur, then ``at least one of
them occurs" is also an event).
\end{enumerate}
When the above properties are satisfied, ${\cal F}$ is said to form a
$\sigma$-algebra on $\Omega$. From the three properties, it follows that $A_1,A_2,A_3, \ldots \in {\cal F} \implies \cap_i A_i
\in {\cal F}$. An element $A \in {\cal F}$ is called a measurable set or
an observable event. The pair $(\Omega,{\cal F})$ forms the measure space.

Given a sample space $\Omega$ and a $\sigma$-algebra ${\cal F}$ on
$\Omega$, a probability measure
is a function that assigns to each event $A \in {\cal F}$ a nonnegative
real number $\le 1$.
Specifically, a probability measure $P$ is a function $P: {\cal F} \to [0,1]$, such
that
\begin{enumerate}
\item $P(A)\ge 0~\forall~A \in {\cal F}$, 
\item $P(\emptyset)=0$ and $P(\Omega)=1$, and
\item For $A_1,A_2,\ldots \in {\cal F}$, if $A_i \cap A_j = \emptyset
~\forall~i \ne j$, then $P(\cup_i A_i)=\sum_i P(A_i)$.
\end{enumerate}
Altogether, the triple $(\Omega,{\cal F},P)$ forms a probability space.

Let us consider an example:
\begin{itemize}
\item Random trial: Tossing a coin two times in a row.
\item Sample space
$\Omega=\{HH,TT,HT,TH\}=\{\omega_1,\omega_2,\omega_3,\omega_4\}$. 
\item Event: could be ``getting
identical result in the two throws": $A=\{\omega_1,\omega_2\}$.  
\item For ${\cal F}$, there are several possibilities:
\begin{enumerate}
\item The smallest $\sigma$-algebra: ${\cal F}_{\rm
min}=\{\emptyset,\Omega\}$ (the events are ``getting nothing" and
``getting something"). ${\cal F}_{\rm min}$ contains what is known before the random trial is
performed. 
\item Another possibility: ${\cal
F}_1=\{\emptyset,\{\omega_1,\omega_2\},\{\omega_3,\omega_4\},\Omega\}$. ${\cal F}_1$ contains what can be observed after the first
trial: whether the random trial gives identical or non-identical
results for the two throws.
\item{Another one:
${\cal
F}_2=\{\emptyset,\{\omega_3\},\{\omega_4\},\{\omega_1,\omega_2\},\{\omega_3,\omega_4\},\{\omega_1,\omega_2,\omega_4\},\{\omega_1,\omega_2,\omega_3\},\Omega\}$. ${\cal F}_2$ contains information on what can be observed after
the second trial. Note that here, e.g., the element $\{\omega_3\}$ refers to the event "Observing $HT$", the element $\{\omega_1,\omega_2,\omega_4\}$ refers to observing the corresponding complement event, i.e., the event "Not observing $HT$."} Here, we
have assumed that ${\cal F}_s \subset {\cal F}_t$ for $s \le t$, since a
natural expectation is that with subsequent throws, we gain new information and do not discard the old ones. 
\item The largest $\sigma$-algebra: 
\begin{eqnarray}
{\cal F}_{\rm max}&=&\{\emptyset,
\{\omega_1\},\{\omega_2\},\{\omega_3\},\{\omega_4\},\{\omega_1,\omega_2\},\{\omega_1,\omega_3\}, \{\omega_1,\omega_4\},\{\omega_2,\omega_3\},\{\omega_2,\omega_4\},\nonumber \\&&\{\omega_3,\omega_4\},\{\omega_1,\omega_2,\omega_3\},\{\omega_1,\omega_2,\omega_4\},\{\omega_1,\omega_3,\omega_4\},\{\omega_2,\omega_3,\omega_4\},\Omega\}.
\end{eqnarray}
                ${\cal F}_{\rm max}$ is the largest possible collection
                of events that can be observed on tossing a coin two
                times in a row.
\end{enumerate}
\end{itemize}
Summarizing, we may think of a $\sigma$-algebra ${\cal F}$ as the amount of information contained in
$\Omega$ that can be observed: The smaller the ${\cal F}$, the lesser is
the amount of information we have of $\Omega$. 

From the above example, we see an illustration of the general result
that the smallest $\sigma$-algebra ${\cal F}_{\rm min}$ consists of the
empty set $\emptyset$ and the sample space $\Omega$, while the largest $\sigma$-algebra
${\cal F}_{\rm max}$ consists of all subsets of $\Omega$ including the empty set and the set 
$\Omega$ itself (${\cal F}_{\rm max}$ would conventionally be called the power set
of $\Omega$); note that the number of elements in ${\cal F}_{\rm max}$
is $2$ raised to the power ``number of elements in $\Omega$", hence, one
writes ${\cal
F}_{\rm max}= 2^\Omega$. A $\sigma$-algebra ${\cal G}$ is a
sub-$\sigma$-algebra of another $\sigma$-algebra ${\cal F}$ if ${\cal G}
\subset {\cal F}$. In the above example of tossing a coin two times in a
row, we have ${\cal F}_1 \subset {\cal F}_{\rm max}$.

%%%%%%%%%%%%%%%%%%%%%%%%%%%%%%%%%%%%%%%%%%%%%%%%%%%%%%%%%%%%%%%%%%%%%%%%%%%%%%%%%%%%%%%%%%%%%%%%%%%%%%%%%%%%%%%
%%%%%%%%%%%%%%%%%%%%%%%%%%%%%%%%%%%%%%%%%%%%%%%%%%%%%%%%%%%%%%%%%%%%%%%%%%%%%%%%%%%%%%%%%%%%%%%%%%%%%%%%%%%%%%%
\subsection{${\cal F}$-measurability, random variables and stochastic
processes}
We now discuss the concept of ${\cal F}$-measurability. Let
$(\Omega,{\cal F},P)$ be a probability space. A function $f: \Omega \to
{\bf R}$ is said to be ${\cal F}$-measurable if to any given interval
$(a,b) \in {\bf R}$ one can associate an event $A \in {\cal F}$.
Consider throwing a die. Here, we have $\Omega=\{1,2,3,4,5,6\}$. Next,
consider the function $X\equiv X(\omega)$ that equals $+1$ if
$\omega$ is either $1$ or $3$ or $5$ and equals $-1$ if $\omega$ is either $2$ or $4$ or $6$. Then, $X$ is
measurable with respect to the $\sigma$-algebra ${\cal
F}_1=\{\emptyset,\Omega,\{1,3,5\},\{2,4,6\}\}$ but is not measurable with
respect to the $\sigma$-algebra ${\cal
F}_2=\{\emptyset,\Omega,\{1,2,3\},\{4,5,6\}\}$ or with respect to the
$\sigma$-algebra ${\cal
F}_3=\{\emptyset,\Omega,\{1,2\},\{3,4\},\{5,6\}\}$.
A random variable $X$ on a probability space $(\Omega, {\cal F}, P )$ is
an ${\cal F}$-measurable function. A collection of random variables
$\{X_t\}$ defined on the probability space $(\Omega,{\cal
F},P)$ and parametrized by the variable $t$ is called a stochastic
process. Taking~$t$ to be time, the stochastic process may be denoted as~$\{X_t\}_{t \ge 0}$, or, when no confusion may arise, by simply~$X_t$ as in the Main Text.  

%%%%%%%%%%%%%%%%%%%%%%%%%%%%%%%%%%%%%%%%%%%%%%%%%%%%%%%%%%%%%%%%%%%%%%%%%%%%%%%%%%%%%%%%%%%%%%%%%%%%%%%%%%%%%%%
%%%%%%%%%%%%%%%%%%%%%%%%%%%%%%%%%%%%%%%%%%%%%%%%%%%%%%%%%%%%%%%%%%%%%%%%%%%%%%%%%%%%%%%%%%%%%%%%%%%%%%%%%%%%%%%
\subsection{Filtration and Adaptation} \label{subsec:FiltrationAndAdaptation}%Ken added a new label
Given a probability space $(\Omega,{\cal F},P)$, a filtration is a collection
$\{{\cal F}_t\}_{t\ge 0}$ of nested sub-$\sigma$-algebras of ${\cal F}$ such that ${\cal F}_s \subset {\cal F}_t$ for $0 \le s \le t$. The probability space with filtration $\{{\cal F}_t\}_{t \ge
0}$ is called the filtered probability space $(\Omega, {\cal F}, \{{\cal
F}_t\},P)$. A stochastic process $\{X_t\}_{t \ge 0}$ defined on $(\Omega, {\cal F}, P)$ whose
 values can be completely determined
from $\{{\cal F}_t\}_{t \ge 0}$ is said to be adapted to the
filtration $\{{\cal F}_t\}_{t\ge 0}$. In other words, the process
$\{X_t\}_{t \ge 0}$ is adapted to the filtration
$\{{\cal F}_t\}_{t \ge 0}$ if $X_t$ is ${\cal F}_t$-measurable for every $t \ge 0$. The natural filtration $\{{\cal F}_t^X\}_{t \ge 0}$ associated to a stochastic process $\{X_t\}_{t \ge 0}$ is a
filtration that records the past behaviour of the stochastic process at each time, i.e., the
information contained in the trajectories $\{X_t\}$ up to time $t$. Thus, all
information related to the process, and only that information, is
available in the natural filtration. Note
that $\{X_t\}_{t\ge 0}$ is obviously adapted to its natural filtration. {The reader is referred to Ref.~\cite{Knill-book} in which several illustrative examples of filtration and adaptation are discussed.}

\subsection{Conditional expectation}
\label{sec:condexpshamik}
Given a random variable $X$ on a probability space $(\Omega,{\cal F},P)$
and a sub-$\sigma$-algebra ${\cal F}' \subset {\cal F}$, one may
define a new random variable as the conditional expectation of $X$:
\begin{equation}
Z \equiv \langle X|{\cal F}'\rangle,
\end{equation}
namely, the expected value of $X$, given the information contained in
${\cal F}'$, i.e., the conditional expectation. The conditional expectation satisfies $\langle X|{\cal
F}\rangle=X$ if $X$ is ${\cal F}$-measurable, and the property of
iterated conditioning~\cite{Shreve:2000} given by $\langle \langle
X|{\cal F}\rangle\rangle=\langle X \rangle$. 

{In more practical terms, for two discrete random variables $X$ and $Y$, the conditional probability distribution of $X$ given $Y$ is the probability distribution of $X$ when $Y$ is known to  have a particular value. Thus, the conditional probability distribution of $X$ given $Y=y$ is given by the Bayes' theorem from probability theory:
\begin{align}
P(X=x|Y=y)=\frac{P_{X,Y}(x,y)}{P_Y(y)}\,. \label{eq:chap-rv-cond-prob}
\end{align}
Here, $P_{X,Y}(x,y)$ is the joint distribution of the random variables $X$ and $Y$, while $P_Y(y)$ is the probability distribution of the random variable $Y$ alone.
The definition in Eq.~(\ref{eq:chap-rv-cond-prob}) holds also for continuous random variables. Considering the case of continuous random variables, we then have the conditional expectation
\begin{equation}
    \langle X|Y\rangle=\int \mathrm{d}x~xP(X=x|Y=y),
\end{equation}
so that 
\begin{eqnarray}
    \langle \langle X|Y\rangle \rangle&&=\int \mathrm{d}y~P_Y(y)\int \mathrm{d}x~xP(X=x|Y=y)\nonumber \\
    &&=\int \mathrm{d}x~x\int \mathrm{d}y~P_{X,Y}(x,y)\nonumber \\
    &&=\int \mathrm{d}x~xP_X(x)\nonumber \\
    &&=\langle X\rangle. \label{eq:towerb5}
\end{eqnarray}
Here, in obtaining the second step, we have used Eq.~(\ref{eq:chap-rv-cond-prob}).}

%%%%%%%%%%%%%%%%%%%%%%%%%%%%%%%%%%%%%%%%%%%%%%%%%%%%%%%%%%%%%%%%%%%%%%%%%%%%%%%%%%%%%%%%%%%%%%%%%%%%%%%%%%%%%%%

%%%%%%%%%%%%%%%%% below by KS %%%%%%%%
\section{Tower Rule} \label{app:towerproperty}

In Sec.~\ref{sec:Examples} we introduced the {\bf conditional-expectation process} as a key example of martingale. For this route to the martingale the core is the {\bf tower property},
$$\langle \langle Z | X_{[0,n]}
 \rangle | X_{[0,m]} \rangle  =  \langle Z | X_{[0,m]} \rangle \qquad 0\le m\le n.$$
In the main text $Z=X_q$ ($q\ge n$) has been taken, but $Z$ can be any random variable whose statistical character is given once 
$X_{[0,m]}$ is known. See, for example, Ch.~\ref{sec:PQ}.

\subsection{Elementary tower rule}
We first {recall Eq.~\eqref{eq:towerb5} in Sec.~\ref{sec:condexpshamik}, that we write,}
\begin{equation} \label{eq:tower0}
\la \,\la{Z}|{X}\ra\,\ra= \la{Z}\ra.
\end{equation}
The fact that the conditional expectation $\la Z|X\ra$ is found at the inside of another expectation implies that $X$ is also a random variable.
In other words, the value of the condition $X=x$ occurs according to the probability of $X,$ that is $\rho_X(x).$ The outer expectation is taken according to such probability distribution. %which has the effect of lifting the condition "$|X$".  This idea is verified by the direct calculation:
%\begin{eqnarray} 
%\la \la Z|X\ra\ra
%&=& \sum_x \la Z|X\ra \rho_X(x)
%\cr &=& \sum\left(\sum_z z \rho_{Z|X}(z|x)\right) \rho_X(x)
%\cr &=& \sum_x\sum_z z \rho_{X,Z}(x,z)
%\cr &=& \sum_z z \rho_Z (z) =\la Z\ra,
%\end{eqnarray}
%where we have used $\rho_{Z|X}(z|x)\rho_X(x)=\rho_{X,Z}(x,z)$ and $\sum_x \rho_{X,Z}(x,z)=\rho_Z(z).$

\subsection{Higher order  Tower Rule}
We can immediately extend the above rule to a higher order {Tower Rule},
\begin{equation}
\la \left. \la{Z}\right|X_{[0,n]}\ra  |  X_{[0,m]}\ra 
= \la{Z}| X_{[0,m]}\ra, \qquad 0\le m\le n,
\label{eq:higherordertower}
\end{equation}
where we have used the abbreviation $X_{[0,m]}=X_0,X_1,\ldots, X_m,$ etc. for the sequence of random variables with consecutive discrete time. Below we shall also abuse this notation for $x_{[0,m]}=x_0,x_1,\ldots, x_m,$ etc. 
The demonstration is done in the same line as (\ref{eq:tower0}):
\begin{eqnarray}
\la \left. \la{Z}|X_{[0,n]}\ra  \right|X_{[0,m]} \ra 
&=&
\sum_{x_{[m+1,n]} 
}
\la Z|  X_{[0,m]}, x_{[m+1,n]}\ra \mP_{ X_{[m+1,n]}|X_{[0,m]}}(
x_{[m+1,n]}|X_{[0,m]})
\cr &=&
\sum_{x_{[m+1,n]} }
\left(\sum_z z \mP_{Z| X_{[0,n]}}(z| X_{[0,m]}, x_{[m+1,n]} )\right) 
\mP_{ X_{[m+1,n]}|X_{[0,m]}}(  x_{[m+1,n]} | X_{[0,m]})
\cr &=&
\sum_{x_{[m+1,n]} } \sum_z z \mP_{ X_{[m+1,n]} ,Z| X_{[0,m]}}( x_{[m+1,n]},z| X_{[0,m]})
\cr &=&
 \sum_z z \mP_{Z| X_{[0,m]} }(z| X_{[0,m]})
=
 \la {Z}|  X_{[0,m]}\ra,
\end{eqnarray}
where we have used
\begin{eqnarray}
&&
\mP_{Z| X_{[0,n]} 
}(z| X_{[0,m]} , x_{[m+1,n]} ) \mP_{ X_{[m+1,n]} | X_{[0,m]} }( x_{[m+1,n]} | X_{[0,m]} )
\cr
 &=&
\frac{\mP_{ X_{[0,n]} ,Z}( X_{[0,m]} ,  x_{[m+1,n]} ,z)}{\mP_{ X_{[0,n]} }(X_{[0,m]},x_{[m+1,n]})}\,
\frac{\mP_{ X_{[0,n]} }( X_{[0,m]} ,  x_{[m+1,n]} )}{\mP_{ X_{[0,m]} }( X_{[0,m]} )}
\cr &=&
\mP_{ X_{[m+1,n]} ,Z| X_{[0,m]} }(  x_{[m+1,n]} ,z| X_{[0,m]} )
\end{eqnarray}
and 
$\sum_{  x_{[m+1,n]} } \mP_{ X_{[0,n]} ,Z}( X_{[0,m]} ,  x_{[m+1,n]} ,z)
=\mP_{ X_{[0,m]} ,Z}( X_{[0,m]} ,z).$ 
{It is worth noting the similarity of this derivation to the one for the martingality of the ratio of path probability densities, see (\ref{eq:RnMartingale}).
In fact both have the common origin in the inclusively ordered series of conditional probabilities, or, the ordered structure of the filtration, see \ref{subsec:FiltrationAndAdaptation}.
In other words, behind these generic ways to make martingale 
    processes, i.e., by the path probability ratios and by the higher 
    order tower rule, there lies the tower rule for the conditional 
    probability {\it function}.
}

{We  illustrate intuitively the tower property or tower-rule of the conditional expectation,
(\ref{eq:higherordertower}).
We hope this illustration helps a little for demystifying the martingale.
Let $z$ be a random variable (RV), that is, a function of the elementary event which we regard to be a sample history.
In Fig.\ref{fig:tower-rule-view} we schematize by the 3D space the functional space
on the elementary events. For example the RV, $z,$ is a vector.
 When $X_k$'s represents the value of an observable $X$ at time $k,$ it is also a function of the history, therefore, of the elementary event. Then the expectation $\langle z|X_{[0,n]}\rangle$ is also the function of the elementary event but through $X_1, \ldots, X_n.$ However, being different from $z$ this expectation spans only a subspace of the whole
functional space, which we symbolize by the 2D bottom plane in Fig.\ref{fig:tower-rule-view}.
Then $\langle z|X_{[0,n]}\rangle$ is said to be the orthogonal projection of $z$ onto the sub-space associated with $X_{[0,n]}$. 
Then it is understandable that $\langle z|X_{[0,m]}\rangle$ with $m<n$ as function of elementary event   finds itself in the (further) sub-space associated with $(X_{[0,m]}),$ which we schematize by an 1D edge in Fig.\ref{fig:tower-rule-view}.%
\begin{figure}[!ht]
\includegraphics[width=\textwidth]{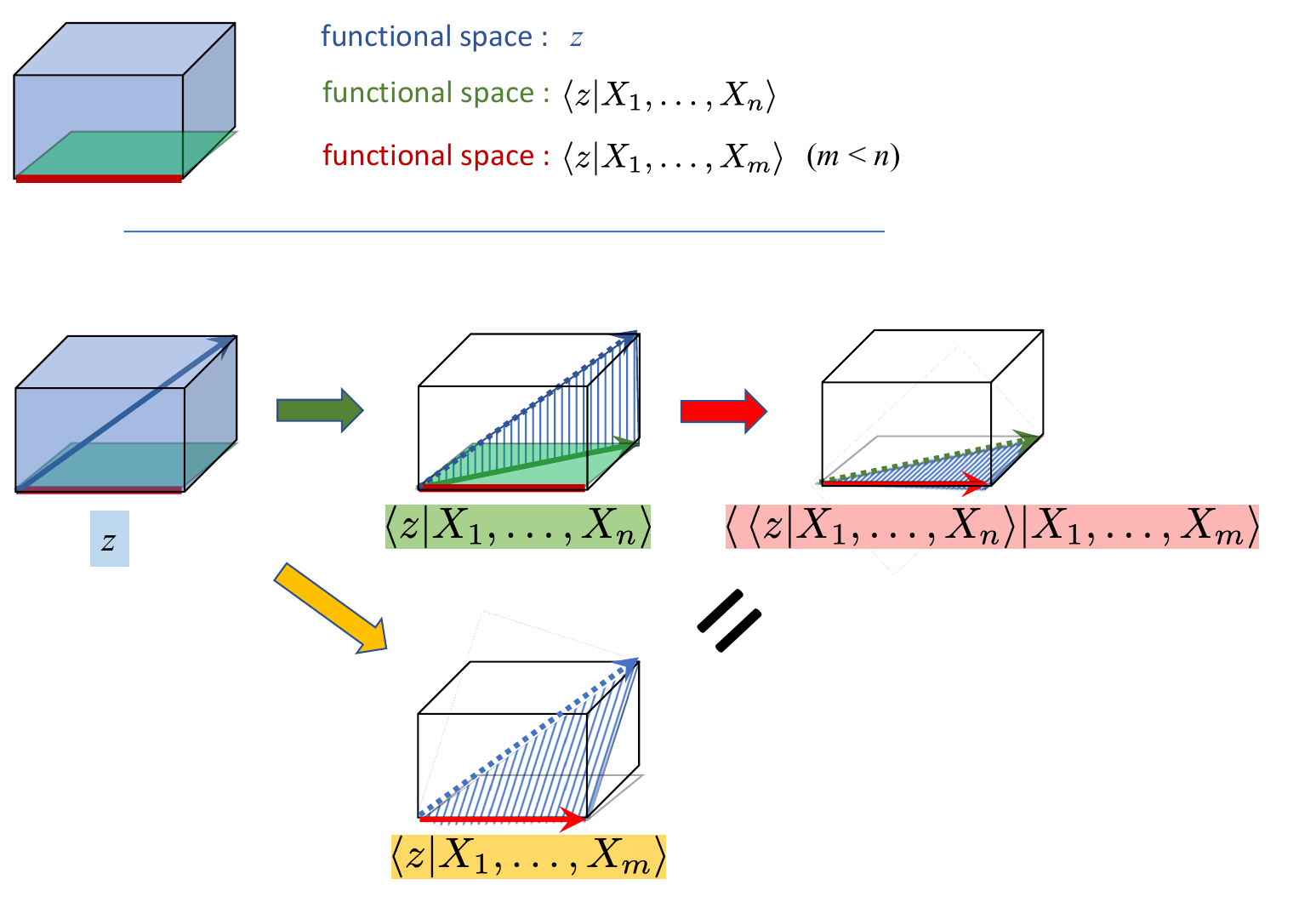}
\caption{
Schematic illustration of the tower property of the conditional expectation.
}
\label{fig:tower-rule-view}
\end{figure}
Physically speaking we interpret $\langle z|X_{[0,n]}\rangle$ as a coarse-grained version of $z$ as function of sample history such that its value is determined only through $n$ data, $X_{[0,n]}$.\footnote{We might appreciate this meaning from different facets:
(1) When a pair of histories, $\omega$ and $\omega',$ realizes the identical set of data $X_{[0,n]},$ therefore also identical $\langle z|X_{[0,n]}\rangle,$ it can occur that $z(\omega)\neq z(\omega').$ (2) When a history $\omega$ is given, $\langle z|X_{[0,n]}\rangle$ takes the average of $z$ over all the histories $\{\omega'\}$ which share the same tata $X_{[0,n]}.$ (3) $z$ {\it can} be any function of $n$ variables, $X_{[0,n]}.$ Nevertheless, each of $X_{[0,n]}$ are {\it prefixed functions} of the history, being independent of $z$. While $z$ is an object of observation, $X_{[0,n]}$ are the measureing apparatus for that. (4) Yet, $z$ is not restricted to a linear combination of $X_{[0,n]}$ and, therefore, the functional subspace spanned by $\langle z|X_{[0,n]}\rangle$ is not $n$-dimensional.} It is understandable that $\langle z|X_{[0,m]}\rangle$ with $m<n$ is even more coarse-grained than $\langle z|X_{[0,n]}\rangle.$
Now the tower property or tower-rule is nothing but an elementary extension of the {\it theorem of three perpendiculars} in Euclidean geometry, which claims that the orthogonal projection through an intermediate orthogonal projection is identical to the one obtained by the direct projection. In the present context the coarse-grained observation of $z$ through the data set, $X_{[0,m]},$ can be either obtained directly, $\langle z|X_{[0,m]}\rangle,$ or passing through an intermediate version, $\langle z|X_{[0,n]}\rangle$ with $t>s.$
In equation,
$$\langle z|X_{[0,m]}\rangle = \langle\,\langle z|X_{[0,n]}\rangle\,|\,X_{[0,m]}\rangle,$$
the martingale $M_m=\langle M_n |\,X_{[0,m]}\rangle$ emerges if we regard
$M_n=\langle z|X_{[0,n]}\rangle$ as a process associated to the process $X_{[0,n]}$.}
%%%%%%%%%%%%%%%%%%%%%%%%%%%%%%%%%%%%%%%%%%%%%%%%%%%%%%%%%%%%%%%%%%%%%%%%%%%%%%%%%%%%%%%%%%%%%%%%%%%%%%%%%%%%%%%
\section{Basics of stochastic  calculus}
\label{app:Ito-lemma}

Let $X_t$ be a stochastic process that obeys a stochastic differential equation.   What is the stochastic differential equation of the process $Y_t = g(X_t)$, where $g$ is a twice continuously, differentiable function?   The rules of stochastic calculus, which we review here, provide a solution to this problem.   

We first review the rules of stochastic calculus for the simplest case of a one-dimensional It\^{o} process in Sec.~\ref{app:ItoFormula}, and subsequently we consider the case of multi-dimensional It\^{o} processes and semimartingales, which is loosely defined as any stochastic process that is a good integrator for the It\^{o} integral, in Secs.~\ref{app:ItoFormula2} and \ref{app:ItoFormula3}.    Lastly, in Sec.~\ref{app:strat}, we review how to express an It\^{o} integral in terms of a Stratonovich integral.  We follow the references~\cite{oksendal2003stochastic, protter2005stochastic}.

%\subsection{Wiener process}
%\label{sec:Wiener}
%A Wiener process
%$B_t$ is a stochastic process in time that satisfies the following
%%properties:
%\begin{enumerate}
%\item $B_0=0$.
%\item For $t>s$, the possible increments $B_t-B_s$ are stationary
%(i.e., function of only the difference $t-s$) and are moreover
%distributed according to a %Gaussian distribution ${\cal N}(0,t-s)$ with zero mean and
%variance equal to $\sqrt{t-s}$. In particular, the increment ${\rm
%d}B_t$ in an infinitesimal time betwen $t$ and $t+{\rm d}t$ is ${\rm d}B_t \sim {\cal
%N}(0,\sqrt{{\rm d}t})$; Here, and in the following, the symbol $\sim$
%represents that the left and the right hand side are equal in distribution.
%\item The trajectories $B_t$ as a function of time are everywhere
%continuous with no jumps.
%\end{enumerate}

\subsection{It\^{o}'s formula}\label{app:ItoFormula}
Let $X_t\in \mathbb{R}$ be a stochastic process that solves a stochastic  differential equation of the form 
\begin{equation}
    \dot{X}_t = b_t(X_{[0,t]}) + \sigma_t(X_{[0,t]})\dot{B}_t, \label{eq:XDot}
\end{equation}
where $B_t$ is the one-dimensional Brownian motion, as defined in Sec.~\ref{sec:examplescont}, and 
where 
\begin{equation}
    \mathcal{P}\left(\int^t_0 ds \sigma^2_s(X_{[0,s]})<\infty, \quad \forall t\geq 0\right) = 1, \label{eq:sigma}
\end{equation}
and 
\begin{equation}
    \mathcal{P}\left(\int^t_0 ds |b_s(X_{[0,s]})|<\infty, \quad \forall t\geq 0\right) = 1. \label{eq:b}
\end{equation}

Let $g(t,x)$ be a twice continously differentiable function in $t\in \mathbb{R}^+$ and $x\in\mathbb{R}$, then the process 
\begin{equation}
    Y_t = g(t,X_t)
\end{equation}
solves the stochastic differential equation \cite{oksendal2003stochastic}
\begin{equation}
    \dot{Y}_t = \frac{\partial g}{\partial t}(t,X_t) +  \frac{\partial g}{\partial x}\dot{X}_t + D_t(X_{[0,t]}) \frac{\partial^2 g}{\left(\partial x\right)^2}(t,X_t)  \label{eq:itoFormula}
\end{equation}
where 
\begin{equation}
    D_t = \frac{\sigma^2_t}{2}.
\end{equation}

It\^{o}'s formula can be understood from a Taylor expansion of $g(t+dt,X_{t+dt})$, viz., \begin{equation}
g(t+dt,X_{t+dt})-g(t,X_t)=\frac{\partial g}{\partial t} dt+\frac{\partial
g}{\partial x}  dX_t+\frac{1}{2}\frac{\partial^2 g}{\partial t^2} (dt)^2+\frac{1}{2}\frac{\partial^2 g}{(\partial x)^2}(
d X_t)^2+\frac{1}{2}\frac{\partial^2 g}{\partial t \partial x}  dt 
dX_t+\ldots,
\label{appeq:dF}
\end{equation}
Neglecting contributions of the order $O((dt)^2)$, and using $dX_t = \dot{X}_tdt$ with $X_t$ solving Eq.~(\ref{eq:XDot}), we obtain 
\begin{equation}
g(t+dt,X_{t+dt})-g(t,X_t)=\frac{\partial g}{\partial t} dt + \frac{\partial
g}{\partial x}  dX_t + \frac{\partial^2 g}{(\partial x)^2} b_t \sigma_t dt dB_t +  \frac{1}{2}\frac{\partial^2 g}{(\partial x)^2}  \sigma^2_t (dB_t)^2 + O((dt)^2).  \label{eq:gPart}
\end{equation}
Using in Eq.~(\ref{eq:gPart}) that $dtdB_t \in o(dt)$ and $(dB_t)^2 = dt$, we readily obtain the It\^{o} formula Eq.~(\ref{eq:itoFormula}) after neglecting $o(dt)$ terms.   

To show that  $(dB_t)^2 = dt$, we  determine the probability  distribution   of $(dB_t)^2 $, see also Ref.~\cite{jacobs2010stochastic}.    The distribution of $dB_t$ is a normal distribution with zero mean and variance $dt$, i.e., 
\begin{equation}
\rho_{dB_t}(x) = \frac{1}{\sqrt{2\pi dt}} \exp\left(-\frac{x^2}{2dt}\right) .     
\end{equation}
Consequently, we obtain for the distribution of $(dB_t)^2$,  
\begin{equation}
   \rho_{(dB_t)^2}(y)=  \frac{1}{\sqrt{2\pi dt}} \int^{\infty}_{-\infty} dx  \exp\left(-\frac{x^2}{2dt}\right) \delta(y-x^2)= \frac{1}{\sqrt{2\pi dt}} \frac{1}{\sqrt{y}} \exp\left(-\frac{y}{2dt}\right).
\end{equation}
In the limit of $dt\rightarrow \infty$ it holds that $(dB_t)^2 =dt$.   Indeed, the average $\langle (dB_t)^2\rangle = dt$ and  $\langle ({\rm d}B_t)^4\rangle=3({\rm
d}t)^2$, so that its variance is negligible.

\subsection{Multidimensional It\^{o} formula}\label{app:ItoFormula2}
We review the generalisation of the It\^{o} formula Eq.~(\ref{eq:itoFormula})  to the multidimensional case.   
 
Consider now 
\begin{equation}
    \dot{X}_t = b_t(X_{[0,t]}) + \sigma_t(X_{[0,t]})\dot{B}_t \label{eq:multiX}
\end{equation} 
where  $b_{t}= (b_t^1, b_t^2, \dots , b_t^d)^\dagger\in \mathbb{R}^d$; where $\sigma_t\in \mathbb{R}^d\times \mathbb{R}^m$ is a matrix with entries $\sigma^{ij}_t$ where $i=1,2,\ldots, d$ and $j=1,2,\ldots, m$; and where $B_t = (B^1_t, B^2_t,\ldots, B^m_t)$ is a vector of $m$ independent   Brownian motions.   

We require that each of the individual $b^i_t$ satisfy Eq.~(\ref{eq:b}) and each of the individual $\sigma^{ij}_t$ satsify Eq.~(\ref{eq:sigma}).  

Let  
\begin{equation}
    Y_t = g(t,X_t),
\end{equation}
where $Y_t\in\mathbb{R}$ and where $g$ is twice, continuously differentiable.   It then holds that 
\begin{equation}
    \dot{Y}_t = \frac{\partial g}{\partial t}(t,X_t) +  \sum^d_{i=1}\frac{\partial g}{\partial x_i}\dot{X}^{i}_t + \frac{1}{2} \sum_{i,j} \frac{\partial^2 g}{\partial x_i \partial x_j} \dot{X}^i_t\dot{X}^j_t, \label{eq:itoFormula2}
\end{equation}
where $\dot{X}^{i}_t\dot{X}^j_t$ follows from applying the rules
\begin{equation}
     \dot{B}_i\dot{B}_j= \delta_{i,j} dt ,\quad \dot{B}_idt= 0, \quad (dt)^2 = 0
\end{equation}
to Eq.~(\ref{eq:multiX}).

\subsection{ Meyer-It\^{o} formula for semimartingales}\label{app:ItoFormula3}
We review the It\^{o} formula for so-called semimartingales $X$, which are    stochastic processes that form  good integrators of   the It\^{o} integral, see Ref.~\cite{protter2005stochastic}.  According to the Bichteler-Dellacherie Theorem a semimartingale can be decomposed into  a local martingale ($L$) and a finite variation process ($A$)~\cite{protter2005stochastic}, viz.,   
\begin{equation}
    X_t = A_t+ L_t
\end{equation}
where a finite variation process, i..e., if with probability one the paths of $A$ have a finite total variation ${\rm sup}_{P}\sum^{n-1}_{i=0}|A_{t_i}-A_{t_{i-1}}|$ on each compact interval $[0,t]$, where $P$ is a finite partition of $[0,t]$, as defined in Sec.~\ref{sec:examplescont}.  
Note that differently from It\^{o} processes, semi-martingales may contain jumps; examples of semi-martingales are It\^{o} processes, (inhomogeneous) Poisson processes, L\'{e}vy processes \cite{ken1999levy}, and  c\'{a}dl\'{a}g (right-continuous in $t$ and with existing left limits) martingales and submartingales.   The fractional Brownian motion, is an example of a stochastic process that is not a semi-martingale, and hence the It\'{o} integral does not exist for the latter~\cite{biagini2008stochastic}.  

Let us assume for simplicity that $X\in \mathbb{R}$, and let $g$ be again a twice, continuously differentiable function, and consider 
\begin{equation}
    Y_t = g(t,X_t). 
\end{equation}

It then holds that 
  \begin{eqnarray}
Y_t-Y_0 &=&\int^t_0 (\partial_t g)(X_s) ds +  \int^t_{0^+}\frac{dg}{dx}(X_{s^{-}})dX_s + \frac{1}{2} \int^t_0 \frac{d^2g}{(dx)^2}(X_{s^-})d[X,X]^{\rm c}_s
\nonumber\\ 
&& + \sum^{N_t}_{j=1} \left(g(X_{\mathcal{T}^{+}_j}) - g(X_{\mathcal{T}^{-}_j}) - \frac{dg}{dx}\left(X_{\mathcal{T}^{-}_j}\right) \left(X_{\mathcal{T}^+_j}-X_{\mathcal{T}^-_j}\right)\right), \label{eq:Itoformula3}
  \end{eqnarray}   
where $[X,X]^{\rm c}_s$ is the continuous part of the quadratic variation $[X,X]$, defined in Eq.~(\ref{eq:quadrVar}), $N_t$ denotes the number of jumps in the interval $[0,t]$, and $\mathcal{T}_j$ are the jump times (this is the same notation as used for Markov jump processes in Sec.~\ref{eq:jumpprocconta}).  

In the particular case of an It\^{o} process of the form Eq.~(\ref{eq:XDot}), $N_t=0$ and 
\begin{equation}
[X,X]^{c}_t = \int^{t}_0 D_s ds, 
\end{equation}
and we recover It\^{o}'s formula Eq.~(\ref{eq:itoFormula}). 

On the other hand, for a pure jump process, 
\begin{equation}
[X,X]^{c}_t = 0,       
\end{equation}
and 
\begin{equation}
    \int^t_{0^+}\frac{dg}{dx}(X_{s^{-}})dX_s = \sum^{N_t}_{j=1}\frac{dg}{dx}\left(X_{\mathcal{T}^{-}_j}\right) \left(X_{\mathcal{T}^+_j}-X_{\mathcal{T}^-_j}\right)
\end{equation}
so that 
 \begin{eqnarray}
Y_t-Y_0 = \int^t_0 (\partial_t g)(X_s) ds  + \sum^{N_t}_{j=1} \left(g(X_{\mathcal{T}^{+}_j}) - g(X_{\mathcal{T}^{-}_j}) \right). \label{eq:pureJumpIto}
  \end{eqnarray}

\subsection{Stratonovich integrals}\label{app:strat} 
{We revise here a generalization of Theorem~\ref{th:0} to semimartingales.}
Let $Y_t\in \mathbb{R}$ and $Z_t\in \mathbb{R}$ represent two semimartingales.  Then, the following conversion formula holds \cite{protter2005stochastic}
\begin{equation}
    \int^t_0 Z_{s^-} \circ dY_s =   \int^t_0 Z_{s^-}  dY_s  + \frac{1}{2}[Z,Y]^{\rm c}_t \label{eq:stratoToIto}
\end{equation}
where the right-hand side contains a Stratonovich integral and the left-hand side an It\^{o} integral, see Eqs.~(\ref{eq:Strat})  and (\ref{eq:Ito2}) for  definitions, and where $[X,Y]^{\rm c}_t$ is the continuous part of the covariation 
\begin{equation}
[Z_t,Y_t] \equiv \lim_{\|P\|\rightarrow0}\sum^{n-1}_{i=0} \left(Z_{t_i}-Z_{t_{i-1}}\right) \left(Y_{t_i}-Y_{t_{i-1}}\right). 
\end{equation}  

Let us consider the example for which $Y$ and $Z$ are It\^{o} processes of the form 
   \begin{equation}
\dot{Y} = b^{(Y)}_t(X_{[0,t]}) + \sigma^{(Y)}(X_{[0,t]})\dot{B}_t   
\end{equation}  
and 
  \begin{equation}
\dot{Z} = b^{(Z)}_t(X_{[0,t]}) + \sigma^{(Z)}(X_{[0,t]})\dot{B}_t,    
\end{equation}  
where $X$ solves Eq.~(\ref{eq:XDot}).   In this case, we obtain the quadratic covariation process by using the rules $(dB_t)^2=dt$, $dtdB_t=0$, and $(dt)^2$, yielding 
\begin{equation}
[Y_t,Z_t]^{\rm c} = [Y_t,Z_t] = \int^{t}_0 \sigma^{(Y)}(X_{[0,s]})\sigma^{(Z)}(X_{[0,s]})  ds . \label{eq:quadraticCov}
\end{equation}
On the other hand, if $Y$ and $Z$ are pure jump processes, then 
\begin{equation}
[Y_t,Z_t]^{\rm c}  =0
\end{equation}
and the Stratonovich integral equals the It\^{o} integral.

%   For example, in the case where    $D_t = f_t(B_t)$, 
%\begin{equation}
% [D_t,B_t]  = \int^t_{0}f'_s(B_s){\rm d}s.
%\end{equation}
%Just as for the quadratic variation, the covariation can be determined with the simple rules $ {\rm d}t {\rm d}B(t) = 0$ and  $({\rm d}B(t))^2 = {\rm d}t$. 

\section{Stochastic exponential for a simple random walk}
We show that the processes Eqs.~(\ref{eq:stochExpExample1}) and (\ref{eq:BExp}) are martingales.  

\subsection{Discrete time}
To show that the process 
$\mathcal{E}_n(z)$  given by Eq.~(\ref{eq:stochExpExample1}) is a martingale, we use  that $\mathcal{E}_n(z)$  is a ratio of two probability densities $R_n$ of the form (\ref{eq:RForm}).

The probability density of a trajectory $X_{[0,n]}$ is given by 
\begin{eqnarray}
\mP(X_{[0,n]}) = \prod^n_{i=1} \left((1-q) \delta_{X_i-X_{i-1},-1}   + q\delta_{X_i-X_{i-1},1}\right).  
\end{eqnarray} 
Analogously, we can define the density   
\begin{eqnarray}
\mQ(X_{[0,n]})  = \prod^n_{i=1} \left((1-\tilde{q}) \delta_{X_i-X_{i-1},-1}   + \tilde{q}\delta_{X_i-X_{i-1},1}\right).  
\end{eqnarray}   
Hence, the ratio  of $\mP(X_{[0,n]})$ and $\mQ(X_{[0,n]})$ is given by 
\begin{eqnarray}
R_n= \frac{\mQ(X_{[0,n]})}{\mP(X_{[0,n]})} = \exp(y(q,\tilde{q}) X_n  + z(q,\tilde{q}) n  ), \label{eq:discreExp}
\end{eqnarray}
with 
\begin{eqnarray}
y(q,\tilde{q})\equiv  \frac{1}{2}\ln\left(\frac{(1-q)\tilde{q}}{q(1-\tilde{q})}\right), \quad z(q,\tilde{q}) \equiv \frac{1}{2} \ln  \left(\frac{\tilde{q}(1-\tilde{q})}{q(1-q)}\right).
\end{eqnarray}   
Solving the first equation towards $\tilde{q}$ we obtain 
\begin{eqnarray}
\tilde{q} = \frac{q\exp(2y)}{1-q + q\exp(2y)} ,
\end{eqnarray} 
and thus 
\begin{eqnarray}
 z(q,\tilde{q}(y))  =   \frac{1}{2} \ln  \left(\frac{\exp(2y)}{[(1-q) + q\exp(2y)]^2}\right) = y -  \ln[(1-q) + \exp(2y) q].
\end{eqnarray}   
Substituting $z$ in (\ref{eq:discreExp}) and writing everything as a function of  $y$ we obtain (\ref{eq:stochExpExample1}).

\subsection{Continuous time}
Using that \begin{eqnarray}
\langle   \exp(z(B_t-B_s)) | B_{[0,s]}  \rangle  = \langle   \exp(z(B_t-B_s))  \rangle  = \exp\left(\frac{z^2}{2}(t-s) \right),
\end{eqnarray}
we obtain 
\begin{eqnarray}
\left\langle \exp\left( zB_t - \frac{z^2t}{2}  \right) \Big| B_{[0,s]} \right\rangle =  \exp\left( zB_s - \frac{z^2s}{2}  \right). 
\end{eqnarray}
\chapter{Appendix to Chapter~\ref{sec:martST1}}\label{appendix:5New} 

\section{Derivation of Eq.~(\ref{forGirs-1-1-1aa})}\label{app:derStot}

%of the stochastic differential equation for $\dot{S}^{\rm tot}_t$ given by  Eq.~(\ref{forGirs-1-1-1aa})}\label{app:derStot}

The stochastic differential equation for $\dot{S}^{\rm tot}_t$, given by Eq.~(\ref{eq:StotStrato}), contains the Stratonovich integral $S_t$ that solves
\begin{equation}
\dot{S}_t = \frac{J_{t,\rho}(X_t)}{\mu T \rho_{t}\left(X_t\right)}\circ \dot{X}_t,  \label{eq:stratS}
\end{equation}
where $\dot{X}_t$ solves Eq.~(\ref{eq:lang2}), and thus \begin{equation}
\dot{S}_t =\frac{F_t(X_t)J_{t,\rho}(X_t)}{T \rho_{t}\left(X_t\right)}  +  \left(\sqrt{\frac{2}{\mu T}}\frac{J_{t,\rho}(X_t)}{ \rho_{t}\left(X_t\right)}\right)\circ \dot{B}_t,  \label{eq:stratS2}
\end{equation}
where $F_t$ is the total force, as defined in Eq.~(\ref{eq:ftot}).

Using Eq.~(\ref{eq:stratoToIto}), Eq.~(\ref{eq:stratS2}) can be expressed as an It\^{o} stochastic differential equation, 
\begin{equation}
\dot{S}_t =\frac{F_t(X_t)J_{t,\rho}(X_t)}{T \rho_{t}\left(X_t\right)}  +  \left(\sqrt{\frac{2}{\mu T}}\frac{J_{t,\rho}(X_t)}{ \rho_{t}\left(X_t\right)}\right) \dot{B}_t + \frac{1}{2}\frac{d}{dt}[Z, B]_{t}, \label{eq:strattoIto2}
\end{equation}
where 
\begin{equation}
    Z_t =\sqrt{\frac{2}{\mu T}} \frac{J_{t,\rho}(X_t)}{ \rho_{t}\left(X_t\right)} 
\end{equation}
and we have used that for a continuous process $[Z,B]^{\rm c}_t =[Z,B]_t$.     The quadratic covariation is given by  Eq.~(\ref{eq:quadraticCov}), where $\sigma^{(B)}_t = 1$ and 
    $\sigma^{(Z)}_t$ is the coefficient in front of the noise term of $\dot{Z}_t$.    We obtain $\dot{Z}_t$ by applying It\^{o}'s formula  Eq.~(\ref{eq:itoFormula}) to $Z$, yielding
    \begin{equation}
        \dot{Z}_t =  \sqrt{\frac{2}{\mu T}} \frac{\left(\partial_xJ_{t,\rho}\right)(X_t)}{ \rho_{t}\left(X_t\right)}\dot{X} -\sqrt{\frac{2}{\mu T}} \frac{\left(\partial_x \rho_t\right)(X_t)J_{t,\rho}(X_t)}{ \rho^2_{t}\left(X_t\right)}\dot{X} + \ldots \label{eq:ZtApp}
    \end{equation}
    where we omitted the $\partial_tg$  and $D\partial^2_xg$ terms in It\^{o}'s formula as they do not contain a noise term and hence do not contribute to $\sigma^{(Z)}_t$.   Using Eq.~(\ref{eq:lang2}) in Eq.~(\ref{eq:ZtApp})     we find 
    \begin{equation}
        \dot{Z}_t =2\left(\frac{\left(\partial_xJ_{t,\rho}\right)(X_t)}{\rho_{t}\left(X_t\right)}  - \frac{\left(\partial_x \rho_t\right)(X_t)J_{t,\rho}(X_t)}{ \rho^2_{t}\left(X_t\right)}\right)\dot{B}_t +  \ldots 
    \end{equation}
   where we identify 
   \begin{equation}
       \sigma^{(Z)}_t = 2\left(\frac{\left(\partial_xJ_{t,\rho}\right)(X_t)}{ \rho_{t}\left(X_t\right)}  - \frac{\left(\partial_x \rho_t\right)(X_t)J_{t,\rho}(X_t)}{ \rho^2_{t}\left(X_t\right)}\right).  
   \end{equation}
   Further using Eq.~(\ref{eq:fpe1d}) and Eq.~(\ref{eq:JFP}), this yields 
     \begin{equation}
       \sigma^{(Z)}_t = 2\left(-\frac{\left(\partial_t\rho_t\right)(X_t)}{ \rho_{t}\left(X_t\right)}  + \frac{J^2_{t,\rho}(X_t)}{\mu T \rho^2_{t}\left(X_t\right)} - \frac{F_t(X_t)J_{t,\rho}(X_t)}{ T \rho_{t}\left(X_t\right)} \right) .  
   \end{equation}
Hence, 
\begin{equation}
 \frac{1}{2} \frac{d}{dt} [Z, B]_{t} =-\frac{\left(\partial_t\rho_t\right)(X_t)}{ \rho_{t}\left(X_t\right)}  + \frac{J^2_{t,\rho}(X_t)}{\mu T \rho^2_{t}\left(X_t\right)} - \frac{F_t(X_t)J_{t,\rho}(X_t)}{ T \rho_{t}\left(X_t\right)} 
\end{equation}
and substituting in Eq.~(\ref{eq:strattoIto2}) gives 
\begin{equation}
\dot{S}_t =  \frac{J_{t,\rho}(X_t)}{\mu T \rho_{t}\left(X_t\right)}\circ \dot{X}_t =   -\frac{\left(\partial_t\rho_t\right)(X_t)}{ \rho_{t}\left(X_t\right)}    + \frac{J^2_{t,\rho}(X_t)}{\mu T \rho^2_{t}\left(X_t\right)} + \left(\sqrt{\frac{2}{\mu T}}\frac{J_{t,\rho}(X_t)}{\rho_{t}(X_t)}\right) \dot{B}_t .  \label{eq:convStratI}
\end{equation} 
Using the latter Eq.~(\ref{eq:convStratI}) in Eq.~(\ref{eq:StotStrato}), we readily obtain  Eq.~(\ref{forGirs-1-1-1aa}).  

\section{Derivation of the inequality in  Eq.~(\ref{eq:SDotResult}) }\label{appendix:5NewSub}
We derive the inequality in Eq.~(\ref{eq:SDotResult}), namely, we show that 
\begin{equation}
 \sum_{(x,y)\in\mathcal{X}^2}\rho_t(x)\omega_{t}(x,y) \ln \left( \frac{\rho_t(x)\omega_{t}(x,y)}{\rho_t(y)\omega_{t}(y,x)}\right)\geq0, \label{eq:SDotResultxx}
\end{equation}

The inequality follows from the nonnegativity of the Kullback-Leibler divergence 
\begin{equation}
 D(p||q) =    \sum_{x\in \mathcal{S}} p(x) \ln \frac{p(x)}{q(x)} \geq 0 
\end{equation}
where $q(x),p(x)\geq0$,  
\begin{equation}
  \sum_{x\in \mathcal{X}} p(x) =   \sum_{x\in \mathcal{X}} q(x) = 1,
\end{equation}
and $\mathcal{X}$ is a finite set, 
see for example Ref.~\cite{cover1999elements}.       

Defining 
\begin{equation}
 \mathcal{N} =    \sum_{(x',y')\in\mathcal{X}^2}\rho_{t}(x')\omega_t(x',y')>0 ,
\end{equation} 
The left-hand side of Eq.~(\ref{eq:SDotResultxx}) can be written as 
\begin{equation}
 \sum_{(x,y)\in\mathcal{X}^2}\rho_t(x)\omega_{t}(x,y) \ln \left( \frac{\rho_t(x)\omega_{t}(x,y)}{\rho_t(y)\omega_{t}(y,x)}\right) = \mathcal{N} \sum_{(x,y)\in\mathcal{X}^2}\frac{\rho_{\rm st}(x)\omega(x,y)}{\mathcal{N}} \ln \left( \frac{\rho_{t}(x)\omega_t(x,y)/\mathcal{N}}{\rho_{t}(y)\omega_t(y,x)/\mathcal{N}}  \right).  \label{eq:SDotResultxx}
\end{equation}
Identifying the two  functions 
\begin{equation}
    p(x,y) = \frac{\rho_{t}(x)\omega_t(x,y)}{\mathcal{N}}, \quad {\rm and} \quad   q(x,y) = \frac{\rho_{t}(y)\omega_t(y,x)}{\mathcal{N}},
\end{equation}
Eq.~(\ref{eq:SDotResultxx}) reads 
\begin{equation}
 \sum_{(x,y)\in\mathcal{X}^2}\rho_t(x)\omega_{t}(x,y) \ln \left( \frac{\rho_t(x)\omega_{t}(x,y)}{\rho_t(y)\omega_{t}(y,x)}\right) = \mathcal{N}D(p||q) \geq 0. 
 \end{equation}

\section{Time independence of the time-reversed Lagrangian in the case of Markov jump processes}\label{app:indepMarkov}
We complete the derivation of the martingale property of $\exp(-S^{\rm tot}_t)$ in Sec.~\ref{sec:radNikoMarkovJump} by showing that  $\mathcal{P}[\Theta_t(X_{[0,t]})]$ is not explicitly dependent on time, and hence we can write $\mathcal{P}[\Theta_t(X_{[0,t]})] = \mathcal{Q}[X_{[0,t]}]$ for a certain measure $\mathcal{Q}$.   To this purpose, we show that the Lagrangian of $\mathcal{P}[\Theta_t(X_{[0,t]})]$ contains no explicit time dependency on $t$ ---see Eq.~ \eqref{eq:LagrangianFormula} for the definition of a Lagrangian.  

Indeed, Eq.~\eqref{eq:RFormContTR} can be written as  
\begin{align*}
\hspace{-0.5cm}\mathcal{P}\left(\Theta_{t}\left(X_{\left[0,t\right]}\right)\right) & =\frac{1}{\mathcal{N}}\rho_{st}\left(X_{t}\right)\exp\left(\sum_{i=1}^{N_{t}}\ln\left(w\left(X_{\mathcal{T}_{i}^{+}},X_{\mathcal{T}_{i}^{-}}\right)\right)-\int_{0}^{t}\lambda(X_{s})ds\right)\\
 & =\frac{1}{\mathcal{N}}\rho_{st}\left(X_{0}\right)\exp\left(\sum_{i=1}^{N_{t}}\ln\left(\frac{\rho_{st}\left(X_{\mathcal{T}_{i}^{+}}\right)w\left(X_{\mathcal{T}_{i}^{+}},X_{\mathcal{T}_{i}^{-}}\right)}{\rho_{st}\left(X_{\mathcal{T}_{i}^{-}}\right)}\right)-\int_{0}^{t}\lambda(X_{s})ds\right)\\
 & =\frac{1}{\mathcal{N}}\rho_{st}\left(X_{0}\right)\exp\left(-\int_{0}^{t}\left\{ -\sum_{x,y}\ln\left(\frac{\rho_{st}\left(y\right)w\left(y,x\right)}{\rho_{st}\left(x\right)}\right)\dot{N}_{s}(x,y)+\lambda(X_{s})\right\} ds\right),
\end{align*}
where in the last line we have used $\dot{N}_s(x,y)$ to denote the rate of change of the jump process $N_s(x,y)$, as defined in Eq.~(\ref{eq:NtDef}).

 Then, the Lagrangian transforms under time reversal as 
\[
\left(\Theta_{t}\mathcal{L}\right)\left[X_{s},\dot{N}_{s}\right]=-\sum_{x,y}\ln\left(\frac{\rho_{st}\left(y\right)w\left(y,x\right)}{\rho_{st}\left(x\right)}\right)\dot{N}_{s}(x,y)+\lambda(X_{s}).
\]
The absence of an explicit $t-$dependence in the right-hand side of the last relation implies that the measure $\mathcal{P}\circ \theta_{t}$  has no explicit $t$-dependency.

\section{Exponentiated negative entropy production as an It\^{o} integral for stationary Markov jump processes}\label{sec:itoMarkovJumpApp}
We derive the stochastic differential Eq.~(\ref{eq:stochExponJump}) presented in Sec.~\ref{sec:ItoIntegralAppraochSec} that describes the evolution in time of  $\exp(-S^{\rm tot}_t)$,  with $M_t$ given by Eq.~(\ref{eq:MtMarkovJumpStoch}).

Since $X$ is a jump process, the rules for stochastic calculus as discussed in Appendix~\ref{app:Ito-lemma} apply, in particular Eq.~(\ref{eq:pureJumpIto}) implies in a differential form that    
\begin{eqnarray}
\frac{d\emStot}{dt} &=&\sum_{x\in \mathcal{X}\setminus\left\{  X_{t^-}\right\}} \left(\emStot-\exp(-S^{\rm tot}_{t-})\right) \dot{N}_t(X_{t^-},x)
\nonumber\\  &=& \exp(-S^{\rm tot}_{t-})\sum_{x\in \mathcal{X}\setminus\left\{  X_{t^-}\right\}} \left(\frac{\rho_{\rm st}(x) \omega(x,X_{t^-})}{\rho_{\rm st}(X_{t^-}) \omega(X_{t^-},x)}-1\right)  \dot{N}_t(X_{t^-}
,x).  \nonumber\\ \label{eq:dExpStotDiff}
\end{eqnarray} 
Subsequently, we write the stationarity condition Eq.~(\ref{eq:statcond}) as 
\begin{eqnarray} 
\sum_{x\in \mathcal{X};x\neq y}\left(\rho_{\rm st}(x)\omega(x,y)-\rho_{\rm st}(y)\omega(y,x)\right) = \rho_{\rm st}(y)\sum_{x\in \mathcal{X};x\neq y}\left(\frac{\rho_{\rm st}(x)\omega(x,y)}{\rho_{\rm st}(y)\omega(y,x)}-1\right)\omega(y,x) = 0 .\nonumber\\
\end{eqnarray} 
Using the latter equation for $y=X_{t^{-}}$, and substracting it from Eq.~(\ref{eq:dExpStotDiff}), we get 
\begin{eqnarray}
\frac{d\emStot}{dt}  &=&\exp(-S^{\rm tot}_{t-}) \sum_{x\in \mathcal{X}\setminus\left\{  X_{t^-}\right\}}  \left(\frac{\rho_{\rm st}(x) \omega(x,X_{t^-})}{\rho_{\rm st}(X_{t^-}) \omega(X_{t^-},x)}-1\right) \left(\dot{N}_t(X_{t^-},x) -\omega(X_{t^-},x)  \right) , \nonumber\\
\end{eqnarray} 
  which can also be written as 
\begin{eqnarray}
\lefteqn{\frac{d\emStot}{dt}  }&& \nonumber\\ 
&=&\exp(-S^{\rm tot}_{t-}) \sum_{x\in \mathcal{X}\setminus\left\{  X_{t^-}\right\}}  \left(\frac{\rho_{\rm st}(x) \omega(x,X_{t^-})}{\rho_{\rm st}(X_{t^-}) \omega(X_{t^-},x)}-1\right) \left(\dot{N}_t(X_{t^-},x) -\dot{\tau}(X_{t^-})\omega(X_{t^-},x)  \right) . \nonumber\\
\end{eqnarray}
Integrating over $t$, we obtain Eqs.~(\ref{eq:stochExponJump}) and (\ref{eq:MtMarkovJumpStoch}) in Sec.~\ref{sec:ItoIntegralAppraochSec}, which we were meant to show. 

 \section{Novikov's condition for Markov jump processes}\label{app:novikov}

We derive  Novikov's condition Eq.~(\ref{eq:novikovMJ}) in Sec.~\ref{sec:novikovMJ}  for the exponentiated negative entropy production of a Markov jump process.  

We use Novikov's condition  for Markov jump processes, which we repeat here for convenience, 
 \begin{eqnarray}
 \Bigg\langle  \exp\left(\frac{1}{2} \langle M^c,M^c\rangle_t + \langle M^d,M^d\rangle_t\right)\Bigg\rangle<\infty, \quad \forall t\geq 0 ,  \label{eq:novikov2repeat}
 \end{eqnarray}
 and where for our purpose here $M$ is the martingale of Eq.~(\ref{eq:MtMarkovJumpStoch}), i.e., 
 \begin{equation}
M_t =    \sum_{x,y\in \mathcal{X}^2} \left(\frac{\rho_{\rm st}(y) \omega(y,x)}{\rho_{\rm st}(x) \omega(x,y)}-1\right) (N_{t}(x,y) -  \tau_t(x) \omega(x,y)) . \label{eq:MtMarkovJumpStochRpeat}
\end{equation}

The predictable quadratic variation $\langle M^c,M^c\rangle_t$ of the continuous part $M^c$ of $M$ equals zero, as also $[M^c,M^c] = 0$.    Let us therefore determine  the predictable quadratic variation $\langle M^d,M^d\rangle_t$ of the discontinuous component 
\begin{equation} 
M^d =  \sum_{x,y\in \mathcal{X}^2} \left(\frac{\rho_{\rm st}(y) \omega(y,x)}{\rho_{\rm st}(x) \omega(x,y)}-1\right) N_{t}(x,y). 
  \end{equation}   
  The quadratic variation of $M^d$ is the process
  \begin{equation}
      [M^d,M^d]  =\sum_{x,y\in\mathcal{X}^2} \left(\frac{\rho_{\rm st}(y) \omega(y,x)}{\rho_{\rm st}(x) \omega(x,y)}-1\right)^2N_t(x,y)
  \end{equation}
  and its compensator 
    \begin{equation}
      \langle M^d,M^d\rangle = \sum_{x,y\in\mathcal{X}^2} \left(\frac{\rho_{\rm st}(y) \omega(y,x)}{\rho_{\rm st}(x) \omega(x,y)}-1\right)^2 \omega(x,y)\tau_t(x). \label{eq:compMd}
  \end{equation}
  Substituting Eq.~(\ref{eq:compMd}) in Eq.~(\ref{eq:novikov2repeat}), we get Eq.~(\ref{eq:novikovMJ}) that we were meant to show.

%%%%%%%
%\include{chap_app6}

\chapter[Appendix to Chapter 6]{Appendix to Chapter~\ref{ch:thermo2}}\label{appendix:6}

We give here alternative expression to the excess stochastic entropy production \eqref{eq:excessentropy} and explicit expression of the housekeeping stochastic entropy production  \eqref{eq:Sehk}, by restricting the class of Markov process.  
\begin{enumerate}

\item For a pure jump process given by transition rates $\omega_t (x, y)$, we obtain from the generic formulae \eqref{eq:excessentropy} the alternative  expression for the excess stochastic entropy production~\cite{harris2007fluctuation,esposito2007fluctuation} :  
\begin{equation}
S^{ex}_{t}  =
-\ln\left(\rho_{t}(X_t)\right)+\ln\left(\rho_0 (X_0)\right)+
\sum_{j=1}^{N_t} \ln \left[\frac{\pi_{\T_{j}}(X_{\T_{j}^+})}{\pi_{\T_{j}}(X_{\T_{j}^-})}  \right]
   %\epsilon_{\T_{j}}(X_{\T^-_{j}},X_{\T^+_{j+1}})
%\displaystyle\sum_{s=0  | \{X_{s^-}\neq X_{s^+}\}}^{t} \ln \left[\frac{\pi_s(X_{s^+})}{\pi_s(X_{s^-})}  \right]
\quad.
\label{SenexJP}
\end{equation}
Moreover, the process with path probability $\mathcal{P}^{\rm ex}$ in the relation \eqref{eq:excessentropy}, or equivalently with Markovian generator given by the dual generator  $\mathcal{L}^{\rm ex}$ \eqref{AFS}, is the pure jump process given by the transition rates  
\[
\omega_{s}^{\rm ex}(x,y) \equiv \frac{\omega_{t-s}(y,x)\pi_{t-s}(y)}{\pi_{t-s}(x)}.
\]

In another side, we have the alternative  expression for the housekeeping stochastic entropy production  \cite{harris2007fluctuation,esposito2007fluctuation} :
\begin{equation*}
S^{\rm hk}_t  =
%\displaystyle\sum_{s=0  | \{X_{s^-}\neq X_{s^+}\}}^{t}
\sum_{j=1}^{N_t}
\ln\left[\frac{\pi_{\T_{j}}(X_{\T_{j}^{-}})\omega_{\T_{j}}(X_{\T_{j}^{-}},X_{\T_{j}^{+}})}{\pi_{\T_{j}}(X_{\T_{j}^{+}}) \omega_{\T_{j}}(X_{s^{+}},X_{s^{-}})}\right]\quad.
\end{equation*}
These last expressions can be obtained from the generic formula~\eqref{eq:Sehk} or, more simply, directly from the explicit expressions given before  \eqref{fatigue}, \eqref{SenexJP} and the Oono-Paniconi decomposition  \eqref{OOP}. Similarly of the total $\Sigma$-entropic functional, this quantity is finite only if for all $x,y$, $\omega_t (x, y)>0$  implies $\omega_t (y, x)>0$, condition which is sometimes call microreversibilty.
Moreover,  the process with path probability $\mathcal{P}^{\rm hk}$ in the relation \eqref{eq:Sehk}, or equivalently with Markovian generator given by the dual generator  $\mathcal{L}^{\rm hk}$ \eqref{DG}, is the pure jump process given by the transition rates  
\[
\omega_{s}^{\rm hk}(x,y) \equiv \frac{\omega_{s}(y,x)\pi_{s}(y)}{\pi_{s}(x)}. 
\]

\item For multidimensional Langevin equation \eqref{eq:LE} (even without Einstein relation \eqref{ER}), we obtain from \eqref{eq:excessentropy} the alternative  expression  for the excess stochastic entropy production~\cite{hatano2001steady,chetrite2008fluctuation,chernyak2006path} : 

\begin{equation}
S^{\rm ex}_t =-\ln\left(\rho_{t}(X_t)\right)+\ln\left(\rho_0 (X_0)\right)+\int_0^t\,[\nabla \ln\left(\pi_s\right)]  (X_s)\circ \dot{X}_{s}ds\quad.
\label{eq:exdiff}
\end{equation}

Moreover, the process with path probability $\mathcal{P}^{\rm ex}$ in the relation \eqref{eq:excessentropy}, or equivalently with Markovian generator given by the dual generator  $\mathcal{L}^{\rm ex}$ \eqref{AFS}, is the Langevin equation in the It\^{o} convention
\begin{equation}
\frac{dX^{\rm ex}_{s}}{ds}= (-\bmu_{t-s}F_{t-s}+2D_{t-s}\nabla\left(\ln\pi_{t-s}\right))(X^{ex}_{s})+\left(\nabla\, \mathbf{D}_{t-s}\right)(X^{ex}_{s})+\sqrt{2\mathbf{D}_{t-s}(X^{ex}_{s})}\dot{B}_{s}\;.
\end{equation}

In another side,  $S_{t}^{\rm hk} $ exists only if the diffusion matrix $D_t$ is invertible and is given by the explicit expression ~\cite{hatano2001steady,chetrite2008fluctuation}: 
\begin{equation*}
S^{\rm hk}_t =\int_{0}^{t}  \left(\left(\bmu_s F_{s}\right) \textbf{D}_{s}^{-1} - \nabla\ln \pi_s\right)(X_{s})\circ \dot{X}_{s}ds\quad.
\end{equation*}
Again, these  expressions can be obtained from the generic formula~\eqref{eq:Sehk} or, more simply, directly from the explicit expressions given before, \eqref{eq:exdiff} and the Oono-Paniconi decomposition  \eqref{OOP}. 
Moreover,  the process with path probability $\mathcal{P}^{\rm hk}$ in the relation \eqref{eq:Sehk}, or equivalently with Markovian generator given by the dual generator  $\mathcal{L}^{\rm hk}$ \eqref{DG}, is the Langevin equation in the It\^{o} convention
\begin{equation}
\frac{dX^{\rm hk}_{s}}{ds}= (-\bmu_{s}F_{s}+2D_{s}\nabla\left(\ln\pi_{s}\right))(X^{hk}_{s})+\left(\nabla\, \mathbf{D}_{s}\right)(X^{hk}_{s})+\sqrt{2\mathbf{D}_{s}(X^{hk}_{s})} \dot{B}_{s}\;.
\end{equation}

\end{enumerate}

%%%%%%%
%\include{chap_app5}

\chapter[Appendix to Chapter~7]{Appendix to Chapter~\ref{sec:martST2}}\label{appendix:5}

\section{Modified fluctuation relation and second law when  $\emStot$ is a  strict local martingale }
As mentioned in Sec.~\ref{sec:martingaleExpS}, we cannot exclude the possibility that there exist  processes $X_{[0,t]}$ for which  $\emStot$ is a  strict local martingale, i.e., a local martingale that is not a martingales.     

Therefore, we analyse here the implication of local martingality on the properties of $S^{\rm tot}_t$.     
\begin{leftbar}
Since $\emStot$ is bounded from below, it is a supermartingale (see in Ref.~\cite{protter2005stochastic}), yielding the following {\bf modified martingale fluctuation relation}
\begin{equation}
    \langle \emStot | X_{[0,s]} \rangle \leq \exp(-S^{\rm tot}_s)
\end{equation}
for all $t>s>0$.   
\end{leftbar}
 Note that for a supermartingale with  $S^{\rm tot}_0 = 0$, 
\begin{equation}
    \langle \emStot\rangle \leq 1  \label{eq:modI}
\end{equation}
and the equality is attained when $\emStot$ is  a martingale. 

Using Jensen's inequality $\exp(-\langle x\rangle)\leq \langle \exp(-x) \rangle $ together with $S^{\rm tot}_0 = 0$, we obtain
\begin{equation}
\langle S^{\rm tot}_t|X_{[0,s]} \rangle \geq0     \label{eq:secondLawStrict}
\end{equation}
which is the martingale version of the second law of thermodynamics. Hence, strict local martingales $ \emStot$ are compatible with the second law of thermodynamics, which is one more indication that strict local martingales $ \emStot$ are physical admissible.     

Since the random time transformation of Sec.~\ref{sec:randomtime} applies to strict local martingales,  also the universal properties of entropy production, such as, the infimum law Eq.~(\ref{eq:inflaw}) and the universal splitting probabilities Eqs.~(\ref{eq:pprob2}-\ref{eq:pprob}),  hold for processes $\emStot$ that are strict local martingales and continuous.

Although the above arguments show that local martingales are compatible with physical laws, it will be interesting to find concrete examples of processes $X$ for which  $\emStot$ is a local martingale.  A possible example are absolutely irreversible processes \cite{murashita2014nonequilibrium}, as such processes obey a   modified integral fluctuation relation of the form Eq.~(\ref{eq:modI}), although this possible connection between absolute irreversibility and local martingales requires s more careful study.

\section[Evaluation of $\hat{s}_{\rm FPR}$ and $\hat{s}_{\rm TUR}$]{Evaluation of the estimators $\hat{s}_{\rm FPR}$ and $\hat{s}_{\rm TUR}$ for a  random walk on a two-dimensional lattice }\label{sec:MFPT}
We derive the equations Eqs.~(\ref{eq:SFPR}) and (\ref{eq:STUR})  for the estimators $\hat{s}_{\rm FPR}$ and $\hat{s}_{\rm TUR}$, respectively, of  the random walk process on the two-dimensional lattice, as  illustrated in Fig.~\ref{fig:tradeoff2}.    To this purpose,  we derive an explicit expression for the quantities $P_-$, $\langle \T\rangle$, and $\langle \T^2\rangle$ appearing in the definitions of $\hat{s}_{\rm FPR}$ and $\hat{s}_{\rm TUR}$ in  (\ref{eq:defSFPR}).     The expressions we require are  first-passage quantities associated with the first passage time  $\T$ of the current $J_t$, as defined in Eqs.~(\ref{eq:TJDef}) and (\ref{eq:JDef}), respectively.

 As will become soon evident, we can use the martingale theory of Sec.~\ref{sec:stop} to derive expressions for  first-passage quantities associated with $\T$.     In this appendix we sketch this approach, and we refer  for details to the   Appendices D and E of Ref.~\cite{neri2022universal}.   \begin{figure}[!ht]
 \begin{center}
\includegraphics[width=80mm]{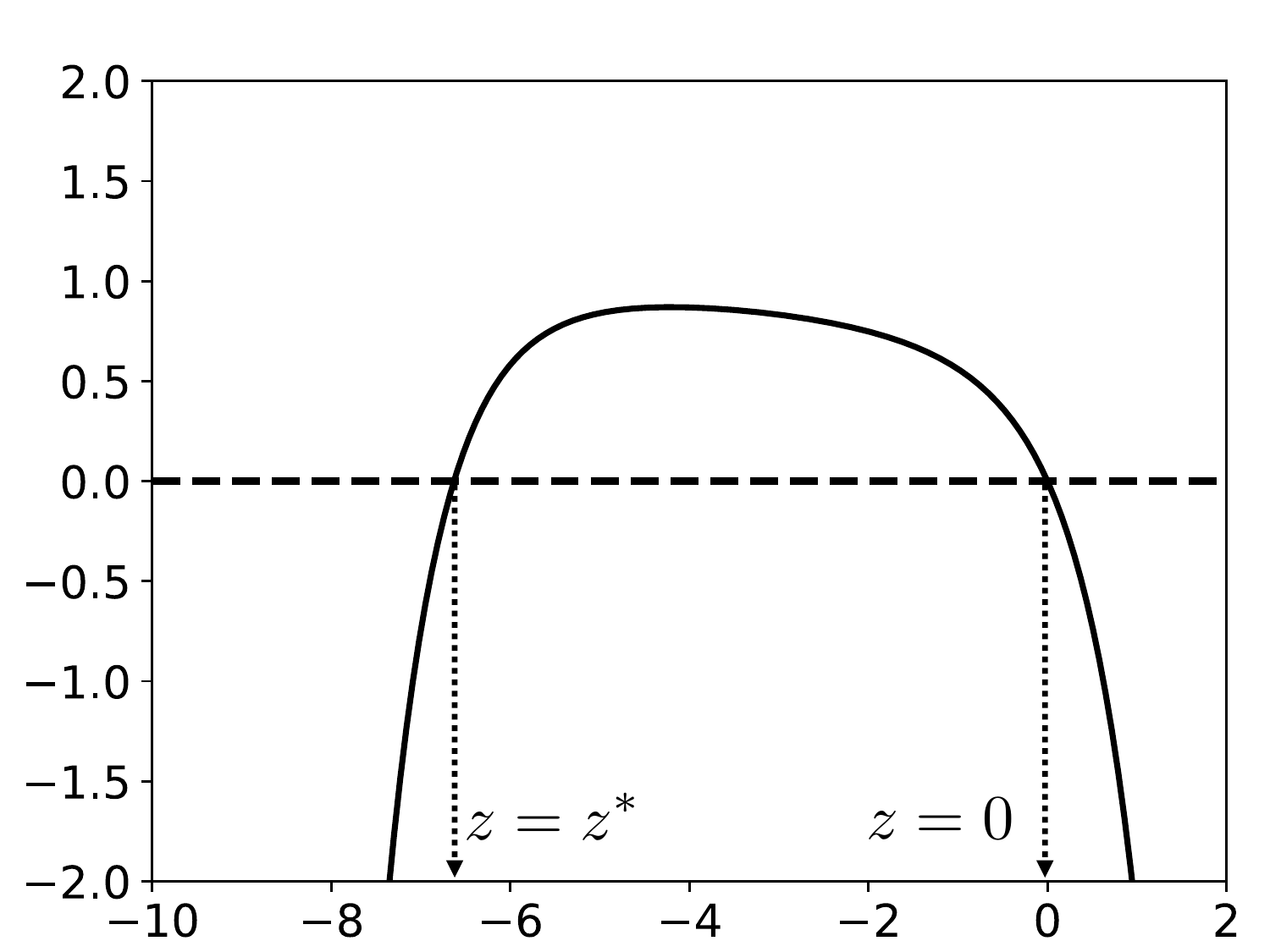}
 \put(-240,80){\Large$f$} 
  \put(-110,-20){\Large$z$} 
 \end{center}
\caption{Plot of the function of $f$, as defined in Eq.~(\ref{eq:fDef}), for $\Delta = 0.6$, and parameters $\omega^+_1 =  \exp(\nu/2)/(4\cosh(\nu/2))$,  $\omega^-_1 = \exp(-\nu/2)/(4\cosh(\nu/2))$, $\omega^+_2 =  \exp(\nu \rho/2)/(4\cosh(\nu/2))$, and $\omega^-_2 =  \exp(-\nu \rho/2)/(4\cosh(\nu/2))$ with $\rho=2$ and $\nu=5$, as in Panel (c) of Fig.~\ref{fig:tradeoff2}.   }
\label{fig:f}
\end{figure}
 
 \subsection{A martingale in the 2D random walk process}
The key insight of the present derivations for $\langle \T\rangle$, $\langle \T^2\rangle$ and $P_-$ is that the  process
 \begin{equation}
     Z_t = \exp\left(z J_t + t f(z)\right)
 \end{equation}
 where
 \begin{eqnarray}
     f(z) &\equiv&  \left[1-\exp\left(z (1-\Delta)\right)\right]\omega^{+}_1  + \left[1-\exp\left(-z (1-\Delta)\right)\right]\omega^{-}_1  
     \nonumber \\ 
     && 
     +  \left[1-\exp\left(z (1+\Delta)\right)\right]\omega^{+}_2 + \left[1-\exp\left(-z (1+\Delta)\right)\right]\omega^{-}_2 , \label{eq:fDef}
 \end{eqnarray}
 is a martingale for all values of $z\in \mathbb{R}$. 
 
  We plot the function $f$ in Fig.~\ref{fig:f} for the same parameters in as Panel(c) of Fig.~\ref{fig:tradeoff2}.   Observe that $f$  has two roots, the trivial root $z=0$ and a nontrivial root $z^\ast$  that  solves 
   \begin{equation}
     f(z^\ast) = 0. \label{eq:fZast}
 \end{equation} 
 The nontrivial root 
  is negative when $\langle J_t\rangle>0$ and is positive when $\langle J_t\rangle<0$. In what follows, we assume that $\langle J_t\rangle>0$ and hence $z^\ast<0$.

  Using Eq.~(\ref{eq:doobStop1}) from Doob's optional stopping theorem, we obtain that for all values $z\in \mathbb{R}$ for which $f(z)<0$ (see Ref.\cite{neri2022universal}), 
 \begin{equation}
     1 = \langle \mathbf{1}_{J_{\T}\geq \ell_+} \exp\left(z\ell_+ [1+o_{\ell_{\rm min}(1)}] + \T f(z)\right) + \mathbf{1}_{J(\T)\leq -\ell_-} \exp\left(-z\ell_- [1+o_{\ell_{\rm min}(1)}] + \T f(z)\right)  \rangle , \label{eq:StopEssential}
 \end{equation}
 where the factors $(1+o_{\ell_{\rm min}(1)})$ take care of the overshoot $J_{\T}- \ell_{\pm}$.

 Equation~(\ref{eq:StopEssential}) is central in the following derivations.  Indeed,    we obtain from this equation  the splitting probabilities $P_-$ and $P_+$, and the moments $\langle \T\rangle$ and $\langle \T^2\rangle$.

  \subsection{Splitting probabilities }

Using Eq.~(\ref{eq:StopEssential}) for the nonzero value of  $z^\ast$ that solves Eq.~(\ref{eq:fZast}), together with 
\begin{equation}
    P_- + P_+ = 1,
\end{equation}
 we obtain, see also Appendix E of Ref.~\cite{neri2022universal},
 \begin{equation}
    P_+ = \frac{1-\exp\left(-\ell_-|z^\ast| \: [1+o_{\ell_{\rm min}}(1)]\right)}{1-\exp\left(-(\ell_-+\ell_+)|z^\ast| \: [1+o_{\ell_{\rm min}}(1)]\right)} \label{eq:pP}
 \end{equation}
 and 
 \begin{equation}
 P_-  =\exp\left(-\ell_-|z^\ast|\:  [1+o_{\ell_{\rm min}}(1)]\right) \frac{1-\exp\left(-\ell_+|z^\ast| \:  [1+o_{\ell_{\rm min}}(1)]\right)}{1-\exp\left(-(\ell_-+\ell_+)|z^\ast| \: [1+o_{\ell_{\rm min}}(1)]\right)} . \label{eq:pM}
 \end{equation}
 Again, we used the factors $[1+o_{\ell_{\rm min}}(1)]$  in the exponentials, as in general  $J_{\T}$  is not equal to either  $\ell_+$ or $\ell_-$ when $J_t$ crosses one of the two threshold.  
 
Hence, in the limit of $\ell_{\rm min}\rightarrow \infty$, we get
\begin{equation}
   P_- = \exp\left(-\ell_- |z^\ast| \: [1+o_{\ell_{\rm min}}(1)]\right) .  \label{eq:pMApp}
\end{equation}
 
   \subsection{Generating function of $\T$}
The generating function is defined as 
\begin{equation}
    g(y) \equiv \langle \exp\left(-y \T\right) \rangle = P_+ g_+(y) + P_- g_-(y), \label{eq:gResult}
\end{equation}
where 
\begin{equation}
g_+(y)  \equiv  \langle \exp\left(-y \T\right) |\T\geq \ell_+\rangle \quad {\rm and} \quad g_-(y)  \equiv  \langle \exp\left(-y \T\right)  |\T\leq -\ell_-\rangle.
\end{equation}

To obtain an expression for $g_+(y)$ and $g_-(y)$, we use  the central Eq.~(\ref{eq:StopEssential}).
 In the range $z\notin [z^\ast, 0]$, for which $f(z)<0$, we set 
 \begin{equation}
     y = -f(z) . 
 \end{equation} 
 Taking the functional inverse of $f$, we obtain two solution branches, 
 \begin{equation}
     z_+(y) \in (-\infty, z^\ast] \quad {\rm and} \quad    z_-(y)\in [0, \infty),
 \end{equation}
  so that 
  \begin{equation}
    f(z_{\pm}(y)) = y. \label{eq:funcInverse}
  \end{equation}
  Selecting these two solution in Eq.~(\ref{eq:StopEssential}), we obtain the equations  
 \begin{equation}
1 = P_+  \exp\left( z_+(y) \ell_+ \: [1+o_{\ell_{\rm min}}(1)] \right)  g_+(y) + P_-  \exp\left(-z_+(y) \ell_-\:[1+o_{\ell_{\rm min}}(1)]\right)  g_-(y)  
 \end{equation}
and
  \begin{equation}
1 = P_+   \exp\left(z_-(y) \ell_+ \: [1+o_{\ell_{\rm min}}(1)] \right)   g_+(y) + P_- \exp\left(-z_-(y) \ell_- \: [1+o_{\ell_{\rm min}}(1)]\right)   g_-(y),
 \end{equation}
  respectively.   Solving the above two equations towards $g_+(y)$ and $g_-(y)$, we obtain 
   \begin{equation}
   g_+(y) = \frac{1}{P_+}  \frac{ 1-\exp\left(-[z_+(y)-z_-(y)] \ell_-\:[1+o_{\ell_{\rm min}}(1)]\right)}{\exp\left(z_+(y) \ell_+ \: [1+o_{\ell_{\rm min}}(1)]\right) -\exp\left(-[z_+(y)\ell_--z_-(y )(\ell_-+\ell_+) ]\:[1+o_{\ell_{\rm min}}(1)]\right)   }    \label{eq:gResultP}
 \end{equation}
 and
 \begin{equation}
  g_-(y)= \frac{1}{P_-} \frac{1-\exp\left(-[z_-(y)-z_+(y)]\ell_+ [1+o_{\ell_{\rm min}}(1)]\right)}{\exp\left(-z_+(y)\ell_-[1+o_{\ell_{\rm min}}(1)]\right) - \exp\left(-[z_-(y)(\ell_-+\ell_+)-z_+(y)\ell_+]\:[1+o_{\ell_{\rm min}}(1)]\right)}.\label{eq:gResultM}
 \end{equation}
 
 Taking the limit $\ell_{\rm min}\rightarrow \infty$, it follows from   Eqs.~(\ref{eq:pP}), (\ref{eq:pM}), (\ref{eq:gResult}), (\ref{eq:gResultP}) and (\ref{eq:gResultM})  that the generating function of $\T$ is given by  
 \begin{equation}
     g(y) = g_+(y)(1+O(\exp\left(\ell_-z^\ast\right))) \label{eq:G}
 \end{equation}
 with 
 \begin{equation}
 g_+(y) = \exp\left(-z_-(y) \ell_+\: [1+o_{\ell_{\rm min}}(1)]\right) ,  \label{eq:gP}
 \end{equation}
and where we have used that $z_+<0$ and $z_- >0$.

\subsection{First moment and second moment of $\T$}
The Eqs.~(\ref{eq:G}) and (\ref{eq:gP}) determine the generating function $g$ of $\T$ in terms of the function $z_{-}(y)$ that solves Eq.~(\ref{eq:funcInverse}) for $z_-(y)\in[0,\infty)$.    Solving Eq.~(\ref{eq:funcInverse})  is not an easy task, but since we only  need  the first two moments of $\T$, we can simplify the problem further.  Indeed, expanding $g$ in small values of $y$ we obtain up to second order in $y$,   
%\begin{equation}
%    g_+(y) = 1 + y g_{+,1} + y^2 g_{+,2} + O(y^3)
%\end{equation}
%and 
\begin{equation}
    g(y) = 1 - y \langle \T\rangle + \frac{y^2}{2} \langle \T^2\rangle + O(y^3), \label{eq:gPTaylor}
\end{equation}
and hence it is sufficient to solve Eq.~(\ref{eq:funcInverse}) up to second order in $y$.    In addition, using that $z_-(y)\approx 0$ for $y\approx 0$ and expanding Eq.~(\ref{eq:funcInverse}) up to second order yields the equation
\begin{eqnarray}
   \lefteqn{ \left[ z (1-\Delta) + \frac{z^2 (1-\Delta)^2}{2} \right] \omega^{+}_1 - \left[z (1-\Delta) -\frac{z^2 (1-\Delta)^2}{2} \right] \omega^{-}_1 } &&  \nonumber\\  
  &&   + \left[z (1+\Delta) + \frac{z^2 (1+\Delta)^2}{2} \right] \omega^{+}_2 - \left[z (1+\Delta) -\frac{z^2 (1+\Delta)^2}{2} \right] \omega^{-}_2 + O(z^3) = y, \label{eq:quadr}\nonumber\\ 
\end{eqnarray}
whose positive   solution determines $z_-$.

\subsubsection*{Mean first-passage time $\langle \T\rangle$}
The solution  of Eq.~(\ref{eq:quadr}) up to linear order in $y$ is, 
\begin{equation}
z_- =  \frac{y}{(1-\Delta)(\omega^+_1 -\omega^-_1)  + (1+\Delta)(\omega^+_2 - \omega^-_2)  } + O(y^2). \label{eq:ZMLinaer}
\end{equation}
Substituting Eq.~(\ref{eq:ZMLinaer}) in  Eqs.~(\ref{eq:G}) and (\ref{eq:gP}) gives 
\begin{equation}
g(y) = \exp\left(-\frac{\ell_+ \left(y + O(y^2)\right)}{(1-\Delta)(k^+_1 -k^-_1)  + (1+\Delta)(k^+_2 - k^-_2) }(1+o_{\ell_{\rm min}}(1))\right). \label{eq:gPResultTaylorx}
\end{equation}
Expanding the latter equation up to linear order in $y$ and comparing with Eq.~(\ref{eq:gPTaylor}) gives
\begin{equation}
\langle \T \rangle =  \frac{\ell_+}{(1-\Delta)(\omega^+_1 -\omega^-_1)  + (1+\Delta)(\omega^+_2 - \omega^-_2) }(1+o_{\ell_{\rm min}}(1)). \label{eq:MeanFPTResult}
\end{equation}

\subsubsection*{Second moment $\langle \T^2\rangle$}
Solving Eq.~(\ref{eq:quadr}) up to quadratic order yields the solution
%\begin{eqnarray}
%z_- = -\frac{(1-\Delta)(k^+_1 -k^-_1) + (1+\Delta)(k^+_2 - k^-_2)}{ (1-\Delta)^2\left( k^+_1 +  k^-_1\right) + (1+\Delta)^2\left( k^+_2 + k^-_2\right)}  
%\nonumber\\ 
%\pm \frac{  \sqrt{[(1-\Delta)(k^+_1 -k^-_1) + (1+\Delta)(k^+_2 - k^-_2)]^2 + 2y \left[ (1-\Delta)^2\left( k^+_1 +  k^-_1\right) + (1+\Delta)^2\left( k^+_2 + k^-_2\right)\right]}}{(1-\Delta)^2\left( k^+_1 +  k^-_1\right) + (1+\Delta)^2\left( k^+_2 + k^-_2\right)}
%\end{eqnarray}

\begin{eqnarray}
z_- &=& \frac{y}{(1-\Delta)(\omega^+_1 -\omega^-_1)  + (1+\Delta)(\omega^+_2 - \omega^-_2) }  
\nonumber \\ 
&&- \frac{y^2}{2} \left[ \frac{(1-\Delta)^2\left( \omega^+_1 +  \omega^-_1\right) + (1+\Delta)^2\left( \omega^+_2 + \omega^-_2\right)}{\left((1-\Delta)(\omega^+_1 -\omega^-_1) + (1+\Delta)(\omega^+_2 - \omega^-_2)\right)^3}\right] + O(y^3). \label{eq:ZM2}
\end{eqnarray}
Substituting Eq.~(\ref{eq:ZM2}) in  Eqs.~(\ref{eq:G}) and (\ref{eq:gP}) and expanding up to second order in $y$ gives 
\begin{eqnarray}
\lefteqn{g_+(y) = \exp\left(-\frac{\ell_+ y}{(1-\Delta)(\omega^+_1 -\omega^-_1)  + (1+\Delta)(\omega^+_2 - \omega^-_2) }(1+o_{\ell_{\rm min}}(1)) \right) }&& \nonumber\\ 
&&\times  \exp\left(\frac{y^2
\ell_+ + O(y^3)}{2} \left[ \frac{(1-\Delta)^2\left( \omega^+_1 +  \omega^-_1\right) + (1+\Delta)^2\left( \omega^+_2 + \omega^-_2\right)}{\left((1-\Delta)(\omega^+_1 -\omega^-_1) + (1+\Delta)(\omega^+_2 - \omega^-_2)\right)^3} \right](1+o_{\ell_{\rm min}}(1)) \right). \nonumber\\
\label{eq:gPResultTaylor2}
\end{eqnarray}
Expanding the latter equation up to second order in $y$, and comparing with Eq.~(\ref{eq:gPTaylor}), we obtain 
\begin{eqnarray}
\lefteqn{\langle \T^2\rangle = \frac{\ell^2_+}{\left[(1-\Delta)(\omega^+_1 -\omega^-_1)  + (1+\Delta)(\omega^+_2 - \omega^-_2)\right]^2}  }&& 
\nonumber\\ 
&& +\ell_+ \frac{(1-\Delta)^2\left( \omega^+_1 +  \omega^-_1\right) + (1+\Delta)^2\left( \omega^+_2 + \omega^-_2\right)}{\left[(1-\Delta)(\omega^+_1 -\omega^-_1) + (1+\Delta)(\omega^+_2 - \omega^-_2)\right]^3},
\end{eqnarray}
and thus  
\begin{equation}
    \langle \T^2\rangle - \langle \T\rangle^2 =  \frac{(1-\Delta)^2\left( \omega^+_1 +  \omega^-_1\right) + (1+\Delta)^2\left( \omega^+_2 + \omega^-_2\right)}{\left[(1-\Delta)(\omega^+_1 -\omega^-_1) + (1+\Delta)(\omega^+_2 - \omega^-_2)\right]^3}. \label{eq:varFinal}
\end{equation}

\subsection{Estimators $\hat{s}_{\rm FPR}$ and $\hat{s}_{\rm TUR}$}
Using the expressions Eqs.~(\ref{eq:pMApp})  and (\ref{eq:MeanFPTResult}) for $P_-$ and $\langle \T\rangle$,  respectively, in the definition of $\hat{s}_{\rm FPR}$,  Eq.~(\ref{eq:defSFPR}), we obtain
\begin{equation}
    \hat{s}_{\rm FPR} =  |z^\ast|\left((1-\Delta)(\omega^+_1-\omega^-_1) + (1+\Delta)(\omega^+_2-\omega^-_2)\right)(1+o_{\ell_{\rm min}}(1)),  \label{eq:SFPRApp}
 \end{equation} 
which is the Eq.~(\ref{eq:SFPR}) we were meant to derive.  

Analogously, using the expressions (\ref{eq:MeanFPTResult}) and (\ref{eq:varFinal}) for $\langle \T\rangle$ and $\langle \T^2\rangle-\langle \T\rangle^2$, respectively,  in the definition of $\hat{s}_{\rm TUR}$, Eq.~(\ref{eq:defSFPR}), we obtain
 \begin{equation}
  \hat{s}_{\rm TUR} =  \frac{ 2\langle T\rangle}{\langle \T^2\rangle - \langle \T\rangle^2} =  \frac{2\left[(1-\Delta)(\omega^+_1 -\omega^-_1) + (1+\Delta)(\omega^+_2 - \omega^-_2)\right]^2}{(1-\Delta)^2\left( \omega^+_1 +  \omega^-_1\right) + (1+\Delta)^2\left( \omega^+_2 + \omega^-_2\right)}(1+o_{\ell_{\rm min}}(1)),
\end{equation}
which is Eq.~(\ref{eq:STUR}) that we were meant to derive. 
%$a = -y$;
%$b = (1-\Delta)(k^+_1 -k^-_1) + %(1+\Delta)(k^+_2 - k^-_2)$; 
%$c = \frac{(1-\Delta)^2}{2}\left( k^+_1 +  %k^-_1\right) + \frac{(1+\Delta)^2}{2}\left( %k^+_2 + k^-_2\right)$

%Expanding further around $y\approx 0$, we bet 
%\begin{eqnarray}
%z_- =  \frac{y}{(1-\Delta)k^+_1 - (1-\Delta)k^-_1 + (1+\Delta)k^+_2 - (1+\Delta)k^-_2  }  
%\nonumber\\ 
%+ O(y^3)
%\end{eqnarray}

%%%%%%%
%\include{chap_app8}

\chapter[Appendix to Chapter 8]{Appendix to Chapter~\ref{sec:martST3}}\label{appendix:8}

\section{Derivation of Eq.~(\ref{eq:StVSNonStat}) demonstrating the exponential martingale for nonstationary processes}\label{app:chapt8}

The derivation of Eq.~(\ref{eq:StVSNonStat})  is similar to the derivation of Eq.~(\ref{eq:entropy}) in Chapter~\ref{sec:martST1}. 

First, we use the definition of $\tilde{\rho}$ as the solution to the Fokker-Planck equation given by Eqs.~(\ref{eq:FPR1}-\ref{eq:FPR2}) to write,  
\begin{eqnarray}
  \lefteqn{\frac{d\left(-\ln \left(\tilde{\rho}_{\tau-s}\left(X_s\right)\right)\right)}{ds}}&& \\
    &=&  -\frac{\left(\partial_s \tilde{\rho}_{\tau-s}\right)\left(X_s\right)}{\tilde{\rho}_{\tau-s}\left(X_s\right)} - \frac{\left(\partial_x \tilde{\rho}_{\tau-s}\right)\left(X_s\right)}{\tilde{\rho}_{\tau-s}\left(X_s\right)}\circ \dot{X}_s\label{eq:us2}\\ 
    &=& \underbrace{-\frac{\left(\partial_s \tilde{\rho}_{\tau-s}\right)\left(X_s\right)}{\tilde{\rho}_{\tau-s}\left(X_s\right)}+ \frac{\tilde{J}_{\tau-s,\tilde{\rho}}(X_s)}{\mu T \tilde{\rho}_{\tau-s}\left(X_s\right)}\circ \dot{X}_s}_{\displaystyle \frac{d\hat{\Sigma}_{s}}{ds}} - \underbrace{\frac{\partial_xV(X_s;\tilde{\lambda}_{\tau-s})}{T}\circ \dot{X}_s}_{\displaystyle \dot{Q}_s/T}, \label{eq:us3}
\end{eqnarray}
where we have used that $\tilde{\lambda}_{\tau-s} = \lambda_s$.  

Hence, in Stratonovich convention 
\begin{eqnarray}
\frac{d\hat{\Sigma}_{s}}{ds} = -\frac{\left(\partial_s \tilde{\rho}_{\tau-s}\right)\left(X_s\right)}{\tilde{\rho}_{\tau-s}\left(X_s\right)}+\frac{1}{\mu T} \frac{\tilde{J}_{\tau-s,\tilde{\rho}}(X_s)}{ \tilde{\rho}_{\tau-s}\left(X_s\right)}\circ \dot{X}_s.\label{eq:appG1}
\end{eqnarray}
Substituting $\dot{X}_s$, given by Eq.~(\ref{eq:noneqLang}), in Eq.~(\ref{eq:appG1}), and using 
\begin{equation}
  \partial_x V(x,\lambda_s) =   \partial_x V(x,\tilde{\lambda}_{\tau-s})   =  -\frac{\tilde{J}_{\tau-s,\tilde{\rho}}}{\mu \tilde{\rho}_{\tau-s}} - T \frac{\partial_x\tilde{\rho}_{\tau-s}}{\tilde{\rho}_{\tau-s}},
\end{equation}
we get
\begin{eqnarray}
\frac{d\hat{\Sigma}_{s}}{ds} = -\frac{\left(\partial_s \tilde{\rho}_{\tau-s}\right)\left(X_s\right)}{\tilde{\rho}_{\tau-s}\left(X_s\right)}+\frac{\tilde{J}_{\tau-s,\tilde{\rho}}\left(X_s\right)\left(\partial_x\tilde{\rho}_{\tau-s}\right)\left(X_s\right)}{\left(\tilde{\rho}_{\tau-s}\left(X_s\right)\right)^2} 
+ v^S_s   +\sqrt{2v^S_s}\circ  \dot{B}_s , \label{eq:AppG2}
\end{eqnarray}  
where we have used Eq.~(\ref{eq:vSs}) to identify $v^S_s$.  Next, we use Theorem~\ref{th:0}
to write the last term in the latter equation in   the It\^{o} convention, obtaining 
\begin{eqnarray}
\sqrt{v^S_s}\circ  \dot{B}_s = \sqrt{v^S_s} \dot{B}_t 
+    \frac{\left(\partial_x \tilde{J}_{\tau-s,\tilde{\rho}}\right)(X_s)}{\tilde{\rho}_{\tau-s}(X_s)} - \frac{\tilde{J}_{\tau-s,\tilde{\rho}}\left(X_s\right)\left(\partial_x\tilde{\rho}_{\tau-s}\right)\left(X_s\right)}{\left(\tilde{\rho}_{\tau-s}\left(X_s\right)\right)^2} \label{eq:STrattoIto}. 
\end{eqnarray}
Lastly, using Eq.~(\ref{eq:STrattoIto}) into Eq.~(\ref{eq:AppG2}) together with  
\begin{equation}
\partial_s \tilde{\rho}_{\tau-s} = -\partial_x \tilde{J}_{\tau-s,\tilde{\rho}},
\end{equation}
we obtain 
\begin{eqnarray}
\frac{d\hat{\Sigma}_{s}}{ds} = v^S_s  + \sqrt{2 v^S_s} \dot{B}_s ,
\end{eqnarray}
which  is the Eq.~(\ref{eq:StVSNonStat}) that  we were meant to derive.

\section{Origin of time reversal in the definition of $\hat{\Sigma}_s$}\label{app:chapt82} 
In the definition Eq.~(\ref{eq:Ss}) of $\hat{\Sigma}_s$  we have set $t/2$, the origin of time reversal, equal to $\tau/2$.   Here, we show that the process $\hat{\Sigma}_s$ is independent of the choice $t/2$ for the origin of time-reversal.

Let us therefore define the  stochastic process 
\begin{equation}
\hat{\Sigma}^{(t)}_{s} \equiv  -\frac{Q_s}{T} + \ln  \rho_{\rm eq}(X_0;\lambda_{\rm i}) -\ln \tilde{\rho}^{(t)}_{t-s}(X_s),  \label{eq:Ssx}
\end{equation}
where $\dot{Q}_s$ is the heat Eq.~(\ref{eq:heat83}) as before, and    
where $\tilde{\rho}^{(t)}_{t-s}$ is the solution to the Fokker-Planck equation 
\begin{equation}
    \partial_{s}\tilde{\rho}^{(t)}_{s}+\partial_x  \tilde{J}^{(t)}_{s,\tilde{\rho}}=0, \label{eq:FPHere1a}
\end{equation} 
with
\begin{equation}
\tilde{J}^{(t)}_{s,\tilde{\rho}}  = -\mu \partial_x V(x;\tilde{\lambda}^{(t)}_s) \tilde{\rho}^{(t)}_s(x)-\mu  T \partial_{x}\tilde{\rho}^{(t)}_{s}(x),   \label{eq:FPHere2a}
\end{equation} 
and 
with  
\begin{equation}
    \tilde{\lambda}^{(t)}_s \equiv \left\{\begin{array}{ccc} \lambda_{\rm f} &{\rm if}& s\leq t-\tau, \\ \lambda_{t-s} &{\rm if}& s\in [t-\tau,t], \\ 
    \lambda_{\rm i} &{\rm if}& s\geq  t, 
    \end{array}\right. 
    \end{equation}  
 the time-reversed protocol.    Note that in the time-reversed protocol $\tilde{\lambda}^{(t)}_s$ the time-reversal reflection point is $t/2$, and not $\tau/2$  as in Eq.~(\ref{eq:tildelambda}).    To complete the definition of $\tilde{\rho}^{(t)}_{s}$  we specify the initial state of the time-reversal dynamics, which for 
 $t\geq \tau$ given by 
\begin{equation}
\tilde{\rho}^{(t)}_0(x) = \rho_{\rm eq}(x;\lambda_{\rm f}),    \label{eq:AppGInit1}   
\end{equation}
and for  $t\in[0,\tau]$ by
\begin{equation}
\tilde{\rho}^{(t)}_0 = \tilde{\rho}^{(\tau)}_{\tau-t}. \label{eq:AppGInit2}   
\end{equation}
Note that in the case $t=\tau$, it holds that $\hat{\Sigma}^{(t)}_s$, as defined in Eq.~(\ref{eq:Ss}), equals $\hat{\Sigma}_s$, as defined in Eq.~(\ref{eq:Ssx}).

It follows  from the definition Eqs.~(\ref{eq:FPHere1a}-\ref{eq:FPHere2a}) for the Fokker-Planck equation with initial condition Eq.~(\ref{eq:AppGInit1}) or (\ref{eq:AppGInit2}) that  
\begin{equation}
    \tilde{\rho}^{(t)}_{t-s} =  \tilde{\rho}^{(\tau)}_{\tau-s}  ,
\end{equation}
and hence  
\begin{equation}
    \hat{\Sigma}^{(t)}_{s} =  \hat{\Sigma}^{(\tau)}_{s}  .
\end{equation}
In other words, the origin $t/2$ of time reversal is not relevant in the definition of    $\hat{\Sigma}^{(t)}_{s}$, as all processes are the same.   For this reason, we have set  $t=\tau$ as in Ref.~\cite{neri2020second}, and we removed the index $(\tau)$ from the definition $\hat{\Sigma}^{(\tau)}_s$ in Eq.~(\ref{eq:Ss}).

%%%%%%%
%\include{chap_app2}
\chapter[Appendix to Chapter~12]{Appendix to Chapter~\ref{chapter:finance}}

%%%%%%%%%%%%%%%%%%%%%%%%%%%%%%%%%%%%%%%%%%%%%%%%%%%%%%%%%%%%%%%%%%%%%%%%%%%%%%%%%%%%%%%%%%%%%%%%%%%%%%%%%%%%%%%
%\section{Options, arbitrage, and all that}
%\label{app:option}

\section{Derivation of Eq.~(\ref{eq:heat-equation}) from
Eq.~(\ref{eq:BS})}
\label{app:heat-derivation}

We start with considering the transformation~(\ref{eq:BS-transformation}) that
implies the following identities:
\begin{equation}
\alpha+\beta =\frac{2r}{\sigma^2},~~\alpha+1=\beta.
\end{equation}
Using $C(S_t,t)=Ku(x,\tau)\exp(-\alpha x -\beta^2 \tau)$ and $\partial
\tau/\partial t=-\sigma^2/2$, we get
\begin{equation}
\frac{\partial C}{\partial t}=-\frac{\sigma^2}{2}K \exp(-\alpha x
-\beta^2\tau)\left(\frac{\partial u}{\partial \tau}-\beta^2 u\right).
\label{eq:A}
\end{equation}
Next, using $\partial x/\partial S_t=1/S_t$, we get
\begin{equation}
\frac{\partial C}{\partial S_t}=\frac{K}{S_t} \exp(-\alpha x
-\beta^2\tau)\left(\frac{\partial u}{\partial x}-u\alpha\right),
\label{eq:delCdelS}
\end{equation}
so that $r=(\alpha+\beta)\sigma^2/2$ yields
\begin{equation}
rS_t\frac{\partial C}{\partial S_t}=\frac{(\alpha+\beta)\sigma^2}{2}K \exp(-\alpha x
-\beta^2\tau)\left(\frac{\partial u}{\partial x}-u\alpha\right).
\label{eq:B}
\end{equation}
Using Eq.~(\ref{eq:delCdelS}) and $K/S_t=\exp(-x)$, we get 
\begin{equation}
\frac{\partial^2C}{\partial
S_t^2}=\frac{1}{S_t}\exp(-(\alpha+1) x
-\beta^2\tau)\left[-(2\alpha+1)
\frac{\partial u}{\partial x}+\frac{\partial^2 u}{\partial
x^2}+\alpha(\alpha+1)u\right].
\end{equation}
It then follows that 
\begin{equation}
\frac{\sigma^2S_t^2}{2}\frac{\partial^2C}{\partial
S_t^2}=\frac{\sigma^2K}{2}\exp(-\alpha x
-\beta^2\tau)\left[-(2\alpha+1)
\frac{\partial u}{\partial x}+\frac{\partial^2 u}{\partial
x^2}+\alpha(\alpha+1)u\right].
\label{eq:C}
\end{equation}
Finally, we have
\begin{equation}
rC=\frac{(\alpha+\beta)\sigma^2 Ku}{2}\exp(-\alpha x
-\beta^2\tau).
\label{eq:D}
\end{equation}
We now substitute Eqs.~(\ref{eq:A},~(\ref{eq:B}),~(\ref{eq:C}),
and~(\ref{eq:D}) in Eq.~(\ref{eq:BS}), and use 
\begin{equation}
\beta^2+\alpha
\beta-(\alpha+\beta)\alpha-(\alpha+\beta)=\beta(\alpha+1)+\alpha\beta-(\alpha+\beta)(\alpha+1)=0,
\end{equation}
where we have used the result $\alpha+1=\beta$; we finally get our
desired result, namely, Eq.~(\ref{eq:heat-equation}):
\begin{equation}
\frac{\partial u}{\partial \tau}=\frac{\partial^2 u}{\partial x^2}.
\end{equation}

%%%%%%%

\bibliographystyle{naturemag}

%%%%%%%
\printindex
\end{document}